\documentclass[fleqn,11pt,oneside,openany]{scrbook}
\pdfoutput=1

\usepackage{feynmp-pdf}               % Feynman diagrams.
\usepackage{amsfonts}                 % AMS font support.
\usepackage{amsthm, amssymb, amsmath} % AMS math support.
\usepackage{latexsym}                 % Symbols support.
\usepackage{graphicx}                 % Graphics support.
\usepackage{array}                    % Define column atributes for tables.
\usepackage{booktabs}                 % Fancy table support.
\usepackage{fp}                       % Allows for simple math manipulation.
\usepackage{calc}                     % Width calculations.
\usepackage[colorlinks=true,linkcolor=black,urlcolor=black,citecolor=black]
{hyperref}                            % Hyperreferencing package.
\usepackage{lmodern}                  % Nice font.
\usepackage{svg}                      % Include SVGs natively.
\usepackage{fancyhdr}                 % Use fancy headers.
\usepackage{bm}                       % Bold math fonts.
\usepackage{grffile}                  % File names with full stops.
\usepackage{relsize}                  % Relative sizes between packages.
\usepackage{xspace}                   % Allow variable space.
\usepackage{ifthen}                   % Conditional package.
\usepackage{mathrsfs}                 % Script math fonts.
\usepackage{multirow}                 % Package for multirow tables.
\usepackage{pbox}                     % Allows for wrapping paragraph boxes.
\usepackage{mciteplus}                % Advanced citing for LHCb style.
\usepackage{patchcmd}                 % Simple command patching.
\usepackage{braket}                   % Expandable braket notation.
\usepackage{tensor}                   % Nicely formatted tensor notation.
\usepackage{slashed}                  % Dirac slash notation.
\usepackage{setspace}                 % Line spacing package.
\usepackage{rotating}                 % Rotating package.
\usepackage[a4paper,top=3cm,bottom=3cm,left=2.5cm,right=2.5cm]
{geometry}                            % Govern the margins.
\usepackage{morefloats}               % Solve the ArXiv float problem.

\allowdisplaybreaks                            % Equation breaks over lines.

\DeclareSubrefFormat{thesis}{\textit{\textbf{#2}}}  % Change subref style.
\makeatletter
\def\gnewcommand{\g@star@or@long\new@command}
\def\grenewcommand{\g@star@or@long\renew@command}
\def\g@star@or@long#1{% 
  \@ifstar{\let\l@ngrel@x\global#1}{\def\l@ngrel@x{\long\global}#1}}
\makeatother

\newcommand{\fmppath}{./}
\newcommand{\executeiffilenewer}[3]{\ifnum\pdfstrcmp{\pdffilemoddate{#1}}
  {\pdffilemoddate{#2}}>0{\immediate\write18{#3}}\fi}

\newcommand{\fmp}[2][]{\ifthenelse{\equal{#1}{}}%
  {\subfloat[]{\begin{fmffile}{\fmppath#2}%
    \setlength{\unitlength}{1mm}\input{\fmppath#2.fmp}%
  \end{fmffile}\executeiffilenewer{\fmppath#2.fmp}%
  {\fmppath#2.1}{cd \fmppath; mpost #2.mp}\labelfig{#2}}}%
  {\begin{fmffile}{\fmppath#2}%
    \setlength{\unitlength}{1mm}\input{\fmppath#2.fmp}%
  \end{fmffile}\executeiffilenewer{\fmppath#2.fmp}%
  {\fmppath#2.1}{cd \fmppath; mpost #2.mp}\global\let\sidecaptiontext\captiontext%
  \grenewcommand{\captiontext}{\sidecaptiontext\labelfig{#2}}}}
\newcommand{\svg}[2][]{\ifthenelse{\equal{#1}{}}%
  {\subfloat[]{\includesvg[pretex=\hspace{-0.43cm}\relsize{-3},%
      width=0.5\columnwidth]{#2}\labelfig{#2}}}%
  {\ifthenelse{\equal{#1}{1}}{\includesvg[pretex=\hspace{-0.43cm}\relsize{-3},%
    width=0.5\columnwidth]{#2}\global\let\sidecaptiontext\captiontext%
    \grenewcommand{\captiontext}{\sidecaptiontext\labelfig{#2}}}%
  {\ifthenelse{\equal{#1}{2}}{}{\rotatebox{90}{#1}&&}%
     \includesvg[pretex=\relsize{-5},width=0.34\columnwidth]%
     {#2}}}}

\newcommand{\lgdsep}{\\[0.4cm]}

\newcommand{\barsep}{&\\[-0.2cm]}
\newcommand{\barend}{&\\[-0.2cm]}

\newcommand{\svgbeg}{\\[-1cm]}
\newcommand{\svgsep}{&\\[-0.8cm]}
\newcommand{\svgend}{&\\[-0.2cm]}
\newcommand{\effbeg}{\\[-1cm]}
\newcommand{\effsep}{&\\[-0.6cm]}
\newcommand{\effend}{&\\[0.0cm]}
\newcommand{\fmpbeg}{\\[-0.2cm]}
\newcommand{\fmpsep}{&\\\\[0.4cm]}
\newcommand{\fmpend}{&\\[0.4cm]}
\newcommand{\captionshow}{true}

\newcommand{\captiontext}{}
\newcounter{sidecapfig}
\newcounter{sidecaptab}
\newcounter{sidefig}
\newcounter{sidetab}
\newcommand{\sidecaption}{%
  \setcounter{sidecapfig}{\value{subfigure}}%
  \setcounter{sidecaptab}{\value{subtable}}%
  \setcounter{sidefig}{\value{figure}}%
  \setcounter{sidetab}{\value{table}}%
  \caption{\captiontext}%
  \setcounter{subfigure}{\value{sidecapfig}}%
  \setcounter{subtable}{\value{sidecaptab}}%
  \setcounter{figure}{\value{sidefig}}%
  \setcounter{table}{\value{sidetab}}%
  \grenewcommand{\captionshow}{false}}
\newlength{\tabcolwidth}
\newlength{\oldtabcolsep}
\newlength{\newtabcolsep}
\newenvironment{subfigures}[3][!ht]{\FPeval{\result}{clip(1/#2)}%
  \begin{figure}[#1]\centering\renewcommand{\captiontext}{#3}%
    \setlength{\tabcolsep}{\newtabcolsep}%
    \setlength{\tabcolwidth}{\result\columnwidth}%
  \ifthenelse{#2 = 1}{\begin{tabular}%
      {M{\tabcolwidth}M{0pt}}}{%
  \ifthenelse{#2 = 2}{\begin{tabular}%
      {M{\tabcolwidth}M{\tabcolwidth}M{0pt}}}{%
  \ifthenelse{#2 = 3}{\begin{tabular}%
      {M{\tabcolwidth}M{\tabcolwidth}M{\tabcolwidth}M{0pt}}}{%
  \ifthenelse{#2 = 4}{\setlength{\tabcolwidth}{0.30\columnwidth}%
    \begin{tabular}{M{0.05\columnwidth}|M{0.02\columnwidth}%
        M{\tabcolwidth}M{\tabcolwidth}M{\tabcolwidth}M{0.4cm}}}{}}}}}
{\end{tabular}\ifthenelse{\equal{\captionshow}{true}}{\caption{\captiontext}}%
{\addtocounter{figure}{1}}\end{figure}\grenewcommand{\captionshow}{true}}
\newenvironment{subtables}[3][!ht]{\FPeval{\result}{clip(1/#2)}%
  \begin{table}[#1]\centering\renewcommand{\captiontext}{#3}%
    \setlength{\tabcolsep}{0pt}\setlength{\tabcolwidth}{\result\columnwidth}%
  \ifthenelse{#2 = 1}{\begin{tabular}%
      {M{\tabcolwidth}M{0pt}}}{%
  \ifthenelse{#2 = 2}{\begin{tabular}%
      {M{\tabcolwidth}M{\tabcolwidth}M{0pt}}}{%
  \ifthenelse{#2 = 3}{\begin{tabular}%
      {M{\tabcolwidth}M{\tabcolwidth}M{\tabcolwidth}M{0pt}}}{}}}}
{\end{tabular}\ifthenelse{\equal{\captionshow}{true}}{\caption{\captiontext}}%
{\addtocounter{table}{1}}\end{table}\grenewcommand{\captionshow}{true}}

\definecolor{red}{rgb}{1,0,0}
\definecolor{blue}{rgb}{0,0,1}
\definecolor{green}{rgb}{0,0.4,0}
\definecolor{orange}{rgb}{1,0.6,0}
\definecolor{magenta}{rgb}{1,0,1}
\definecolor{cyan}{rgb}{0,1,1}

\newcommand{\lep}{\ensuremath{\ell}\xspace}
\newcommand{\had}{\ensuremath{h}\xspace}

\newcommand{\z}{\ensuremath{Z}\xspace}
\newcommand{\w}{\ensuremath{W}\xspace}
\newcommand{\vb}{\ensuremath{V}\xspace}
\newcommand{\ww}{\ensuremath{{WW}}\xspace}
\newcommand{\ttbar}{\ensuremath{{t\bar{t}}}\xspace}
\newcommand{\bbbar}{\ensuremath{{b\bar{b}}}\xspace}
\newcommand{\dy}{\ensuremath{\gamma^*/Z}\xspace}
\newcommand{\jpsi}{\ensuremath{J/\psi}\xspace}
\newcommand{\dip}{\ensuremath{pp}\xspace}
\newcommand{\die}{\ensuremath{ee}\xspace}
\newcommand{\dimu}{\ensuremath{\mu\mu}\xspace}
\newcommand{\dilep}{\ensuremath{\lep\lep}\xspace}
\newcommand{\ditau}{\ensuremath{{\tau\tau}}\xspace}
\newcommand{\mumu}{\ensuremath{{\tau_\mu\tau_\mu}}\xspace}
\newcommand{\mue}{\ensuremath{{\tau_\mu\tau_e}}\xspace}
\newcommand{\emu}{\ensuremath{{\tau_e\tau_\mu}}\xspace}
\newcommand{\muh}{\ensuremath{{\tau_\mu\tau_\had}}\xspace}
\newcommand{\eh}{\ensuremath{{\tau_e\tau_\had}}\xspace}
\newcommand{\h}{\ensuremath{{\phi^0}}\xspace}
\newcommand{\hz}{\ensuremath{{\Phi^0}}\xspace}
\newcommand{\hH}{\ensuremath{H}\xspace}
\newcommand{\hhz}{\ensuremath{{h^0}}\xspace}
\newcommand{\hAz}{\ensuremath{{A^0}}\xspace}
\newcommand{\hHz}{\ensuremath{{H^0}}\xspace}

\newcommand{\hHpm}{\ensuremath{{H^\pm}}\xspace}
\newcommand{\tauto}{\ensuremath{{\tau^- \to \nu_\tau}}\xspace}

\newcommand{\wtl}{\ensuremath{\tau~\mathrm{lepton}}\xspace}
\newcommand{\wtls}{\ensuremath{\tau} leptons\xspace}

\newcommand{\whb}{Higgs boson\xspace}
\newcommand{\whbs}{Higgs bosons\xspace}

\newcommand{\Whbs}{Higgs bosons\xspace}
\newcommand{\wzb}{\z boson\xspace}
\newcommand{\wzbs}{\z bosons\xspace}

\newcommand{\wwb}{\w boson\xspace}
\newcommand{\wwbs}{\w bosons\xspace}

\newcommand{\wuq}{$u$-quark\xspace}
\newcommand{\wuqs}{$u$-quarks\xspace}

\newcommand{\wdq}{$d$-quark\xspace}
\newcommand{\wdqs}{$d$-quarks\xspace}

\newcommand{\wtq}{$t$-quark\xspace}
\newcommand{\wtqs}{$t$-quarks\xspace}

\newcommand{\wbq}{$b$-quark\xspace}
\newcommand{\wbqs}{$b$-quarks\xspace}

\newcommand{\wsq}{$s$-quark\xspace}

\newcommand{\wcq}{$c$-quark\xspace}

\newcommand{\cp}{\ensuremath{\mathcal{CP}}\xspace}
\newcommand{\sm}{\ensuremath{\mathrm{SM}}\xspace}
\newcommand{\MC}[1][]{\ensuremath{\texttt{MC#1}}\xspace}
\newcommand{\mssm}{\ensuremath{\mathrm{MSSM}}\xspace}
\newcommand{\susy}{\ensuremath{\mathrm{SUSY}}\xspace}
\newcommand{\mhmax}{\ensuremath{{m_\hhz^\mathrm{max}}}\xspace}
\newcommand{\PDF}{\ensuremath{\mathrm{PDF}}\xspace}
\newcommand{\pdf}{\ensuremath{\mathit{pdf}}\xspace}
\newcommand{\cdf}{\ensuremath{\mathit{cdf}}\xspace}
\newcommand{\equcomma}{\ensuremath{,\quad\quad}}
\newcommand{\equsep}{\\[0.2cm]}

\newcommand{\pb}{\ensuremath{\mathrm{pb}}\xspace}
\newcommand{\ipb}{\ensuremath{\mathrm{pb}^{-1}}\xspace}

\newcommand{\ifb}{\ensuremath{\mathrm{fb}^{-1}}\xspace}
\newcommand{\tev}{\ensuremath{\mathrm{Te\kern -0.1em V}}\xspace}
\newcommand{\tevc}{\ensuremath{\mathrm{Te\kern -0.1em V\!/}c}\xspace}
\newcommand{\tevcc}{\ensuremath{\mathrm{Te\kern -0.1em V\!/}c^2}\xspace}
\newcommand{\gev}{\ensuremath{\mathrm{Ge\kern -0.1em V}}\xspace}
\newcommand{\gevc}{\ensuremath{\mathrm{Ge\kern -0.1em V\!/}c}\xspace}
\newcommand{\gevcc}{\ensuremath{\mathrm{Ge\kern -0.1em V\!/}c^2}\xspace}
\newcommand{\mev}{\ensuremath{\mathrm{Me\kern -0.1em V}}\xspace}
\newcommand{\mevc}{\ensuremath{\mathrm{Me\kern -0.1em V\!/}c}\xspace}
\newcommand{\mevcc}{\ensuremath{\mathrm{Me\kern -0.1em V\!/}c^2}\xspace}
\newcommand{\ev}{\ensuremath{\mathrm{e\kern -0.1em V}}\xspace}
\newcommand{\rad}{\ensuremath{\mathrm{rad}}\xspace}

\newcommand{\tauola}{\ensuremath{\textsc{Tauola}}\xspace}
\newcommand{\pythia}[1]{\ensuremath{\textsc{Pythia}~#1}\xspace}
\newcommand{\citepythiasix}{sjostrand.06.1, pythia.13.1}
\newcommand{\citepythiaeight}{sjostrand.06.1, sjostrand.08.1, pythia.13.2}
\newcommand{\powheg}{{\sc Powheg}\xspace}
\newcommand{\herwig}[1]{{\sc Herwig\ensuremath{#1}}\xspace}
\newcommand{\citeherwigpp}{bahr.08.1, herwig.13.1}
\newcommand{\sherpa}{{\sc Sherpa}\xspace}
\newcommand{\citesherpa}{gleisberg.08.1, sherpa.13.1}
\newcommand{\mstw}{{MSTW$08$}\xspace}
\newcommand{\cteq}{{CTEQ$6$L$1$}\xspace}

\newcommand{\geant}{{\sc Geant4}\xspace}
\newcommand{\gauss}{{\sc Gauss}\xspace}
\newcommand{\brunel}{{\sc Brunel}\xspace}
\newcommand{\davinci}{{\sc DaVinci}\xspace}
\newcommand{\moore}{{\sc Moore}\xspace}
\newcommand{\boole}{{\sc Boole}\xspace}
\newcommand{\evtgen}{{\sc EvtGen}\xspace}
\newcommand{\gaudi}{{\sc Gaudi}\xspace}
\newcommand{\ganga}{{\sc Ganga}\xspace}
\newcommand{\dirac}{{\sc Dirac}\xspace}

\newcommand{\feynhiggs}{\ensuremath{\textsc{\href{http://www.feynhiggs.de/%
      }{FeynHiggs}}}\xspace}
\newcommand{\citefeynhiggs}{heinemeyer.98.1, heinemeyer.98.2,%
  degrassi.03.1, frank.07.1, feynhiggs.13.1}
\newcommand{\hdecay}{\ensuremath{\textsc{\href{http://people.web.psi.ch/%
        spira/hdecay/}{HDecay}}}\xspace}
\newcommand{\citehdecay}{djouadi.97.1, spira.97.1, butterworth.10.1,
  hdecay.13.1}
\newcommand{\prophecy}{\ensuremath{\textsc{\href{http://omnibus.uni-freiburg.de%
        /~sd565/programs/prophecy4f/prophecy4f.html}{Prophecy4f}}}\xspace}
\newcommand{\citeprophecy}{bredenstein.06.1, bredenstein.06.2,
  bredenstein.06.3, prophecy.13.1}

\newcommand{\lo}{\ensuremath{\textrm{LO}}\xspace}
\newcommand{\nlo}{\ensuremath{\textrm{NLO}}\xspace}
\newcommand{\nnlo}{\ensuremath{\textrm{NNLO}}\xspace}

\newcommand{\dynnlo}{\ensuremath{\textsc{\href{http://theory.fi.infn.it/%
        grazzini/dy.html}{Dynnlo}}}\xspace}
\newcommand{\citedynnlo}{catani.07.1, catani.09.1, dynnlo.13.1}
\newcommand{\bbh}{\ensuremath{\textsc{\href{http://particle.uni-wuppertal.de/%
        harlander/software/bbh@nnlo/}{bbH@NNLO}}}\xspace}
\newcommand{\citebbh}{harlander.03.1, bbh.13.1}
\newcommand{\ggh}{\ensuremath{\textsc{\href{http://particle.uni-wuppertal.de/%
        harlander/software/ggh@nnlo/}{ggH@NNLO}}}\xspace}
\newcommand{\citeggh}{harlander.02.1, harlander.03.2, ggh.13.1}
\newcommand{\higlu}{\ensuremath{\textsc{\href{http://people.web.psi.ch/spira/%
        higlu/}{Higlu}}}\xspace}
\newcommand{\citehiglu}{graudenz.93.1, spira.95.1, spira.95.2, higlu.13.1}
\newcommand{\dfg}{\ensuremath{\textsc{\href{http://theory.fi.infn.it/grazzini/%
        hcalculators.html}{dFG}}}\xspace}
\newcommand{\citedfg}{catani.03.1, florian.09.1, dfg.13.1}
\newcommand{\vbfh}{\ensuremath{\textsc{\href{http://vbf-nnlo.phys.ucl.ac.be/%
        vbf.html}{vbf@NNLO}}}\xspace}
\newcommand{\citevbfh}{bolzoni.10.1, bolzoni.11.1, vbfh.13.1}
\newcommand{\vh}{\ensuremath{\textsc{\href{http://particle.uni-wuppertal.de/%
        harlander/software/vh@nnlo/}{vH@NNLO}}}\xspace}
\newcommand{\citevh}{brein.12.1, vh.13.1}

\newcommand{\ggf}{gluon-gluon fusion\xspace}
\newcommand{\Ggf}{Gluon-gluon fusion\xspace}
\newcommand{\vbf}{vector-boson fusion\xspace}

\newcommand{\aqp}{associated heavy quark production\xspace}
\newcommand{\abp}{associated \wbq production\xspace}

\newcommand{\avp}{associated vector boson production\xspace}
\newcommand{\awp}{associated \wwb production\xspace}
\newcommand{\azp}{associated \wzb production\xspace}

\newcommand{\tanb}[1][]{\ensuremath{{\tan #1\beta}}\xspace}

\newcommand{\sinb}[1][]{\ensuremath{{\sin #1\beta}}\xspace}
\newcommand{\cosb}[1][]{\ensuremath{{\cos #1\beta}}\xspace}

\newcommand{\sina}[1][]{\ensuremath{{\sin #1\alpha}}\xspace}
\newcommand{\cosa}[1][]{\ensuremath{{\cos #1\alpha}}\xspace}
\newcommand{\cosbma}[1][]{\ensuremath{{\cos #1\left(\beta -%
        \alpha\right)}}\xspace}
\newcommand{\sinbma}[1][]{\ensuremath{{\sin #1\left(\beta -%
        \alpha\right)}}\xspace}
\newcommand{\tw}{\ensuremath{\theta_w}\xspace}
\newcommand{\sintw}[1][]{\ensuremath{{\sin #1\tw}}\xspace}
\newcommand{\costw}[1][]{\ensuremath{{\cos #1\tw}}\xspace}
\newcommand{\gE}[1][]{\ensuremath{{g_e #1}}\xspace}
\newcommand{\gW}[1][]{\ensuremath{{g_w #1}}\xspace}
\newcommand{\gS}[1][]{\ensuremath{{g_s #1}}\xspace}
\newcommand{\aE}[1][]{\ensuremath{{\alpha_e #1}}\xspace}

\newcommand{\aS}[1][]{\ensuremath{{\alpha_s #1}}\xspace}
\newcommand{\un}[1][]{\ensuremath{{\mathit{U}%
      \ifthenelse{\equal{#1}{}}{}{(#1)}}}\xspace}
\newcommand{\su}[1][]{\ensuremath{{\mathit{SU}%
      \ifthenelse{\equal{#1}{}}{}{(#1)}}}\xspace}
\newcommand{\so}[1][]{\ensuremath{{\mathit{SO}%
      \ifthenelse{\equal{#1}{}}{}{(#1)}}}\xspace}
\newcommand{\msbar}{\ensuremath{\overline{\textrm{MS}}}\xspace}
 
\newcommand{\gm}[2]{\ensuremath{\gamma_{#1}^{#2}}\xspace}

\newcommand{\vc}[1][]{\ensuremath{{v#1}}\xspace}
\newcommand{\ac}[1][]{\ensuremath{{a#1}}\xspace}
\newcommand{\Q}{\ensuremath{Q}\xspace}

\newcommand{\me}{\ensuremath{\mathcal{M}}\xspace}
\newcommand{\ld}{\ensuremath{\mathcal{L}}\xspace}
\newcommand{\xf}{\ensuremath{\mathit{f}}\xspace}

\newcommand{\termstrut}[1]{\vrule width0pt height0pt depth#1}
\newcommand{\termlabel}[3][4mm]{\ensuremath{
    \underbrace{\termstrut{#1}\smash{#2}}_{\mbox{#3}}}}

\newcommand{\collab}{collaboration\xspace}
\newcommand{\lhc}{LHC\xspace}
\newcommand{\lhcb}{LHCb\xspace}
\newcommand{\alice}{ALICE\xspace}
\newcommand{\atlas}{ATLAS\xspace}
\newcommand{\cms}{CMS\xspace}
\newcommand{\tevatron}{Tevatron\xspace}
\newcommand{\hera}{HERA\xspace}
\newcommand{\dlep}{LEP\xspace}
\newcommand{\stt}{\ensuremath{\mathrm{ST}}\xspace}
\newcommand{\ttt}{\ensuremath{\mathrm{TT}}\xspace}
\newcommand{\itt}{\ensuremath{\mathrm{IT}}\xspace}
\newcommand{\ott}{\ensuremath{\mathrm{OT}}\xspace}
\newcommand{\velo}{\ensuremath{\mathrm{VELO}}\xspace}
\newcommand{\ecal}{\ensuremath{\mathrm{ECAL}}\xspace}
\newcommand{\hcal}{\ensuremath{\mathrm{HCAL}}\xspace}
\newcommand{\prs}{\ensuremath{\mathrm{PRS}}\xspace}
\newcommand{\rich}[1]{\ensuremath{\mathrm{RICH}#1}\xspace}
\newcommand{\spd}{\ensuremath{\textrm{SPD}}\xspace}

\newcommand{\hlt}{\ensuremath{\mathrm{HLT}}~trigger\xspace}
\newcommand{\inelastic}{\ensuremath{\mathrm{inelastic}}\xspace}

\newcommand{\m}{\ensuremath{m}\xspace}

\newcommand{\pt}{\ensuremath{{p_\mathrm{T}}}\xspace}
\newcommand{\qt}{\ensuremath{{q_\mathrm{T}}}\xspace}
\newcommand{\et}{\ensuremath{{E_\mathrm{T}}}\xspace}
\newcommand{\met}{\ensuremath{{{\not\mathrel{E}}_\mathrm{T}}}\xspace}
\newcommand{\prc}[1]{\ensuremath{{(#1)\%}}\xspace}
\newcommand{\abs}[1]{\ensuremath{\left|#1\right|}\xspace}

\newcommand{\eff}[1][]{\ensuremath{{\varepsilon#1}}\xspace}
\newcommand{\iso}{\ensuremath{{I_{p_\mathrm{T}}}}\xspace}
\newcommand{\dphi}{\ensuremath{{|\Delta \phi|}}\xspace}

\newcommand{\ips}{\ensuremath{{\mathrm{IPS}}}\xspace}
\newcommand{\apt}{\ensuremath{{A_{p_\mathrm{T}}}}\xspace}
\newcommand{\kin}{\ensuremath{\mathrm{kin}}\xspace}
\newcommand{\gec}{\ensuremath{\mathrm{GEC}}\xspace}
\newcommand{\rec}{\ensuremath{\mathrm{rec}}\xspace}
\newcommand{\gen}{\ensuremath{\mathrm{gen}}\xspace}
\newcommand{\sel}{\ensuremath{\mathrm{sel}}\xspace}

\newcommand{\bkg}{\ensuremath{\mathrm{bkg}}\xspace}

\newcommand{\trg}{\ensuremath{\mathrm{trg}}\xspace}
\newcommand{\trk}{\ensuremath{\mathrm{trk}}\xspace}
\newcommand{\id}{\ensuremath{\mathrm{id}}\xspace}
\newcommand{\ewk}{\ensuremath{\mathrm{EWK}}\xspace}
\newcommand{\qcd}{\ensuremath{\mathrm{QCD}}\xspace}
\newcommand{\simu}{\ensuremath{\mathrm{sim}}\xspace}

\newcommand{\fwidth}{\ensuremath{{f_\mathrm{width}}}\xspace}
\newcommand{\ffit}{\ensuremath{{f_\mathrm{fit}}}\xspace}
\newcommand{\fshift}{\ensuremath{{f_\mathrm{shift}}}\xspace}
\newcommand{\fscale}{\ensuremath{{f_\mathrm{scale}}}\xspace}
\newcommand{\rqcd}{\ensuremath{{r_\mathrm{\qcd}}}\xspace}
\newcommand{\rewk}{\ensuremath{{r_\mathrm{\ewk}}}\xspace}
\newcommand{\sqcd}{\ensuremath{{S_\mathrm{\qcd}}}\xspace}
\newcommand{\sewk}{\ensuremath{{S_\mathrm{\ewk}}}\xspace}

\newcommand{\br}[1][]{\ensuremath{{\mathcal{B}#1}}\xspace}
\newcommand{\lum}{\ensuremath{\mathscr{L}}\xspace}
\newcommand{\acc}[1][]{\ensuremath{{\mathcal{A}#1}}\xspace}
\newcommand{\sys}{\ensuremath{\mathrm{sys}}\xspace}

\newcommand{\cor}{\ensuremath{\mathcal{C}}\xspace}

\newcommand{\sdif}{\ensuremath{\textrm{\,d}}\xspace}
\newcommand{\dif}{\ensuremath{\textrm{d}}\xspace}
\newcommand{\wpv}{\ensuremath{p\textrm{\,-value}}\xspace}
\newcommand{\wpvs}{\ensuremath{p\textrm{\,-values}}\xspace}

\newcommand{\lcrit}[1][]{\ensuremath{{c #1}}\xspace}
\newcommand{\crit}[1][]{\ensuremath{{\mathcal{C} #1}}\xspace}

\newcommand{\hypo}[1][]{\ensuremath{{\mathcal{H} #1}}\xspace}
\newcommand{\stat}[1][]{\ensuremath{{t #1}}\xspace}
\newcommand{\unc}[1][]{\ensuremath{\theta #1}\xspace}
\newcommand{\uncs}[1][]{\ensuremath{\vec{\theta} #1}\xspace}
\newcommand{\var}[1][]{\ensuremath{x #1}\xspace}
\newcommand{\vars}[1][]{\ensuremath{\vec{x} #1}\xspace}
\newcommand{\asi}[1][]{\ensuremath{a #1}\xspace}
\newcommand{\asis}[1][]{\ensuremath{\vec{a} #1}\xspace}
\newcommand{\pval}[1][]{\ensuremath{p #1}\xspace}

\newcommand{\mval}[2][]{\ensuremath{M#1[#2]}\xspace}
\newcommand{\lh}{\ensuremath{{\mathit{L}}}\xspace}
\newcommand{\llh}{\ensuremath{{\mathit{LL}}}\xspace}

\newcommand{\cls}{\ensuremath{{\mathrm{CL_s}}}\xspace}
\newcommand{\clb}{\ensuremath{{\mathrm{CL_b}}}\xspace}
\newcommand{\clsb}{\ensuremath{{\mathrm{CL_{s+b}}}}\xspace}

\newcommand{\obs}{\ensuremath{{\mathrm{obs}}}\xspace}
\newcommand{\sig}{\ensuremath{{\mathrm{sig}}}\xspace}

\newcommand{\bins}{\ensuremath{{\mathrm{bins}}}\xspace}
\newcommand{\p}{\ensuremath{\phantom{1}}}

\newlength{\backspace}
\newcolumntype{L}{>{$}l<{$}}
\newcolumntype{R}{>{$}r<{$}}
\newcolumntype{C}{>{$}c<{$}}
\newcolumntype{B}{>{\centering\arraybackslash}b}
\newcolumntype{P}{>{\centering\arraybackslash}p}
\newcolumntype{M}{>{\centering\arraybackslash}m}
\newcolumntype{E}{>{$}r<{\:\pm$}@{\:}>{$}r<{$}}
\newcolumntype{F}{>{$}r<{\:\pm$}@{$\:$\hspace{-\backspace}}>{$}r<{$}}

\newcommand{\chp}[1]  {\mbox{Chap.~\ref{chp:#1}\xspace}}

\newcommand{\chps}[1] {\mbox{Chaps.~\ref{chp:#1}\xspace}}

\renewcommand{\sec}[1]{\mbox{Sect.~\ref{sec:#1}\xspace}}

\newcommand{\secs}[1] {\mbox{Sects.~\ref{sec:#1}\xspace}}

\newcommand{\sap}[1]{\mbox{App.~\ref{sec:#1}\xspace}}

\newcommand{\app}[1]  {\mbox{App.~\ref{app:#1}\xspace}}

\newcommand{\fig}[1]  {\mbox{Fig.~\ref{fig:#1}\xspace}}
\newcommand{\Fig}[1]  {\mbox{Figure~\ref{fig:#1}\xspace}}
\newcommand{\figs}[1] {\mbox{Figs.~\ref{fig:#1}\xspace}}
\newcommand{\Figs}[1] {\mbox{Figures~\ref{fig:#1}\xspace}}
\newcommand{\equ}[1]  {\mbox{Eq.~\ref{equ:#1}\xspace}}
\newcommand{\Equ}[1]  {\mbox{Equation~\ref{equ:#1}\xspace}}
\newcommand{\equs}[1] {\mbox{Eqs.~\ref{equ:#1}\xspace}}

\newcommand{\tab}[1]  {\mbox{Table~\ref{tab:#1}\xspace}}
\newcommand{\Tab}[1]  {\mbox{Table~\ref{tab:#1}\xspace}}
\newcommand{\tabs}[1] {\mbox{Tables~\ref{tab:#1}\xspace}}

\newcommand{\rfr}[1]  {\mbox{Ref.~\cite{#1}\xspace}}
\newcommand{\Rfr}[1]  {\mbox{Reference~\cite{#1}\xspace}}
\newcommand{\rfrs}[1] {\mbox{Refs.~\cite{#1}\xspace}}

\newcommand{\labelchp}{}
\newcommand{\labelsec}[1]{\label{sec:\labelchp:#1}}
\newcommand{\labelfig}[1]{\label{fig:\labelchp:#1}}
\newcommand{\labelequ}[1]{\label{equ:\labelchp:#1}}
\newcommand{\labelali}[1]{\stepcounter{equation}\tag{\theequation}%
  \ifthenelse{\equal{#1}{}}{}{\labelequ{#1}}}
\newcommand{\labeltab}[1]{\label{tab:\labelchp:#1}}
\newcommand{\labelite}[1]{\label{ite:\labelchp:#1}}
\newcommand{\subfig}[2][]{\ifthenelse{\equal{#1}{}}{\subref*{fig:\labelchp:#2}}%
  {\subref*{fig:\labelchp:#1}\textit{\textbf{--}}\subref*{fig:\labelchp:#2}}}
\newcommand{\subtab}[2][]{\ifthenelse{\equal{#1}{}}{\subref*{tab:\labelchp:#2}}%
  {\subref*{tab:\labelchp:#1}\textit{\textbf{--}}\subref*{tab:\labelchp:#2}}}

\newcommand{\newchapter}[2]{\chapter{#1}\label{chp:#2}%
  \setsvg{svgpath=Figures/#2/}\renewcommand{\fmppath}{Figures/#2/}%
  \renewcommand{\labelchp}{#2}}
\newcommand{\newappendix}[2]{\chapter{#1}\label{app:#2}%
  \setsvg{svgpath=Figures/#2/}\renewcommand{\fmppath}{Figures/#2/}%
  \renewcommand{\labelchp}{#2}}
\newcommand{\newsection}[2]{\section{#1}\labelsec{#2}}
\newcommand{\newsubsection}[2]{\subsection{#1}\labelsec{#2}}
\newcommand{\newsubsubsection}[2]{\subsubsection{#1}}
\newcommand{\ltext}{\hspace{-0.05\unitlength}}
\newcommand{\ptext}[1]{\pbox{2cm}{~\\\phantom{1}#1}}

\begin{document}
\setsvg{svgpath=Figures/Ttl/}

% PDF meta tag.
\newcommand{\infotitle}{Electroweak and Higgs Measurements Using Tau
  \texorpdfstring{\\[0.4cm]}{} Final States with the LHCb Detector}
\newcommand{\infoauthor}{Philip Ilten}
\hypersetup{
  pdftitle    = {\infotitle},
  pdfauthor   = {\infoauthor},
  pdfsubject  = {high energy particle physics},
  pdfkeywords = {HEP, tau lepton, Z boson, Higgs}
}

% Title page.
\thispagestyle{empty}
\title{\infotitle}
\author{\infoauthor}
\begin{center}
  {\LARGE \bf \infotitle}
  \\[1cm]
  {\LARGE by \infoauthor}
  \\[2cm]
  \includesvg[width=4cm]{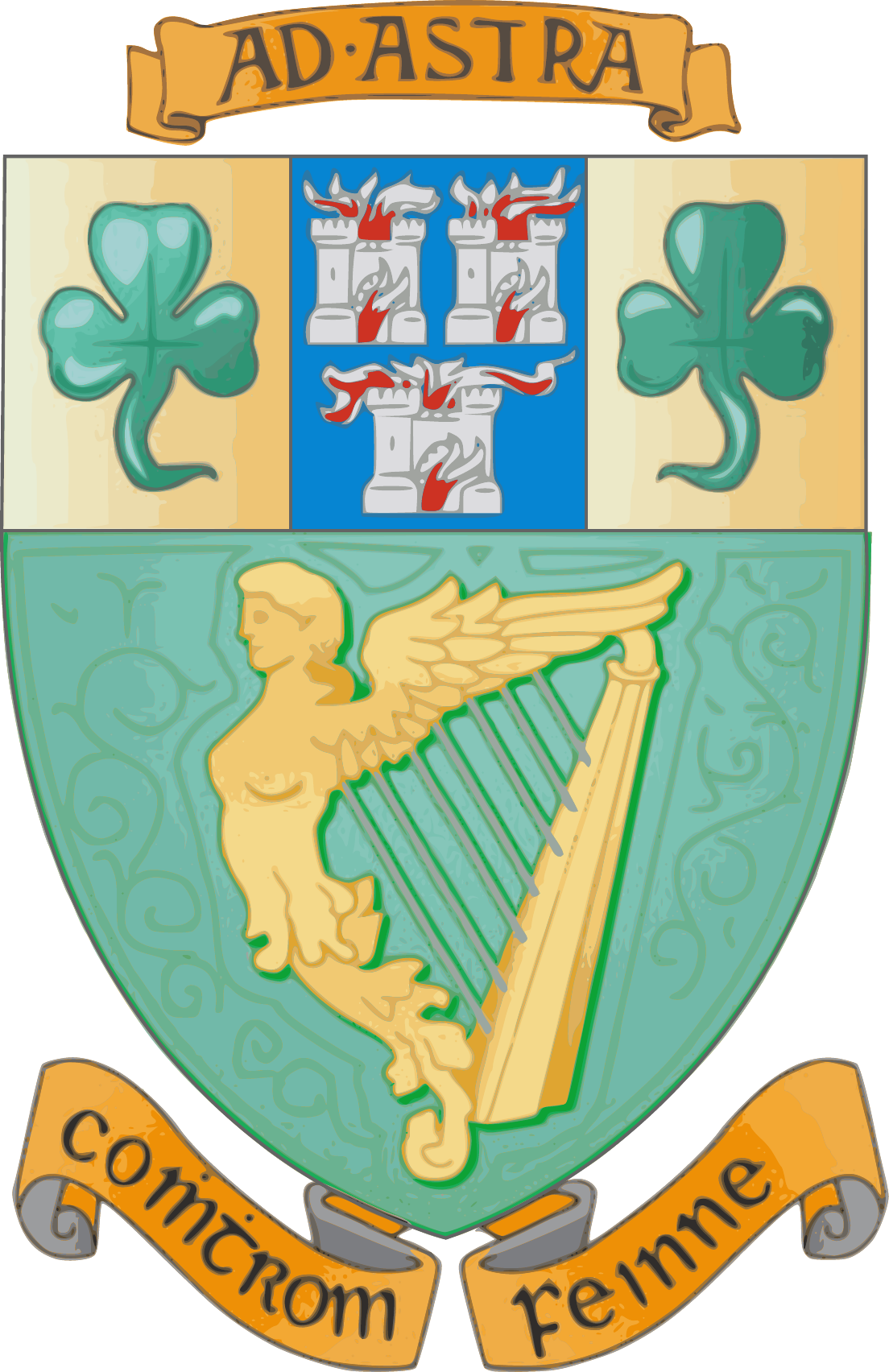} \\[2cm]
  {\Huge UNIVERSITY COLLEGE DUBLIN}
  \\[1cm]
  {\Huge School of Physics} \\
\end{center}
\vfill
\noindent
This thesis is submitted to University College Dublin in fulfilment
of the requirements for the degree of Doctor of Philosophy. \\[0.5cm]
\makebox[4cm][l]{Head of School:}
\makebox[6cm][l]{Prof.\ Padraig Dunne} \\
\makebox[4cm][l]{Supervisor:}
\makebox[6cm][l]{Dr.\ Ronan McNulty} \\
\makebox[4cm][l]{Submitted:}
\makebox[6cm][l]{September, $2013$} \\
\makebox[4cm][l]{Examined:}
\makebox[6cm][l]{November, $2013$} \\[0.5cm]
\makebox[4cm][l]{{\it Viva Voce} Panel} \\
\makebox[4cm][l]{Chairperson:}
\makebox[6cm][l]{Prof.\ Padraig Dunne} \\
\makebox[4cm][l]{Internal Examiner:}
\makebox[6cm][l]{Prof.\ Martin Gr\"unewald} \\
\makebox[4cm][l]{External Examiner:}
\makebox[6cm][l]{Prof.\ Eilam Gross} \\
\newpage

% Copyright page.
\thispagestyle{empty}
{\it \noindent I, Philip Ilten, hereby certify that the submitted work
  is my own work, was completed while registered as a candidate for
  the degree of Doctor of Philosophy at University College Dublin, and
  I have not obtained a degree elsewhere on the basis of the research
  presented in the submitted work.} \\

\noindent This thesis is licensed under the
\href{http://creativecommons.org}{Creative Commons}
\href{http://creativecommons.org/licenses/by/3.0/legalcode}{Attribution
  3.0} license. The contents of this thesis, in full or in part, can
be copied, distributed, transmitted, or remixed if the material is
attributed to the author. \\

\begin{center}
  $\vcenter{\hbox{\includesvg[width=2cm]{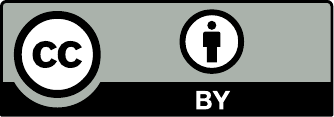}}}$
  \copyright~2013 by Philip Ilten
\end{center}
\newpage

% Table of contents.
\pagestyle{fancy}
\pagenumbering{roman}
\tableofcontents
\newpage

% Abstract page.
\pagestyle{plain}
\chapter*{Abstract}
\addcontentsline{toc}{chapter}{\it Abstract}

Spin correlations for \wtl decays are included in the \pythia{8} event
generation software with a framework which can be expanded to include
the decays of particles other than the \wtl. The spin correlations for
the decays of \wtls produced from electroweak and \whbs are
calculated. Decays of the \wtl using sophisticated resonance models
are included in \pythia{8} for all channels with experimentally
observed branching fractions greater than $0.04\%$. The mass
distributions for the decay products of these channels calculated with
\pythia{8} are validated against the equivalent distributions from the
\herwig{++} and \tauola event generators. The technical
implementation of the \wtl spin correlations and decays in
\pythia{8} is described.

A measurement of the inclusive ${Z \to \ditau}$ cross-section using
${1.0~\ifb}$ of data from \dip collisions at $\sqrt{s} = 7~\tev$
collected with the \lhcb detector is presented. Reconstructed final
states containing two muons, a muon and an electron, a muon and a
charged hadron, or an electron and a charged hadron are selected as
${\z \to \ditau}$ candidates. The cross-section for \wzbs with a mass
between $60$ and $120~\gev$ decaying into \wtls with pseudo-rapidities
between $2.0$ and $4.5$ and transverse momenta greater than $20~\gev$
is measured to be ${72.3 \pm 3.5 \pm 2.9 \pm 2.5~\pb}$. The first
uncertainty is statistical, the second uncertainty is systematic, and
the third is to due the integrated luminosity uncertainty. The ${\z
  \to \ditau}$ to ${\z \to \dimu}$ cross-section ratio is found to be
${0.94 \pm 0.09}$ and the ${\z \to \ditau}$ to ${\z \to \die}$
cross-section ratio is found to be ${0.95 \pm 0.07}$. The uncertainty
on these ratios is the combined statistical, systematic, and
luminosity uncertainties.

Limits on the production of neutral \whbs decaying into \wtl pairs
with pseudo-rapidities between $2.0$ and $4.5$ are set at a $95\%$
confidence level using the same \lhcb dataset. A model independent
upper limit on the production of neutral \whbs decaying into \wtls is
set and ranges between $8.6~\pb$ for a \whb mass of $90~\gev$ to
$0.7~\pb$ for a \whb mass of $250~\gev$. This limit is compared to the
expected standard model cross-section. An upper limit on \tanb in the
$\m_\hAz$ and \tanb plane is set for the \mhmax scenario of the
minimal supersymmetric model and varies from $34$ for a \cp-odd \whb
mass of $90~\gev$ to $70$ for a \cp-odd \whb mass of $140~\gev$.

% Authorship page.
\chapter*{Authorship}
\addcontentsline{toc}{chapter}{\it Authorship}

I undertook the work presented in this thesis as a member of the \lhcb
collaboration at CERN, sharing responsibility with the rest of the
collaboration for the collection and integrity of the data used in
this thesis. The analysis of \chp{Zed} was performed independently by
me and then combined with that of another University College Dublin
student~\cite{farry.12.1} in order to produce a paper~\cite{lhcb.13.1}
whose results I have shown at
conference~\cite{ilten.12.3,*ilten.12.4}. The analysis of \chp{Hig} is
my own and extends the results of \chp{Zed} to a search for the
\whb. I wrote the paper~\cite{lhcb.13.3} and presented the results at
conference~\cite{ilten.13.1,*ilten.13.2}. The theoretical work in
\chp{Tau} was undertaken at the University of Lund, under the
supervision of Prof. Torbj\"orn Sj\"ostrand, and was presented at the
{\it 12th International Workshop on Tau Physics}~\cite{ilten.12.1}
with proceedings to appear in \rfr{ilten.12.2}.

% Acknowledgements page.
\chapter*{Acknowledgements}
\addcontentsline{toc}{chapter}{\it Acknowledgements}

This thesis would not have been possible without the support from a
set of extraordinary people.

I would first like to acknowledge my supervisor, Ronan McNulty. His
scientific integrity and rigour are unparallelled and his work speaks
volumes, but his dedication to education, whether of his PhD students,
undergraduates, or the general public, is uncompromising. He allowed
me free reign with my ideas, but always made sure I understood the
practicalities of the situation as well. He supported me in every step
of my PhD, even if this meant sending me to Sweden for a semester.

My time in Sweden was an incredible opportunity which I owe entirely
to Torbj\"orn Sj\"ostrand, my supervisor away from home. Not only is
he an outstanding physicist, he is also one of the kindest people I
have had the pleasure of meeting. Despite his busy schedule of
teaching, theoretical research, and developing {\sc Pythia},
Torbj\"orn took the time to make me feel at home in Lund. I owe a debt
of gratitude to Torbj\"orn, MCNet, and everyone in Lund. Thank you for
the experience.

My PhD would not have been possible without funding from the UCD
physics department, Science Foundation Ireland, and MCNet through
Marie Curie grant MRTN-CT-2006-035606. I sincerely hope that these
funding sources will be able to continue to support particle physics
research in the future, especially in Ireland.

Joining a collaboration, especially one as large as \lhcb is never
easy, but I had the support of some amazing colleagues in the
electroweak, exotica, and Monte Carlo groups. I am especially indebted
to Tara Shears, Roger Barlow, and Gloria Corti.

The UCD particle physics group also provided me with critical support
over my four years at UCD. Thank you Steve, Simone, James, Dermot,
Ronan (Wallace), Wenchao, Zoltan, Francesco, Shane, and Sara. I would
like to especially thank Simone Bifani who always had the answer I
needed, and gently notified me when the cluster was not working. I
cannot thank Steve Farry enough. His patience in dealing with me is
nearly unrivalled (except for my wife), and despite his attempts to
fool people with his knowledge of football, he is a particle physicist
at heart. He is a true scholar and a gentleman.

I would like to also thank everyone else in the UCD physics department
who has helped me through the years. Thank you Marian, Bairbre, and
John for making the physics department not just a loose affiliation of
students and teachers but a real community. Thank you Padraig and
Lorraine for guiding the school over the years and supporting my
research, and thank you Peter for trusting me to tutor your classes.

To all my friends at UCD over the years, thank you. The karate club
kept me sane and fit, and was my home away from home. There is nothing
like some light sparring to take your mind off particle physics. Thanks
to the CASL and Settlers of Catan crews, my Thursday evenings were
always enjoyable. I also am indebted to Dominic's group for letting me
drop by and chat.

As for my housemates, past and present, I could not ask for
better. Majella, thank you for the wine, food, and stories. I always I
could come home looking forward to a (usually) good story and an
interesting conversation. Noel and Eibhlin, thank you for the hikes,
dives, and barbeques; they shall continue!

To all my friends I have not yet mentioned, thanks for helping me
become the person I am today. I would like to especially thank John
Wuestneck, Andrew Hulke, and the Core: Matt, Mary, Sarah, and
Dan. Everyone who made it out to my and \'Eadaoin's wedding, thank
you. To all my Irish friends, thank you for welcoming me with open
arms. Of course I cannot neglect $Z\Psi$ and especially the
$A\Gamma$s, $\tau\kappa\phi$. Coblenz, thank you for the statistics
discussions, they were very illuminating, and Akil, thank you for
understanding why the eagles could not just fly into Mordor.

Of course, I would not have reached this point without incredible
teachers over the years. I would especially like to thank Markus for
sending me off to Ireland, Tim Corcoran for teaching me to love
science, and John Rubel for helping me truly understand the scientific
method.

Finally, I would like to thank my family immediate and extended. Mom,
Dad, Jen, and Nathan, I would be nothing without you. Karen and Adam,
thank you for joining the Ilten family, I cherish your love and
support. Zoe and Henry, thanks for remembering who crazy uncle Phil
is! Peter and Caitr\'iona, thank you for unconditionally accepting me
into your family. I am proud to be your son-in-law. D\'onal, Andrew,
Deirdre, and D\'onal, thank you for all the good times, past, present
and future. \'Eadaoin, there are so many things I should thank you for
that I will just thank you for being my better half.\\[1cm]

\begin{center}
  Thank you.
\end{center}

% Dedication page.
\newpage
\topskip0pt
\begin{center}
  \vspace*{\fill}
  \begin{center}
    {\it To \'{E}adaoin Ilten, my wife and the love of my life.}
  \end{center}
  \vspace*{\fill}
\end{center}\newpage

\pagestyle{fancy}
\pagenumbering{arabic}
\newchapter{Introduction}{Int}

The natural world is complex, and humankind attempts to understand
this complexity through the search for the underlying laws by which
nature is governed. This search takes many forms, but at the forefront
is particle physics, which endeavours to describe the mechanisms
whereby the fundamental constituents of nature interact. The
predictive power of particle physics spans from the early formation of
the universe to the structure of the proton, yet remains incomplete.

The theoretical framework for the standard model (\sm) of particle
physics is relativistic quantum field theory where fundamental
particles, thought to be the indivisible constituents of matter, are
represented as the quanta of relativistic fields. The quantum nature
of the \sm dictates that its predictions are not certainties but
rather probabilities: the probabilities for particles to interact
through collisions or the probabilities for particles to decay. Many
of these probabilities can be directly calculated with the \sm to
produce theoretical predictions which can be confirmed or rejected
with experiment. A review of the \sm and how these calculations are
made is provided in \chp{Thr} of this thesis.

Within the \sm the interactions of the fundamental particles are
through the electromagnetic, weak, and strong forces. These forces are
carried through fundamental particles which are labelled bosons. By
the symmetry of the theory, these bosons are expected to be
massless. However, the two carriers of the weak force, the \w and
\wzbs, are massive. To rectify this in the \sm, an additional field,
the Higgs field, is introduced. This field can be used not only to
generate the masses of the \w and \wzbs, but also the masses for all
remaining fundamental particles with non-zero mass. The Higgs field
also has an associated particle, the \whb, with a mass not fixed by
theory and whose interaction strength with other particles is
proportional to their mass.

A particle with a mass near $125~\gev$ has been
discovered~\cite{atlas.12.2, cms.13.1} which exhibits many of the
properties of the expected \whb from the \sm. To verify this particle
is indeed consistent with the \sm \whb, its decay probabilities must
be fully measured. An \sm \whb with a mass of $125~\gev$ is expected
to decay into a pair of \wtls in approximately $6\%$ of all its
decays. The \wtl is the heaviest of the charged leptons, fundamental
particles which interact through only the electromagnetic and weak
forces, and can decay into a large variety of complex final states
that can be experimentally detected. Consequently, to measure the
probability of the \whb decaying into \wtl pairs, the decays of the
\wtl must first be theoretically understood.

In \chp{Tau}, all \wtl decays with a probability greater than $0.04\%$
are implemented using a variety of theoretical models in the
open-source software \pythia{8}~\cite{\citepythiaeight}, which
performs theoretical calculations by simulating events where particles
interact and decay. The type of particle producing the \wtl influences
its intrinsic spin. This spin in turn influences the kinematics of the
\wtl decay products. To ensure the kinematics of the \wtl decays are
properly modelled, full spin correlations are also included in the
decay models of the \wtl in \chp{Tau}.

The probabilities of particle interactions and decays are
experimentally measured by using accelerators where particles are
collided together at high energies within particle detectors. The
detectors measure the passage of particles produced from the
collision, and the output from the detector is processed using
specialised software which reconstructs the particles from the
collision. The reconstructed particles can then be used to determine
the interactions and decays occurring within the collision and measure
the probabilities predicted by the \sm.

The Large Hadron Collider (\lhc) is a particle accelerator which
collides two protons head-on, each with an energy of $3.5~\tev$. The
Large Hadron Collider Beauty detector (\lhcb) is built around one of
the four \lhc collision points and measures the production of
particles along the forward direction of the beam. The \lhcb detector
is designed to detect $B$-mesons, composite particles whose decays
might help explain the observed asymmetry between matter and
anti-matter in the universe.  In \chp{Exp} the \lhc and \lhcb are
described, as well as the techniques used to reconstruct particles
passing through the \lhcb detector.

The detector characteristics which allow \lhcb to identify $B$-mesons
are also well suited for the identification and reconstruction of \wtl
decays. Consequently, \lhcb can be used to search for the production
of \whbs decaying into \wtl pairs. However, these \whb events are
expected to be rare, and cannot be easily separated from events where
a \wtl pair is produced from a \wzb decay. In \chp{Zed} the
cross-section, or probability per particle flux and time, is
measured for the production of \wzbs which decay into \wtl pairs
within \lhcb. This cross-section can be used to refine current
knowledge of the proton structure. Additionally, the \wzb is expected
to decay with equal probability into the three types of charged
leptons: electrons, muons, and \wtls. This prediction is tested by
comparing the ${\z \to \ditau}$ cross-section to the cross-sections of
\wzbs decaying into muon and electron pairs.

An excess in the ${\z \to \ditau}$ cross-section might indicate the
presence of the \whb or some other new physics contaminating the
number of observed events of \chp{Zed}. In \chp{Hig} a statistical
analysis is performed to determine upper limits on the cross-section
for the production of \whbs decaying into \wtls within \lhcb that is
consistent with the number of observed events. This limit is compared
to the cross-section expected from the \sm, as well as from the
alternative minimal supersymmetric model (\mssm). A conclusion
summarising all the results from \chp{Tau}, \chp{Zed}, and \chp{Hig}
of this thesis is provided in \chp{Con}.

\newchapter{Theory}{Thr}

A review is given in this chapter which provides the necessary
theoretical framework for this thesis. The chapter is split into two
sections, \secs{Thr:Sta} and~\ref{sec:Thr:Mon}. In \sec{Thr:Sta}, an
overview of the standard model of particle physics is given. This
includes an introduction to the perturbative methods used to calculate
experimental observables, the underlying Lagrangian densities used in
these perturbative calculations, and the experimental observables
themselves. Additionally, an outline of alternatives and extensions to
the standard model is given. The results from this section are used in
\chp{Tau} to model \wtl decays, in \chp{Zed} to calculate the ${\dip
  \to \z \to \ditau}$ cross-section, and in \chp{Hig} to place limits
on \whb production. In \sec{Thr:Mon} numerical analysis techniques
used to calculate the experimental observables introduced in
\sec{Thr:Sta} are presented. These include the methods necessary for
the modelling of \wtl decays in \chp{Tau}, as well as the simulation
of background and signal events for \chps{Zed} and~\ref{chp:Hig}.

\newsection{Standard Model}{Sta}

The standard model (\sm) of particle
physics~\cite{weinberg.67.1,salam.68.1} describes the interactions
between all the experimentally observed particles of \fig{Thr:Sm}
through the electromagnetic, weak, and strong forces, but not
gravity. In \fig{Thr:Sm} the symbol, electromagnetic charge quantum
number, spin quantum number, and mass for each particle is given. The
particles are grouped by their spin quantum numbers into fermions,
half-integer spin particles, and bosons, integer spin particles. The
fermions are the constituents of matter and are further grouped into
quarks, particles which interact through all three forces, and
leptons, particles which interact through only the electromagnetic and
weak forces. Both quarks and leptons have a spin quantum number of
$1/2$.

\begin{subfigures}{1}{The \sm particles: the spin-$1/2$ fermions
    (solid boxes) divided into quarks (magenta) and leptons (orange),
    and the spin-$1$ gauge bosons (dashes and cyan). The spin-$0$ \whb
    has not been included. The fermions are within dotted lines
    representing their interactions with the strong (green),
    electromagnetic (red), and weak (blue) forces, with the force
    mediating bosons included in the grouping. The top line for each
    particle is its symbol, the left middle its electromagnetic
    charge, the right middle its spin, and the bottom its mass. The
    masses are reported up to an uncertainty on three significant
    digits and are taken from \rfr{pdg.12.1}. No masses are given for
    the neutrinos as they are not mass eigenstates.\labelfig{Sm}}
  \svgbeg
  \includesvg[width=\columnwidth]{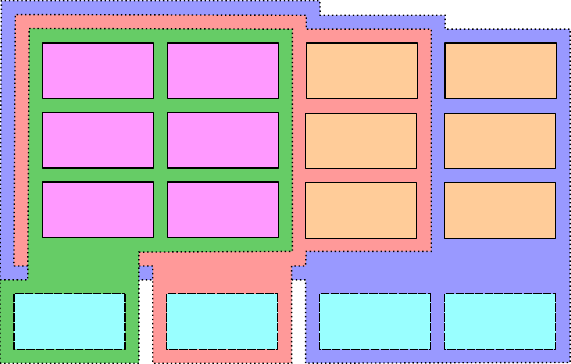} \svgend
\end{subfigures}

Each quark has a colour charge quantum number of either red, blue, or
green. The up ($u$), charm ($c$), and top quarks ($t$) all have an
electromagnetic charge quantum number of $+2/3$ while the down ($d$),
strange ($s$), and bottom ($b$) quarks have an electromagnetic charge
quantum number of $-1/3$. Each quark type also has a corresponding
anti-quark type which carries an anti-colour of anti-red, anti-blue,
or anti-green, and the opposite sign electromagnetic charge. The
quarks are grouped into three generations of up/down, charm/strange,
and top/bottom. While a quark can interact with another quark outside
its generation, this type of behaviour is suppressed in the \sm by the
Cabibbo-Kobayashi-Maskawa (CKM) matrix which relates the mass and
flavour eigenstates of the quarks. The quarks listed in \fig{Thr:Sm}
are mass eigenstates with well defined masses, but because all the
quarks except the massive top quark have only been observed in bound
states, their masses have not been directly measured experimentally.

The leptons are grouped into charged leptons which have an
electromagnetic charge quantum number of $-1$, and neutrinos with an
electromagnetic charge quantum number of $0$. There are three flavours
of charged lepton, the electron ($e$), the muon ($\mu$), and the \wtl
($\tau$), each grouped with a neutrino of the same flavour, the
electron ($\nu_e$), muon ($\nu_\mu$), and \wtl ($\nu_\tau$)
neutrinos. Each charged lepton has an anti-lepton partner with
anti-flavour and an electromagnetic charge of $+1$, while each
neutrino has an anti-neutrino partner with anti-flavour. Within the
\sm, lepton flavour is approximately conserved with the exception of
oscillations of the neutrinos between their flavour eigenstates. The
charged leptons of \fig{Thr:Sm} are mass eigenstates with well defined
masses, while the neutrinos are flavour eigenstates. The neutrinos are
also known to have mass eigenstates $\nu_1$, $\nu_2$, and $\nu_3$, but
only the mass differences between these states have been measured.

In the \sm there are four gauge bosons with a spin quantum number of
$1$. The gluon ($g$) is the massless mediator of the strong force and
carries one of eight colour/anti-colour charge quantum number
combinations. The photon ($\gamma$) is the massless mediator of the
electromagnetic force, while the \w and \wzbs are the massive
mediators of the weak force. The gluons only interact with fermions
with colour charge, photons only interact with fermions with non-zero
electromagnetic charge, and the weak bosons interact with all the
fundamental fermions. The gauge bosons can also interact amongst
themselves, with the details of these interactions given in
\sec{Thr:Lag}. The \w and \wzbs acquire their masses through the Higgs
mechanism which requires the presence of at least one spin-$0$ boson,
the \whb ($H$). Recently, a Higgs-like boson has been
observed~\cite{atlas.12.2,cms.13.1}, but further measurements to fully
understand its nature are needed.

The \sm requires $18$ experimentally measured parameters, excluding
the parameters for the neutrino sector which are not yet sufficiently
understood. One possible representation for these $18$ parameters is
the $9$ masses of the fundamental fermions with the neutrino masses
excluded, $3$ angles and $1$ phase describing the mixing of the quark
generations with the CKM matrix, $2$ couplings and $1$ mixing angle
describing the strengths of the three forces, and $1$ vacuum
expectation value and $1$ mass describing the Higgs sector. A
comprehensive review of most experimental measurements to date of
these parameters, as well as world averages and theoretical reviews
can be found in \rfr{pdg.12.1}.

The mathematical framework describing the interactions between the
fundamental particles of the \sm is a relativistic quantum field
theory (QFT), where particles are associated with continuous physical
fields that are invariant under the Poincar\'e group. In \sec{Thr:Sca}
the scattering matrix, which is used to calculate observables of the
\sm, is introduced. The Lagrangian densities used to construct
scattering matrices for the \sm are then provided in \sec{Thr:Lag}, as
well as the underlying gauge symmetries used to build them. In
\sec{Thr:Exp} the experimental observables that can be calculated from
the scattering matrix are presented, as well as issues arising from
these calculations. Finally, in \sec{Thr:Alt}, alternative models and
extensions to the \sm are explored. All these sections are intended to
provide a broad overview of how QFT is used with the \sm to produce
observable predictions, and is not intended as a rigorous treatment of
QFT; indeed most of the subtleties behind QFT are omitted from these
sections. However, many excellent QFT textbooks exist, including
\rfrs{bjorken.65.1}, \cite{veltman.94.1}, \cite{peskin.95.1}, and
\cite{weinberg.95.1} which are used as references for these sections.

\newsubsection{Scattering Matrix}{Sca}

In a typical high energy particle physics experiment a set of initial
particles is collided and the momentum and energy of the resultant
final particles are then measured. The experiment is repeated a large
number of times and the probability of observing a specific final
state of particles, given an initial state, is measured. The initial
particles are separated by a large length-scale, as are the final
particles, and only during the collision do the length-scales between
the particles become sufficiently small for the particles to
interact. Consequently, both the initial and final particles are
considered as free states and so using the standard Dirac notation of
\rfr{dirac.39.1} the probability of observing a final state $\ket{B}$
after the interaction of an initial state $\ket{A}$ is,
\begin{equation}
  \abs{\tensor*[_{\mathrm{+\infty}}]{\Braket{B | A}}{_{\mathrm{-\infty}}}}^2 =
  \abs{\tensor*[_{\mathrm{+\infty}}]{\Braket{b_1 \ldots b_m | a_1
        \ldots
        a_n}}{_{\mathrm{-\infty}}}}^2
  \labelequ{Probability}
\end{equation}
where $\ket{A}$ and $\bra{B}$ consist of $n$ fully specified free
particles $a_1$ through $a_n$ and $m$ free particles $b_1$ through
$b_m$, respectively. Here the initial state is at a time in the far
past, $t = -\infty$, while the final state is at a time in the far
future, $t = +\infty$, as the time-scale of the particle interactions
is very small.

To calculate \equ{Thr:Probability} either $\ket{A}$ must be evolved to
$t = +\infty$ or $\bra{B}$ to $t = -\infty$. The scattering-matrix $S$
is defined such that,
\begin{equation}
  \tensor*{\ket{\Psi}}{_{+\infty}} = S^\dagger
  \tensor*{\ket{\Psi}}{_{-\infty}} \equcomma S^\dagger S = 1
  \labelequ{S}
\end{equation}
for the state $\tensor*{\ket{\Psi}}{_t}$ at time $t$ and so the
probability of \equ{Thr:Probability} becomes,
\begin{equation}
  \abs{\tensor*[_{\mathrm{+\infty}}]{\Braket{B | A}}{_{\mathrm{-\infty}}}}^2
  = \abs{\tensor*[_{\mathrm{-\infty}}]{\Braket{B | S | A}}{_{\mathrm{-\infty}}}}^2
\end{equation}
where the final state is now also at ${t = -\infty}$. The
scattering-matrix can be divided into a non-interacting and
interacting term such that the probability amplitude can be written
as,
\begin{align*}\labelali{MatrixElement}
  \tensor*[_{\mathrm{-\infty}}]{\Braket{B | S |
      A}}{_{\mathrm{-\infty}}} =
  &\tensor*[_{\mathrm{-\infty}}]{\Braket{B | A}}{_{\mathrm{-\infty}}} +
  i (2\pi)^4\delta\left(\sum_i^n q_{a_i} - \sum_j^m q_{b_j}\right) \\
  &\me_{A \to B} \prod_i^n
  \frac{1}{\sqrt{2E_{a_i}\mathcal{V}}}
  \prod_j^m \frac{1}{\sqrt{2E_{b_j}\mathcal{V}}} 
\end{align*}
where the first term is one if the final state is the same as the
initial, and zero otherwise, and the second term describes the
interactions between the initial particles. The delta-function in the
second term imposes conservation of energy and momenta, $q$, between
the initial and final particles, while the matrix element \me provides
the interactions between the initial particles for the production of
the given final state. The normalisation is given by the two products
over $i$ and $j$, where $\mathcal{V}$ is the unit volume for the
particles with the particle energy $E$ included, as the term
$\mathcal{V}E$ is Lorentz invariant.

In the following two sections, the time-dependent perturbation and
functional integration methods for calculating \equ{Thr:Probability}
will be outlined. This is followed by the formulation of the Feynman
rules used to build \me, which can be derived from either the
time-dependent perturbation method or functional integration method.

% See soper.12.1.pdf for more details.
\newsubsubsection{Time-dependent Perturbation Theory}{}

From the definition of \equ{Thr:S} the scattering-matrix can be
interpreted as a time-evolution operator $U(t, t_0)$ where ${t_0 =
  -\infty}$ and ${t = +\infty}$ or $U(+\infty,
-\infty)$. Consequently, the calculation of the time-evolution
operator will yield the scattering-matrix. The Schr\"odinger picture
wave-function $\tensor*{\ket{\Psi_S}}{_{t_0}}$ for a state at time
$t_0$ can be evolved to an arbitrary time $t$ by,
\begin{equation}
  \tensor*{\ket{\Psi_S}}{_t} = U_S(t, t_0) \tensor*{\ket{\Psi_S}}{_{t_0}}
  \labelequ{SchrodingerU}
\end{equation}
where $U_S(t, t_0)$ is the Schr\"odinger picture time-evolution
operator with the conditions ${U_S(t_0, t_0) = 1}$ and ${U^\dagger(t,
  t_0) U(t, t_0) = 1}$. Given a time-dependent Hamiltonian in the
Schr\"odinger picture $H_S(t)$, the Schr\"odinger equation of
\rfr{schrodinger.26.1} applied to $\tensor*{\ket{\Psi_S}}{_t}$ is,
\begin{equation}
  i \frac{\partial U_S(t, t_0)
    \tensor*{\ket{\Psi_S}}{_{t_0}}}{\partial t} = H_S(t) U_S(t, t_0)
  \tensor*{\ket{\Psi_S}}{_{t_0}}
  \labelequ{SchrodingerEquation}
\end{equation}
but since $\tensor*{\ket{\Psi_S}}{_{t_0}}$ is constant then $U_S(t,
t_0)$ must fulfil,
\begin{equation}
  \frac{\partial U_S(t, t_0)}{\partial t} = -i H_S(t) U_S(t, t_0)
  \labelequ{SchrodingerDu}
\end{equation}
which, when
$H_S$ is time-independent, results in the solution, 
\begin{equation}
  U_S(t, t_0) = e^{-iH_S(t - t_0)}
  \labelequ{SchrodingerSu}
\end{equation}
where $H_S(t-t_0)$ is the product of $H_S$ and $t-t_0$, and not $H_S$
evaluated at time $t-t_0$.

When $H_S(t)$ can be split into,
\begin{equation}
  H_S(t) = {H_S}_0 + {H_S}_1(t)
  \labelequ{SchrodingerH}
\end{equation}
where ${H_S}_0$ is a Schr\"odinger picture time-independent
Hamiltonian without interaction terms and ${H_S}_1(t)$ is a
Schr\"odinger picture time-dependent Hamiltonian with interaction
terms, working in the Dirac picture rather than the Schr\"odinger
picture is oftentimes more convenient. In the Dirac picture, both the
states and observables are time-dependent, unlike the Schr\"odinger
picture where only the states are time-dependent. The transformations,
\begin{equation}
  \tensor*{\ket{\Psi_D}}{_t} = e^{i{H_S}_0t}
  \tensor*{\ket{\Psi_S}}{_t} \equcomma O_D(t) = e^{i{H_S}_0 t} O_S(t)
  e^{-i{H_S}_0 t}
  \labelequ{DiracTransform}
\end{equation}
take the state $\tensor*{\ket{\Psi_S}}{_t}$ and the operator $O_S(t)$
from the Schr\"odinger picture to the Dirac picture. Using
\equ{Thr:SchrodingerU} once, and the transformation of
\equ{Thr:DiracTransform} twice, the state $\tensor*{\ket{\Psi_D}}{_t}$
can be written as,
\begin{equation}
  \tensor*{\ket{\Psi_D}}{_t} = e^{i{H_S}_0t} U_S(t,
  t_0)\tensor*{\ket{\Psi_S}}{_{t_0}} = e^{i{H_S}_0t} U_S(t,
  t_0) e^{-i{H_S}_0t_0}\tensor*{\ket{\Psi_D}}{_{t_0}}
\end{equation}
and so,
\begin{equation}
  U_D(t, t_0) = e^{i{H_S}_0t} U_S(t,t_0) e^{-i{H_S}_0t_0}
  \labelequ{DiracU}
\end{equation}
is the Dirac time-evolution operator which takes the state
$\tensor*{\ket{\Psi_D}}{_{t_0}}$ to the state
$\tensor*{\ket{\Psi_D}}{_t}$.

The derivative of the Dirac time-evolution operator is,
\begin{align*}\labelali{DiracDu}
  \frac{\partial U_D(t, t_0)}{\partial t} &=
  e^{i{H_S}_0t} \left(\frac{\partial U_S(t,t_0)}{\partial t} \right)
  e^{-i{H_S}_0t_0} + i{H_S}_{0} e^{i{H_S}_0t} U_S(t,t_0)
  e^{-i{H_S}_0t_0} \\
  &= -i \left(e^{i{H_S}_0t} {H_S}_1(t) e^{-i{H_S}_0t}\right)
  \left(e^{i{H_S}_0t} U_S(t,t_0) e^{-i{H_S}_0t_0} \right) \\
  &= -i {H_D}_1(t) U_D(t, t_0) \phantom{\left(\frac{\partial
        U_S(t,t_0)}{\partial t} \right)} 
\end{align*}
where \equs{Thr:SchrodingerDu} and \ref{equ:Thr:SchrodingerH} are used
in the second line and \equs{Thr:DiracTransform}, and
\ref{equ:Thr:DiracU} in the third. Here, ${H_D}_1(t)$ is the
interaction Hamiltonian in the Dirac picture. \Equ{Thr:DiracDu} was
first proposed in a covariant formulation of quantum electrodynamics
by Tomonaga in \rfr{tomonaga.46.1} and Schwinger in
\rfr{schwinger.48.1}.

The power series method for solving differential equations of
\rfr{frobenius.73.1} can be used to solve \equ{Thr:DiracDu} for ${U_D(t,
  t_0)}$,
\begin{equation}
  U_D(t,t_0) = 1 + \sum_n^\infty \left( (-i)^n \int_{t_0}^t \ldots
  \int_{t_0}^{t_{n-1}} {H_D}_1(t_1') \ldots {H_D}_1(t_n') \sdif{t_n'}
  \ldots \dif{t_1'} \right)
\end{equation}
where the power series has been expanded about the Dirac picture
interaction Hamiltonian ${H_D}_1(t)$, beginning with ${n = 1}$. The
scattering-matrix in the Dirac picture can then be found by setting
${t_0 = -\infty}$ and ${t = +\infty}$,
\begin{equation}
  S_D = 1 + \sum_n^\infty \left( \frac{(-i)^n}{n!} \int_{-\infty}^{+\infty}
    \ldots \int_{-\infty}^{+\infty} T\left\{ {H_D}_1(t_1) \ldots
      {H_D}_1(t_n) \right\} \sdif{t_n}
    \ldots \dif{t_1} \right)
  \labelequ{DiracS}
\end{equation}
where the integration has been simplified by applying the
time-ordering operator $T$ to the operators ${H_D}_1(t)$ such that
${t_n > t_{n+1}}$. This expansion is the Dyson series of
\rfr{dyson.49.1} which can be used in conjunction with the contractions
of Wick's theorem from \rfr{wick.50.1} to calculate the probability of
\equ{Thr:Probability} in the Dirac picture.

% See gardi.12.1.pdf for more details.
\newsubsubsection{Functional Integration}{}

An alternative, yet equivalent method to using the time-perturbation
derived scattering-matrix of \equ{Thr:DiracS} to determine the
probability of \equ{Thr:Probability}, is the method of functional
integration first introduced by Dirac in \rfr{dirac.32.1} and more
fully realised by Feynman in \rfr{feynman.48.1}. The method of
functional integration provides several advantages to the
time-perturbation method: a clear graphical interpretation via Feynman
diagrams, a general method to determine Feynman rules for complex
interactions, a non-perturbative calculation of \equ{Thr:Probability},
and a manifestly Poincar\'e invariant formalism. However, the
mathematical derivation of the functional integration formalism is
more involved than time-perturbation theory, and so only a short
overview is given here.

In the Schr\"odinger picture the probability of \equ{Thr:Probability}
can be calculated using the probability amplitude,
\begin{equation}
  \tensor*[_{t_B}]{\Braket{B_S | U_S(t_B, t_A) | A_S}}{_{t_A}}
  \labelequ{ProbabilityAmplitude}
\end{equation}
where $U_S(t,t_0)$ is the Schr\"odinger picture time-evolution
operator given by the solution of \equ{Thr:SchrodingerDu}. The time
between $t_A$ and $t_B$ can be broken down into ${n+1}$ elements of
time $\varepsilon$, and given a complete set of $n$ coordinate-space
states $X_i$,
\begin{equation}
  \int \tensor{\ket{X_i}}{_t}\tensor[_t]{\bra{X_i}}{} \sdif{X_i}  = 1
\end{equation}
such that the probability amplitude of \equ{Thr:ProbabilityAmplitude}
can be written as,
\begin{align*}\labelali{}
  \tensor*[_{t_B}]{\Braket{B_S | U_S(t_B, t_A) | A_S}}{_{t_A}}
  \labelequ{ProbabilityAmplitude} = & \int \ldots \int
  \tensor*[_{t_B}]{\Braket{B_S | U_S(t_B, t_B - \varepsilon) |
      X_n}}{_{t_B - \varepsilon}} \ldots \\
  & \tensor*[_{t_A + \varepsilon}]{\Braket{X_1 | U_S(t_A +
      \varepsilon, t_A | A_S}}{_{t_A}}
  \sdif{X_1} \ldots \dif{X_n} 
\end{align*}
which can be thought of as the probability amplitude of $\ket{A_S}$
transitioning through the intermediate states $\ket{X_i}$ to the state
$\bra{B_S}$. In \rfr{feynman.48.1} Feynman showed that this can be
written as,
\begin{equation}
  \tensor*[_{t_B}]{\Braket{B_S | U_S(t_B, t_A) | A_S}}{_{t_A}} = \int
  \ldots \int e^{\int_{t_A}^{t_B} \int \ld \sdif{\vec{x}\sdif{t}}}
  \,\mathcal{D}X_1 \ldots \,\mathcal{D}X_n
  \labelequ{Functional}
\end{equation}
where $\mathcal{D}X_i$ indicates the functional integral over the path
$X_i$. Here $\vec{x}$ is a space three-vector and \ld is the
Lagrangian density, which following the convention of
\rfr{peskin.95.1}, will be labelled as just the Lagrangian. The result
of \equ{Thr:Functional} can be interpreted as $n$ paths contributing
equally to the probability amplitude of
\equ{Thr:ProbabilityAmplitude}, but each with a phase given by the
classical action for that path.

\newsubsubsection{Feynman Rules}{}

The paths of the functional integration method can be graphically
depicted by Feynman diagrams where each diagram represents a component
matrix element of the total matrix element $\mathcal{M}$ introduced in
\equ{Thr:MatrixElement}. The total matrix element is then the sum of
the component matrix elements. Each component matrix element also
corresponds to a term from the Dyson series of \equ{Thr:DiracS} after
applying Wick contractions. Example Feynman diagrams are given in
\fig{Thr:FeynmanDiagrams} where each line represents a fully specified
particle, which for the \sm corresponds to a given momentum and
energy, electromagnetic charge, colour charge, and spin. The diagrams
proceed from left to right, with the time axis given along the
$x$-axis and the spacial axis along the $y$-axis. Consequently, the
leftmost lines correspond to the initial state particles, and all
remaining lines reaching the edge of the diagram are final state
particles. Every Feynman diagram consists of the following three
components which are derived from the Lagrangian describing the free
particles and their interactions.

\begin{subfigures}{2}{Example Feynman diagrams for
    \subfig{LO}~leading-order, \subfig{NLO}~next-to-leading-order, and
    \subfig{NLL}~one-leg scattering processes. The $x$-axis is the
    time axis while the $y$-axis is the spacial
    axis.\labelfig{FeynmanDiagrams}}
  \fmpbeg 
  \fmp{LO}  & \fmp{NLO}    \fmpsep 
  \fmp{NLL} & \sidecaption \fmpend
\end{subfigures}

\begin{itemize}
  \settowidth{\itemindent}{external lines}
\item[{\makebox[\itemindent][l]{vertices:}}] The junction of $n$ lines
  corresponds to an interaction between the $n$ particles represented
  by the lines. Each vertex is given by an interaction term from the
  Lagrangian, where the fields contained in the term dictate the lines
  of the vertex and the remainder of the term yields the vertex
  factor. The vertex factors are translated into momentum-space from
  the Lagrangian, where each ${i \partial_\mu}$ is replaced with a
  $p_\mu$. The incoming electromagnetic charge, colour charge, energy,
  and momentum are conserved in the corresponding outgoing quantities
  of the vertex.
\item[{\makebox[\itemindent][l]{propagators:}}] All internal lines
  connecting two vertices are propagators, or virtual particles
  through which the scattering process proceeds. Each propagator is
  given by the product of $i$ and the inverse of the free field
  equation without the field term for that propagator type, derived
  from the corresponding free Lagrangian. Again, the propagator is
  translated into momentum-space.
\item[{\makebox[\itemindent][l]{external lines:}}] All external lines
  correspond to real initial and final state particles that can be
  observed. These lines represent the scalar or vector pre-factors to
  the plane-wave solutions of the free field equation for the field of
  the corresponding particle type.
\end{itemize}

The component matrix element for a Feynman diagram is then the product
of the ordered external lines, internal lines, and vertex factors. In
the diagram of \fig{Thr:LO}, a two-to-two scattering process occurs,
for which a minimum of two vertices is necessary. Any diagrams with
the minimum number of vertices are the leading-order terms of the
total matrix element, while diagrams containing the next-to-minimum
number of vertices, such as the diagram of \fig{Thr:NLO}, are the
next-to-leading order terms of the total matrix element. If an
additional final state line is added to a diagram, like the
two-to-three scattering process of \fig{Thr:NLL}, the diagram is
typically categorised as a one-leg diagram. The $n$-leg terminology is
used when explicitly calculating hard radiation from the initial or
final state with the matrix element. Any diagram without internal
loops, {\it i.e.}  \figs{Thr:LO} and \ref{fig:Thr:NLL}, are tree-level
diagrams, while a diagram with $n$ loops is an $n$-loop level diagram,
like the one-loop level diagram of \fig{Thr:NLO}.

\newsubsection{Lagrangians}{Lag}

In order to derive the Feynman rules introduced in \sec{Thr:Sca} and
calculate the matrix elements for a given theory, the Lagrangian must
be known. Beginning with the Lagrangians for the free fields of the
theory provides the rules for both the propagators and external
lines. The Euler-Lagrange equation is given for a field $\phi$ by,
\begin{equation}
  \partial_\mu \left(\frac{\partial \ld}{\partial \left(\partial_\mu
        \phi\right)}\right) - \frac{\partial \ld}{\partial \phi} = 0
  \labelequ{EulerLagrange}
\end{equation}
and is applied to the free Lagrangians to determine the equations of
motion for the free fields, which can then be translated to
propagators. The solutions to the equations of motion produce the
rules for the external lines. Within the \sm, three fundamental
particle types of spin-$0$, spin-$1/2$, and spin-$1$ have been
observed, corresponding to scalar, spinor, and vector
fields. Consequently, the free Lagrangians for these fields provide
the propagators and external lines for the \sm.

In this section, first the free Lagrangians for the \sm particles are
introduced, as well as their corresponding propagators and external
line factors. A summary of these Feynman rules are provided in
\tab{Thr:Rules}. Next, local gauge invariances of the free Lagrangian
for fermions is imposed, producing the full \sm Lagrangian,
\begin{equation}
  \ld_\sm = \ld_\qcd + \ld_\ewk + \ld_\mathrm{MHM}
\end{equation}
consisting of the quantum chromodynamics (\qcd), electroweak (\ewk),
and minimal Higgs mechanism (MHM) Lagrangians. From these Lagrangians
the vertex rules for the \sm are provided.

\newsubsubsection{Free Fields}{}

The free Lagrangian for spin-$0$ fields is given by the Klein-Gordon
equation,
\begin{equation}
  \ld_{\textrm{spin-}0} = \frac{1}{2}\left(\partial_\mu
    \phi \partial^\mu \phi - m^2\phi^\dagger\phi\right)
  \labelequ{Spin0L}
\end{equation}
where $m$ is the mass of the particle corresponding to the scalar
field $\phi$. Applying \equ{Thr:EulerLagrange} results in the equation
of motion,
\begin{equation}
  \left(\partial_\mu\partial^\mu + m^2\right)\phi = 0
  \labelequ{Spin0M}
\end{equation}
for scalar fields. Using the prescription of \sec{Thr:Sca} the scalar
propagator is,
\begin{equation}
  \frac{i}{q^2 - m^2}
  \labelequ{Spin0P}
\end{equation}
where $q$ is the four-momentum of the particle. Since the solution to
\equ{Thr:Spin0M} is just a plane-wave, the external line for a scalar
is $1$.

The free Lagrangian for spin-$1/2$ fields can be found from the Dirac
equation of \rfr{dirac.28.1},
\begin{equation}
  \ld_{\textrm{spin-}1/2} = i\bar{\psi}\gamma^\mu\partial_\mu\psi -
  m\bar{\psi}\psi
  \labelequ{Spin1L}
\end{equation}
for a spinor field $\psi$ and adjoint spinor field $\bar{\psi}$, where
$\gamma^\mu$ are the Dirac matrices. The equation of motion for the
field is then
\begin{equation}
  \left(i\gamma^\mu\partial_\mu - m\right)\psi = 0
  \labelequ{Spin1M}
\end{equation}
which produces
\begin{equation}
  \frac{i}{\gamma^\mu q_\mu-m}
  \labelequ{Spin1P}
\end{equation}
as the propagator for a spin-$1/2$ particle. The external lines for
spin-$1/2$ particles are given by the spinors $u(q,\lambda)$ for
particles and the anti-spinors $v(q,\lambda)$ for anti-particles,
where $\lambda$ is the helicity $\pm 1$ and $q$ is the momentum
four-vector. For the purposes of calculating the helicity matrix
elements necessary for \chp{Tau}, these spinors are defined using the
conventions of \rfr{hagiwara.86.1}.

In the Weyl basis the two vectors,
\begin{equation}
  \kappa(q,\lambda) = \begin{cases}
    \frac{1}{2\left(\vec{q}^2 + \abs{\vec{q}}q_z\right)}
    \begin{pmatrix}
      iq_y - q_x \\
      \abs{\vec{q}} + q_z \\
    \end{pmatrix} & \textrm{for}~\lambda=+1, ~
    \begin{pmatrix} 0 \\ 1 \\ \end{pmatrix}\phantom{-}
    \quad\textrm{as}~{q_z \rightarrow -\abs{\vec{q}}} \\
    \frac{1}{2\left(\vec{q}^2 + \abs{\vec{q}}q_z\right)}
    \begin{pmatrix}
      \abs{\vec{q}} + q_z \\
      iq_y + q_x \\
    \end{pmatrix} & \textrm{for}~\lambda=-1, ~
    \begin{pmatrix} -1 \\ 0 \\ \end{pmatrix}
    \quad\textrm{as}~{q_z \rightarrow -\abs{\vec{q}}} \\
  \end{cases}
\end{equation}
are eigenvectors to the helicity operator,
\begin{equation}
  \frac{\sigma_j q^j}{\abs{\vec{q}}} \kappa(q,\lambda) = \lambda
  \kappa(q, \lambda)
\end{equation}
where $\sigma_i$ are the Pauli matrices,
\begin{equation}
  \sigma_0 = 
  \begin{pmatrix}
    1 & 0 \\
    0 & 1 \\
  \end{pmatrix}\equcomma \sigma_1 = 
  \begin{pmatrix}
    0 & 1 \\
    1 & 0 \\
  \end{pmatrix}\equcomma \sigma_2 =
  \begin{pmatrix}
    0 & -i \\
    i &  0 \\
  \end{pmatrix}\equcomma \sigma_3 =
  \begin{pmatrix}
    1 &  0  \\
    0 & -1  \\
  \end{pmatrix}
  \labelequ{Su2G}
\end{equation}
and the index $j$ indicates the spatial components of $q$: $x$,
$y$, and $z$. The spinors and anti-spinors can then be written as,
\begin{equation}
  u(q,\lambda) = \begin{pmatrix}
    \kappa(q,\lambda)\sqrt{E - \lambda\abs{\vec{q}}} \\
    \kappa(q,\lambda)\sqrt{E + \lambda\abs{\vec{q}}} \\
  \end{pmatrix}\equcomma
  v(q,\lambda) = \begin{pmatrix}
    -\lambda\kappa(q,-\lambda)\sqrt{E + \lambda\abs{\vec{q}}} \\
    \phantom{-}\lambda\kappa(q,-\lambda)\sqrt{E - \lambda\abs{\vec{q}}} \\
  \end{pmatrix}
  \labelequ{Spin1E}
\end{equation}
where $E$ is the energy component of $q$. These solutions of
\equ{Thr:Spin1M} provide the Feynman rules for spin-$1/2$ external
lines and require the Dirac matrices to be defined as,
\begin{equation}
  \gamma^0 = 
  \begin{pmatrix}
    0 & 1 \\
    1 & 0 \\
  \end{pmatrix}\equcomma \gamma^i =
  \begin{pmatrix}
    0         & \sigma_i \\
    -\sigma_i & 0        \\
  \end{pmatrix}\equcomma \gamma_5 =
  \begin{pmatrix}
    -1 & 0 \\
    0 & 1 \\
  \end{pmatrix}
  \labelequ{DiracMatrices}
\end{equation}
where the bar of a spinor or anti-spinor is given by $\bar{u} =
u^\dagger \gamma^0$ if $u^\dagger$ is the Hermitian adjoint of $u$.

The free Lagrangian for spin-$1$ fields can be found from the Proca
equation, resulting in,
\begin{equation}
  \ld_{\textrm{spin-}1} = \left(\partial^\mu
    \omega^\nu - \partial^\nu \omega^\mu \right) \left(\partial_\mu
    \omega_\nu - \partial_\nu \omega_\mu \right) - \frac{m^2}{2}
  \omega^\nu\omega_\nu
  \labelequ{Spin2L}
\end{equation} 
for a vector field $\omega^\mu$. Applying \equ{Thr:EulerLagrange}
yields the equation of motion,
\begin{equation}
  \partial_\mu \left(\partial^\mu
    \omega^\nu - \partial^\nu \omega^\mu \right) - m^2\omega^\nu = 0
  \labelequ{Spin2M}
\end{equation}
which provides the propagator,
\begin{equation}
  \frac{-i}{q^2 - m^2} \left(g_{\mu\nu} - \frac{q_\mu q_\nu}{m^2}\right)
  \labelequ{Spin2P}
\end{equation}
for massive spin-$1$ particles. If the mass term from \equ{Thr:Spin2L}
is removed, {\it i.e.} a massless spin-$1$ particle like the photon or
gluon, and the Lorentz condition ${\partial_\mu\omega^\mu = 0}$ is
imposed, then the equation of motion becomes,
\begin{equation}
  \partial^2 \omega^\nu = 0
  \labelequ{Spin2MM}
\end{equation}
which yields the propagator,
\begin{equation}
  \frac{-ig_{\mu\nu}}{q^2}
  \labelequ{Spin2MP}
\end{equation}
for massless spin-$1$ particles.

\begin{table}\centering
  \captionabove{A summary of the Feynman rules
    for the particles of the \sm, excluding vertices but including
    the symbols used when drawing the Feynman
    diagrams.\labeltab{Rules}}
  \begin{tabular}{l|c|c|c|c|M{2cm}}
    \toprule
    & propagator
    & incoming line
    & outgoing line & equ.
    & symbols \\
    \midrule
    spin-$0$ 
    & $\frac{i}{q^2 - m^2}$ 
    & $1$ 
    & $1$ 
    &
    & \begin{tabular}{l} \begin{fmffile}{\fmppathSpin0}%
    \setlength{\unitlength}{1mm}\input{\fmppathSpin0.fmp}%
  \end{fmffile}\executeiffilenewer{\fmppathSpin0.fmp}%
  {\fmppathSpin0.1}{cd \fmppath; mpost Spin0.mp} \\ \end{tabular} \\ &&&&&\\
    spin-$1/2$
    & $\frac{i}{\gamma^\mu q_\mu-m}$ 
    & $u(q,\lambda)$, $\bar{v}(q,\lambda)$ 
    & $\bar{u}(q,\lambda)$, $v(q,\lambda)$
    & \ref{equ:Thr:Spin1E} 
    & \begin{tabular}{l} \begin{fmffile}{\fmppathSpin1}%
    \setlength{\unitlength}{1mm}\input{\fmppathSpin1.fmp}%
  \end{fmffile}\executeiffilenewer{\fmppathSpin1.fmp}%
  {\fmppathSpin1.1}{cd \fmppath; mpost Spin1.mp} \\ \end{tabular} \\ &&&&&\\
    spin-$1$ ($\m = 0$)
    & $\frac{-ig_{\mu\nu}}{q^2}$ 
    & $\varepsilon(q,\lambda)$
    & $\varepsilon^\dagger(q,\lambda)$
    & \ref{equ:Thr:Spin2E} 
    & \begin{tabular}{l} \begin{fmffile}{\fmppathSpin2Photon}%
    \setlength{\unitlength}{1mm}\input{\fmppathSpin2Photon.fmp}%
  \end{fmffile}\executeiffilenewer{\fmppathSpin2Photon.fmp}%
  {\fmppathSpin2Photon.1}{cd \fmppath; mpost Spin2Photon.mp} \\ 
      \begin{fmffile}{\fmppathSpin2Gluon}%
    \setlength{\unitlength}{1mm}\input{\fmppathSpin2Gluon.fmp}%
  \end{fmffile}\executeiffilenewer{\fmppathSpin2Gluon.fmp}%
  {\fmppathSpin2Gluon.1}{cd \fmppath; mpost Spin2Gluon.mp} \\ \end{tabular} \\ &&&&&\\
    spin-$1$ ($\m > 0$) 
    & $\frac{-i\left(g_{\mu\nu} - \frac{q_\mu q_\nu}{m^2}\right)}{q^2 - m^2}$
    & $\varepsilon(q,\lambda)$ 
    & $\varepsilon^\dagger(q,\lambda)$ 
    & \ref{equ:Thr:Spin2E}
    & \begin{tabular}{l} \begin{fmffile}{\fmppathSpin2Massive}%
    \setlength{\unitlength}{1mm}\input{\fmppathSpin2Massive.fmp}%
  \end{fmffile}\executeiffilenewer{\fmppathSpin2Massive.fmp}%
  {\fmppathSpin2Massive.1}{cd \fmppath; mpost Spin2Massive.mp} \\ \end{tabular} \\
    \bottomrule
  \end{tabular}
\end{table}

The external lines for spin-$1$ particles are given by the
polarisation vectors $\varepsilon^\mu$ which satisfy
\equ{Thr:Spin2M}. Again, to calculate the helicity matrix elements
necessary for \chp{Tau} the conventions of \rfr{hagiwara.89.1} which
are consistent with the conventions of \rfr{hagiwara.86.1} are
used. The helicity polarisation vectors are defined as,
\begin{equation}
  \varepsilon(q,\lambda) = 
  \begin{cases}
    \lambda\begin{pmatrix}
      0 \\
      \frac{q_x q_z}{\abs{\vec{q}}\qt} - \frac{iq_y}{\qt} \\ 
      \frac{q_y q_z}{\abs{\vec{q}}\qt} + \frac{iq_x}{\qt} \\ 
      \frac{-\qt}{\abs{\vec{q}}} \\
    \end{pmatrix} & \textrm{for}~\lambda=\pm1, ~
    \lambda\begin{pmatrix}
      0 \\ q_z \\ 0 \\ 0 \\
    \end{pmatrix}
    \quad\textrm{as}~\qt \rightarrow 0 \\
    \frac{1}{\m\abs{\vec{q}}}\begin{pmatrix}
      \vec{q}^2 \\ Eq_x \\ Eq_y \\ Eq_z \\
    \end{pmatrix} & \textrm{for}~\lambda=0~\textrm{and}~\m > 0 \\
  \end{cases}
  \labelequ{Spin2E}
\end{equation}
where there is no $\lambda = 0$ polarisation state for massless
spin-$1$ particles and ${\qt = \sqrt{q_x^2 + q_y^2}}$.

A summary of the Feynman rules derived from the free Lagrangians are
given in \tab{Thr:Rules} for the spin-$0$, spin-$1/2$, and spin-$1$
particles of the \sm. The only spin-$0$ \sm particle is the \whb,
while both leptons and quarks are the spin-$1/2$ \sm particles, and
photons, gluons, \wwbs, and \wzbs are the \sm spin-$1$ particles. Free
Lagrangians for particles with spins not given in \tab{Thr:Rules} can
be built using the Bargmann-Wigner equations of \rfr{bargmann.48.1},
but are not currently necessary in the \sm.

% See griffiths.08.1.pdf for more details.
\newsubsubsection{Quantum Chromodynamics}

A method for generating Lagrangians invariant under non-Abelian local
gauge theory was first proposed in \rfr{yang.54.1} by Yang and Mills,
laying the groundwork for both \qcd and unified electroweak
theory. Yang-Mills theory was then applied to \qcd in \rfr{han.65.1}
requiring the free fields for quarks to be invariant under local
\su[3] gauge transformations, leading to the conservation of three
colour charge quantum numbers. Direct evidence for three colour
charges has been observed by measuring the ratio of hadron production
to muon pair-production in electron-positron collisions, which should
be approximately $11/3$ at off resonance centre-of-mass energies below
the \wtq mass. \Rfr{ezhela.04.1} provides this measurement made using
the experiments on \dlep. Further measurements testing the underlying
group structure of \qcd have also been made by \dlep experiments and
reported in \rfr{dissertori.97.1}, with results consistent with the
\sm and \su[3] theory. Because quarks of different colours with the
same flavour are identical except for colour, the three quark colour
fields can be written as a colour triplet,
\begin{equation}
  q_f = \begin{pmatrix} {q_f}_r \\ {q_f}_b \\ {q_f}_g \\ \end{pmatrix}
  \labelequ{Su3F}
\end{equation}
where $f$ is one of the six flavours of the quark, and $r$, $b$, and
$g$ are the colour charges. Substituting $q_f$ for $\psi$ into the
free Lagrangian for a spin-$1/2$ particle, \equ{Thr:Spin1L}, results
in the free Lagrangian for the quark fields.

This Lagrangian is invariant under global \un[3] transformations of
the field,
\begin{equation}
  U q_f = e^{i\theta}e^{i\theta_a \lambda^a} q_f
\end{equation}
where the first exponential is a \un[1] transformation and the second
exponential is an \su[3] transformation. The \un[1] transformation is
given by a phase $\theta$ while the \su[3] transformation is given by
eight phases $\theta_a$ and Gell-Mann matrices
$\lambda^a$. The Gell-Mann matrices are,
\begin{alignat*}{2}\labelali{Su3G}
  &\lambda^1 = 
  \begin{pmatrix} 0 & 1 & 0 \\ 1 & 0 & 0 \\ 0 & 0 & 0 \\ \end{pmatrix}
  \equcomma && \lambda^2 = 
  \begin{pmatrix} 0 & -i & 0 \\ i & 0 & 0 \\ 0 & 0 & 0 \\ \end{pmatrix}
  \equcomma \lambda^3 = 
  \begin{pmatrix} 1 & 0 & 0 \\ 0 & -1 & 0 \\ 0 & 0 & 0 \\ \end{pmatrix}, \\
  &\lambda^4 = 
  \begin{pmatrix} 0 & 0 & 1 \\ 0 & 0 & 0 \\ 1 & 0 & 0 \\ \end{pmatrix}
  \equcomma && \lambda^5 = 
  \begin{pmatrix} 0 & 0 & -i \\ 0 & 0 & 0 \\ i & 0 & 0 \\ \end{pmatrix}
  \equcomma \lambda^6 = 
  \begin{pmatrix} 0 & 0 & 0 \\ 0 & 0 & 1 \\ 0 & 1 & 0 \\ \end{pmatrix}, \\
  &\lambda^7 = 
  \begin{pmatrix} 0 & 0 & 0 \\ 0 & 0 & -i \\ 0 & i & 0 \\ \end{pmatrix}
  \equcomma && \lambda^8 = \frac{1}{\sqrt{3}} 
  \begin{pmatrix} 1 & 0 & 0 \\ 0 & 1 & 0 \\ 0 & 0 & -2 \\ \end{pmatrix}
\end{alignat*}
where their commutation relations are,
\begin{equation}
  \left[\lambda^a, \lambda^b \right] = i2f^{abc}\lambda^c
  \labelequ{Su3F}
\end{equation}
and $f^{abc}$ are the $512$ anti-symmetric \su[3] structure
constants.

Requiring that the free Lagrangian for the quark fields remains
invariant under a local \su[3] transformation,
$e^{i\theta_a(x)\lambda^a}$, necessitates the introduction of eight
vector fields $G^\mu$, corresponding to the eight gluons of the \sm,
by replacing $\partial_\mu$ in \equ{Thr:Spin1L} with,
\begin{equation}
  D_\mu = \left( \partial_\mu + \frac{i\gS}{2} \lambda_a G^a_\mu
  \right)
  \labelequ{Thr:Su3D}
\end{equation}
the \qcd covariant derivative. Including the free Lagrangian for the
gluon fields, given by \equ{Thr:Spin2L} without the mass term, results
in the Lagrangian,
\begin{align*}\labelali{QcdL}
  \ld_\qcd =
  &
  % Free gluon.
  \termlabel{-\frac{1}{2}\left(\partial^\mu G_a^\nu - \partial^\nu
      G_a^\mu\right)\left(\partial_\mu G_\nu^a - \partial_\nu
      G_\mu^a\right)}
  {\equs{Thr:Spin2MP}, \ref{equ:Thr:Spin2E}}
  % G2QQ interaction.
  \termlabel{-\frac{i\gS}{2}\bar{q}_i^f \gamma^\mu \lambda_{ij}^a q_f^j G_\mu^a}
  {\fig{Thr:G2QQ}}
  \lgdsep &
  % G2GG interaction.
  \termlabel{-i\gS f^{abc}\left(\partial^\mu G_a^\nu - \partial^\nu
      G_a^\mu\right) G_\mu^b G_\nu^c}
  {\fig{Thr:G2GG}}
  % GG2GG interaction.
  \termlabel{-\frac{i}{4}\gS[^2] f^{abc} f^{adc} G_\mu^b G_\nu^c
    G_\mu^d G_\nu^e}
  {\fig{Thr:GG2GG}}
\end{align*}
for \qcd, where the Feynman diagram for each interaction term is given
in \fig{Thr:Vertices.Qcd}. The first term is the gluon propagator, the
second term is a coupling of the gluon with two quarks, the third term
a coupling of three quarks, and the fourth term a coupling of four
quarks. Here, \gS is the strong coupling constant. In \equ{Thr:QcdL}
the free Lagrangian for the quark fields has been explicitly omitted,
as this Lagrangian will be included in the electroweak Lagrangian for
all fermions. However, when this term is included, the \qcd Lagrangian
is invariant under local \su[3] gauge transformations.

\begin{subfigures}{3}{\qcd vertices for the \subfig{G2QQ}~gluon
    with quarks, \subfig{G2GG}~cubic gluon, and
    \subfig{GG2GG}~quartic gluon couplings.\labelfig{Vertices.Qcd}}
  \fmpbeg
  \fmp{G2QQ}    & \fmp{G2GG}  & \fmpsep
  \sidecaption  & \fmp{GG2GG} & \fmpend
\end{subfigures}

Because the \su[3] colour group is non-Abelian, {\it i.e.} the
Gell-Mann matrices of \equ{Thr:Su3G} do not commute, as evidenced by
non-zero structure constants $f^{abc}$ in \equ{Thr:Su3F}, cubic and
quartic self-interacting gluon terms are introduced into the
Lagrangian. These terms in conjunction with the number of quark
flavours, $N_f$, and colour charges, $N_c$, dictate the range of the
force. If $2N_f - 11N_c$ is greater than $0$ the strong force
increases at small length-scales, otherwise the strong force decreases
at small length-scales~\cite{griffiths.08.1}. The number of quark
flavours in the \sm is $6$ and the number of colours $3$, and so the
strong force decreases at small length-scales or high energies. This
phenomena, or asymptotic freedom, was first proposed in
\rfrs{gross.73.1} and \cite{politzer.73.1} and allows for the
perturbative calculation of interactions involving the strong force at
high energies using the methods of \sec{Thr:Sca}. Conversely, at large
length-scales and low energies the strong force becomes large,
resulting in the confinement of quarks in colourless states, {\it
  i.e.} hadrons. While confinement has been observed both in
experiment and in lattice \qcd calculations, see \rfr{haymaker.98.1},
no theoretical proof accepted by the particle physics community has
yet been made, primarily due to the non-perturbative nature of this
energy regime.

% See pich.07.1.pdf for more details.
\newsubsubsection{Electroweak Theory}{}

A unification of the electromagnetic and weak forces was first
proposed by Glashow in \rfr{glashow.61.1}, where the masses of the
fermions and gauge bosons are neglected, and a local
\un[2] gauge invariance is required. The quark
and lepton fields can be decomposed into left-handed and right-handed
components via the chirality operators,
\begin{equation}
  \psi_L = \frac{1 - \gamma^5}{2}\psi\equcomma
  \psi_R = \frac{1 + \gamma^5}{2}\psi
  \labelequ{Chirality}
\end{equation}
where $\psi_L$ indicates a left-handed field and $\psi_R$ a
right-handed field. These two types of fields are assigned a weak
isospin quantum number $T$, which is $1/2$ for left-handed fields and
$0$ for right-handed fields. The third component of isospin, $T_3$, is
$+1/2$ for the left-handed neutrino and $u$-type quark fields, and
$-1/2$ for the charged leptons and $d$-type quark fields. Both the
left-handed and right-handed fields are also assigned a weak
hypercharge quantum number, $Y$.

The left-handed fields can be combined into weak isospin doublets,
\begin{equation}
  t_f = \begin{pmatrix} \nu_f \\ \lep_f \\ \end{pmatrix}_L ,~
  \begin{pmatrix} u_f \\ d_f' \\ \end{pmatrix}_L
  \labelequ{Su2F}
\end{equation}
each consisting of a ${T_3 = +1/2}$ and ${T_3 = -1/2}$ left-handed
field. Here, the left-handed fields $\nu_f$ and $u_f$ correspond to
neutrinos and $u$-type quarks of generation $f$ and ${T_3 = +1/2}$,
while the left-handed fields $\lep_f$ and $d_f'$ correspond to charged
leptons and weak eigenstate $d$-type quarks of generation $f$ and
${T_3 = -1/2}$.

The field $d_f'$ is not an observable mass eigenstate, but rather a
flavour eigenstate that is a superposition of the mass eigenstates
$d_f$. The quark flavour eigenstates are related to their mass
eigenstates by the Cabibbo-Kobayashi-Maskawa (CKM) matrix $V$ of
\rfr{kobayashi.72.1},
\begin{equation}
  \begin{pmatrix} d' \\ s' \\ b' \\ \end{pmatrix} = 
  \begin{pmatrix}
    c_{12} c_{13}
    & s_{12}c_{13}
    & s_{13} e^{-i\delta} 
    \\ -s_{12}c_{23} -
    c_{12}s_{23}s_{13}e^{i\delta}
    & c_{12}c_{23} -
    s_{12}s_{23}s_{13}e^{i\delta}
    & s_{23}c_{13}
    \\ s_{12}s_{23} - c_{12}c_{23}s_{13}e^{i\delta}
    & -c_{12}s_{23} - s_{12}c_{23}s_{13}e^{i\delta}
    & c_{23}c_{13} \\
  \end{pmatrix}
  \begin{pmatrix} d \\ s \\ b \\ \end{pmatrix}
  \labelequ{Ckm}
\end{equation}
where $V$ is fully specified by the experimentally determined mixing
angles $\theta_{12}$, $\theta_{13}$, and $\theta_{23}$, and the
\cp-violating phase angle $\delta$. Here, $c_{ij}$ indicates
$\cos\theta_{ij}$ and $s_{ij}$ indicates $\sin\theta_{ij}$.

Unlike the left-handed fields, the right-handed fields have a weak
isospin of zero and must be written as weak hypercharge singlets,
\begin{equation}
  y_f = \begin{pmatrix} \lep_f \end{pmatrix}_R ,~
  \begin{pmatrix} u_f \end{pmatrix}_R ,~
  \begin{pmatrix} d_f \end{pmatrix}_R
  \labelequ{U1F}
\end{equation}
where the neutrinos have been assumed to be massless and only
left-handed. Evidence for neutrino oscillations, see {\it e.g.}
\rfr{cleveland.98.1}, indicates the neutrinos must have mass with
eigenstates $\nu_1$, $\nu_2$, and $\nu_3$, but indirect measurements
constrain the sum of these three masses with an upper limit of
approximately ${0.5~\ev}$~\cite{pdg.12.1}. In this thesis the
neutrinos are assumed to be massless, but the \sm can be extended to
include either Majorana or Dirac mass terms for the
neutrinos~\cite{bilenky.80.1}.

Substituting the weak isospin doublets $t_f$ of \equ{Thr:Su2F} and
weak hypercharge singlets $y_f$ of \equ{Thr:U1F} into the free
Lagrangian of \equ{Thr:Spin1L} for a spin-$1/2$ field $\psi$, without
the mass term, provides the free Lagrangians for $t_f$ and $y_f$. The
weak isospin doublet Lagrangian is invariant under global \un[2]
transformations of the field,
\begin{equation}
  U t_f = e^{i\theta} e^{\frac{i}{2}\theta_j\sigma^j} t_f
  \labelequ{U2D}
\end{equation}
where the first exponential is a \un[1] transformation with phase
$\theta$, and the second exponential is a \su[2] transformation with
three phases $\theta_j$ and the Pauli matrices $\sigma^j$ of
\equ{Thr:Su2G}. The weak hypercharge singlet Lagrangian is also
invariant under,
\begin{equation}
  U y_f = e^{i\theta} y_f
\end{equation}
or global \un[1] transformations of the fields $y_f$. Requiring the
free Lagrangians to be invariant under local transformations of these
types with $\theta(x)$ and $\theta_i(x)$ necessitates the replacement
of $\partial_\mu$ with the covariant derivative,
\begin{equation}
  D_\mu = \left(\partial_\mu + \frac{ig_1Y_f}{2}B_\mu +
    ig_2\abs{{T_3}_f} \sigma_iW^i_\mu\right)
  \labelequ{U2D}
\end{equation}
for the free Lagrangians of $y_f$ and $t_f$, where $Y_f$ is the weak
hypercharge for fermion $f$. Here, one vector field $B_\mu$ and three
vector fields $W_\mu^i$ with gauge coupling strengths $g_1$ and $g_2$,
respectively, have been introduced. These four fields can be
transformed to physical fields by,
\begin{equation}
  \begin{pmatrix} W_\mu^+ \\ W_\mu^- \\ Z_\mu \\ A_\mu
    \\ \end{pmatrix} =
  \begin{pmatrix}
    \frac{1}{\sqrt{2}} & \frac{i}{\sqrt{2}} & 0 & 0 \\
    \frac{1}{\sqrt{2}} & \frac{-i}{\sqrt{2}} & 0 & 0 \\
    0 & 0 & \costw & -\sintw \\
    0 & 0 & \sintw & \costw  \\
  \end{pmatrix}
  \begin{pmatrix} W_\mu^1 \\ W_\mu^2 \\ W_\mu^3 \\ B_\mu \\ \end{pmatrix}
\end{equation}
where $A_\mu$ is the field for a photon, $W_\mu^{\pm}$ for \wwbs, and
$Z_\mu$ for \wzbs, and the couplings are related by ${g_1 = g_e /
  \costw}$ and ${g_2 = g_w = g_e / \sintw}$. The fermion charge can be
written as,
\begin{equation}
  Q = T_3 + Y/2
  \labelequ{Charge}
\end{equation}
in terms of the third component of weak isospin $T_3$ and the weak
hypercharge $Y$.

Introducing the covariant derivatives of \equ{Thr:U2D} into the free
Lagrangians for $y_f$ and $t_f$, the Lagrangian for unified
electroweak theory $\ld_\ewk$ can then be written, and is supplied in
\equ{Tvr:EwkL} of \sap{Tvr:Ewk} due to its length. However, the vertex
factors from the interactions of the electroweak Lagrangian, used in
\chp{Tau}, are given in \fig{Thr:Vertices.Ewk}. The vector and axial
couplings of the fermions with the \wzb are given in
\tab{Thr:VaCouplings}.

\begin{subtables}{2}{Vector and axial couplings of the fermions with
    the \wzb used in the vertex of
    \fig{Thr:Z2FF}.\labeltab{VaCouplings}}
  \setlength{\tabcolsep}{\oldtabcolsep}
  \begin{tabular}{l|c|r}
    \toprule
    fermion & $v$ & \multicolumn{1}{c}{$a$} \\
    \midrule
    $\nu_e$, $\nu_\mu$, $\nu_\tau$ & $1$ & $1$ \\
    $e$, $\mu$, $\tau$            & $\sintw{^2} - 1$ & $-1$ \\
    $u$, $c$, $t$                 & $1 - \frac{8}{3}\sintw{^2}$ & $1$ \\
    $d$, $s$, $b$                 & $\frac{4}{3}\sintw{^2} - 1$ & $-1$ \\
    \bottomrule
  \end{tabular}
  \setlength{\tabcolsep}{\newtabcolsep} & \sidecaption \\
\end{subtables}

\begin{subfigures}[p]{2}{Electroweak vertices for the
    \subfig[Gm2FF]{W2QQ}~gauge bosons with fermions,
    \subfig[Gm2WW]{Z2WW}~cubic gauge boson, and
    \subfig[WW2WW]{WW2GmZ}~quartic gauge boson
    couplings.\labelfig{Vertices.Ewk}}
  \fmpbeg
  \begin{tabular}{M{0.9\tabcolwidth}M{0.9\tabcolwidth}M{0pt}}
    \fmp{Gm2FF}   & \fmp{Z2FF} \fmpsep
    \fmp{W2LL}    & \fmp{W2QQ} \fmpsep
    \fmp{Gm2WW}   & \fmp{Z2WW} \fmpsep
    \fmp{WW2WW}   & \fmp{WW2ZZ} \fmpsep
    \fmp{WW2GmGm} & \fmp{WW2GmZ}
  \end{tabular} \fmpend
\end{subfigures}

% See pich.07.1.pdf for more details.
\newsubsubsection{Higgs Mechanism}{}

The local \un[2] gauge symmetry of unified electroweak theory is
broken with the introduction of mass terms for the fermions and
bosons. However, Weinberg and Salam in \rfrs{weinberg.67.1} and
\cite{salam.68.1} introduced mass terms into the electroweak
Lagrangian via the Higgs mechanism of \rfrs{englert.64.1},
\cite{higgs.64.1}, and \cite{guralnik.64.1}, which allows for
spontaneous symmetry breaking of electroweak theory. Consider a
scalar weak isospin doublet with,
\begin{equation}
  h = \begin{pmatrix} h_+ \\ h_0 \\ \end{pmatrix} = \frac{1}{\sqrt{2}}
  \begin{pmatrix} h_3 + ih_4 \\ h_1 + ih_2 \\ \end{pmatrix}
  \labelequ{HiggsF}
\end{equation}
which is made up of four real scalar fields $h_1$ through $h_4$ that
can be written in terms of two complex scalar fields $h_+$ and
$h_0$. Let both complex scalar fields $h_+$ and $h_0$ have hypercharge
$+1$. Then using \equ{Thr:Charge}, the upper field has charge $+1$
while the lower field has charge $0$, hence the $+$ and $0$ subscript
notation. The free Lagrangian for this doublet is then given by
substituting $h$ for $\phi$ into \equ{Thr:Spin0L} and changing
$\partial_\mu$ to the covariant derivative $D_\mu$ given by
\equ{Thr:U2D}. Furthermore, let the mass term $m^2\phi^\dagger\phi/2$
be replaced with the potential,
\begin{equation}
  V = \mu^2\phi^\dagger\phi + \lambda\left(\phi^\dagger \phi\right)^2
  \labelequ{SmPotential}
\end{equation}
where $\mu^2$ and $\lambda$ are free real parameters. This potential
as a function of the norm of the scalar field is plotted in
\fig{Thr:Potential}. Here, the potential is given for the four
possible sign combinations of the parameters $\mu^2$ and $\lambda$. As
can be seen, for a stable potential $\lambda$ must be greater than
zero, and for a non-zero potential minimum, $\mu^2$ must be less than
zero.

\begin{subfigures}{2}{The potential $V$ of \equ{Thr:SmPotential} as
    a function of the complex scalar field $\phi$. The units on $V$
    are in terms of $4\mu^4/\lambda$ while the units on $\abs{\phi}$
    are given in terms of $\abs{\mu}/\lambda$.}
  \svgbeg
  \svg[1]{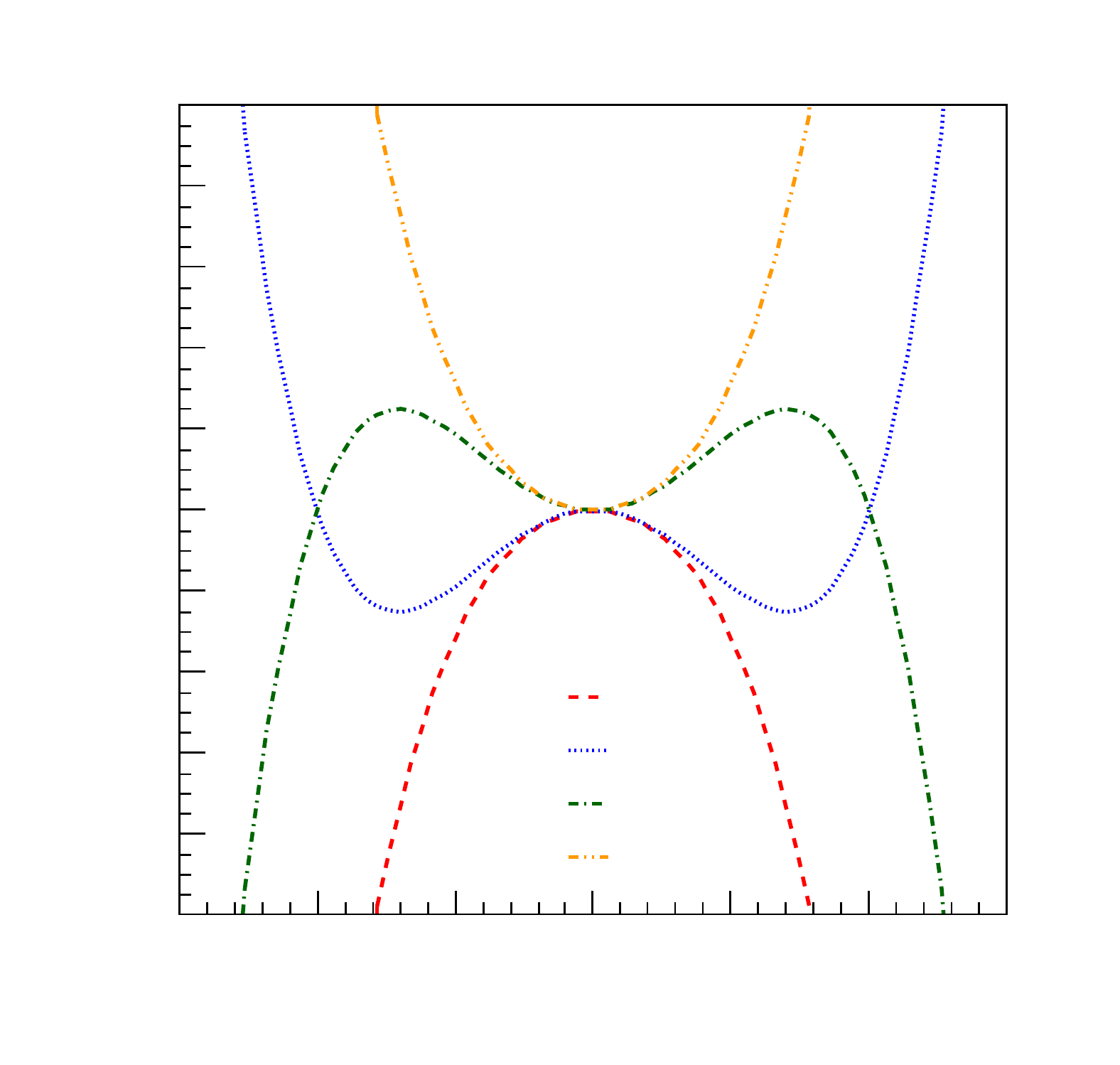} & \sidecaption \svgend
\end{subfigures}

This choice of signs for the parameters $\mu^2$ and $\lambda$ results
in the minimum potential,
\begin{equation}
  V_0 = -\frac{4\mu^4}{\lambda} = -\frac{\lambda v^4}{4}
\end{equation}
for the complex scalar field doublet of \equ{Thr:HiggsF} at,
\begin{equation}
  \abs{h} = \sqrt{\frac{-\mu^2}{\lambda}} = \frac{v}{\sqrt{2}}
\end{equation}
where $v = \sqrt{-\mu^2/\lambda}$ is the vacuum expectation value for
the ground state of $h$. The symmetry of the potential can then be
spontaneously broken by choosing the unitary gauge, ${h_2 = h_3 = h_4
  = 0}$, where $h_1$ is the only non-zero field. \Equ{Thr:HiggsF} can
then be written as,
\begin{equation}
  h = \frac{1}{\sqrt{2}} \begin{pmatrix} 0 \\ v + H \\ \end{pmatrix}
\end{equation}
where $h_1$ has been expanded as the neutral scalar field $H$ about
the vacuum expectation value $v$.

The Lagrangian density of the Higgs field $h$, using the unitary
gauge, results in the terms,
\begin{equation}
  -\frac{v^2\gW[^2]}{4}W_\mu^+W^{-\mu}\equcomma
  -\frac{v^2\gW[^2]}{8\costw[^2]}Z_\mu Z^{\mu}\equcomma
  -\frac{\lambda v^2}{2} HH
\end{equation}
where the first two terms are from introducing the covariant
derivative of \equ{Thr:U2D} and the last term is from the potential of
\equ{Thr:SmPotential}. These terms, however, are just mass terms for the
\w, \z, and \whbs where,
\begin{equation}
  \m_\w = \frac{v\gW}{2}\equcomma
  \m_\z = \frac{v\gW}{2\costw}\equcomma
  \m_\hH = v\sqrt{\lambda}
\end{equation}
gives their masses in terms of \gW, \costw, and $v$. Notice that
before requiring the unitary gauge the Higgs doublet $h$, as well as
the massless fields $W_\mu^i$ and $B_\mu$, constitute twelve free
fields as the four scalar fields of $h$ take on only one polarisation
state, while the massless vector fields $W_\mu^i$ and $B_\mu$ take on
two polarisation states each. After choosing the unitary gauge there
are still twelve fields; three scalar fields have been eliminated but
three of the massless vector fields now are massive, taking on an
additional polarisation state each.

The vacuum expectation value $v$ can be measured from the decays of
muons assuming a four-point Fermi function, and so the masses of both
the \w and \wzbs are fully predicted, whereas the mass of the \whb
remains a free variable which must be measured experimentally. Yukawa
couplings of the \whb with the fundamental fermions of the form
${-\m_f^2\bar{\psi}_f\psi_f}$ can also be added to the Higgs
Lagrangian without breaking the underlying symmetry and provide mass
terms for the fermions~\cite{gunion.90.1}. Note that all of these
masses are also free parameters, and must be measured
experimentally. The full Higgs Lagrangian $\ld_\mathrm{MHM}$ is
extensive, like the electroweak Lagrangian, and consequently is also
supplied in \equ{Tvr:MhmL} of \sap{Tvr:Mhm}. However, the vertices for
the interaction terms, used in \chp{Tau}, are provided in
\fig{Thr:Vertices.Sm.Higgs}. Further details on \whb phenomenology are
explored in \chp{Hig}.

\begin{subfigures}{3}{\sm \whb vertices for the
    \subfig[H02H0H0]{H02ZZ}~cubic gauge boson couplings,
    \subfig[H0H02H0H0]{H0H02ZZ}~quartic gauge boson couplings, and
    \subfig{H02FF}~coupling with fermions. The vertices follow the
    conventions of \rfr{gunion.90.1}.\labelfig{Vertices.Sm.Higgs}}
  \fmpbeg
  \fmp{H02H0H0}   & \fmp{H02WW}   & \fmp{H02ZZ}   \fmpsep
  \fmp{H0H02H0H0} & \fmp{H0H02WW} & \fmp{H0H02ZZ} \fmpsep
  \fmp{H02FF}     & \multicolumn{2}{M{2\tabcolwidth}}\sidecaption \fmpend
\end{subfigures}

\newsubsection{Experimental Observables}{Exp}

The theory of \secs{Thr:Sca} and \ref{sec:Thr:Lag} provides a method
for calculating the probability of observing a free initial state
transitioning into a free final state. Consequently, to test the
theory, these probabilities must be related to viable experimental
measurements. Colliding single particles and observing the outcome is
experimentally challenging, and so oftentimes bunches of particles are
collided and the result is measured. In this type of experiment, the
cross-sections, either for all possible momentum configurations of a
specific final state or differentially with respect to some
experimental observable, are measured. From these measurements more
complex analyses can be applied to extract measurements of the theory.

Within this section the cross-section is defined, as well as the decay
widths for particles. The technical difficulties in calculating the
perturbative scattering-matrix used for higher order predictions of
cross-sections and decay widths are then introduced, as well as an
overview on the methods used to overcome these difficulties. Finally,
a procedure for calculating cross-sections from bound states such as
the proton is given.

% See anastasiou.10.1.pdf and anastasiou.10.2.pdf for more details.
\newsubsubsection{Cross-sections and Decay Widths}{}

In experimental particle physics either the scattering of two
particles or the decay of a single particle is typically
measured. Both of these measurements can be theoretically described
using the probability from \equ{Thr:Probability}. From
\equ{Thr:MatrixElement} the differential probability, with respect to
the momenta of the outgoing particles, for observing a transition from
the initial state $\ket{A}$ to a different final state $\ket{B}$ per
unit time $\mathcal{T}$, is given by,
\begin{align*}\labelali{}
  \dif{\mathcal{W}_{A \to B}} &=
  \frac{\abs{\tensor*[_{\mathrm{-\infty}}]{\Braket{B | S |
          A}}{_{\mathrm{-\infty}}}}^2}{\mathcal{T}} \prod_{i}^m
  \frac{\mathcal{V}}{(2\pi)^3}\sdif{\vec{q}} \\
  &= \frac{\mathcal{V}^{1-n}}{(2\pi)^{3m-4}} \delta \left(\sum_i^n
    q_{a_i} - \sum_j^m q_{b_j} \right) \abs{\me_{A \to B}}^2 \prod_i^n
  \frac{1}{2E_{a_i}}
  \prod_j^m \frac{\dif{\vec{q}_{b_j}}}{2E_{b_j}} 
\end{align*}
where the square of the delta-function yields $\delta(q)^2 =
\mathcal{V}\mathcal{T}\delta(q)$. Integrating over all final state
momenta provides,
\begin{equation}
  \mathcal{W}_{A \to B} = \frac{\mathcal{V}^{1-n}}{(2\pi)^{3m-4}} \prod_i^n
  \frac{1}{2E_{a_i}}
  \int \ldots \int \delta \left(\sum_i^n
    q_{a_i} - \sum_j^m q_{b_j} \right)  \abs{\me_{A \to B}}^2 \prod_j^m
  \frac{\dif{\vec{q}_{b_j}}}{2E_{b_j}} 
\end{equation}
or the transition probability per unit time.

% See veltman.94.1.pdf, page 57, for cross-section definition.
The cross-section can then be defined as,
\begin{equation}
  \sigma_{A \to B} = \frac{\mathcal{W}_{A \to B}}{\mathcal{F}}
\end{equation}
which is the transition probability per unit time over the particle
flux $\mathcal{F}$. Consider observing the process of two initial
state particles $a_1$ and $a_2$ scattering into $m$ final state
particles $b_j$ in the centre-of-mass frame ${\vec{q}_{a_1} =
  -\vec{q}_{a_2}}$. The particle flux is given by,
\begin{equation}
  \mathcal{F} = \frac{\abs{\vec{v}_{a_1} -
      \vec{v}_{a_2}}}{\mathcal{V}} = \left(E_{a_1} + E_{a_2}\right)
  \frac{\abs{\vec{q}_{a_1}}}{\mathcal{V}E_{a_1}E_{a_2}} =
  \sqrt{(q_{a_1}q_{a_2})^2 - (m_{a_1}m_{a_2})^2}
  \frac{1}{\mathcal{V}E_{a_1}E_{a_2}}
\end{equation}
and so the cross-section for this two-to-$m$ process is,
\begin{align*}\labelali{CrossSection}
  \sigma_{a_1a_2 \to B} =\:
  &\frac{1}{\mathcal{S}!4(2\pi)^{3m-4}\sqrt{(q_{a_1}q_{a_2})^2 -
        (m_{a_1}m_{a_2})^2}} \\
  & \int \ldots \int \delta \left(q_{a_1} + q_{a_2} - \sum_j^m q_{b_j}
  \right) \abs{\me_{a_1a_2 \to B}}^2 \prod_j^m
  \frac{\dif{\vec{q}_{b_j}}}{2E_{b_j}} 
\end{align*}
where $\mathcal{S}$ is the number of sets of identical particle types
in the final state. Both the normalisation and integral of the
cross-section are Lorentz invariant, and so the cross-section itself
must also be Lorentz invariant. The cross-section has units of area
which are typically given in barns where ${1~\mathrm{b} =
  10^{-28}~\mathrm{m}^2}$.

In a typical scattering experiment the cross-section for a specific
two-to-$m$ scattering process, given by \equ{Thr:CrossSection}, is
determined by
\begin{equation}
  \sigma_{a_1a_2 \to B} = \frac{N}{\lum}
\end{equation}
where $N$ is the number of scattering events observed, and \lum is the
integrated luminosity for the experiment. The luminosity is defined as
the number of particles per unit area and time, so the
integrated luminosity is the number of particles per unit area. The
luminosity can be written as,
\begin{equation}
  \frac{\partial \lum}{\partial t} = \rho
  \abs{\vec{v}_{a_1} - \vec{v}_{a_2}} = \frac{N\abs{\vec{v}_{a_1} -
      \vec{v}_{a_2}}}{\mathcal{V}} = N\mathcal{F}
  \labelequ{Luminosity}
\end{equation}
where $\rho$ is the particle density.

Following a similar process to the cross-section formula
determination, the decay width or decay rate for a particle $a_1$
decaying into $m$ particles $b_j$ can be written as,
\begin{equation}
  \Gamma_{a_1 \to B} = 
  \frac{1}{2E_{a_1}(2\pi)^{3m - 4}} \int \ldots \int \delta
  \left(q_{a_1} - \sum_j^m q_{b_j}
  \right) \abs{\me_{a_1 \to B}}^2 \prod_j^m
  \frac{\dif{\vec{q}_{b_j}}}{2E_{b_j}}
  \labelequ{DecayWidth}  
\end{equation}
which is not Lorentz invariant. Typically, the decay width is defined
in the rest frame of the decaying particle and so ${E_{a_1} =
  m_{a_1}}$. Conceptually, the decay width is not consistent with the
definition of the scattering matrix because an unstable particle
cannot be a free state, but the optical theorem, see {\it e.g.}
\rfr{peskin.95.1}, validates the relation of
\equ{Thr:DecayWidth}. Decay widths are usually given in units of
\ev and are related directly to the mean lifetime of the particle
by $1/\Gamma$. Consequently, decay widths can be experimentally
determined by measuring the mean lifetimes of particles at rest.

\newsubsubsection{Renormalisation}{}

One of the issues in comparing theory and experiment is determining
theoretical predictions to the same level of precision as the
experimental results. Because predictions such as cross-sections must
be calculated perturbatively, the precision of the theoretical
prediction is dependent not only on the order at which it was
calculated, but also on the rate of convergence of the perturbation
series. Consider the example from \rfr{griffiths.08.1} for the
$t$-channel electron-muon scattering cross-section calculated at
leading order using only the diagram of \fig{Thr:MuE2MuE.LO} and at
next-to-leading order using both the diagram of \fig{Thr:MuE2MuE.LO}
and of \fig{Thr:MuE2MuE.NLO}.

\begin{subfigures}{2}{Feynman diagrams for $t$-channel muon-electron
    scattering at \subfig{MuE2MuE.LO}~leading-order, and
    \subfig{MuE2MuE.NLO}~next-to-leading-order with one electron loop,
    corresponding to vacuum polarisation.\labelfig{Renormalisation}}
  \fmpbeg 
  \fmp{MuE2MuE.LO} & \fmp{MuE2MuE.NLO} \fmpend
\end{subfigures}

The matrix element for the leading-order calculation can be determined
by applying the Feynman rules of \tab{Thr:Rules} and the interaction
vertex of \fig{Thr:Gm2FF} to \fig{Thr:MuE2MuE.LO}, resulting in,
\begin{equation}
  \me_{\mu e \to \mu e}^\lo = -\gE[^2]\left(\bar{u}(q_3,\lambda_3)\gamma^\mu
    u(q_1,\lambda_1)\right) \frac{g_{\mu\nu}}{q^2}
  \left(\bar{u}(q_4,\lambda_4)\gamma^\nu
    u(q_2,\lambda_2)\right)
  \labelequ{RenormLO}
\end{equation}
which is finite. Consequently, when \equ{Thr:RenormLO} is used in
\equ{Thr:CrossSection}, the resulting cross-section is finite. The
matrix element from the next-to-leading order diagram of
\fig{Thr:MuE2MuE.NLO} is given by,
\begin{align*}\labelali{RenormNLO}
  \me_{\mu e \to \mu e}^\nlo &= -\gE[^2]
  \left(\bar{u}(q_3,\lambda_3)\gamma^\mu
    u(q_1,\lambda_1)\right) \mathcal{I}_{\mu\nu}
  \left(\bar{u}(q_4,\lambda_4)\gamma^\nu
    u(q_2,\lambda_2)\right) \\
  \mathcal{I}_{\mu\nu} &= \frac{i\gE[^2]}{q^4}\int
  \frac{\textrm{Tr}\left(\gamma_\mu(\slashed{q}_0 +
      m_e)\gamma_\nu(\slashed{q}_0 - \slashed{q} +
      m_e)\right)}{(2\pi)^4\left(q_0^2 - m_e^2\right)\left((q_0 - q)^2
      - m_e^2)\right)} \sdif{q_0}
\end{align*}
where the integral $\mathcal{I}_{\mu\nu}$ is due to the internal loop
and ${\slashed{q} \equiv q^\mu \gamma_\mu}$ is Dirac slash
notation. The trace, $\textrm{Tr}$, in the integral arises from
applying Casimir's trick~\cite{casimir.33.1} to sum over all spin
states of the internal loop. This matrix element is not finite; the
integral $\mathcal{I}_{\mu\nu}$ between the two fermion lines diverges
and approaches $\ln\abs{q_0}$ as $\abs{q_0}$ approaches
infinity. Consequently, a cross-section calculated with this matrix
element included is also not finite. This type of divergence, when the
momentum approaches infinity, is an ultraviolet divergence, as opposed
to an infrared divergence which occurs when the momentum approaches
zero.

The divergent integral of \equ{Thr:RenormNLO} can be rewritten, or
regularised, in such a way that it diverges for only a single cut-off
parameter $\Lambda$ as,
\begin{equation}
  \mathcal{I}_{\mu\nu} =
  -g_{\mu\nu} q^2
  \frac{\gE[^2]}{12\pi^2}\left(\ln\left(\frac{\Lambda^2}{m_e^2}\right)
    - \mathcal{R} \left(\frac{-q^2}{m^2}\right) \right)
\end{equation}
where,
\begin{equation}
\mathcal{R}(x) = \frac{4}{x} - \frac{5}{3} +
\frac{2(x-2)}{x}\sqrt{\frac{x+4}{x}} \tanh^{-1} \sqrt{\frac{x}{x+4}}
\end{equation}
is finite for the limits of $x$ approaching both zero and
infinity. For further details, see \rfr{griffiths.08.1}. The infinite
cut-off term can then be absorbed into a renormalised coupling
constant,
\begin{equation}
  g_r^2 \equiv \gE[^2]\left(1 - \frac{\gE[^2]}{12\pi^2}
    \ln\left(\frac{\Lambda^2}{m_e^2}\right) \right)\equcomma
  g_r^2(q) \equiv g_r^2\left(1 - \frac{g_r^2}{12\pi^2}
    \mathcal{R}\left(\frac{-q^2}{m_e^2}\right) \right)
  \labelequ{RenormG}
\end{equation}
so the matrix element of the sum of the leading-order and
next-to-leading-order terms can be written as,
\begin{equation}
    \me_{\mu e \to \mu e} = -g_r^2(q) \left(\bar{u}(q_3,\lambda_3)\gamma^\mu
      u(q_1,\lambda_1)\right) \frac{g_{\mu\nu}}{q^2}
    \left(\bar{u}(q_4,\lambda_4)\gamma^\nu u(q_2,\lambda_2)\right)
\end{equation}
where the divergence has been absorbed by the renormalised momentum
dependent coupling. This renormalised coupling is the experimentally
measured coupling.

Any Lagrangian where the infinities from higher-order diagrams can be
absorbed into constants of the Lagrangian, {\it i.e.} masses and
couplings, is renormalisable. In \rfr{thooft.72.1} t'Hooft and Veltman
demonstrated that all gauge theories are renormalisable using the
process of dimensional regularisation. This is a method by which
integrals similar to that of \equ{Thr:RenormNLO} can be rewritten in
terms of a single cut-off parameter. Since the Lagrangian of the \sm
is an ${\su[3]_C \otimes \su[2]_T \otimes \un[1]_Y}$ gauge theory, the
Lagrangian for the \sm is fully renormalisable, and the ultraviolet
divergences are absorbed in the masses and couplings of the \sm
Lagrangian. The minimal subtraction (MS) scheme for renormalisation
was proposed by t'Hooft in \rfr{thooft.73.1} and Weinberg in
\rfr{weinberg.73.1} from which emerged the modified minimal
subtraction (\msbar) scheme. This renormalisation scheme is currently
used for most \sm calculations.

One of the consequences of renormalisation is the running of the
couplings as a function of momentum. Oftentimes the couplings \gE and
\gS with the Weinberg angle \tw, are expressed as~\cite{amaldi.91.1},
\begin{alignat*}{3}\labelali{AlphaData}
  \alpha_1 &= \left(\frac{5}{3}\right)
  \frac{\gE[^2]}{(4\pi)\costw[^2]}\equcomma
  & \alpha_1^{-1}(\m_\z) &= 58.9 \pm 0.3
  \\ \alpha_2 &= \frac{\gE[^2]}{(4\pi)\sintw[^2]}\equcomma
  & \alpha_2^{-1}(\m_\z) &= 29.7 \pm 0.2
  \\ \alpha_3 &= \frac{\gS[^2]}{4\pi}\equcomma
  & \alpha_3^{-1}(\m_\z) &= 8.47 \pm 0.5 \\
\end{alignat*}
where the factor of $5/3$ for $\alpha_1$ is a normalisation from the
\su[5] unification theory of \rfr{georgi.74.1} and the numerical
values are from \rfr{kim.93.1}.

Performing renormalisation at the one-loop level, like that of
\equ{Thr:RenormG}, results in the renormalised couplings fulfilling
the differential equation,
\begin{equation}
  \frac{\partial \alpha_i^{-1}}{\partial \ln\left(\frac{Q}{Q_0}\right)}
  = \frac{-b_i}{2\pi}\equcomma b_i = \begin{pmatrix}\frac{41}{10},&
    \frac{-19}{6},& -7 \end{pmatrix}
  \labelequ{AlphaDiff}
\end{equation}
where $Q$ is the renormalisation group scale and the values $b_i$
arise from the \sm Lagrangian~\cite{martin.11.1}. Solving this for
$\alpha_i^{-1}$ results in the relation,
\begin{equation}
  \alpha_i^{-1}(Q) = \alpha_i^{-1}(Q_0) - \frac{b_i}{2\pi}
  \ln\left(\frac{Q}{Q_0}\right)
  \labelequ{AlphaRun}
\end{equation}
for the running of the couplings, which is plotted in
\fig{Thr:Couplings.Sm} using the $\alpha_i^{-1}(\m_\z)$ from
\equ{Thr:AlphaData}. From this figure the confinement and asymptotic
freedom properties of \qcd are apparent. At low $Q$ the \qcd coupling
\gS increases, while for high $Q$ it decreases. As $Q$ becomes
smaller, perturbative \qcd calculations begin to no longer converge,
and at around ${Q = 1~\gev}$ perturbative \qcd fails completely.

\begin{subfigures}{2}{Running of the couplings $\alpha_i^{-1}(Q)$
    given by \equs{Thr:AlphaDiff} and \ref{equ:Thr:AlphaRun}. The
    couplings are run upwards from their experimental values at
    ${\m_\z \approx 10^2~\gev}$ from \equ{Thr:AlphaData}.}
  \svgbeg 
  \svg[1]{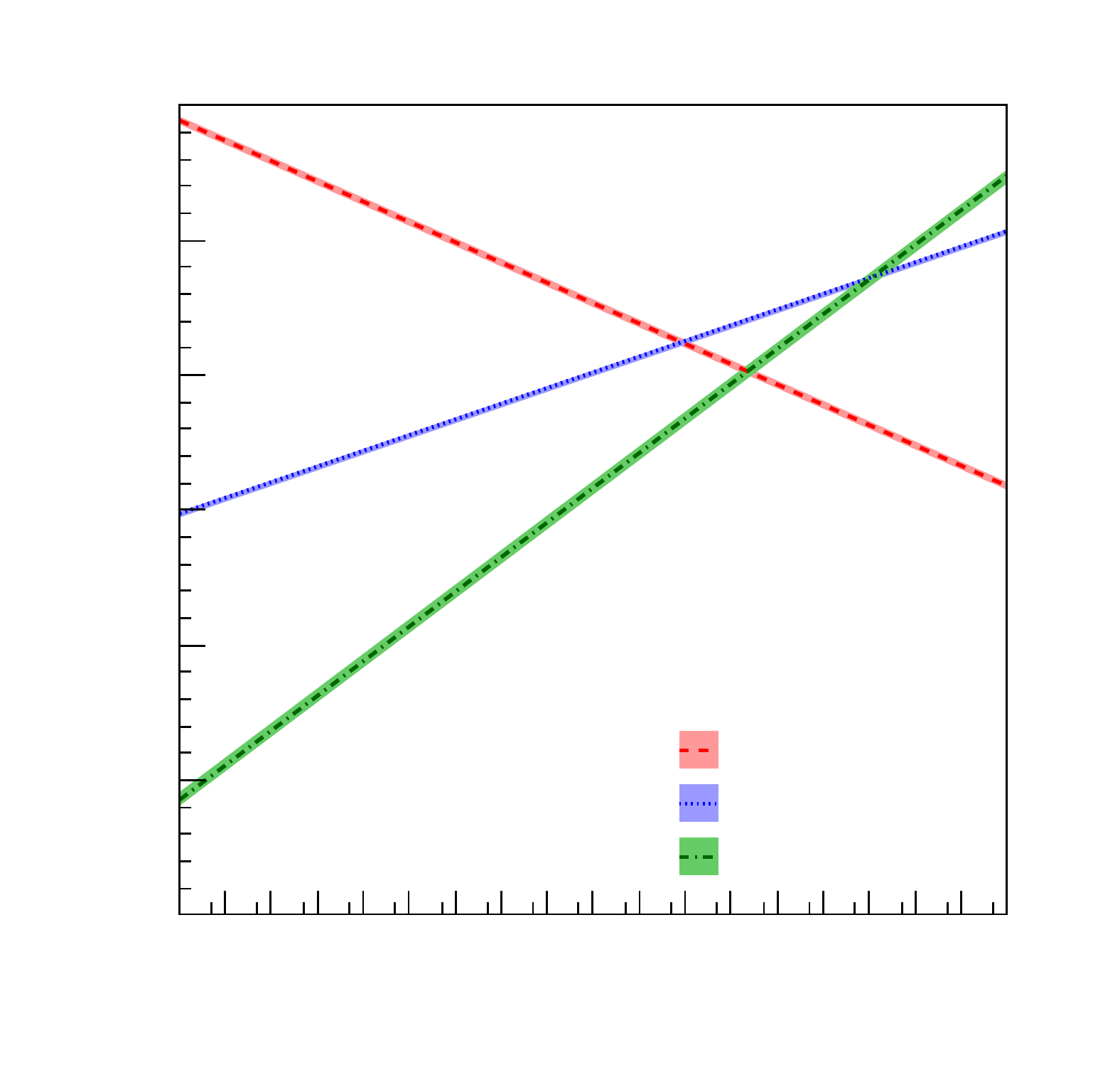} & \sidecaption \svgend
\end{subfigures}

\newsubsubsection{Factorisation}

In the formulation of the scattering matrix from \sec{Thr:Sca}, the
initial and final states are assumed to be free states. When colliding
bound states, such as two protons at the \lhc, or a proton and an
anti-proton at the Tevatron, the initial states are no longer free,
and so a method is needed to theoretically calculate cross-sections
for these processes. Consider scattering an electron off a proton at
low energies. Here, the proton is resolved as a point charge and the
scattering of the electron can be treated classically. However, at
higher energies the electron begins to probe the inner structure of
the proton and will scatter off partons such as the valence \wuqs and
\wdqs. At higher energies yet, the electron will be able to resolve
even more structure of the proton, including gluons and sea quarks
such as anti-quarks and heavy-flavour quarks.

The cross-section for an electron scattering off a quark can be
calculated using \equ{Thr:CrossSection}, but the momentum of both the
electron and quark is needed. Consequently, if the fractional momentum
of the proton carried by the quark is known, then the cross-section
for an electron scattering off a quark within the proton can be
calculated. Similarly, if two protons are collided, and the momentum
fractions of the two interacting partons of the protons are known,
then the cross-section can again be calculated with
\equ{Thr:CrossSection}. The factorisation theorem generalises this
concept; the cross-section for two colliding bound states $a_1$ and
$a_2$ with interacting partons $p_1$ and $p_2$ producing a final state
$B$ which can be written as,
\begin{equation}
  \sigma_{a_1a_2 \to B} = \int \int \xf_{a_1}\left(x_{p_1},
    Q^2, p_1\right)
  \xf_{a_2}\left(x_{p_2}, Q^2, p_2\right)
  \sigma_{p_1 p_2 \to B} \sdif{x_{p_1}}\sdif{x_{p_2}}
  \labelequ{Factorisation}
\end{equation} 
where $\xf_{a_i}\left(x_{p_i}, Q^2, p_i\right)$ is the parton
distribution function (\PDF) at the energy scale $Q^2$ for parton type
$p_i$ of the bound state $a_i$, and $\sigma_{p_1p_2 \to B}$ is the
partonic cross-section which can be calculated with
\equ{Thr:CrossSection}.

The \PDF, at leading-order for a given bound state, is the joint
probability density function for finding a parton of type $p_i$ with a
longitudinal momentum fraction $x_{p_i}$. Since the parton cannot have
a momentum larger than the momentum of the bound state, the momentum
fraction $x_{p_i}$ must range between $0$ and $1$. The \PDF evaluated
for a given parton type $p_i$ as a function of momentum fraction
$x_{p_i}$ is typically notated as ${\xf\left(x_{p_i}, Q^2,
    p_i\right)}$. Because the \PDF is a probability density function,
the \xf for a given parton type, integrated over $x$, is the
probability of observing that parton type in the bound
state. Consequently, the sum of the integrated \xf over all parton
types is unity. The \PDF{s} for bound states must be determined from
fits of experimental data, and are energy scale dependent. In
\fig{Thr:Pdf.All} the \xf, with uncertainties, for \wuqs, \wdqs, their
corresponding anti-quarks, and gluons are given for an energy scale of
${Q^2 = 100~\gev^2}$ using the \mstw leading-order proton \PDF set of
\rfr{martin.09.1}. For low momentum fractions the gluon dominates the
structure of the proton, while at very high momentum fractions, the
valence \wuq dominates. At this energy scale, gluons carry
approximately $46\%$ of the proton momentum while \wuqs carry $26\%$
and \wdqs carry $12\%$.

\begin{subfigures}{2}{\subfig{Pdf.All}~The \xf as a function
    of momentum fraction $x$ for the \wuq, \wdq, $\bar{u}$-quark,
    $\bar{d}$-quark, and gluon partons of the leading-order \mstw
    proton \PDF set at an energy scale of ${Q^2 =
      100~\gev^2}$. \subfig{Pdf.U}~The \xf for the \wuq
    parton of the proton as a function of momentum fraction $x$ and
    energy scale $Q^2$.}
  \svgbeg 
  \svg{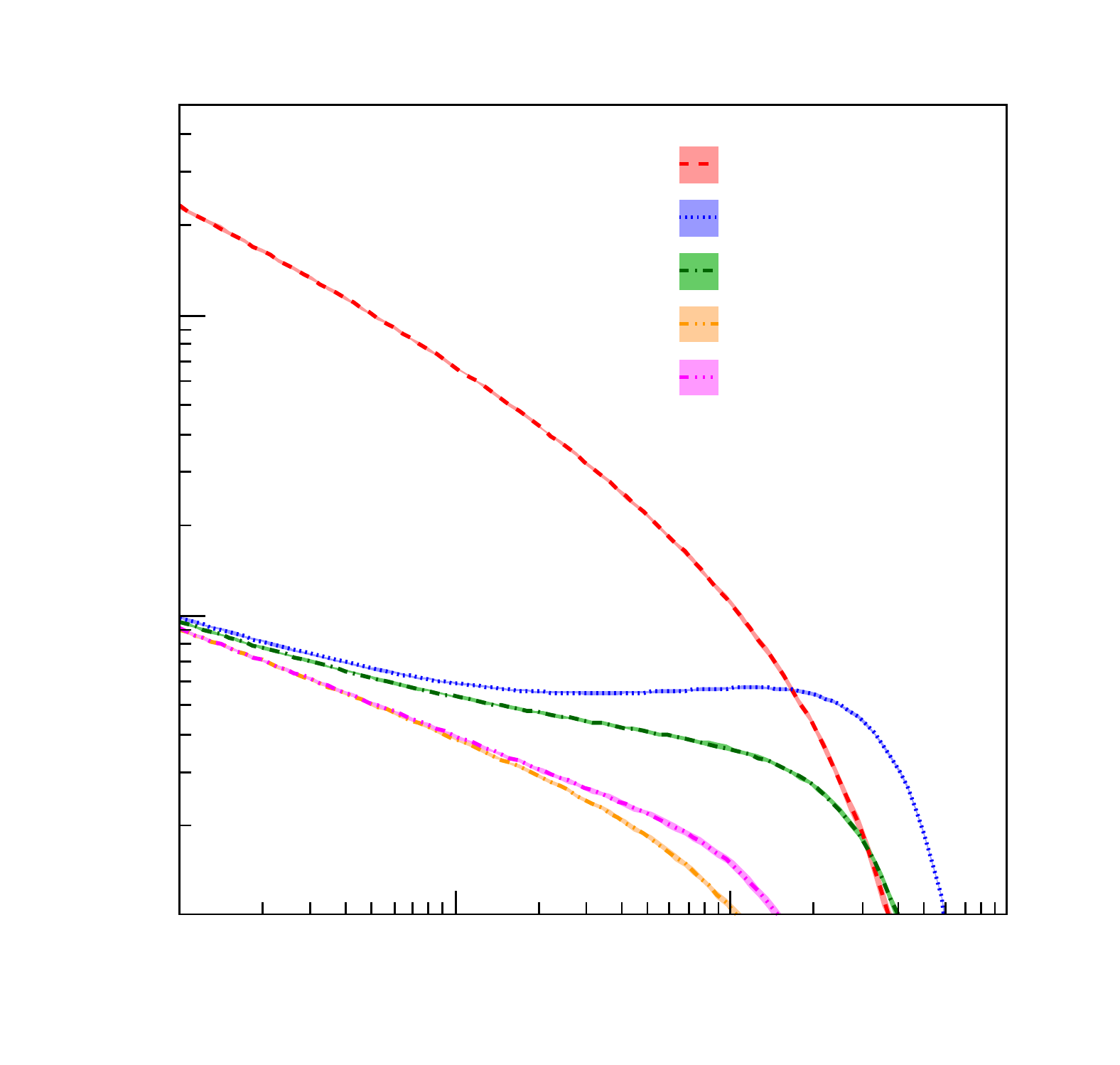} & \svg{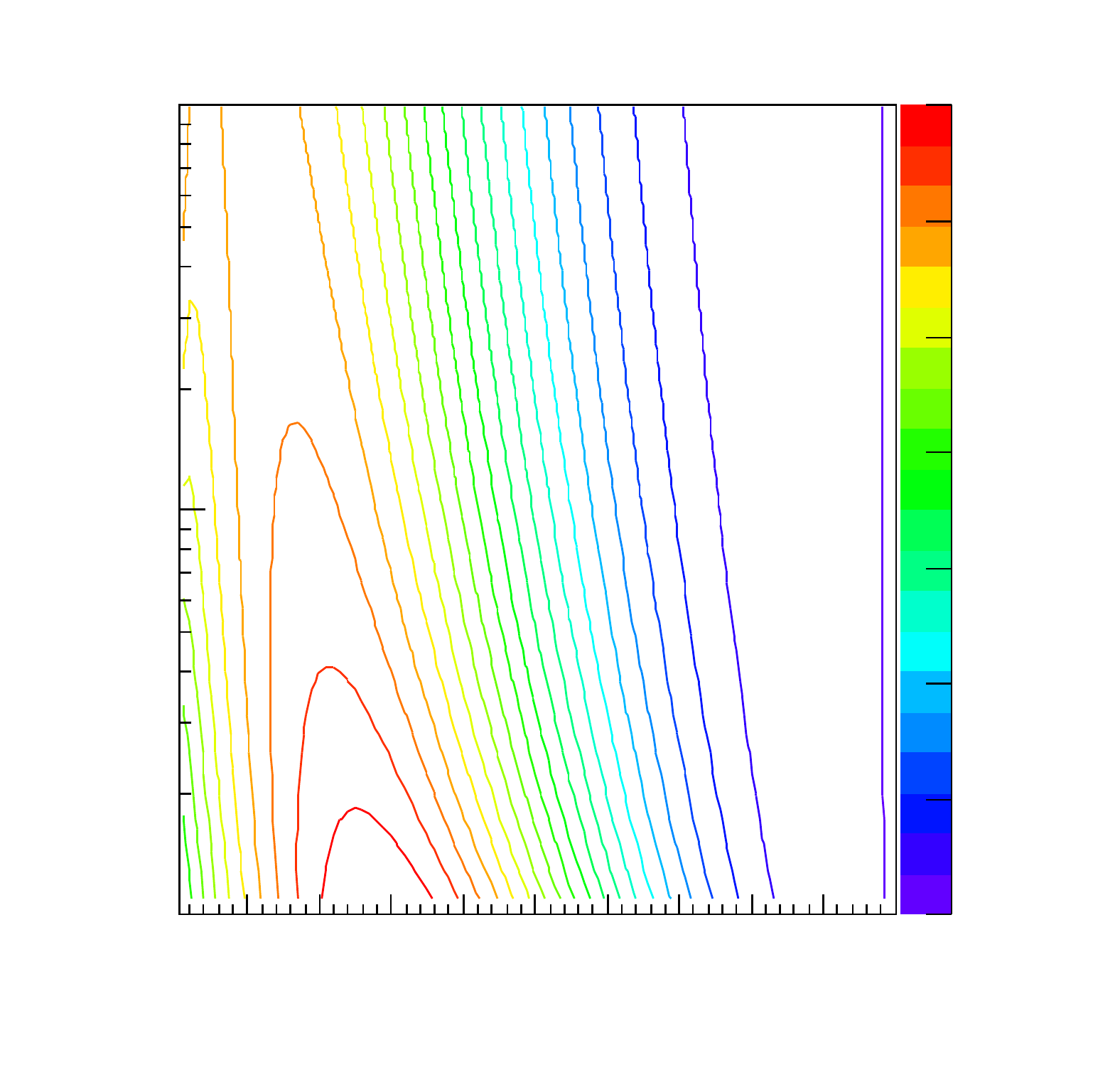} \svgend
\end{subfigures}

The \PDF{s}, just like the couplings of \equ{Thr:AlphaRun}, are both
energy scale dependent and renormalisation scheme dependent. The
Dokschitzer Gribov-Lipatov Altarelli-Parisi (DGLAP) equation of
\rfrs{dokshitzer.77.1}, \cite{gribov.72.1}, and \cite{altarelli.77.1}
is an analogue to \equ{Thr:AlphaRun}, evolving the \PDF measured at
some energy scale $Q_0^2$ to the energy scale $Q^2$. The \xf for the
partons of a hadron must satisfy,
\begin{align*}\labelali{Dglap}
  Q^2 \frac{\partial}{\partial \left(Q^2\right)}
  &\begin{pmatrix} \xf(x,Q^2,q_i) \\ \xf(x,Q^2,g) \\ \end{pmatrix}
  = \frac{\aS(Q^2)}{2\pi} \sum_j \int_x^1 \frac{1}{x'} \\
  &
  \begin{pmatrix}
    \mathcal{P}_{q_iq_j}\left(\frac{x}{x'}, \aS(Q^2)\right) &
    \mathcal{P}_{q_ig}\left(\frac{x}{x'}, \aS(Q^2)\right) \\
    \mathcal{P}_{gq_j}\left(\frac{x}{x'}, \aS(Q^2)\right) &
    \mathcal{P}_{gg}\left(\frac{x}{x'}, \aS(Q^2)\right) \\
  \end{pmatrix} 
  \begin{pmatrix} \xf(x',Q^2,q_j) \\ \xf(x',Q^2,g) \\ \end{pmatrix}
  \sdif{x'}
\end{align*}
where $q_i$ is a quark of flavour $i$ and $g$ is a gluon. The
splitting functions $\mathcal{P}_{p_ip_j}$ encode the splitting of a
parton into two additional partons via the \qcd vertices of either
\fig{Thr:G2QQ} or \fig{Thr:G2GG}. These splitting functions are also
used for the evolution of parton showers, and are described at leading
order in \sec{Thr:Par} where $\mathcal{P}_{q_iq_j}$ is zero unless the
flavour of $q_i$ and $q_j$ are the same. For further details on these
splitting functions as well as a derivation of \equ{Thr:Dglap}, see
\rfr{ellis.96.1}. In \fig{Thr:Pdf.U} the \xf for the \wuq parton of
the proton is given as a function of momentum transfer $x$ and energy
scale $Q^2$ as evolved with the DGLAP equation. Generally, at lower
energy scales the \wuq partons carry a larger fraction of the proton
momentum than at higher energies. For example, the \wuqs carry $38\%$
of the proton momentum at ${Q^2=1~\gev^2}$ and only $26\%$ at ${Q^2 =
  100~\gev^2}$.

\newsubsection{Alternatives and Extensions}{Alt}

While the \sm provides predictions which have been experimentally
validated to better than $1\%$, there remain a variety of outstanding
issues, both experimental and theoretical. The following list is by no
means exhaustive, but is intended to provide a general overview of
some of the issues currently confronting the particle physics
community.
\begin{itemize}
  \settowidth{\itemindent}{dark matter:}
\item[{\makebox[\itemindent][l]{gravity:}}] Currently, the \sm does
  not account for gravity, as the gravitational forces acting on
  particles at the scattering energies of most particle physics
  experiments is nearly forty times weaker then the electroweak or
  strong forces. In the development of the Lagrangians in
  \sec{Thr:Lag}, only special relativity was accounted for, and not
  general relativity. Introducing fields for gravity into the \sm
  Lagrangian results in a non-renormalisable theory and so the
  question of how to incorporate gravity into the \sm remains.
\item[{\makebox[\itemindent][l]{unification:}}] In the \sm the
  electromagnetic and weak forces are unified into the electroweak
  force via a ${\su[2]_T\otimes\un[1]_Y}$ gauge symmetry. Is it
  possible to unify the strong force with the electroweak force, and
  possibly gravity as well? In a unified theory the couplings
  $\alpha_i$ of \equ{Thr:AlphaData} would converge at some grand
  unified theory (GUT) scale. As can be seen for the \sm in
  \fig{Thr:Couplings.Sm}, this is not the case, but if the forces do
  unify, the GUT scale is expected to be on the order of $\approx
  10^{16}~\gev$.
\item[{\makebox[\itemindent][l]{parameters:}}] There are $18$
  parameters within the \sm which must be experimentally determined,
  excluding the neutrino sector, which introduces an additional $7$
  parameters. For a fundamental theory, one might assume only a single
  parameter must be experimentally determined, or even better, no
  parameters. Furthermore, within the \sm the observed mass of the
  \whb is the result of large cancellations from renormalisation and
  requires fine-tuning, which many theorists think is unnatural and is
  indicative of a more fundamental theory.
\item[{\makebox[\itemindent][l]{dark matter:}}] Measurements of the
  rotational velocities of galaxies as a function of the galactic
  radius do not match the expected velocities determined from the
  visible mass of the galaxies, see {\it e.g.}
  \rfr{trimble.87.1}. This discrepancy implies the presence of a new
  type of weakly interacting massive particles (WIMP) not accounted
  for within the \sm.
\item[{\makebox[\itemindent][l]{anti-matter:}}] According to the
  standard cosmological big-bang theory of \rfr{lemaitre.31.1} both
  matter and anti-matter should have been created in equal parts
  during the early formation of the universe. However, precision
  observations of the cosmic microwave background from \rfr{wmap.12.1}
  indicate that the universe is primarily made of matter. Currently,
  the \lhcb collaboration is exploring aspects as to why this
  matter/anti-matter asymmetry in the universe exists. Additionally,
  the AMS experiment of \rfr{ams.13.1} is searching for cosmic sources
  of anti-matter.
\item[{\makebox[\itemindent][l]{neutrinos:}}] As stated in
  \sec{Thr:Lag}, the observation of neutrino oscillations implies
  neutrinos have mass. However, the details on how massive neutrinos
  should be incorporated into the \sm have not yet been fully
  determined by experiment. For a review of current neutrino
  experiment results see \rfr{drexlin.13.1}.
\item[{\makebox[\itemindent][l]{Higgs boson:}}] The details of the
  Higgs mechanism introduced in \sec{Thr:Lag} have not yet been fully
  confirmed by experiment. However, the observation of a Higgs-like
  boson by the \atlas \cite{atlas.12.2} and \cms \cite{cms.13.1}
  collaborations with a mass of approximately $125~\gev$ are the
  beginnings of detailed measurements of the Higgs
  mechanism. Furthermore, this thesis investigates the forward
  production of \whbs within \lhcb.
\end{itemize}
Within the remainder of this section the supersymmetry extension to
the \sm is presented, where in \chp{Hig} parts of this theory are
tested. Alternative models such as \su[5]~\cite{georgi.74.1},
\so[10]~\cite{fritzsch.75.1}, technicolour~\cite{susskind.78.1}, and string
theory~\cite{susskind.69.1} have also been proposed, but are not
discussed here.

\newsubsubsection{Supersymmetry}{}

% See pietro.05.1.pdf and martin.11.1.pdf for more details.
Consider a diagram with an internal fermion loop similar to the
diagram of \fig{Thr:MuE2MuE.NLO}, but replacing the photon mediator with
the neutral \sm \whb. This loop also produces a divergent integral,
that is no longer proportional to \gE[^2] as in the case of the photon
with the vertex of \fig{Thr:Gm2FF}, but rather is proportional to the
mass of the fermion squared, given by the vertex factor of
\fig{Thr:H02FF}. Consequently, the divergence from this one-loop
correction is now absorbed in the renormalised \whb mass of the form,
\begin{equation}
  \m_r^2 = \m_H^2 - \frac{g_f^2}{8\pi^2}\left(\Lambda^2 +
    3m_f^2\ln\left(\frac{\Lambda^2}{m_f^2}\right) - 2m_f^2 +
    \frac{m^4}{\Lambda^2 + m_f^2} \right)
\end{equation}
where $\m_H$ is the bare \whb mass, $m_f$ is the mass of the fermion,
$g_f$ is the coupling of the fermion to the \whb, and $\Lambda$ is
some cut-off parameter. Notice that this renormalisation contains a
quadratically divergent term, $\Lambda^2$, unlike the renormalised
coupling $g_r$ of \equ{Thr:RenormG} which is only logarithmically
divergent. The measured value of the Higgs vacuum expectation value
$v$ requires the renormalised \whb mass to be near $100~\gev$ which
is consistent with the Higgs-like boson observed by \atlas and
\cms. Consequently, if the cut-off parameter is on the order of the
Planck scale, where $\Lambda \approx 10^{19}~\gev$ and effects from
quantum gravity become significant, then the bare mass must also be on
the same order and a cancellation of at least $17$ digits must
occur to produce the renormalised \whb mass. This problem is known as
the naturalness or fine-tuning problem.

Massive scalars which couple to the \whb, such as the self-coupling of
\fig{Thr:H02H0H0}, also contribute to the renormalised \whb mass at the
one-loop level,
\begin{equation}
  m_r^2 = m_H^2 + \frac{g_\phi^2}{16\pi^2}\left(\Lambda^2 - m_\phi^2
    \ln \left(\frac{\Lambda^2}{m_\phi^2} \right) \right)
\end{equation}
where $g_\phi$ is the coupling of the scalar with the \whb and
$m_\phi$ is the mass of the scalar. These contributions also contain a
quadratically divergent term, but with the opposite sign to the term
from the fermion, and so the quadratic divergences of the renormalised
\whb mass could be cancelled if there is a symmetry between fermions
and bosons. This type of symmetry is a supersymmetry (\susy) and was
introduced by Wess and Zumino in \rfr{wess.74.1} where the
translations,
\begin{equation}
  \delta\phi = 2\bar{\epsilon}\psi\equcomma \delta\psi =
  -i\gamma^\mu\epsilon\left(\partial_\mu \phi\right)
\end{equation}
mix scalar fields $\phi$ with spinor fields $\psi$, and $\delta$ and
$\epsilon$ describe the transformations~\cite{griffiths.08.1}. A
Lagrangian invariant under this translation can be constructed by
combining the spin-$0$ Lagrangian of \equ{Thr:Spin0L} with the
spin-$1/2$ Lagrangian of \equ{Thr:Spin1L}, where both fields
correspond to particles with the same mass.

A supersymmetric Lagrangian is built by requiring a fermionic or
bosonic superpartner for every particle of the \sm and combining them
into supermultiplets which preserve the gauge groups of the \sm. For
every fermion/boson pair the number of bosonic degrees of freedom must
equal the number of fermionic degrees of freedom. For each spin-$1/2$
fermion of the \sm, a complex scalar field or sfermion must be
introduced. The sfermion partners to leptons are sleptons, while the
sfermion partners to quarks are squarks. The gauge bosons give rise to
gaugino fermionic superpartners: gluinos, winos, zinos, photinos, and
higgsinos. This however, is only an introduction to some of the
creative naming conventions of \susy models. None of the plethora of
new particles introduced by \susy has been experimentally observed,
and this requires that \susy is also a broken symmetry, similar to
electroweak symmetry. Methods for breaking \susy are not explored
here, but an excellent introduction is given in \rfr{martin.11.1}.

The minimal supersymmetric model (\mssm) is the \susy model with the
simplest Higgs sector: a two Higgs doublet model consisting of a weak
isospin doublet with hypercharge $+1$ and two complex fields, and a
weak isospin doublet with hypercharge $-1$, also with two complex
fields. The two doublets, analogous to the single \sm doublet of
\equ{Thr:HiggsF}, are typically written as,
\begin{equation}
  h_u = \begin{pmatrix} h_u^+ \\ h_u^0 \\ \end{pmatrix}\equcomma
  h_d = \begin{pmatrix} h_d^0 \\ h_d^- \\ \end{pmatrix}\equcomma
\end{equation}
where $h_u$ has hypercharge $+1$ and $h_d$ has hypercharge $-1$. Just
as a potential was introduced for the \sm Higgs doublet with
\equ{Thr:SmPotential}, the potential,
\begin{align*}\labelali{MssmPotential}
  V =
  &\left(\abs{\mu}^2 + \m_{h_u}^2\right)
  \left(h_u^{0\dagger}h_u^{0} + h_u^{+\dagger}h_u^{+}\right) + 
  \left(\abs{\mu}^2 + \m_{h_d}^2\right)
  \left(h_d^{0\dagger}h_d^{0} + h_d^{-\dagger}h_d^{-}\right) \\
  &+ m_{h_{ud}}^2\left(h_u^+h_d^- - h_u^0h_d^0 +
    h_u^{+\dagger}h_d^{-\dagger} - h_u^{0\dagger}h_d^{0\dagger}
  \right) \\
  &+ \frac{\left(g_1^2 + g_2^2\right)}{8}
  \left(h_u^{0\dagger}h_u^{0} + h_u^{+\dagger}h_u^{+}
    - h_d^{0\dagger}h_d^{0} - h_d^{-\dagger}h_d^{-} \right)^2 \\
  &+ \frac{g_2}{2}
  \left(h_u^+h_d^{0\dagger}+h_u^0h_d^{-\dagger}\right)^\dagger
  \left(h_u^+h_d^{0\dagger}+h_u^0h_d^{-\dagger}\right)
\end{align*}
must be introduced for the \mssm Higgs
Lagrangian~\cite{gunion.90.1,martin.11.1}. Here, $\mu$ is the Higgs
mixing parameter and $\m_{h_u}$, $\m_{h_u}$, and $\m_{h_{ud}}$ are
additional constants which can be related to more physical
constants. The field $h_u^0$ has a vacuum expectation value of $v_u$
while the field $h_d^0$ has a vacuum expectation value of $v_d$. These
two vacuum expectation values are specified by,
\begin{equation}
  v_u^2 + v_d^2 = \frac{2\m_\z^2}{\left(g_1^2 + g_2^2\right)}
\end{equation}
and so only the ratio of the two is unknown, which is normally written
as ${\tanb \equiv v_u/v_d}$.

The eight real fields of the two doublets can be written as mass
eigenstates by the relations,
\begin{alignat*}{3}\labelali{HiggsMass}
  &\begin{pmatrix} h_u^0 \\ h_d^0 \\ \end{pmatrix} &=& 
  \begin{pmatrix} v_u \\ v_d \end{pmatrix} + \frac{1}{\sqrt{2}}
  \begin{pmatrix} 
    \cos\alpha & \sin\alpha \\ -\sin\alpha & \cos\alpha \\
  \end{pmatrix}
  \begin{pmatrix} \hhz \\ \hHz \\ \end{pmatrix} + \frac{i}{\sqrt{2}}
  \begin{pmatrix} 
    \sin\beta & \cos\beta \\ -\cos\beta & \sin\beta \\
  \end{pmatrix}
  \begin{pmatrix} \phi^0 \\ \hAz \\ \end{pmatrix} \\
  &\begin{pmatrix} h_u^+ \\ h_d^{-\dagger} \\ \end{pmatrix} &=&
  \begin{pmatrix} 
    \sin\beta & \cos\beta \\ -\cos\beta & \sin\beta \\
  \end{pmatrix}
  \begin{pmatrix} \phi^+ \\ H^+ \\ \end{pmatrix}
\end{alignat*}
for the neutral and charged mass eigenstates, assuming $v_u$ and $v_d$
minimise the tree-level potential. At tree-level the \hhz is a light
\cp-even \whb, the \hHz a heavy \cp-even \whb, and the \hAz a \cp-odd
\whb. There are two real charged \whbs where ${H^{+\dagger} = H^-}$
and two charged Goldstone bosons, ${\phi^{+\dagger} = \phi^-}$, which
are absorbed into the longitudinal polarisations of the \wwbs with an
appropriate gauge transformation. Similarly, $\phi^0$ is a neutral
Goldstone boson which is absorbed into the longitudinal polarisation
of the \wzb with the proper transformation. The masses are related via
both $\beta$ and the \whb mass mixing angle $\alpha$ which is given by
the relations,
\begin{equation}
  \frac{\sin2\alpha}{\sin2\beta} = \frac{-\m_\hHz^2 -
    \m_\hhz^2}{\m_\hHz^2 - \m_\hhz^2}\equcomma
  \frac{\tan2\alpha}{\tan2\beta} = \frac{\m_\hAz^2 +
    \m_\z^2}{\m_\hAz^2 - \m_\z^2}
\end{equation}
in terms of $\beta$, the masses of the neutral mass eigenstates, and
the mass of the \wzb.

\begin{subfigures}[p]{3}{Neutral \mssm \whb couplings with
    \subfig[H12FuFu]{H32FuFu}~$u$-type fermions ($u$, $c$, $t$, $\nu$)
    and \subfig[H12FdFd]{H32FdFd}~$d$-type fermions ($d$, $s$, $b$,
    $\lep$). \subfig[H4p2FF]{H4m2FF}~Charged \whb couplings with
    fermions. \subfig[H12WW]{H22ZZ}~Couplings of the neutral \mssm
    \whbs with two gauge bosons. The vertices follow the conventions
    of \rfr{gunion.90.1}.\labelfig{Vertices.Mssm.Higgs}}
  \fmpbeg
  \fmp{H12FuFu} & \fmp{H22FuFu} & \fmp{H32FuFu} \fmpsep
  \fmp{H12FdFd} & \fmp{H22FdFd} & \fmp{H32FdFd} \fmpsep
  \begin{tabular}{M{\tabcolwidth}M{0.5\tabcolwidth}M{\tabcolwidth}M{0pt}}
    \fmp{H4p2FF}  && \fmp{H4m2FF} \fmpsep            
    \fmp{H12WW}   && \fmp{H22WW}  \fmpsep
    \fmp{H12ZZ}   && \fmp{H22ZZ}  \fmpsep
  \end{tabular}\fmpend
\end{subfigures}

The parameters of the potential given by \equ{Thr:MssmPotential} can
then be related by the equations,
\begin{alignat*}{3}\labelali{}
  &\m_{h_u}^2 &=\:& \m_\hAz^2\cos^2\beta + \frac{\m_\z^2}{2}\cos2\beta -
  \abs{\mu}^2 \\
  &\m_{h_d}^2 &=\:& \m_\hAz^2\sin^2\beta + \frac{\m_\z^2}{2}\cos2\beta -
  \abs{\mu}^2 \\
  &\m_{h_{ud}}^2 &=\:& \frac{\m_\hAz^2}{2}\sin2\beta \\
\end{alignat*}
where typically the free parameters are taken as \tanb, $\m_\hAz$, and
$\abs{\mu}^2$. The masses of the remaining \whbs can also be specified
in terms of these parameters, and is done so in the \mssm \whb
phenomenology presented in \chp{Hig}. Vertices for the \mssm \whbs
coupling with $u$-type and $d$-type fermions as well as for the
neutral \whbs coupling with \w and \wzbs pairs are given in
\fig{Thr:Vertices.Mssm.Higgs}. Note that the \cp-odd \whb does not
have \w and \wzb pair couplings. These couplings are used in \chp{Tau}
when determining the spin correlations for \wtls produced from \mssm
\whbs, and in \chp{Hig} for the \mssm \whb phenomenology used to
produce \whb limits.

\begin{subfigures}{2}{Running of the couplings ${\alpha_i^{-1}(Q)}$
    given by \equs{Thr:AlphaDiff} and \ref{equ:Thr:AlphaRun}, using the
  \mssm coefficients of \equ{Thr:MssmRun}. The couplings are run
  upwards from their experimental values at ${\m_\z \approx
    10^2~\gev}$ from \equ{Thr:AlphaData}.}
  \svgbeg
  \svg[1]{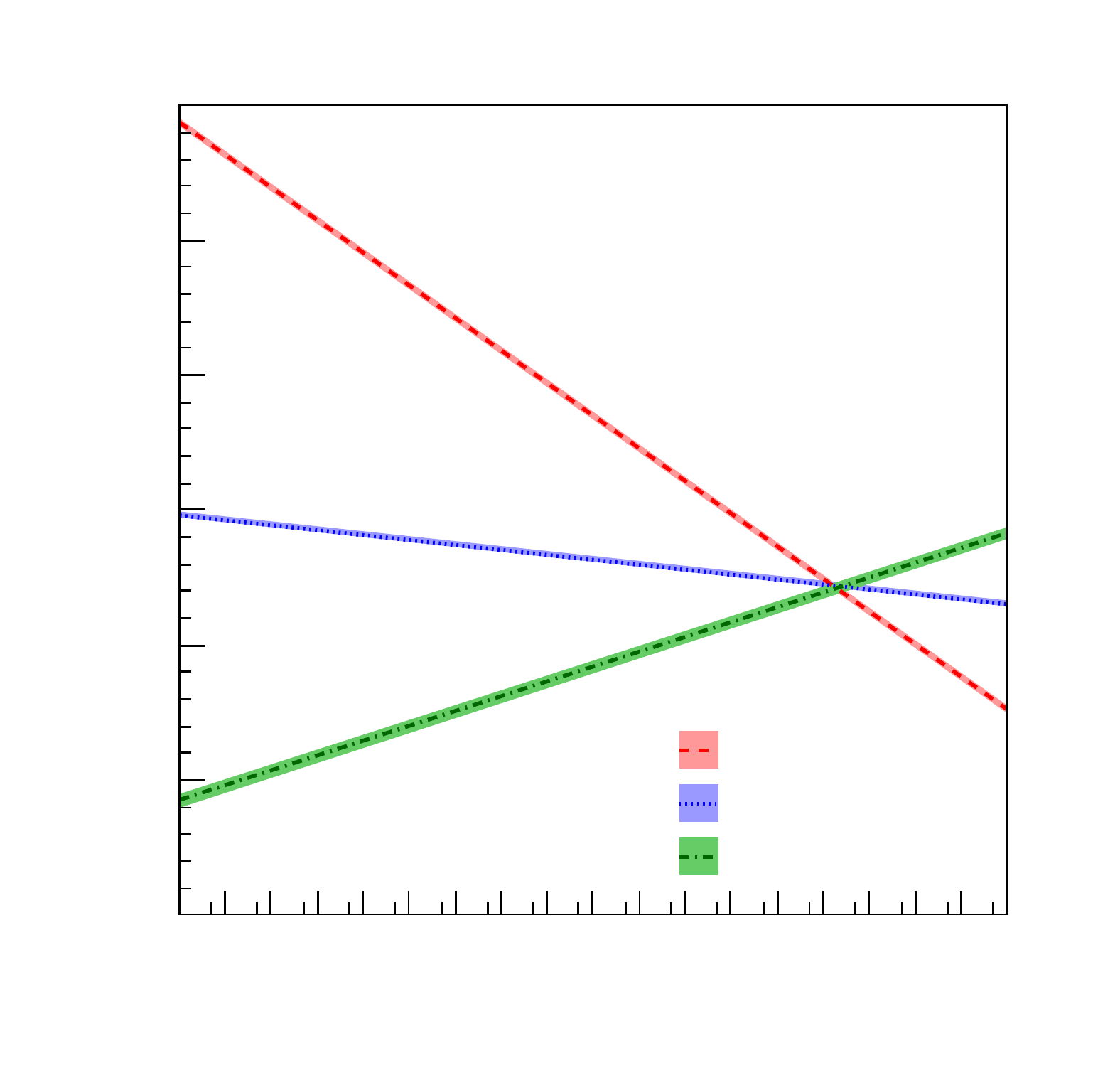} & \sidecaption \svgend
\end{subfigures}

The \mssm, and \susy in general, could resolve some of the current
issues with the \sm. Because \susy maintains the gauge symmetries of
the \sm, it can also be unified with \su[5] theory. When this is done
for the \mssm, the coefficients $b_i$ for the couplings $\alpha_i$ of
\equ{Thr:AlphaRun} at the one-loop level become,
\begin{equation}
  b_i = \begin{pmatrix} \frac{33}{5}, & 1, & -3 \end{pmatrix}
  \labelequ{MssmRun}
\end{equation}
which results in the forces unifying at an energy of $\approx
2\times10^{16}~\gev$, as shown in
\fig{Thr:Couplings.Mssm}. Additionally, the introduction of the
superpartners provides the possibility of weakly interacting massive
particles which could be dark matter candidates. Finally, the
introduction of these superpartners stabilises the observable mass of
the \whbs, and solves the fine-tuning problem. However, the addition
of superpartners requires even more model parameters then the \sm,
with a minimum of $124$ parameters required for the \mssm, nearly five
times the number of the \sm~\cite{pdg.12.1}. Currently, no experiments
have observed any evidence for \susy, and with results from the \lhc,
\susy is rapidly being excluded as a viable theory.

\newsection{Monte Carlo Techniques}{Mon}

The groundwork for performing perturbative calculations of
experimental observables such as the cross-section of
\equ{Thr:CrossSection} and the decay width of \equ{Thr:DecayWidth} has
been laid in \sec{Thr:Sta} for both the \sm and the \whb sector
of the \mssm. However calculating these observables can be non-trivial
and sometimes impossible using only perturbative methods, and so
advanced numerical methods, many of which rely on random sampling, must be
employed. These numerical methods are often broadly classified as
Monte Carlo. Within this section the general usage of Monte Carlo
within high energy particle physics will be introduced with an
emphasis given to the techniques required in \chp{Tau}.

Calculating a two-to-$n$ cross section or the width of an $n$-body
decay requires a $3n-4$ dimensional integral over the momentum
three-vectors of the $n$ final particles, not including integration
over quantum numbers such as helicity, flavour, or colour. One of the
most robust methods to numerically perform large multi-dimensional
integrals is Monte Carlo integration. In the simplest form of Monte
Carlo integration for an $n$-dimensional integral, $N$ random points
are uniformly selected from some $n$-dimensional volume bounded by the
limits of the integral, and the integrand is evaluated for each
point. The running average of the integrand is computed, and the
integral is then the product of the sampling volume and the average
integrand.

The precision of Monte Carlo integration converges on the order of
$N^{-1/2}$, where $N$ is the number of integration points, and is not
dependent upon the dimensionality of the integral. Comparatively, if
the desired precision for a one-dimensional integral using quadrature
methods is obtained with $N$ points, then roughly $N^n$ sampling
points will be needed to maintain the same precision for an
$n$-dimensional integral. Monte Carlo integration also can be
terminated whenever a sufficient precision is reached and does not
require complicated boundary conditions, unlike quadrature methods
which require pre-determined boundaries and sampling points, and
cannot be prematurely terminated~\cite{press.07.1}.

Because particle scattering events are truly random by nature, Monte
Carlo integration also has the advantage that each sampling point is a
simulated scattering event, where the final state particles are fully
specified. The correct distribution of random events, according to
theory, can then be obtained by using the accept-and-reject method. An
additional uniform random number is selected for each point; if this
number is less than ratio of the integrand at that point over the
maximum possible integrand, the event is accepted, otherwise it is
rejected and another point is chosen. Consequently, the accepted
sampling points from Monte Carlo integration can be passed directly
through material simulations of experimental detectors. These
simulations can then be used to estimate detector effects which are
necessary for detector design and calibration, as well as for many
physics analyses. Within the experimental high energy particle physics
community, the term Monte Carlo is often synonymous with detector
simulation.

Monte Carlo techniques are also well suited for extending the
perturbative calculations of \sec{Thr:Sca} to non-perturbative
regimes, primarily for \qcd. Specifically, Monte Carlo can handle
infrared and collinear divergences in the radiation of massless
particles, such as gluons and photons, as well provide models to
combine quarks and gluons into hadrons. These techniques are known as
parton showering and hadronisation respectively. Additionally, Monte
Carlo can provide methods for calculating the soft \qcd interactions
of the underlying events from hadron collisions like those at the
\lhc.

\begin{subfigures}{1}{Schematic of an example proton-proton to \sm
    \whb event produced by a general purpose Monte Carlo generator
    such as \pythia. The process begins with a ${q\bar{q} \to \hH \to
      \w\w}$ hard process and then proceeds with resonance decays,
    FSR, ISR, the underlying event, hadronisation, and finally,
    particle decays.\labelfig{MonteCarlo}}
  \fmpbeg
  \begin{fmffile}{\fmppathMonteCarlo}%
    \setlength{\unitlength}{1mm}\input{\fmppathMonteCarlo.fmp}%
  \end{fmffile}\executeiffilenewer{\fmppathMonteCarlo.fmp}%
  {\fmppathMonteCarlo.1}{cd \fmppath; mpost MonteCarlo.mp} \fmpend
\end{subfigures}

General purpose Monte Carlo generators are programs that combine all
of the techniques outlined above, and more, into a single coherent
generation of particle physics events. A variety of generators are
publicly available, each with advantages and disadvantages, but the
three primary general purpose generators are
\pythia{8}~\cite{\citepythiaeight}, \herwig{++}~\cite{\citeherwigpp},
and \sherpa~\cite{\citesherpa}. A schematic of an example event
produced by a general purpose Monte Carlo generator is provided in
\fig{Thr:MonteCarlo}. This schematic is a simplification of the
process, but attempts to provide all the salient features. The event
generation begins with the calculation of the hard process by
performing Monte Carlo integration of the cross-section formula of
\equ{Thr:CrossSection}, where the matrix element is built from the
elements of \sec{Thr:Lag}. In this example, the hard process is the
production of an \sm \whb from a quark pair decaying into two \wwbs.

Next, resonance decays are performed, again using perturbative
QFT and Monte Carlo integration. Resonance decays occur on a
time-scale shorter than the hadronisation of quarks and gluons, and
are primarily decays of \w, \z, or \whbs, or \wtqs. In
\fig{Thr:MonteCarlo}, the $\w^-$ from the hard process decays into a
quark pair, and the $\w^+$ into a \wtl and neutrino. After the hard
process and resonance decays are simulated, the initial and final
state quarks and gluons are dressed with parton showers which
probabilistically simulate the radiation of gluons and quarks as
determined by perturbative theory. The parton shower on the final
state particles is labelled final state radiation (FSR) and the shower
on the initial state particles is initial state radiation (ISR). Here,
FSR is only performed on the decay products of the $W^-$ as the $W^+$
has not decayed to quarks or gluons. At this point electromagnetic
final state radiation may also be added, but is not included in
\fig{Thr:MonteCarlo}.

In hadron-hadron collisions, interactions besides just the hard
process will also occur between the partons of the hadrons and are
categorised as underlying event. These interactions are typically via
soft \qcd and are simulated in Monte Carlo generators using
non-perturbative models which must be tuned to data. Both FSR and ISR
must also be applied to the underlying event, although they have not
been included in \fig{Thr:MonteCarlo}. Following this, the partons
from the resonance decays, parton showers, and underlying event are
combined into bound hadrons using phenomenological models. This
process is typically called hadronisation but is sometimes referred to
as fragmentation. In the final step of the event, all unstable
particles are decayed, using either perturbative QFT or models
determined using non-perturbative theories. In \chp{Tau} sophisticated
models for the decays of \wtls are implemented in \pythia{8}.

In the remainder of this section further details are given on parton
showers in \sec{Thr:Par}, hadronisation in \sec{Thr:Had}, and particle
decays in \sec{Thr:Dec}. The particle decay techniques are further used
in \chp{Tau} for the modelling of \wtl decays. For an excellent
overview on all aspects of Monte Carlo event generation see
\rfr{buckley.11.1}, while for more specifics on the \qcd aspects of
parton showering and hadronisation see \rfr{ellis.96.1}.

\newsubsection{Parton Showers}{Par}

Following the example of \rfrs{buckley.11.1} and \cite{ellis.96.1},
consider the leading-order tree-level production of a quark pair from
electron-positron annihilation as shown in \fig{Thr:EE2QQ}. The
cross-section for this diagram, $\sigma_{e^+e^-\to q\bar{q}}$, is
finite and can be calculated using \equ{Thr:CrossSection},
\tab{Thr:Rules}, and the vertices of \figs{Thr:Gm2FF} and
\ref{fig:Thr:Z2FF}. However, because of confinement, the quarks of
\fig{Thr:EE2QQ} must somehow interact through the strong force to
produce stable hadrons which are then experimentally observable. The
idea behind the parton shower is to evolve the quarks of
\fig{Thr:EE2QQ} from the energy scale at which they are produced,
using perturbative \qcd, to the non-perturbative \qcd regime where
hadronisation can then be applied to create bound final states.

\begin{subfigures}{3}{\subfig{EE2QQ}~Feynman diagram for
    electron-positron annihilation producing a quark
    pair. \subfig[EE2QGQ]{EE2QQG}~One-leg diagrams contributing to the
    leading-order diagram. \subfig{Angles}~Coordinate system in the
    centre-of-mass frame for the one-leg diagrams.}
  \fmpbeg
  \fmp{EE2QQ}  & \fmp{EE2QGQ}  & \fmp{EE2QQG} \fmpsep
  \fmp{Angles} & \multicolumn{2}{M{2\tabcolwidth}}{\sidecaption} \fmpend
\end{subfigures}

In \fig{Thr:EE2QQ} the quarks can interact through the strong force
via either a real or virtual emission of a gluon. The real gluon
emission results in an additional leg added on to the tree-level
diagram in the two configurations of \figs{Thr:EE2QGQ} and
\ref{fig:Thr:EE2QQG}. The differential cross-section for the sum of
the three diagrams from \figs{Thr:EE2QQ} through \ref{fig:Thr:EE2QQG}
can be approximated as,
\begin{equation}
  \dif{\sigma_{e^+e^-\to q\bar{q}g}}
  \approx \sigma_{e^+e^-\to q\bar{q}}
  \left(\frac{2\sdif{\cos\theta_q}}{\sin^2\theta_q}\right)
  \left(\frac{\aS}{2\pi}\right)
  \left(\frac{N_c^2-1}{2N_c}\right)
  \left(\frac{1+(1-z)^2}{z}\right)\sdif{z}
  \labelequ{XsBoth}
\end{equation}
where the coordinates are summarised in \fig{Thr:Angles} and only the
divergent terms are included but with full interference. Here, $z$ is
the fractional energy of the gluon with respect to the quark energy
${E_g/E_q}$, $\theta_q$ is the opening angle between the quark and
gluon in the centre-of-mass frame, and ${N_c = 3}$ is the number of
colour charges. This differential cross-section diverges for three
limits of the phase-space: when $z$ approaches zero, when $\theta_q$
approaches zero, and when $\theta_q$ approaches $\pi$.

The first divergence is an infrared divergence where the energy of the
emitted gluon is much less than that of the emitting quark, ${E_g \ll
  E_q}$ and the divergence does not depend on
$\theta_q$. Consequently, this divergence can be attributed to the
interference between the two diagrams of \figs{Thr:EE2QGQ} and
\ref{fig:Thr:EE2QQG}. The second divergence occurs when a gluon is
emitted in the same direction as the quark, while the third divergence
occurs when a gluon is emitted in the same direction as the
anti-quark. Both of these emissions are collinear divergences which
can be interpreted as independent emissions of the gluon by either the
quark or the anti-quark and so \equ{Thr:XsBoth} can be factorised as,
\begin{equation}
  \dif{\sigma_{e^+e^-\to q\bar{q}g}}
  \approx \sigma_{e^+e^-\to q\bar{q}}
  \sum_i\left(
    \left(\frac{\dif{\theta_{p_i}^2}}{\theta_{p_i}^2}\right)
    \left(\frac{\aS}{2\pi}\right)
    \left(\frac{N_c^2-1}{2N_c}\right)
    \left(\frac{1+(1-z)^2}{z}\right)\sdif{z}
  \right)
  \labelequ{XsSplit}
\end{equation}
where $\theta_{p_i}$ is the opening angle between parton $p_i$ and the
gluon. Here the summation is over the two partons, the quark and the
anti-quark. Note that this cross-section only accounts for the
collinear divergences.

The relation of \equ{Thr:XsSplit} can be generalised to any scattering
process resulting in final state quarks or gluons as,
\begin{equation}
  \dif{\sigma_{A \to B b_j}} \approx \sigma_{A \to B}
  \sum_i\left(
    \left(\frac{\dif{\theta_{b_i}^2}}{\theta_{b_i}^2}\right)
    \mathcal{P}_{b_jb_i}\left(z,\aS\right)\sdif{z}
  \right)
  \labelequ{XsGeneral}
\end{equation}
where $b_i$ are the final state partons of $B$ with flavour $i$ which
can emit a parton of flavour $j$. The splitting functions
$\mathcal{P}_{b_jb_i}$ are the same splitting functions used for the
DGLAP evolution of \equ{Thr:Dglap} and are given by,
\begin{alignat*}{2}\labelali{SplittingFunctions}
  \mathcal{P}_{gq}(z,\aS) &= \left(\frac{\aS}{2\pi}\right)
  \left(\frac{N_c^2-1}{2N_c}\right)
  \left(
    \frac{1+(1-z)^2}{z}
  \right)\quad && (q \to g q) \\
  \mathcal{P}_{gg}(z,\aS) &= \left(\frac{\aS}{2\pi}\right)
  \left(2N_c\right)
  \left( 
    \frac{1-z}{z} + \frac{z}{1-z} + z(1-z)
  \right)\quad && (g \to g g) \\
  \mathcal{P}_{qg}(z,\aS) &= \left(\frac{\aS}{2\pi}\right)
  \left(\frac{1}{2}\right)
  \left( 
    z^2 + (1-z)^2
  \right)\quad && (g \to q \bar{q}) \\
  \mathcal{P}_{qq}(z,\aS) &= \left(\frac{\aS}{2\pi}\right)
  \left(\frac{N_c^2-1}{2N_c}\right)
  \left( 
    \frac{1+z^2}{1-z}
  \right)\quad && (q \to q g) \\
\end{alignat*}
in their helicity-averaged form~\cite{altarelli.77.1}. See
\rfr{ellis.96.1} for their helicity-dependent form,
$\mathcal{P}_{b_jb_i}(z,\phi,\aS)$, where $\phi$ is the azimuthal
angle of the splitting and the helicity dependent $\mathcal{P}_{gq}$
can be found by ${\mathcal{P}_{qq}(1-z,\phi,\aS)}$. The splitting
functions of \equ{Thr:SplittingFunctions} correspond to the splitting
process given in brackets after each function.

\newsubsubsection{Final State Showers}{}

The term $\theta^2$ of \equ{Thr:XsGeneral} can be replaced by the
virtuality $q^2$, where $q$ is the momentum of the parton being
split. Because the splitting parton is from some hard process, $q^2$
must be less than than some maximum virtuality $Q^2$ given by the hard
process scale. Additionally there is some virtuality scale $Q_0^2$
where the emission of a parton cannot be physically resolved from the
emitting parton, typically on the order of $1~\gev^2$. The probability
of a parton not emitting between the virtualities $q_1^2$ and $q_2^2$
is given by a Sudakov form factor,
\begin{equation}
  \Delta_{ij}(q_1^2,q_2^2) = \exp\left(
    -\int_{q_2^2}^{q_1^2} \frac{1}{q^2} \int_{Q_0^2/q^2}^{1-Q_0^2/q^2}
    \mathcal{P}_{ji}(z,\aS) \sdif{z}\sdif{q^2}
  \right)
  \labelequ{SudakovFsr}
\end{equation}
where the probability of not emitting a resolvable parton is then
$\Delta_{ij}(Q^2,Q_0^2)$.

A final state parton shower (FSR) is performed by uniformly picking a
random number $r$ between $0$ and $1$, and solving
${\Delta_{ij}(Q^2,q^2) = r}$ for $q^2$. If $q^2$ is above $Q_0^2$ an
emission is generated at the scale $q^2$, otherwise the shower is
terminated. This process is iteratively applied until the condition
${q^2 < Q_0^2}$ is met and terminates the process for each showerable
parton. The parton shower includes not only the effect of tree-level
diagrams with unresolvable collinear emissions, but also the effect of
one-loop diagrams, as a non-emission can either be an unresolvable
parton or a virtual emission. A large number of issues regarding final
state parton showers have not been addressed here, including soft
gluon emission, and further details can be found in
\rfr{buckley.11.1}.

\newsubsubsection{Initial State Showers}{}

Just as final state partons are expected to radiate, initial state
partons leading up to the hard process are also expected to
radiate. Consequently, a process similar to the final state shower
could be applied to determine these emissions, and ultimately, the
partons used in the hard process. Using the same prescription as
forward evolution for final state showers, partons could be randomly
selected using the incoming \PDF{s} of the hadrons and evolved
downwards until the parton shower is terminated with the condition
${q^2 < Q_0^2}$ for each branch. However, this would result in a large
number of events generated where the partons after showering would not
be suitable for use in the hard process of choice, {\it e.g.} \w, \z,
or \whb production. A more efficient method is to choose the partons
of the hard process first using ${x_{a_1}x_{a_2}Q^2 \approx s}$ where
$x_{a_1}$ and $x_{a_2}$ are the longitudinal momentum fractions of the
two incoming partons, $\Q^2$ is the centre-of-mass energy of the hard
process, and $s$ is the centre-of-mass energy of the two colliding
hadrons. The emissions are then evolved backward, starting with a
large $Q^2$ and small $x$, and move towards a smaller $q^2$ and larger
$x$~\cite{sjostrand.85.1, marchesini.88.1}.

The Sudakov form factor of \equ{Thr:SudakovFsr} must be modified for
backwards evolution by,
\begin{align*}\labelali{}
  \Delta_{ij}(q_1^2,q_2^2,x) = 
  \exp\bigg(&
    -\int_{q_2^2}^{q_1^2} \frac{1}{q^2} \int_{Q_0^2/q^2}^{1-Q_0^2/q^2}
    \mathcal{P}_{ij}(z,\aS) \\
    &\left(\frac{x}{zx}\right)
    \left(\frac{\xf(x/z,q^2,j)}{\xf(x,q^2,i)}\right)
    \sdif{z}\sdif{q^2}
  \bigg) 
\end{align*}
where $\xf(x,q^2,i)$ is the parton distribution function from
\sec{Thr:Exp} for parton $i$ at momentum transfer $x$ and energy
scale $q^2$. The initial state parton shower (ISR) is then performed
by picking a uniform random number $r$ between $0$ and $1$ and solving
${\Delta_{ij}(Q^2,q^2,x) = r}$ for $q^2$, where $x$ is the momentum
transfer for the parton from the hard process being showered. If $q^2$
is above the cut-off $Q_0^2$ then an emission is generated, and the
process is iteratively continued, just as is done for a final state
parton shower. Now, however, the momentum fraction $x$ must be
recalculated for each step in the parton shower at the new lower
$q^2$. By this process, the partons are evolved backwards to a low
energy scale with high momentum transfer. Further details on how final
and initial state parton showers are matched with the hard process can
be found in \rfr{buckley.11.1}.

\newsubsection{Hadronisation}{Had}

Hadronisation is the process by which coloured quarks and gluons from
the initial and final state showers are combined to produce colourless
hadrons which are then either decayed or observable in the final
state. The hadronisation step of a general purpose Monte Carlo
generator involves all quarks and gluons from the event, including
particles from the underlying event and beam remnants, which are not
discussed here. Because of \qcd confinement, hadronisation must be
performed via phenomenological models, although these models are based
on behaviour observed in non-perturbative \qcd such as lattice
\qcd. Currently two major hadronisation models are used in general
purpose Monte Carlo generators, the Lund string fragmentation model of
\rfr{andersson.83.1} and the cluster model of \rfr{fox.79.1}. The
\pythia{8} event generator uses the string model, while \herwig{++}
and \sherpa use the cluster model. Early versions of both models were
originally introduced in \rfr{artru.74.1} but hadronisation studies
did not begin in earnest until an iterative process for jet production
was outlined in \rfr{field.77.1}. Note that in many of these papers
the terms hadronisation and fragmentation are used interchangeably,
although occasionally fragmentation is used to denote both parton
showers and hadronisation.

\begin{subfigures}{2}{Schematic of the final state parton shower from
    \fig{Thr:MonteCarlo} with \subfig{String}~string hadronisation and
    \subfig{Cluster}~cluster hadronisation models applied. For the
    string model the lower gluon creates a kink in the
    string, while for the cluster model, the gluon is forced to
    split. The coloured lines indicate the colour flow of the
    event.\labelfig{Hadronisation}}
  \svgbeg
  \svg{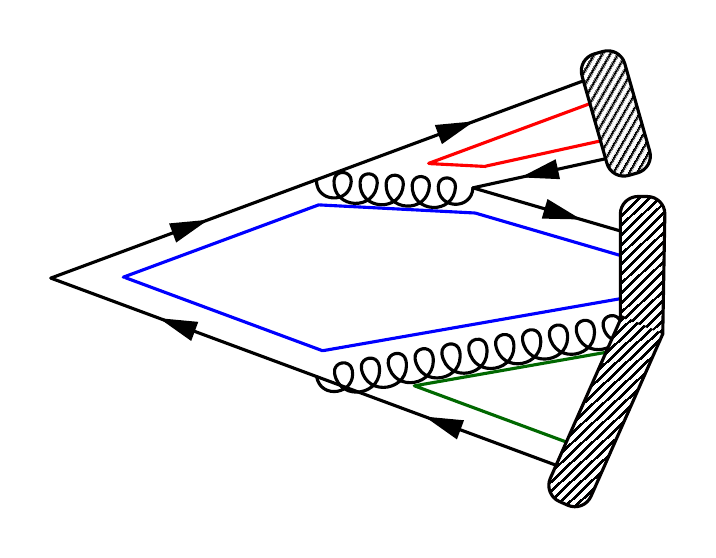} & \svg{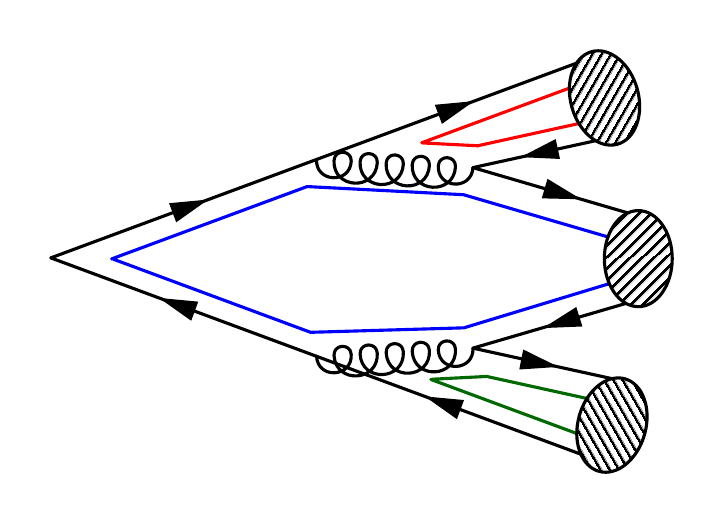} \svgend
\end{subfigures}

\newsubsubsection{String Model}{}

The string model is based on the linear confinement of quarks at large
distances observed from lattice \qcd results; for a study of three
quark systems see {\it e.g.} \rfr{suganuma.11.1}. These results have
shown that the potential for a static multi-quark system can be
modelled with the sum of a linear and Coulomb potential,
\begin{equation}
  V \approx \kappa r - \frac{4\aS}{3r}
\end{equation}
where the linear potential is thought to arise from the
self-interaction of the gluon. Here, $r$ is the distance between the
quarks and $\kappa$ is the string constant. The Coulomb potential is
only significant at small distances, and in the string hadronisation
model, is neglected, as it is expected only to affect the properties
of the hadrons and not their production distribution. In the string
model, quark/anti-quark pairs are connected via colour flux tubes or
strings where the potential arising from the string is the linear
portion of the potential, ${V = \kappa r}$. Here, the string constant
$\kappa$ is estimated to have an energy per unit length of
approximately ${0.2~\gev^2}$ from experimental measurements of hadron
masses~\cite{sjostrand.06.1}.

After the hard process, parton shower, and underlying event steps of a
general purpose Monte Carlo generator, the quarks, anti-quarks, and
gluons are divided into colour singlets dictated by the colour flow of
the event, assuming the large $N_c$ limit. In this limit, a unique
colour is introduced for each \qcd splitting within the event. For
simplicity, consider a colour singlet consisting of a quark and
anti-quark, $q_0\bar{q}_0$, and connected by a colour string. As the
two partons separate from each other, the potential increases linearly
until the string splits into another quark/anti-quark pair,
$q_{1}\bar{q}_1$, as dictated by a fragmentation function. This split
produces two new colour singlets, $q_0\bar{q}_1$ and $q_1\bar{q}_0$,
each connected by their own colour string. The process can be then
iterated until the invariant masses of the colour singlets are small
enough to form mesons. A system which splits $n$ times will produce
$n+1$ final state mesons, $q_0\bar{q}_1$ through $q_n\bar{q}_0$.
Baryons can also be produced in the string model by the creation of
di-quark/anti-di-quark pairs in addition to the creation of just
quark/anti-quark pairs, although the popcorn models of
\rfrs{andersson.85.1} and~\cite{eden.96.1} are more commonly used to
handle baryon production.

Gluons are incorporated into the string model by attaching them to
colour strings between quark/anti-quark pairs. This produces kinks in
the strings that influence the kinematics of the string
splittings. With the inclusion of gluons in the string model, only a
few parameters are required to produce the final state kinematics of
the hadrons, and these kinematics are well matched by experimental
observation, including the string effect first observed by the JADE
experiment in \rfr{jade.81.1}. However, a large number of parameters
are required to describe the flavours of the quark/anti-quark pairs
produced in the string splittings, and so the flavour composition of
the final state hadrons is not predictive and oftentimes does not
match well with experimental observation. An example schematic of the
string model applied to the final state parton shower of
\fig{Thr:MonteCarlo} is given in \fig{Thr:String}.

\newsubsubsection{Cluster Model}{}

The cluster model is based on the property of pre-confinement from
\rfr{amati.79.1}, where the colour singlet combinations of partons, or
clusters, have an invariant mass distribution that does not depend
upon the scale of the hard process, $Q^2$, but rather is only
dependent on the \qcd scale and the parton shower cut-off scale
$Q_0^2$. The large $N_c$ limit is used to determine the colour flow of
the partons and form the colour singlet clusters of quarks where the
gluons carry colour/anti-colour lines and the quarks and anti-quarks
carry either colour or anti-colour. In the cluster hadronisation
model, the clusters are formed by forcing all gluons to split into
quark/anti-quark pairs which enhances the ${g \to q\bar{q}}$ splitting
of \equ{Thr:SplittingFunctions}. Some measurements of jet profiles,
such as \rfr{dasgupta.07.1}, indicate that perhaps this gluon
splitting enhancement is experimentally supported.

After the clusters are formed from the final parton colour singlets,
they are decayed via a series of two-body decays until stable final
state hadrons are reached. The kinematics of the hadrons from the
simple cluster hadronisation model do not match experiment well, and
require a variety of additional phenomenological mechanisms and
parameters to be introduced. However, the flavour composition of the
final state hadrons is controlled by only a few parameters, and
produces a flavour spectrum that provides a better match to experiment
than the string hadronisation model. An example schematic of the
cluster model applied to the final state parton shower of
\fig{Thr:MonteCarlo} is given in \fig{Thr:Cluster}.

\newsubsection{Particle Decays}{Dec}

After hadronisation, a set of colourless final state particles have
been produced, but these particles themselves might still decay. For
example, a pion produced from the hadronisation process might decay
into a muon and neutrino, or a muon produced from the hard process
might decay into an electron or neutrino. These decays can then be
continued until stable particles terminate the decay chain. However,
in typical high energy particle physics experiments the particles
produced are travelling near the speed of light, so particles that
would quickly decay in their own rest frame, such as muons, are
considered stable in the laboratory frame. Consequently, most general
purpose Monte Carlo generators provide a variety of options for when
particles should be decayed. By default, all particles with a mean
lifetime times the speed of light $c\tau$ less than $1~\mathrm{m}$ are
decayed by \pythia{8}, and all other particles are set as stable. With
this default behaviour, particles such as the \wtl and neutral pion
are decayed, while particles such as the muon and charged pion are
not. For full detector simulation, the final state particles after
decays by the Monte Carlo generator are passed to the simulation
software, where material interactions are then modelled.

In the simplest form of particle decays, a list of available decay
channels is supplied for each particle. Each decay channel specifies
the decay products of the channel, as well as the relative decay width
for the channel with respect to the total decay width of the particle,
{\it i.e.} the branching fraction for that channel. Typically the
branching fractions for the decay channels are set from experimental
observation and are not calculated from theory, unlike for resonance
decays, which must be calculated for the invariant mass of the
resonance. The decay channel for the particle is randomly selected
from the available decay channels, weighted by their branching
fractions. The $n$-body decay of the particle is then performed, where
the decay products are kinematically distributed according to
isotropic phase-space.

If the matrix element for the decay is known from either perturbative
theory or some phenomenological model, a more realistic distribution
of the kinematics for the decay products can be produced by weighting
the decays by the matrix element for the decay channel. Additionally,
if the kinematics of the decay are influenced by helicity correlations
from the hard process, further weighting can be applied. In the
remainder of this section, the method used for determining isotropic
phase-space for $n$-body decays used in \chp{Tau}, as well as the
helicity correlation algorithm used in \chp{Tau} are described.

\newsubsubsection{Phase-Space}{}

% See sjostrand.11.1.pdf for more details.
One of the requirements of any decay algorithm is the ability to
distribute an $n$-body decay with isotropic phase-space, {\it i.e.}
the kinematic distribution of the decay products assuming the decay
matrix element, \me, is unity.  For decays involving only massless
decay products, the random momenta and boosts (RAMBO) algorithm of
\rfr{kleiss.85.1} is completely efficient. The RAMBO approach randomly
generates the momenta for the $n$ decay products and then rescales
their invariant mass to match the decaying particle mass. However,
this rescaling is not possible for massive decay products, and so the
RAMBO approach is not suitable for \wtls. Instead, the phase-space
generator for \wtl decays implemented in \pythia{8} is based on the
\m-generator algorithm of \rfr{james.68.1}.

The principle behind the \m-generator algorithm is that two-body
phase-space can be distributed in the rest frame of the decaying
particle by uniformly sampling $\phi$ and $\cos\theta$ where $\phi$
and $\theta$ are the azimuthal and polar angles of the decay
products. Consequently, an $n$-body decay can be written as a series
of intermediate $n-1$ two-body decays, where each two-body decay is
performed in the rest frame of the decaying intermediate particle.

The phase-space element for a two-body decay, using
\equ{Thr:DecayWidth}, is given by,
\begin{equation}
  \dif{\Phi_2(q_0, q_1, q_2)} = \left(\frac{1}{(2\pi)^2 2^2}\right)
  \delta(q_0-q_1-q_2) 
  \frac{\dif{\vec{q}_1}}{E_1}
  \frac{\dif{\vec{q}_2}}{E_2}
  \labelequ{TwoBody}
\end{equation}
where $q_0$, $q_1$, and $q_2$ are the momentum four-vectors for the
decay particle and its two decay products, and $E_1$ and $E_2$ are the
energies of the two decay products. In a similar fashion the
three-body phase-space element can be written as,
\begin{equation}
  \dif{\Phi_3(q_0,q_1,q_2,q_3)} =
  \left(\frac{1}{(2\pi)^5 2^3}\right)
  \delta(q_0-q_1-q_2-q_3) 
  \frac{\dif{\vec{q}_1}}{E_1}
  \frac{\dif{\vec{q}_2}}{E_2}
  \frac{\dif{\vec{q}_3}}{E_3}
\end{equation}
where $q_3$ is the momentum four-vector for the third decay
product. An intermediate decay product with momentum ${q_{12} =
  q_1 + q_2}$ can be introduced and the three-body
phase-space element can be rewritten as,
\begin{alignat*}{2}\labelali{}
  \dif{\Phi_3(q_0,q_1,q_2,q_3)}
  &=&\:&\left(\frac{1}{(2\pi)^5 2^3}\right)
  \delta(q_0-q_1-q_2-q_3) 
  \frac{\dif{\vec{q}_1}}{E_1}
  \frac{\dif{\vec{q}_2}}{E_2}
  \frac{\dif{\vec{q}_3}}{E_3} \\ &&
  &\delta(q_{12}-q_1-q_2)\delta(q_{12}^2-\m_{12}^2)\sdif{q_{12}}\sdif{\m_{12}^2}
  \\
  &=&\:&\left(\frac{2}{\pi}\right)
  \dif{\Phi_2(q_0,q_{12},q_3)}\m_{12}\sdif{\m_{12}}\sdif{\Phi_2(q_{12},q_1,q_2)}
\end{alignat*}
where in the second step, the three-body phase-space element has been
expressed in terms of the two-body decay ${p_0 \to p_{12}p_3}$, the
two-body decay ${p_{12} \to p_1p_2}$, and the intermediate mass
$\m_{12}^2$. Here $p_i$ indicates particle $i$. The process can then
be recursively applied to $n$-body phase-space,
\begin{align*}\labelali{NBody}
  \dif{\Phi_n(q_0,\ldots,q_n)} = &\left(\frac{2}{\pi}\right)^{n-2}
  \dif{\Phi_2(q_0,q_{1 \ldots n-1},q_n)}
  \m_{1 \ldots n-1} 
  \sdif{\m_{1 \ldots n-1}} \\
  &\ldots \dif{\Phi_2(q_{123},q_{12},q_3)}
  \m_{12} \sdif{\m_{12}}
  \sdif{\Phi_2(q_{12},q_1,q_2)} 
\end{align*}
where $n-1$ two-body decays are performed with $n-2$ intermediate
decay products.

The two-body phase-space element of \equ{Thr:TwoBody} can be
rewritten in terms of invariant masses,
\begin{equation}
  \dif{\Phi_2(\m_0^2,\m_1^2,\m_2^2)} = \left(\frac{1}{(2\pi)^22^2}\right)
  \frac{\sqrt{\mathcal{K}(\m_0^2,\m_1^2,\m_2^2)}}{2\m_0^2} \sdif{\Omega_0}
\end{equation}
using the centre-of-mass frame of the two-body decay where
$\dif{\Omega_0}$ is the solid angle element for the rest frame of the
decaying particle $p_0$ and,
\begin{equation}
  \mathcal{K}(\m_0^2,\m_1^2,\m_2^2) = (\m_0^2 - \m_1^2 - \m_2^2)^2 - 4\m_1^2\m_2^2
\end{equation}
is the K\"all\'en triangle function. The $n$-body phase-space element
of \equ{Thr:NBody} then becomes,
\begin{align*}\labelali{NBodyMass}
  \dif{\Phi_n(\m_0^2,\ldots,\m_n^2)} =
  &\left(\frac{1}{\m_02^{4n-2}\pi^{3n-4}}\right)
  \frac{\sqrt{\mathcal{K}(\m_0^2,\m_{1 \ldots n-1}^2, m_n^2)}}{\m_0}
  \sdif{m_{1 \ldots n-1}} \sdif{\Omega_0} \\
  &\ldots \frac{\sqrt{\mathcal{K}(\m_{123}^2,\m_{12}^2, m_3^2)}}{\m_{123}}
  \sdif{m_{12}} \sdif{\Omega_{123}} \\
  &\frac{\sqrt{\mathcal{K}(\m_{12}^2,\m_{1}^2, m_2^2)}}{\m_{12}}
  \sdif{\Omega_{12}} 
\end{align*}
which is differential in terms of the solid angle elements
$\dif{\Omega}$ and the intermediate invariant mass elements
$\dif{\m}$.

The relationship of \equ{Thr:NBodyMass} can then be translated into a
Monte Carlo algorithm for an $n$-body decay where first the
intermediate masses are randomly sampled, and then the solid
angles. To sample the intermediate masses, select $n-2$ random numbers
$r_1$ to $r_{n-2}$ and order them so that ${r_i < r_{i+1}}$. The $n-2$
intermediate masses $\m_{12}$ through $\m_{1 \ldots n-1}$ can then be
generated by,
\begin{equation}
  \m_{1 \ldots i} = \sum_j^{i} \m_j + r_{i-1}\Delta \m \equcomma 
  \Delta \m = \m_0 - \sum_j^n \m_j
  \labelequ{IntermediateMasses}
\end{equation}
where all summation indices begin at $1$ and $i$ runs between $2$ and
$n-1$. This set of intermediate masses then covers the intermediate
mass space with a weight,
\begin{equation}
  \mathcal{J} = \frac{\sqrt{\mathcal{K}(\m_0^2,\m_{1 \ldots n-1}^2, m_n^2)}}{\m_0}
  \ldots \frac{\sqrt{\mathcal{K}(\m_{123}^2,\m_{12}^2,
      m_3^2)}}{\m_{123}}
  \frac{\sqrt{\mathcal{K}(\m_{12}^2,\m_{1}^2, m_2^2)}}{\m_{12}}
  \labelequ{MassWeight}
\end{equation}
which is proportional to the integrand of \equ{Thr:NBodyMass}. The
invariant masses can then be randomly sampled by using the
accept-and-reject method where an additional uniform random number
$r_0$ between $0$ and $1$ is selected, and if $r_0$ is less than
$\mathcal{J}/\mathcal{J}_\mathrm{max}$, the masses are accepted. Here,
$\mathcal{J}_\mathrm{max}$ is the maximum $\mathcal{J}$ which can be
determined empirically.

\begin{subfigures}{2}{Example of the \m-generator algorithm for a
    five-body decay where the particle $p_0$ decays into its products
    $p_1$ through $p_5$ via the intermediate particles $p_{1234}$,
    $p_{123}$, and $p_{12}$ indicated by the dashed lines. Each vertex
    represents a two-body decay and the solid lines represent the real
    particles.}
  \fmpbeg 
  \fmp[1]{Decays} & \sidecaption \fmpend
\end{subfigures}

After the intermediate masses have been sampled, the individual
two-body decays can then be performed. The decay ${p_0 \to p_{1 \ldots
    n-1} p_n}$ is first performed, in the rest frame of $p_0$. This is
done by uniformly sampling $\phi$ between $0$ and $2\pi$ and
$\cos\theta$ between $-1$ and $1$. The momenta for $q_{1 \ldots n-1}$
and $n$ are then fully determined by these two angles. Next the decay,
${p_{1 \ldots n-1} \to p_{1 \ldots n-2} p_{n-1}}$ is performed in the
rest frame of $p_{1 \ldots n-1}$ using the same method for sampling
$\phi$ and $\cos\theta$. The decay products are then boosted back into
the rest frame of $p_0$. This process continues until all $n-1$
two-body decays have been performed. A schematic of the process for a
five-body decay is shown in \fig{Thr:Decays}, where each solid line
indicates either the decaying particle $p_0$ or one of its decay
products, and each dashed line indicates an intermediate
particle. Each vertex represents a two-body decay.

\newsubsubsection{Helicity Correlations}{}

Using the \m-generator algorithm of \equs{Thr:IntermediateMasses}
and~\ref{equ:Thr:MassWeight}, an $n$-body isotropic decay can be
performed. Furthermore, the decay matrix element for the decay, \me,
can be calculated and used to weight the isotropic decays. However,
this does not take into account helicity correlations between the
decays of particles within the event. This is of particular importance
for \wtl decays where the helicities of the \wtls play an important
role in their decays. The helicity correlation algorithm implemented
in \pythia{8} for the \wtl decays of \chp{Tau}, is described here and
is based on the algorithm expanded by Richardson in
\rfr{richardson.01.1} and proposed in its original form by Collins and
Knowles in \rfrs{collins.87.1} and~\cite{knowles.90.1}.

First, the momenta of the two-to-$n$ hard process is calculated using
the matrix element weight,
\begin{equation}
  \mathcal{W} = {\rho_1}_{\kappa_1\kappa_1'}{\rho_2}_{\kappa_2\kappa_2'}
  \me_{\kappa_1\kappa_2 \lambda_1 \ldots \lambda_n}
  \me_{\kappa_1'\kappa_2' \lambda_1' \ldots \lambda_n'}^\dagger
  \prod_{i}^{n} {D_i}_{\lambda_i\lambda_i'}
  \labelequ{WeightHard}
\end{equation}
where $\kappa_i$ is the helicity of incoming particle $p_i$,
$\lambda_i$ is the helicity of outgoing particle $p_i$, $\rho_i$ is
the helicity density matrix for the incoming particle $p_i$, \me is
the helicity matrix element for the hard process, and $D_i$ is the
decay matrix for the outgoing particle $p_i$. Here, repeated indices
are summed over and the product of $D_i$ is over all $n$ outgoing
particles. The helicity density matrix for a two helicity state
incoming particle is given by,
\begin{equation}
  \rho_{\kappa\kappa'} =
  \begin{pmatrix}
    \frac{1}{2}(1 + \mathcal{P}_z) & 0 \\
    0 & \frac{1}{2}(1 - {\mathcal{P}_z}) \\
  \end{pmatrix}
\end{equation}
where $\mathcal{P}_z$ is the longitudinal polarisation of the incoming
particle with respect to the beam axis. The decay matrices $D_i$ for
the outgoing particles begin as the identity matrix.

After the hard process is generated, one of the outgoing particles
$p_j$ is chosen, and its helicity density matrix $\rho_j$ is calculated by,
\begin{equation}
  {\rho_j}_{\lambda_j \lambda_j'} = 
  {\rho_1}_{\kappa_1\kappa_1'}{\rho_1}_{\kappa_2\kappa_2'}
  \me_{\kappa_1\kappa_2; \lambda_1 \ldots \lambda_n}
  \me_{\kappa_1'\kappa_2'; \lambda_1' \ldots \lambda_n'}^\dagger
  \prod_{i \neq j}^n {D_i}_{\lambda_i\lambda_i'}
  \labelequ{RhoHard}
\end{equation}
and normalised, ${\rho_j = \rho_j/\mathrm{Tr}\left(\rho_j \right)}$,
such that its trace is one. An $m$-body decay channel for particle
$p_j$ is selected, and then $p_j$ is decayed according to the matrix
element weight,
\begin{equation}
  \mathcal{W} = 
  {\rho_j}_{\lambda_0\lambda_0'}
  \me_{\lambda_0\lambda_1 \ldots \lambda_m}
  \me_{\lambda_0'\lambda_1' \ldots \lambda_m'}^\dagger
  \prod_k^m {D_k}_{\lambda_k\lambda_k'}
  \labelequ{WeightDecay}
\end{equation}
where $\mathcal{M}$ is now the helicity matrix element for the decay
of $p_j$, $\lambda_0$ is the helicity of $p_j$, and $\lambda_k$ is the
helicity of decay product $p_k$. Given that the decay channel of $p_j$
is an $m$-body decay, the product of $D_k$ runs from $1$ to $m$.

After decaying $p_j$, one of its $m$ outgoing particles $p_l$ is
picked at random and the helicity density matrix for this particle is
calculated by,
\begin{equation}
  {\rho_l}_{\lambda_l\lambda_l'} = {\rho_j}_{\lambda_0,\lambda_0'} 
  \mathcal{M}_{\lambda_0;\lambda_1 \ldots \lambda_m}
  \mathcal{M}_{\lambda_0';\lambda_1' \ldots \lambda_m'}^\dagger \prod_{k
    \neq l}^m {D_k}_{\lambda_k\lambda_k'}
  \labelequ{RhoDecay}
\end{equation}
and then normalised such that the trace of the helicity density matrix
is one. Here \me is still the helicity matrix element for the decay of
$p_j$ and $D_k$ is the decay matrix for outgoing particle $p_k$ from
the decay of $p_j$. The process of selecting a particle from the
decay, calculating the helicity density matrix using
\equ{Thr:RhoDecay}, and then decaying the particle using the weight of
\equ{Thr:WeightDecay}, continues until a particle is reached where all
the decay products are stable. Up until this point, all decay matrices
$D$ have been initialised as the identity matrix. Assuming that all
the decay products $p_k$ of particle $p_j$ have either already been
decayed or are stable, the decay matrix for $p_j$ is then updated by,
\begin{equation}
  {D_j}_{\lambda_0\lambda_0'} = 
  \mathcal{M}_{\lambda_0;\lambda_1 \ldots \lambda_m}
  \mathcal{M}_{\lambda_0';\lambda_1' \ldots \lambda_m'}^\dagger \prod_{k}^m
  {D_k}_{\lambda_k\lambda_k'}
  \labelequ{DecayMatrix}
\end{equation}
which must be normalised such that the trace of $D_j$ is one.

After the end of a decay chain is reached, a new undecayed particle
from the decay prior to the terminating decay is randomly selected. If
this particle is from the hard process, \equ{Thr:RhoHard} is used to
calculate its helicity density matrix, otherwise \equ{Thr:RhoDecay} is
used. Note that at this point, the decay matrices $D$ used in the
calculation of the helicity density matrix are no longer necessarily
identity matrices, and for particles that have already been decayed,
are given by \equ{Thr:DecayMatrix}. The process of performing decays
using the weight of \equ{Thr:WeightDecay} until reaching a terminating
decay with only stable particles, tracing backwards and calculating
the decay matrix of the prior decay with \equ{Thr:DecayMatrix}, and
then moving forward again, is continued until all particles in the
event have been decayed.

\newchapter{Tau Leptons}{Tau}

The \wtl, unlike the lighter muon and electron, can decay both
leptonically and hadronically. Consequently, the \wtl provides an
important bridge between electroweak and \qcd theory. Additionally, the
low mass of the \wtl with respect to the energy regime of perturbative
\qcd places the hadronic decay of the \wtl at the border of
non-perturbative and perturbative regimes, allowing \wtl decays to be
modelled through a variety of theories including both lattice \qcd and
chiral perturbation theory~\cite{wingate.95.1, colangelo.96.1}.

While the hadronic decay of the \wtl is important for low energy \qcd,
\wtl decays are also important for both direct and indirect \whb
searches. Within the \sm, the minimal Higgs mechanism of \sec{Thr:Lag}
predicts a branching fraction of $\approx 10\%$ for the \whb decaying
into a \wtl pair within the mass window of $115$ to $140~\gev$. For
large portions of parameter space in the \mssm, the neutral \whbs are
predicted to have a branching fraction to \wtl pairs of $\approx 10\%$
and the charged \whb can decay almost exclusively to a \wtl and \wtl
neutrino~\cite{djouadi.97.1}. Further \whb phenomenology is explored
in \chp{Hig}.

Because of the role the \wtl plays in new physics searches, it is
important that current Monte Carlo event generators handle the decay
of the \wtl using the best possible models. The phase-space
distribution of the \wtl decay products can be heavily influenced by
the polarisation of the \wtl, and so it is also important to ensure
that proper spin correlations between the \wtls and their decay
products are modelled. Currently, the \herwig{++}~\cite{\citeherwigpp}
event generator incorporates \wtl decays with full correlation
effects~\cite{grellscheid.07.1} while the \sherpa~\cite{\citesherpa}
event generator incorporates approximate correlation effects
\cite{laubrich.06.1}. The \wtl specific event generator \tauola
performs only \wtl decays but provides approximate correlation effects
for a variety of \wtl decays and can be interfaced with many current
event generators~\cite{golonka.03.1}.

For \pythia{6} and previous versions of \pythia{8} (at or below
$8.145$), leptonic \wtl decays were distributed by the standard
vector-axial matrix element while hadronic decays were distributed by
phase-space weighted by a factor of $2E_{\nu_\tau} / \m_\tau (3 -
2E_{\nu_\tau})$, where $E_{\nu_\tau}$ is the energy of the neutrino in
the rest frame of the \wtl \cite{sjostrand.06.1}. No polarisation
information in the decays of the \wtl are included in \pythia{6}. In
this chapter the implementation of well modelled matrix elements with
full spin correlations using the algorithm of \sec{Thr:Dec} for \wtl
decays in \pythia{8} is described.

The spin correlation algorithm does not implement the full recursion
of the algorithm of \sec{Thr:Dec}, as this is not necessary for \wtl
decays, and assumes that the \wtls are produced directly from the hard
process of the event. The isotropic phase-space of the \wtl decays is
distributed using an implementation of the \m-generator outlined in
\sec{Thr:Dec}. The matrix elements used in both the the helicity
correlation algorithm and decay calculations are built from the
Feynman rules of \tab{Thr:Rules} and the vertices from
\figs{Thr:Vertices.Qcd}, \ref{fig:Thr:Vertices.Ewk},
\ref{fig:Thr:Vertices.Sm.Higgs}, and
\ref{fig:Thr:Vertices.Mssm.Higgs}.

In \sec{Tau:Pro} the \wtl production matrix elements available in
\pythia{8} are outlined and in \sec{Tau:Dec} the matrix elements used
for \wtl decays are documented. Finally, in \sec{Tau:Imp} the
implementation of \wtl decays within \pythia{8} is summarised.

\newsection{Tau Lepton Production}{Pro}

In \pythia{8} the hard process is already calculated using helicity
averaged matrix elements and so the first step of using
\equ{Thr:WeightHard} from the spin correlation algorithm of
\sec{Thr:Dec} is not necessary. However, the helicity matrix element
of the hard process is still needed to calculate the helicity density
matrices of the outgoing \wtls from the hard process in
\equ{Thr:RhoHard}. For the case of \wtl production from particles that
are not spin zero, such as ${\z \to \ditau}$, the full information of
the hard process is required, whereas for \wtls produced from spin
zero particles, such as \whbs, the incoming particles of the hard
process are not required. In either case it is important that the
helicity matrix elements match the helicity averaged matrix elements
used in the hard process machinery of \pythia{8}.

Most helicity matrix elements which produce \wtls within \pythia{8}
have been implemented and fall into three broad categories which are
presented here. In \sec{Tau:ProEwk} the available electroweak matrix
elements are outlined, while in \sec{Tau:ProHig} the \whb matrix
elements are presented, and in \sec{Tau:ProOth} approximated helicity
matrix elements from other mechanisms such as $B$-hadron decays, where
the production of the \wtl is not a hard process, are
considered. Additionally, the method used for handling initial state
and final state radiation in the calculation of the hard process
helicity matrix elements is outlined in \sec{Tau:ProRad}.

\newsubsection{Electroweak Production}{ProEwk}

A \wtl can be produced from a hard electroweak process in \pythia{8}
through either a Drell-Yan process or a \wwb. For these two hard
processes the momenta and types of the incoming particles must be
known to fully calculate the helicity matrix element and ensure the
correct polarisation of the outgoing \wtls. The helicity matrix
element for Drell-Yan fermion production mediated by an excited photon is
given by,
\begin{equation}
  \me_\gamma = \frac{\gE{^2}Q_0Q_2}{s}\big(\bar{v}_1 \gamma_\mu
  u_0\big) \big(\bar{u}_3 \gamma^\mu v_2 \big)
  \labelequ{MeGamma}
\end{equation}
using the vertex of \fig{Thr:Gm2FF} and the spinors of
\equ{Thr:Spin1E} where $s$ is the mass of the propagator squared,
$p_0$ and $p_1$ are the incoming fermions, $p_2$ and $p_3$ are the
outgoing fermions, and $Q_i$ is the charge of particle $p_i$. Here the
numerical subscripts indicate the corresponding particle to which the
quantity belongs, {\it i.e.} $\bar{v}_1$ is the anti-spinor for
$p_1$. Additionally, the explicit dependence of the spinors on momentum
and helicity are omitted for brevity. This convention is maintained
throughout the chapter.

The Drell-Yan process can also be mediated by a \wzb for which the
helicity matrix element is given by the vertex of \fig{Thr:Z2FF},
\begin{align*}\labelali{MeZ}
  \me_\z =\, &\frac{\gE{^2}}{16\costw{^2}\sintw{^2}
    \left(s - \m_\z^2 + i\frac{\Gamma_\z}{\m_\z}\right)} \bigg(\bar{v}_1
  \gamma^\mu \big(v_0 - a_0\gamma^5\big) u_0\bigg) \\
  & \bigg(g_{\mu\nu} - \frac{q_\mu q_\nu}{\m_\z^2}\bigg)
  \bigg(\bar{u}_3 \gamma^\nu \big(v_2 -
  a_2\gamma^5\big)v_2 \bigg)
\end{align*}
where $a_i$ and $v_i$ are the axial and vector couplings of fermion
$p_i$ to the \wzb given in \tab{Thr:VaCouplings}. When both incoming
fermions are oriented along the $z$-axis and their helicities are
equal, ${\lambda_0 = \lambda_1}$, the full matrix element is
zero. This simplification is used for numerical speed.

In \pythia{8} it is possible to produce \wtls from a Drell-Yan hard
process with either the excited photon matrix element or the \wzb
matrix element, or full interference through both the excited photon
and \wzb matrix elements. When full interference is requested the
helicity matrix element for the process is,
\begin{equation}
  \me_\mathrm{DY} = \me_\gamma + \me_\z
  \labelequ{MeDy}
\end{equation}
where the simplification for the \wzb matrix element is still made
when the incoming fermions are oriented along the $z$-axis and
$\lambda_0 = \lambda_1$.

The matrix element for fermions produced from a \wwb created in the
hard process is given by,
\begin{equation}
  \me_\w \propto \big( \bar{v}_1 \gamma_\mu(1 - \gamma_5)
  u_0 \big) \big( \bar{u}_3 \gamma^\mu(1 - \gamma_5) v_2 \big)
  \labelequ{MeW}
\end{equation}
where the $s$-channel has been assumed. Here $p_0$ and $p_1$ are the
incoming particles and $p_2$ and $p_3$ are the outgoing
fermions. Because $\me_\w$ is not combined with any other matrix
elements for interference effects, unlike the $\gamma$ and \z matrix
elements, the additional factors of proportionality from the \su[2]
gauge coupling have been omitted for numerical simplicity.

The effects of helicity correlations in \wtls from the electroweak
matrix elements outlined above are most readily observed in the rest
frame of the decaying electroweak boson where the \wtl decays into a
\wtl neutrino and a charged pion. In this frame the number of events,
$N$, is proportional to,
\begin{equation}
  N \propto 1 + 2\mathcal{P}_\tau \left( \frac{2E_{\pi^-}}{\sqrt{s}}
    - \frac{1}{2} \right)
  \labelequ{FractionalE}
\end{equation}
where $E_{\pi^-}$ is the energy of the charged pion in the rest frame
of the electroweak boson, $\sqrt{s}$ is the interaction energy, and
$\mathcal{P}_\tau$ is the average polarisation of the \wtls produced
from this type of event. For on-shell \wwb production the average \wtl
polarisation should be $\mathcal{P}_\tau \approx -1$, while the
average \wtl polarisation from an on-shell \wzb depends upon the type
of incoming fermions, but for proton-proton collisions is
$\mathcal{P}_\tau \approx -0.15$. For the case of the photon, the
average \wtl polarisation is zero.

\begin{subfigures}{2}{Comparisons of the fractional energy of a
    charged pion from a \wtl decay in the rest frame of the
    electroweak boson producing the \wtl. The distributions are
    sensitive to the average \wtl polarisations and are given for
    \pythia{8}, \herwig{++}, and \tauola with \wtl
    production from a \subfig{q_q.Z.16_211.E_211}~\wzb hard process
    and a \subfig{q_qp.W.16_211.E_211}~\wwb hard
    process.\labelfig{EwkPolarisation}}
  \svgbeg
  \svg{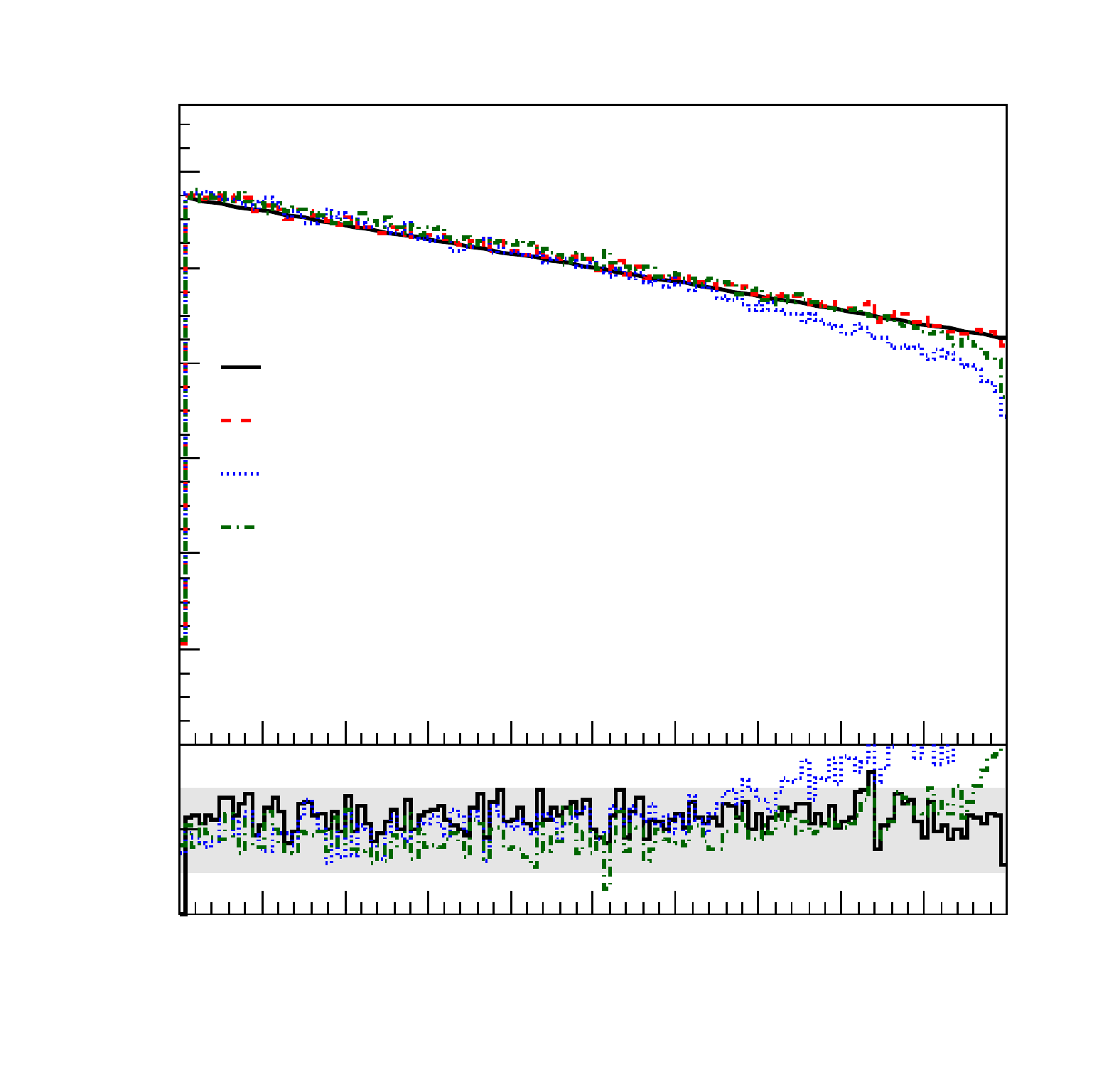} & \svg{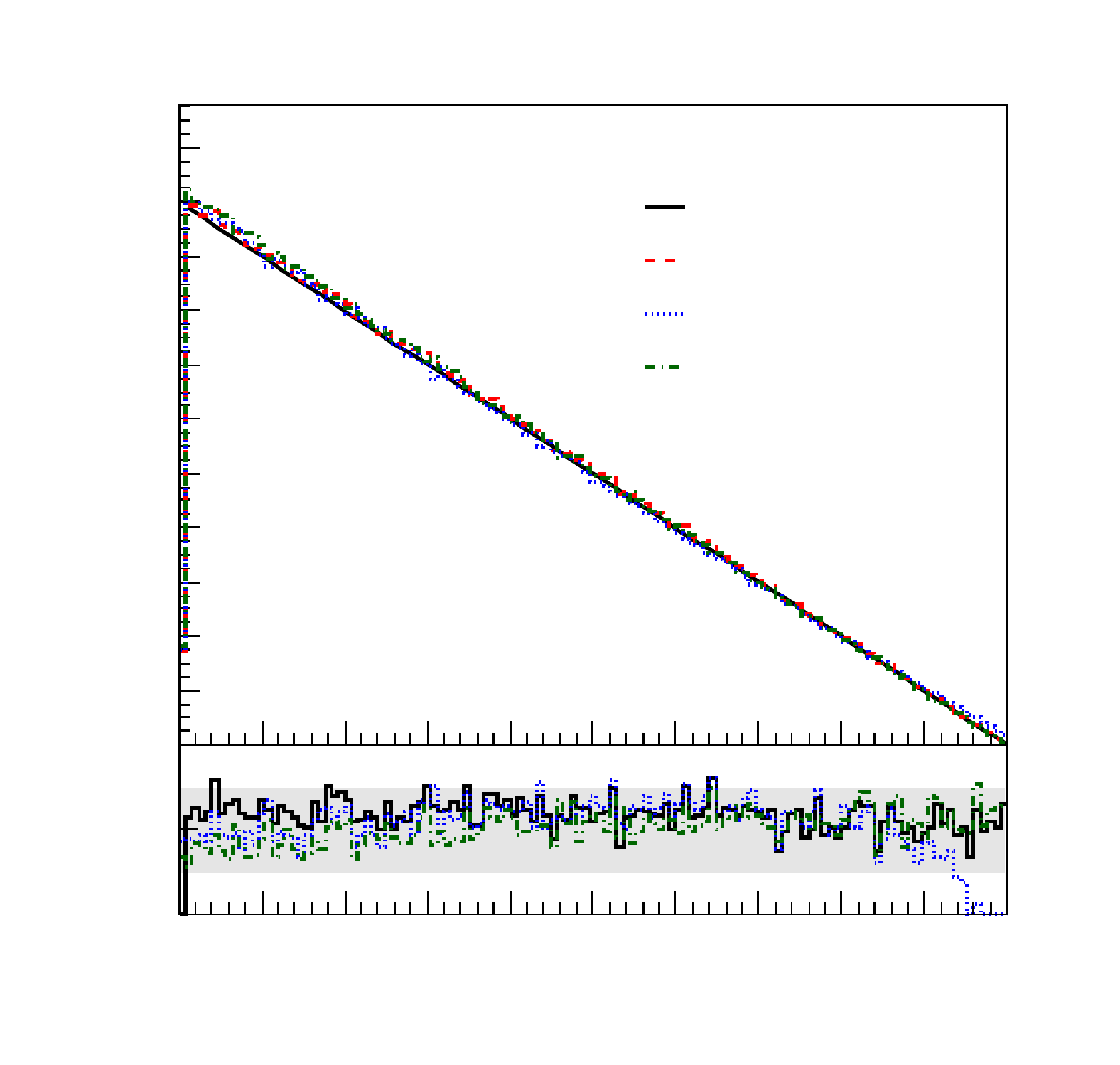} \svgend
\end{subfigures}

In \fig{Tau:EwkPolarisation} the energy of the charged pion, in the
rest frame of the intermediate electroweak boson, is plotted for both
a \wzb hard process and \wwb hard process. For the \wzb hard process
the \z is produced with all photon interference turned off. All plots
within this chapter, unless specified otherwise, are generated from
proton-proton simulations at ${\sqrt{s} = 14~\tev}$, although this
does not affect most distributions. A total of $10^6$ events are
generated for each sample. Additionally, the lower subplot of each
figure provides the difference divided by the statistical uncertainty
between the \pythia{8} result and any other generator distributions or
analytic functions in the plot. The grey band indicates a difference
range within three standard deviations. As can be seen, \pythia{8}
agrees well with both \herwig{++} and \tauola. Additionally,
\pythia{8} matches the expected distribution of \equ{Tau:FractionalE},
assuming $\mathcal{P}_\tau = -0.15$ for \fig{Tau:q_q.Z.16_211.E_211}
and $\mathcal{P}_\tau = -1$ for \fig{Tau:q_qp.W.16_211.E_211}.

\newsubsection{Higgs Boson Production}{ProHig}

Because \whbs, either \sm or beyond, are spin zero, the production of
the \whb does not influence the decay correlation of the \wtls, and
only the vertex factor for the \whb coupling with the two fermions is
needed. For \cp-even \whbs the \whb is predicted by the \sm and \mssm
as a scalar and so the helicity matrix element is proportional to,
\begin{equation}
  \mathcal{M}_{\cp-\mathrm{even}} \propto \left(\frac{i g_w \m_2}{2 \m_\w}\right)
  \bar{u}_3 F_{\h p_2 p_3} v_2
  \labelequ{MeEven}
\end{equation}
for a neutral \whb coupling with two fermions, where $g_w$ is the
\su[2] gauge coupling, $\m_2$ is the mass of one of the outgoing
fermions, and $F_{\h p_2 p_3}$ is the vector coupling of the Higgs to
the outgoing fermions $p_1$ and $p_2$ where $p_0$ is the incoming
\whb. Within \pythia{8} these couplings are not included in the
calculation of the helicity matrix element for numerical simplicity
and speed but are given for the \sm \whb, \hH, in the vertex of
\fig{Thr:H02FF} and for the \mssm light \whb, \hhz, and heavy \whb,
\hHz, in the vertices of \figs{Thr:H12FuFu} and \ref{fig:Thr:H22FuFu}
for $u$-type fermions and \figs{Thr:H12FdFd} and \ref{fig:Thr:H22FdFd}
for $d$-type fermions.

For the \cp-odd \mssm \whb the vertex factor for the
coupling of the \whb to fermions is proportional to,
\begin{equation}
  \mathcal{M}_{\cp-\mathrm{odd}} \propto -\left(\frac{g_w \m_2}{2
      \m_\w}\right) \bar{u}_3 F_{\h p_2 p_3} \gamma^5 v_2
  \labelequ{MeOdd}
\end{equation}
where the coupling constant $F_{\h p_2 p_3}$ is given in the vertices
of \figs{Thr:H32FuFu} and \ref{fig:Thr:H32FdFd} for the \mssm \cp-odd
\whb, \hAz. The vertex factors of \figs{Thr:H4p2FF} and
\ref{fig:Thr:H4m2FF} for the charged \whb, \hHpm, gives the matrix
element,
\begin{equation}
  \mathcal{M}_{\hHpm} \propto \left(\frac{i g_w}{2 \sqrt{2}
      \m_\w}\right) \bar{u}_3
  \Big((\m_2\tan\beta + \m_3\cot\beta) \pm (\m_2\tan\beta - \m_3\cot\beta)
  \gamma^5 \Big) v_2
  \labelequ{MeCharged}
\end{equation}
where $p_1$ is a $d$-type fermion and $p_2$ is a $u$-type
fermion. Note the change in sign between the vector and axial portions
of the vertex depend upon the charge of the \whb.

\begin{subfigures}{2}{The same distributions as
    \fig{Tau:EwkPolarisation} but for \wtl production from a
    \subfig{q_q.H1.16_211.E_211}~\cp-even light \whb hard process and
    a \subfig{q_qp.H4.16_211.E_211}~charged \whb hard
    process.\labelfig{HiggsPolarisation}}
  \svgbeg 
  \svg{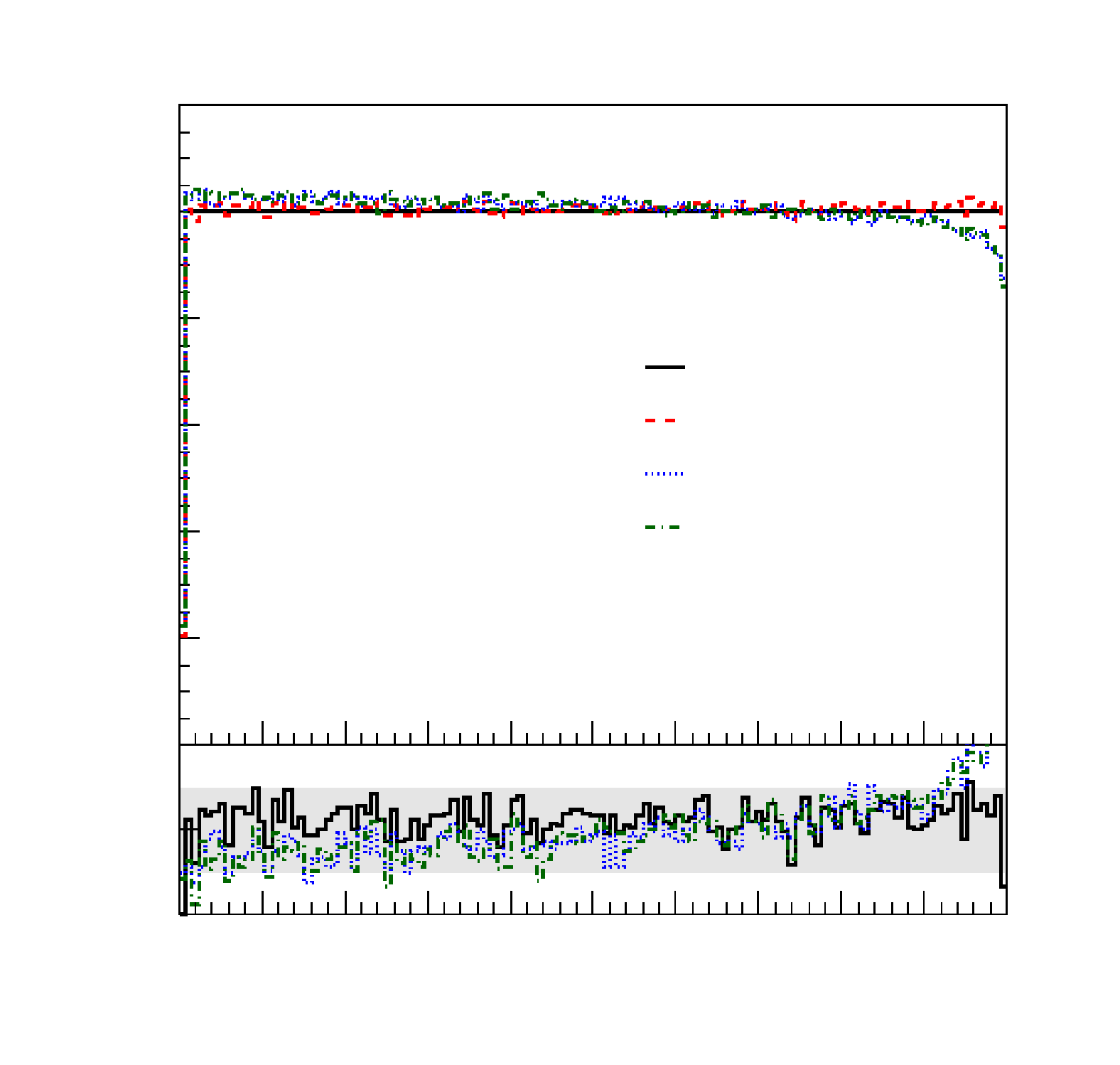} & \svg{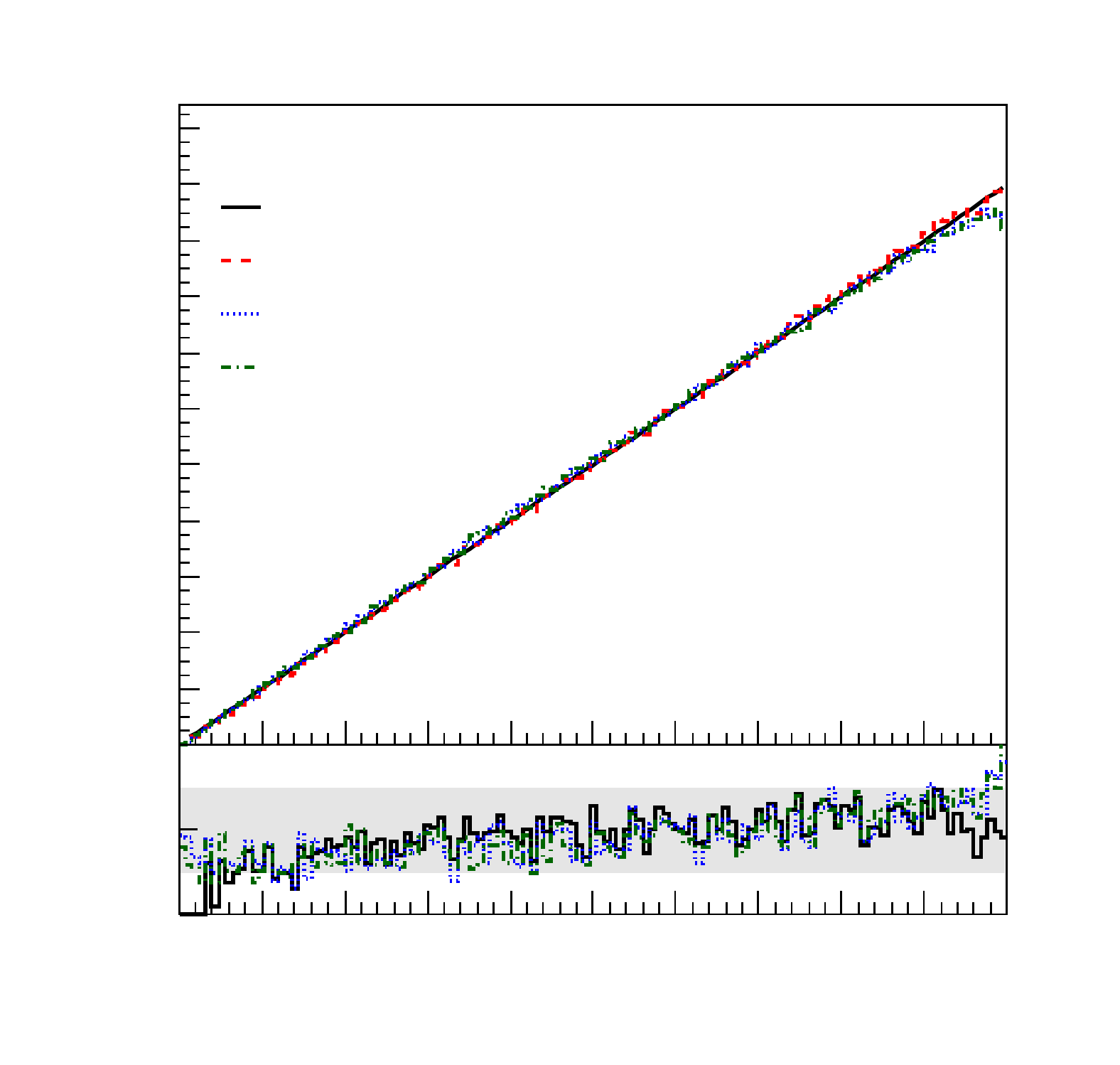}
  \svgend
\end{subfigures}

Because the \whbs are spin zero, the equivalent distributions of
\fig{Tau:EwkPolarisation} yield flat distributions, shown in
\fig{Tau:q_q.H1.16_211.E_211} where $\mathcal{P}_\tau = 0$ is used for
\equ{Tau:FractionalE}. However, \wtls produced from the charged \mssm
\whb have an average polarisation of $P_\tau \approx +1$ as shown in
\fig{Tau:q_qp.H4.16_211.E_211} where \equ{Tau:FractionalE} is plotted
for $P_\tau = +1$. As can be seen, there is good agreement between
\pythia{8}, \herwig{++}, \tauola, and the expected distribution of
\equ{Tau:FractionalE} in both of these plots.

The \cp-even \whb of \equ{Tau:MeEven} and the \cp-odd \whb of
\equ{Tau:MeOdd} produce a correlation in the polarisation of \wtls
that can be seen by plotting the acoplanarity angle $\phi^*$ between
the two pions from \wtl decays into a \wtl neutrino and single charged
pion~\cite{kramer.94.1}. The acoplanarity angle is defined as the
azimuthal angle between the two decay planes of the \wtl in the rest
frame of the \whb and can be written as,
\begin{equation}
  \phi^* = \mathrm{acos}(\vec{n}_2 \cdot \vec{n}_3)
  \labelequ{Acoplanarity}
\end{equation}
where the vector $\vec{n}_i$ is given by,
\begin{equation}
  \vec{n}_i = \frac{\vec{q}_i \times \vec{q}_1}{\abs{\vec{q}_i \times
      \vec{q}_1}}
\end{equation}
and $q_0$ is the $\tau^+$ three-momentum, $q_1$ is the $\tau^-$
three-momentum, $q_2$ is the $\pi^+$ three-momentum, and $q_3$ is the
$\pi^-$ three-momentum~\cite{was.02.1}.

\begin{subfigures}{2}{Comparisons of the acoplanarity angle, $\phi^*$,
    defined in \equ{Tau:Acoplanarity} for \wtls produced from a
    \subfig{q_q.H1.16_211.phi_211_211}~\cp-even \whb and a
    \subfig{q_q.H3.16_211.phi_211_211}~\cp-odd \whb where the \wtl
    decays through the ${\tauto \pi^-}$ channel.\labelfig{HiggsCorrelation}}
  \svgbeg 
  \svg{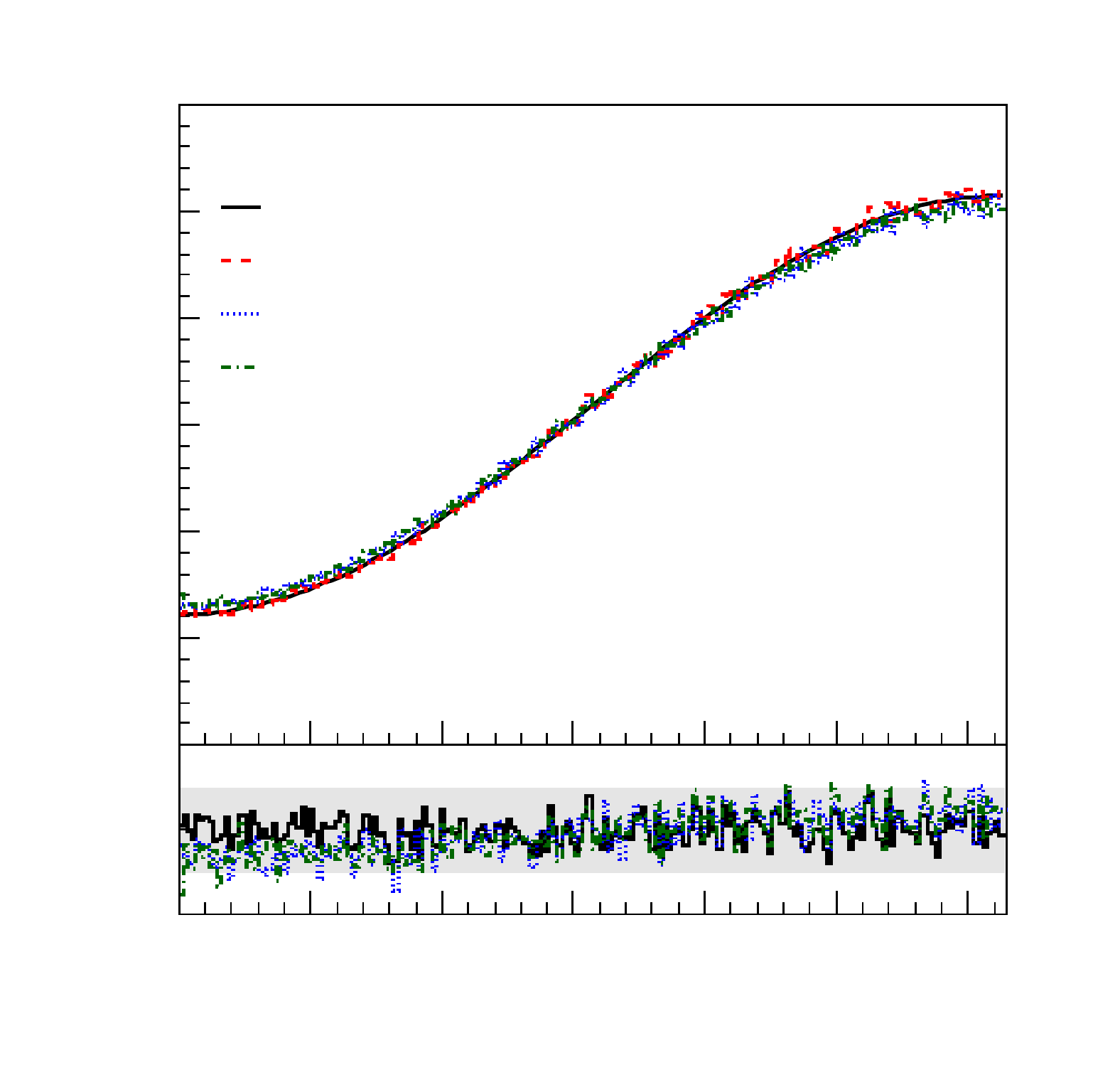} & \svg{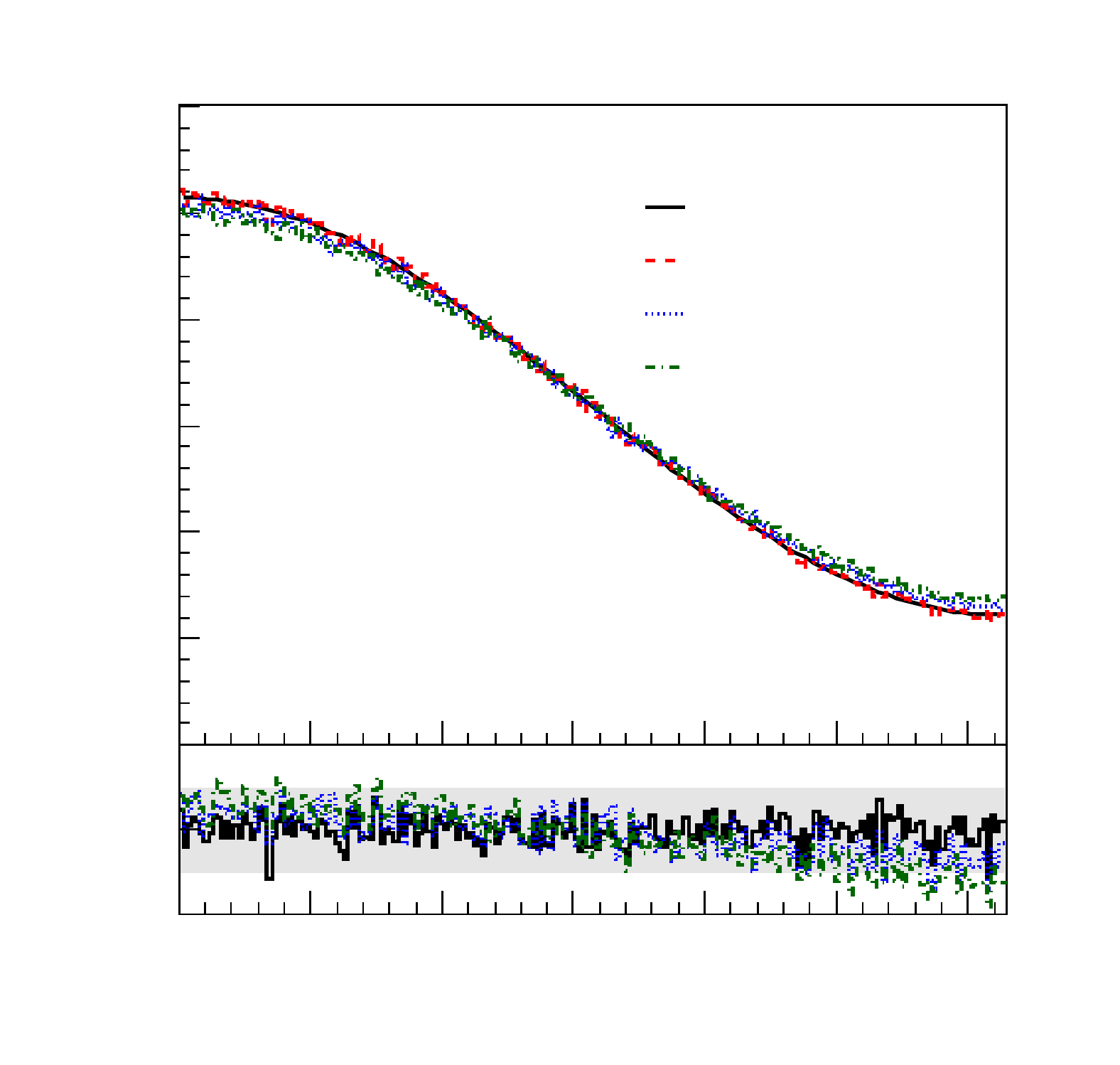}
  \svgend
\end{subfigures}

\Fig{Tau:HiggsCorrelation} provides distributions of the acoplanarity
angle for the \cp-even \whb in \fig{Tau:q_q.H1.16_211.phi_211_211} and
\cp-odd \whb in \fig{Tau:q_q.H3.16_211.phi_211_211}, where the \whbs
are decaying into \wtl pairs which are further decaying into final
states with a \wtl neutrino and charged pion. The number of events,
$N$, is proportional to,
\begin{equation}
  N \propto 1 \pm \frac{\pi^2}{16} \cos \phi^*
\end{equation}
for the \cp-even and \cp-odd \whbs
respectively~\cite{kramer.94.1}. All three generators match the
theoretically expected distribution of \equ{Tau:Acoplanarity} well.

\begin{subfigures}[t]{2}{Distribution of ${2E_{\pi^-}/\sqrt{s}}$ from
    ${\tauto \pi^-}$ decays where the \wtl is produced from a ${\hH
      \to \z\z}$ hard process. The subplot compares the difference
    divided by uncertainty between the \pythia{8} distribution and the
    \herwig{++}, \tauola, and \equ{Tau:FractionalE}
    distributions.\labelfig{ZzPolarisation}}
  \svgbeg 
  \svg[1]{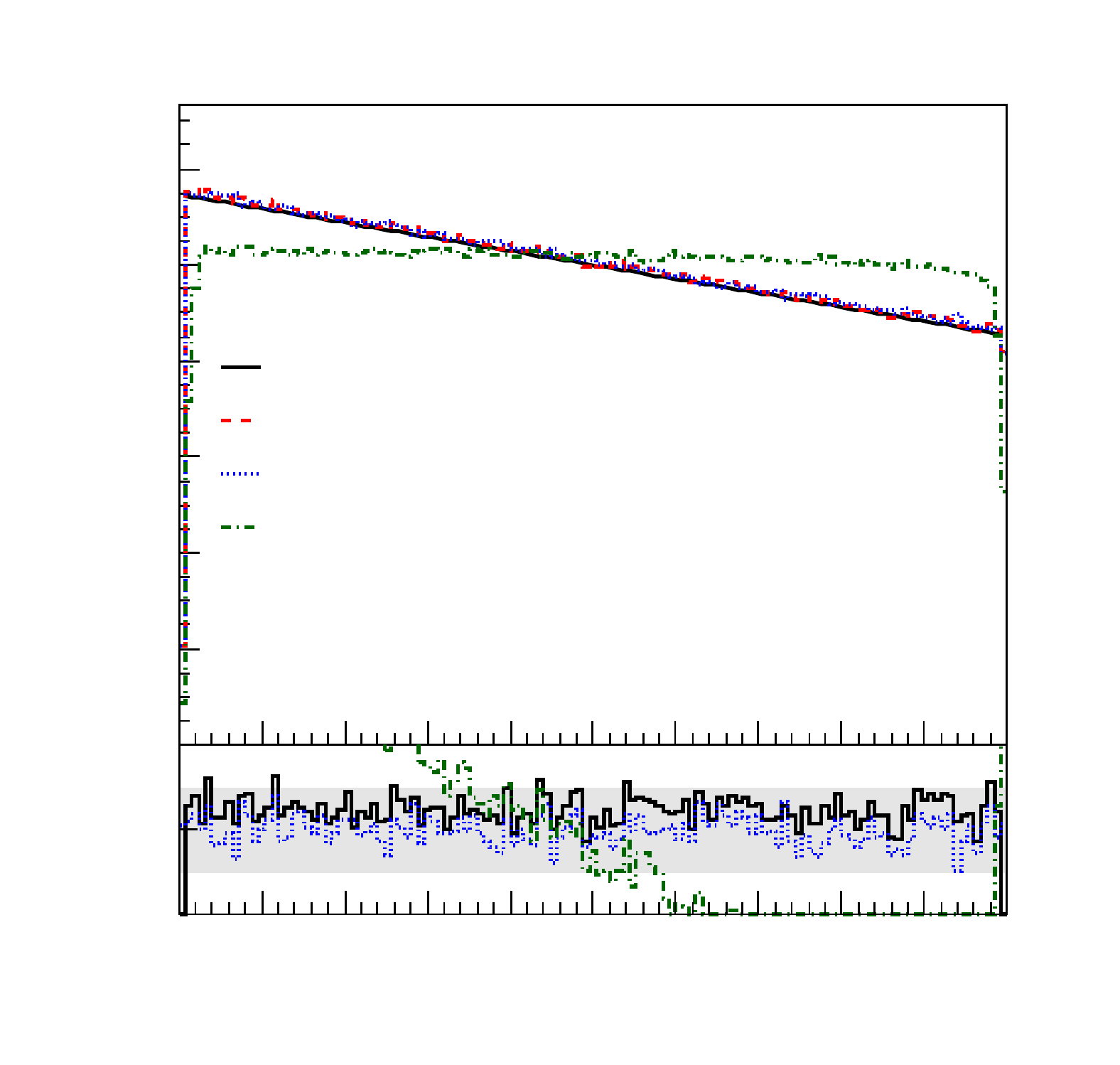} & \sidecaption
  \svgend
\end{subfigures}

\Whbs can also decay into \wzb and \wwb pairs which then subsequently
decay into final states with \wtls. These decays are treated in
\pythia{8} by assuming the electroweak bosons produced from the \whb
are unpolarised and so the matrix element,
\begin{equation}
  \me_{\z\z} \propto {\varepsilon_0}_\mu \bigg(\bar{u}_3 \gamma^\mu \big(v_2 -
    a_2\gamma^5\big)v_2 \bigg)
  \labelequ{MeZZ}
\end{equation}
is used for \wtls produced from a \wzb where $\varepsilon_0$ is the
polarisation vector of the \wzb given by \equ{Thr:Spin2E} and $p_1$
and $p_2$ are the two \wtls. The analytic distribution of
\equ{Tau:FractionalE} also applies for \wtls produced with this
mechanism, where $\mathcal{P}_\tau \approx -0.15$. In
\fig{Tau:ZzPolarisation} the distribution of ${2E_{\pi^-}/\sqrt{s}}$
for \wtls produced from a \whb decaying into a \wzb pair hard process,
where the \wtls decay via ${\tauto \pi^-}$, is given. Both \pythia{8}
and \herwig{++} model this correctly, while \tauola does not, as
compared to \equ{Tau:FractionalE} using $\mathcal{P}_\tau =
-0.15$. For \wtls produced from ${\phi \to \w\w}$, the polarisation of
the \wtl is assumed to be ${\mathcal{P}_\tau = -1}$, as the \wwb will
have a mass much larger than the \wtl, and so no specialised matrix
element is needed.

\newsubsection{Other Production}{ProOth}

Within \pythia{8}, \wtls can also be produced from a variety of decays
which are not the hard process or do not match the electroweak or \whb
hard processes of \secs{Tau:ProEwk} and \ref{sec:Tau:ProHig}. A
special case is top quark production and decay, which is considered as
a hard process and is handled using the \wwb helicity matrix element
of \equ{Tau:MeW} as the top quark, bottom quark, \wtl, and \wtl
neutrino from the process are all well defined within the \pythia{8}
event record.

For the case of \wtls produced from $D$ or $B$-mesons, however, the
constituent quarks are not recorded in the \pythia{8} event record,
and it is necessary to approximate the incoming particles of the \w
helicity matrix element. When the meson decays into another meson plus
a \wtl and \wtl neutrino, the incoming quark momentum is approximated
as the momentum of the decaying meson, while the momentum of the
outgoing quark is approximated as the incoming meson momentum less the
\wtl and \wtl neutrino momentum. The properties of the incoming and
outgoing quarks are set to that of a \wbq. For the decay of a
$D$ or $B$-meson to only a \wtl and \wtl neutrino, the two incoming
quark momenta are approximated as half the momentum of the decaying
meson.

For the production of \wtls from virtual photons, the excited photon
matrix element of \equ{Tau:MeGamma} is used. The incoming fermions are
set as \wdqs and the momenta for each quark is set as half the
momentum of the virtual photon. For this decay there are two \wtls
produced, and so full correlations between the decays of the two \wtls
are calculated.

For any process where the production of the \wtl is unknown or does
not match the scenarios described above, the \wtl is assumed to be
uncorrelated, and the helicity density of the \wtl is set as a
normalised identity matrix, {\it i.e.} the two on-diagonal elements
are one half and the off-diagonal elements are zero.

\newsubsection{Initial and Final State Radiation}{ProRad}

Both initial state and final state radiation can effect the helicity
density matrix for \wtls and can be handled through a variety of
methods. The most complete solution would be to treat each radiation
as a decay and perform the full correlation algorithm of
\sec{Thr:Dec}. This method, however, is not possible to implement for
initial state radiation in \pythia{8} without significant changes.

Because the helicity density matrices for the hard processes are used
only for the calculation of \wtl helicity density matrices, it is
possible to approximate the incoming and outgoing particles using
particles before or after initial and final state radiation have been
applied without a significant change in the calculation of the
helicity density matrices. Within \pythia{8} the incoming particles
used in the helicity density matrices for the hard process are taken
before initial state radiation is applied, and the outgoing particles
are taken after final state radiation has been applied.

\newsection{Tau Lepton Decays}{Dec}

The \wtl can decay through a large variety of channels, and so while
an attempt has been made to implement as many channels as possible in
\pythia{8}, some very rare and high multiplicity channels are missing,
and modelled using only isotropic phase-space. Currently, all known
\wtl decays with a branching fraction greater than $0.04\%$ are
implemented with fully modelled helicity matrix
elements. Additionally, the implementation of the helicity matrix
elements within the \pythia{8} code is intended to be both transparent
and easily extensible so the implementation of new channels in the
future, or the modification of old channels, is possible. Many of the
hadronic currents presented in this section are based on the hadronic
currents implemented in \herwig{++}~\cite{grellscheid.07.1} and
\tauola~\cite{golonka.03.1}.

The \wtl decays through a weak interaction and so all \wtl decays take
the form of $p_0 \rightarrow p_1 + p_2 \ldots + p_n$ where $p_0$ is a
\wtl, $p_1$ is a \wtl neutrino, and $p_2$ through $p_n$ are the
remaining leptonic or hadronic children of the decay. The matrix
element for the \wtl decay can be written as,
\begin{equation}
  \mathcal{M} = \frac{g_w^2}{8 \m_\w^2} L_\mu J^\mu
  \labelequ{MeTau}
\end{equation}
where $L_\mu$ is the leptonic current of the \wtl and \wtl neutrino
fermion line and $J^\mu$ is the hadronic current. Here the propagator
for the \w has been approximated as $1/\m_\w^2$ as ${\m_\tau \ll
  \m_\w}$ and \gW is the \su[2] gauge coupling. The leptonic current
  is given by,
\begin{equation}
  L_\mu = \bar{u}_1 \gamma_\mu (1 - \gamma^5) u_0
\end{equation}
where $\bar{u}_1$ is the spinor for the outgoing \wtl neutrino and
$u_0$ is the spinor for the decaying \wtl.

In many of the hadronic currents for the \wtl decays of this section,
Breit-Wigner distributions are used to model hadronic resonances. The
fixed width Breit-Wigner used is,
\begin{equation}
  BW(s, m, \Gamma) = \frac{i\m\Gamma - \m^2}{s - m^2 + i\m\Gamma}
  \labelequ{Bw}
\end{equation}
while the $s$-wave Breit-Wigner,
\begin{equation}
  BW_s(\m_0, \m_1, s, m, \Gamma) = \frac{m^2}{\m^2 - s - \left( \frac{i \Gamma
        \m^2}{\sqrt{s}} \right) \left(\frac{g\left(\m_0, \m_1, s
        \right)}{g\left(\m_0, \m_1, \m^2 \right)}\right)}
  \labelequ{BwS}
\end{equation}
is used for spin-$0$ systems, the $p$-wave Breit Wigner,
\begin{equation}
  BW_p(\m_0, \m_1, s, \m, \Gamma) = \frac{\m^2}{\m^2 - s - \left( \frac{i \Gamma
        \m^2}{\sqrt{s}} \right) \left(\frac{g\left(\m_0, \m_1, s
        \right)}{g\left(\m_0, \m_1, \m^2 \right)}\right)^3}
  \labelequ{BwP}
\end{equation}
is used for spin-$1$ systems, and the $d$-wave Breit Wigner,
\begin{equation}
  BW_d(\m_0, \m_1, s, \m, \Gamma) = \frac{\m^2}{\m^2 - s - \left( \frac{i \Gamma
        \m^2}{\sqrt{s}} \right) \left(\frac{g\left(\m_0, \m_1, s
        \right)}{g\left(\m_0, \m_1, \m^2 \right)}\right)^5}
  \labelequ{BwD}
\end{equation}
is used for spin-$2$ systems where,
\begin{equation}
  g\left(\m_0, \m_1, s\right) = \frac{\sqrt{\big(s - (\m_0 + \m_1)^2\big)
      \big(s - (\m_0 - \m_1)^2\big)}}{2\sqrt{s}}
  \labelequ{BwG}
\end{equation}
and $\m_0$ and $\m_1$ are the masses of the particles from the
resonance decay while $s$ is the square of the centre-of-mass energy
of the resonance. Here, $m$ is the on-shell mass of the resonance and
$\Gamma$ is the width of the resonance. For some decay matrix elements
more sophisticated running widths for the Breit-Wigners are necessary,
but these running widths are explicitly defined for each hadronic
current as required.

In this section the implemented \wtl decay channels are described and
grouped by the multiplicity of the decay. The helicity matrix elements
for each channel are given by \equ{Tau:MeTau} and so only the hadronic
current $J^\mu$ is given for each channel. The hadronic current is
given only to a constant of proportionality, as the branching
fractions of the \wtl are given {\it a priori} within \pythia{8} and
not calculated from the helicity matrix elements.

For each channel the \wtl is designated by $p_0$, the \wtl neutrino as
$p_1$, and the remaining particles of the decay as $p_i$, where the
order is specified. The four-momentum of each particle is denoted by
$q_i$, and the mass by $m_i$. Plots comparing the important invariant
mass distributions for each decay channel are given. When possible,
results from \pythia{8}, \herwig{++}, and \tauola are plotted. In some
cases where the specific decay channel is not available in \tauola,
the \tauola distribution has been omitted.

\newsubsection{Two-Body Decays}{DecTwo}

The \wtl has only two known decay channels available with two decay products,
$\tauto \pi^-$ and $\tauto
K^-$, both of which are decays to a pseudoscalar meson. The hadronic
current for the decay is given by \cite{jadach.90.1},
\begin{equation}
  J^\mu = f_2 q_2^\mu
  \labelequ{Meson}
\end{equation}
where $f_2$ is given by ${130.41 \pm 0.20~\mev}$ for the charged pion
channel and ${156.1 \pm 0.85~\mev}$ for the charged kaon
channel~\cite{pdg.12.1}. The maximum helicity averaged matrix element
amplitude for this channel is given by,
\begin{equation}
  \langle \abs{\mathcal{M}}^2 \rangle_\mathrm{max} = \frac{g_W^4}{16 m_W^4}
  m_0^2 \left(m_0^2 - m_2^2 \right)
  \labelequ{Meson.Max}
\end{equation}
where \gW is the \su[2] gauge coupling and $\m_\w$ the mass of the
\wwb.

\newsubsection{Three-Body Decays}{DecThree}

The \wtl can decay through a variety of three-body channels, and
currently, three models have been implemented in \pythia{8} for
three-body decays. These models are for \wtl decays into a \wtl
neutrino and two leptons via a leptonic current, a neutrino and two
mesons via a vector current, and a neutrino and two mesons via a
vector and scalar current.

\newsubsubsection{Three Leptons}{}

The three lepton current for the decays $\tauto e^- \bar{\nu}_e$ and
$\tauto \mu^- \bar{\nu}_\mu$ is given by,
\begin{equation}
  J^\mu = \bar{u}_2 \gamma^\mu (1 - \gamma^5) v_3
  \labelequ{TwoLeptons}
\end{equation}
where $p_2$ is the charged lepton and $p_3$ is the corresponding
neutrino. The maximum helicity average matrix element amplitude is,
\begin{equation}
  \langle \abs{\mathcal{M}}^2 \rangle_\mathrm{max} = \frac{g_W^4}{4 m_W^4}
  \left(m_0^2 - m_2^2 \right)^2
  \labelequ{TwoLeptons.Max}
\end{equation}
where the masses of the neutrinos are assumed to be zero and $m_2$ is
the mass of the outgoing electron or muon.

\begin{subfigures}{2}{Distributions of the combined $p_2$ and $p_3$
    invariant mass, $m_{23}$ for
    \subfig{q_qp.W.16_11_12.m_11_12}~$\tauto e^- \bar{\nu}_{e^-}$ and
    \subfig{q_qp.W.16_13_14.m_13_14}~$\tauto \mu^- \bar{\nu}_{\mu^-}$
    decays.\labelfig{TwoLeptons}}
  \svgbeg 
  \svg{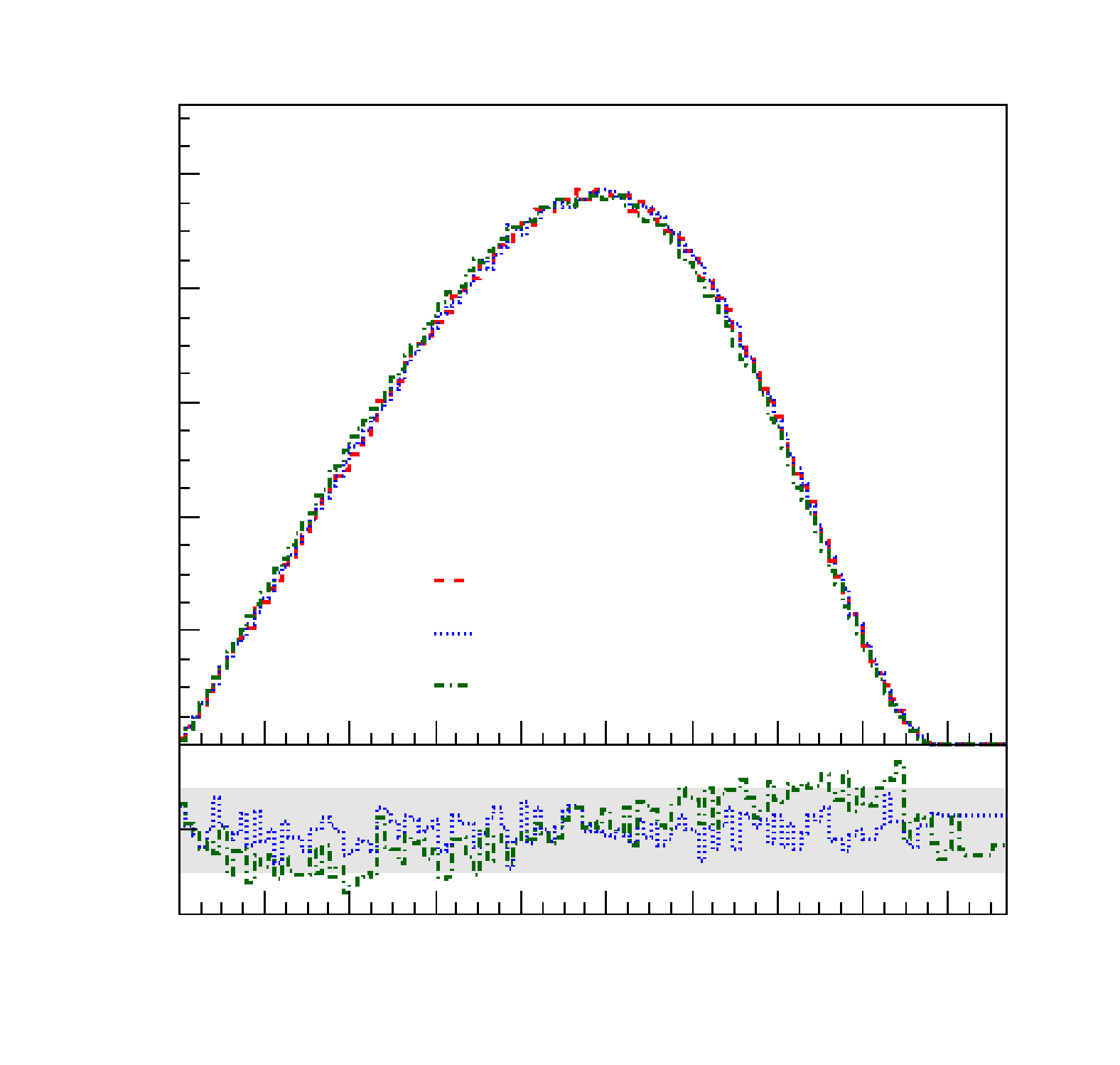} & \svg{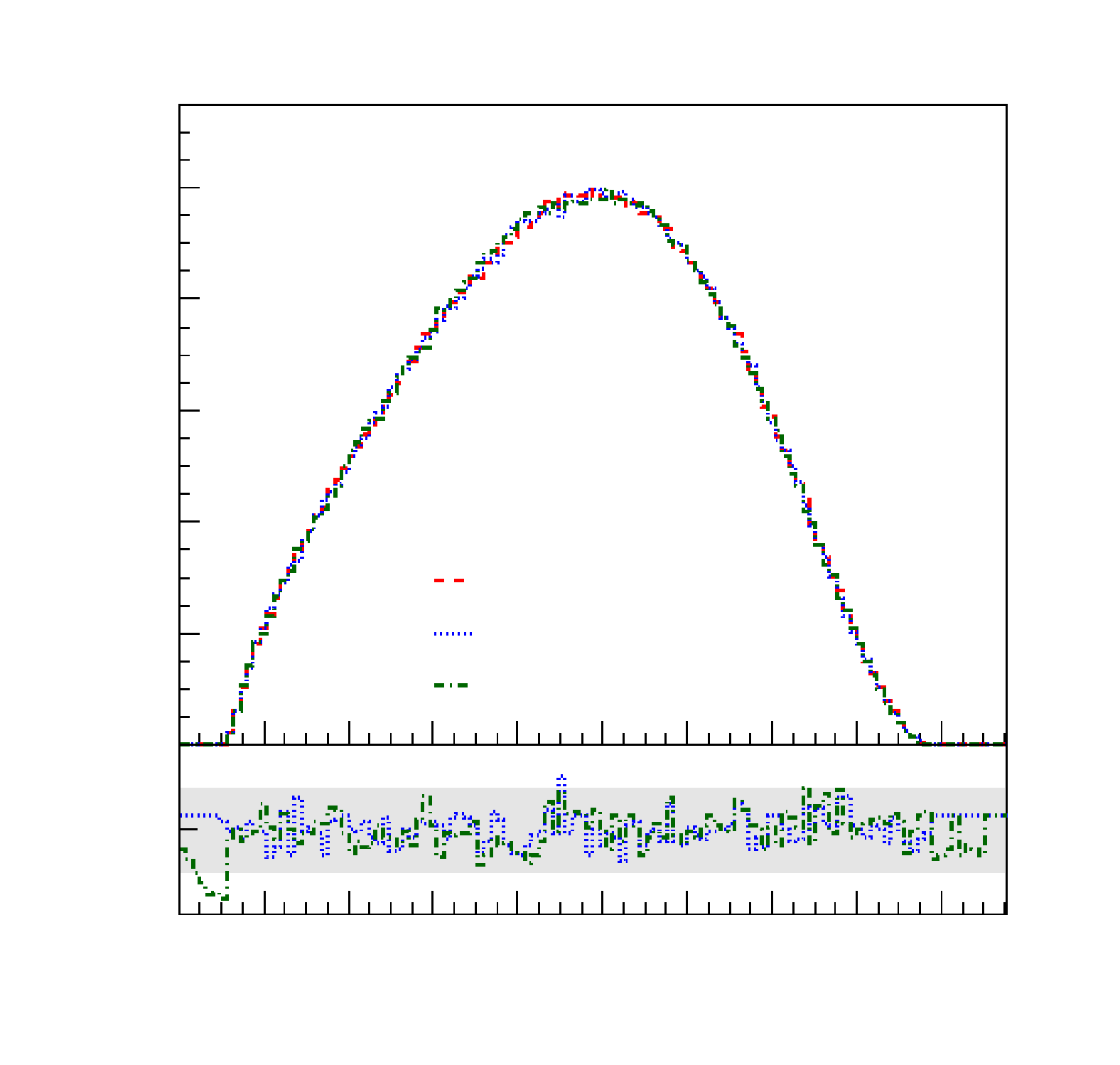} \svgend
\end{subfigures}

\Fig{Tau:TwoLeptons} gives the invariant mass distributions for
the lepton and anti-lepton neutrino system, $m_{23}$, for \pythia{8},
\herwig{++}, and \tauola. As expected, there is good agreement between
the three generators.

\newsubsubsection{Two Mesons Via a Vector Current}{}

The \wtl can decay into two mesons via a vector meson resonance for the
$\tauto \pi^0 \pi^-$, $\tauto K^0 K^-$, and $\tauto \eta K^-$
channels. The hadronic current for these decays is given by the
K\"uhn and Santamaria model~\cite{kuhn.90.1},
\begin{equation}
  J^\mu \propto \left((q_3 - q_2)^\mu -
    \frac{s_1}{s_2} (q_2 + q_3)^\mu \right) \sum_i {w_v}_i BW_p(\m_2, \m_3,
  s_2, {\m_v}_i, {\Gamma_v}_i)
  \labelequ{TwoMesonsViaVector}
\end{equation}
where $s_1$ is given by $(q_3 - q_2)_\mu(q_2 + q_3)^\mu$, $s_2$ is
given by $(q_2 + q_3)_\mu(q_2 + q_3)^\mu$, ${w_V}_i$ are complex
weights for the vector resonances, $BW_p$ is the $p$-wave Breit-Wigner
of \equ{Tau:BwP}, ${\m_v}_i$ are the on-shell masses of the vector
resonances, and ${\Gamma_v}_i$ are the on-shell widths of the vector
resonances. The $\tau^- \rightarrow \nu_\tau \pi^0 \pi^-$ and $\tauto
K^0 K^-$ channels proceed through $\rho$ resonances, while the $\tau^-
\rightarrow \eta K^-$ channel proceeds through $K^*$
resonances. Because the $\rho$ resonances are dominated by the $\pi^0
\pi^-$ decay, the masses used in calculating the $p$-wave Breit-Wigner
are set at $\m_2 = \m_{\pi^0}$ and $\m_3 = \m_{\pi^-}$. For the $K^*$
resonances, the $\pi^- \bar{K}^0$ channel dominates and so the masses
used in calculating the $p$-wave Breit-Wigner are set at $\m_2 =
\m_{\pi^-}$ and $\m_3 = \m_{K^0}$. The $\tau^- \rightarrow \nu_\tau \pi^-
\bar{K}^0$ channel is not modelled using the hadronic current of
\equ{Tau:TwoMesonsViaVector}, but rather through the model of
Finkemeier and Mirkes in \rfr{finkemeier.96.1}, where a scalar meson
resonance is included. The hadronic current for this model is given in
the following section by \equ{Tau:TwoMesonsViaVectorScalar}.

\begin{table}\centering
  \captionabove{Parameters used for the \wtl decay into two mesons via a
    vector meson
    resonance.\labeltab{TwoMesonsViaVector}}
  \begin{tabular}{L|L|L|L|L}
    \toprule
    \multicolumn{1}{c}{resonance} 
    & \multicolumn{1}{c|}{$m~[\gev]$}
    & \multicolumn{1}{c|}{$\Gamma~[\gev]$} 
    & \multicolumn{1}{c|}{$\phi$}
    & \multicolumn{1}{c}{$A$} \\
    \midrule
    \rho(770)  & 0.7746 & 0.149  & 0   & 1     \\
    \rho(1450) & 1.408  & 0.502  & \pi & 0.167 \\
    \rho(1700) & 1.7    & 0.235  & 0   & 0.050 \\
    K^*(892)   & 0.8921 & 0.0513 & 0   & 1     \\
    K^*(1680)  & 1.7    & 0.235  & \pi & 0.038 \\
    \bottomrule
  \end{tabular}
\end{table}

The parameters used for the hadronic current of
\equ{Tau:TwoMesonsViaVector} are given in
\tab{Tau:TwoMesonsViaVector} and are the same as those used in
\herwig{++} which are based on the fits of CLEO \cite{cleo.99.1}. The
complex weights for the vector resonances are calculated from the
phases, ${\phi_v}_i$, and amplitudes, ${A_v}_i$, as a position vector
in the complex plane.
\begin{equation}
  w_{i} = A_{i} \left(\cos\phi_i + i\sin\phi_i \right)
  \labelequ{ComplexWeights}
\end{equation}
Three $\rho$ resonances are used, $\rho(770)$, $\rho(1450)$, and
$\rho(1700)$ while only two $K^*$ resonances are used, $K^*(892)$
and $K^*(1680)$.

\begin{subfigures}{2}{Comparisons of the $\m_{23}$ invariant mass
    distributions for the \subfig{q_qp.W.16_111_211.m_111_211}~$\tauto
    \pi^0 \pi^-$ decay channel, the
    \subfig{q_qp.W.16_311_321.m_311_321}~$\tau^- \rightarrow \nu_\tau
    K^0 K^-$ decay channel, and the
    \subfig{q_qp.W.16_221_321.m_221_321}~$\tau^- \rightarrow \nu_\tau
    \eta K^-$ decay channel for the two meson hadronic current through
    a vector resonance of
    \equ{Tau:TwoMesonsViaVector}.\labelfig{TwoMesonsViaVector}}
  \svgbeg
  \svg{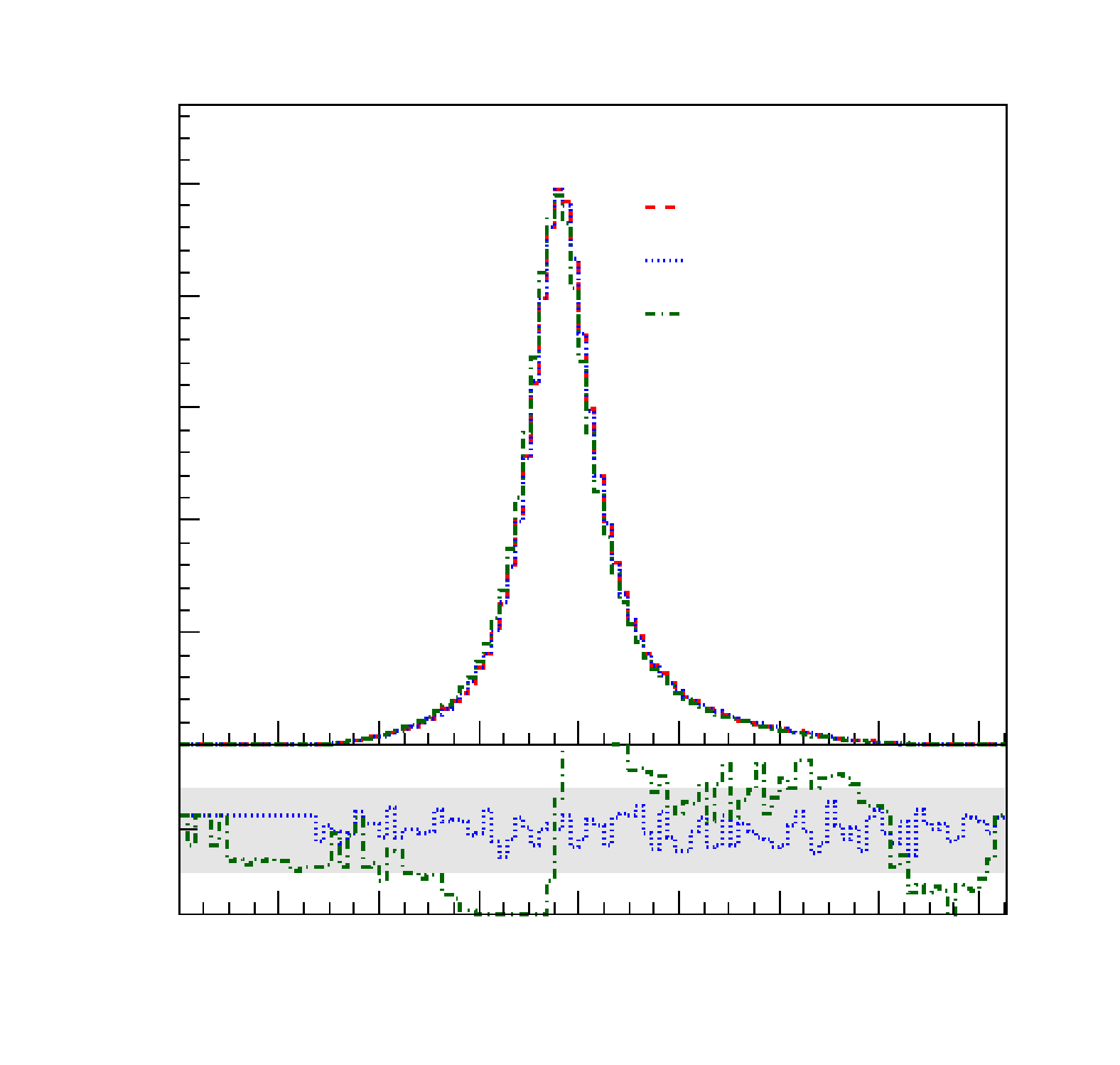} &
  \svg{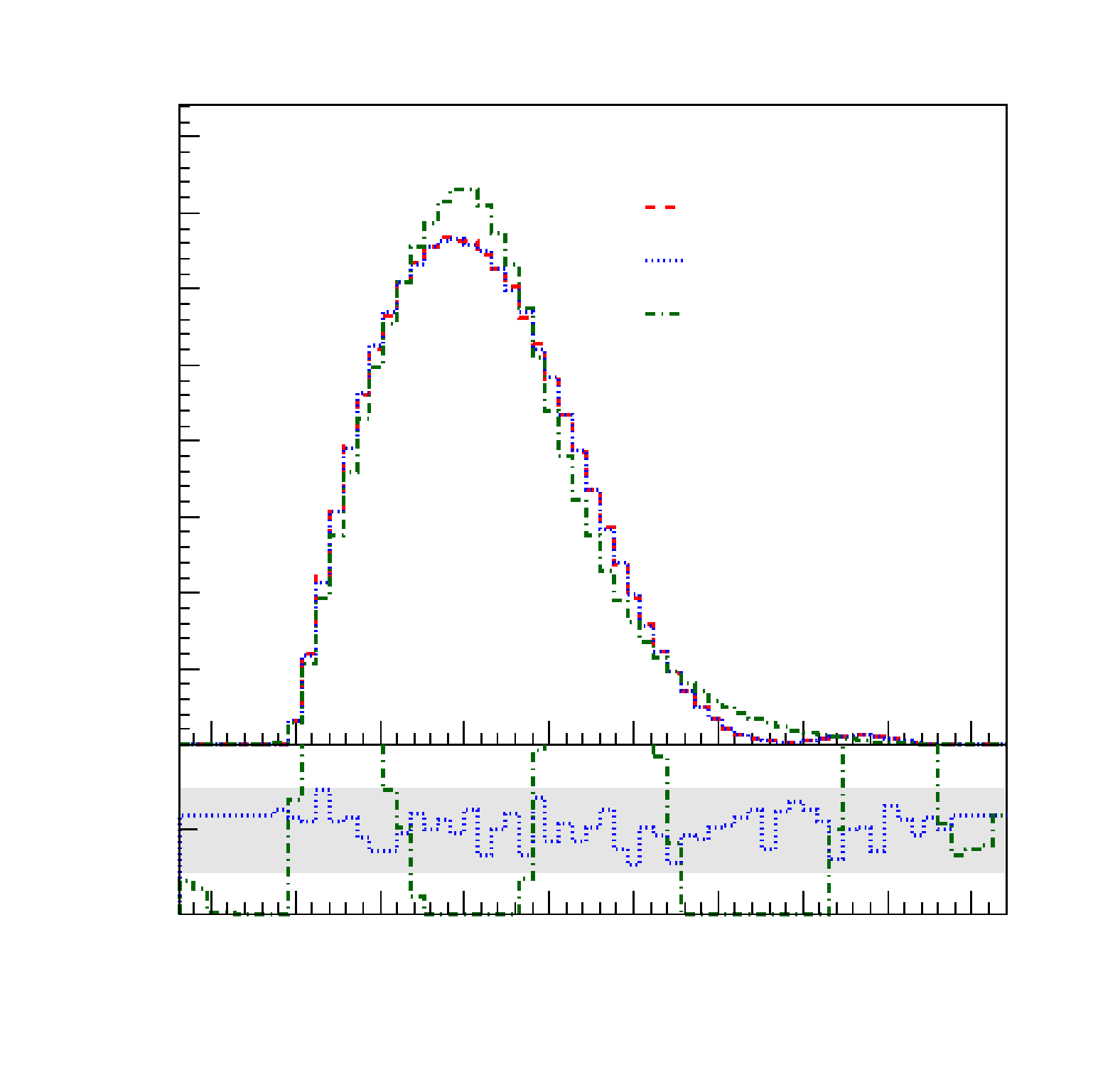} \svgsep
  \svg{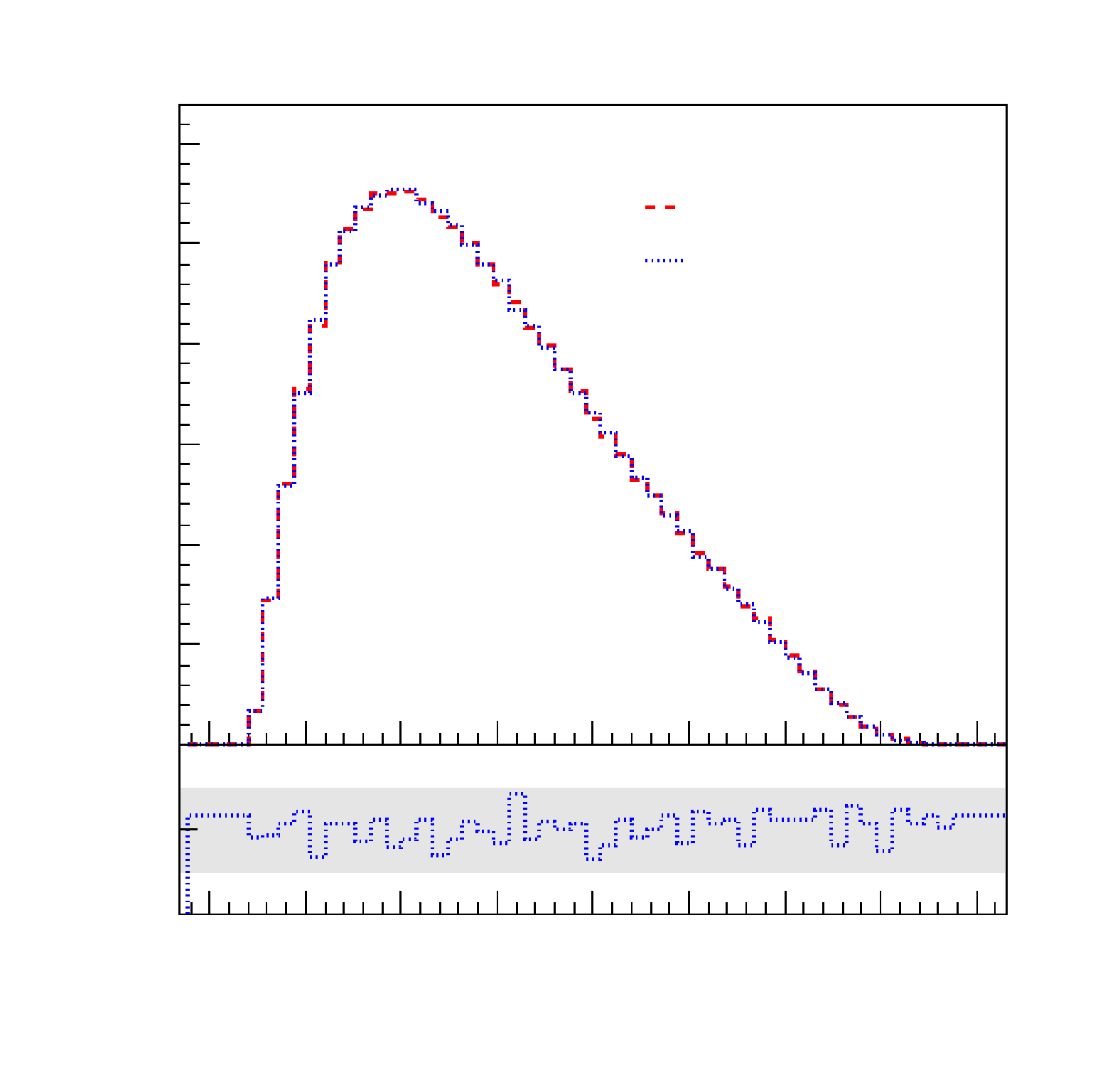} & \sidecaption \svgend
\end{subfigures}

Plots comparing the hadronic invariant masses of the $\pi^0 \pi^-$,
$K^0 K^-$, and $\eta K^-$ for the three decay channels modelled with
\equ{Tau:TwoMesonsViaVector} are given in Figure
\fig{Tau:TwoMesonsViaVector}. For both the $\pi^0 \pi^-$ and $K^0
K^-$ there is good agreement between \pythia{8} and \herwig{++} while
the vector resonance is slightly sharper in \tauola for the $K^0 K^-$
invariant mass. The $\tauto \eta K^-$ decay channel is not available
in \tauola, but the $\eta K^-$ invariant mass distributions generated
by \herwig{++} and \pythia{8} match well.

\newsubsubsection{Two Mesons Via Vector and Scalar Current}{}

For the rare decays $\tauto \pi^- \bar{K}^0$ and $\tauto \pi^0 K^-$,
the \wtl decay can proceed through both vector and scalar meson
resonances. These channels have been modelled by Finkemeier and Mirkes
in \rfr{finkemeier.96.1} with the hadronic current,
\begin{align*}\labelali{TwoMesonsViaVectorScalar}
  J^\mu \propto~ & \frac{c_v}{\sum_i {w_v}_i} \bigg( (q_3 -
  q_2)^\mu\sum_i {w_v}_i
  BW_p(\m_2, \m_3, s_2, {\m_v}_i, {\Gamma_v}_i) \\
  & - s_1 (q_2 + q_3)^\mu \sum_i \frac{{w_v}_i
    BW_p(\m_2, \m_3, s_2, {\m_v}_i, {\Gamma_v}_i)}{{\m_v}_i^2}
  \bigg) \\ 
  & + \frac{c_s }{\sum_j {w_s}_j} (q_2 + q_3)^\mu \sum_j {w_s}_j
  BW_s(\m_2, \m_3, s_2, {\m_s}_j,
  {\Gamma_s}_j)
\end{align*}
where $c_v$ and $c_s$ are the couplings to the vector and scalar
resonances, ${w_v}_i$ and ${w_s}_j$ are the complex vector and scalar
weights, ${\m_v}_i$ and ${m_s}_j$ are the on-shell resonance masses,
and ${\Gamma_v}_i$ and ${\Gamma_s}_j$ are the on-shell resonance
widths. The energies $s_1$ and $s_2$ are the same as for the vector
resonance current of \equ{Tau:TwoMesonsViaVector}. Again, the
complex weights are calculated from a phase, $\phi$, and amplitude,
$A$, using \equ{Tau:ComplexWeights}. The propagator for the vector
resonances is modelled using a $p$-wave Breit-Wigner, given by
\equ{Tau:BwP}, while the propagator for the scalar resonances is
modelled using an $s$-wave Breit-Wigner, given by \equ{Tau:BwS}.

\begin{table}\centering
  \captionabove{Parameters used for the \wtl decay into two mesons via
    both vector and scalar meson
    resonances.\labeltab{TwoMesonsViaVectorScalar}}
  \begin{tabular}{L|L|L|L|L}
    \toprule
    \multicolumn{1}{c|}{resonance} 
    & \multicolumn{1}{c|}{$\m~[\gev]$} 
    & \multicolumn{1}{c|}{$\Gamma~[\gev]$} 
    & \multicolumn{1}{c|}{$\phi$} 
    & \multicolumn{1}{c}{$A$} \\
    \midrule
    K^*(892)   & 0.89547 & 0.04619 & 0      & 1     \\
    K^*(1410)  & 1.414   & 0.232   & 1.4399 & 0.075 \\
    K^*_0(800) & 0.878   & 0.499   & 0      & 1     \\
    \bottomrule
  \end{tabular}
\end{table}

\Tab{Tau:TwoMesonsViaVectorScalar} gives the parameters used in
\pythia{8} for the hadronic current of
\equ{Tau:TwoMesonsViaVectorScalar}. The vector resonances proceed
through a $K^*(892)$ and a $K^*(1410)$ while the scalar resonance
proceeds through only a $K^*_0(800)$. Both the $\tauto \pi^0 K^-$
and $\tauto \pi^- K^0$ channels proceed through all three
resonances. The vector resonance coupling is $c_v = 1$ and the scalar
resonance coupling is $c_s = 0.465$. These parameters match those used
in \herwig{++} and are taken from the fits of Belle in
\rfr{belle.07.1}.

\begin{subfigures}{2}{Distributions of the $m_{23}$ invariant mass for
    the \subfig{q_qp.W.16_111_321.m_111_321}~$\tauto \pi^0 K^-$ decay
    channel and the \subfig{q_qp.W.16_211_311.m_211_311}~$\tauto \pi^-
    K^0$ decay channel for the two meson hadronic current via vector
    and scalar resonances using
    \equ{Tau:TwoMesonsViaVectorScalar}.
    \labelfig{TwoMesonsViaVectorScalar}}
    \svgbeg
    \svg{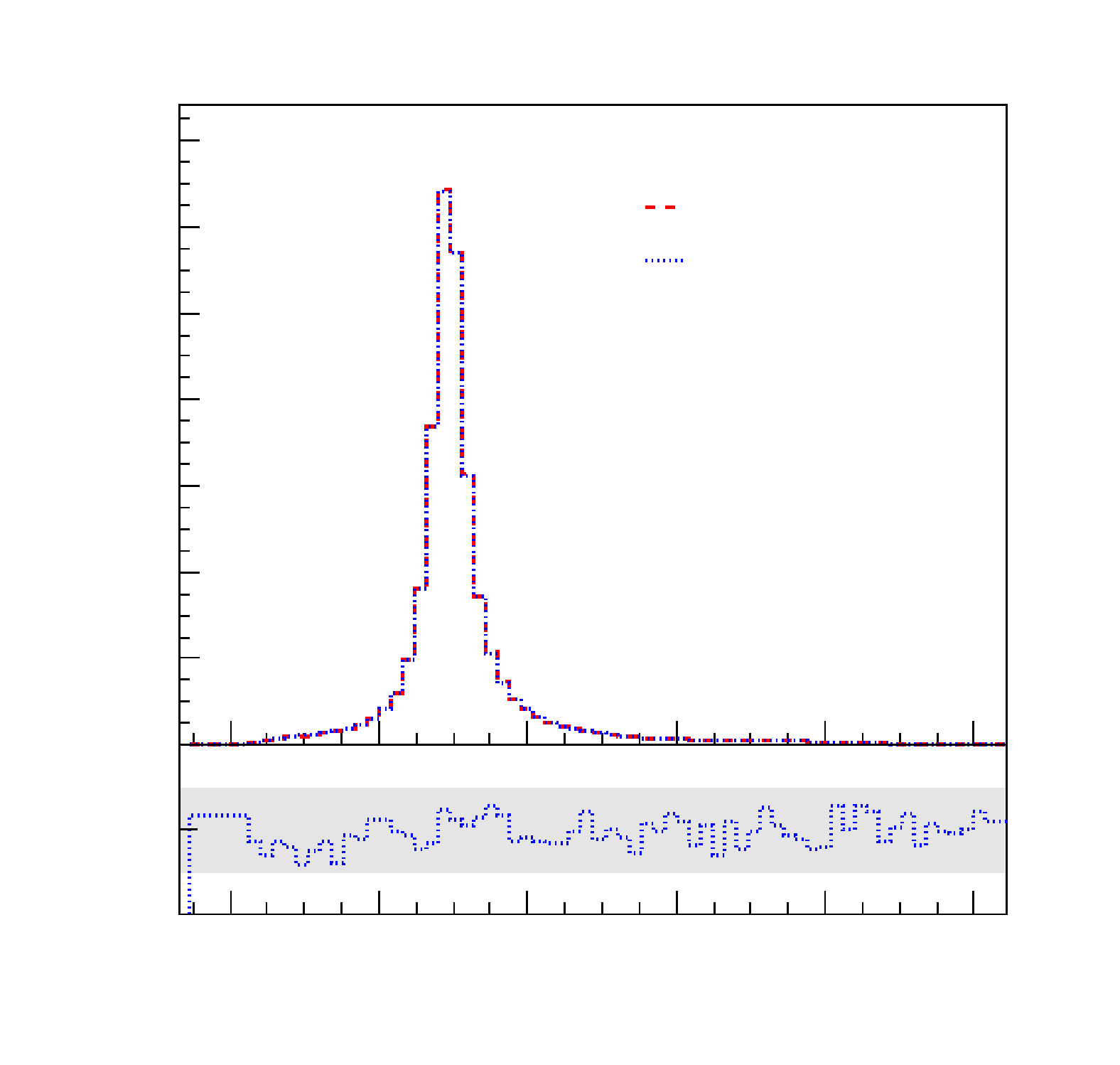} &
    \svg{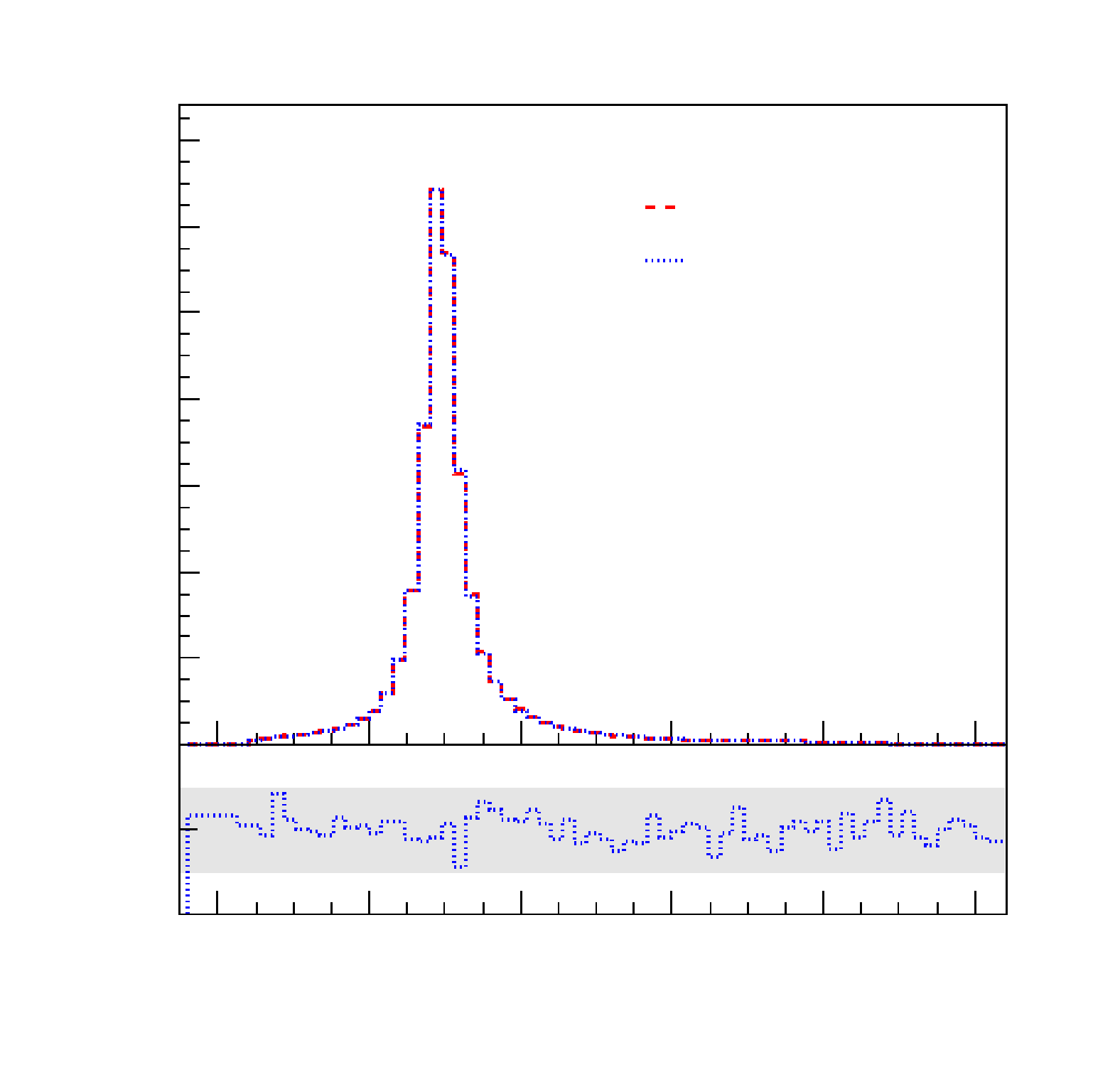} \svgend
\end{subfigures}

\Fig{Tau:TwoMesonsViaVectorScalar} provides the comparison plots
between \pythia{8} and \herwig{++} for $\tauto \pi^0 K^-$ and $\tauto
\pi^- K^0$ decays. The $m_{\pi^0 K^-}$ and $m_{\pi^- K^0}$
distributions are similar, and the distributions from \pythia{8} and
\herwig{++} agree well for both decay channels.

\newsubsection{Four-Body Decays}{DecFour}

Currently four models are implemented in \pythia{8}, which provide
four-body decays of the \wtl into three mesons for a total of twelve
\wtl decay channels. The first hadronic current is used specifically
for final states with three pions, while the second two hadronic
currents are generalised currents which can be used for final states
with both pions and kaons. The final four-body hadronic current
includes an explicit photon for the $\pi^0 \pi^-$ channel. The
general hadronic current is given by,
\begin{align*}\labelali{ThreeMesons}
  J^\mu \propto &\left(g^{\mu\nu} - \frac{q^\mu q^\nu}{s_1}\right)
  \bigg((F_3 - F_2) q_2 + (F_1 - F_3) q_3 + (F_2 - F_1) q_4
  \bigg)^\mu \\
  & + F_4 q^\mu + iF_5 \epsilon^\mu(q_2, q_3, q_4)
\end{align*}
from \rfr{jadach.93.1}, where $F_i$ are form factors specific to the
model, $\epsilon$ is the permutation operator, and $q$ is $q_2 + q_2 +
q_3$, where $q_2$, $q_3$, and $q_4$ are the four-momenta of the three
mesons. The centre-of-mass energies are defined as,
\begin{align*}\labelali{ThreeMesons.S}
  s_1 &= (q_2 + q_3 + q_4)^\mu (q_2 + q_3 + q_4)_\mu \\
  s_2 &= (q_3 + q_4)^\mu (q_3 + q_4)_\mu \\
  s_3 &= (q_2 + q_4)^\mu (q_2 + q_4)_\mu \\
  s_4 &= (q_2 + q_3)^\mu (q_2 + q_3)_\mu
\end{align*}
where $s_1$ gives the centre-of-mass energy for the three meson
system, while $s_2$, $s_3$, and $s_4$ provide the centre-of-mass
energies for the combinations of meson pairs.

\newsubsubsection{CLEO Current}{}

The $\pi^0 \pi^0 \pi^-$ and $\pi^- \pi^- \pi^+$ decay
channels are modelled using the current from the CLEO collaboration fit
of \rfr{cleo.99.2} which first proceed through an $a_1$ resonance, and
then an additional scalar or tensor resonance. Both the $\pi^0
\pi^0 \pi^-$ and $\pi^- \pi^- \pi^+$ decay channels can proceed
through scalar $f_0$, tensor $f_2$, or scalar $\sigma$ resonances. The
$F_4$ and $F_5$ form factors for the CLEO current are zero, and so
only the first three form factors need to be specified.

The form factor $F_1$ for the $\pi^- \pi^- \pi^+$ channel is
given by,
\begin{align*}\labelali{ThreePions.F1c}
  F_1^- =~ & BW_{a_1}(s_1)\sum_i \Bigg( -{w_\rho}_i^p BW_p(m_3, m_4, s_2,
  {\m_\rho}_i,
  {\Gamma_\rho}_i) \\
  &- \frac{{w_\rho}_i^d}{3} BW_p(m_2, m_4, s_3,
  {\m_\rho}_i,
  {\Gamma_\rho}_i)(s_2 - s_4) \Bigg) \\
  & - \frac{2}{3} \Bigg( w_\sigma BW_s(m_2, m_4, s_3, \m_\sigma,
  \Gamma_\sigma) + w_{f_0} BW_s(m_2,
  m_4, s_3, \m_{f_0}, \Gamma_{f_0}) \Bigg) \\
  & + w_{f_2} \Bigg(\frac{s_4 - s_3}{2} BW_d(m_3, m_4, s_2, \m_{f_2},
  \Gamma_{f_2}) \\
  &-  \frac{1}{18 s_3} (4 m_2^2 - s_3) (s_1 + s_2 -
  m_2^2) BW_d(m_2, m_4, s_3, \m_{f_2}, \Gamma_{f_2}) \Bigg)
\end{align*}
while the first form factor for the $\pi^0
\pi^0 \pi^-$ channel is given by,
\begin{align*}\labelali{ThreePions.F1n}
  F_1^0 =~ &  BW_{a_1}(s_1)\sum_i \Bigg( {w_\rho}_i^p BW_p(m_3, m_4, s_2,
  {\m_\rho}_i,
  {\Gamma_\rho}_i)\\
  &  - \frac{{w_\rho}_i^d}{3} BW_p(m_2, m_4, s_3,
  {\m_\rho}_i, {\Gamma_\rho}_i) (s_4 - s_2 - m_4^2 - m_2^2) \Bigg) \\
  & + \frac{2}{3} \Bigg( w_\sigma BW_s(m_2, m_3, s_4 \m_\sigma,
  \Gamma_\sigma) + w_{f_0} BW_s(m_2, m_3, s_4, \m_{f_0},
  \Gamma_{f_0}) \Bigg) \\
  & + \frac{w_{f_2}}{18 s_4} (s_1 - m_4^2 + s_4) (4 m_2^2 - s_4)
  BW_d(m_2, m_3, s_4, \m_{f_2}, \Gamma_{f_2})
\end{align*}
where ${w_\rho}_i^p$ are the complex couplings of the $p$-wave $\rho$
resonances to the $a_1$, ${w_\rho}_i^d$ are the $d$-wave $\rho$
couplings, $w_{f_0}$ is the $f_0$ coupling, $w_{f_2}$ is the $f_2$
coupling, and $w_\sigma$ is the $\sigma$ coupling.

The second form factor, $F_2$, is given by,
\begin{align*}\labelali{ThreePions.F2c}
  F_2^- =~ &  BW_{a_1}(s_1) \sum_i \Bigg( {w_\rho}_i^p BW_p(m_2, m_4, s_3,
  {\m_\rho}_i, {\Gamma_\rho}_i) \\
  &  - \frac{{w_\rho}_i^d}{3} BW_p(m_3,
  m_4, s_2, {\m_\rho}_i, {\Gamma_\rho}_i) (s_3 - s_4) \Bigg) \\
  & - \frac{2}{3} \Bigg( w_\sigma BW_s(m_3, m_4, s_2, \m_\sigma,
  \Gamma_\sigma) + w_{f_0} BW_s(m_3, m_4, s_2, \m_{f_0},
  \Gamma_{f_0}) \Bigg) \\
  & + w_{f_2} \Bigg( \frac{s_4 - s_2}{2} BW_d(m_2, m_4, s_3,
  \m_{f_2},
  \Gamma_{f_2}) \\
  & - \frac{1}{18 s_2} (4 m_2^2 - s_2) (s_1 + s_2 -
  m_2^2) BW_d(m_3, m_4, s_2, \m_{f_2}, \Gamma_{f_2}) \Bigg) 
\end{align*}
for the $\pi^- \pi^- \pi^+$ channel while,
\begin{align*}\labelali{ThreePions.F2n}
    F_2^0 =~ &  BW_{a_1}(s_1) \sum_i \Bigg( - \frac{{w_\rho}_i^p}{3}
    BW_p(m_2, m_4, s_3, {\m_\rho}_i, {\Gamma_\rho}_i) \\
    & - {w_\rho}_i^d BW_p(m_3, m_4, s_2,
    {\m_\rho}_i, {\Gamma_\rho}_i) (s_4 - s_3 - m_4^2 - m_3^2) \Bigg) \\
    & + \frac{2}{3} \Bigg( w_\sigma BW_s(m_2, m_3, s_4, \m_\sigma,
    \Gamma_\sigma) + w_{f_0} BW_s(m_2, m_3, s_4, \m_{f_0},
    \Gamma_{f_0}) \Bigg) \\
    & + \frac{w_{f_2}}{18 s_4} (s_1 - m_4^2 + s_4) (4 m_2^2 - s_4)
    BW_d(m_2, m_3, s_4, \m_{f_2}, \Gamma_{f_2}) 
\end{align*}
is the second form factor for the $\pi^0
\pi^0 \pi^-$ channel.

The third form factor is given by,
\begin{align*}\labelali{ThreePions.F3c}
  F_3^- =~ &  BW_{a_1}(s_1) \sum_i - {w_\rho}_i^d \Bigg( \frac{1}{3}
  (s_3 - s_4) BW_p(m_2, m_4, s_2, {\m_\rho}_i, {\Gamma_\rho}_i) \\
  & - \frac{1}{3} (s_2 - s_4) BW_p(m_2,
  m_4, s_3, {\m_\rho}_i, {\Gamma_\rho}_i) \Bigg) \\
  & - \frac{2}{3} \Bigg( w_\sigma BW_s(m_3, m_4, s_2, \m_\sigma,
  \Gamma_\sigma) + w_{f_0} BW_s(m_3, m_4, s_2, \m_{f_0},
  \Gamma_{f_0}) \Bigg) \\
  & + \frac{2}{3} \Bigg( w_\sigma BW_s(m_2, m_4, s_3, \m_\sigma,
  \Gamma_\sigma) + w_{f_0} BW_s(m_2, m_4, s_3, \m_{f_0},
  \Gamma_{f_0}) \Bigg) \\
  & + w_{f_2} \Bigg( - \frac{1}{18 s_2} (4 m_2^2 - s_2) (s_1 + s_2 -
  m_2^2) BW_d(m_3, m_4, s_2, \m_{f_2}, \Gamma_{f_2}) \\
  & + \frac{1}{18 s_3} (4 m_2^2 - s_3) (s_1 + s_3 -
  m_2^2) BW_d(m_2, m_4, s_3,
  \m_{f_2}, \Gamma_{f_2}) \Bigg) 
\end{align*}
for the $\pi^- \pi^- \pi^+$ channel and,
\begin{align*}\labelali{ThreePions.F3n}
  F_3^0 =~ & BW_{a_1}(s_1) \sum_i {w_\rho}_i^d \Bigg( - \frac{1}{3}
  (s_4 - s_3 - m_4^2 + m_3^2) \\
  & BW_p(m_3, m_4, s_2, {\m_\rho}_i, {\Gamma_\rho}_i) \\
  & + \frac{1}{3} (s_4 - s_2 - m_4^2 + m_2^2)
  BW_p(m_2, m_4, s_3, {\m_\rho}_i,
  {\Gamma_\rho}_i) \Bigg) \\
  & - \frac{w_{f_2}}{2} (s_2 - s_3) BW_d(m_2, m_3, s_4, \m_{f_2},
  \Gamma_{f_2}) 
\end{align*}
for the $\pi^0 \pi^0 \pi^-$ channel.

All the complex couplings are calculated from a phase and amplitude
using \equ{Tau:ComplexWeights}. The on-shell widths and masses along
with the phases and couplings used in \pythia{8} for the $\rho$
resonances are given in \tab{Tau:ThreePions.RhoParameters}, while the
scalar $f_0$ and $sigma$, and tensor $f_2$ parameters are given in
\tab{Tau:ThreePions.ScalarParameters}. Both sets of parameters are
based on the fits performed by the CLEO
collaboration~\cite{cleo.99.2}.

\begin{table}\centering
  \captionabove{Parameters for the $\rho$ resonances used by the \wtl
    decay into three pions.\labeltab{ThreePions.RhoParameters}}
  \begin{tabular}
    {L|L|L|L|L|L|L}
    \toprule
    \multicolumn{1}{c|}{resonance} 
    & \multicolumn{1}{c|}{$\m~[\gev]$} 
    & \multicolumn{1}{c|}{$\Gamma~\gev$} 
    & \multicolumn{1}{c|}{$\phi_p$} 
    & \multicolumn{1}{c|}{$A_p$} 
    & \multicolumn{1}{c|}{$\phi_d$} 
    & \multicolumn{1}{c}{$A_d$} \\
    \midrule
    \rho(770) & 0.7743 & 0.1491 & 0 & 1 & -0.471239 & 3.7 \times 10^{-7} \\
    \rho(1450) & 1.37 & 0.386 & 3.11018 & 0.12 & \phantom{1}1.66504 & 8.7
    \times 10^{-7} \\
    \rho(1700) & 1.72 & 0.25 & 0 & 0 & \phantom{-}0 & 0 \\
    \bottomrule
  \end{tabular} \\ ~ \\ ~ \\
  \captionabove{Parameters used by the three pion \wtl decay channels for the
    secondary $f_0$, $f_2$, and $\sigma$
    resonances.\labeltab{ThreePions.ScalarParameters}}
  \begin{tabular}{L|L|L|L|L}
    \toprule
    \multicolumn{1}{c|}{resonance}
    & \multicolumn{1}{c|}{$\m~[\gev]$} 
    & \multicolumn{1}{c|}{$\Gamma~[\gev]$} 
    & \multicolumn{1}{c|}{$\phi$} 
    & \multicolumn{1}{c}{$A$} \\
    \midrule
    f_0(980)   & 1.186 & 0.350 & -1.69646 & 0.77 \\
    f_2(1270)  & 1.275 & 0.185 & 1.75929  & 7.1 \times 10^{-7} \\
    \sigma     & 0.86  & 0.88  & 0.722466 & 2.1 \\
    \bottomrule
  \end{tabular}
\end{table}

The Breit-Wigners used in the form factors $F_1$, $F_2$, and $F_3$ are
given by the $s$, $p$, and $d$-wave Breit-Wigners of \equs{Tau:BwS},
\ref{equ:Tau:BwP}, and \ref{equ:Tau:BwD}. The Breit-Wigner for the
$a_1$ is given by,
\begin{equation}
  BW_{a_1}(s) = \frac{\m_{a_1}^2}{\m_{a_1}^2 - s - \Gamma_{a_1}(s)}
\end{equation}
where $\m_{a_1}$ is the on-shell mass of the $a_1$. The running width
$\Gamma_{a_1}(s)$ is the weighted sum of the three partial widths
$\Gamma_{\pi^0 \pi^0 \pi^-}(s)$, $\Gamma_{\pi^- \pi^- \pi^+}(s)$, and
the $s$-wave contribution $\Gamma_{K K^*}(s)$,
\begin{equation}
  \Gamma_{a_1}(s) = w_\pi \big(\Gamma_{\pi^0 \pi^0 \pi^-}(s) +
  \Gamma_{\pi^- \pi^- \pi^0}(s) + w_{K} \Gamma_{K K^*}(s) \big)
  \labelequ{ThreePions.A1Width}
\end{equation}
where $w_{\pi}$ and $w_{K}$ are the pion and kaon weights given in
\tab{Tau:ThreePions.A1Parameters}, and $\Gamma_{\pi^0 \pi^0
  \pi^-}(s)$, $\Gamma_{\pi^- \pi^- \pi^+}(s)$, and $\Gamma_{K K^*}(s)$
are piece-wise fitted functions given in \tab{Tau:ThreePions.A1Fit}
with the parameters given in \tabs{Tau:ThreePions.A1Parameters} and
\ref{tab:Tau:ThreePions.A1FitParameters}.

\begin{table}\centering
  \captionabove{Functions used to fit the partial widths of the running
    $a_1$ width used in the hadronic current for the decay of the
    \wtl into three pions.\labeltab{ThreePions.A1Fit}}
  \begin{tabular}{L|L|L}
    \toprule
    \multicolumn{1}{c|}{$\Gamma_{\mathrm{partial}}$} 
    & \multicolumn{1}{c|}{limits~$\left[\gev^2\right]$}
    & \multicolumn{1}{c}{$\Gamma(s)~[\gev]$} \\
    \midrule
    \multirow{3}{*}{$\Gamma_{\pi^-\pi^-\pi^+}$}
    & 0 \leq s < \m_{3\pi^-} & 0 \\
    & \m_{3\pi^-} \leq s < \m_{\rho\pi^0}
    & P_0 (s - \m_{3\pi^-})^3 \left( 1 - P_1 (s - \m_{3\pi^-}) + P_2 (s -
      \m_{3\pi^-})^2 \right) \\
    & \m_{\rho\pi^0} \leq s & P_0 + P_1 s + P_2 s^2 + P_3 s^3 + P_4
    s^4 \\
    \midrule
    \multirow{4}{*}{$\Gamma_{\pi^0\pi^0\pi^-}$}
    & 0 \leq s < \m_{2\pi^0\pi^-} & 0 \\
    & \m_{2\pi^0\pi^-} \leq s \leq \m_{\rho\pi^0}
    & \begin{aligned} &P_0 (s - \m_{2\pi^0\pi^-})^3 \\ &\left( 1 - P_1 (s -
        \m_{2\pi^0\pi^-}) + P_2 (s - \m_{2\pi^0\pi^-})^2 \right) \\ 
    \end{aligned} \\
    & \m_{\rho\pi^0} \leq s & P_0 + P_1 s + P_2 s^2 + P_3 s^3 + P_4
    s^4 \\
    \midrule
    \multirow{2}{*}{$\Gamma_{K K^*}$}
    & 0 \leq s < \m_{K K^*} & 0 \\
    & \m_{K K^*} \leq s & \sqrt{(s - (\m_K + \m_{K^*})^2) (s - (\m_K
      - \m_{K^*})^2)} / (2 s) \\
    \bottomrule
  \end{tabular}
\end{table}

\begin{subtables}[t]{2}{Parameters used for the $a_1$ Breit-Wigner and
    running width of \equ{Tau:ThreePions.A1Width} and
    \tab{Tau:ThreePions.A1Fit} for the three pion hadronic current of
    the \wtl.\labeltab{ThreePions.A1Parameters}}
  \setlength{\tabcolsep}{\oldtabcolsep}
  \begin{tabular}{L|L}
    \toprule
    \multicolumn{1}{c|}{parameter} 
    & \multicolumn{1}{c}{value} \\
    \midrule
    w_\pi & 0.2384^2/1.0252088 \\
    w_K   & 4.7621 \\
    \m_{3\pi^-} & (\m_{\pi^-} + \m_{\pi^-} + \m_{\pi^+})^2 \\
    \m_{2\pi^0\pi^-} & (\m_{\pi^0} + \m_{\pi^0} + \m_{\pi^-})^2 \\
    \m_{\rho\pi^0} & (\m_\rho + \m_{\pi^0})^2 \\
    \m_{K K^*} & (\m_K + \m_{K^*})^2 \\
    \bottomrule
  \end{tabular} & \sidecaption \\
\end{subtables}

\begin{table}\centering
  \captionabove{Parameters used in the $a_1$ running partial width fits
    of \tab{Tau:ThreePions.A1Fit} for the three pion
    hadronic current.\labeltab{ThreePions.A1FitParameters}}
  \begin{tabular}{L|L|LLL}
    \toprule
    \multicolumn{1}{c|}{$\Gamma_{\mathrm{partial}}$}
    & \multicolumn{1}{c|}{limits} 
    & \multicolumn{3}{c}{parameters} \\
    \midrule
    \multirow{3}{*}{$\Gamma_{\pi^-\pi^-\pi^+}$}
    & \m_{3\pi^-} \leq s < \m_{\rho\pi^0}
    & P_0 = 5.8090 & P_1 = 3.0098 & P_2 = 4.5792 \\
    & \m_{\rho\pi^0} \leq s 
    & P_0 = -13.914 & P_1 = 27.69    & P_2 = -13.393 \\
    & & P_3 = 3.1924  & P_4 = -0.10487 & \\
    \midrule
    \multirow{3}{*}{$\Gamma_{\pi^0\pi^0\pi^-}$}
    & \m_{2\pi^0\pi^-} \leq s \leq \m_{\rho\pi^0}
    & P_0 = 6.28450 & P_1 = 2.9595 & P_2 = 4.3355 \\
    & \m_{\rho\pi^0} \leq s
    & P_0 = -15.411 & P_1 = 32.088   & P_2 = -17.666 \\
    & & P_3 = 4.9355  & P_4 = -0.37498 & \\
    \bottomrule
  \end{tabular}
\end{table}

\begin{subfigures}{2}{Comparisons of the $m_{23}$ invariant mass
    distributions for the
    \subfig{q_qp.W.16_111_111_211.m_111_211}~$\pi^0 \pi^0 \pi^-$ decay
    channel and the \subfig{q_qp.W.16_211_211_211.m_211n_211p}~$\pi^-
    \pi^- \pi^+$ decay channel for the three pion hadronic
    current.\labelfig{ThreePions}}
  \svgbeg
  \svg{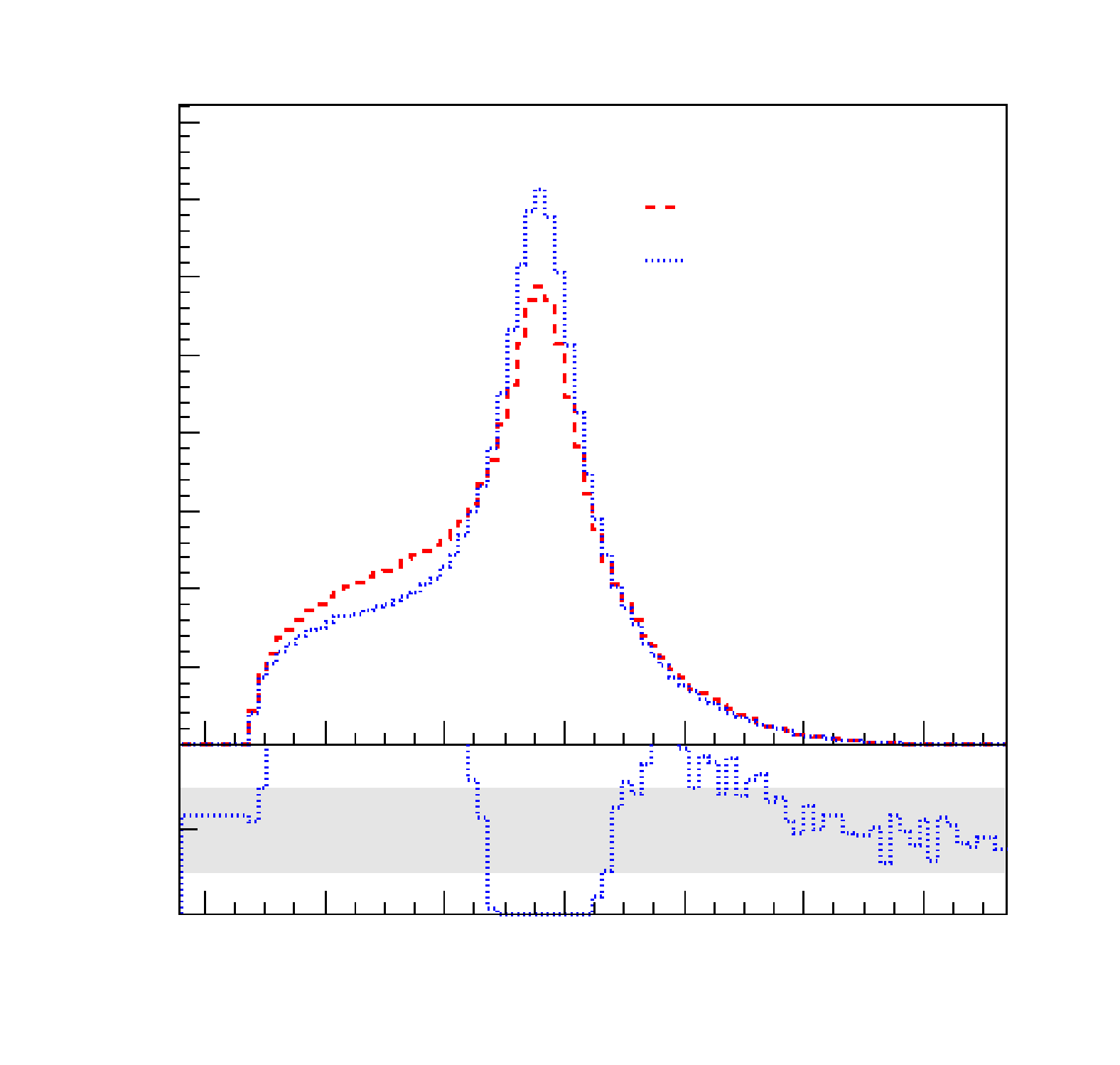} &
  \svg{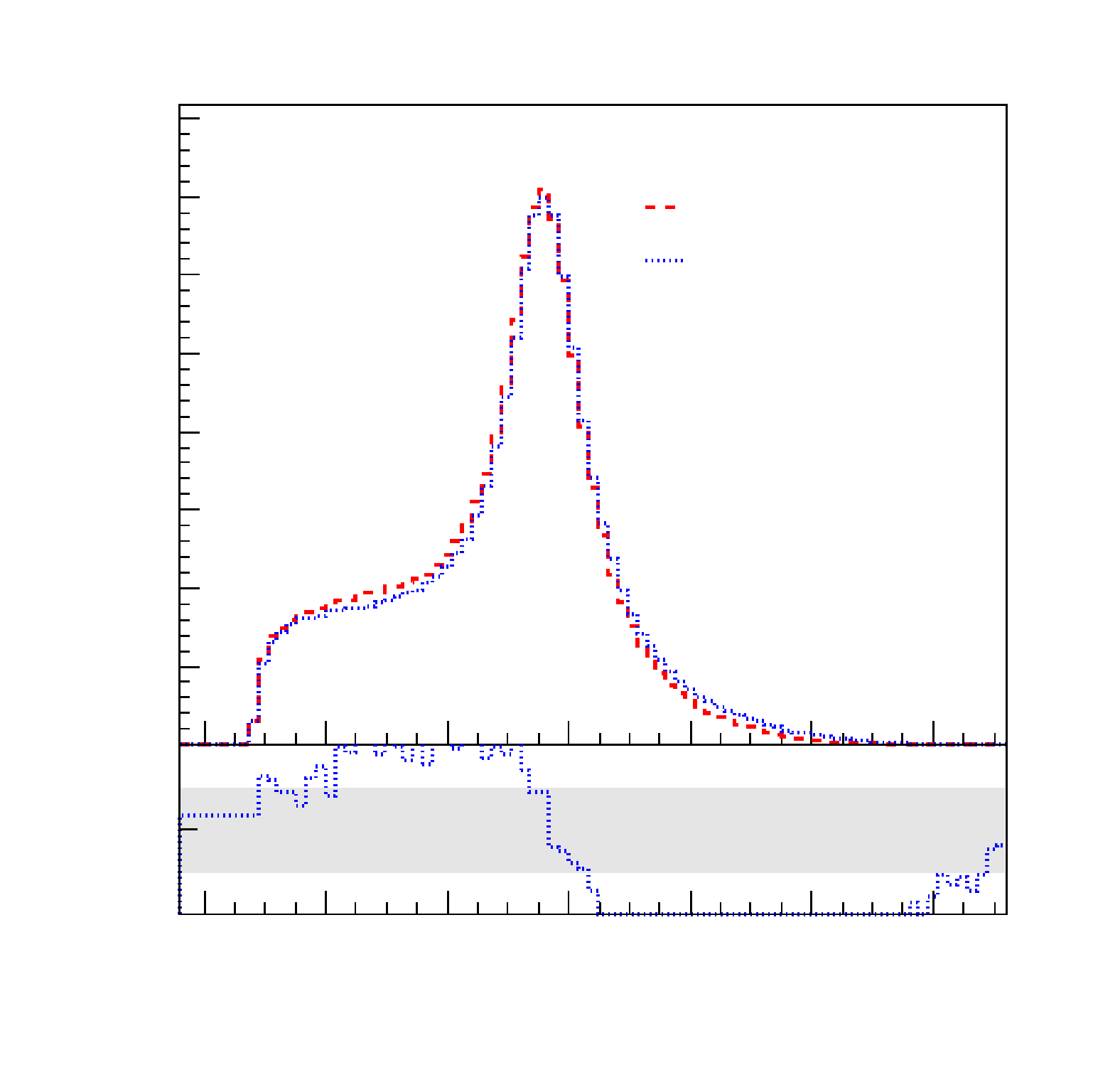} \svgend
\end{subfigures}

\Fig{Tau:ThreePions} gives the $\pi^0\pi^-$ and $\pi^-\pi^+$ invariant
mass distributions for the $\pi^- \pi^- \pi^+$ and $\pi^0 \pi^0 \pi^-$
decay channels. There is good general agreement between \herwig{++}
and \pythia{8} for the $\pi^- \pi^- \pi^+$ channel but a sizable
discrepancy for the $\pi^0 \pi^0 \pi^-$ channel due to a difference in
the implementation of the second form factor, $F_2^0$, of
\equ{Tau:ThreePions.F2n}. Additional differences between the two
distributions for both decay channels arise from the choice of
parameters used.

\newsubsubsection{Three Mesons with Kaons}{}

The model of \rfr{finkemeier.95.1} by Finkemeier and Mirkes is
implemented in \pythia{8} to provide $K^- \pi^- K^+$, $K^0 \pi^-
\bar{K}^0$, $K_S^0 \pi^- K_S^0$, $K_L^0 \pi^- K_L^0$, $K_S^0 \pi^-
K_L^0$,$K^- \pi^0 K^0$, $\pi^0 \pi^0 K^-$, $K^- \pi^- \pi^+$, and
$\pi^- \bar{K}^0 \pi^0$ decays for the \wtl. For this model the $F_1$,
$F_2$, and $F_4$ form factors are non-zero, while the $F_3$ and $F_5$
form factors are zero. The form factors for this model are given in
\tab{Tau:ThreeMesonsWithKaons.F1} for $F_1$,
\tab{Tau:ThreeMesonsWithKaons.F1} for $F_2$, and
\tab{Tau:ThreeMesonsWithKaons.F4} for $F_4$. These form factors use
the weighted sums of the intermediate resonances,
\begin{align*}\labelali{ThreeMesonsWithKaons.T}
  & t(s, \vec{m}, \vec{\Gamma}, \vec{w}) = \frac{1}{\sum_i w_i} w_i BW(s,
  m_i, \Gamma_i) \\
  & t_p(m_0, m_1, s, \vec{m}, \vec{\Gamma}, \vec{w}) = \frac{1}{\sum_i w_i}
  w_i BW_p(m_0, m_1, s, m_i, \Gamma_i) 
\end{align*}
where $\vec{m}$ is a vector of the resonance masses, $\vec{\Gamma}$
are their widths, and $\vec{w}$ are their real weights. The lengths of
these three vectors are the same, and correspond to the number of
resonances being modelled. Here, $t$ is a sum of fixed width
Breit-Wigners, given by \equ{Tau:Bw} and $t_p$ is the sum of $p$-wave
Breit-Wigners, given by \equ{Tau:BwP}.

\begin{table}\centering
  \captionabove{Vectors of masses, widths, and real weights for the
    resonances used in the form factors of
    \tabs{Tau:ThreeMesonsWithKaons.F1} through
    \ref{tab:Tau:ThreeMesonsWithKaons.F4} for \wtl decays into three
    mesons where one or more of the mesons are
    kaons.\labeltab{ThreeMesonsWithKaons.Resonances}}
  \begin{tabular}{L|L|L|L}
    \toprule
    \multicolumn{1}{c|}{resonances}
    & \multicolumn{1}{c|}{$\vec{\m}~[\gev]$}
    & \multicolumn{1}{c|}{$\vec{\Gamma}~[\gev]$}
    & \multicolumn{1}{c}{$\vec{w}$} \\
    \midrule
    \rho_a 
    & \begin{pmatrix}  0.773, &  1.370      \end{pmatrix}
    & \begin{pmatrix}  0.145, &  \p\p0.510  \end{pmatrix}
    & \begin{pmatrix}  1    , & -\frac{29}{200} \end{pmatrix}
    \\
    \rho_v 
    & \begin{pmatrix}  0.773, &  1.500,   &  1.750   \end{pmatrix}
    & \begin{pmatrix}  0.145, &  \p\p0.220, &  0.120 \end{pmatrix}
    & \begin{pmatrix}  1    , & -\frac{13}{52}, & -\frac{1}{26} \end{pmatrix}
    \\
    {K^*}_a 
    & \begin{pmatrix}  0.892, &  1.412      \end{pmatrix}
    & \begin{pmatrix}  0.050, &  \p\p0.227  \end{pmatrix}
    & \begin{pmatrix}  1    , & -\frac{27}{200} \end{pmatrix}
    \\
    {K^*}_v 
    & \begin{pmatrix}  0.892, &  1.412,     &  1.714 \end{pmatrix}
    & \begin{pmatrix}  0.050, &  \p\p0.227, &  0.323 \end{pmatrix}
    & \begin{pmatrix}  1    , & -\frac{13}{52}, & -\frac{1}{26} \end{pmatrix}
    \\
    {K_1}_a 
    & \begin{pmatrix}  1.402, &  1.270      \end{pmatrix}
    & \begin{pmatrix}  0.174, &  \p\p0.090  \end{pmatrix}
    & \begin{pmatrix}  1,     &  \phantom{-}\frac{33}{100} \end{pmatrix}
    \\
    {K_1}_b 
    & \begin{pmatrix}  1.270 \end{pmatrix}
    & \begin{pmatrix}  0.090 \end{pmatrix}
    & \begin{pmatrix}  1    \end{pmatrix}
    \\
    \omega 
    & \begin{pmatrix}  0.782,   &  1.020   \end{pmatrix}
    & \begin{pmatrix}  0.00843, &  0.0443  \end{pmatrix}
    & \begin{pmatrix}  1,       &  \phantom{-}\frac{1}{20} \end{pmatrix}
    \\
   \bottomrule
  \end{tabular}
\end{table}

The channels containing a single pion can proceed through an initial
$a_1$ resonance, similar to the three pion channels modelled with the
CLEO current. The Breit-Wigner for the $a_1$ is now defined as,
\begin{equation}
  BW_{a_1}(s) = \frac{\m_{a_1}^2}{\m_{a_1}^2 - s - i\m_{a_1}
    \Gamma_{a_1} \frac{g_{a_1}(s)}{g_{a_1}(\m_{a_1}^2)}}
  \labelequ{ThreeMesonsWithKaons.A1BW}
\end{equation}
where,
\begin{equation}
  g_{a_1}(s) = \begin{cases}
    0 
    & \mbox{if } s < (3\m_{\pi^-})^2 \\
    \begin{aligned} & 4.1(s - 9\m_{\pi^-}^2)^3 (1 -3.3 (s -
      9\m_{\pi^-}^2) \\ & + 5.8(s - 9\m_{\pi^-}^2)^2 \end{aligned}
    & \mbox{else if } s < (\m_\rho + \m_{\pi^-})^2 \\
    s(1.623 + \frac{10.38}{s} - \frac{9.32}{s^2} + \frac{0.65}{s^3})
    & \mbox{else} \\
  \end{cases}
\end{equation}
is an $a_1$ phase-space factor as given in \rfr{kuhn.90.1}. The
channels with a single pion can also proceed through initial $\rho$
resonances which are modelled with $p$-wave Breit-Wigners. The channels
containing two pions proceed through either initial $K_1$ resonances
modelled with fixed width Breit-Wigners or $K^*$ resonances modelled
with $p$-wave Breit-Wigners.

All the channels with a single pion can proceed through secondary
$K^*$ or $\rho$ resonances, except the $K_S^0 \pi^- K_S^0$ and $K_L^0
\pi^- K_L^0$ channels which can only proceed through secondary $K^*$
resonances. Additionally, the $K^- \pi^- K^+$, $K^0 \pi^- \bar{K}^0$,
and $K_S^0 \pi^- K_L^0$ channels can proceed via secondary $\omega$
resonances. The channels with two pions can proceed through secondary
$K^*$ resonances, and the $K^- \pi^- \pi^+$ and $\pi^- \bar{K}^0
\pi^0$ can proceed through additional secondary $\rho$ resonances.

\begin{table}[p]\centering
  \captionabove{The $F_1$ form factors for the three mesons with kaons
    decay model given for the relevant \wtl decay
    channels.\labeltab{ThreeMesonsWithKaons.F1}}
  \begin{tabular}{L|L}
    \toprule
    \multicolumn{1}{c|}{channel} & \multicolumn{1}{c}{$F_1$} \\
    \midrule
    K^- \pi^- K^+ 
    & -\frac{1}{6}BW_{a_1}(s_1) t_p(\m_{\pi^-}, \m_{K^0}, s_2,
    \vec{\m}_{{K^*}_a}, \vec{\Gamma}_{{K^*}_a}, \vec{w}_{{K^*}_a})
    \\ \midrule
    K^0 \pi^- \bar{K}^0 
    & -\frac{1}{6}BW_{a_1}(s_1) t_p(\m_{\pi^-}, \m_{K^0}, s_2,
    \vec{\m}_{{K^*}_a}, \vec{\Gamma}_{{K^*}_a}, \vec{w}_{{K^*}_a})
    \\ \midrule
    K_{S/L}^0 \pi^- K_{S/L}^0 
    & \begin{aligned}&\tfrac{1}{6}BW_{a_1}(s_1)
      (t_p(\m_{\pi^-}, \m_{K^0}, s_2, \vec{\m}_{{K^*}_a},
      \vec{\Gamma}_{{K^*}_a}, \vec{w}_{{K^*}_a})\\ &- t_p(\m_{\pi^-}, \m_{K^0},
      s_4, \vec{\m}_{{K^*}_a}, \vec{\Gamma}_{{K^*}_a}, \vec{w}_{{K^*}_a})) 
    \end{aligned}
    \\ \midrule
    K_S^0 \pi^- K_L^0 
    & \begin{aligned}&-\tfrac{1}{6}BW_{a_1}(s_1)
      (t_p(\m_{\pi^-}, \m_{K^0}, s_2, \vec{\m}_{{K^*}_a},
      \vec{\Gamma}_{{K^*}_a}, \vec{w}_{{K^*}_a})\\ &- t_p(\m_{\pi^-}, \m_{K^0},
      s_4, \vec{\m}_{{K^*}_a}, \vec{\Gamma}_{{K^*}_a}, \vec{w}_{{K^*}_a})) 
    \end{aligned}
    \\ \midrule
    K^- \pi^0 K^0 
    & \begin{aligned}&-\tfrac{1}{6}BW_{a_1}(s_1)
      (t_p(\m_{\pi^-}, \m_{K^0}, s_2, \vec{\m}_{{K^*}_a},
      \vec{\Gamma}_{{K^*}_a}, \vec{w}_{{K^*}_a})\\ &- t_p(\m_{\pi^-}, \m_{K^0},
      s_4, \vec{\m}_{{K^*}_a}, \vec{\Gamma}_{{K^*}_a}, \vec{w}_{{K^*}_a})) 
    \end{aligned}
    \\ \midrule
    \pi^0 \pi^0 K^-
    & -\frac{1}{3} t(s_1, \vec{\m}_{{K_1}_a}, \vec{\Gamma}_{{K_1}_a},
    \vec{w}_{{K_1}_a}) t_p(\m_{\pi^-}, \m_{K^0}, s_2,
    \vec{\m}_{{K^*}_a}, \vec{\Gamma}_{{K^*}_a}, \vec{w}_{{K^*}_a})
    \\ \midrule
    K^- \pi^- \pi^+
    & -\frac{1}{3} t(s_1, \vec{\m}_{{K_1}_b}, \vec{\Gamma}_{{K_1}_b},
    \vec{w}_{{K_1}_b}) t_p(\m_{\pi^-}, \m_{\pi^-}, s_2,
    \vec{\m}_{\rho_a}, \vec{\Gamma}_{\rho_a}, \vec{w}_{\rho_a})
    \\ \midrule
    \pi^- \bar{K}^0 \pi^0
    & \begin{aligned}&-\tfrac{1}{3} t(s_1, \vec{\m}_{{K_1}_a},
      \vec{\Gamma}_{{K_1}_a}, \vec{w}_{{K_1}_a}) (t_p(\m_{\pi^-},
      \m_{K^0}, s_2, \vec{\m}_{{K^*}_a}, \vec{\Gamma}_{{K^*}_a},
      \vec{w}_{{K^*}_a}) \\ & -t_p(\m_{\pi^-}, \m_{K^0}, s_4,
      \vec{\m}_{{K^*}_a}, \vec{\Gamma}_{{K^*}_a},
      \vec{w}_{{K^*}_a})) \end{aligned}
    \\
    \bottomrule
  \end{tabular} \\ ~ \\ ~ \\
  \captionabove{The $F_2$ form factors for the three mesons with kaons
    decay model given for the relevant \wtl decay
    channels.\labeltab{ThreeMesonsWithKaons.F2}}
  \begin{tabular}{L|L}
    \toprule
    \multicolumn{1}{c|}{channel} & \multicolumn{1}{c}{$F_2$} \\
    \midrule
    K^- \pi^- K^+ 
    & \frac{1}{6}BW_{a_1}(s_1) t_p(\m_{\pi^-}, \m_{\pi^-}, s_3,
    \vec{\m}_{\rho_a}, \vec{\Gamma}_{\rho_a}, \vec{w}_{\rho_a})
    \\ \midrule
    K^0 \pi^- \bar{K}^0 
    & \frac{1}{6}BW_{a_1}(s_1) t_p(\m_{\pi^-}, \m_{\pi^-}, s_3,
    \vec{\m}_{\rho_a}, \vec{\Gamma}_{\rho_a}, \vec{w}_{\rho_a})
    \\ \midrule
    K_{S/L}^0 \pi^- K_{S/L}^0 
    & \frac{1}{6}BW_{a_1}(s_1) (t_p(\m_{\pi^-}, \m_{K^0}, s_4,
    \vec{\m}_{{K^*}_a}, \vec{\Gamma}_{{K^*}_a}, \vec{w}_{{K^*}_a})
    \\ \midrule
    K_S^0 \pi^- K_L^0 
    & \begin{aligned}&\tfrac{1}{6}BW_{a_1}(s_1)
      (2 t_p(\m_{\pi^-}, \m_{\pi^-}, s_3, \vec{\m}_{\rho_a},
      \vec{\Gamma}_{\rho_a}, \vec{w}_{\rho_a})\\ &+ t_p(\m_{\pi^-}, \m_{K^0},
      s_4, \vec{\m}_{{K^*}_a}, \vec{\Gamma}_{{K^*}_a}, \vec{w}_{{K^*}_a})) 
    \end{aligned}
    \\ \midrule
    K^- \pi^0 K^0 
    & \begin{aligned}&\tfrac{1}{6}BW_{a_1}(s_1)
      (2 t_p(\m_{\pi^-}, \m_{\pi^-}, s_3, \vec{\m}_{\rho_a},
      \vec{\Gamma}_{\rho_a}, \vec{w}_{\rho_a})\\ &+ t_p(\m_{\pi^-}, \m_{K^0},
      s_4, \vec{\m}_{{K^*}_a}, \vec{\Gamma}_{{K^*}_a}, \vec{w}_{{K^*}_a})) 
    \end{aligned}
    \\ \midrule
    \pi^0 \pi^0 K^-
    & \frac{1}{3} t(s_1, \vec{\m}_{{K_1}_a}, \vec{\Gamma}_{{K_1}_a},
    \vec{w}_{{K_1}_a}) t_p(\m_{\pi^-}, \m_{K^0}, s_3,
    \vec{\m}_{{K^*}_a}, \vec{\Gamma}_{{K^*}_a}, \vec{w}_{{K^*}_a})
    \\ \midrule
    K^- \pi^- \pi^+
    & \frac{1}{3} t(s_1, \vec{\m}_{{K_1}_a}, \vec{\Gamma}_{{K_1}_a},
    \vec{w}_{{K_1}_a}) t_p(\m_{\pi^-}, \m_{K^0}, s_3,
    \vec{\m}_{{K^*}_a}, \vec{\Gamma}_{{K^*}_a}, \vec{w}_{{K^*}_a})
    \\ \midrule
    \pi^- \bar{K}^0 \pi^0
    & \begin{aligned} &\tfrac{2}{3} t(s_1, \vec{\m}_{{K_1}_b},
      \vec{\Gamma}_{{K_1}_b}, \vec{w}_{{K_1}_b}) t_p(\m_{\pi^-},
      \m_{\pi^-}, s_3, \vec{\m}_{\rho_a}, \vec{\Gamma}_{\rho_a},
      \vec{w}_{\rho_a}) \\ & + \tfrac{1}{3} t(s_1, \vec{\m}_{{K_1}_a},
      \vec{\Gamma}_{{K_1}_a}, \vec{w}_{{K_1}_a}) t_p(\m_{\pi^-},
      \m_{K^0}, s_4, \vec{\m}_{{K^*}_a}, \vec{\Gamma}_{{K^*}_a},
      \vec{w}_{{K^*}_a}) \end{aligned}
    \\
    \bottomrule
  \end{tabular}
\end{table}

\begin{table}[p]\centering
  \captionabove{The $F_4$ form factors for the three mesons with kaons
    decay model given for the relevant \wtl decay
    channels.\labeltab{ThreeMesonsWithKaons.F4}}
  \begin{tabular}{L|L}
    \toprule
    \multicolumn{1}{c|}{channel} & \multicolumn{1}{c}{$F_4$} \\
    \midrule
    K^- \pi^- K^+ 
    & \begin{aligned} &
      \tfrac{\sqrt{2} - 1}{8\pi^2 w_\pi^2} t_p(\m_{\pi^-},
      \m_{\pi^-}, s_1, \vec{\m}_{\rho_v}, \vec{\Gamma}_{\rho_v},
      \vec{w}_{\rho_v}) (\sqrt{2} t(s_3, \vec{\m}_\omega,
      \vec{\Gamma}_\omega, \vec{w}_\omega) \\ & + t_p(\m_{\pi^-},
      \m_{K^0}, s_2, \vec{\m}_{{K^*}_a}, \vec{\Gamma}_{{K^*}_a},
      \vec{w}_{{K^*}_a})) \end{aligned}
    \\ \midrule
    K^0 \pi^- \bar{K}^0 
    & \begin{aligned} &
      \tfrac{1 - \sqrt{2}}{8\pi^2 w_\pi^2} t_p(\m_{\pi^-},
      \m_{\pi^-}, s_1, \vec{\m}_{\rho_v}, \vec{\Gamma}_{\rho_v},
      \vec{w}_{\rho_v}) (\sqrt{2} t(s_3, \vec{\m}_\omega,
      \vec{\Gamma}_\omega, \vec{w}_\omega) \\ & + t_p(\m_{\pi^-},
      \m_{K^0}, s_2, \vec{\m}_{{K^*}_a}, \vec{\Gamma}_{{K^*}_a},
      \vec{w}_{{K^*}_a})) \end{aligned}
    \\ \midrule
    K_{S/L}^0 \pi^- K_{S/L}^0 
    & \begin{aligned} &
      \tfrac{\sqrt{2} - 1}{8\pi^2 w_\pi^2} t_p(\m_{\pi^-},
      \m_{\pi^-}, s_1, \vec{\m}_{\rho_v}, \vec{\Gamma}_{\rho_v},
      \vec{w}_{\rho_v}) \\ & (t_p(\m_{\pi^-}, \m_{K^0}, s_2,
      \vec{\m}_{{K^*}_a}, \vec{\Gamma}_{{K^*}_a},
      \vec{w}_{{K^*}_a}) \\ & - t_p(\m_{\pi^-}, \m_{K^0}, s_4,
      \vec{\m}_{{K^*}_a}, \vec{\Gamma}_{{K^*}_a},
      \vec{w}_{{K^*}_a})) \end{aligned}
    \\ \midrule
    K_S^0 \pi^- K_L^0 
    & \begin{aligned} &
      \tfrac{1 - \sqrt{2}}{8\pi^2 w_\pi^2} t_p(\m_{\pi^-},
      \m_{\pi^-}, s_1, \vec{\m}_{\rho_v}, \vec{\Gamma}_{\rho_v},
      \vec{w}_{\rho_v}) (2\sqrt{2}t(s_3, \vec{\m}_\omega, \vec{\Gamma}_\omega,
      \vec{w}_\omega) \\ & + t_p(\m_{\pi^-}, \m_{K^0}, s_2,
      \vec{\m}_{{K^*}_a}, \vec{\Gamma}_{{K^*}_a},
      \vec{w}_{{K^*}_a}) \\ & + t_p(\m_{\pi^-}, \m_{K^0}, s_4,
      \vec{\m}_{{K^*}_a}, \vec{\Gamma}_{{K^*}_a},
      \vec{w}_{{K^*}_a})) \end{aligned}
    \\ \midrule
    K^- \pi^0 K^0 
    & \begin{aligned} &
      \tfrac{1 - \sqrt{2}}{8\pi^2 w_\pi^2} t_p(\m_{\pi^-},
      \m_{\pi^-}, s_1, \vec{\m}_{\rho_v}, \vec{\Gamma}_{\rho_v},
      \vec{w}_{\rho_v}) \\ & (t_p(\m_{\pi^-}, \m_{K^0}, s_4,
      \vec{\m}_{{K^*}_a}, \vec{\Gamma}_{{K^*}_a}, \vec{w}_{{K^*}_a})
      \\ & - t_p(\m_{\pi^-}, \m_{K^0}, s_2, \vec{\m}_{{K^*}_a},
      \vec{\Gamma}_{{K^*}_a}, \vec{w}_{{K^*}_a})) \end{aligned}
    \\ \midrule
    \pi^0 \pi^0 K^-
    & \begin{aligned} &
      \tfrac{1}{8\pi^2 w_\pi^2} t_p(\m_{\pi^-},
      \m_{K^0}, s_1, \vec{\m}_{{K^*}_v}, \vec{\Gamma}_{{K^*}_v},
      \vec{w}_{{K^*}_v}) \\ &(t_p(\m_{\pi^-}, \m_{K^0}, s_2,
      \vec{\m}_{{K^*}_a}, \vec{\Gamma}_{{K^*}_a}, \vec{w}_{{K^*}_a})
      \\ & - t_p(\m_{\pi^-}, \m_{K^0}, s_3, \vec{\m}_{{K^*}_a},
      \vec{\Gamma}_{{K^*}_a}, \vec{w}_{{K^*}_a})) \end{aligned}
    \\ \midrule
    K^- \pi^- \pi^+
    & \begin{aligned} &
      -\tfrac{1}{8\pi^2 w_\pi^2} t_p(\m_{\pi^-},
      \m_{K^0}, s_1, \vec{\m}_{{K^*}_v}, \vec{\Gamma}_{{K^*}_v},
      \vec{w}_{{K^*}_v}) \\ &(t_p(\m_{\pi^-}, \m_{\pi^-}, s_2,
      \vec{\m}_{\rho_a}, \vec{\Gamma}_{\rho_a}, \vec{w}_{\rho_a})
      \\ & + t_p(\m_{\pi^-}, \m_{K^0}, s_3, \vec{\m}_{{K^*}_a},
      \vec{\Gamma}_{{K^*}_a}, \vec{w}_{{K^*}_a})) \end{aligned}
    \\ \midrule
    \pi^- \bar{K}^0 \pi^0
    & \begin{aligned} &
      \tfrac{1}{8\pi^2 w_\pi^2} t_p(\m_{\pi^-},
      \m_{K^0}, s_1, \vec{\m}_{{K^*}_v}, \vec{\Gamma}_{{K^*}_v},
      \vec{w}_{{K^*}_v}) \\ & (2t_p(\m_{\pi^-}, \m_{\pi^-}, s_3,
      \vec{\m}_{\rho_a}, \vec{\Gamma}_{\rho_a}, \vec{w}_{\rho_a})
      \\ & + t_p(\m_{\pi^-}, \m_{K^0}, s_2, \vec{\m}_{{K^*}_a},
      \vec{\Gamma}_{{K^*}_a}, \vec{w}_{{K^*}_a}) \\ &
      + t_p(\m_{\pi^-}, \m_{K^0}, s_4, \vec{\m}_{{K^*}_a},
      \vec{\Gamma}_{{K^*}_a}, \vec{w}_{{K^*}_a})) \end{aligned}
    \\
    \bottomrule
  \end{tabular}
\end{table}

The $\m_{23}$ distributions for the five channels modelled with this
current containing a single pion are given in
\fig{Tau:ThreeMesonsWithKaons.OnePion} while the distributions for the
three channels containing two pions are given in
\fig{Tau:ThreeMesonsWithKaons.TwoPions}. For all single pion channels,
the \pythia{8} and \herwig{++} distributions match well. However, he
\tauola distributions, particularly in the $K^0 \pi^- \bar{K}^0$ and
$K^- \pi^0 K^0$ channels, do not match. This is expected, as the
\tauola implementation for these channels uses an older model, which
among other differences, does not include the $K^*(1410)$
resonance. This older model is used for general three meson decays and
is introduced in the following section. A similar level of agreement
between the \pythia{8} and \herwig{++} distributions for the two pion
channels can also be seen, with the same expected discrepancies with
the \tauola distributions.

\begin{subfigures}[p]{2}{Distributions of $\m_{23}$ for the
    \subfig{q_qp.W.16_211_321_321.m_211_321n}~$K^- \pi^- K^+$,
    \subfig{q_qp.W.16_211_311_311.m_211_311}~$K^0 \pi^- \bar{K}^0$,
    \subfig{q_qp.W.16_211_310_310.m_211_310}~$K_S^0 \pi^- K_S^0$,
    \subfig{q_qp.W.16_130_211_310.m_211_310}~$K_S^0 \pi^- K_L^0$, and
    \subfig{q_qp.W.16_111_311_321.m_111_321}~$K^- \pi^0 K^0$ \wtl
    decay channels using the three mesons with kaons model by
    Finkemeier and
    Mirkes~\cite{finkemeier.95.1}.\labelfig{ThreeMesonsWithKaons.OnePion}}
  \svgbeg
  \svg{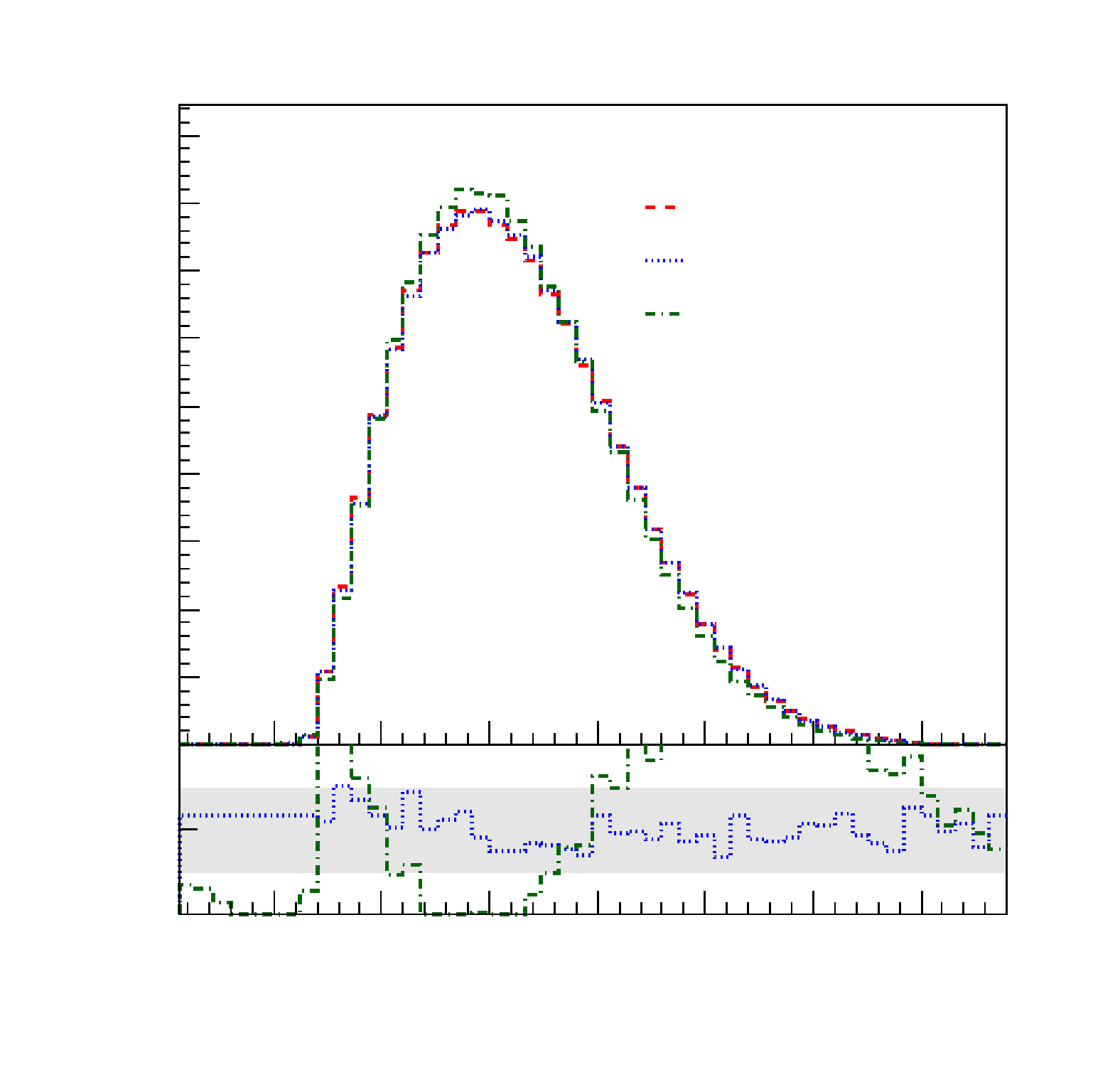} &
  \svg{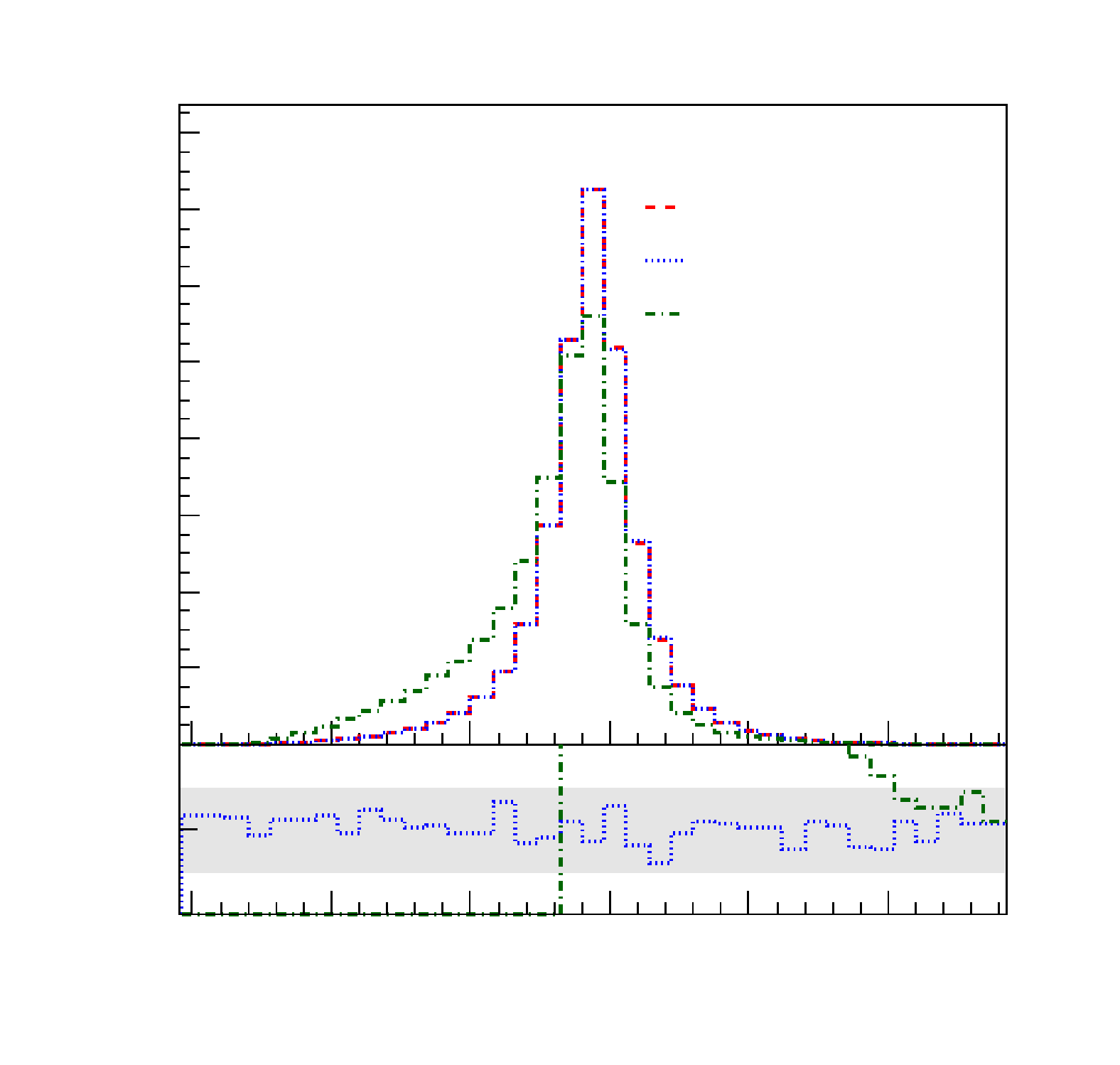} \svgsep
  \svg{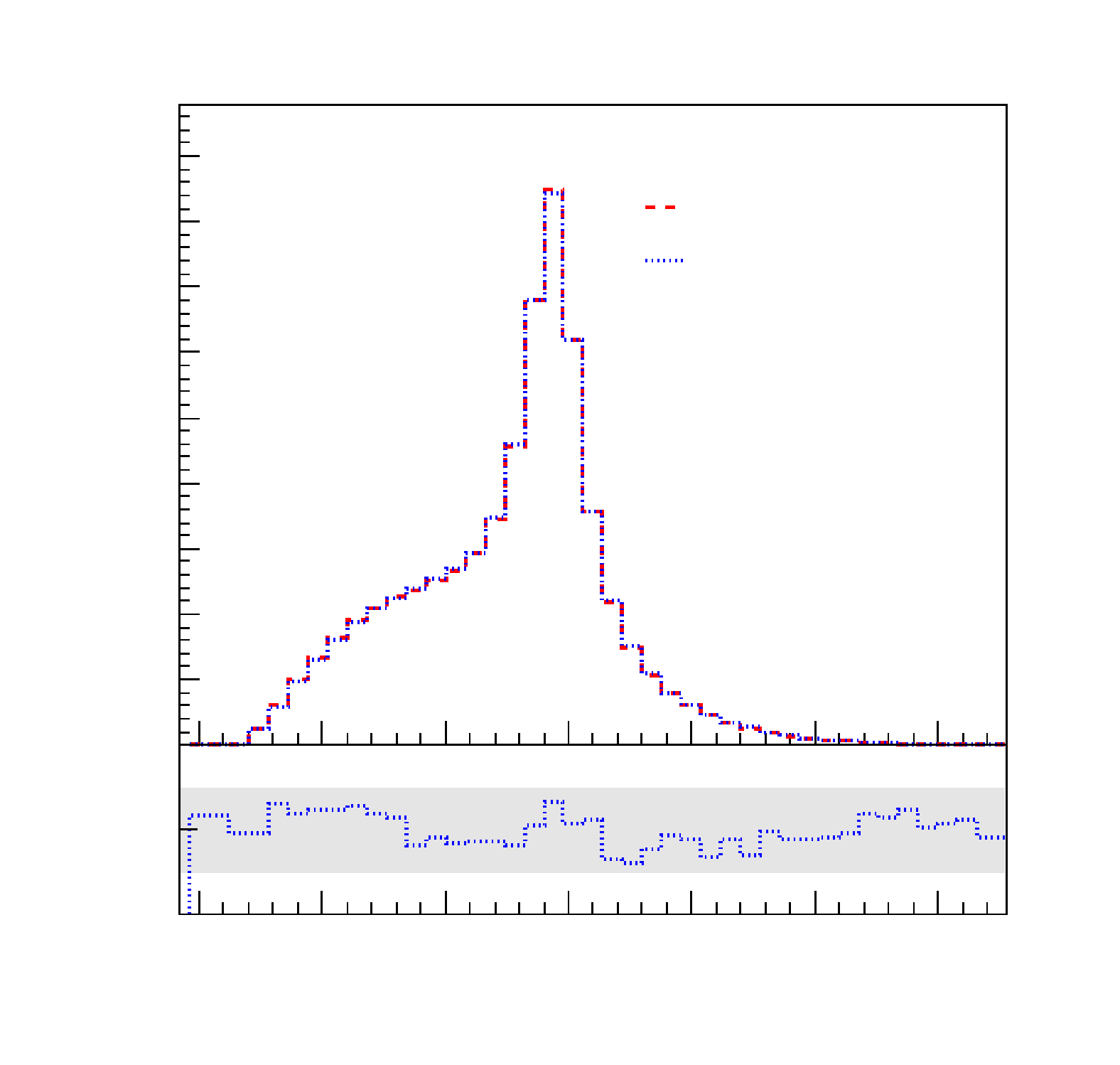} &
  \svg{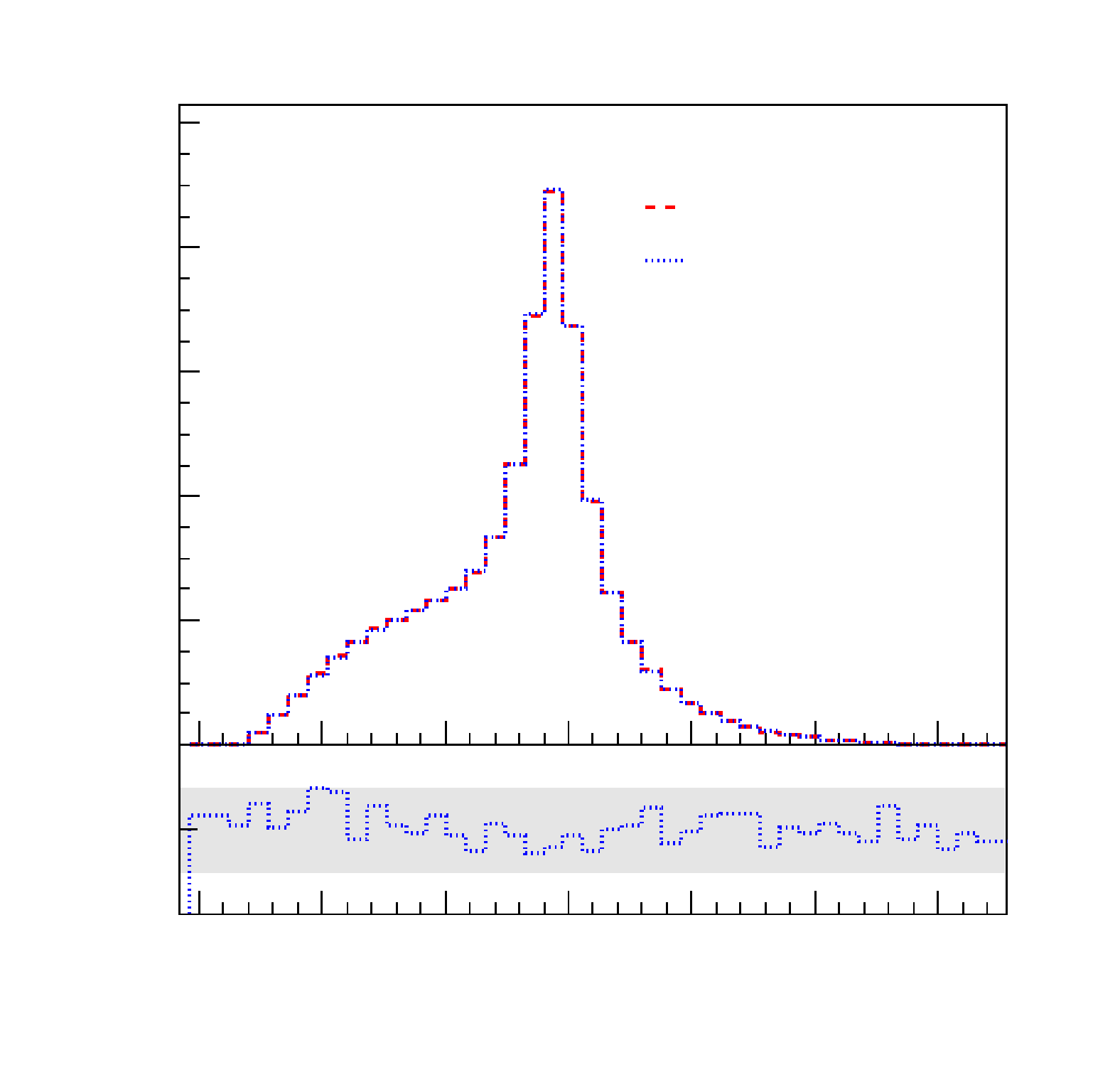} \svgsep
  \svg{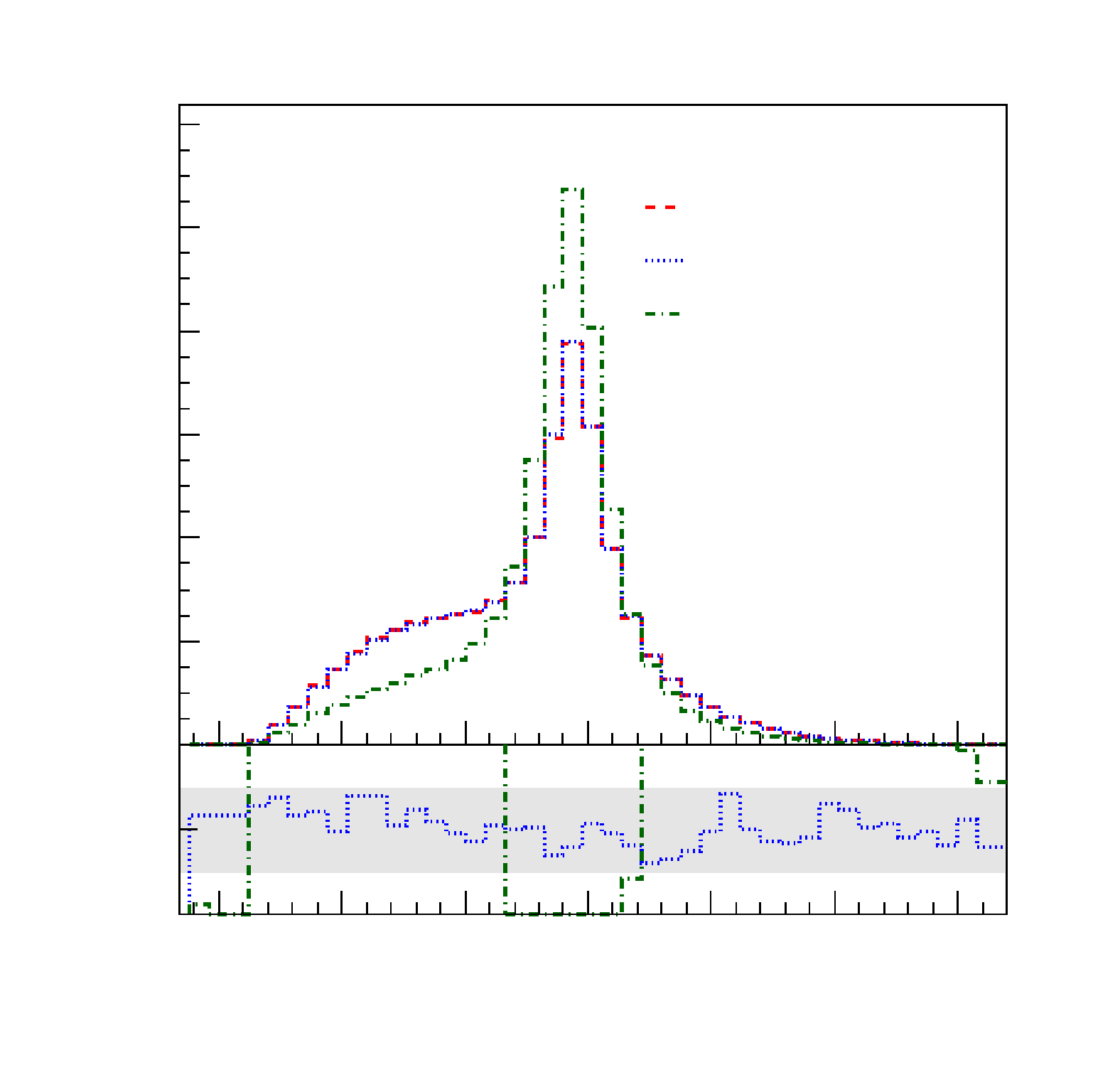} & \sidecaption \svgend
\end{subfigures}

\begin{subfigures}{2}{Distributions of $\m_{23}$ for the
    \subfig{q_qp.W.16_111_111_321.m_111_321}~$\pi^0 \pi^0 K^-$,
    \subfig{q_qp.W.16_211_211_321.m_211n_321}~$K^- \pi^- \pi^+$, and
    \subfig{q_qp.W.16_111_211_311.m_211_311}~$\pi^- \bar{K}^0 \pi^0$
    decay channels using the three mesons with kaons model by
    Finkemeier and
    Mirkes~\cite{finkemeier.95.1}.\labelfig{ThreeMesonsWithKaons.TwoPions}}
  \svgbeg
  \svg{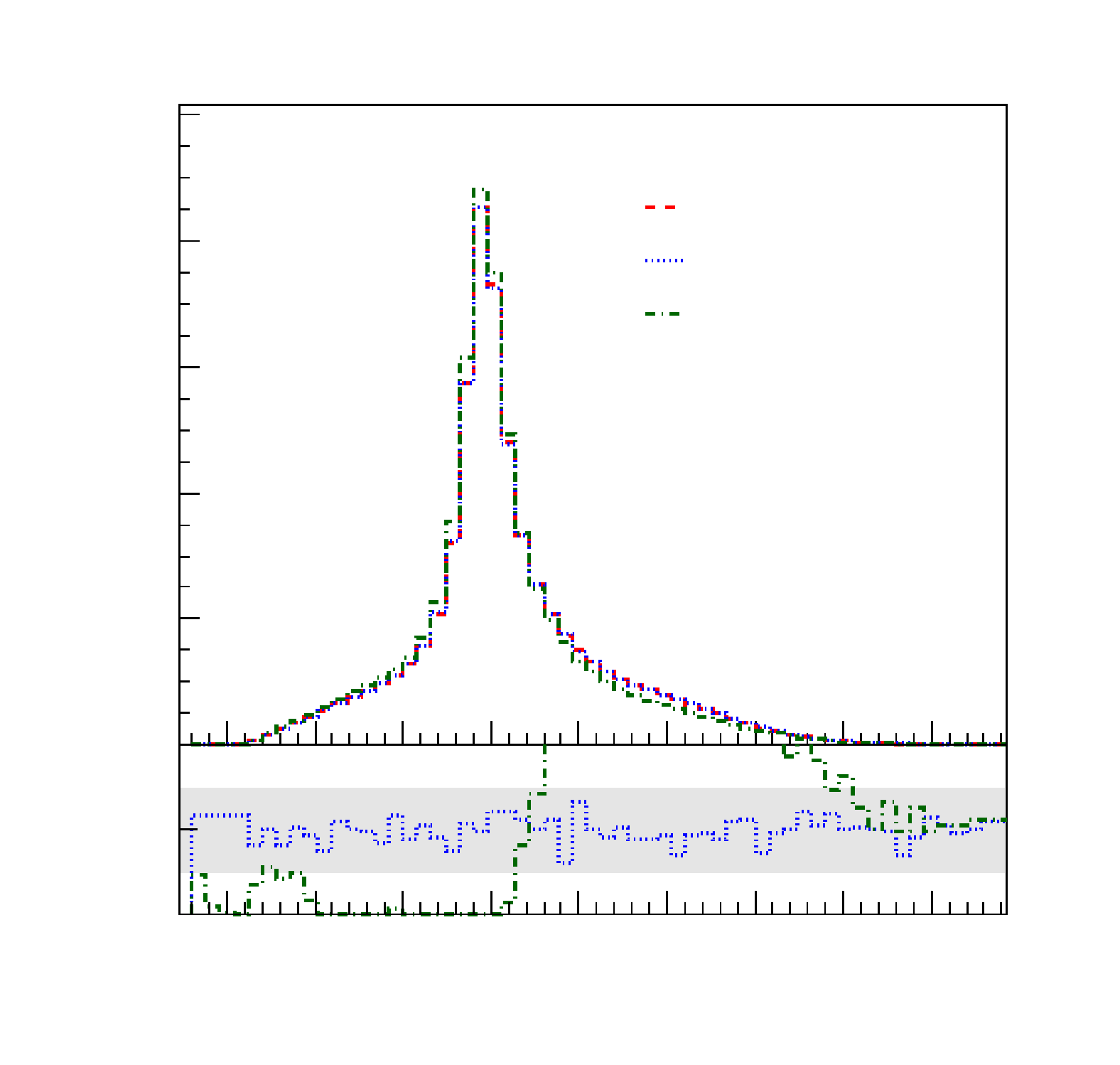} &
  \svg{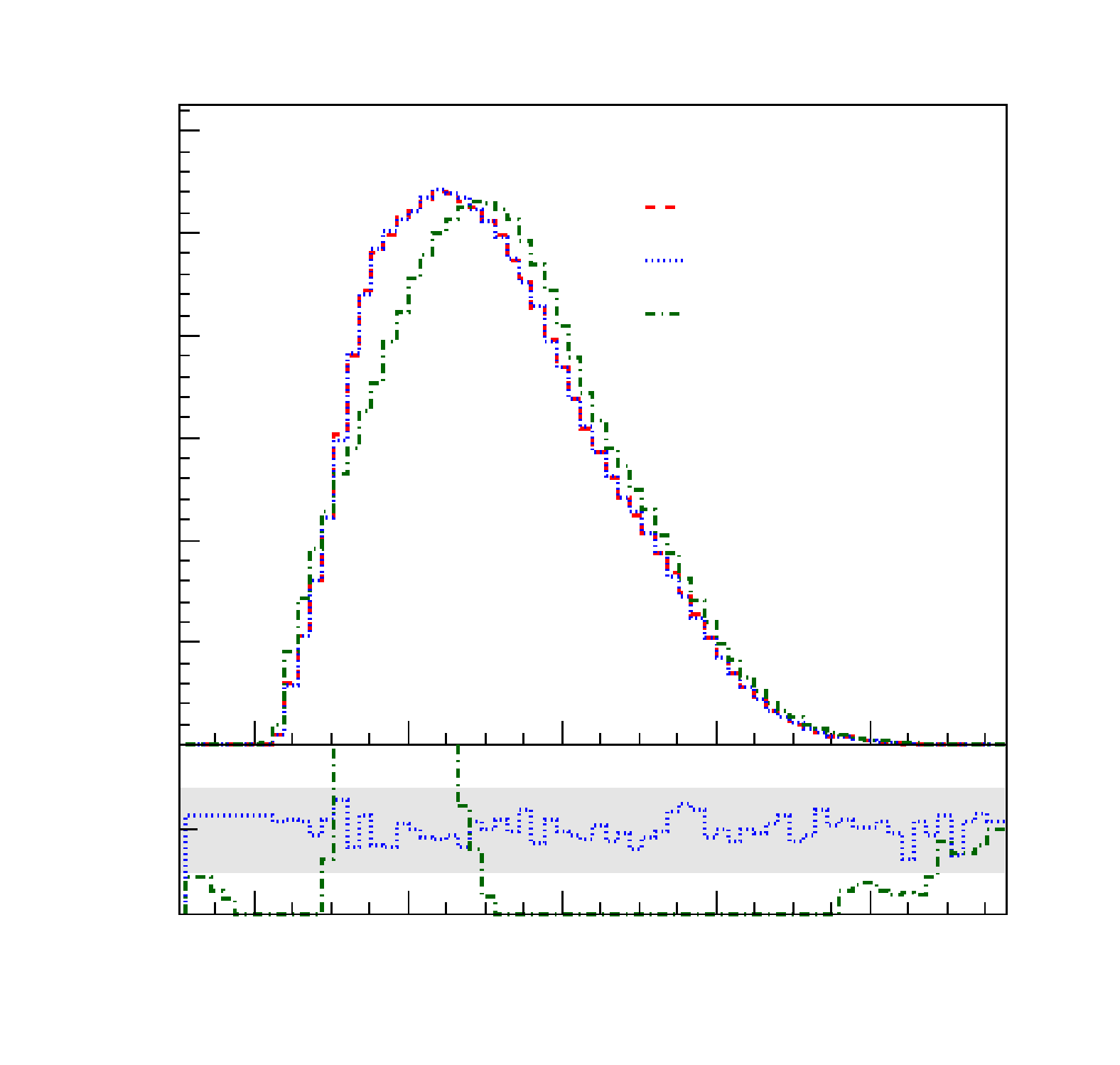} \svgsep
  \svg{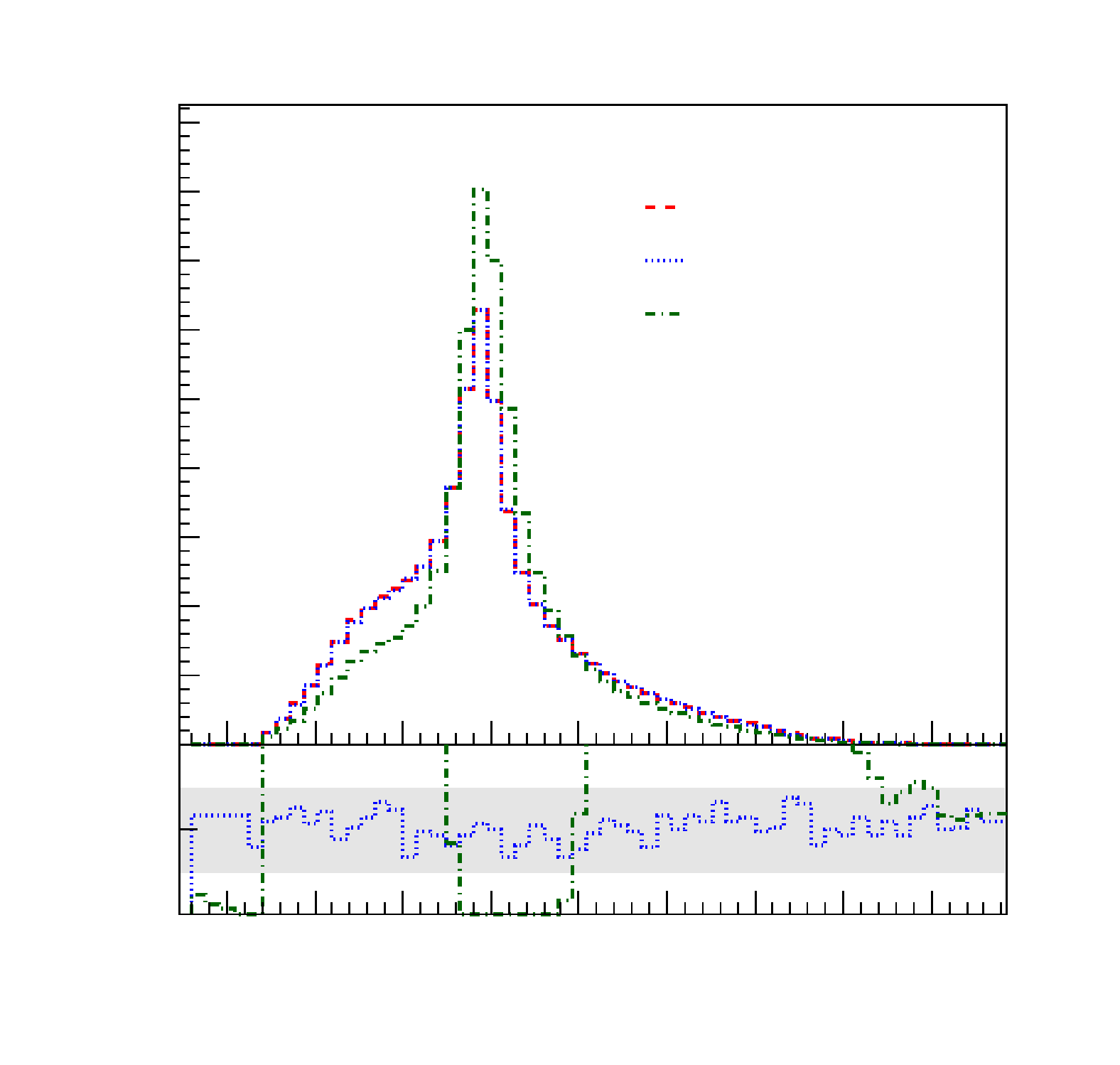} & \sidecaption \svgend
\end{subfigures}

\newsubsubsection{General Three Mesons}{}

The older and more general model of Decker, {\it et al.} of
\rfr{decker.93.1} is also implemented in \pythia{8} and can be used to
perform the $\pi^0 \pi^0 \pi^+$, $\pi^- \pi^- \pi^+$, $K^- \pi^- K^+$,
$K^0 \pi^- \bar{K}^0$, $K^- \pi^0 K^0$, $\pi^0 \pi^0 K^-$, $K^- \pi^-
\pi^+$, $\pi^- \bar{K}^0 \pi^0$, and $\pi^- \pi^0 \eta$ decays of the
\wtl. However, by default, only the $\pi^- \pi^0 \eta$ decay of the
\wtl is performed using this model. The same summation of fixed width
and $p$-wave Breit-Wigners using $t$ and $t_p$ of
\equ{Tau:ThreeMesonsWithKaons.T} is used, as well as the same $a_1$
Breit-Wigner of \equ{Tau:ThreeMesonsWithKaons.A1BW}. However, the
masses, widths, and real weights of the resonances differ from those
of \tab{Tau:ThreeMesonsWithKaons.Resonances} and are given in
\tab{Tau:ThreeMesonsGeneric.Resonances}. Note that no $\omega$
resonances are introduced.

\begin{table}\centering
  \captionabove{Vectors of masses, widths, and real weights for the
    resonances used in the form factors of
    \tabs{Tau:ThreeMesonsGeneric.F1} through
    \ref{tab:Tau:ThreeMesonsGeneric.F4} for \wtl decays into three
    mesons using the general three meson
    model.\labeltab{ThreeMesonsGeneric.Resonances}}
  \begin{tabular}{L|L|L|L}
    \toprule
    \multicolumn{1}{c|}{resonances}
    & \multicolumn{1}{c|}{$\vec{\m}~[\gev]$}
    & \multicolumn{1}{c|}{$\vec{\Gamma}~[\gev]$}
    & \multicolumn{1}{c}{$\vec{w}$} \\
    \midrule
    \rho_a 
    & \begin{pmatrix}  0.773, &  1.370 \end{pmatrix}
    & \begin{pmatrix}  0.145, &  0.510 \end{pmatrix}
    & \begin{pmatrix}  1    , & -\frac{29}{200} \end{pmatrix}
    \\
    \rho_v 
    & \begin{pmatrix}  0.773, &  1.500, &  1.750 \end{pmatrix}
    & \begin{pmatrix}  0.145, &  0.220, &  0.120 \end{pmatrix}
    & \begin{pmatrix}  1    , &  -\frac{13}{52}, & -\frac{1}{26} \end{pmatrix}
    \\
    K^*
    & \begin{pmatrix}  0.892 \end{pmatrix}
    & \begin{pmatrix}  0.0513 \end{pmatrix}
    & \begin{pmatrix}  1     \end{pmatrix}
    \\
    K_1 
    & \begin{pmatrix}  1.402 \end{pmatrix}
    & \begin{pmatrix}  0.174 \end{pmatrix}
    & \begin{pmatrix}  1     \end{pmatrix}
    \\
   \bottomrule
  \end{tabular}
\end{table}

\begin{table}[p]\centering
  \captionabove{The $F_1$ form factors for the general three meson
    model given for the relevant \wtl decay
    channels.\labeltab{ThreeMesonsGeneric.F1}}
  \begin{tabular}{L|L}
    \toprule
    \multicolumn{1}{c|}{channel} & \multicolumn{1}{c}{$F_1$} \\
    \midrule
    \pi^{-/0} \pi^{-/0} \pi^{+/-} 
    & BW_{a_1}(s_1) t_p(\m_{\pi^-}, \m_{\pi^-}, s_2,
    \vec{\m}_{\rho_a}, \vec{\Gamma}_{\rho_a}, \vec{w}_{\rho_a})
    \\ \midrule
    K^- \pi^- K^+ 
    & -\frac{1}{3}BW_{a_1}(s_1) t_p(\m_{\pi^-}, \m_{K^0}, s_2,
    \vec{\m}_{K^*}, \vec{\Gamma}_{K^*}, \vec{w}_{K^*})
    \\ \midrule
    K^0 \pi^- \bar{K}^0 
    & -\frac{1}{3}BW_{a_1}(s_1) t_p(\m_{\pi^-}, \m_{K^0}, s_2,
    \vec{\m}_{K^*}, \vec{\Gamma}_{K^*}, \vec{w}_{K^*})
    \\ \midrule
    K^- \pi^0 K^0
    & 0 
    \\ \midrule
    \pi^0 \pi^0 K^-
    & t(s_1, \vec{\m}_{K_1}, \vec{\Gamma}_{K_1}, \vec{w}_{K_1})
    t_p(\m_{\pi^-}, \m_{K^0}, s_2,
    \vec{\m}_{K^*}, \vec{\Gamma}_{K^*}, \vec{w}_{K^*})
    \\ \midrule
    K^- \pi^- \pi^+
    & -\frac{1}{3}t(s_1, \vec{\m}_{K_1}, \vec{\Gamma}_{K_1}, \vec{w}_{K_1})
    t_p(\m_{\pi^-}, \m_{\pi^-}, s_2,
    \vec{\m}_{\rho_a}, \vec{\Gamma}_{\rho_a}, \vec{w}_{\rho_a})
    \\ \midrule
    \pi^- \bar{K}^0 \pi^0
    & 0
    \\ \midrule
    \pi^- \pi^0 \eta
    & 0 \\
    \bottomrule
  \end{tabular} \\ ~ \\ ~ \\
  \captionabove{The $F_2$ form factors for the general three meson
    model given for the relevant \wtl decay
    channels.\labeltab{ThreeMesonsGeneric.F2}}
  \begin{tabular}{L|L}
    \toprule
    \multicolumn{1}{c|}{channel} & \multicolumn{1}{c}{$F_2$} \\
    \midrule
    \pi^{-/0} \pi^{-/0} \pi^{+/-} 
    & -BW_{a_1}(s_1) t_p(\m_{\pi^-}, \m_{\pi^-}, s_3,
    \vec{\m}_{\rho_a}, \vec{\Gamma}_{\rho_a}, \vec{w}_{\rho_a})
    \\ \midrule
    K^- \pi^- K^+ 
    & \frac{1}{3}BW_{a_1}(s_1) t_p(\m_{\pi^-}, \m_{\pi^-}, s_3,
    \vec{\m}_{\rho_a}, \vec{\Gamma}_{\rho_a}, \vec{w}_{\rho_a})
    \\ \midrule
    K^0 \pi^- \bar{K}^0 
    & \frac{1}{3}BW_{a_1}(s_1) t_p(\m_{\pi^-}, \m_{\pi^-}, s_3,
    \vec{\m}_{\rho_a}, \vec{\Gamma}_{\rho_a}, \vec{w}_{\rho_a})
    \\ \midrule
    K^- \pi^0 K^0
    & BW_{a_1}(s_1) t_p(\m_{\pi^-}, \m_{\pi^-}, s_3,
    \vec{\m}_{\rho_a}, \vec{\Gamma}_{\rho_a}, \vec{w}_{\rho_a})
    \\ \midrule
    \pi^0 \pi^0 K^-
    & -t(s_1, \vec{\m}_{K_1}, \vec{\Gamma}_{K_1}, \vec{w}_{K_1})
    t_p(\m_{\pi^-}, \m_{K^0}, s_3,
    \vec{\m}_{K^*}, \vec{\Gamma}_{K^*}, \vec{w}_{K^*})
    \\ \midrule
    K^- \pi^- \pi^+
    & \frac{1}{3}t(s_1, \vec{\m}_{K_1}, \vec{\Gamma}_{K_1}, \vec{w}_{K_1})
    t_p(\m_{\pi^-}, \m_{K^0}, s_3,
    \vec{\m}_{K^*}, \vec{\Gamma}_{K^*}, \vec{w}_{K^*})
    \\ \midrule
    \pi^- \bar{K}^0 \pi^0
    & t(s_1, \vec{\m}_{K_1}, \vec{\Gamma}_{K_1}, \vec{w}_{K_1})
    t_p(\m_{\pi^-}, \m_{\pi^-}, s_3,
    \vec{\m}_{\rho_a}, \vec{\Gamma}_{\rho_a}, \vec{w}_{\rho_a})
    \\ \midrule
    \pi^- \pi^0 \eta
    & 0 \\
    \bottomrule
  \end{tabular}
\end{table}

\begin{table}\centering
  \captionabove{The $F_4$ form factors for the general three meson
    model given for the relevant \wtl decay
    channels.\labeltab{ThreeMesonsGeneric.F4}}
  \begin{tabular}{L|L}
    \toprule
    \multicolumn{1}{c|}{channel} & \multicolumn{1}{c}{$F_4$} \\
    \midrule
    \pi^{-/0} \pi^{-/0} \pi^{+/-} 
    & 0
    \\ \midrule
    K^- \pi^- K^+ 
    & \begin{aligned} & \tfrac{5}{16\pi^2w_\pi^2} 
      t_p(\m_{\pi^-}, \m_{\pi^-}, s_1,
      \vec{\m}_{\rho_v}, \vec{\Gamma}_{\rho_v}, \vec{w}_{\rho_v})
      (t_p(\m_{\pi^-}, \m_{\pi^-}, s_3,
      \vec{\m}_{\rho_a}, \vec{\Gamma}_{\rho_a}, \vec{w}_{\rho_a})
      \\ & - \tfrac{2}{10}t_p(\m_{\pi^-}, \m_{K^0}, s_2,
      \vec{\m}_{K^*}, \vec{\Gamma}_{K^*}, \vec{w}_{K^*})) \end{aligned}
    \\ \midrule
    K^0 \pi^- \bar{K}^0 
    & \begin{aligned} & -\tfrac{5}{16\pi^2w_\pi^2} 
      t_p(\m_{\pi^-}, \m_{\pi^-}, s_1,
      \vec{\m}_{\rho_v}, \vec{\Gamma}_{\rho_v}, \vec{w}_{\rho_v})
      \\ & (t_p(\m_{\pi^-}, \m_{\pi^-}, s_3,
      \vec{\m}_{\rho_a}, \vec{\Gamma}_{\rho_a}, \vec{w}_{\rho_a})
      \\ & - \tfrac{2}{10}t_p(\m_{\pi^-}, \m_{K^0}, s_2,
      \vec{\m}_{K^*}, \vec{\Gamma}_{K^*}, \vec{w}_{K^*})) \end{aligned}
    \\ \midrule
    K^- \pi^0 K^0
    & 0 
    \\ \midrule
    \pi^0 \pi^0 K^-
    & 0
    \\ \midrule
    K^- \pi^- \pi^+
    & \begin{aligned} & -\tfrac{5}{16\pi^2w_\pi^2} 
      t_p(\m_{\pi^-}, \m_{K^0}, s_1,
      \vec{\m}_{K^*}, \vec{\Gamma}_{K^*}, \vec{w}_{K^*})
      \\ &(t_p(\m_{\pi^-}, \m_{\pi^-}, s_2,
      \vec{\m}_{\rho_a}, \vec{\Gamma}_{\rho_a}, \vec{w}_{\rho_a})
      \\ & - \tfrac{2}{10}t_p(\m_{\pi^-}, \m_{K^0}, s_3,
      \vec{\m}_{K^*}, \vec{\Gamma}_{K^*}, \vec{w}_{K^*})) \end{aligned}
    \\ \midrule
    \pi^- \bar{K}^0 \pi^0
    & \begin{aligned} & \tfrac{5}{8\pi^2w_\pi^2} 
      t_p(\m_{\pi^-}, \m_{K^0}, s_1,
      \vec{\m}_{K^*}, \vec{\Gamma}_{K^*}, \vec{w}_{K^*})
      (t_p(\m_{\pi^-}, \m_{\pi^-}, s_3,
      \vec{\m}_{\rho_a}, \vec{\Gamma}_{\rho_a}, \vec{w}_{\rho_a})
      \\ & - \tfrac{2}{10}t_p(\m_{\pi^-}, \m_{K^0}, s_2,
      \vec{\m}_{K^*}, \vec{\Gamma}_{K^*}, \vec{w}_{K^*})) \end{aligned}
    \\ \midrule
    \pi^- \pi^0 \eta
    & \tfrac{1}{4\pi^2w_\pi^2}  t_p(\m_{\pi^-}, \m_{\pi^-}, s_1,
    \vec{\m}_{\rho_v}, \vec{\Gamma}_{\rho_v}, \vec{w}_{\rho_v})
    t_p(\m_{\pi^-}, \m_{\pi^-}, s_4,
    \vec{\m}_{\rho_a}, \vec{\Gamma}_{\rho_a}, \vec{w}_{\rho_a})
    \\
    \bottomrule
  \end{tabular}
\end{table}

\begin{subfigures}{2}{Distributions of $\m_{23}$ for the default $\pi^-
    \pi^0 \eta$ decay channel using the general three meson model by
    Decker {\it et
      al.}~\cite{decker.93.1}.\labelfig{ThreeMesonsGeneric.Eta}}
  \svgbeg
  \svg[1]{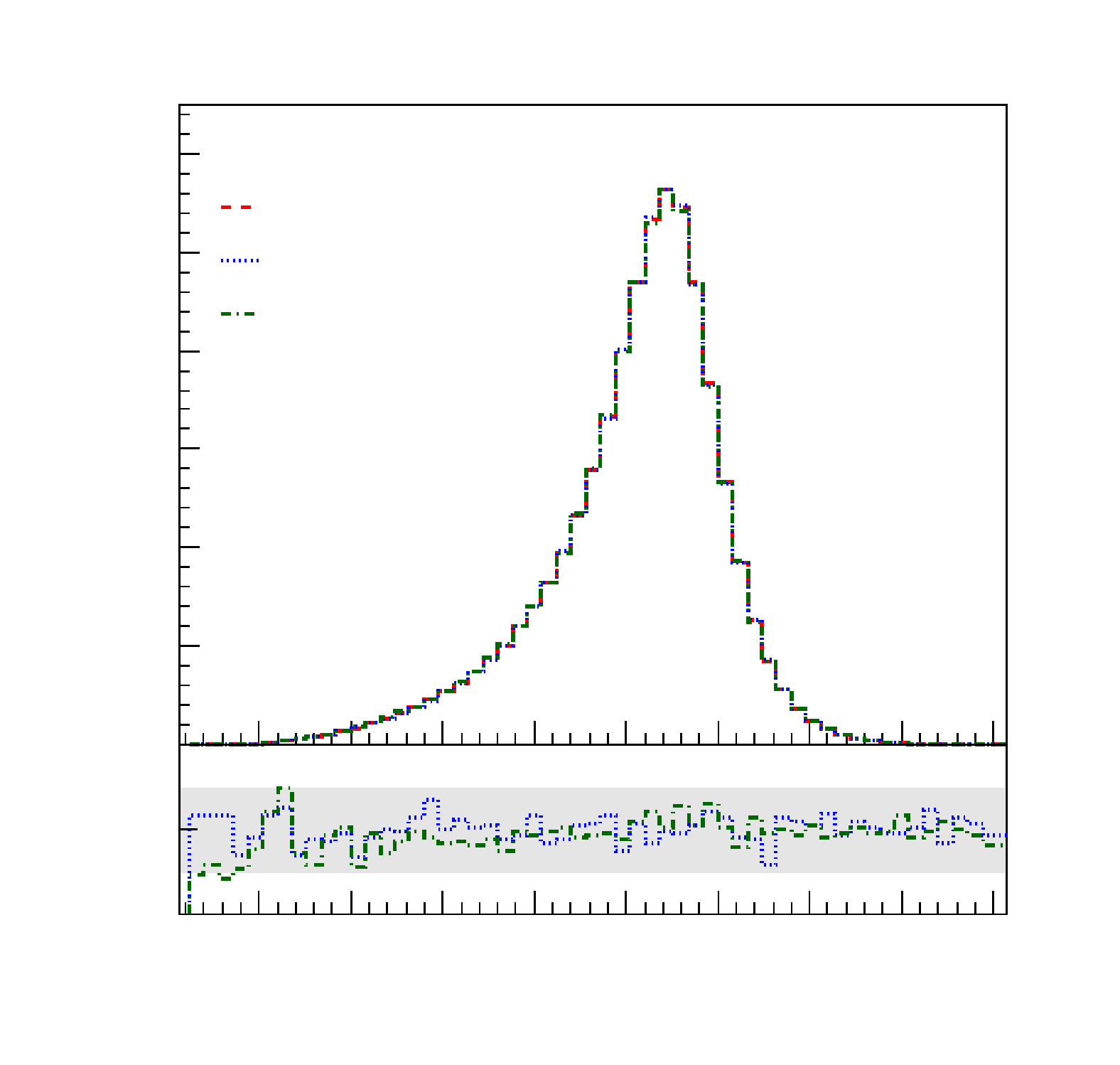} & \sidecaption \svgend
\end{subfigures}

The form factors for this model are provided in
\tab{Tau:ThreeMesonsGeneric.F1} for $F_1$,
\tab{Tau:ThreeMesonsGeneric.F1} for $F_2$, and
\tab{Tau:ThreeMesonsGeneric.F4} for $F_4$. Here $F_3$ and $F_5$ are
zero for all channels. The three pion channels decay through an
initial $a_1$ resonance followed by a decay through secondary $\rho$
resonances. The $K^- \pi^- K^+$ and $K^- \pi^- \bar{K}^0$ channels can
decay through an initial $a_1$ or $\rho$ resonance and through
secondary $K^*$ and $\rho$ resonances. The $K^- \pi^0 K^0$ channel can
proceed through an initial $a_1$ resonance followed by secondary $K^*$
resonances. The $\pi^0 \pi^- K^-$ channel proceeds through initial
$K_1$ resonances and secondary $K^*$ resonances, while the $K^- \pi^-
\pi^+$ and $\pi^- \bar{K}^0 \pi^0$ channels proceed through initial
$K_1$ or $K^*$ resonances and secondary $K^*$ or $\rho$ resonances.

In \fig{Tau:ThreeMesonsGeneric.Eta} the $\m_{23}$ distribution from
\pythia{8}, \herwig{++}, and \tauola for the $\pi^- \pi^0 \eta$ decay
channel is given. There is good agreement between all three
generators. Distributions for the additional decays that can be
performed with the general three meson model but are not used by
default in \pythia{8} are provided in \app{Dvr} for brevity, where
\fig{Dvr:ThreeMesonsGeneric.OnePion} are the $\m_{23}$ distributions
for decays with only one pion and
\fig{Dvr:ThreeMesonsGeneric.TwoPions} are the distributions for decays
with two or more pions.

\newsubsubsection{Two Pions with a Photon}{}

The $\omega$ meson can decay into $\pi^0 \pi^- \pi^+$ or $\gamma
\pi^0$ final states. Consequently, the decay $\tauto \omega \pi^0
\pi^-$ can result in a four-body or five-body final state. The
five-body final state is included in the intermediate resonances of
the five-body decays in \sec{Tau:DecFive}, but the four-body final
state decay $\tauto \gamma \pi^0 \pi^-$ needs to be accounted for with
an independent model. In \pythia{8} the model of Jadach {\it et al.}
from \rfr{jadach.93.1} is implemented for this decay, and is given by
the hadronic current,
\begin{align*}\labelali{TwoPionsGamma}
  J^\mu \propto~ & F(s_1, \vec{m}_\rho, \vec{G}_\rho, \vec{w}_\rho) 
  F(0, \vec{m}_\rho, \vec{G}_\rho, \vec{w}_\rho) 
  F(s_4, \vec{m}_\omega, \vec{G}_\omega, \vec{w}_\omega) \\
  & \bigg(
  \varepsilon_2^\mu\left(\m_{\pi^-}^2{q_4}_\nu q_2^\nu -
    {q_3}_\nu q_2^\nu({q_4}_\nu q_3^\nu - {q_4}_\nu q_2^\nu)\right) \\ &
  - {q_3}^\mu \left(({q_3}_\nu \varepsilon_2^\nu)({q_4}_\nu q_2^\nu)
    - ({q_4}_\nu \varepsilon_2^\nu)({q_3}_\nu q_2^\nu) \right) \\ &
  - {q_2}^\mu \left(({q_3}_\nu \varepsilon_2^\nu)({q_4}_\nu q_3^\nu)
    - ({q_4}_\nu \varepsilon_2^\nu)(\m_{\pi^-}^2 + {q_3}_\nu q_2^\nu) \right)
  \bigg) 
\end{align*}
where $\varepsilon_2$ is the polarisation vector of the photon given
by \equ{Thr:Spin2E}. Unlike the other hadronic currents of
\sec{Tau:Dec}, this current takes on two spin states due to the photon
in the final state. The form factor $F$ is given by,
\begin{equation}
  F(s, \vec{m}, \vec{\Gamma}, \vec{w}) = \sum_j \frac{w_j}{\m_j^2 - s
    - i \m_j \Gamma_j}
  \labelequ{TwoPionsGamma.F}
\end{equation}
where $s$ is the centre-of-mass energy, $\vec{m}$ is the vector of
resonance masses, $\vec{\Gamma}$ the vector of widths, and $\vec{w}$
the vector of real weights. These vectors are given in
\tab{Tau:TwoPionsGamma.Resonances}, where the channel proceeds through
initial $\rho$ resonances followed by an $\omega$ resonance.

\begin{table}\centering
  \captionabove{Vectors of masses, widths, and real weights for the
    resonances used in the form factor of \equ{Tau:TwoPionsGamma.F}
    for the $\gamma \pi^0 \pi^-$
    channel.\labeltab{TwoPionsGamma.Resonances}}
  \begin{tabular}{L|L|L|L}
    \toprule
    \multicolumn{1}{c|}{resonances}
    & \multicolumn{1}{c|}{$\vec{\m}~[\gev]$}
    & \multicolumn{1}{c|}{$\vec{\Gamma}~[\gev]$}
    & \multicolumn{1}{c}{$\vec{w}$} \\
    \midrule
    \rho 
    & \begin{pmatrix}  0.773, &  1.7 \end{pmatrix}
    & \begin{pmatrix}  0.145, &  0.26 \end{pmatrix}
    & \begin{pmatrix}  1    , & -\frac{1}{10} \end{pmatrix}
    \\
    \omega
    & \begin{pmatrix}  0.782  \end{pmatrix}
    & \begin{pmatrix}  0.0085 \end{pmatrix}
    & \begin{pmatrix}  1      \end{pmatrix}
    \\
   \bottomrule
  \end{tabular}
\end{table}

\begin{subfigures}[t]{2}{Distributions of $\m_{23}$ for the default $\gamma
    \pi^0 \pi^-$ decay channel using the model
    of \equ{Tau:TwoPionsGamma}.\labelfig{TwoPionsGamma}}
  \svgbeg
  \svg[1]{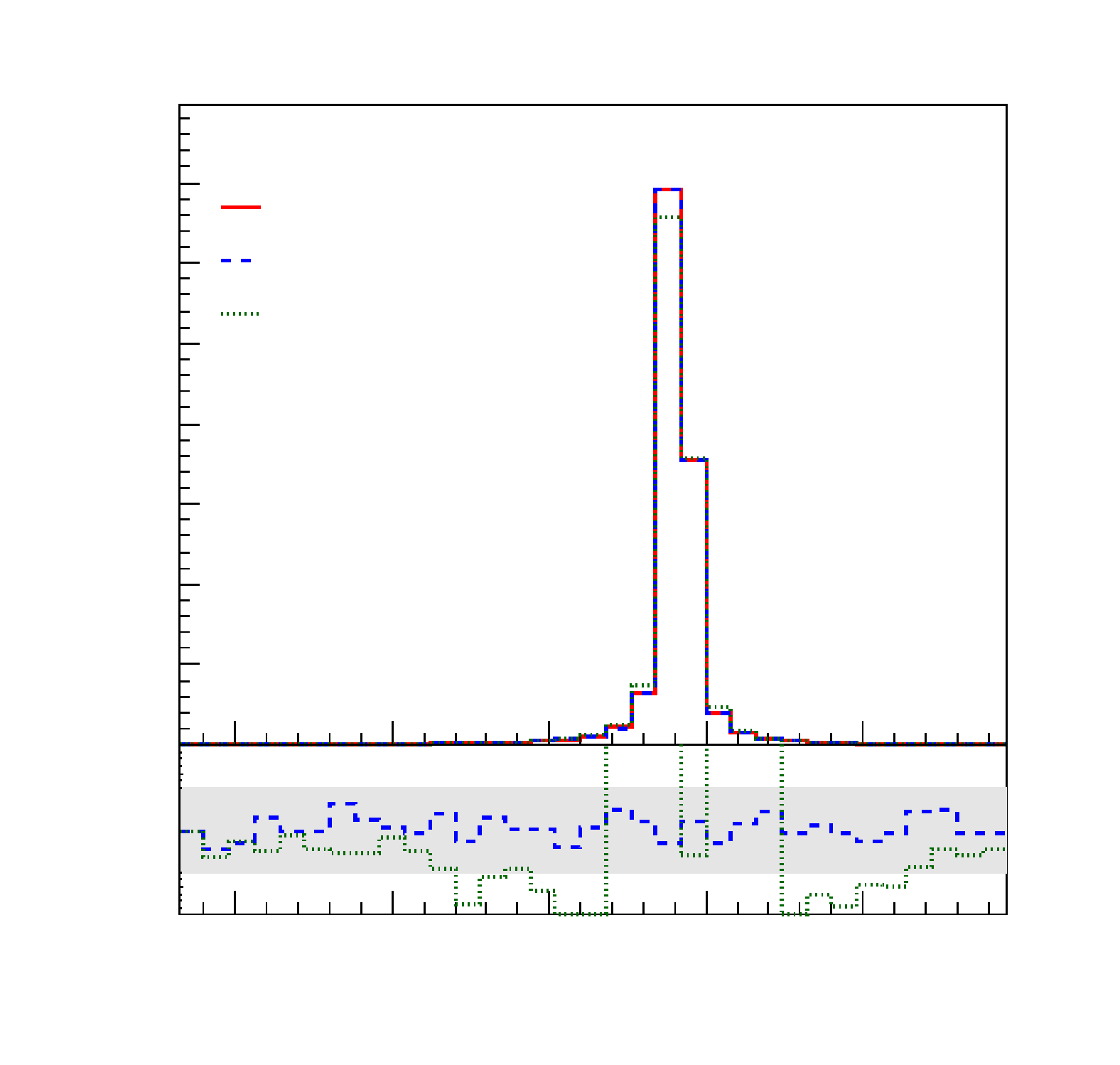} & \sidecaption \svgend
\end{subfigures}

In \fig{Tau:TwoPionsGamma} the $\m_{23}$ distributions of the photon
and neutral pion are given for \pythia{8}, \herwig{++}, and
\tauola. All three distributions match well, although the \tauola
distribution deviates slightly from the \pythia{8} and \herwig{++}
distributions due to a difference in the parameters used in
implementing the channel.

\newsubsection{Five-Body Decays}{DecFive}

Five-body decays of the \wtl into four pions and a \wtl neutrino have
been observed experimentally, but due to the low branching ratio and
complexity of the decay, experimental observations do not strongly
constrain the model \cite{weinstein.01.1}. Consequently, the hadronic
current for the $\pi^0 \pi^0 \pi^0 \pi^-$ and $\pi^0 \pi^- \pi^-
\pi^+$ decays in \pythia{8} is based on the Novosibirsk model of
\rfrs{bondar.02.1} and~\cite{golonka.03.1}, a phenomenological model
fitted to four pion production from electron-positron annihilation
within the energy range of $1.0$ to $2.5~\gev$. Within this model the
hadronic decay of the \wtl occurs through an initial excited $\rho$
resonance and secondary $a_1$, $\rho$, and $\sigma$ resonances. For
the $\pi^0 \pi^- \pi^- \pi^+$ decay an additional secondary $\omega$
resonance is also included. The four pion hadronic currents are the
summation of the resonance subcurrents,
\begin{align*}\labelali{FourPions}
  J^\mu_{\pi^0\pi^0\pi^0\pi^-} & \propto J^\mu_{0,a_1 \rightarrow
    \rho
    \pi} + J^\mu_{0,a_1 \rightarrow \sigma \pi} \\
  J^\mu_{\pi^0\pi^-\pi^-\pi^+} & \propto J^\mu_{-,a_1 \rightarrow
    \rho \pi} + J^\mu_{-,a_1 \rightarrow \sigma \pi} +
  J^\mu_{-,\omega
    \rightarrow \rho \pi} 
\end{align*}
where $J_{0,X}$ are the subcurrents for the $\pi^0 \pi^0 \pi^0 \pi^-$
decay and $J_{-,X}$ are the subcurrents for the $\pi^0 \pi^- \pi^-
\pi^+$ decay.

The hadronic subcurrent for the decay of the $a_1$ to a $\rho \pi$ pair
for the three neutral pion current is given by the combinatorics of
the $\pi^0 \pi^0 \pi^0 \pi^-$ final state,
\begin{align*}\labelali{FourPions.RhoNeutralCurrent}
  J^\mu_{0,a_1 \rightarrow \rho \pi} =~ & G_1(s) \bigg(
  t_1^\mu(q_3,q_4,q_5,q_2) + t_1^\mu(q_3,q_2,q_5,q_4) +
  t_1^\mu(q_4,q_3,q_5,q_2) \\
  & + t_1^\mu(q_4,q_2,q_5,q_3) +
  t_1^\mu(q_2,q_3,q_5,q_4) +
  t_1^\mu(q_2,q_4,q_5,q_3) \bigg) 
\end{align*}
where $t_1$ is a four-vector based on the $a_1$ and $\rho$ resonance
propagators, and $G_1$ is a phenomenological fit of the four pion
invariant mass distribution given later in \tabs{Tau:FourPions.GFit}
through \ref{tab:Tau:FourPions.GFitParameters}. The Mandelstam
variable $s$ as used within this context is given by, $q_\mu q^\mu$
where,
\begin{equation}
  q^\mu = (q_2 + q_3 + q_4 + q_5)^\mu
  \labelequ{FourPions.Q}
\end{equation}
is the four-momentum of the four pion system.

The hadronic subcurrent for the decay of the $a_1$ into a $\sigma \pi$
pair for the $\pi^0 \pi^0 \pi^0 \pi^-$ pion final state is given by,
\begin{align*}\labelali{FourPions.SigmaNeutralCurrent}
  J^\mu_{0,a_1 \rightarrow \sigma \pi} =~ & G_1(s) \bigg(
  t_2^\mu(q_3,q_5,q_4,q_2) + t_2^\mu(q_4,q_5,q_3,q_2) +
  t_2^\mu(q_2,q_5,q_4,q_3) \\
  &- t_2^\mu(q_5,q_3,q_4,q_2) -
  t_2^\mu(q_5,q_4,q_3,q_2) -
  t_2^\mu(q_5,q_2,q_4,q_3) \bigg) 
\end{align*}
where $t_2$ is a four-vector based on the $a_1$ and $\sigma$ resonance
propagators.

For the $\pi^0 \pi^- \pi^- \pi^+$ decay,
the hadronic subcurrent for the decay of the $a_1$ into $\rho\pi$ is
given by,
\begin{align*}\labelali{FourPions.RhoChargedCurrent}
  J^\mu_{-,a_1 \rightarrow \rho \pi} =~ & G_2(s) \bigg(
  t_1^\mu(q_3,q_5,q_4,q_2) + t_1^\mu(q_4,q_5,q_3,q_2) +
  t_1^\mu(q_3,q_4,q_5,q_2) \\
  &+ t_1^\mu(q_4,q_3,q_5,q_2) +
  t_1^\mu(q_2,q_4,q_3,q_5) +
  t_1^\mu(q_2,q_3,q_4,q_5) \bigg) 
\end{align*}
where $G_2$ is another phenomenological fit of the four pion invariant
mass also given in \tabs{Tau:FourPions.GFit} through
\ref{tab:Tau:FourPions.GFitParameters}. The $a_1$ to $\sigma \pi$
subcurrent is similar,
\begin{align*}\labelali{FourPions.SigmaChargedCurrent}
  J^\mu_{-,a_1 \rightarrow \sigma \pi} =~
  & G_2(s) \bigg( t_2^\mu(q_2,q_4,q_3,q_5) + t_2^\mu(q_2,q_3,q_4,q_5) \\
  &- t_2^\mu(q_3,q_2,q_4,q_5) -
  t_2^\mu(q_4,q_2,q_3,q_5) \bigg) 
\end{align*}
but the $\sigma$ propagator is accounted for by $t_2$. Finally, the
additional $\omega$ to $\rho \pi$ subcurrent for the $\pi^0 \pi^-
\pi^- \pi^+$ channel is given by,
\begin{align*}\labelali{FourPions.OmegaChargedCurrent}
  J^\mu_{-,\omega \rightarrow \rho \pi} =~ & G_3(s) \bigg(
  t_3^\mu(q_3,q_5,q_4,q_2) + t_3^\mu(q_4,q_5,q_3,q_2) -
  t_3^\mu(q_3,q_4,q_5,q_2) \\
  &- t_3^\mu(q_4,q_3,q_5,q_2) -
  t_3^\mu(q_3,q_2,q_4,q_5) -
  t_3^\mu(q_4,q_2,q_3,q_5) \bigg) 
\end{align*}
where $t_3$ is a four-vector based on the $\omega$ and $\rho$
propagators, and $G_3$ is a phenomenological fit of the four pion
invariant mass given in \tabs{Tau:FourPions.GFit} through
\ref{tab:Tau:FourPions.GFitParameters}.

The the four-vector $t_1$ is given by,
\begin{align*}\labelali{FourPions.T1}
  t_1^\mu(q_i,q_j,q_k,q_l) =~ & - F_{a_1}(s_{a_1}) \frac{\m_{a_1}^2
    (\m_\rho^2
    + \m_\rho \Gamma_\rho dm(0))}{D_{a_1}(s_{a_1}) D_{\rho}(s_{\rho})} \\
  & \bigg( (q_\nu q_{a_1}^\nu) \big(({q_k}_\nu q_{a_1}^\nu)
  q_l^\mu -
  ({q_l}_\nu q_{a_1}^\nu) q_k^\mu \big) \\
  & + \big( (q_\nu q_l^\nu) ({q_i}_\nu q_k^\nu) - (q_\nu
  q_k^\nu)({q_i}_\nu q_l^\nu) \big)
  q_{a_1}^\mu \bigg)
\end{align*}
where $D_{a_1}$ and $D_{\rho}$ are the denominators for the $a_1$ and
$\rho$ propagators respectively, and are given later in
\equs{Tau:FourPions.A1D} and \ref{equ:Tau:FourPions.RhoD}. These
propagators differ from those used by Bondar {\it et al.} in
\rfr{bondar.02.1} in the numerator where the $a_1$ and corrected
$\rho$ masses have been added such that the forms of the propagators
are that of Breit-Wigners, as is done in \herwig{++}. The $\rho$ mass
is corrected by a running mass correction, $dm(s)$, which is defined
later in \equ{Tau:FourPions.RhoDm}. The Mandelstam variables and
propagator four-momenta are,
\begin{align*}\labelali{FourPions.Q}
  q_{a_1}^\mu &= (q_j + q_k + q_l)^\mu & q_\rho^\mu &= (q_k +
  q_l)^\mu & q_\sigma^\mu &= (q_k + q_l)^\mu &
  q_\omega^\mu &= (q_j + q_k + q_l)^\mu \\
  s_{a_1} &= {q_{a_1}}_\mu q_{a_1}^\mu & s_\rho &= {q_\rho}_\mu
  q_\rho^\mu & s_\sigma &= {q_\sigma}_\mu q_\sigma^\mu &
  s_\omega &= {q_\omega}_\mu q_\omega^\mu
\end{align*}
for the $a_1$, $\rho$, $\sigma$ and $\omega$. The form factor for the
$a_1$, $F_{a_1}$, is given by,
\begin{equation}
  F_{a_1}(s) = \left(\frac{\Lambda^2 + \m_{a_1}^2}{\Lambda^2 + s}\right)^2
\end{equation}
where the cutoff value $\Lambda$ is taken as $1.2~\gev$ from
\rfr{bondar.99.1}.

\begin{table}\centering
  \captionabove{Parameters used by the four pion current for
    the $a_1$, $\rho$, $\sigma$, and $\omega$ resonances.
    \labeltab{FourPions.Parameters}}
  \begin{tabular}{L|L|L|L|L}
    \toprule
    \multicolumn{1}{c|}{resonance} 
    & \multicolumn{1}{c|}{$\m~[\gev]$} 
    & \multicolumn{1}{c|}{$\Gamma~[\gev]$} 
    & \multicolumn{1}{c|}{$\phi$} 
    & \multicolumn{1}{c}{$A$} \\
    \midrule
    a_1(1260)   & 1.23   & 0.45    &         &         \\
    \rho(770)   & 0.7761 & 0.1445  &         &         \\
    \sigma      & 0.8    & 0.88    & 0.43585 & 1.39987 \\
    \omega(782) & 0.782  & 0.00841 & 0       & 1       \\
    \bottomrule
  \end{tabular}
\end{table}

The four-vector $t_2$ is similar to $t_1$ but with different
combinatorics and propagators,
\begin{align*}\labelali{FourPions.T2}
  t_2^\mu(q_i,q_j,q_k,q_l) =~ & w_\sigma F_{a_1}(s_{a_1})
  \frac{\m_{a_1}^2 \m_\sigma^2}{D_{a_1}(s_{a_1})
    D_{\sigma}(s_{\sigma})} \\ &\bigg( (q^\nu {q_{a_1}}_\nu) s_{a_1}
  q_j^\mu - (q^\nu {q_j}_\nu)
  s_{a_1} q_{a_1}^\mu \bigg)
\end{align*}
where $w_\sigma$ is a complex weight for the $\sigma$ resonance
calculated from an amplitude and phase given in
\tab{Tau:FourPions.Parameters}, and $D_\sigma$ is the denominator of
the propagator for the $\sigma$ given later in
\equ{Tau:FourPions.SigmaD}. Again, the form of $t_2$ used in
\pythia{8} differs from Bondar {\it et al.} in the propagator with the
addition of $\m_{a_1}^2$ and $\m_\sigma^2$ in the numerator.

The final four-vector, $t_3$, is used only in the $\pi^0 \pi^- \pi^-
\pi^+$ decay, where the second resonance can occur through either an
$\omega$ or $a_1$,
\begin{align*}\labelali{FourPions.T3}
  t_3^\mu(q_i,q_j,q_k,q_l) =~ & w_\omega F_\omega(s_\omega)
  \frac{\m_{\omega}^2 (\m_\rho^2 + \m_\rho \Gamma_\rho
    dm(0))}{D_\omega(s_{\omega})
    D_{\rho}(s_{\rho})} \\
  & \bigg( \big( (q_\nu q_k^\nu)({q_i}_\nu q_l^\nu) -
  (q_\nu q_l^\nu)({q_i}_\nu q_k^\nu) \big) q_j^\mu \\
  & + \big( (q_\nu q_l^\nu)({q_i}_\nu q_j^\nu) -
  (q_\nu q_j^\nu)({q_i}_\nu q_l^\nu) \big) q_k^\mu \\
  & + \big( (q_\nu q_j^\nu)({q_i}_\nu q_k^\nu) - (q_\nu
  q_k^\nu)({q_i}_\nu q_j^\nu) \big) q_l^\mu
  \bigg)
\end{align*}
where $w_\omega$ is a complex weight calculated from a phase and
amplitude given in \tab{Tau:FourPions.Parameters} using
\equ{Tau:ComplexWeights}, $F_\omega$ is the $\omega$ form factor, and
$D_\omega$ is the denominator of the propagator for the $\omega$. Note
that the $\omega$ mass and corrected $\rho$ mass have been added to
the numerator of the propagator. Currently, the $\omega$ form factor
is taken as $F_\omega = 1$.

The energy dependent denominator of the propagator for the $a_1$ is
given by,
\begin{equation}
  D_{a_1}(s) = s - \m_{a_1}^2 + i \sqrt{s} \Gamma_{a_1}(s)
  \labelequ{FourPions.A1D}
\end{equation}
where the running width for the $a_1$, $\Gamma_{a_1}$, is calculated
from integrating over phase-space for the $a_1 \rightarrow \pi^0 \pi^-
\pi^+$ and $a_1 \rightarrow \pi^0 \pi^0 \pi^0$ decays taking into
account the combinatorics of the $\rho$ and $\sigma$ propagators. In
\tauola this phase-space integration is performed using Monte Carlo
integration and an interpolation table is built at initialisation. In
\pythia{8}, a fit of the \tauola interpolation table, given in
\tab{Tau:FourPions.A1WidthFit} with parameters in
\tab{Tau:FourPions.A1WidthFitParameters} and plotted in
\fig{Tau:FourPions.A1WidthFit}, is used for $\Gamma_{a_1}$.

\begin{table}\centering
  \captionabove{Functions used to fit the running $a_1$ width of
    \equ{Tau:FourPions.A1D} where $s_{3\pi^-} = 0.16960~\gev^2$ and
    $s_{\rho\pi^0} = 0.83425~\gev^2$ for the four pion
    current.\labeltab{FourPions.A1WidthFit}}
  \begin{tabular}{L|L}
    \toprule
    \multicolumn{1}{c|}{limits~$\left[\gev^2\right]$} 
    & \multicolumn{1}{c}{$\Gamma_{a_1}(s)~[\gev]$} \\
    \midrule
    0 \leq s < s_{3\pi^-} & 0 \\
    s_{3\pi^-} \leq s < s_{\rho\pi^0}
    & P_0 (s - s_{3\pi^-})^3 \left( 1 - P_1 (s - s_{3\pi^-}) + P_2 (s -
      s_{3\pi^-})^2 \right) \\
    s_{\rho\pi^0} \leq s & P_0 + P_1 s + P_2 s^2 + P_3 s^3 + P_4
    \frac{s + P_5}{s} \\
    \bottomrule
  \end{tabular}
\end{table}

\begin{table}\centering
  \captionabove{Parameters used in the $a_1$ running width fits of
    \tab{Tau:FourPions.A1WidthFit} for the four pion
    current.\labeltab{FourPions.A1WidthFitParameters}}
  \begin{tabular}{L|LLL}
    \toprule
    \multicolumn{1}{c|}{limits~$\left[\gev^2\right]$}  
    & \multicolumn{3}{c}{parameters} \\
    \midrule
    s_{3\pi^-} \leq s < s_{\rho\pi^0}
    & P_0 = 0.003052   & P_1 = -151.088  & P_2 = 174.495  \\ \midrule
    \multirow{2}{*}{$s_{\rho\pi^0} \leq s$} 
    & P_0 = 2.60817    & P_1 = -2.47790 & P_2 = 0.66539  \\
    & P_3 = -0.0678183 & P_4 = 1.66577  & P_5 = -1.23701 \\
    \bottomrule
  \end{tabular}
\end{table}

\begin{subfigures}{2}{Fit of the $a_1$ running width, using
    \tabs{Tau:FourPions.A1WidthFit} and
    \ref{tab:Tau:FourPions.A1WidthFitParameters}, compared to \tauola.
    This width is used in the four pion
    current.\labelfig{FourPions.A1WidthFit}}
  \svgbeg
  \svg[1]{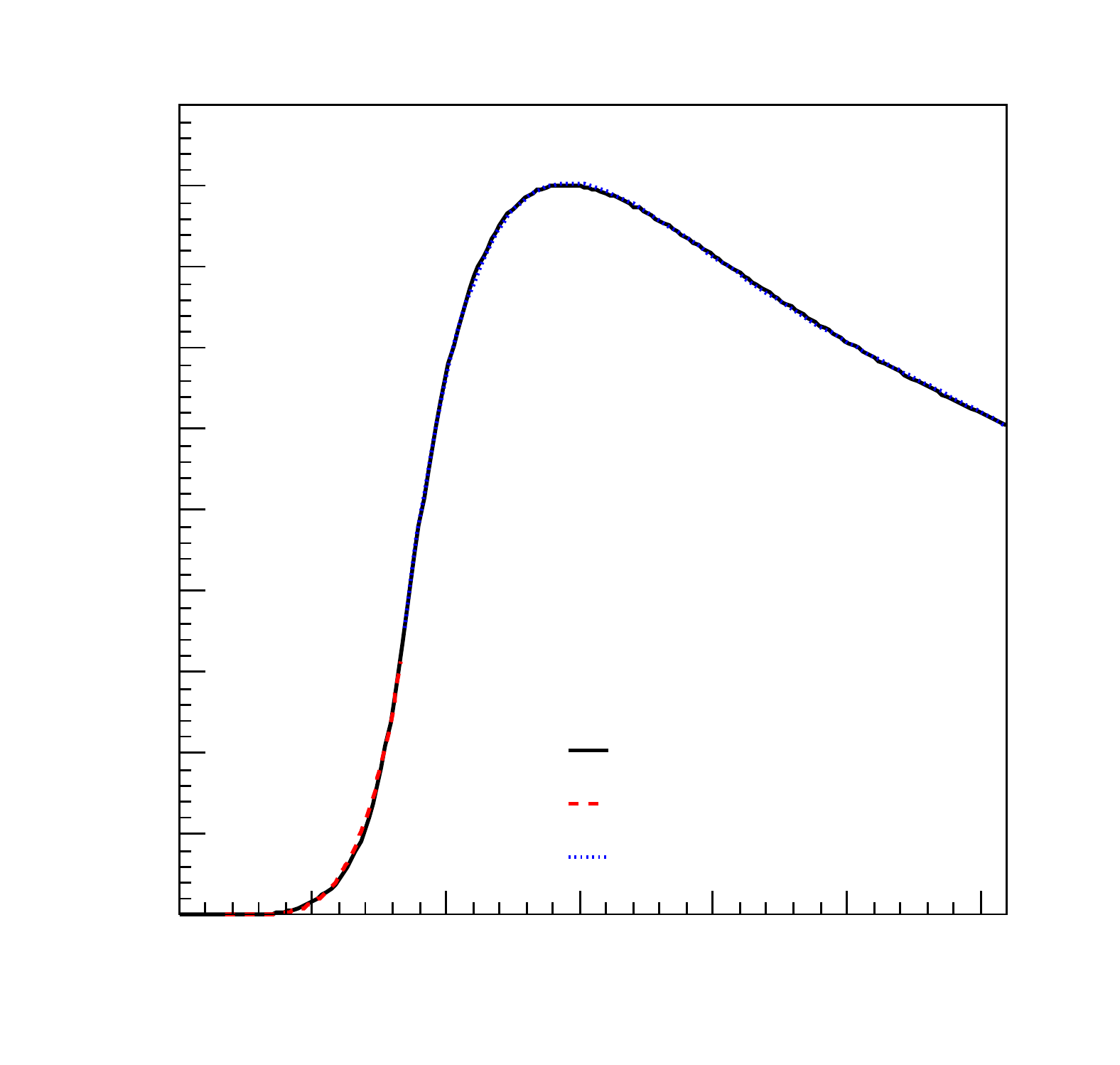} & \sidecaption \svgend
\end{subfigures}

The denominator for the $\rho$ resonance propagator includes a running
mass correction, $dm(s)$, similar to that of Gounaris and Sakurai from
\rfr{gounaris.68.1} which is also applied to $\m_\rho$ in the
numerators of the propagators from \equs{Tau:FourPions.T1} to
\ref{equ:Tau:FourPions.T3}. Consequently, the denominator to the
$\rho$ propagator is,
\begin{equation}
  D_\rho(s) = s - \m_\rho^2 - \m_\rho \Gamma_\rho dm(s) + i \m_\rho
  \Gamma_\rho \left( \frac{\m_\rho^2(s - 4 m_{\pi^-}^2)^3} {s(\m_\rho^2 - 4
      m_{\pi^-}^2)^3} \right)^{1/2}
  \labelequ{FourPions.RhoD}
\end{equation}
where the running mass correction is given as,
\begin{equation}
  dm(s) = \frac{\m_\rho \big( h(s) - h(\m_\rho^2) - (s - \m_\rho^2)
    dh(\m_\rho^2) \big) }{(\m_\rho^2 - 4m_{\pi^-}^2)^{3/2}}
  \labelequ{FourPions.RhoDm}
\end{equation}
and $h(s)$ is defined by Golonka {\it et al.} in \rfr{golonka.03.1} as,
\begin{equation}
  h(s) = 
  \begin{cases}
    \frac{x}{\pi} \left( \frac{\ln(1 + x)}{\ln(1 - x)} \right) (s - 4
    m_{\pi^-}^2) & \mbox{if $s > 4 m_{\pi^-}^2$} \\
    - \frac{8 m_{\pi^-}^2}{\pi} & \mbox{if $s = 0$} \\
    0 & \mbox{else} \\
  \end{cases}
\end{equation}
where $x = \sqrt{1 - 4 m_{\pi^-}^2 / s}$. The derivative of $h(s)$
with respect to $s$ is given by,
\begin{equation}
  dh(s) =
  \begin{cases}
    \frac{x}{\pi s} (s x + (2m_{\pi^-}^2 + s) \ln \left( \frac{1 +
        x}{1 - x} \right) & \mbox{if $s > 4 m_{\pi^-}^2$} \\
    0 & \mbox{else} \\
  \end{cases}
\end{equation}
which is used in the calculation of $dm(s)$ in
\equ{Tau:FourPions.RhoDm}.

The denominator of the $\sigma$ propagator, $D_\sigma(s)$, is given by
the denominator of the $s$-wave Breit-Wigner of \equ{Tau:BwS},
\begin{equation}
  D_\sigma(s) = s - \m_\sigma^2 - \left( \frac{i \Gamma_\sigma
      \m_\sigma^2}{\sqrt{s}} \right) \left( \frac{g(m_{\pi^-},
      m_{\pi^-}, s)} {g(m_{\pi^-}, m_{\pi^-}, \m_\sigma^2)}
  \right)
  \labelequ{FourPions.SigmaD}
\end{equation}
where $g(m_{\pi^-},m_{\pi^-},s)$ is given by \equ{Tau:BwG}. For the
$\omega$ resonance,
\begin{equation}
  D_\omega(s) = s - \m_\omega^2 + i \m_\omega \Gamma_\omega(s)
  \labelequ{FourPions.OmegaD}
\end{equation}
where the running width of the $\omega$ is taken from Golonka {\it et
al.} of \rfr{golonka.03.1} and is given by the fit of
\tab{Tau:FourPions.OmegaWidthFit} with the parameters of
\tab{Tau:FourPions.OmegaWidthFitParameters}.

\begin{table}\centering
  \captionabove{Running width of the $\omega$ from
    \equ{Tau:FourPions.OmegaD} where $x = \sqrt{s} - \m_\omega$ 
    for the four pion
    current.\labeltab{FourPions.OmegaWidthFit}}
  \begin{tabular}{L|L}
    \toprule
    \multicolumn{1}{c|}{limits$~[\gev^2]$} 
    & \multicolumn{1}{c}{$\Gamma_\omega(s)~[\gev]$} \\
    \midrule
    \sqrt{s} < 1  & P_0 + P_1
    x + P_2 x^2 + P_3 x^3 + P_4 x^4 + P_5 x^5 + P_6 x^6 \\
    \sqrt{s} < 1,~ \Gamma_\omega(s) \leq 0  & 0 \\
    s \geq 1 & P_0 + P_1 s^{1/2} + P_2 s + P_3 s^{3/2} \\
    \bottomrule
  \end{tabular} \\ ~ \\ ~ \\
  \captionabove{Parameters used in the $\omega$ running width fits of
    \tab{Tau:FourPions.OmegaWidthFit} for the four pion
    current.\labeltab{FourPions.OmegaWidthFitParameters}}
  \begin{tabular}{L|LLLL}
    \toprule
    \multicolumn{1}{c|}{limits$~[\gev^2]$} 
    & \multicolumn{3}{c}{parameters} \\
    \midrule
    \multirow{2}{*}{$\sqrt{s} < 1$} & P_0 = 1
    & P_1 = 17.56   & P_2 = 141.11  & P_3 = 894.884 \\
    & P_4 = 4977.35 & P_5 = 7610.66 & P_6 = -42524.4 \\ \midrule
    s \geq 1 & P_0 = -1333.26 & P_1 = 4860.19 
    & P_2 = -6000.81 & P_3 = 2504.97 \\
    \bottomrule
  \end{tabular}
\end{table}

The phenomenological $G$ factors of the hadronic currents used in
\equs{Tau:FourPions.RhoNeutralCurrent} through
\ref{equ:Tau:FourPions.OmegaChargedCurrent} are fitted piece-wise with
the functions of \tab{Tau:FourPions.GFit}, where the limits
$s_i$ are given in \tab{Tau:FourPions.GFitLimits} and the fit
parameters $P_i$ are given in
\tab{Tau:FourPions.GFitParameters}. The fits of the three $G$
factors are plotted in \fig{Tau:FourPions.GFit}. Note that the $G$
factors given in \rfr{bondar.02.1} must be corrected using the
functions described in \rfr{golonka.03.1}.

\begin{table}\centering
  \captionabove{Fitting functions used for the phenomenological $G$
    factors where the centre-of-mass limits $s_i$ are given in
    \tab{Tau:FourPions.GFitLimits} and the parameters
    are given in \tab{Tau:FourPions.GFitParameters} for the four pion
    current.\labeltab{FourPions.GFit}}
  \begin{tabular}{L|L}
    \toprule
    \multicolumn{1}{c|}{limits~$\left[\gev^2\right]$} 
    & \multicolumn{1}{c}{$G(s)$} \\
    \midrule
    s < s_0 & 0 \\
    s_0 \leq s < s_1 & P_0 + P_1 s \\
    s_1 \leq s < s_2 & P_0 s^{P_1} + P_2 s^2 + P_3 s^3 + P_4 s^4 \\
    s_2 \leq s < s_3 & P_0 + P_1 s + P_2 s^2 + P_3 s^3 + P_4 s^4 \\
    s_3 \leq s < s_4 & P_0 + P_1 s \\
    s_4 \leq s < s_5 & P_0 + P_1 s \\
    s_5 \leq s & 0 \\
    \bottomrule
  \end{tabular}
\end{table}

\begin{table}[p]\centering
  \captionabove{Limits used in the $G$ factors of
    \tab{Tau:FourPions.GFit} for the four pion
    current.\labeltab{FourPions.GFitLimits}}
  \begin{tabular}{L|LLL}
    \toprule
    \multicolumn{1}{c|}{$G(s)$} 
    & \multicolumn{3}{c}{limits~$\left[\gev^2\right]$} \\
    \midrule
    \multirow{2}{*}{$G_1(s)$} 
    & s_0 = 0.614403 & s_1 = 0.656264 & s_2 = 1.57896 \\
    & s_3 = 3.08198 & s_4 = 3.12825 & s_5 = 3.17488 \\ \midrule
    \multirow{2}{*}{$G_2(s)$} 
    & s_0 = 0.614403 & s_1 = 0.635161 & s_2 = 2.30794 \\
    & s_3 = 3.08198 & s_4 = 3.12825 & s_5 = 3.17488 \\ \midrule
    \multirow{2}{*}{$G_3(s)$} 
    & s_0 = 0.81364 & s_1 = 0.861709 & s_2 = 1.92621 \\
    & s_3 = 3.08198 & s_4 = 3.12825 & s_5 = 3.17488 \\
    \bottomrule
  \end{tabular} \\ ~ \\ ~ \\
  \captionabove{Parameters used in the $G$ factors of
    \tab{Tau:FourPions.GFit} for the four pion
    current.\labeltab{FourPions.GFitParameters}}
  \begin{tabular}{L|L|LLLL}
    \toprule
    \multicolumn{1}{c|}{$G(s)$} 
    & \multicolumn{1}{c|}{limits~$\left[\gev^2\right]$} 
    & \multicolumn{3}{c}{parameters} \\
    \midrule
    \multirow{7}{*}{$G_1(s)$}
    & s_0 \leq s < s_1 & P_0 = -23383.7 & P_1 = 38059.2 & \\ \cmidrule{2-5}
    & \multirow{2}{*}{$s_1 \leq s < s_2$} 
    & P_0 = 230.368 & P_1 = -4.39368 & P_2 = 687.002 \\
    & & & P_3 = -732.581 & P_4 = 207.087 & \\ \cmidrule{2-5}
    & \multirow{2}{*}{$s_2 \leq s < s_3$}
    & P_0 = 1633.92 & P_1 = -2596.21 & P_2 = 1703.08 \\
    & & & P_3 = -501.407 & P_4 = 54.5919 & \\ \cmidrule{2-5}
    & s_3 \leq s < s_4 & P_0 = -2982.44 & P_1 = 986.009 & \\ \cmidrule{2-5}
    & s_4 \leq s < s_5 & P_0 = 6948.99 & P_1 = -2188.74 & \\ \midrule
    \multirow{7}{*}{$G_2(s)$}
    & s_0 \leq s < s_1 & P_0 = -54171.5 & P_1 = 88169.3 & \\ \cmidrule{2-5}
    & \multirow{2}{*}{$s_1 \leq s < s_2$}
    & P_0 = 454.638 & P_1 = -3.07152 & P_2 = -48.7086 \\
    & & & P_3 = 81.9702 & P_4 = -24.0564 & \\ \cmidrule{2-5}
    & \multirow{2}{*}{$s_2 \leq s < s_3$}
    & P_0 = -162.421 & P_1 = 308.977 & P_2 = -27.7887 \\
    & & & P_3 = -48.5957 & P_4 = 10.6168 & \\ \cmidrule{2-5}
    & s_3 \leq s < s_4 & P_0 = -2650.29 & P_1 = 879.776 & \\ \cmidrule{2-5}
    & s_4 \leq s < s_5 & P_0 = 6936.99 & P_1 = -2184.97 & \\ \midrule
    \multirow{7}{*}{$G_3(s)$}
    & s_0 \leq s < s_1 & P_0 = -84888.9 & P_1 = 104332 & \\ \cmidrule{2-5}
    & \multirow{2}{*}{$s_1 \leq s < s_2$}
    & P_0 = 2698.15 & P_1 = -3.08302 & P_2 = 1936.11 \\
    & & & P_3 = -1254.59 & P_4 = 201.291 & \\ \cmidrule{2-5}
    & \multirow{2}{*}{$s_2 \leq s < s_3$}
    & P_0 = 7171.67 & P_1 = -6387.94 & P_2 = 3056.29 \\
    & & & P_3 = -888.635 & P_4 = 108.632 & \\ \cmidrule{2-5}
    & s_3 \leq s < s_4 & P_0 = -5607.47 & P_1 = 1917.27 & \\ \cmidrule{2-5}
    & s_4 \leq s < s_5 & P_0 = 26573 & P_1 = -8369.76 & \\
    \bottomrule
  \end{tabular}
\end{table}

\begin{subfigures}{2}{Fit of the \subfig{FourPions.G1Fit}~$G_1$,
    \subfig{FourPions.G2Fit}~$G_2$, and \subfig{FourPions.G2Fit}~$G_3$
    factors, using \tabs{Tau:FourPions.GFit},
    \ref{tab:Tau:FourPions.GFitLimits}, and
    \ref{tab:Tau:FourPions.GFitParameters}, compared to \tauola.
    These factors are used in the four pion current in
    \equs{Tau:FourPions.RhoNeutralCurrent} through
    \ref{equ:Tau:FourPions.OmegaChargedCurrent}.\labelfig{FourPions.GFit}}
  \svgbeg
  \svg{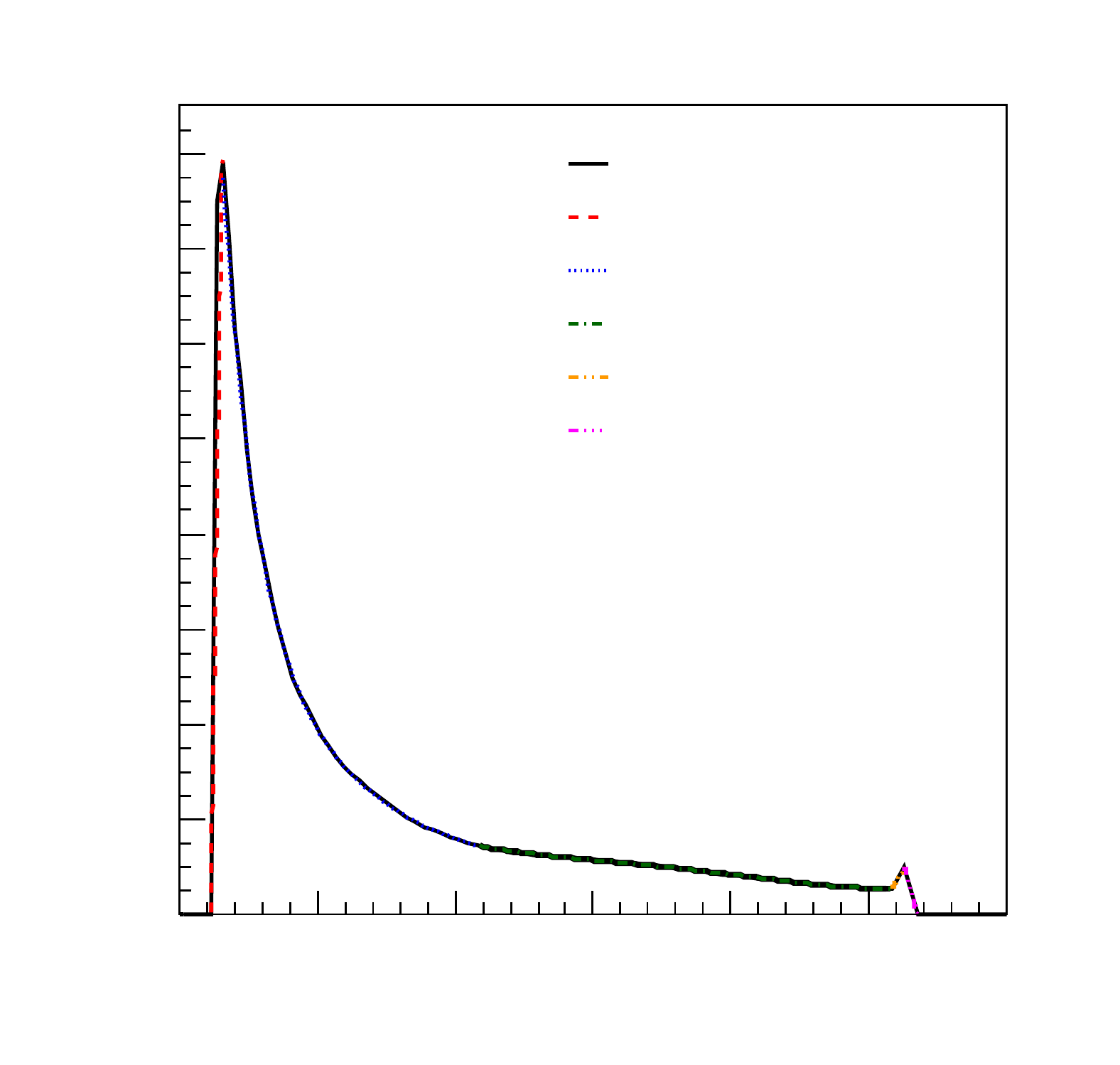} & \svg{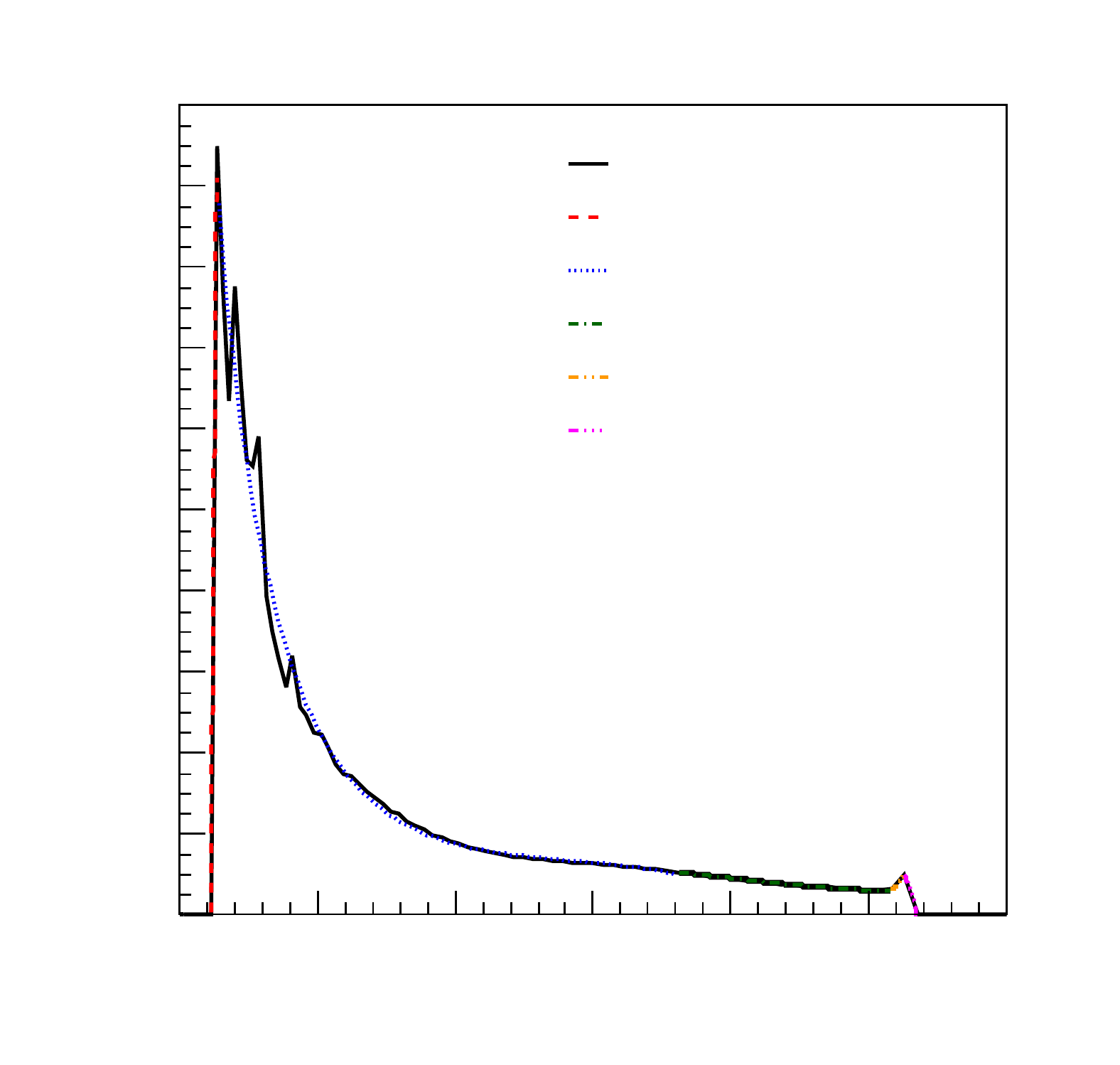} \svgsep
  \svg{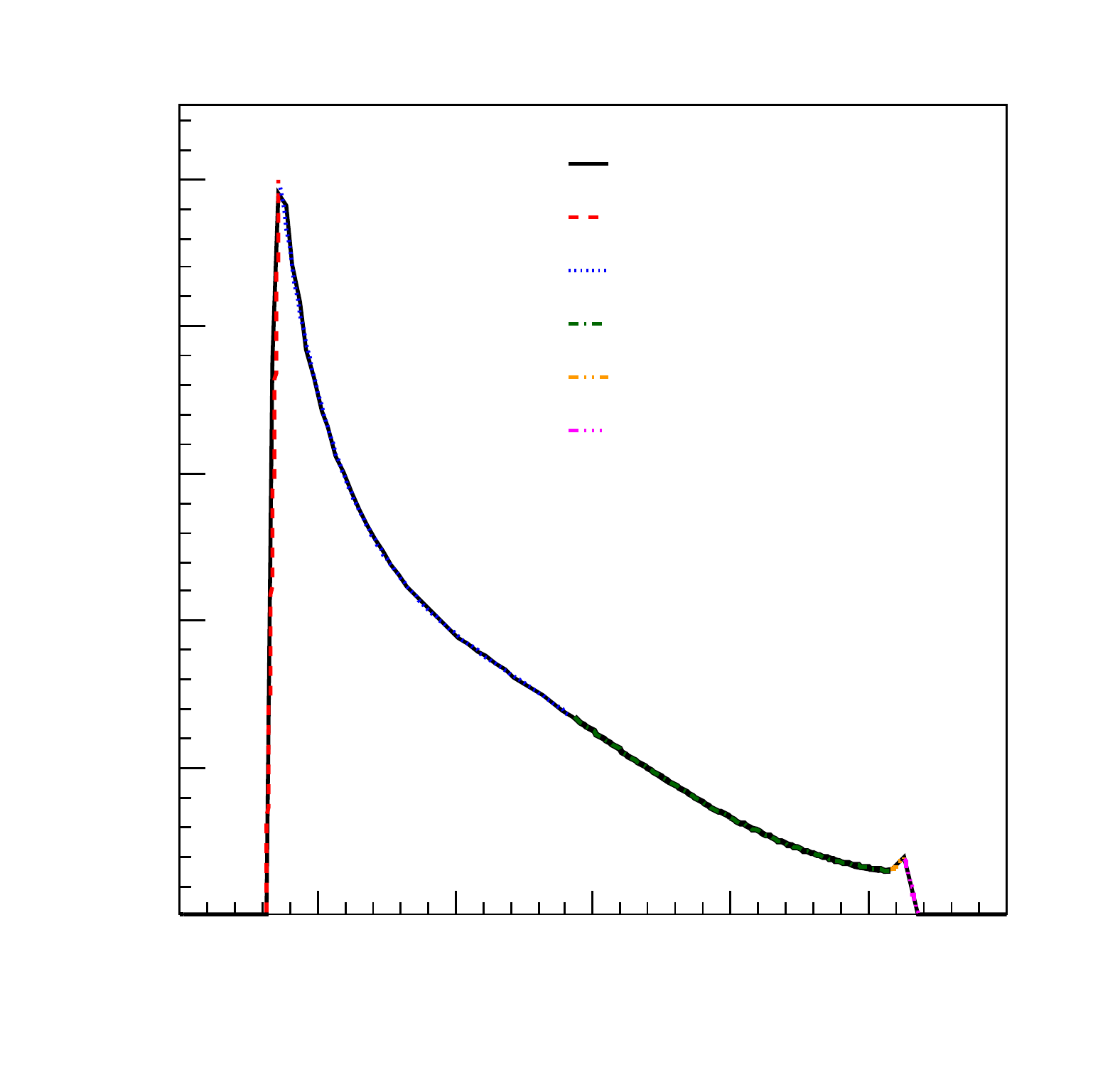} & \sidecaption \svgend
\end{subfigures}

\begin{subfigures}{2}{Invariant mass distributions of $m_{345}$ for
    the \subfig{q_qp.W.16_111_111_111_211.m_111_111_211}~$\pi^0 \pi^0
    \pi^0 \pi^-$ and
    \subfig{q_qp.W.16_111_211_211_211.m_111_211n_211p}~$\pi^0 \pi^-
    \pi^- \pi^+$ decay channels of the \wtl using the four pion
    model.\labelfig{FourPions}}
  \svgbeg
  \svg{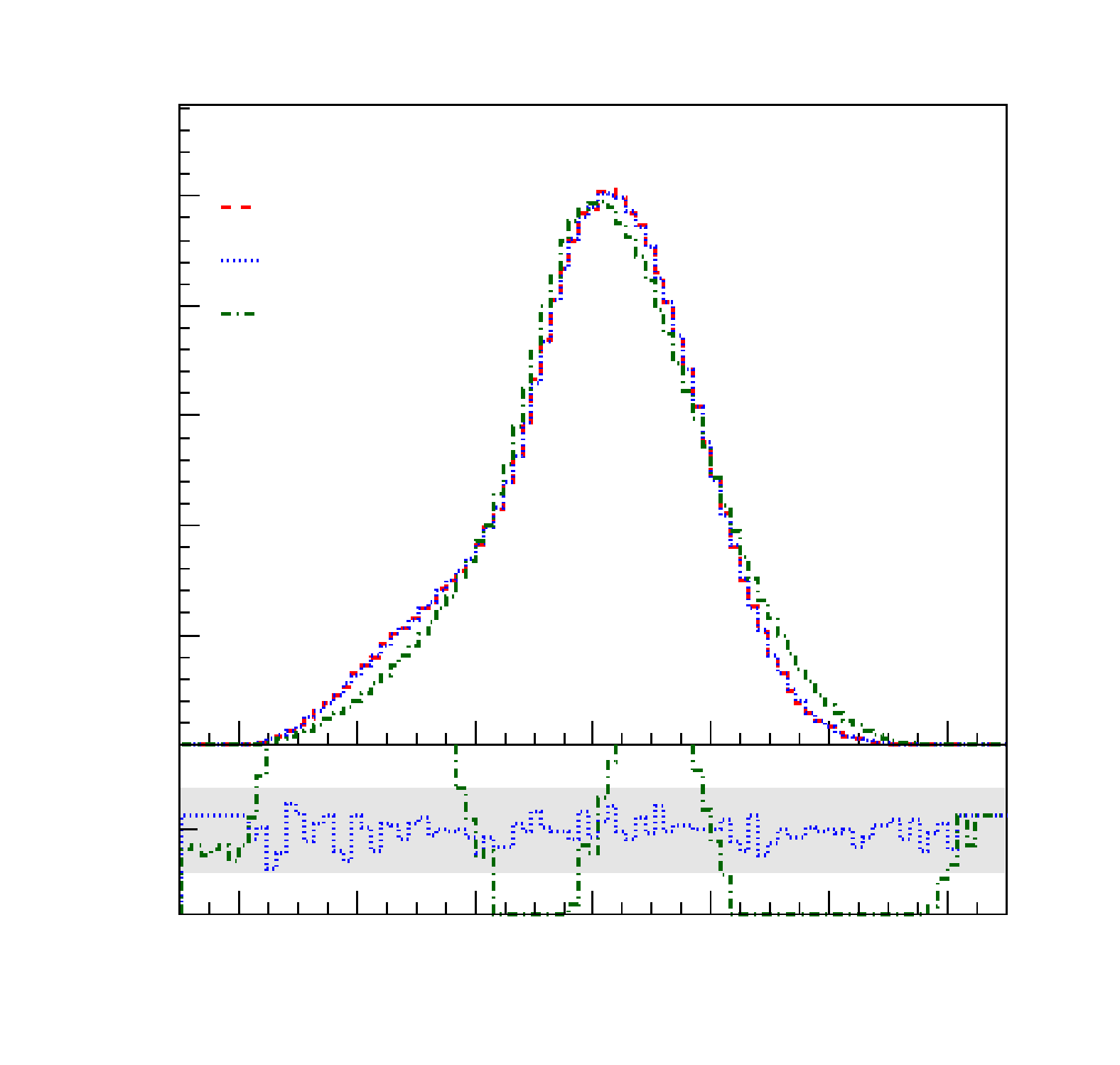} &
  \svg{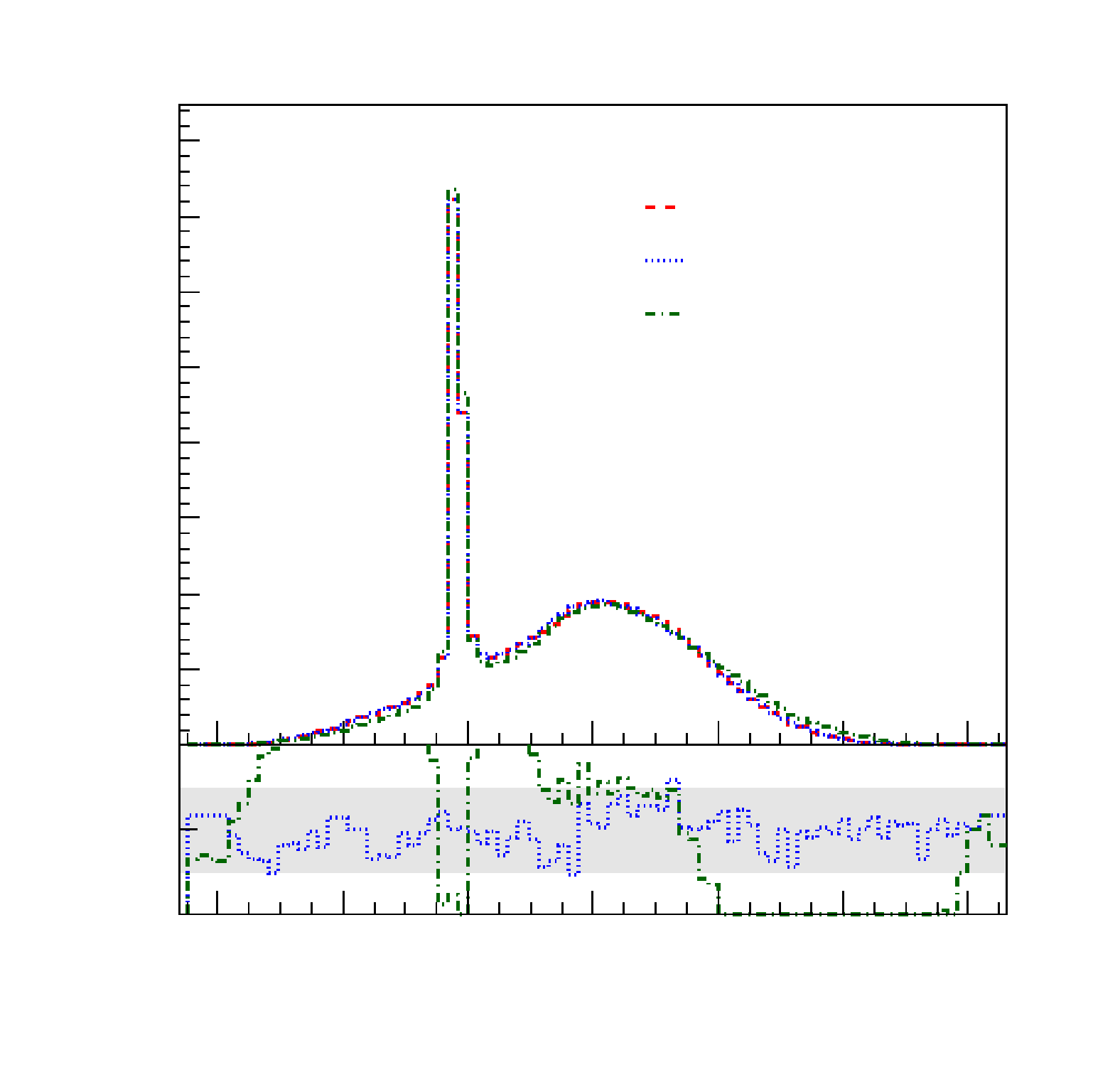} \svgend
\end{subfigures}

\Fig{Tau:FourPions} compares the invariant mass distributions of
$m_{345}$ for the four pion decays produced from \pythia{8},
\herwig{++}, and \tauola. The \pythia{8} and \herwig{++} distributions
match well for both channels with four pions, with the distinct peak at
$0.8~\gev$ in \fig{Tau:q_qp.W.16_111_211_211_211.m_111_211n_211p} due
to the additional $\omega$ resonance of the $\pi^- \pi^- \pi^-
\pi^+$ channel. The \tauola distributions differ slightly from the
\pythia{8} and \herwig{++} due to differences in the implementations
and the parameters used in the hadronic current.

\newsubsection{Six-Body Decays}{DecSix}

Three six-body decays are implemented in \pythia{8} with the hadronic
final states $\pi^0 \pi^0 \pi^- \pi^- \pi^+$, $\pi^0 \pi^0 \pi^0 \pi^0
\pi^-$, and $\pi^- \pi^- \pi^- \pi^+ \pi^+$. These decays are modelled
with the five pion current by K\"uhn and W\c{a}s of
\rfr{kuhn.08.1}. The decays can occur through $a_1$, $\rho$, $\omega$,
and $\sigma$ resonances. The hadronic currents for the three decay
channels are given by, 
\begin{alignat*}{3}\labelali{FivePions}
  & J_{\pi^0 \pi^0 \pi^- \pi^- \pi^+} && \propto ~&&
  J_a^\mu(q_6, q_4, q_2, q_5, q_3) + J_a^\mu(q_6, q_5, q_2, q_4, q_3) \\
  &&&&&+ J_a^\mu(q_6, q_4, q_3, q_5, q_2) + J_a^\mu(q_6, q_5, q_3, q_4, q_2) \\
  &&&&&+ J_b^\mu(q_4, q_5, q_6, q_2, q_3) + J_b^\mu(q_2, q_3, q_4, q_6, q_5) \\
  &&&&&+ J_b^\mu(q_2, q_3, q_5, q_6, q_4) \\
  & J_{\pi^0 \pi^0 \pi^0 \pi^0 \pi^-}^\mu && \propto ~&&
  J_b^\mu(q_2, q_3, q_6, q_4, q_5) + J_b^\mu(q_5, q_3, q_6, q_4, q_2) \\
  &&&&&+ J_b^\mu(q_3, q_4, q_6, q_2, q_5) + J_b^\mu(q_2, q_4, q_6, q_3, q_5) \\
  &&&&&+ J_b^\mu(q_2, q_5, q_6, q_4, q_3) + J_b^\mu(q_4, q_5, q_6, q_2, q_3) \\
  & J_{\pi^- \pi^- \pi^- \pi^+ \pi^+}^\mu && \propto ~&&
  J_b^\mu(q_2, q_3, q_5, q_6, q_4) + J_b^\mu(q_4, q_3, q_5, q_6, q_2) \\
  &&&&&+ J_b^\mu(q_2, q_4, q_5, q_6, q_3) + J_b^\mu(q_2, q_3, q_6, q_5, q_4) \\
  &&&&&+ J_b^\mu(q_4, q_3, q_6, q_5, q_2) + J_b^\mu(q_2, q_4, q_6, q_5, q_3)
\end{alignat*}
where $J_a^\mu$ and $J_b^\mu$ are hadronic subcurrents corresponding
to the decay type. A schematic of the $a$-type decay is given in
\fig{Tau:FivePions.Ja} where the \wtl can decay through an $a_1$
resonance. This resonance then decays into secondary $\rho$ and
$\omega$ resonances.  While the $\omega$ resonance of the $a$-type
decay could be modelled further with a tertiary $\rho$ resonance
decay, its decay is modelled as a contact interaction. A schematic of
the $b$-type decay is shown in \fig{Tau:FivePions.Jb}. Here, an $a_1$
resonance decays into secondary $a_1$ and $\sigma$ resonances. This is
followed with the secondary $a_1$ decaying into a tertiary $\rho$
resonance.

\begin{subfigures}{2}{Schematics of the resonance structure for
    \subfig{FivePions.Ja}~$a$-type and \subfig{FivePions.Jb}~$b$-type
    decays of the \wtl into a final state with five
    pions.\labelfig{FivePions.J}}
  \fmpbeg
  \fmp{FivePions.Ja} & \fmp{FivePions.Jb} \fmpend
\end{subfigures}

The hadronic subcurrent for the $a$-decay resonance structure is given by,
\begin{align*}\labelali{FivePions.Ja}
  J_a^\mu(q_1, q_2, q_3, q_4, q_5) = ~& w_\omega 
  BW(q_\nu q^\nu, \m_{a_1}, \Gamma_{a_1}) \\ &
  BW((q_1 + q_2 + q_3)_\nu(q_1 + q_2 + q_3)^\nu, \m_\omega, \Gamma_\omega) \\ &
  BW((q_4 + q_5)_\nu(q_4 + q_5)^\nu, \m_\rho, \Gamma_\rho) \\ &
  \epsilon^\mu\left(q_4 - q_5, \epsilon(q_1, q_2, q_3), q \right)
  \big(BW((q_2 + q_3)_\nu(q_2 + q_3)^\nu, \m_\rho, \Gamma_\rho) \\ &
  + BW((q_1 + q_3)_\nu(q_1 + q_3)^\nu, \m_\rho, \Gamma_\rho) \\ &
  + BW((q_1 + q_2)_\nu(q_1 + q_2)^\nu, \m_\rho, \Gamma_\rho)\big)
\end{align*}
where $w_\omega$ is a real weight, $BW$ is the Breit-Wigner defined
by,
\begin{equation}
  BW(s, \m, \Gamma) = \frac{\m^2}{\m^2 - s^2 - i\m\Gamma}
\end{equation}
and $\epsilon$ is the permutation operator also used in
\equ{Tau:ThreeMesons} for the three meson channel. Here, $q$ is the
sum of all five momenta, $q_1+q_2+q_3+q_4+q_5$. The parameters of the
resonances used in \pythia{8} are given in
\tab{Tau:FivePions.Parameters}.

\begin{subtables}[t]{2}{Parameters used by the five pion subcurrents for
    the $a_1$, $\rho$, $\sigma$, and $\omega$ resonances.
    \labeltab{FivePions.Parameters}}
  \setlength{\tabcolsep}{\oldtabcolsep}
  \begin{tabular}{L|L|L|L}
    \toprule
    \multicolumn{1}{c|}{resonance} 
    & \multicolumn{1}{c|}{$\m~[\gev]$} 
    & \multicolumn{1}{c|}{$\Gamma~[\gev]$} 
    & \multicolumn{1}{c}{$w$} \\
    \midrule
    a_1(1260)   & 1.26   & 0.4    &      \\
    \rho(770)   & 0.776  & 0.15   &      \\
    \sigma      & 0.8    & 0.6    & 11.5 \\
    \omega(782) & 0.782  & 0.0085 & \p1  \\
    \bottomrule
  \end{tabular} & \sidecaption \\
\end{subtables}

The hadronic subcurrent for the $b$-decay resonance structure is given by,
\begin{align*}\labelali{FivePions.Jb}
  J_b^\mu(q_1, q_2, q_3, q_4, q_5) = ~& w_\sigma
  BW(q_\nu q^\nu, \m_{a_1}, \Gamma_{a_1}) \\ &
  BW((q_1+q_2+q_3)_\nu (q_1+q_2+q_3)^\nu, \m_{a_1}, \Gamma_{a_1}) \\ &
  BW((q_4+q_5)_\nu (q_4+q_5)^\nu, \m_\sigma, \Gamma_\sigma) \\ & \left(
  \left(\frac{{J_c}_\nu(q_1,q_2,q_3) q^\nu}{q_\nu q^\nu}\right) q^\mu
  - J_c^\mu(q_1,q_2,q_3) \right)
\end{align*}
where the subcurrent $J_c^\mu$ is,
\begin{align*}\labelali{FivePions.Jc}
  J_c^\mu(q_1, q_2, q_3) = ~& 
  BW((q_1+q_3)_\nu (q_1+q_3)^\nu, \m_\rho, \Gamma_\rho) \\ &
  \left(\frac{{q_2}_\nu(q_1-q_3)^\nu}{(q_1+q_2+q_3)_\nu(q_1+q_2+q_3)^\nu}
    (q_1+q_2+q_2)^\mu - q_1^\mu + q_3^\mu \right) \\ &
  + BW((q_2+q_3)_\nu (q_2+q_3)^\nu, \m_\rho, \Gamma_\rho) \\ &
  \left(\frac{{q_1}_\nu(q_2-q_3)^\nu}{(q_1+q_2+q_3)_\nu(q_1+q_2+q_3)^\nu}
    (q_1+q_2+q_2)^\mu - q_2^\mu + q_3^\mu \right)
\end{align*}
and the parameters are again given in
\tab{Tau:FivePions.Parameters}. The definitions of $q$ and $BW$ are
the same as for $J_a^\mu$.

\begin{subfigures}{2}{Invariant mass distributions of $m_{234}$ for
    the \subfig{q_qp.W.16_111_111_211_211_211.m_111_111_211n}~$\pi^0 \pi^0
    \pi^- \pi^- \pi^+$,
    \subfig{q_qp.W.16_111_111_111_111_211.m_111_111_111}~$\pi^0 \pi^0
    \pi^0 \pi^0 \pi^+$, and
    \subfig{q_qp.W.16_211_211_211_211_211.m_211n_211n_211n}~$\pi^- \pi^-
    \pi^- \pi^+ \pi^+$ decay channels of the \wtl using the five pion
    model.\labelfig{FivePions}}
  \svgbeg
  \svg{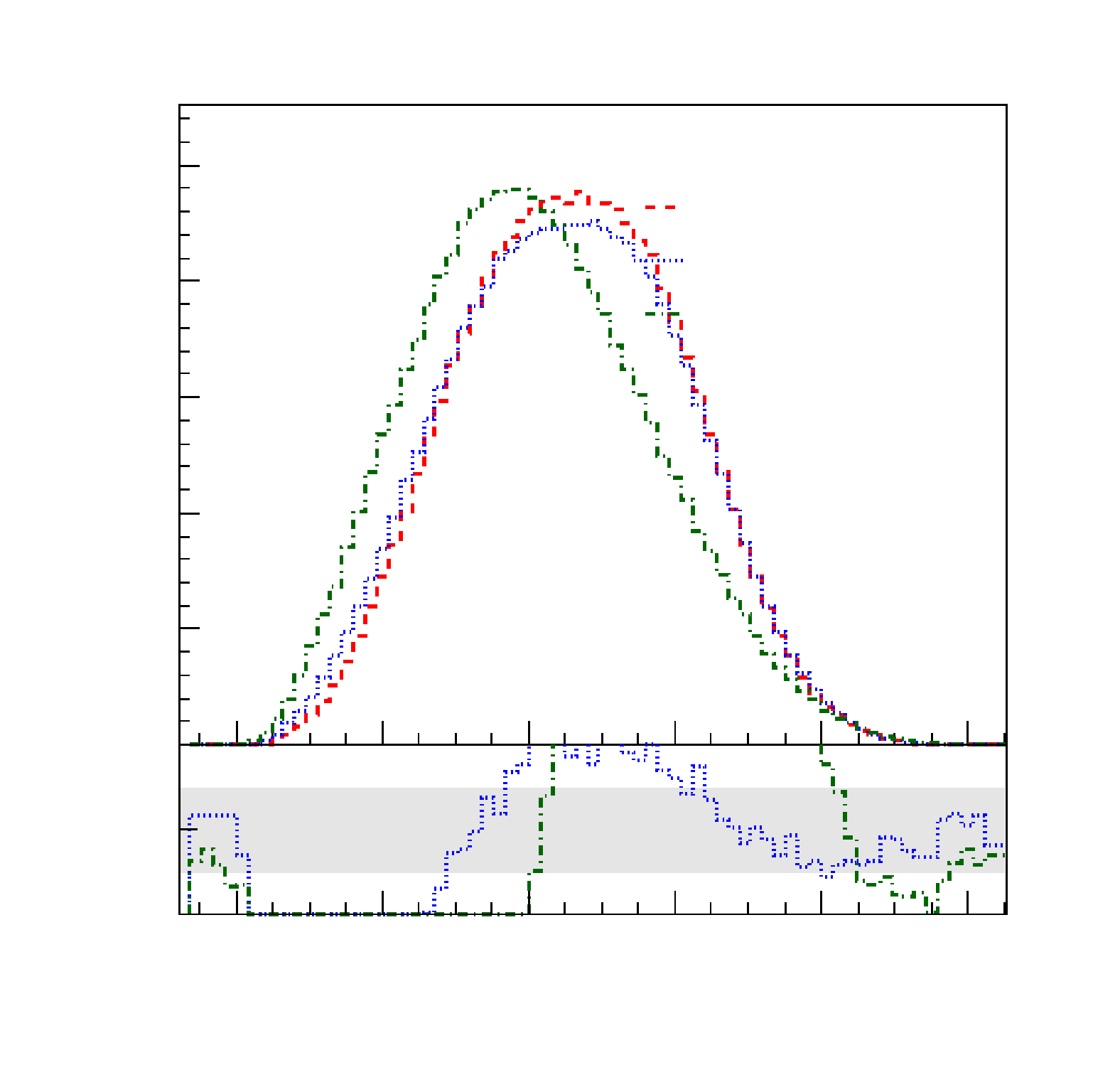} &
  \svg{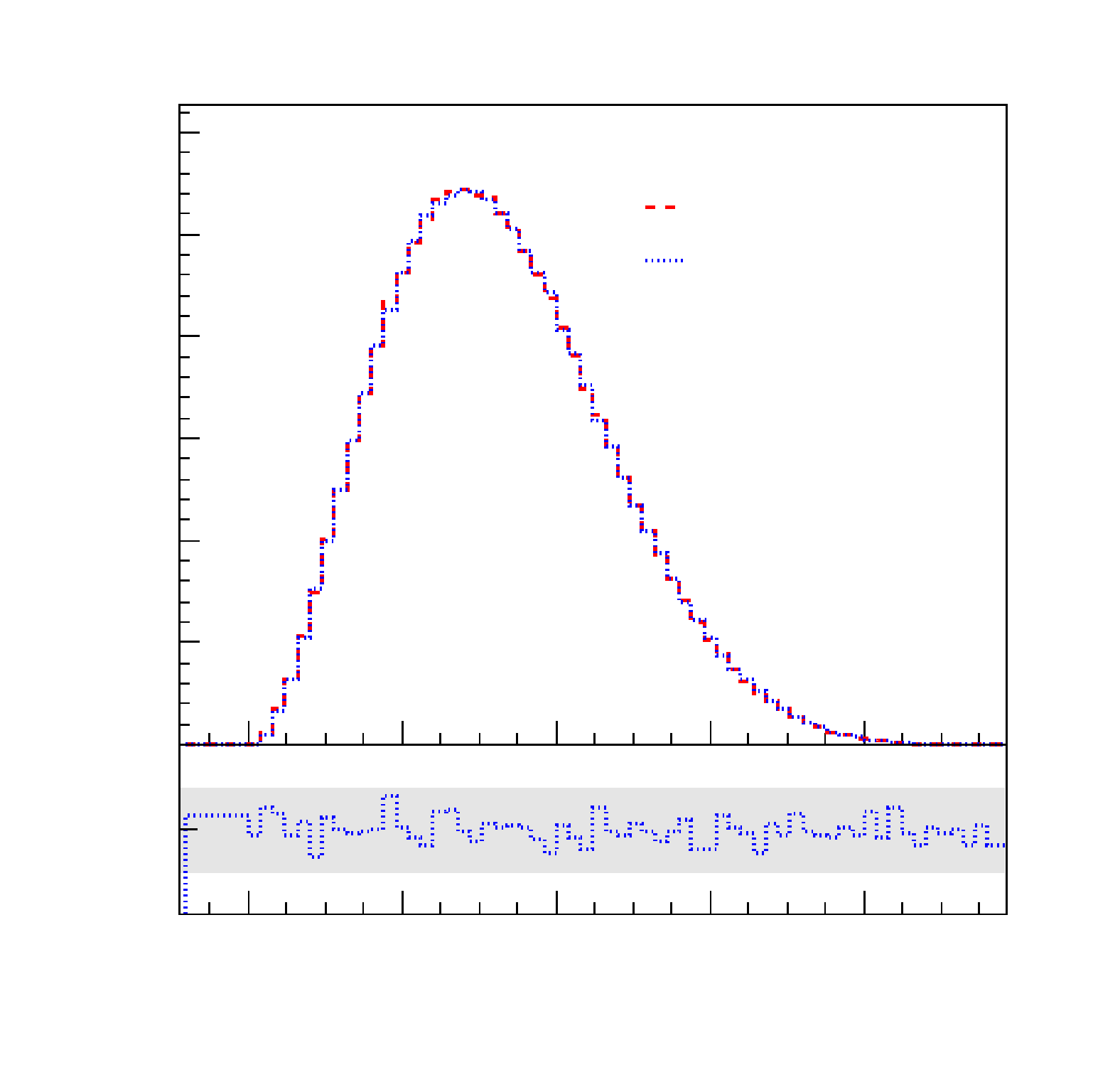} \svgsep
  \svg{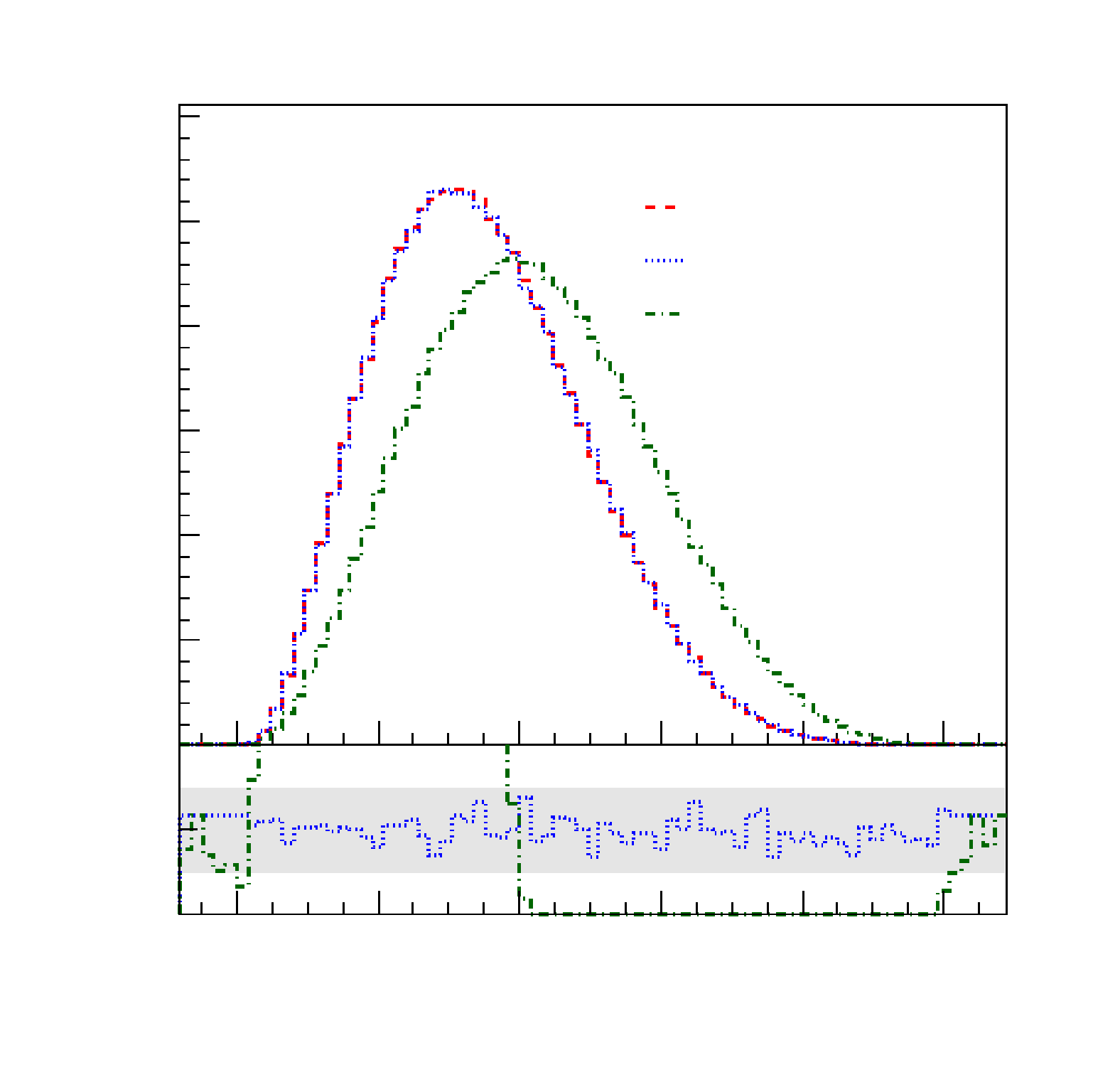} & \sidecaption \svgend
\end{subfigures}

In \fig{Tau:FivePions}, the $m_{234}$ invariant mass distributions are
given for \pythia{8}, \herwig{++}, and \tauola. There is good
agreement between \pythia{8} and \herwig{++} for the $\pi^0 \pi^0
\pi^0 \pi^0 \pi^-$ and $\pi^- \pi^- \pi^- \pi^+ \pi^+$ channels, with
a slight disagreement for the $\pi^0 \pi^0 \pi^- \pi^- \pi^+$
channel. However, the \tauola distributions do not agree well with the
\pythia{8} and \herwig{++} distributions.

\newsection{Implementation}{Imp}

One of the goals of the \wtl decay implementation in \pythia{8} is
that the production processes and decay channels of the \wtls are
easily extensible for new physics. Additionally, the machinery used to
generate helicity correlations for \wtl decays may also be used for
helicity correlation in other processes such as \wtq decays in the
future. This section outlines the \wtl decay machinery architecture
used in \pythia{8} as well as the additional software developed for
validation. The implementation of the \wtl decays with full helicity
correlations is summarised in \sec{Tau:ImpDec}, while the
implementation of the helicity matrix elements used in the \wtl decays
is introduced in \sec{Tau:ImpMe} and the validation procedure is
outlined in \sec{Tau:ImpVal}. A summary of the \wtl production
mechanisms presented in \sec{Tau:Pro} and implemented in \pythia{8}
is given in \tab{Tau:Production}, while a summary of the default \wtl
decays introduced in \sec{Tau:Dec} and implemented in \pythia{8} is
given in \tab{Tau:Decays}.

\begin{table}\centering
  \captionabove{Production mechanisms implemented in \pythia{8} for
    which full spin correlations of \wtl decays are
    calculated.\labeltab{Production}}
  \begin{tabular}{c|l}
    \toprule
    type & 
    \multicolumn{1}{c}{processes} \\
    \midrule
    \multirow{3}{*}{electroweak}
    & $f\bar{f} \to \gamma \to f\bar{f}$, $f\bar{f} \to \z \to f\bar{f}$, \\
    & $f\bar{f} \to \dy \to f\bar{f}$, $f_i\bar{f}_j \to \w \to f_k\bar{f}_l$,\\
    & $\z \to f\bar{f}$, $\w \to f_i\bar{f}_j$ \\
    \midrule
    \multirow{2}{*}{\whb}
    & $\hH \to f\bar{f}$, $\hhz \to f\bar{f}$, $\hHz \to f\bar{f}$ \\
    & $\hAz \to f\bar{f}$, $\hHpm \to f_i\bar{f}_j$ \\
    \midrule
    other
    & $B/D \to f_i\bar{f}_j + X$ \\
    \bottomrule
  \end{tabular}
\end{table}

\begin{table}[p]\centering
  \captionabove{Summary of default \wtl decay models and channels in \pythia{8}
    sorted by multiplicity. For each model the reference, internal \pythia{8}
    matrix element mode identifier {\tt meMode}, default decay
    channels using the model, and branching fractions of the channels
    are given. The implicit \wtl neutrinos are omitted. Additional
    channels are available in \pythia{8} but are decayed using
    isotropic phase-space.\labeltab{Decays}}
  \begin{tabular}{C|ll|C|L|R}
    \toprule
    \multicolumn{1}{c|}{mult.}
    & \multicolumn{2}{c|}{model}
    & \multicolumn{1}{c|}{\tt meMode}
    & \multicolumn{1}{c|}{products}
    & \multicolumn{1}{c}{$\mathcal{B}~[\%]$}
    \\
    \midrule
    \multirow{2}{*}{2}
    & \multirow{2}{*}{single hadron}
    &
    & \multirow{2}{*}{1521}
    & \pi^- 
    & 10.76825
    \\ 
    &&&
    & K^-
    & 0.69601
    \\ \midrule
    \multirow{8}{*}{3}
    & \multirow{2}{*}{leptonic}
    &
    & \multirow{2}{*}{1531}
    & e^- \bar{\nu}_e 
    & 17.72832
    \\
    &&& 
    & \mu^- \bar{\nu}_\mu
    & 17.31072
    \\ \cmidrule{2-6}
    & \multirow{3}{*}{two mesons via vector}
    & \multirow{3}{*}{\cite{kuhn.90.1}}
    & \multirow{3}{*}{1532}
    & \pi^0 \pi^-
    & 25.37447
    \\
    &&&
    & K^0 K^-
    & 0.15809
    \\
    &&&
    & \eta K^-
    & 0.01511
    \\ \cmidrule{2-6}
    & \multirow{2}{*}{two mesons via vector and scalar}
    & \multirow{2}{*}{\cite{finkemeier.96.1}}
    & \multirow{2}{*}{1533}
    & \pi^- \bar{K}^0
    & 0.83521
    \\
    &&&
    & \pi^0 K^-
    & 0.42655
    \\ \midrule
    \multirow{13}{*}{4}
    & \multirow{2}{*}{CLEO three pions}
    & \multirow{2}{*}{\cite{cleo.99.2}}
    & \multirow{2}{*}{1541}
    & \pi^0  \pi^0 \pi^-
    & 9.24697
    \\
    &&&
    & \pi^- \pi^- \pi^+
    & 9.25691
    \\ \cmidrule{2-6}
    & \multirow{8}{*}{three mesons with kaons}
    & \multirow{8}{*}{\cite{finkemeier.95.1}}
    & \multirow{8}{*}{1542}
    & \pi^- \bar{K}^0 \pi^0
    & 0.39772
    \\
    &&&
    & K^- \pi^- \pi^+
    & 0.34701
    \\
    &&&
    & K^0 \pi^- \bar{K}^0
    & 0.14318
    \\
    &&&
    & K^- \pi^0 K^0
    & 0.15809
    \\
    &&&
    & K_S^0 \pi^- K_L^0
    & 0.11932
    \\
    &&&
    & \pi^0 \pi^0 K^-
    & 0.06463
    \\
    &&& 
    & K_S^0 \pi^- K_S^0
    & 0.02386
    \\
    &&&
    & K_L^0 \pi^- K_L^0
    & 0.02386
    \\ \cmidrule{2-6}
    & general three mesons
    & \cite{decker.93.1}
    & 1543
    & \pi^- \pi^0 \eta
    & 0.13821
    \\ \cmidrule{2-6}
    & two pions with photon
    & \cite{jadach.93.1}
    & 1544
    & \gamma \pi^0 \pi^-
    & 0.17520
    \\ \midrule
    \multirow{2}{*}{5}
    & \multirow{2}{*}{four pions}
    & \multirow{2}{*}{\cite{golonka.03.1}}
    & \multirow{2}{*}{1551}
    & \pi^0 \pi^- \pi^- \pi^+
    & 4.59365
    \\
    &&&
    & \pi^0 \pi^0 \pi^0 \pi^-
    & 1.04401
    \\ \midrule
    \multirow{3}{*}{6}
    & \multirow{3}{*}{five pions}
    & \multirow{3}{*}{\cite{kuhn.08.1}}
    & \multirow{3}{*}{1561}
    & \pi^0 \pi^0 \pi^- \pi^- \pi^+
    & 0.49069
    \\
    &&&
    & \pi^0 \pi^0 \pi^0 \pi^0 \pi^-
    & 0.09515
    \\
    &&&
    & \pi^- \pi^- \pi^- \pi^+ \pi^+
    & 0.08342
    \\
    \bottomrule
  \end{tabular}
\end{table}

\newsubsection{Tau Decays}{ImpDec}

The code for \wtl decays is provided in the four files (headers and
source), {\tt Particle\-Decays}, {\tt Tau\-Decays}, {\tt
  Helicity\-Basics}, and {\tt Helicity\-Matrix\-Elements} of the
\pythia{8} source code. When a particle decay is requested by
\pythia{8} the following program flow occurs, where the first step
occurs within {\tt Particle\-Decays::\-decay} and all remaining steps
occur within {\tt Tau\-Decays::\-decay}.
\begin{enumerate}
\item The particle is passed to {\tt ParticleDecays::\-decay}.
\item If the particle is a \wtl, the decay is passed to {\tt
    TauDecays::\-decay}.
\item The hard process is determined.
  \begin{enumerate}
  \item The correlated \wtl is found (if it exists).
  \item The incoming and outgoing particles are set.
  \item The helicity matrix element, \me, from
    {\tt Helicity\-Matrix\-Elements} is set.
  \end{enumerate}
\item The \wtl is selected (if correlated, randomly selected from the two
  \wtls).
\item The helicity density matrix, $\rho$, is calculated by {\tt
    Helicity\-Matrix\-Element::\-calc\-ulate\-Rho} using
  \equ{Thr:RhoHard}.\labelite{Imp.CalculateRho}
\item The \wtl children are created by {\tt
    Tau\-Decays::\-createChildren}.\labelite{Imp.createChildren}
  \begin{enumerate}
  \item The decay channel is selected.
  \item The \wtl children are created.
  \item The \wtl decay matrix element is set from {\tt
      Helicity\-Matrix\-Elements}.
  \end{enumerate}
\item The children momenta, $q_i$, are assigned isotropically until
  the condition $x \in \mathcal{U}(0,1) \geq
  \frac{\mathcal{W}}{\mathcal{W}_\mathrm{max}}$ is met.
  \begin{enumerate}
  \item The variable $x$ is a random number from a uniform distribution,
    $\mathcal{U}(0,1)$.\labelite{Imp.Momenta}
  \item The decay weight $\mathcal{W}$ is calculated by {\tt
      Helicity\-Matrix\-Elements::\-decay\-Weight} using
    \equ{Thr:WeightDecay}.
  \item The maximum weight $\mathcal{W}_\mathrm{max}$ is empirically or
    analytically known.
  \end{enumerate}
\item {\tt TauDecays::\-writeEvent} writes the decay to the event
  record.\labelite{Imp.WriteEvent}
  \begin{enumerate}
  \item If there is no correlated \wtl, {\tt TauDecays::\-decay} returns to
    {\tt Particle\-Decays\-::\-decay}.
  \end{enumerate}
\item If the \wtl is correlated, the decay matrix, $D$, of the decayed
  \wtl is
  calculated by {\tt Helicity\-Matrix\-Elements::\-calculateD} using
  \equ{Thr:DecayMatrix}.
\item The second correlated \wtl is selected, and
  steps~\ref{ite:Tau:Imp.CalculateRho} through
  \ref{ite:Tau:Imp.WriteEvent} are repeated for the second \wtl.
\item {\tt TauDecays::\-decay} returns to {\tt ParticleDecays::\-decay}
  which returns to the main algorithm of \pythia{8}.
\end{enumerate}

\begin{table}\centering
  \captionabove{A list of the methods implemented in the {\tt
      TauDecays} class.\labeltab{TauDecays.Methods}}
  \begin{tabular}{>{\tt}l<{}|p{0.7\columnwidth}}
    \toprule
    \multicolumn{1}{c|}{method} 
    & \multicolumn{1}{c}{description} \\
    \midrule
    decay & this is the main method called by {\tt ParticleDecays}
    and performs the correlated decays of the \wtl \\
    createChildren & selects the \wtl decay channel, assigns the \wtl
    helicity matrix element, and returns a
    vector of {\tt HelicityParticle} children with $q_i = 0$ \\
    isotropicDecay & takes the \wtl children and reassigns their
    momenta using isotropic phase-space \\
    writeEvent & writes the \wtl children to the \pythia{8} event
    record \\
    \bottomrule
  \end{tabular}
\end{table}

\Tab{Tau:TauDecays.Methods} outlines the methods
implemented within the {\tt TauDecays} class and provides a brief
description of each method. The process of decaying a \wtl is very
similar to the decay of a standard particle except for the use of more
sophisticated helicity matrix elements and helicity correlations. As
such, the {\tt TauDecays::\-createChildren} method is very similar to
inline code within the {\tt ParticleDecays} class and {\tt
  TauDecays::\-isotropicDecay} reimplements {\tt
  ParticleDecays::\-mGenerator}, the \m-generator algorithm
introduced in \sec{Thr:Dec}.

\newsubsection{Matrix Elements}{ImpMe}

Within the {\tt Helicity\-Basics} source files, three major classes
are defined: {\tt Wave4}, {\tt Gamma\-Matrix}, and {\tt
  Helicity\-Particle}. The class {\tt Wave4} is intended to store
four-momenta and spinors and is just a complex four-vector with
standard vector operations defined: vector addition/subtraction,
vector multiplication, and scalar
multiplication/division. Additionally, the operator {\tt Wave4::\-(i)}
is defined, where {\tt i} must be in the range $0 \leq {\tt i} \leq
3$, and allows access to the corresponding element of the four-vector,
{\it i.e.} $q(0)$ returns the energy of the momentum $q$.

The {\tt GammaMatrix} class is intended to be used in conjunction with
the {\tt Wave4} class such that helicity matrix elements can be easily
translated from analytic expressions to code. For example, the
helicity matrix element for $\tauto \pi^-$,
\begin{equation}
  \mathcal{M} = \bar{u}_1 \gamma_\mu (1 - \gamma_5) u_0 q_2^\mu 
\end{equation}
can be written in pseudo-code as,
\begin{align*}\labelali{MeExample}
  \mathcal{M} = \sum_\mu ~& {\tt Wave4(}\bar{u}_1{\tt ) *
    GammaMatrix(}\mu{\tt ) * (1 - GammaMatrix(5))} \\ & {\tt *
    Wave4(}u_0{\tt )} {\tt * GammaMatrix(4)(}\mu{\tt ,}\mu{\tt ) *
    Wave4(}q_2{\tt )(}\mu{\tt )}
\end{align*}
in the {\tt HelicityMatrixElements} source code. Again, the operator
{\tt GammaMatrix::(i,j)} can be used to access the Dirac matrix
element of row {\tt i} and column {\tt j} where $0 \leq {\tt i,j} \leq
3$. The constructor {\tt GammaMatrix(i)} returns $\gamma^{\tt i}$.

The Dirac matrices are defined by \equ{Thr:DiracMatrices}, using the
representation of \sec{Thr:Lag}, and $\gamma^4$ is defined as
$g_{\mu\nu}$. Because carrying out the full matrix multiplication of
\equ{Tau:MeExample} is very time consuming, the special sparse
properties of the Dirac matrices are exploited, as well as left to
right order of operations. Accordingly, the {\tt GammaMatrix} class is
represented by four ordered {\tt values} where the first value is the
non-zero element of the first row, the second value is the non-zero
element of the second row, the third is the non-zero element of the
third row, and the fourth is the non-zero element of the fourth
row. Corresponding to each of the {\tt values} is an {\tt index} which
provides the column index for the non-zero value. The left to right
multiplication $w = {\tt Wave4 * GammaMatrix}$ is then given by,
\begin{equation}
  w^\mu = {\tt Wave4(GammaMatrix.index[}\mu{\tt ]) *
    GammaMatrix.values[}\mu{\tt ]}
\end{equation}
which is four multiplication and four assignment operations, rather
than the sixteen multiplication, sixteen addition, and four assignment
operations required for full matrix multiplication.

In \equ{Tau:MeExample}, the term {\tt 1 - GammaMatrix(5)} is included,
where the nonsensical subtraction of a matrix from a scalar is
performed. This is because the addition and subtraction of scalars
with the {\tt GammaMatrix} class is defined as the addition or
subtraction of the scalar applied to the non-zero elements of the
Dirac matrix. Because $\gamma^5$ is on-diagonal in the representation
used, {\tt 1 - GammaMatrix(5)} is just the subtraction of $\gamma^5$
from the identity matrix.

The final class defined in {\tt Helicity\-Basics}, {\tt
  HelicityParticle}, takes the standard {\tt Particle} class of
\pythia{8} and extends the class to include a helicity density matrix,
$\rho$, and a decay matrix, $D$. Additionally, the method {\tt
  HelicityParticle::\-wave(h)} is defined which returns the {\tt
  Wave4} spinor or polarisation vector for the particle with helicity
{\tt h}.

\begin{table}\centering
  \captionabove{A list of the public methods implemented in the {\tt
        HelicityMatrixElement}
      class.\labeltab{MatrixElements.Methods}}

  \begin{tabular}{>{\tt}l<{}|p{0.7\columnwidth}}
    \toprule
    \multicolumn{1}{c|}{method} 
    & \multicolumn{1}{c}{description} \\
    \midrule
    initPointers & initialise the pointers to the \pythia{8} \sm and
    \mssm couplings database and particle properties database \\
    initChannel & takes as an argument a vector of {\tt
      HelicityParticle} which are used to initialise any constants
    used in the matrix element \\
    decayWeight & takes a vector of {\tt HelicityParticle} and
    calculates the decay weight $\mathcal{W}$ for the matrix element \\
    calculateME & calculates the helicity matrix element \\
    calculateRho & calculates the helicity density matrix \me for one of
    the {\tt HelicityParticles} being used in the matrix element \\
    calculateD & calculates the decay matrix $D$ for one of the {\tt
      HelicityParticles} being used in the matrix element \\
    setFermionLine & determines the order to assign a fermion line
    in the matrix element based on direction of
    particle or anti-particle \\
    xBreitWigner & the Breit-Wigners of \equs{Tau:Bw} through
    \ref{equ:Tau:BwD} where {\tt x} is fixed, {\tt s}, {\tt p}, or
    {\tt d} \\
    \bottomrule
  \end{tabular}
\end{table}

The actual helicity matrix elements for both the hard processes and
\wtl decays are defined in the {\tt HelicityMatrixElements} files
which use the classes outlined above from {\tt HelicityBasics}. The
important methods of the {\tt HelicityMatrixElement} class are
outlined in \tab{Tau:MatrixElements.Methods} with brief descriptions
provided for each method. The hard processes are written as classes
that derive directly from the {\tt HelicityMatrixElement} class and
utilise pointer polymorphism within C$++$.

The \wtl decay matrix elements are written as classes that derive from
the {\tt HMETauDecay} class which itself derives from the {\tt
  HelicityMatrixElement} class. The {\tt HMETauDecay} class is very
similar to the {\tt Helicity\-Matrix\-Element} class except that the
additional method {\tt HME\-Tau\-Decay\-::\-init\-Hadronic\-Current}
has been implemented which allows the \wtl decay matrix element to be
calculated using the general form of Equation \equ{Tau:MeTau}, so that
for a new \wtl decay only the hadronic current needs to be defined when
implementing the matrix element.

\newsubsection{Validation}{ImpVal}

Because the matrix elements defined in \secs{Tau:Pro} and
\ref{sec:Tau:Dec} can be complex, it is important to validate
\pythia{8} against the other event generators which provide similar
features for \wtl decays. In order to facilitate validation, a series
of tools were written in addition to the code implemented in
\pythia{8} and are available by request.

Within the validation package are three sets of tools: a generation
tool, an analysis tool, and a plotting tool. The generation tool
provides a common {\sc Root} $n$-tuple output for events generated
with \pythia{8}, \herwig{++}, and \pythia{6} with \tauola. The
analysis package runs over the generated $n$-tuples and applies a
common analysis to the events from the three generators. The
histograms from the analysis tool can then be plotted using the
plotting tool. The tools are written in C$++$ and interfaced with {\sc
  Bash} scripts. For event generation, the \pythia{8}, \herwig{++},
{\sc ThePEG}, \pythia{6}, and \tauola libraries are required as well
as {\sc Root} libraries. The analysis and plotting tools require only
{\sc Root} libraries.

A timing tool is provided within the validation package as a method to
compare the average time per event required by the various generators
to ensure that no significant timing issues arise. Runs of various
sizes, {\it i.e.} $2000$, $6000$, and $12000$ events, are performed
for each decay channel with each generator. The results are aggregated
by the timing tool and a linear fit is performed for each decay type
with each generator. The slope from the linear fit is taken as the
average time per event with associated uncertainty, while the
intercept of the fit is taken as the average initialisation time of
the generator.

\begin{subfigures}{1}{A comparison of the average time per \wtl decay
    between the \pythia{8}, \herwig{++}, and \tauola event generators
    for a selection of the \wtl decays of
    \sec{Tau:Dec}.\labelfig{Time}}
  \svgbeg
  \includesvg[pretex=\hspace{-0.43cm}\relsize{-3},width=\columnwidth]{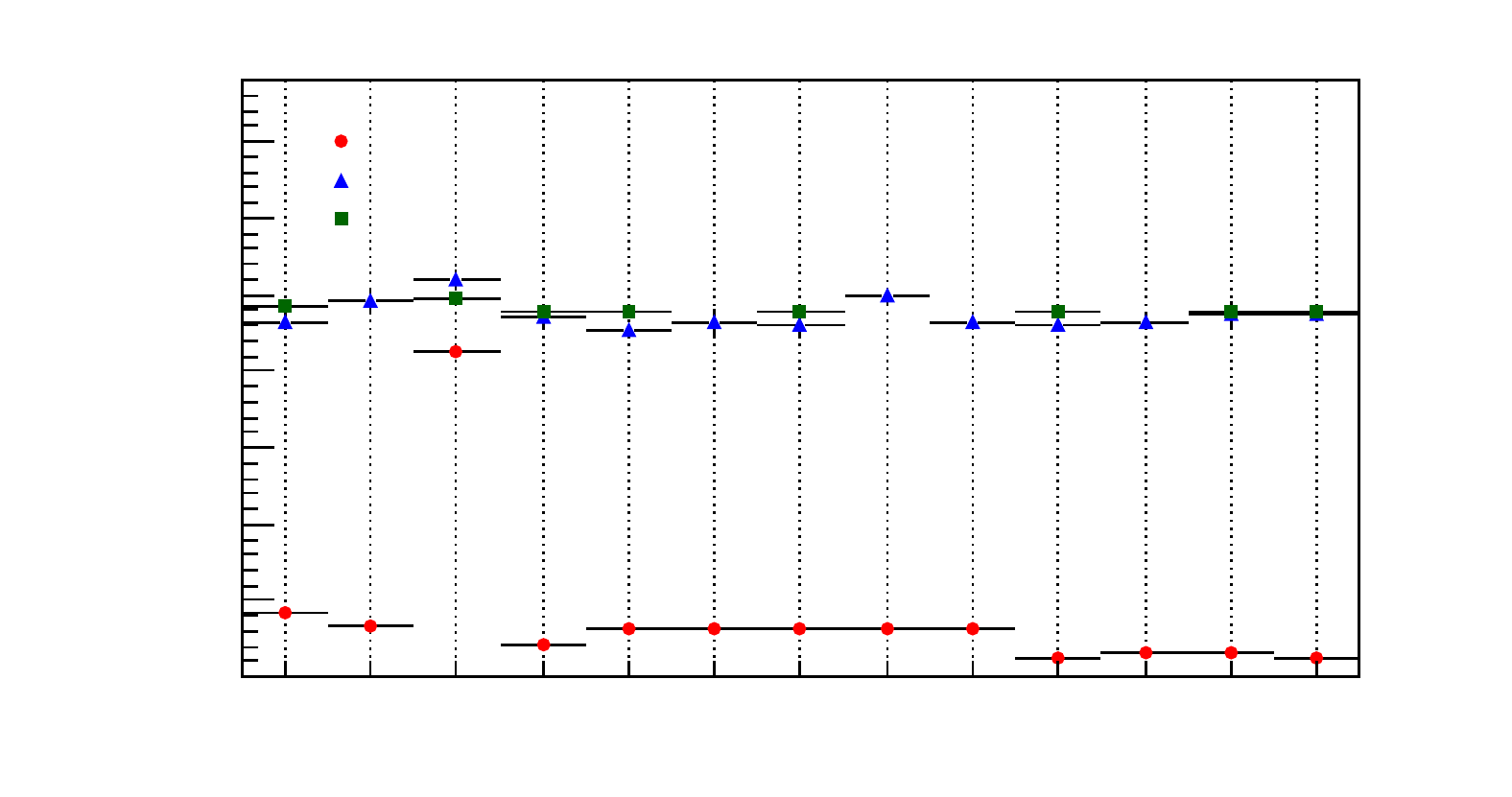}
  \svgend
\end{subfigures}

In \fig{Tau:Time} the average time per event for a selection of the
\wtl decay channels of \sec{Tau:Dec} is given. In these tests
\pythia{8} outperforms \herwig{++} and \tauola for all of the
implemented \wtl decay channels by a factor of $\approx 5$ except for
the $\tauto \pi^0 \pi^- \pi^- \pi^+$ channel. The decrease in speed
for this channel is due to the sharp resonance from the $\omega$ which
is not present in the $\tau^- \rightarrow \nu_\tau \pi^0 \pi^0 \pi^0
\pi^-$ channel. This sharp resonance is not easily sampled using the
\m-generator phase-space algorithm and leads to inefficiencies for
this channel. The timing information given in \fig{Tau:Time} is
dependent upon a variety of factors other than just the \wtl decay
algorithms used, including compile time options, machine architecture,
and most importantly, the configuration used for event generation with
each generator. Consequently, the results of \fig{Tau:Time} provide a
general sense for the timing of the \wtl decay implementation in
\pythia{8}, but will vary from system to system.

\newchapter{Experimental Setup}{Exp}

The data used in the analyses of \chp{Zed} and \chp{Hig} were
collected using the Large Hadron Collider Beauty detector (\lhcb) on
the Large Hadron Collider (\lhc) at the European Organisation for
Nuclear Research (CERN). CERN was founded in 1954 and is situated on
the French-Swiss border near the city of Geneva,
Switzerland. Currently, CERN is run by $20$ European member states
with $7$ observer states and organisations, and $37$ participating
non-member states. Nearly $10,000$ visiting scientists from over $600$
universities and institutes and $113$ countries utilise the research
facilities at CERN. Fundamental advances in particle physics have been
made throughout the years at CERN including the first observation of
the \w~\cite{ua1.83.1,ua2.83.1} and \wzbs~\cite{ua1.83.2,ua2.83.2} by
the UA$1$ and UA$2$ detectors on the Super Proton Synchrotron (SPS),
precision electroweak measurements~\cite{lep.06.1} by the detectors on
the Large Electron-Positron collider (LEP), the creation of
anti-hydrogen~\cite{ps210.96.1}, and most recently the discovery of a
Higgs-like boson~\cite{atlas.12.2,cms.13.1} with the \atlas and \cms
detectors on the \lhc. Within this chapter the collider is introduced
in \sec{Exp:Col}, the \lhcb detector is described in \sec{Exp:Det},
and the the methods used for reconstructing events observed within the
\lhcb detector are outlined in \sec{Exp:Rec}.

\newsection{Large Hadron Collider}{Col}

The \lhc accelerates protons in opposite directions around a
$27~\mathrm{km}$ ring, colliding them at four interaction points
around which four detectors are built. Schematics in the vertical
plane, defined by the beamline and a vector perpendicular to the \lhc
ring, for the general purpose \atlas and \cms detectors are given in
\figs{Exp:Atlas} and \ref{fig:Exp:Cms}. These detectors are fully
instrumented with tracking systems, calorimeters, and muon
chambers. The \lhcb detector, with a schematic shown in
\fig{Exp:Lhcb}, is designed specifically for forward physics, in
particular the physics of $B$-hadrons, and is a forward arm
spectrometer, extending outwards on only one side of the interaction
point. The data used in the analyses of \chps{Zed} and \ref{chp:Hig}
were taken with this detector. Further details on the \lhcb detector
and \lhcb event reconstruction are given in \secs{Exp:Det} and
\ref{sec:Exp:Rec}. The \alice detector is a heavy ion detector, with
its schematic shown in \fig{Exp:Alice}. During nominal \lhc operations
the \lhcb and \alice detectors receive reduced luminosities with
respect to the \atlas and \cms detectors.

\begin{subfigures}{2}{Rough schematics of the \subfig{Atlas}~\atlas,
    \subfig{Cms}~\cms, \subfig{Lhcb}~\lhcb, and \subfig{Alice}~\alice
    detectors of the \lhc. Here, \ecal are electromagnetic
    calorimeters, \hcal are hadronic calorimeters, TRD are transition
    radiation detectors, TOF are time of flight detectors, and TPC are
    time projection chambers. These schematics were modified from
    \rfrs{atlas.08.1}, \cite{cms.08.1}, \cite{lhcb.08.1},
    and~\cite{alice.08.1}.\labelfig{Detectors}}
  \multicolumn{2}{c}{\includesvg[width=\columnwidth]{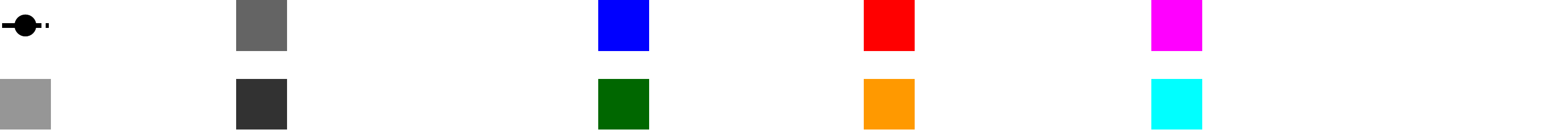}} \\
  \svg{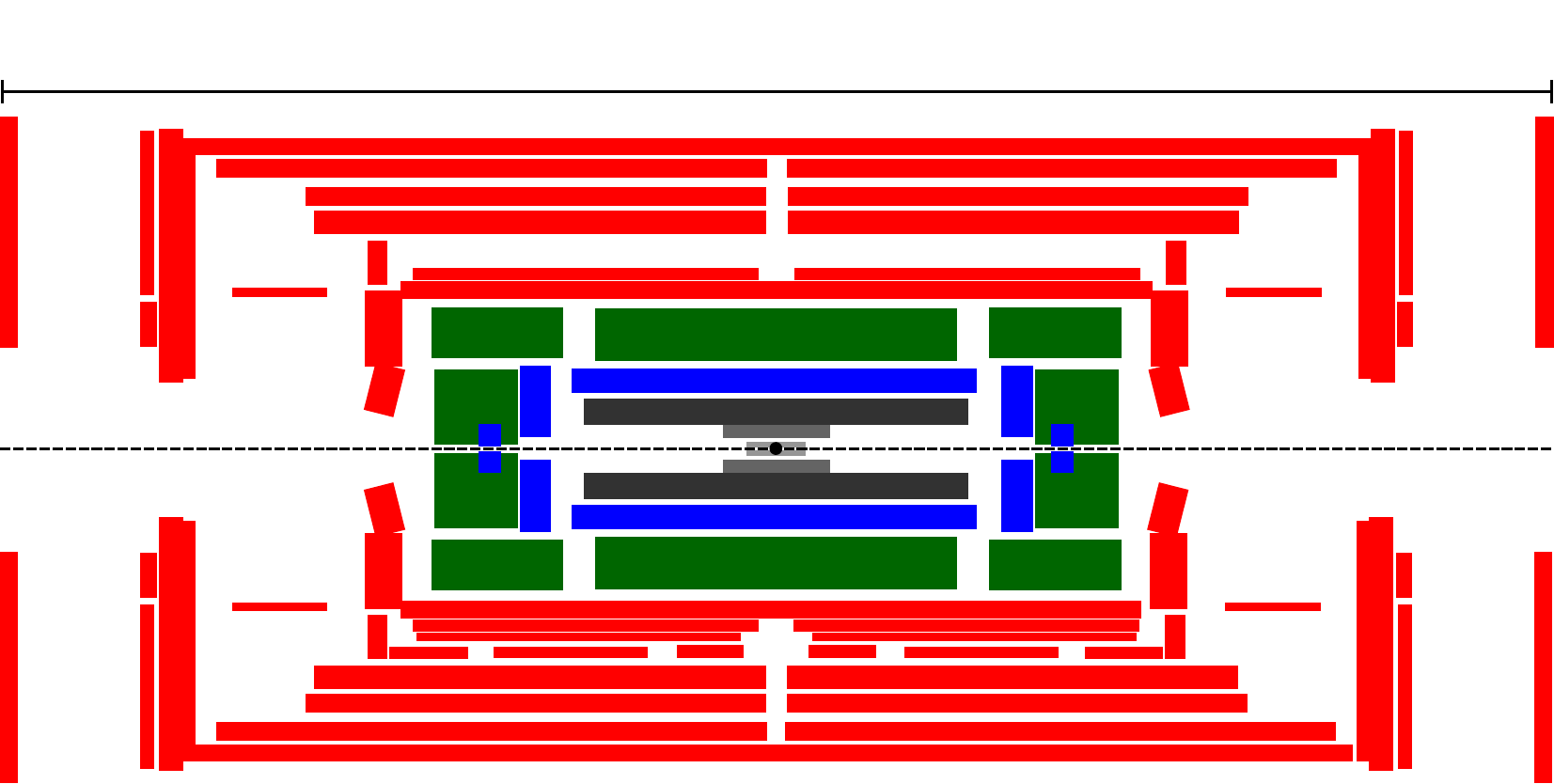} & \svg{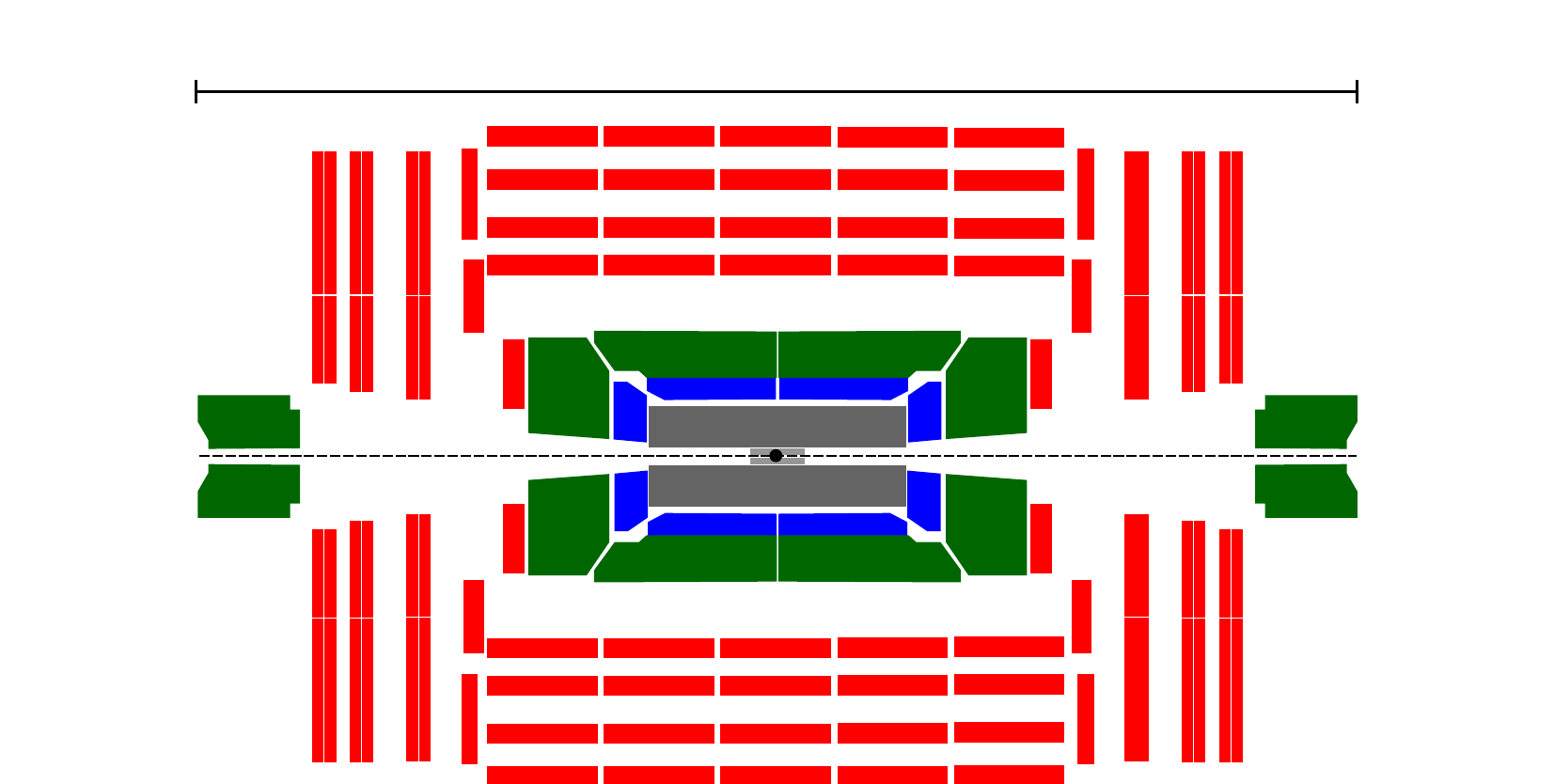}   \\
  \svg{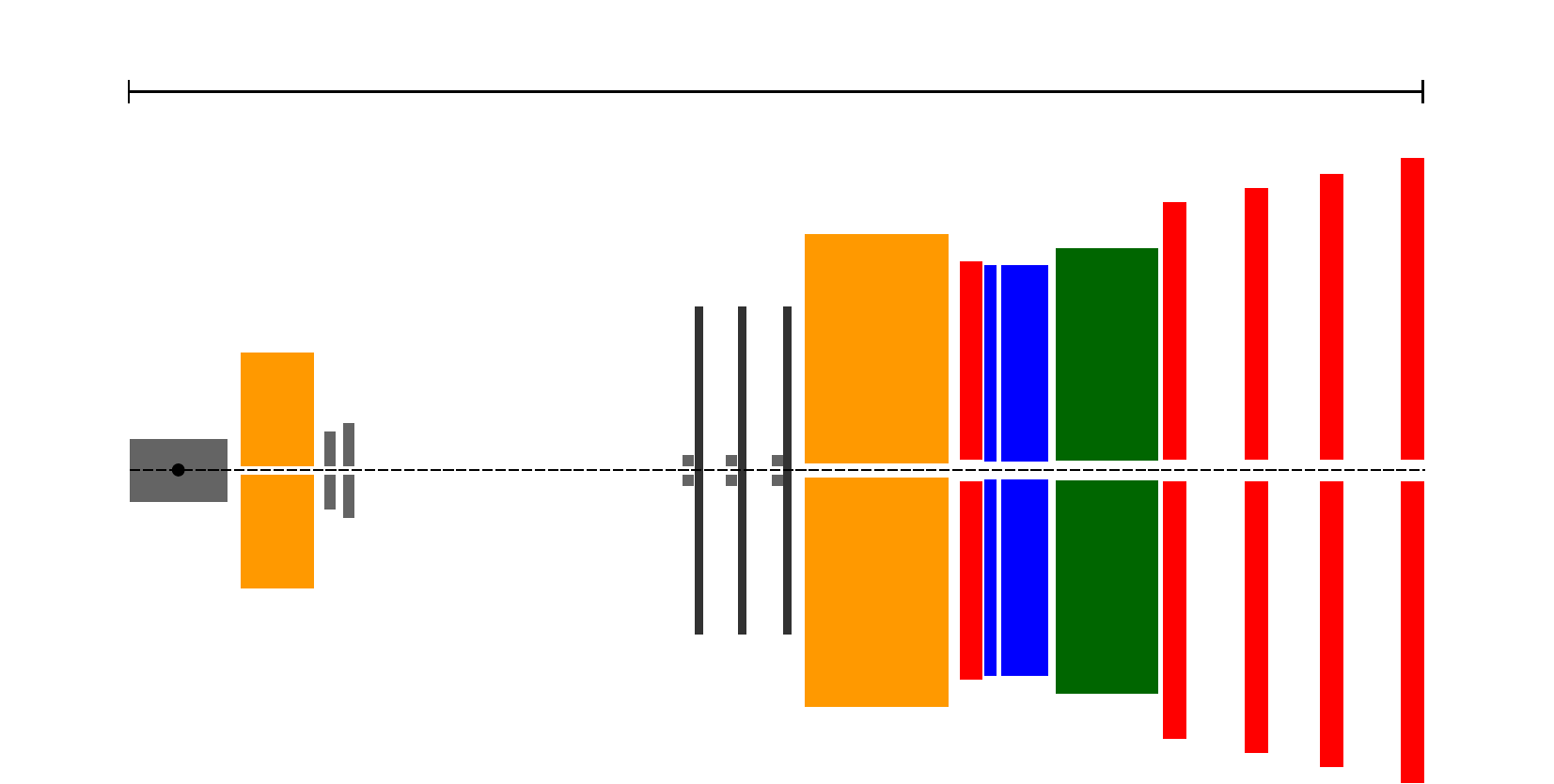}  & \svg{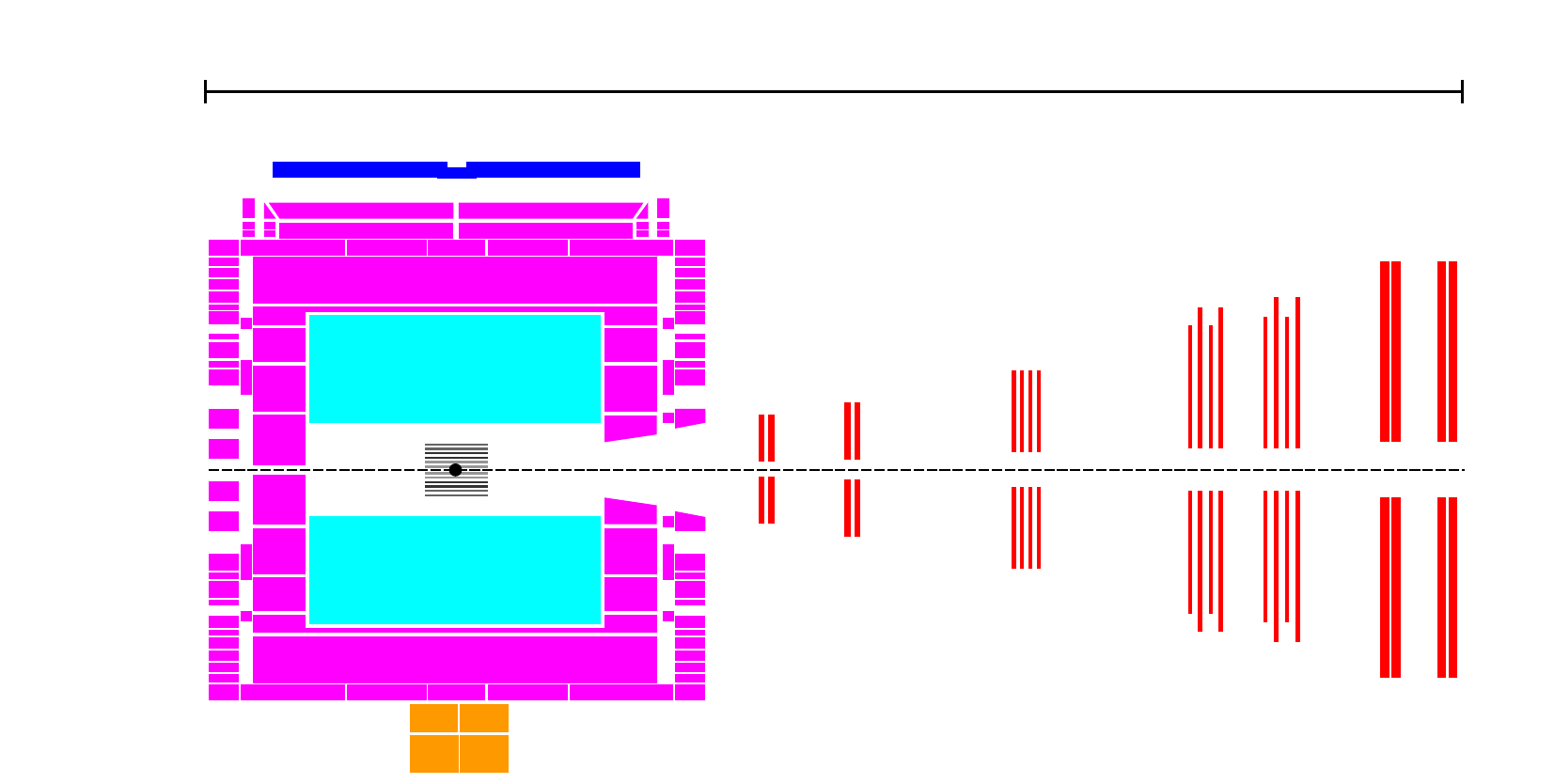} \svgend
\end{subfigures}

The \lhc began operation in September of $2008$, but nine days after
the inaugural start-up, a fault of the busbars in the interconnects
between a dipole and quadrupole magnet system caused a magnet quench
and delayed operations of the \lhc until November of
$2009$~\cite{lhc.09.1}. Full-scale data-taking began in March of
$2010$ at a centre-of-mass energy of $7~\tev$, $3.5~\tev$ per beam,
and continued until November of $2010$ when lead ion beams were
circulated for a month. In March of $2011$ proton-proton collisions
were again begun at a centre-of-mass energy of $7~\tev$ and continued
until November of $2011$.  A plot of the integrated luminosity over
time for the \atlas, \cms, and \lhcb detectors during $2011$
data-taking is given in \fig{Exp:Luminosity.11}. In April of $2012$,
proton-proton data-taking at a centre-of-mass energy of $8~\tev$
began, and continued until February of $2013$. The $2012$ integrated
luminosity is plotted in \fig{Exp:Luminosity.12}. Currently, the \lhc
has entered a long shutdown until sometime in $2014$ after which
operation will recommence at higher luminosities and with a
centre-of-mass energy of $14~\tev$~\cite{heuer.12.1}. The data used in
\chps{Zed} and \ref{chp:Hig} were collected during the $2011$
data-taking operations.

\begin{subfigures}{2}{Integrated luminosity as a function of time
    provided by the \lhc to the \lhcb, \atlas, and \cms detectors for
    the \subfig{Luminosity.11}~$2011$ data-taking at ${\sqrt{s} =
      7~\tev}$ and \subfig{Luminosity.12}~$2012$ data-taking at
    ${\sqrt{s} = 8~\tev}$. The integrated luminosity data are taken
    from the \href{https://lhc-statistics.web.cern.ch/}{\lhc statistics
      web-page}.\labelfig{Luminosity}}
  \svgbeg
  \svg{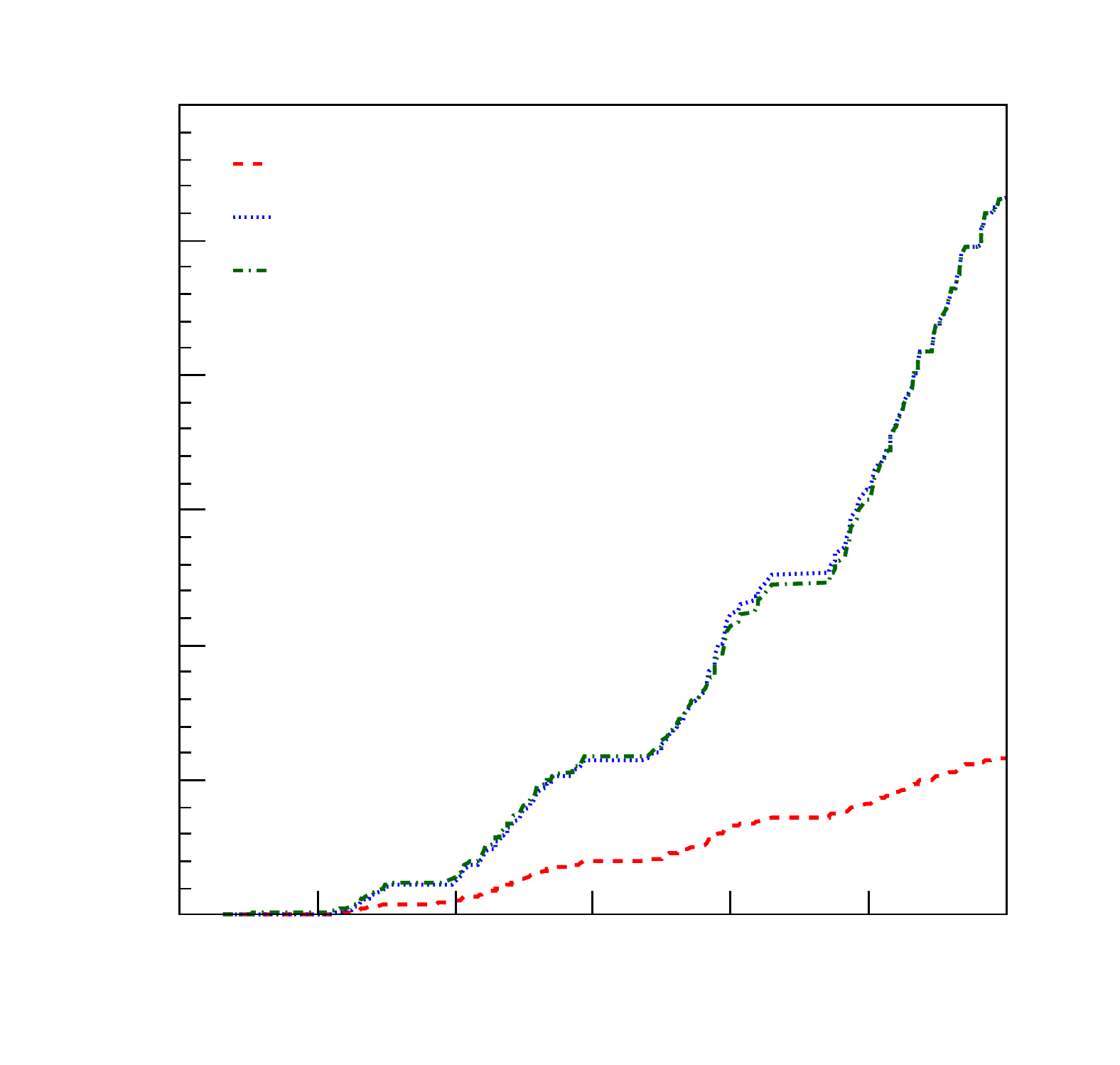} & \svg{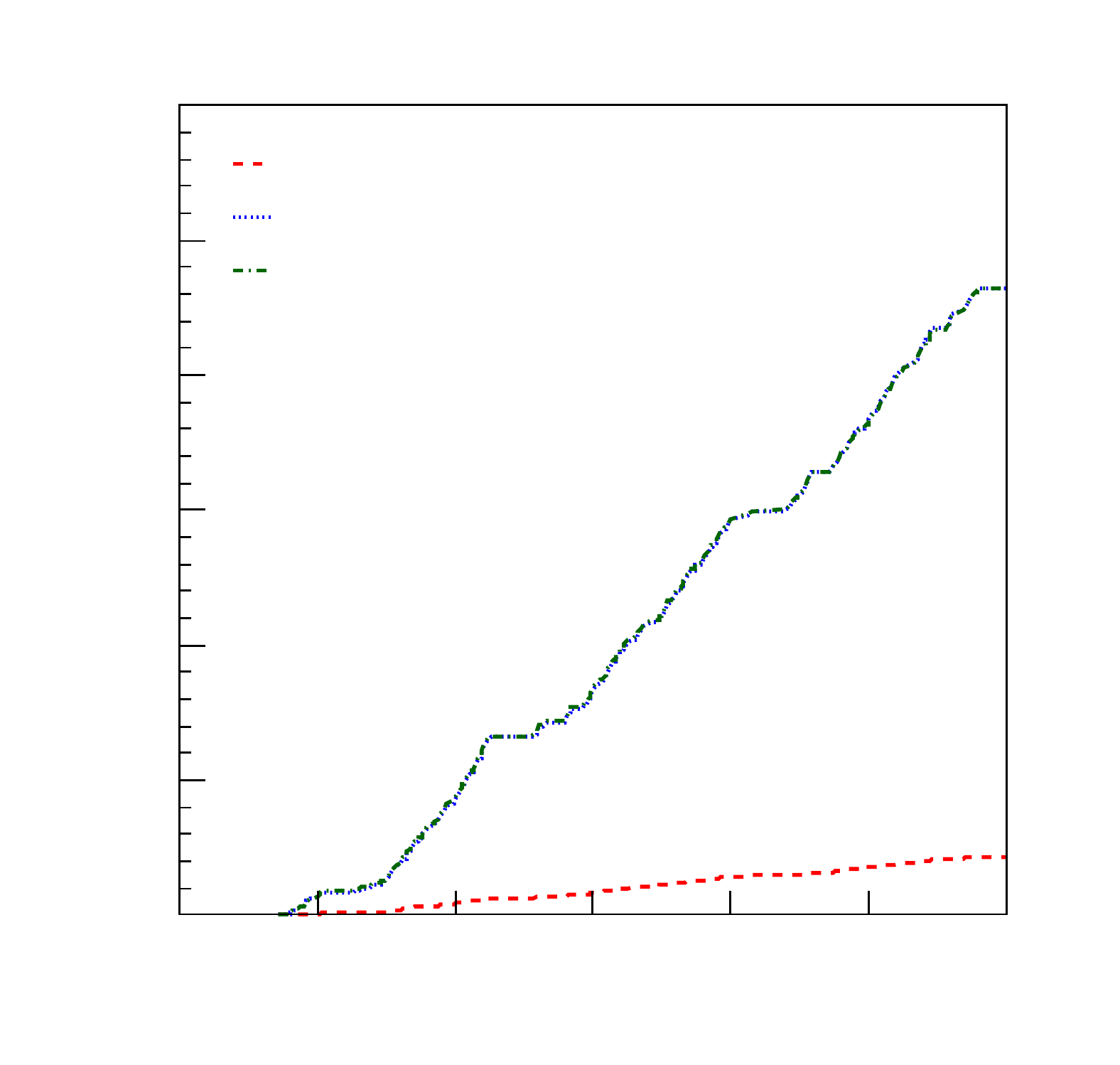} \svgend
\end{subfigures}

The technical design reports for the \lhc can be found in the three
volumes of \rfrs{lhc.04.1}, \cite{lhc.04.2}, and \cite{lhc.04.3},
corresponding to the main ring, infrastructure and general services,
and injector chain for the \lhc. An abridged and updated version of
the technical design reports can be found in \rfr{lhc.08.2}. In the
remainder of this section the layout of the machine is outlined in
\sec{Exp:Lay} and the injector chain is described in \sec{Exp:Inj}.

\newsubsection{Layout}{Lay}

The layout of the main \lhc ring is summarised in the schematic of
\fig{Exp:Lhc}. The ring consists of eight long straight sections (LSS)
each with a length of approximately ${528~\mathrm{m}}$ alternating
with eight arcs (ARC) each with a distance of approximately
${2.8~\mathrm{km}}$. The ring is also divided into octants, with each
octant centred about an LSS, and each ARC divided between two
octants. The collider ring is located between ${45~\mathrm{m}}$ to
${170~\mathrm{m}}$ below the surface, with an access point provided at
every LSS. Each LSS is used for beam utilities or experiments, and the
ARC segments contain the dipoles needed to bend the beam and
quadrupoles used to focus the beam. Between each ARC and LSS is
located a dispersion suppressor (DS) which is used to adapt the \lhc
reference orbit to the tunnel geometry. Additionally, the dispersion
suppressors are used to match the ARC optics with the insertion optics
for each LSS, as well as cancel horizontal dispersion from the dipole
magnets.

\begin{subfigures}{1}{Schematic of the main \lhc ring and the injector
    chain as seen from above. The first proton beam (red) rotates
    clockwise while the second proton beam (blue) rotates
    anti-clockwise.\labelfig{Lhc}}
  \includesvg[width=\columnwidth]{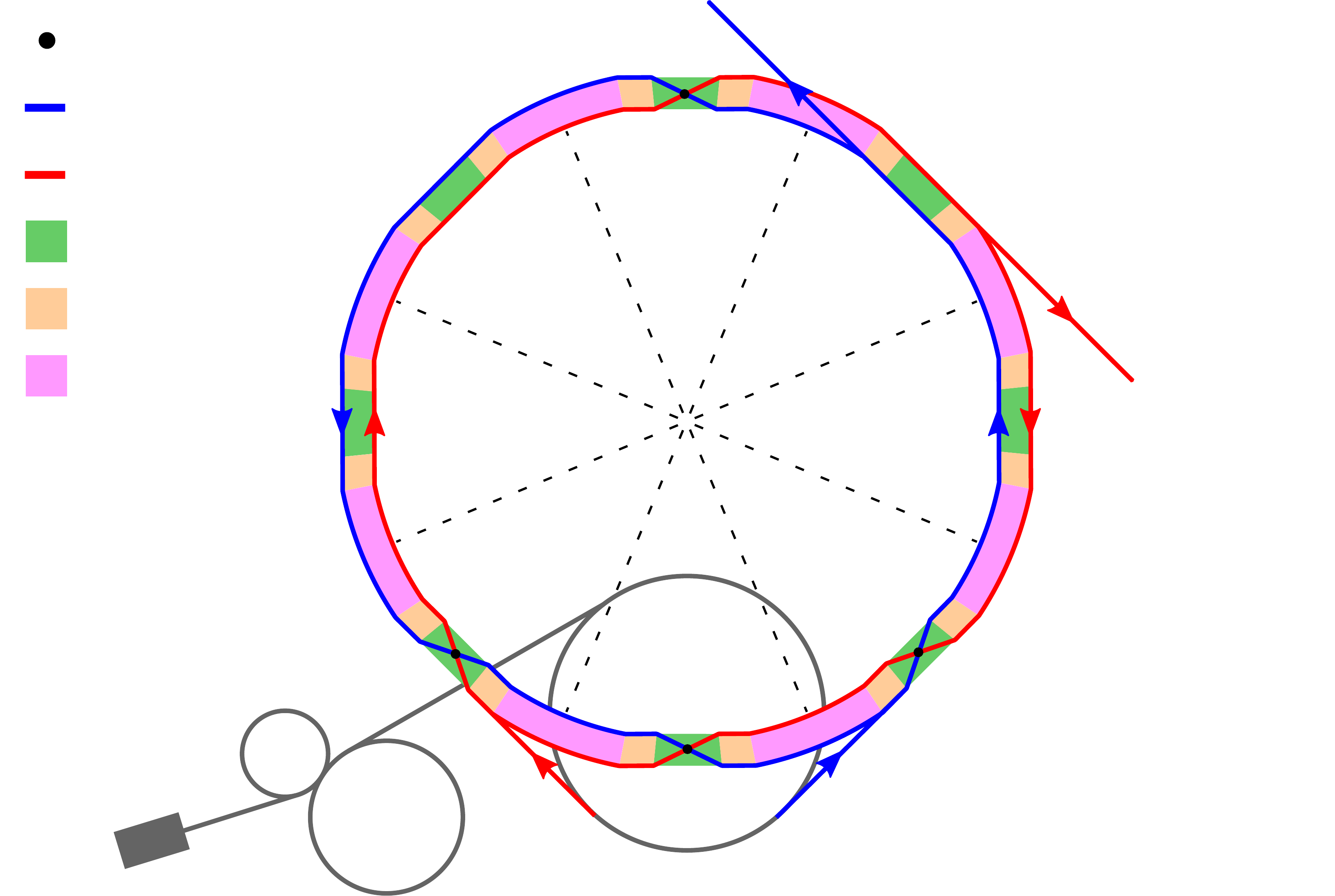} \svgend
\end{subfigures}

The beams cross in the LSSs of the first, second, fifth, and eighth
octants. The high luminosity \atlas and \cms experiments are located
at points $1$ and $5$ respectively. The plane of the beam crossing
angle at point $1$ is vertical while it is horizontal at point
$5$. Point $2$ houses the lower luminosity \alice experiment and point
$8$ contains the \lhcb experiment. Beam $1$, rotating clockwise from
above, is injected at point $2$ while beam $2$, rotating
anti-clockwise from above, is injected at point $8$.

In the LSS of the third octant, momentum cleaning of the beams is
performed while in the LSS of the seventh octant, optical cleaning of
the beam is performed. The radio frequency (RF) cavities for the main
ring are installed in the LSS of the fourth octant and contained in
the old cavern used to house the ALEPH detector on the \dlep
collider. The LSS of the sixth octant contains the two independent
beam dump systems for both beams. These dumps abort the beams by using
kicker magnets to horizontally bump the beams into septum magnets
which then deflect the beams vertically into absorbers within
dedicated tunnels.

\newsubsection{Injection Chain}{Inj}

The injection chain for the \lhc is also shown in \fig{Exp:Lhc}, and
begins with LINAC $2$, an Alvarez linear accelerator, where a plasma
is created from an ionised gas within a duoplasmatron running at
${90~\mathrm{kV}}$. The plasma is formed into an ion beam within a
${750~\mathrm{keV}}$ RF quadrupole which then passes the beam into
three Alvarez tanks, consisting of in-phase drift tubes, which
accelerate the beam up to an energy of ${50~\mev}$. For nominal \lhc
operation, the LINAC $2$ must output a proton beam with a current of
${180~\mathrm{mA}}$ and pulse lengths of ${150~\mu\mathrm{s}}$ at a
rate of ${0.8~\mathrm{Hz}}$ into the Proton Synchrotron Booster
(PSB)~\cite{hill.00.1}. The PSB consists of four rings each with a
radius of ${25~\mathrm{m}}$~\cite{reich.69.1}. One proton bunch is
injected into each of the rings and the four bunches are then
accelerated to an energy of ${1.4~\gev}$. These bunches are then
extracted into the Proton Synchrotron (PS). The PS has a radius of
$100~\mathrm{m}$ and holds $84$ proton bunches with a bunch spacing of
${25~\mu\mathrm{s}}$. These bunches are accelerated to an energy of
${26~\gev}$ after which their bunch length is reduced to
${4~\mathrm{ns}}$ for insertion into the ${200~\mathrm{MHz}}$ RF
cavities of the Super Proton Synchrotron~\cite{benedikt.00.1}. The
SPS, with a radius of ${1.1~\mathrm{km}}$ accelerates the beam to
${250~\gev}$ using two ${200~\mathrm{MHz}}$ RF cavities and injects
the beams into the main \lhc ring at either point $2$ or
$8$~\cite{schindl.99.1}.

\newsection{LHC Beauty Detector}{Det}

The Large Hadron Collider Beauty experiment (\lhcb) is a forward arm
spectrometer located in the experimental cavern of point $8$ on the
\lhc ring. The detector is primarily designed for the identification
of $B$-hadrons in order to further explore their decays and make
precise measurements of $\delta$, the \cp-violating phase angle of the
CKM matrix from the unified electroweak Lagrangian of \sec{Thr:Lag},
as well as to search for new physics through the deviation of rare
$B$-hadron decays from their \sm predictions, {\it e.g.} ${B_s \to
  \dimu}$~\cite{lhcb.13.4}. The production of \wbq pairs within the
\lhc environment, which then hadronise into jets of $B$-hadrons, is
typically close to the direction of either beam, with both \wbqs being
produced in the same direction. Consequently, \lhcb is designed as a
forward arm spectrometer about the beamline in order to maximise the
acceptance of $B$-hadrons while minimising construction and material
costs. The reconstruction and identification of $B$-hadrons within the
hadronic environment of the \lhc requires excellent secondary vertex
reconstruction, good momentum resolution, a fast and robust trigger,
and hadron identification capabilities.

\begin{subfigures}[t]{1}{Schematic of the \lhcb detector in the
    longitudinal $yz$-plane.\labelfig{Detector}}
  \includesvg[width=\columnwidth]{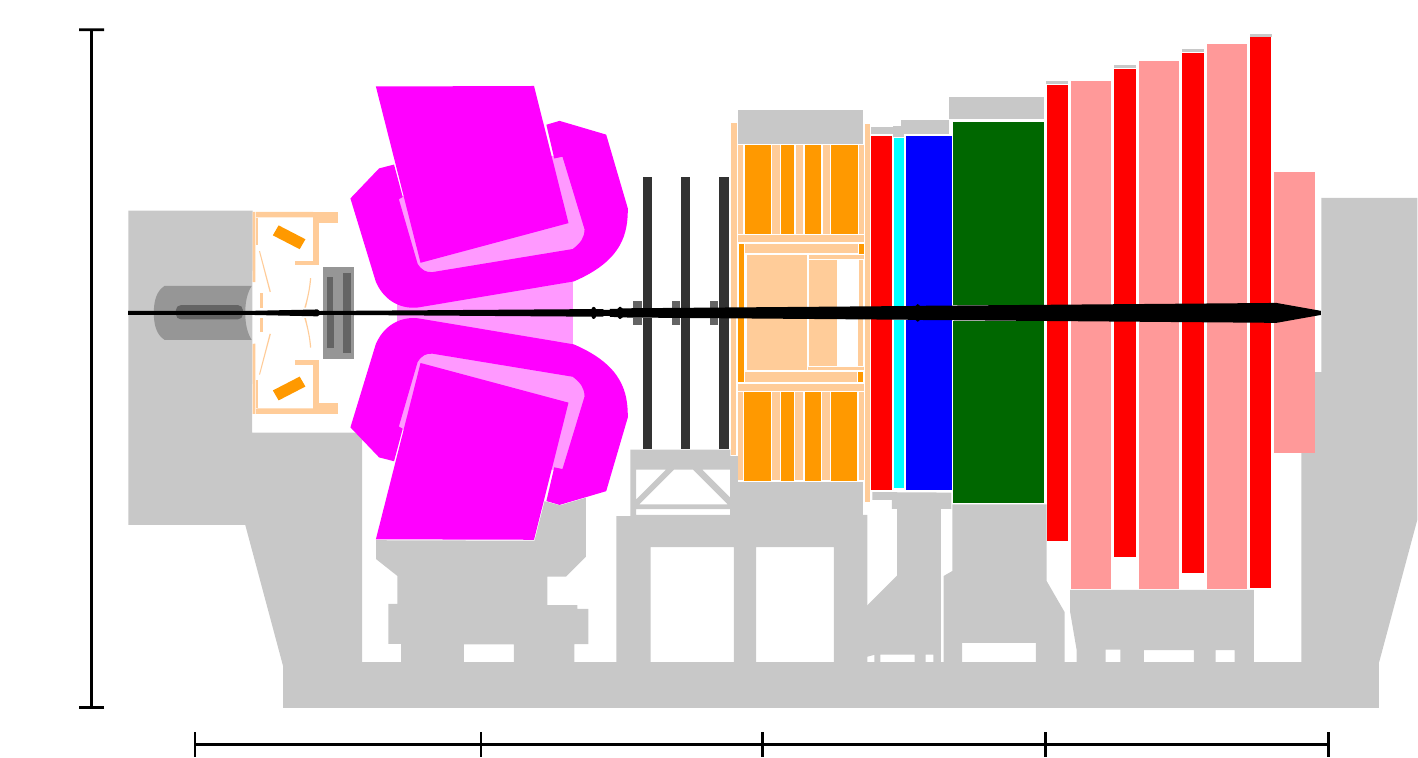}
\end{subfigures}

A detailed schematic of the \lhcb detector and its sub-components in
the longitudinal $yz$-plane is given in \fig{Exp:Detector}. The origin
of the \lhcb coordinate system is defined as the nominal interaction
point within \lhcb. The $x$-axis points outward from the centre of the
\lhc ring and is parallel to the plane of the \lhc ring. The $z$-axis
points along the beam in the direction of the detector, resulting in
an angle of ${\approx 3.6~\mathrm{mrad}}$ with the floor of the \lhcb
cavern. The $y$-axis points upward and is perpendicular with both the
$x$-axis and $z$-axis to form a right-handed Cartesian coordinate
system. The azimuthal angle $\phi$ is defined in the transverse
$xy$-plane with a $\phi$ of zero along the direction of the $x$-axis
and ranges between values of $-\pi$ and $\pi$. The polar angle
$\theta$ is defined as the opening angle with the $z$-axis such that a
particle along the beamline has a $\theta$ of zero. The
pseudo-rapidity $\eta$ and rapidity $y$ are oftentimes used rather
than $\theta$ and are defined as,
\begin{equation}
  \eta \equiv \frac{1}{2}\ln\left(\frac{\abs{\vec{p}} + p_z}{\abs{\vec{p}} -
      p_z}\right) = -\ln\left(\tan\left(\frac{\theta}{2}\right)\right)
  \equcomma
  y = \frac{1}{2}\ln\left(\frac{E + p_z}{E - p_z}\right)
  \labelequ{Rapidity}
\end{equation}
where $p_z$ is the $z$-component of the particle momentum. For
massless particles the pseudo-rapidity and rapidity are
equivalent. The transverse momentum of a particle, \pt, is also
commonly used and is defined as $\sqrt{p_x^2 + p_y^2}$.

Surrounding the interaction point in \fig{Exp:Detector} is a vertex
locator (\velo) which is followed, in the positive $z$-direction, by a
ring imaging Cherenkov detector (\rich{1}) and a tracker turicensis
(\ttt). After the \ttt is a large dipole magnet which provides a
bending field for the tracking systems. The remainder of the tracking
system follows the magnet, consisting of an inner tracker (\itt) about
the beamline and an outer tracker (\ott). This part of the tracking
system is separated into three tracking stations, T$1$, T$2$, and
T$3$. Another ring imaging Cherenkov detector (\rich{2}) is located
after the \itt and \ott, which is larger than \rich{1}. After \rich{2}
is the first station M$1$ of the muon system, followed by a
scintillating pad detector (\spd) and pre-shower calorimeter (\prs),
an electromagnetic calorimeter (\ecal), and a hadronic calorimeter
(\hcal). The remainder of the detector consists of four muon tracking
stations M$2$ through M$5$. In the remainder of this section a
description of the tracking system is given in \sec{Exp:Trk} followed
by an overview of the particle identification systems in
\sec{Exp:Pid}. A full description of the detector can be found in
\rfr{lhcb.08.1} on which this section is based.

\newsubsection{Tracking}{Trk}

The tracking system for \lhcb consists of the \velo, \ttt, and \itt
which are silicon microstrip detectors, and the \ott which is a
straw-tube detector. The \ttt and \itt were developed under the
combined silicon tracker (\stt) project and utilise similar hardware
designs. Both the silicon microstrip and straw-tube technologies
record the passage of charged particles, and when combined with a
magnetic field, the trajectories of the charged particles built from
their hits in the detectors are used to determine the momenta of the
particles. The passage of a charged particle through a silicon
detector ionises atoms, creating a current which is detected, while in
straw-tube detectors the particle ionises the gas within the tube
which then avalanches and creates a detected current. Both silicon
microstrips and straw-tubes are combined in layers to provide a
three-dimensional hit coordinate. During $2011$ data taking the
combined tracking system provided a momentum resolution, $\delta_p/p$,
for charged particles between \prc{0.4 - 0.6}~\cite{tobin.12.1}.

\begin{subfigures}[t]{2}{Magnet-down $y$-component of the magnetic
    field within the \lhcb detector along the $z$-axis, beginning at
    the \velo and moving outwards to the \rich{2} detector. The
    uncertainty on the precision of the measurements is less than
    ${10^{-4}~\mathrm{T}}$. The approximate $z$-axis locations of the
    subdetectors are indicated by the shaded areas. The magnetic field
    data is taken from \rfr{lhcb.08.1}.}
  \svgbeg
  \svg[1]{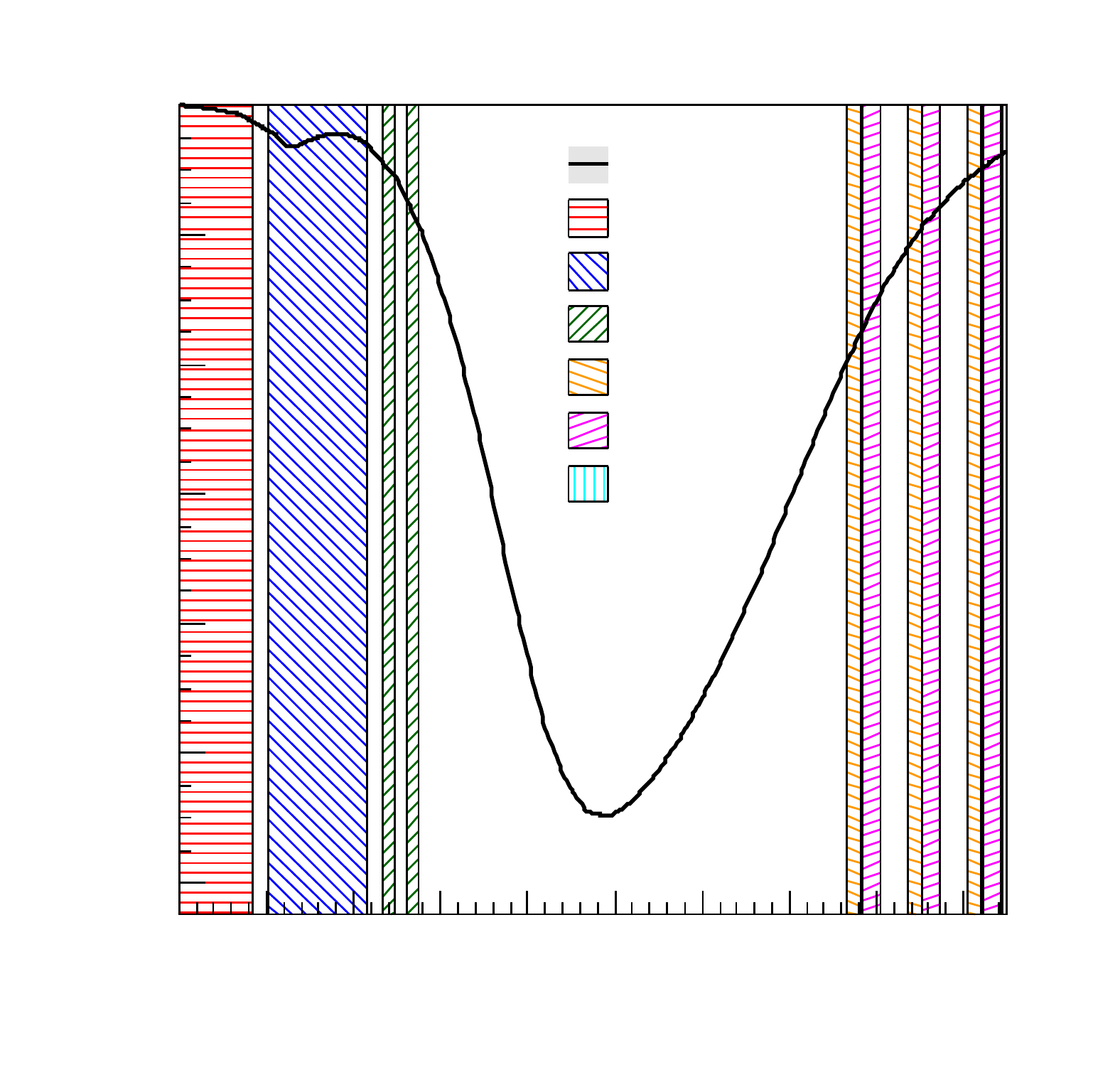} & \sidecaption \svgend
\end{subfigures}

A large dipole magnet located between the \ttt and \ott provides the
bending field for the tracking system. The dipole is a warm magnet
consisting of two saddle-shaped coils, and is designed to meet the
requirements of the tracking system while fitting within the space of
the experimental cavern and minimising costs. The bending plane of the
magnet is in the $xz$-plane of the detector, and is designed to
provide a magnetic field of less than ${50~\mathrm{mT}}$ within the
\rich{} systems and a maximum magnetic field between the \ttt and \ott
with an integrated magnetic field of ${4~\mathrm{Tm}}$ for a
${10~\mathrm{m}}$ track. The dipole can be run with a magnet-down or
magnet-up configuration, corresponding to the direction of the
$y$-component of the magnetic field. Both field configurations have
been mapped between the \velo and \rich{2} using an array of Hall
probes within a precision of better than ${10^{-4}~\mathrm{T}}$. A
report on the field map for the magnet taken during $2011$ is
available in \rfr{lhcb.12.4}, with the $y$-component of the magnetic
field along the $z$-axis from \rfr{lhcb.08.1} given in
\fig{Exp:Magnet}. The initial technical design report for the magnet
system can be found in \rfr{lhcb.07.1}.

\newsubsubsection{Vertex Locator}

The vertex locator is a microstrip detector located about the \lhcb
interaction point and is designed for high resolution reconstruction
of secondary vertices from long-lived particles such as
$B$-hadrons. The \velo is designed to have a pseudo-rapidity coverage
of ${1.6 \leq \eta \leq 4.9}$ for particles produced from vertices
within ${10.7~\mathrm{cm}}$ of the interaction point. A schematic of
the \velo in the $xz$-plane is given in \fig{Exp:Velo}, consisting of
$42$ semi-circular detector modules and $4$ pile-up veto sensors. The
modules have a diameter of ${9~\mathrm{cm}}$ and during stable beam
conditions the active areas of the modules are located
${8~\mathrm{mm}}$ from the centre of the beamline. Because the
transverse beam width during injection of the \lhc beams is larger
than this separation, the \velo modules are retracted during beam
injection so that the active areas of the module sensors are
${29~\mathrm{mm}}$ from the centre of the beamline.

\begin{subfigures}[t]{1}{Schematic of the \velo detector in the
    $xz$-plane. Each of the $42$ modules consists of an $r$-sensor
    (red) and a $\phi$-sensor (blue). The leftmost $r$-sensors are
    pile-up veto sensors. Figure adapted from
    \rfr{lhcb.08.1}.\labelfig{Velo}}
  \includesvg[width=\columnwidth]{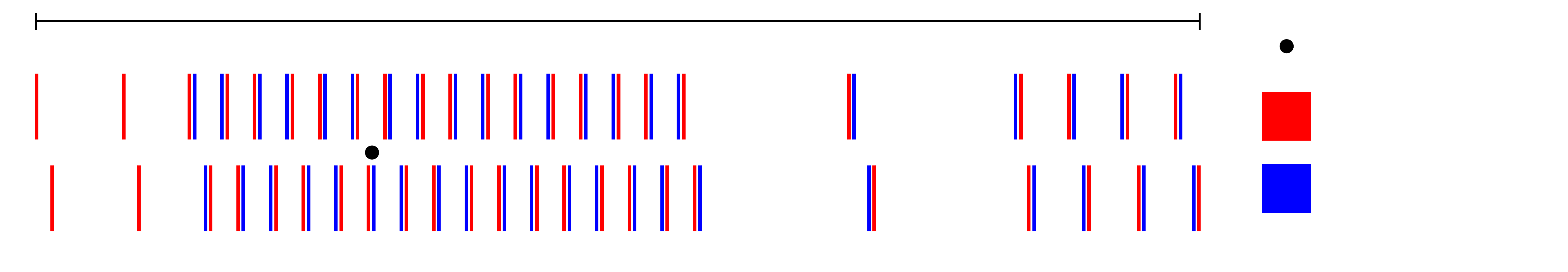}
\end{subfigures}

Each module consists of a back-to-back radial $r$-sensor and azimuthal
$\phi$-sensor, where the $z$-coordinate of each module in conjunction
with the $r$ and $\phi$-measurements provide a full three-dimensional
location for each hit in the module. The $r$-sensors are made up of
quadrants, with each quadrant containing $512$ strips, while the
$\phi$-sensors consist of $683$ inner strips and $1365$ outer strips;
the total number of strips per sensor for both sensor types is
$2048$. The strip pitch of the $r$-sensors increases linearly from
${40~\mu\mathrm{m}}$ to ${102~\mu\mathrm{m}}$ and the strip pitch for
the $\phi$-sensors range between ${38~\mu\mathrm{m}}$ and
${97~\mu\mathrm{m}}$. The use of a cylindrical geometry for the \velo
sensors, rather than a more typical rectilinear geometry, allows fast
reconstruction of track impact parameters with sufficient resolution
using information from the $r$-sensors, as is done by the high level
trigger.

The pitches of the $r$ and $\phi$-sensors were chosen to provide a
single hit resolution of approximately $4~\mathrm{\mu}m$ for a track
with a pseudo-rapidity of $3$ passing through the inner
$\phi$-strips. The impact parameter significance used in \sec{Zed:Sel}
to separate \wtl decays from prompt backgrounds primarily depends upon
the precision which the \velo can reconstruct track positions close to
the interaction region (impact parameter
resolution). The impact parameter resolution is
given by,
\begin{equation}
  \delta_{\mathrm{IP}} = \delta_\mathrm{HIT} \oplus
  \frac{\delta_\mathrm{MSE}}{\pt}
\end{equation}
where $\delta_\mathrm{HIT}$ is the uncertainty due to the intrinsic
precision of the track hits, $\delta_\mathrm{MSE}$ is the uncertainty
due to multiple scattering effects, and \pt is the transverse momentum
of the particle. From $2010$ data, $\delta_\mathrm{HIT}$ was found to
be ${13.1~\mathrm{mm}}$ for the $x$-component and ${12.1~\mathrm{mm}}$
for the $y$-component, while $\delta_\mathrm{MSE}$ was found to be
${23.9~\mathrm{mm}\,\gev}$ and ${23.7~\mathrm{mm}\,\gev}$ for the $x$
and $y$-components of the impact parameter~\cite{farry.12.1}.

\newsubsubsection{Tracker Turicensis}

% See Tracker.m for pseudo-rapidity calculations.
The tracker turicensis is a microstrip silicon detector designed to be
used in conjunction with the \velo, \itt, and \ott to measure the
momenta of charged particles. This sub-detector begins approximately
${2.3~\mathrm{m}}$ down the $z$-axis, ends at approximately
${2.8~\mathrm{m}}$ along the $z$-axis, and covers a pseudo-rapidity
range of ${2.0 \leq \eta \leq 5.0}$ in the $yz$-plane and ${1.9 \leq
  \eta \leq 4.9}$ in the $xz$-plane for particles produced from the
interaction point. The \ttt consists of four layers, with the first
two layers grouped into a single station and separated from the second
station of the final two layers by a distance of ${0.3~\mathrm{m}}$. A
schematic of the first layer in the $xy$-plane is given in
\fig{Exp:Tt}, which includes $16$ strip modules with $8$ modules above
the $xz$-plane and $8$ modules below. The second \ttt layer is similar
to the first, but the modules are rotated by an angle of $+5^\circ$
with respect to the $yz$-plane. The third layer has an additional $4$
modules, $2$ above and $2$ below the $xz$-plane, and is rotated by an
angle of $-5^\circ$, while the fourth and final layer also has an
additional $4$ modules, but is not rotated.

\begin{subfigures}[t]{1}{Schematic of the first layer of the \ttt
    detector in the $xy$-plane. The layer consists of $16$ strip
    modules, each with $7$ sensors and $2$ or $3$ read out
    hybrids. Adapted from \rfr{lhcb.03.1}.\labelfig{Tt}}
  \includesvg[width=\columnwidth]{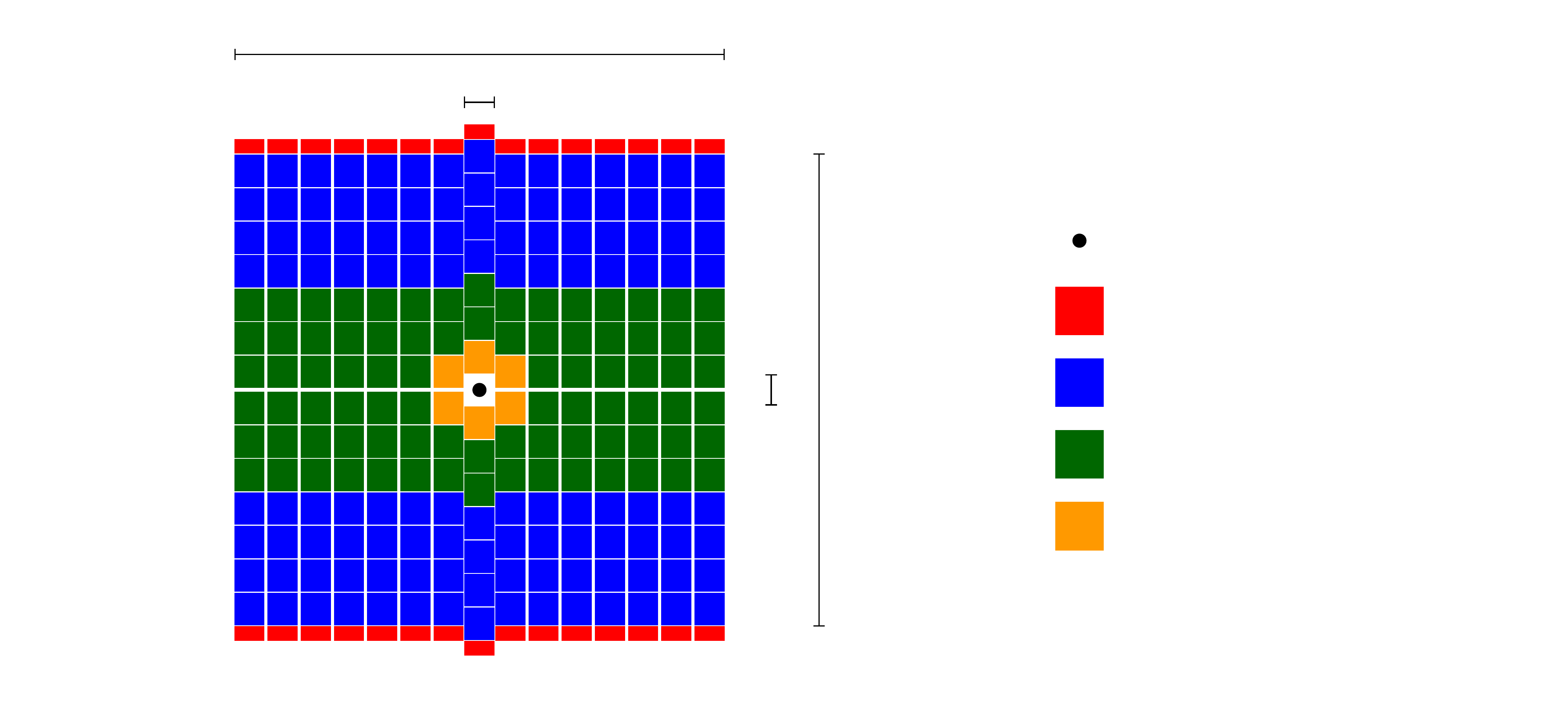}
\end{subfigures}

Every strip module contains $7$ silicon sensors, each with $512$
strips aligned vertically along the length of the module, where the
pitch for each strip is $183~\mu\mathrm{m}$. This vertical alignment
provides maximum resolution in the bending plane of the dipole
magnet. The sensors are divided into three read out sectors, $L$, $M$,
and $K$. The $4$ outermost silicon sensors for each module read out to
the $L$-sector, while the $3$ innermost sensors read out to the
$M$-sector. Modules with high occupancy sensors adjacent to the
beamline contain an $M$-sector subdivided into an additional
$K$-sector, with the two outermost sensors reading out to the
$M$-sector, and the innermost sensor adjacent to the beamline reading
out to the $K$-sector.

This design of the \ttt system provides a minimum single hit
resolution of ${59~\mu\mathrm{m}}$ as found from
data~\cite{tobin.12.1}. The \ttt in combination with the muon system
is used to determine muon track finding efficiencies from data in
\sec{Zed:RecEff}. Further details on the \ttt can be found in the
technical design report of \rfr{lhcb.03.2}.

\newsubsubsection{Inner Tracker}

% See Tracker.m for pseudo-rapidity calculations.
The inner tracker uses the same silicon microstrip sensors as the \ttt
and provides high pseudo-rapidity tracking coverage along the beam
line complimentary to the \ott. There are three \itt stations located
at approximately ${7.7~\mathrm{m}}$, ${8.4~\mathrm{m}}$, and
${9.0~\mathrm{m}}$ along the $z$-axis corresponding to the combined
\itt and \ott tracking stations T$1$, T$2$, and T$3$ of
\fig{Exp:Detector}. The three stations cover a common pseudo-rapidity
range for a track produced at the interaction point of ${4.5 \leq \eta
  \leq 4.9}$ in the $yz$-plane and ${3.4 \leq \eta \leq 5.0}$ in the
$xz$-plane. Despite the small acceptance of the \itt, nearly $20\%$ of
all tracks produced within \lhcb pass through the \itt. Every \itt
station contains four layers, similar to the four layers of the
\ttt. Each layer consists of $28$ modules in a configuration similar
to the schematic of \fig{Exp:It}. The strip modules of the first and
fourth layers are vertical like that of \fig{Exp:It}, while the second
layer is rotated by $+5^\circ$ with respect to the $yz$-plane and the
third layer is rotated by $-5^\circ$. This rotation configuration is
the same as for the \ttt layers and is designed to maximise the \itt
resolution in the bending plane of the dipole.

\begin{subfigures}[t]{1}{Schematic of the first and fourth layer of
    the \itt detector in the $xy$-plane. The layer consists of $28$
    strip modules, each with $1$ or $2$ sensors and $1$ or $2$ read
    out hybrids. Adapted from \rfr{lhcb.02.1}.\labelfig{It}}
  \includesvg[width=\columnwidth]{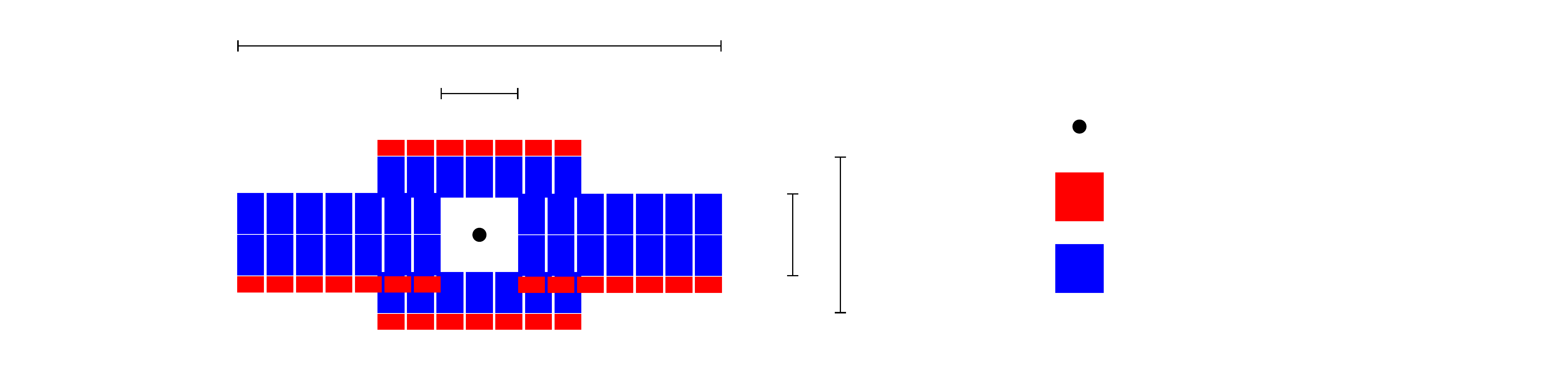}
\end{subfigures}

The $14$ outermost strip modules contain two silicon sensors each,
with a dedicated read out hybrid for every sensor, while the $14$
innermost strip modules only contain one sensor each. These sensors
use the same technology as the \ttt sensors and contain $384$
vertically aligned strips, each with a pitch of
${198~\mu\mathrm{m}}$. The \itt provides a single hit resolution of
${50~\mu\mathrm{m}}$ as found from data~\cite{tobin.12.1}, and is
critical for the reconstruction of the charged particle tracks used in
the analysis of \chp{Zed}. Further details on the \itt can be found in
the technical design report of \rfr{lhcb.02.1}.

\newsubsubsection{Outer Tracker}

The outer tracker is a large straw-tube detector which provides lower
pseudo-rapidity coverage about the higher pseudo-rapidity \itt
stations. There are three \ott stations which are contained in the
combined \itt and \ott stations T$1$, T$2$, and T$3$. The three \ott
stations are located at ${7.8~\mathrm{m}}$, ${8.5~\mathrm{m}}$, and
${9.2~\mathrm{m}}$ along the $z$-axis and cover a common
pseudo-rapidity range of ${2.0 \leq \eta \leq 4.5}$ in the $yz$-plane
and ${1.8 \leq \eta \leq 3.4}$ in the $xz$-plane. Each \ott station
consists of four layers, where the same rotation scheme for the \ttt
and \itt is used; the first and fourth layers are vertical, the second
layer is rotated by $+5^\circ$ with respect to the $yz$-plane, and the
third layer is rotated by $-5^\circ$. A schematic of a vertical \ott
layer is given in \fig{Exp:Ot}, where each layer contains $14$ long
$F$-modules and $8$ short $S$-modules.

\begin{subfigures}[t]{1}{Schematic of the first and fourth layer of
    the \ott detector in the $xy$-plane. The layer consists of $14$
    $F$-modules and $8$ $S$-modules.\labelfig{Ot}}
  \includesvg[width=\columnwidth]{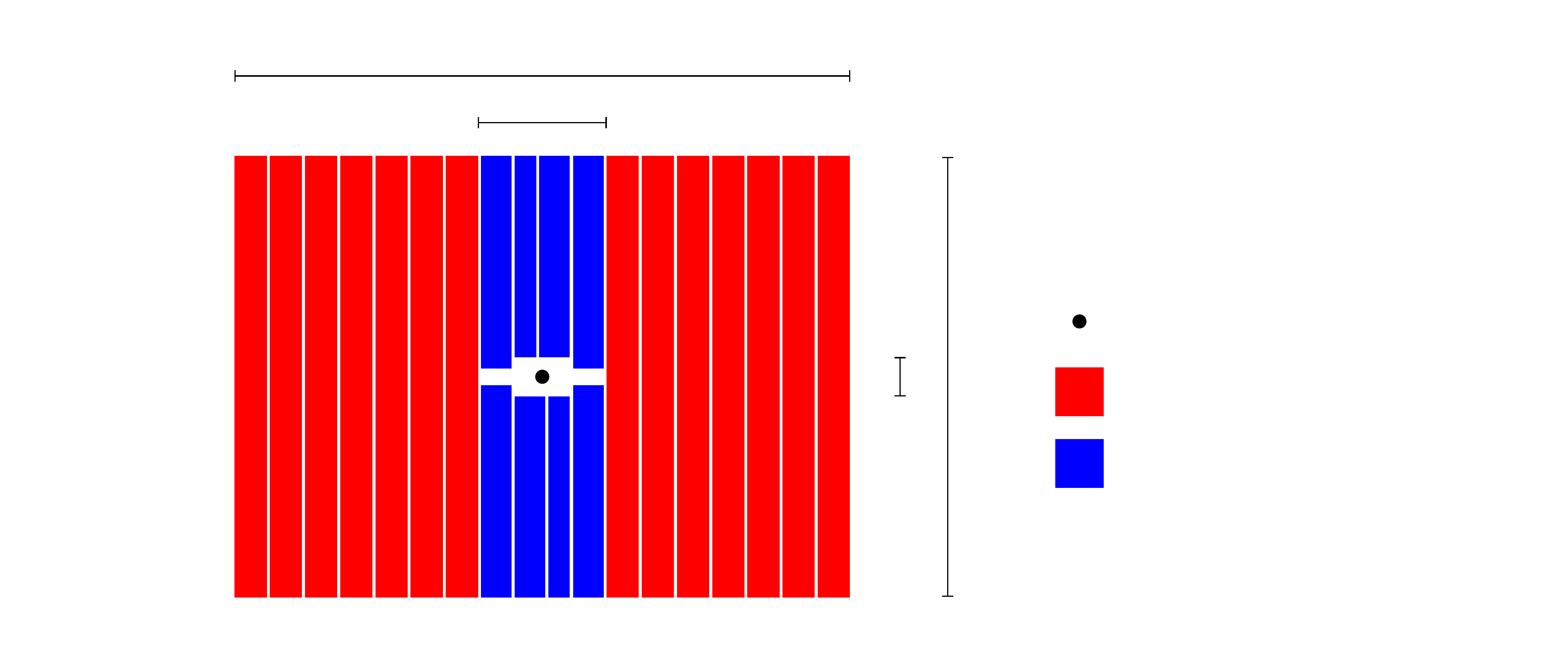}
\end{subfigures}

Every $F$-module contains $256$ straw-tubes, with the tubes divided
into two monolayers in the $xy$-plane, each containing $128$
tubes. The monolayers are divided horizontally in half, with each half
containing $64$ straw-tubes. Each layer half reads out at the
outermost edge of the module. The $S$-modules also consist of two
straw-tube monolayers, but these monolayers are not divided in
half. Each layer contains $128$ tubes, and so every $S$-modules
contains $256$ straw-tubes. Again, each layer reads out to the
outermost edge of the module. The straw-tubes contain a mixture of
$30\%$ argon and $20\%$ carbon dioxide, and with an inner diameter of
$4.9~\mathrm{mm}$ have a drift time of less than
${50~\mathrm{ns}}$. The single hit resolution of the \ott is
${220~\mathrm{\mu}\mathrm{m}}$~\cite{storaci.11.1}. Hits from the \ott
are used in the reconstruction of the tracks used in the analysis of
\chp{Zed}. Further details on the \ott can be found in the technical
design report of \rfr{lhcb.01.2}.

\newsubsection{Particle Identification}{Pid}

The particle identification subdetectors of \lhcb consist of two ring
imaging Cherenkov radiation (\rich{}) detectors, four calorimeters,
and a muon system. The \rich{} detectors use Cherenkov radiation from
charged particles to determine the velocity of particles, and in
conjunction with momentum information from the tracking system, are
used to calculate the mass of charged particles. Information from the
\rich{} detectors is used to differentiate between charged pions and
kaons from $B$-hadron decays. The energy of charged or neutral
particles are determined with the four calorimeters. Here, the
particles interact with scintillating material, producing particle
showers resulting in photons that are measured by photodetectors; the
magnitude and intensity of the photons from the particle shower
correspond to the energy of the particle passing through the
calorimeter. The calorimeters are used to differentiate between
hadrons and leptons, and in combination with the tracking system,
between neutral and charged particles. The muon systems are designed
specifically to identify muons and measure both their momentum and
energy through the use of multi-wire proportional chambers.

\newsubsubsection{Ring Imaging Cherenkov Detectors}{}

Charged particles passing through a medium at a velocity, larger than
the velocity of light propagating through that medium, radiate photons
in a cone about the direction of travel of the particle. The polar
angle $\theta$ between the velocity vector of the particle and the
radiated light is given by,
\begin{equation}
  \theta = \cos^{-1}\left(\frac{1}{n\beta}\right)
\end{equation}
where $n$ is the index of refraction for the medium and ${\beta =
  v/c}$, where $v$ is the velocity of the particle. \rich{} detectors
consist of a radiating medium through which the particle passes, and a
photodetector which measures the rings produced from the radiating
medium. Because \rich{} detectors can only resolve rings with radii
within a given range, the index of refraction for the radiating
material dictates the velocity range of particles observable by the
detector.

The \rich{1}, located approximately ${1.0~\mathrm{m}}$ along the
$z$-axis directly after the \velo, is designed to differentiate
between charged pions and kaons with low momenta within the range $1$
to ${60~\gev}$. Typically, particles produced with low momenta from
$B$-hadron decays provide a broad spread in pseudo-rapidities, and so
the \rich{1} is designed to cover a pseudo-rapidity range of ${2.1
  \leq \eta \leq 4.4}$ in the $yz$-plane and ${1.9 \leq \eta \leq
  4.4}$ in the $xz$-plane for a particle originating from the
interaction point. \rich{1} utilises two radiating materials to
achieve this momentum range, aerogel and $\mathrm{C_4F_{10}}$. The
photodetectors used for measuring the Cherenkov radiation are hybrid
pixel photon detectors which are housed in magnetic shielding,
allowing the photodetectors to operate in magnetic fields with a field
strength below ${50~\mathrm{mT}}$. The photodetectors can detect
photons with wavelengths between $200$ and ${600~\mathrm{nm}}$, and
are made up of $500$ by ${500~\mu\mathrm{m}}$ pixels.

The \rich{2} is located after the \ott at a distance of approximately
${9.5~\mathrm{m}}$ along the $z$-axis and is designed to separate
charged pions and kaons with momenta in the range $15$ to
$100~\gev$. Particles from $B$-hadrons with larger momenta are
expected to also have larger pseudo-rapidities, and so \rich{2} covers
the pseudo-rapidity range ${3.0 \leq \eta \leq 4.9}$ in the $yz$-plane
and ${2.8 \leq \eta \leq 4.9}$ in the $xz$-plane. The \rich{2} uses
the same photodetectors as \rich{1}, but uses the radiating material
$\mathrm{C_4F}$, with an index of refraction approximately $1.7$ times
smaller than $\mathrm{C_4F_{10}}$, to achieve a larger momentum upper limit
than \rich{1}. While the \rich{} detectors are not directly used in
the analysis of \chp{Zed}, they provide an important component in the
detection of $B$-hadron decays. Further details on the \rich{} systems
can be found in the technical design report of \rfr{lhcb.00.1}.

\newsubsubsection{Scintillating Pad Detector and Preshower Calorimeter}{}

The \spd, \prs, and \ecal are aligned along lines of pseudo-rapidity
from the interaction point to provide a one-to-one mapping between
calorimeter cells. The scintillating pad detector is the first
calorimeter along the $z$-axis and is located at
${12.3~\mathrm{m}}$. No absorber material is placed in front of the
\spd and so charged particles shower within it, providing a fast
method to determine the track multiplicity of the event and
differentiate between charged particles and photons. The
pseudo-rapidity coverage of the \spd is designed to match the coverage
of the tracking system and ranges from ${2.1 \leq \eta \leq 4.4}$ in
the $yz$-plane and from ${1.9 \leq \eta \leq 4.4}$ in the $xz$-plane
for particles produced at the interaction point. A schematic for the
layout of a single quadrant of the \spd, \prs, and \ecal is given in
\fig{Exp:Ecal}, with the \ecal dimensions provided. The dimensions for
the \spd are similar but reduced by a factor of approximately
$1.0\%$. The \spd contains $3312$ modules, with $828$ modules per
quadrant. The modules all have the same dimensions and are divided
into outer-modules, middle-modules, and inner-modules.

\begin{subfigures}[t]{1}{Schematic of the lower left quadrant of the
    \spd, \prs, and \ecal in the $xy$-plane, along the direction of
    the $z$-axis. The quadrant consists of $672$ outer-modules with
    $1$ square cell each, $112$ middle-modules with $4$ square cells
    each, and $44$ inner modules with $9$ square cells each. All
    modules have the same dimensions, with the dimensions given here
    for the \ecal.\labelfig{Ecal}}
  \includesvg[width=\columnwidth]{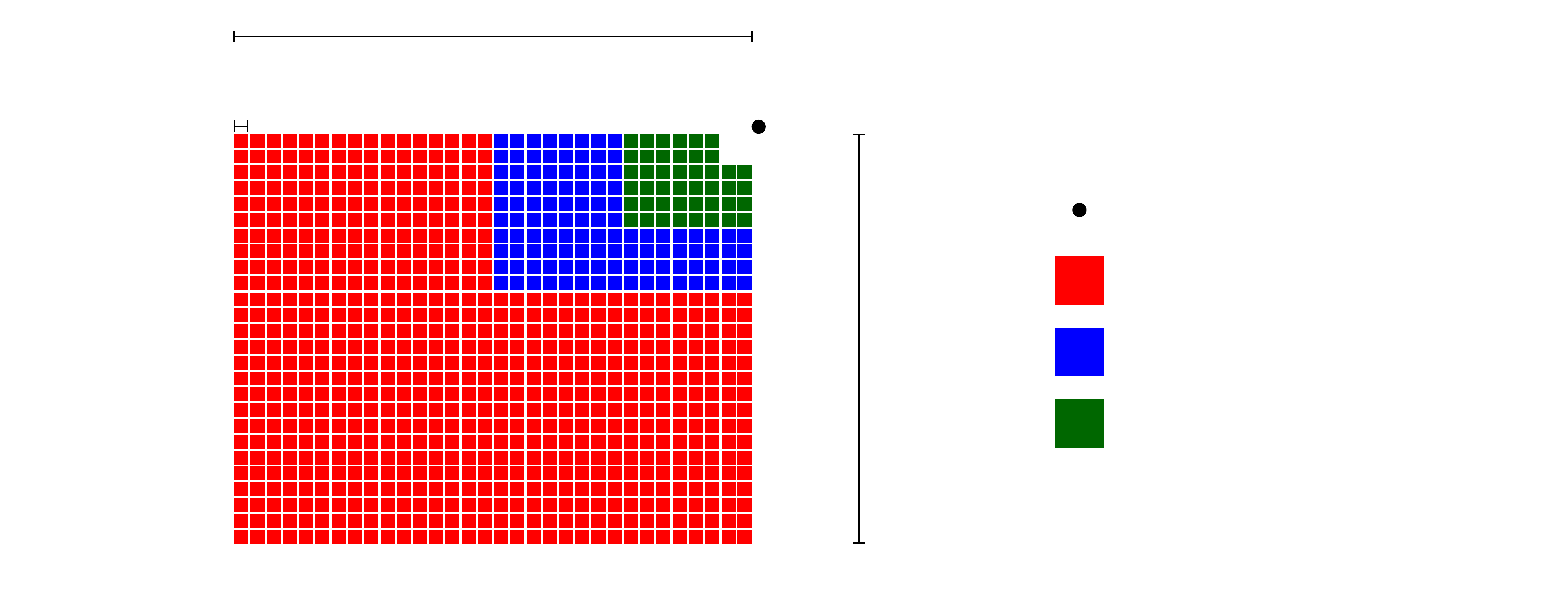}
\end{subfigures}

Each outer-module consists of a single square scintillating tile made
from polystyrene, while every middle-module houses $4$ square tiles,
and every inner-module contains $9$ square tiles. The outer-module
tiles are approximately ${12~\mathrm{cm}}$ to a side in the
$xy$-plane, the middle-module tiles ${6~\mathrm{cm}}$, and the
inner-module tiles ${4~\mathrm{cm}}$. Each tile, which is
${1.5~\mathrm{cm}}$ deep and $0.28$ radiation lengths, is connected
via wavelength-shifting fibres to a multi-anode photomultiplier to
create an \spd calorimeter cell. The segmentation of the \spd in the
$xy$-plane ensures a roughly similar particle occupancy between all
calorimeter cells. The \spd is not used in the reconstruction of
electrons or hadrons in the analysis of \chp{Zed}, but is used to
perform a fast global event cut based on the \spd multiplicity of the
event which must be accounted for in \sec{Zed:RecEff}.

Behind each \spd calorimeter cell, along the $z$-axis, is placed a
${1.5~\mathrm{cm}}$ deep lead absorber of $2.5$ radiation lengths,
followed by another ${1.5~\mathrm{cm}}$ deep scintillating tile of
$0.28$ radiation lengths. These tiles are not part of the \spd, but
rather the \prs, and are slightly smaller than the \ecal cells by
$0.5\%$ in order to maintain the one-to-one correspondence between
\spd, \prs, and \ecal cells along lines of pseudo-rapidity. Every \prs
tile, like \spd tiles, is connected via wavelength-shifting fibres to
a multi-anode photomultiplier tube to create a \prs calorimeter
cell. The cells for the pre-shower calorimeter provide a segmentation
in the $yz$-plane for the \ecal, and have the same layout and
pseudo-rapidity coverage as both the \spd and \ecal. The \prs is used
to identify neutral and charged particles which interact through the
electromagnetic force. In the analysis of \chp{Zed}, the \prs is used
to differentiate between the electrons and charged hadrons of
\sec{Zed:Rec}.

\newsubsubsection{Electromagnetic Calorimeter}{}

The electromagnetic calorimeter is used to identify photons and
electrons and measure their energy. It is located ${12.5~\mathrm{m}}$
along the $z$-axis, just after the \spd and \prs calorimeters. The
\ecal has the same pseudo-rapidity coverage as the \spd and \prs, and
is given by the schematic of \fig{Exp:Ecal}, with the same module
configuration and cell layout in the $xy$-plane as the \spd and
\prs. Each module, however, contains a stack of $132$ layers,
alternating between lead absorbers and calorimeter tiles along the
direction of the $z$-axis; this corresponds to $66$ absorbers and $66$
scintillating tiles. Each absorber is ${0.2~\mathrm{cm}}$ deep and
$0.33$ radiation lengths, and each scintillator tile is
${0.4~\mathrm{cm}}$ deep and $0.08$ radiation lengths. The dimensions
for each absorber and tile from a stack in the $xy$-plane are
approximately the same as those for the \spd and \prs. The photons
from all $66$ tiles in a stack are collected via wavelength-shifting
fibres into a phototube to produce a single \ecal cell. The entire
\ecal covers $25$ radiation lengths along the direction of the
$z$-axis.

The energy resolution for a calorimeter can be written as,
\begin{equation}
  \delta_E = \delta_\mathrm{clb}E \oplus \delta_\mathrm{smp}\sqrt{E}
  \oplus \delta_\mathrm{noi}
  \labelequ{Calorimeter}
\end{equation}
where $E$ is the measured energy of the particle,
$\delta_\mathrm{clb}$ is the calibration uncertainty,
$\delta_\mathrm{smp}$ is the uncertainty due to sampling fluctuations,
and $\delta_\mathrm{noi}$ is the noise uncertainty. For high energy
particles, oftentimes only the first two uncertainties are considered
as the noise uncertainty is negligible. In \rfr{lhcb.07.2} these
uncertainties were measured to be ${\delta_\mathrm{clb} =
  8.3\times10^{-2}}$ and ${\delta_\mathrm{smp} =
  9.5\times10^{-1}~\gev^{1/2}}$ for the \lhcb \ecal. The energy of
particles as measured by the \ecal is used to separate electrons from
hadrons in the analysis of \chp{Zed}.

\newsubsubsection{Hadronic Calorimeter}{}

The hadronic calorimeter is directly after the \ecal,
${13.3~\mathrm{m}}$ along the $z$-axis, and is used to identify
hadrons and measure their energy. The pseudo-rapidity coverage of the
\hcal is ${1.8 \leq \eta \leq 4.2}$ in the $yz$-plane and ${2.1 \leq
  \eta \leq 4.2}$ in the $xz$-plane for a particle originating from
the interaction point. A schematic of the calorimeter cell layout for
the \hcal is given in \fig{Exp:Hcal} for the lower left quadrant in
the $xy$-plane along the direction of the $z$-axis. Each \hcal
quadrant contains $152$ square outer-cells with sides of
${26.26~\mathrm{cm}}$ in the $xy$-plane, and $215$ inner-cells with
sides of ${13.13~\mathrm{cm}}$. The segmentation of the \hcal cells in
the $xy$-plane is larger than the \ecal cells due to the larger size
of hadronic showers.

\begin{subfigures}[t]{1}{Schematic of the lower left quadrant of the
    \hcal in the $xy$-plane, along the direction of the $z$-axis. The
    quadrant consists of $152$ outer-cells and $215$
    inner-cells.\labelfig{Hcal}}
  \includesvg[width=\columnwidth]{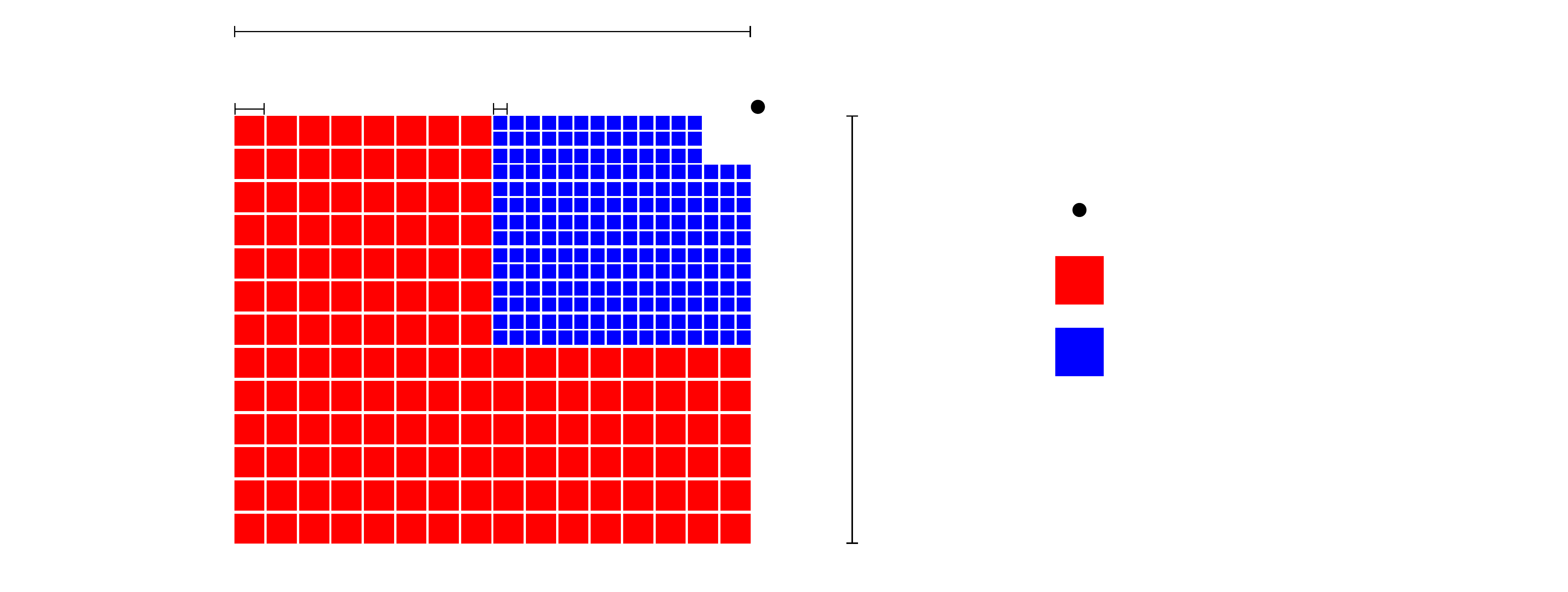}
\end{subfigures}

Each outer-cell is made from a stack of $26$ layers along the
direction of the $x$-axis, opposed to the \ecal cells which are
stacked along the $z$-axis. The width of each layer is
${1.01~\mathrm{cm}}$, with a height of ${26.26~\mathrm{cm}}$ and a
depth of ${128.3~\mathrm{cm}}$. Every layer contains two sublayers, a
${0.6~\mathrm{cm}}$ wide structural plate, followed by a sublayer of
$3$ polystyrene scintillator tiles alternating with iron absorbers
along the direction of the $z$-axis. The scintillator tiles are
${0.3~\mathrm{cm}}$ wide and ${19.7~\mathrm{cm}}$ deep, while the iron
absorbers are ${0.4~\mathrm{cm}}$ wide. Each scintillator tile is
separated by ${20.2~\mathrm{cm}}$ of iron absorber along the direction
of the $z$-axis, which is $1.0$ interaction length of material. The
scintillator-absorber sublayer layout is staggered between the layers
so that the scintillator of one layer is followed by an absorber in
the next layer of the stack. The scintillator tiles are connected to
photomultiplier tubes via wavelength-shifting fibres, with one read
out channel per cell.

The inner-cells have the same stack structure as the outer-cells, but
each stack contains $13$ layers. Additionally, each scintillator tile
is split in two with the same width and depth but a height of
${13.13~\mathrm{cm}}$. Consequently, each inner-cell stack contains
two inner-cells along the direction of the $y$-axis. From this
configuration, each outer-cell houses $78$ scintillator tiles and each
inner-cell contains $39$ scintillator tiles. The entire \hcal,
including structure and absorbers, is $5.6$ interaction lengths along
the $z$-axis. The energy resolution of the \hcal, using the same
parametrisation of \equ{Exp:Calorimeter}, is ${\delta_\mathrm{clb} =
  0.9}$ and ${\delta_\mathrm{smp} = 6.9~\gev^{1/2}}$ from
\rfr{lhcb.08.1}. The energy of particles measured by the \hcal is used
to separate hadrons from electrons in the analysis of
\chp{Zed}. Further information on all of the calorimeter systems can
be found in the technical design report of \rfr{lhcb.00.2}.

\newsubsubsection{Muon System}{}

Because muons are two orders of magnitude more massive than electrons,
the energy lost through bremsstrahlung radiation for muons is ten
orders of magnitude smaller than electrons, and so muons do not shower
in the \ecal. Additionally, muons interact minimally with the \hcal,
as they are leptons and do not interact through the strong
force. Consequently, the majority of the muon system is placed after
the \ecal and \hcal. The muon system is divided into five stations,
M$1$ through M$5$, located at distances of ${12.1~\mathrm{m}}$,
${15.3~\mathrm{m}}$, ${16.5~\mathrm{m}}$, ${17.7~\mathrm{m}}$, and
${18.9~\mathrm{m}}$ along the $z$-axis. The five stations cover a
pseudo-rapidity range of ${2.0 \leq \eta \leq 4.8}$ in the $yz$-plane
and ${1.9 \leq \eta \leq 4.6}$ in the $xz$-plane for particles
produced from the interaction point. The first three muon stations are
built for high resolution transverse momentum measurements, while the
final two stations are designed for primarily particle identification.
To ensure only muons reach the final two stations, three iron
absorbers, each with a depth of ${80~\mathrm{cm}}$, are placed between
the final four muon stations. To reach the final station, a muon will
have traversed over $20$ interaction lengths of material, including
the calorimeters.

\begin{subfigures}[t]{1}{Schematic of the lower left quadrant of a
    muon station in the $xy$-plane, along the direction of the
    $z$-axis. The width and height of the quadrant for each muon
    station is given. Adapted from \rfr{lhcb.08.1}.\labelfig{Muon}}
  \includesvg[width=\columnwidth]{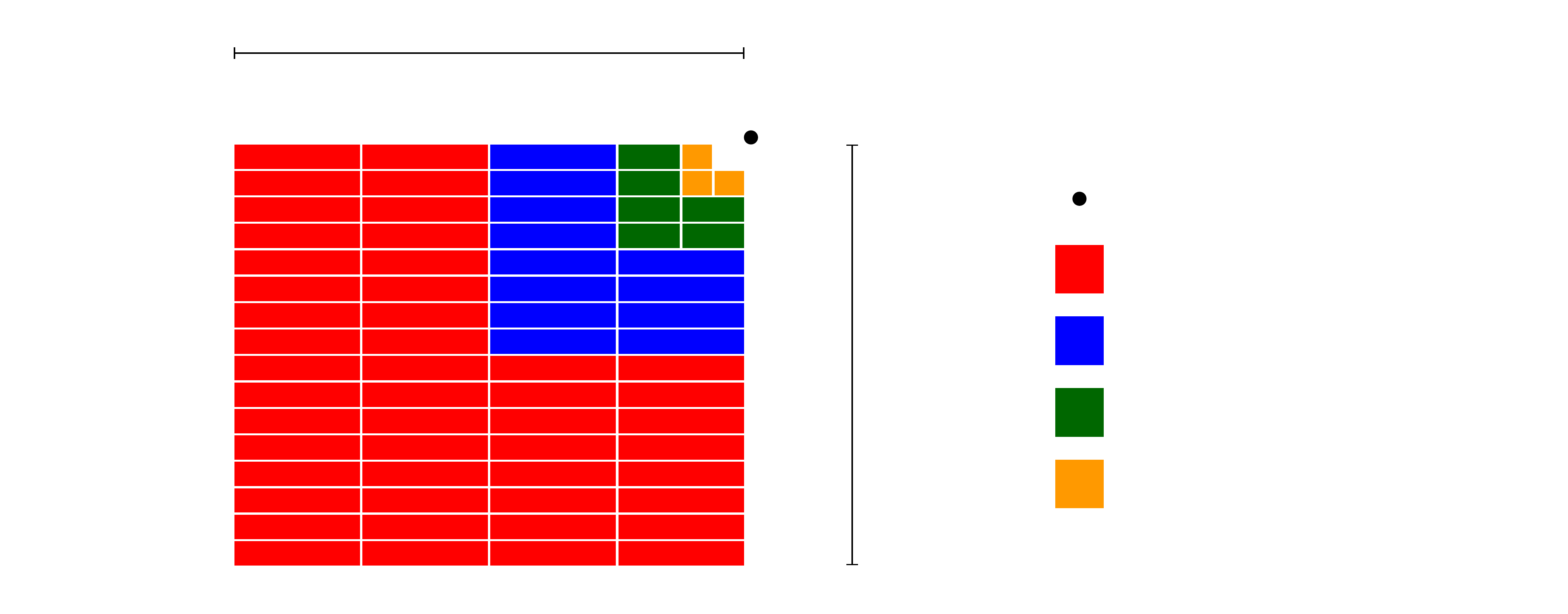}
\end{subfigures}

The five muon stations are segmented into chambers, with the chamber
layout and size designed to provide a one-to-one mapping between the
chambers of each station along lines of pseudo-rapidity. A schematic
of the chamber layout for a quadrant of a muon station in the
$xy$-plane is given in \fig{Exp:Muon}. The quadrant is divided into
four regions, R$1$ through R$4$, where the chamber resolution is lower
for chambers in regions farther from the $z$-axis. Each chamber is a
multi-wire proportional chamber except for the R$1$-chambers of the
first muon station, which are triple-gas electron multiplier
chambers. The chambers are divided into rectangular pads, where each
pad has a read out. For all muon stations the R$1$-chambers are
segmented into $8$ pads along the $y$-axis, while the R$2$-chambers
are segmented into $4$ pads, the R$3$-chambers into $2$ pads, and the
R$1$-chambers into only a single pad. Along the $x$-axis, the R$1$
through R$3$-chambers of the M$1$ station are divided into $24$ pads,
while the R$4$-chambers are divided into $12$ pads. For M$1$ and M$2$,
the number of pads along the $x$-axis for each chamber type are
doubled with respect to the M$1$ segmentation. For M$4$ and M$5$, the
pad segmentation of the chambers along the $x$-axis is halved.

The transverse momentum resolution, $\delta_\pt/\pt$, for muon tracks
using only hits from the muon stations was measured to be
approximately $20\%$ in \rfr{lhcb.10.2}. The use of the muon system in
the analysis of \chp{Zed} is critical. Every decay category of ${\z
  \to \ditau}$ events analysed, except one, requires at least a single
muon, which is identified using M$2$ through M$5$. Additionally, the
track finding efficiency for muons passing through the \velo, \itt,
and \ott is determined using combined hits from the \ttt and muon
system in \sec{Zed:RecEff}. Further information on the \lhcb muon
system can be found in the technical design report of \rfr{lhcb.01.3}.

\newsection{LHCb Event Reconstruction}{Rec}

To interpret the raw output from the \lhcb detector described in
\sec{Exp:Det}, the signals from the subdetectors are combined using
reconstruction to produce high level physics objects that can be used
in analyses. Hits from the tracking system are combined to produce
tracks, particle trajectories of charged particles, while cells from
the calorimeters are clustered to find deposits of energy from either
charged or neutral particles. Matching algorithms between tracks and
calorimeter clusters, as well as \rich{} information, is then run to
identify particles.

Reconstruction of \lhcb data is performed twice, once at the online
level of the trigger system and once at the offline level. The online
triggers perform a fast reconstruction that is not complete, to
quickly filter events from the detector and write them to disk for
later offline reconstruction. Offline reconstruction uses the
information from the entire event to create a more precise
detector-wide reconstruction of the event. The offline \lhcb
reconstruction software is \brunel~\cite{brunel.13.1} which produces
data files that can be processed by the \davinci~\cite{davinci.13.1}
physics analysis software.

Full simulations of events in the detector are performed with the
\gauss~\cite{gauss.13.1} simulation software which utilises
\pythia{6}~\cite{\citepythiasix} as a general purpose Monte Carlo
event generator, \evtgen~\cite{evtgen.13.1} as a $B$-hadron particle
decayer, and \geant as a detector simulator. The simulated events
produced from \gauss are digitised by the \boole~\cite{boole.13.1}
software, with trigger emulation applied by the
\moore~\cite{moore.13.1} software.

All \lhcb software is based on the \gaudi~\cite{gaudi.13.1}
framework. Large distributed processing jobs for any of the \lhcb
software is typically submitted through the \ganga~\cite{ganga.13.1}
software and can be distributed to the CERN and \lhcb computing grids
via the \dirac~\cite{dirac.13.1} software. Further information on
\lhcb computing and software can be found in the technical design
report of \rfr{lhcb.05.1}.

The track reconstruction implemented in \brunel is described in
\sec{Exp:RecTrk}, and the use of calorimeter information in the
\brunel reconstruction process is outlined in \sec{Exp:RecCal}. The
trigger system, which uses similar algorithms to the tracking
algorithms implemented in \brunel, is introduced in
\sec{Exp:RecTrg}. Finally, the methods used for determining the
integrated luminosity for a reconstructed data sample are described in
\sec{Exp:RecLum}.

\newsubsection{Tracking Information}{RecTrk}

Information from the tracking system of \sec{Exp:Trk} provides the
foundation for \lhcb event reconstruction. Hits from the \velo, \ttt,
\itt, and \ott are combined into tracks which represent possible
trajectories of charged particles passing through the detector. These
trajectories, if passing through the dipole magnetic field of
\fig{Exp:Magnet}, can be used to determine the momentum of the
particle. Tracks from a common origin can be used to produce vertices,
{\it i.e.} locations in the detector where two or more particles were
produced. In \lhcb reconstruction, five types of tracks are
considered~\cite{lhcb.07.3}.
\begin{itemize}
  \settowidth{\itemindent}{downstream}
\item[{\makebox[\itemindent][l]{\velo:}}] Contain hits only in the
  \velo and consequently have no associated momenta. Because the \velo
  surrounds the interaction point, these tracks can have either a
  forward or backward direction along the $z$-axis. These tracks are
  particularly useful for vertex reconstruction.
\item[{\makebox[\itemindent][l]{T (\itt/\ott):}}] Have hits from either
  the \itt or \ott, or both, but do not have any associated hits
  from the \velo or \ttt. T-tracks are combined with information from
  the \rich{2} and used for the reconstruction of
  charged pions and kaons.
\item[{\makebox[\itemindent][l]{upstream:}}] Are built from both \velo
  and \ttt hits, but do not contain \velo, \itt, or \ott
  hits. Oftentimes, these are low momentum tracks which are deflected
  from within the detector by the magnetic field of the dipole. If the
  tracks are within \rich{1} acceptance the mass of the particle
  producing the track can typically be determined.
\item[{\makebox[\itemindent][l]{downstream:}}] Consist of hits from
  the \ttt and the \itt, \ott, or both. These tracks can be produced
  from long lived particles that may not decay within the \velo, such
  as the $K_S$.
\item[{\makebox[\itemindent][l]{long:}}] Contain hits from the \velo
  and the \itt, \ott, or both. These tracks may also contain hits from
  the \ttt, but are not required to contain \ttt hits. This track type
  is the most commonly used type in \lhcb analyses.
\end{itemize}
A graphical summary of the \lhcb track types is given in
\fig{Exp:Tracks}. The tracks used in the analysis of \chp{Zed} are
long-tracks, and the efficiency of long-track reconstruction for muons
is measured in \sec{Zed:RecEff}.

\begin{subfigures}{2}{Schematic of the \lhcb tracking system in the
    bending $xz$-plane depicting the different types of reconstructed
    \lhcb tracks.}
  \svgbeg
  \svg[1]{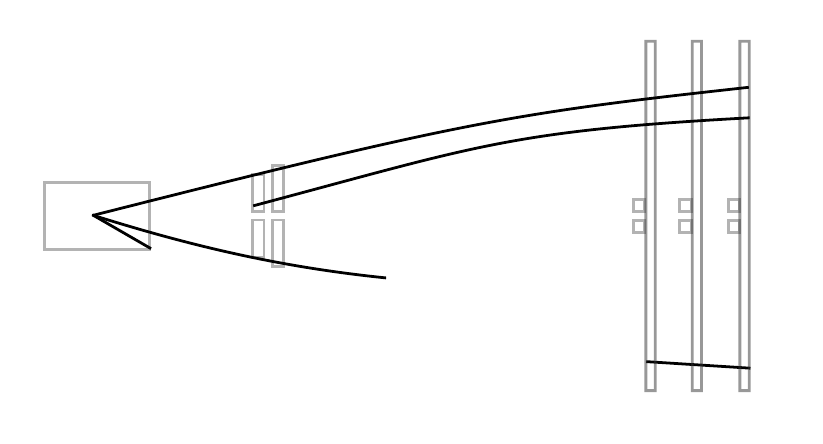} & \sidecaption \svgend
\end{subfigures}

Tracks are reconstructed by first track-seeding in the individual
tracking system detectors, then track-finding where a search is made
outside the seeding detector, and finally track-fitting where the
trajectories from track-finding are refined. After this, reconstructed
long-tracks are extrapolated to the muon system, and if compatible
hits are found, a combined muon track is formed. The tracks produced
from the reconstruction process are then used to seed possible
vertices, which are then further refined through vertex
fitting. Further information on \lhcb track reconstruction can be
found in \rfr{lhcb.04.1, *lhcb.06.1, *lhcb.07.4, *lhcb.07.5,
  *lhcb.07.6, *lhcb.07.7, *lhcb.07.8, *lhcb.07.9, *lhcb.07.10}.

\newsubsection{Calorimeter Information}{RecCal}

After the reconstruction of tracks from \sec{Exp:RecTrk} is performed,
calorimeter and \rich{} information is used to identify the type of
particle from which the track was produced. In \chp{Zed} only muon
system and calorimeter information is used to identify muons,
electrons, and charged hadrons, with no \rich{} information
used. Consequently, \rich{} reconstruction is not introduced in this
thesis, but an overview can be found in \rfr{lhcb.08.2}. \ecal
information is used to create neutral pion and photon candidates, as
well as identify tracks from electrons and correct their
energy. Additionally, the energies from \prs, \ecal, and \hcal cells
associated with tracks extrapolated to the calorimeters are used to
differentiate between muons, electrons, and charged hadrons.

Electrons within \lhcb interact with the detector material, losing
energy through the emission of bremsstrahlung photons, and so the
recovery of these photons is important in reconstructing the full
momentum of electrons. Within \lhcb reconstruction, all long-tracks
are considered as electron candidates and bremsstrahlung recovery is
performed by searching for compatible photon candidates and adding
their momentum to the track momentum. Photon candidates are
reconstructed by grouping \ecal cells into \ecal clusters, matching
\ecal clusters with tracks, and generating candidates from clusters
without matching tracks. Details on the \lhcb calorimeter
reconstruction can be found in \rfrs{lhcb.01.4}, \cite{lhcb.03.3}, and
\cite{lhcb.03.4}.

The calorimeter energy associated with a long-track is found by
extrapolating the track from its last state vector to the start of the
calorimeter. The track is then linearly extrapolated from the start of
the calorimeter through the depth of the calorimeter and points are
sampled along the line of the trajectory. A list of the calorimeter
cells which contain the points is made, with any duplicate cells
removed. The energies from the corresponding digits of the cells from
the list are then summed to produce the associated calorimeter energy
for the track.

\begin{subfigures}{2}{\subfig{Z2EE.Et}~Distribution of transverse
    \ecal energy associated with electrons tracks taken from ${\z \to
      \die}$ data events, illustrating \ecal cell
    saturation. \subfig{Z2EE.Mass}~Invariant mass distributions for
    ${\z \to \dimu}$ and ${\z \to \die}$ events from data,
    demonstrating incomplete bremsstrahlung recovery.}
  \svgbeg
  \svg{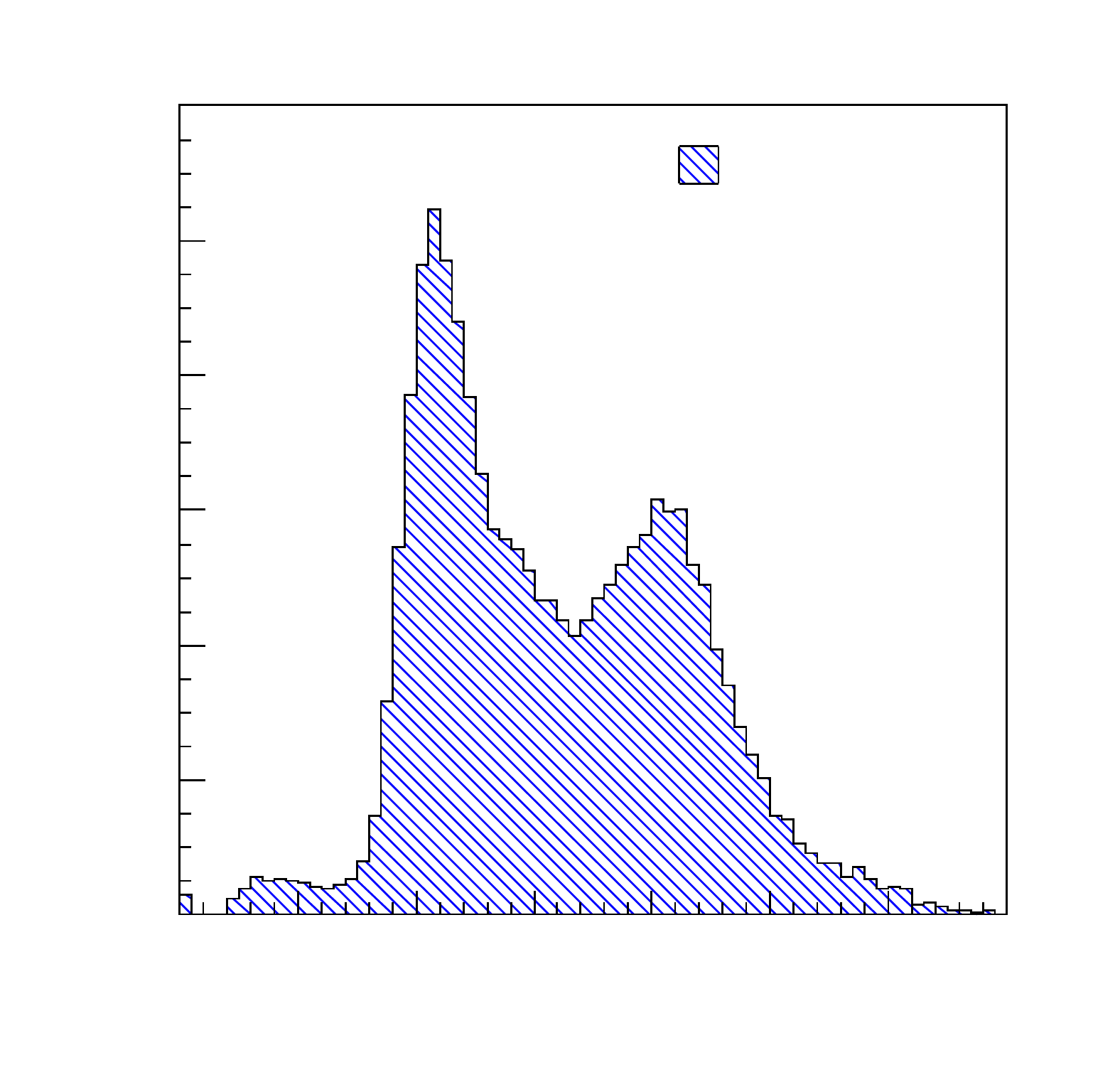} & \svg{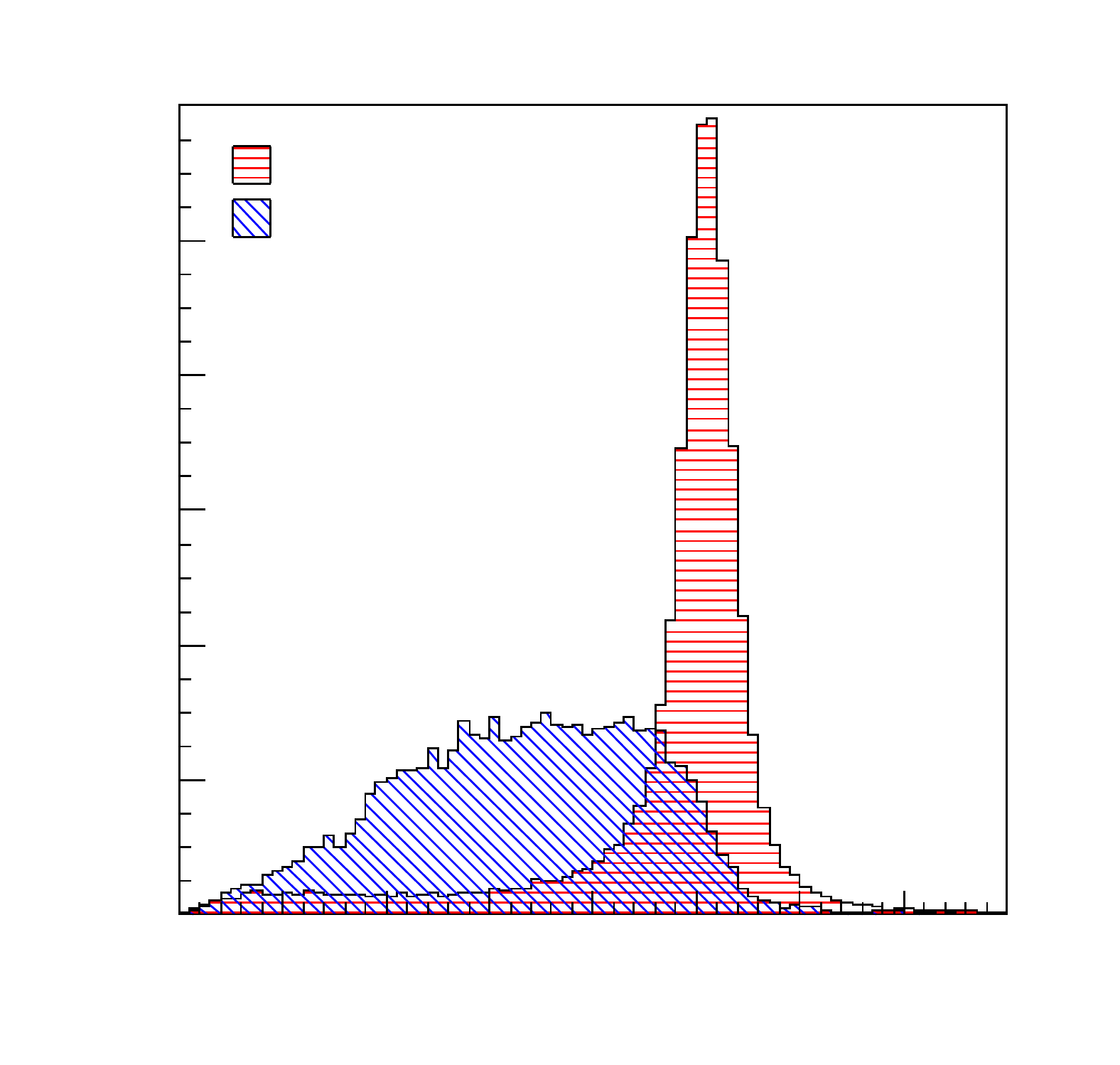} \svgend
\end{subfigures}

The associated \prs, \ecal, and \hcal energies of tracks are used to
identify high \pt electrons and charged hadrons in \chp{Zed}. However,
the \lhcb calorimeter systems were designed for lower \pt $B$-hadron
decay products, and the individual \ecal cells saturate at transverse
energies, \et, greater than $10~\gev$. Consequently, many of the
energies from the \ecal cells associated with an electron track are
fully saturated, and so the \ecal \et for an electron is degraded. In
\fig{Exp:Z2EE.Et} the distribution for the \et of electrons produced
from ${\z \to \die}$ events in data is plotted, where clear peaks can
be seen at multiples of $10~\gev$, the \ecal saturation
\et. Additionally, the energies for recovered bremsstrahlung photons
are also degraded, and so the bremsstrahlung correction of the
momentum for electron tracks is incomplete, resulting in a reduced
momentum resolution. This can be seen in the broadened and shifted
${\z \to \die}$ invariant mass distribution taken from data with
respect to the ${\z \to \dimu}$ distribution plotted in
\fig{Exp:Z2EE.Mass}.

\newsubsection{Triggers}{RecTrg}

The rate of detectable collisions produced within the \lhcb detector
is approximately ${10~\mathrm{MHz}}$ and must be reduced to a rate of
${2~\mathrm{kHz}}$ for storage. Most of the events produced within the
detector are not of interest for \lhcb physics analyses, and
consequently can be discarded. The \lhcb trigger system provides a
fast decision on whether an event should be kept or discarded and is
used to reduce the rate of events from the detector to a rate that can
be written to disk. The \lhcb trigger system consists of both hardware
and software components and is divided into a level $0$ trigger (L$0$)
and a high level trigger (HLT). A full description of the \lhcb
trigger system can be found in \rfr{lhcb.08.1} and the technical
design report of \rfr{lhcb.03.5}.

\newsubsubsection{Level 0 Trigger}{}

The level $0$ trigger is a hardware trigger which must process each
event within $2~\mu\mathrm{s}$ and reduce the event rate from
${10~\mathrm{MHz}}$ to ${1~\mathrm{MHz}}$ using partial detector
information. At this lower event rate the entire detector can be read
out for each event and passed on to the HLT for further
refinement. The L$0$ is divided into a calorimeter trigger and a muon
trigger. With these two components, the L$0$ attempts to reconstruct
the highest \et photon, electron, neutral pion, hadron, and the two
highest \pt muons in the event. The information from each component is
passed to a decision unit which then produces a combined decision on
whether the event should be passed to the HLT.

The calorimeter trigger attempts to reconstruct the highest transverse
energy photon, electron, and hadron in the event using $19420$ read
out channels from the \spd, \prs, \ecal, and \hcal. Portions of the
\ecal and \hcal are read out, and the sum of the transverse energy for
every two-by-two group of cells is calculated. The group of cells with
the highest \et in the \hcal is passed on as a hadron candidate. The
two-by-two groups of \ecal cells are merged with the \spd and \prs
cells to produce photon, electron, and neutral pion candidates. The
candidate with the largest \et of each type is passed on to the
decision unit along with the total \hcal \et and \spd multiplicity.

The muon trigger uses $25920$ read out channels and searches for muon
tracks in the event. The tracks are seeded from hits in the third muon
station, M$3$, which are linearly extrapolated to the interaction
point. A search for hits in a field of interest along the $x$-axis is
made in M$2$ and along both the $x$ and $y$-axis in M$4$ and M$5$. If
hits are found in all three of these stations the line from the M$3$
hit to the M$2$ hit is extrapolated to M$1$ and the closest hit within
a field of interest is added to the track. The momentum of the track
is estimated and the two highest \pt tracks are passed on to the
decision unit.

The decision unit combines the information from the two L$0$ triggers,
and either accepts the event and passes on the information to the HLT,
or discards the event. The L$0$ will accept an event if a hadron
candidate with ${\et >5~\gev}$ is found. The L$0$ will also accept an
event if an electron, photon, or neutral pion candidate with ${\et >
  2.5~\gev}$ is found or if ${\pt > 1.2~\gev}$ for any muon
track. Additionally, an event will be accepted if ${\pt_1 + \pt_2 >
  1~\gev}$ for the two muon tracks. All calorimeter candidates and
tracks that pass the requirements above are sent to the \hlt as L$0$
objects.

\newsubsubsection{High Level Trigger}{}

The high level trigger is a software trigger which runs on an event
filter farm consisting of $2000$ computing nodes and is divided into
an HLT$1$ and HLT$2$ level. The HLT$1$ must reduce the
${1~\mathrm{MHz}}$ event rate from the L$0$ to the ${30~\mathrm{kHz}}$
event rate required by HLT$2$, and begins by confirming the L$0$
objects. L$0$ calorimeter objects are confirmed by accessing tracking
system information and attempting to reconstruct a track associated
with the object; the L$0$ decision is confirmed if no track is found
for neutral objects and if a track is found for charged objects.

The HLT$2$ takes the ${30~\mathrm{kHz}}$ event rate from HLT$1$, and
reduces this to a rate of ${2~\mathrm{kHz}}$ where the raw data from
the event can be written to disk for later full reconstruction and
analysis. The HLT$2$ performs full track fitting. Because the HLT$1$
and HLT$2$ are both software based triggers, the algorithms and
selections applied can be modified to accommodate improved techniques
or changing physics interests. To ensure consistency between the
algorithms and requirements used between data taking periods, a
trigger configuration key (TCK) is assigned to each unique trigger
setup which can be later accessed during analysis.

\newsubsection{Luminosity Determination}{RecLum}

Many of the physics analyses performed using \lhcb data require a
precise measurement of the integrated beam luminosity for the dataset
being analysed. The luminosity can be determined with,
\begin{equation}
  \frac{\dif{\lum}}{\dif{t}} = \frac{\mu N_b f}{\sigma_\inelastic}
\end{equation}
where $\mu$ is the average number of visible proton-proton
interactions per bunch crossing, $\sigma_\inelastic$ is the
proton-proton inelastic cross-section, $N_b$ is the number of proton
bunches which is well known, and $f$ is the revolution frequency which
is also well known. Only the values of $\mu$ and $\sigma_\inelastic$
are unknown; $\mu$ must be measured per bunch crossing while
$\sigma_\inelastic$ remains constant and need only be measured once.

The value of $\mu$ can be measured per bunch crossing by recording
observables, or luminosity counters, that are proportional to
$\mu$. For \lhcb luminosity determination, five luminosity counters
are recorded: the number of \velo vertices, the number of \velo
$rz$-tracks, the number of hits in the \velo pile-up sensors, the
number of \spd hits, and the \et deposition within the
calorimeters. Of these counters, the number of \velo $rz$-tracks is
found to be the most reliable measure of $\mu$. The inelastic
cross-section is measured using two different methods, a Van de Meer
scan and a beam-gas imaging technique. Both are described in the
remainder of this section. The combination of the uncertainties from
the measurements on $\mu$ and $\sigma_\inelastic$ result in a $3.5\%$
uncertainty on the integrated luminosity measurement used for any
\lhcb analysis with $2011$ data. Full details on the \lhcb luminosity
determination can be found in \rfr{lhcb.11.1}.

\newsubsubsection{Van de Meer Scan}{}

The Van de Meer scan was first proposed by Van de Meer in
\rfr{meer.68.1}, and is performed by scanning the two colliding beams
across each other in the transverse plane. The visible inelastic
cross-section is given by,
\begin{equation}
  \sigma_\inelastic = \frac{\int \mu(\Delta_x,{\Delta_y}_0)\sdif{\Delta_x}
    \int \mu({\Delta_x}_0, \Delta_y) \sdif{\Delta_y}}
  {{N_p}_1{N_p}_2\mu({\Delta_x}_0,{\Delta_y}_0)\cos\alpha_1}
  \labelequ{Vdm}
\end{equation}
where ${N_p}_1$ is the number of protons per bunch in the first beam,
${N_p}_2$ is the number of protons per bunch in the second beam, and
$\alpha_1$ is the polar angle of the first beam velocity with respect
to the $z$-axis~\cite{balagura.11.1}. Here $\Delta_x$ is the
$x$-coordinate offset between the two beams with a nominal offset of
${\Delta_{x}}_0$, and $\Delta_y$ is the $y$-coordinate offset with a
nominal value of ${\Delta_y}_0$.

The beams are scanned over a series of approximately $15$ to $30$
$x$-offset steps with a constant $y$-offset of ${\Delta_y}_0$, and the
average $\mu$ per step is measured. These measurements are used to
calculate the first integral of \equ{Exp:Vdm}. The same process is
repeated, but now with $y$-offsets and a constant $x$-offset of
${\Delta_x}_0$. These measurements of $\mu$ are then used to calculate
the second integral. The systematic uncertainty on $\sigma_\inelastic$
is determined from the uncertainty from the $\mu$ measurements of the
scans, the offsets of the scans, and the uncertainty on the number of
protons per bunch for both beams. The primary source of uncertainty is
from the number of protons per bunch, and is on the order of $2.7\%$.

\newsubsubsection{Beam-Gas Imaging}{}

The beam-gas imaging method, first proposed in \rfr{luzzi.05.1}, uses
the vertices reconstructed within the \velo from the interactions of
the beams with residual gas in the beam pipes to measure the profile
of the beams and determine $\sigma_\inelastic$. The distribution of
vertices provide a transverse image of the beams from which their
angles, profiles, and positions can be extracted. While the rate of
interactions is much smaller than for a Van de Meer scan, the beams do
not need to be moved, and so many uncertainties of the Van de Meer
scan method are mitigated. However, the beam-gas imaging method
requires a vertex resolution smaller than the transverse beam width
and a well understood uncertainty, as this contributes to the overall
systematic uncertainty.

Beam-gas interaction measurements are made both when the bunches
collide and do not collide. Only the measurements from colliding
bunches can be used to make luminosity measurements, but the
measurements from non-colliding bunches can be used to further
understand the beams and perform cross-checks with the measurements
from the colliding bunches. The transverse beams widths are determined
with beam-gas interactions from colliding bunches, but must be
measured away from the interaction point so beam-gas interactions can
be separated from proton-proton interactions. The same beam width
measurement can be made from non-colliding bunches, but the
measurement can also be made at the interaction point. These two
widths from non-colliding bunches are compared to ensure the offset
width does not differ within uncertainty from the interaction point
width. The beam crossing angles are also measured using beam-gas
interactions from non-colliding beams.

The systematic uncertainty on the luminosity determination using the
beam-gas imaging method depends upon the vertex resolution, time
stability, beam sizes and offsets, gas pressure gradient, and
crossing-angle effects as well as the number of protons per bunch. The
luminosity measurements made with the beam-gas imaging method for
$2011$ data are consistent with the Van de Meer measurements, and also
have a systematic uncertainty dominated by the number of protons per
bunch.

\newchapter{\textit{Z} Boson Cross-Section}{Zed}

Within this chapter, a measurement of the cross-section for \wzbs
decaying into a \wtl pair using data from \lhcb with ${\sqrt{s} =
  7~\tev}$ is presented. The analysis is described in \sec{Zed:Ana},
while the cross-section measurement is performed in \sec{Zed:Sig} and
the results are presented in \sec{Zed:Res}. In this introduction, the
${\dip \to \z \to \ditau}$ cross-section measurement is motivated by
describing how it can be used to test lepton universality, search for
new physics, and further constrain the proton \PDF.

In the standard model of particle physics (\sm), the \wzb couples to
fermions with a vertex given by \fig{Thr:Z2FF} which is a factor of
${\frac{-ig_w}{2\costw}\gamma^\mu(v_f - a_f\gamma^5)}$ where the
vector and axial couplings, $v_f$ and $a_f$, are for neutral leptons,
charged leptons, $u$-type quarks, or $d$-type quarks, and are given in
\tab{Thr:VaCouplings}. The charged leptons, $e$, $\mu$, and $\tau$,
all have the same factors, and so the \wzb couples with an identical
strength to each charged lepton. This property of the SM, lepton
universality, can be experimentally tested by comparing the ratios of
${\z \to \die}$, ${\z \to \dimu}$, and ${\z \to \ditau}$
production. Previous measurements by \dlep \cite{lep.06.1} have been
made,
\begin{equation}
  \frac{\sigma_{\die \to \z \to \dimu}}{\sigma_{\die \to \z \to \die}}
  = 1.0009 \pm 0.0028 \equcomma
  \frac{\sigma_{\die \to \z \to \ditau}}{\sigma_{\die \to \z \to \die}}
  = 1.0019 \pm 0.0032
  \labelequ{Lep}
\end{equation}
which verify lepton universality in \wzb decays to a precision of
better than $1\%$. The measurement of $\sigma_{pp \to \z \to \ditau}$
within this chapter, when compared to the \lhcb measurements of
$\sigma_{\dip \to \z \to \die}$ \cite{lhcb.13.2} and $\sigma_{\dip \to
  \z \to \dimu}$ \cite{lhcb.12.1}, can test lepton universality to the
level of $4\%$ at best, assuming the precision of $\sigma_{\dip \to \z
  \to \ditau}$ is similar to the precisions of the $\sigma_{\dip \to
  \z \to \die}$ and $\sigma_{\dip \to \z \to \dimu}$
measurements. However, the \lhcb tests of lepton universality are
unique; the \wzbs are produced from
proton-proton collisions, unlike the electron-positron collisions at
\dlep, and the leptons produced from the \wzbs are observed in the
forward pseudo-rapidity range of $2.0 \leq \eta \leq 4.5$.

Measuring a ratio different from unity might be indicative of new
physics, rather than a failure of lepton universality. In most models
with \whbs, described in \sec{Thr:Lag} for the \sm and \sec{Thr:Alt}
for the minimal supersymmetric model (\mssm), the \whbs are expected
to couple to fermions with a strength proportional to the mass of the
fermion. This can be seen by the factor of the fermion mass, $m_f$, in
the vertices of \fig{Thr:H02FF} for the \sm and \figs{Thr:H12FuFu}
through \ref{fig:Thr:H32FdFd} for the \mssm. The \whbs can also couple
to the electroweak vector bosons with strengths proportional to the
mass of the electroweak bosons. The relevant vertices are given in
\figs{Thr:H02WW} and \ref{fig:Thr:H02ZZ} for the \sm and
\figs{Thr:H12WW} through \ref{fig:Thr:H22ZZ} for the \mssm.

Consequently, for a neutral \whb with a mass less than $2\m_\w$, the
dominant decays of the \whb at leading order are into \wtl and \wbq
pairs. For the case of neutral \whbs decaying into \wtl pairs, this
process will be similar to the ${\z \to \ditau}$ signal, except for
the invariant mass and spin correlations of the \wtl pair, and will
provide an unaccounted background to the ${\z \to \ditau}$ signal,
resulting in an excess of observed events. Using the measured ${\dip
  \to \z \to \ditau}$ cross-section, \whb searches are performed in
\chp{Hig} by looking for an unexpected excess of events.

\begin{subfigures}[t]{2}{\subfig{Stau2TauMuUD} An example \susy decay
    chain from a stau LSP, producing a \wtl in the final state. Dashed
    lines indicate bosons and solid line
    fermions. \subfig{Pdf.Kinematics} Rough outline in momentum
    fraction and momentum transfer space of the measurements currently
    used to constrain the proton \PDF. The $x$ and $Q^2$ limits on the
    data available from \tevatron (blue), \hera (green), and fixed
    target experiment (yellow) are taken from \rfr{ball.11.1}.}
  \svgbeg
  \fmp{Stau2TauMuUD} & \svg{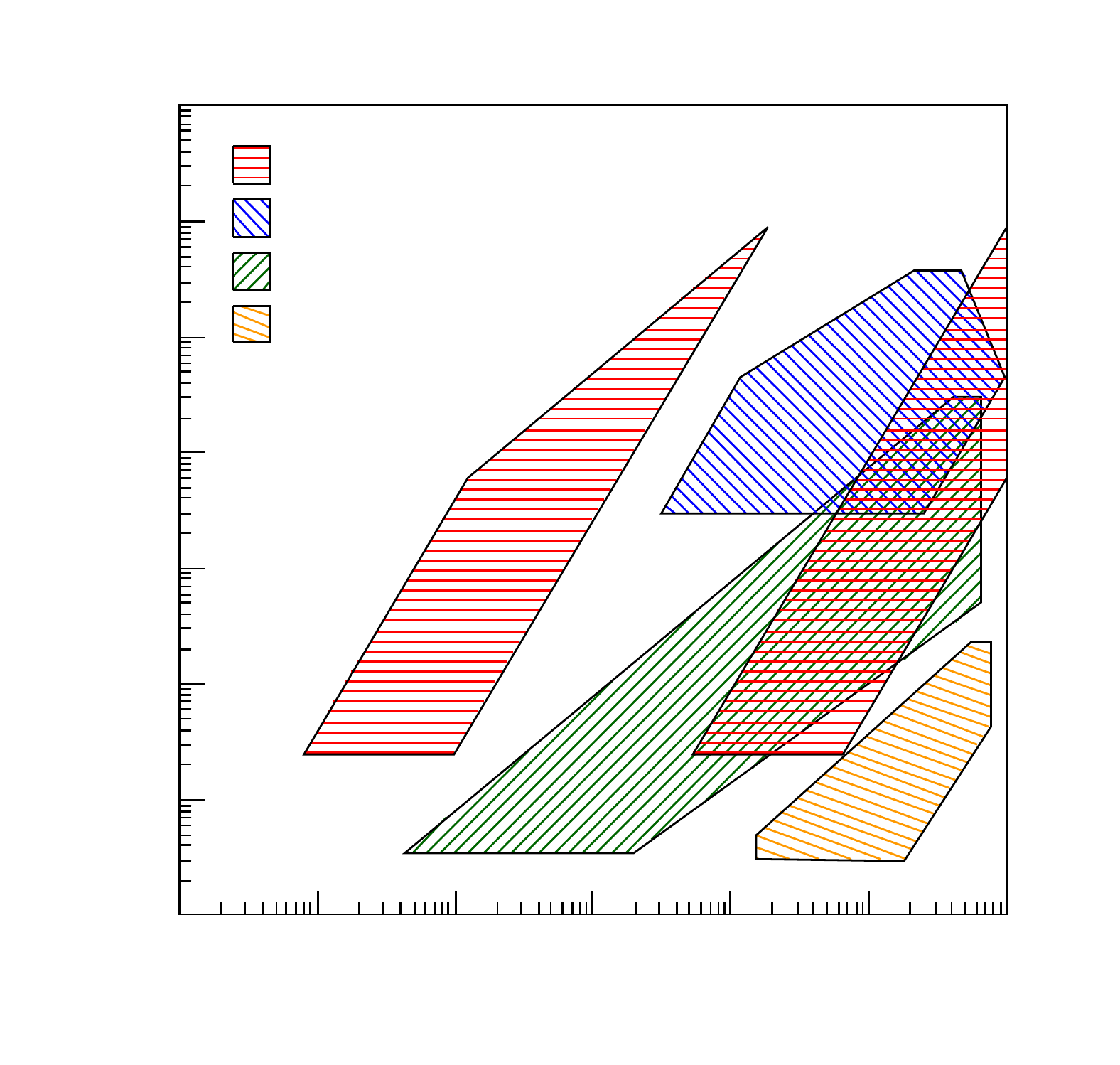} \svgend
\end{subfigures}

In addition to excesses from \whb decays, many \susy models, described
in \sec{Thr:Alt}, predict a lightest supersymmetric partner (LSP) such
as the stau. The decay of the stau depends upon the \susy parameter
space, but oftentimes the stau will decay via chains containing \wtls,
such as the decay of \fig{Zed:Stau2TauMuUD}~\cite{dreiner.09.1}. Here,
the stau decays into a virtual lightest neutralino, $\tilde{N}_1$, and
a \wtl. The neutralino then decays into a muon and left-handed smuon,
$\tilde{\mu}_L$, which further decays into a \wuq and \wdq pair. The
dashed lines in this diagram indicate bosons, while the solid lines
indicate fermions and the arrows indicate the flow of electromagnetic
charge. These types of events can also produce unaccounted backgrounds
to the ${\z \to \ditau}$ signal, resulting in an excess in the ratio
of the measured $\sigma_{\dip \to \z \to \ditau}$ to $\sigma_{\dip \to
  \z \to \dimu}$ and $\sigma_{\dip \to \z \to \die}$.

Unlike at \dlep, \wzbs at \lhcb are produced from composite particles;
typically a valence quark and sea anti-quark from two protons
annihilate, producing a \wzb. Consequently, the measurement of the \z
cross-section at \lhc can be used to refine the structure of the
proton \PDF, described in \sec{Thr:Exp}. The proton \PDF is
constrained by previous results from \tevatron, \hera, and fixed
target experiments at values of the parton momentum fraction, $x$, and
momentum transfer, $Q^2$, in the regions of parameters space shown in
\fig{Zed:Pdf.Kinematics}. The $7~\tev$ centre-of-mass energy of the
\lhc combined with the forward coverage of \lhcb provides \wzb data,
which can be used to further constrain the proton \PDF. The region in
which \lhcb data can be used is also shown in
\fig{Zed:Pdf.Kinematics}. This estimated coverage for \lhcb \wzb data
is bounded by,
\begin{equation}
  Q^2(x) = 
  \begin{cases}
    xs & \mbox{upper bound} \\
    e^{\pm 2 \abs{\eta}} x^2 \sqrt{s} & \mbox{left-right bounds} \\
    \m_\mathrm{min}^2 & \mbox{lower bound} \\
  \end{cases}
  \labelequ{Pdf.Kinematics}
\end{equation}
where $\sqrt{s}$ is the centre-of-mass energy of the two colliding
beams, $\eta$ is defined in \equ{Exp:Rapidity} and is the minimum or
maximum pseudo-rapidity of the detector, and $\m_\mathrm{min}$ is the
minimum measured invariant mass of the two leptons produced from the
\wzb. The relations of the upper and left-right bounds from
\equ{Zed:Pdf.Kinematics} are determined by setting the momentum of one
colliding parton as $x\sqrt{s}/2$, the momentum of the other parton as
$Q^2/(2x/\sqrt{s})$, and assuming both partons have only longitudinal
momentum.

\newsection{Analysis}{Ana}

Within this section the methods used for selecting the ${\z \to
  \ditau}$ events used in the cross-section calculation of
\sec{Zed:Sig} are described. First the final states of the ${\z \to
  \ditau}$ signal events and possible backgrounds are outlined. In
\sec{Zed:Rec} the details of the particle identification criteria used
to obtain \wtl decay product candidates are given. The event selection
which uses these candidates is then described in
\sec{Zed:Sel}. Finally, in \sec{Zed:Bkg} the methods used to estimate
the backgrounds to the ${\z \to \ditau}$ signal are introduced.

The cross-section for producing a \wtl pair from the partons of two
protons can be perturbatively expanded into terms represented by
diagrams mediated by an excited photon or \wzb with their leading
order matrix elements given by \equs{Tau:MeGamma} and
\ref{equ:Tau:MeZ} respectively. Experimentally, these terms and their
interference, given by \equ{Tau:MeDy}, cannot be
separated. Consequently, for the remainder of this chapter, the symbol
\z is used to indicate contributions from an excited photon or \wzb
with an invariant mass about the on-shell mass of the
\wzb. Additionally, the symbol \lep for a charged lepton is used to
indicate only an electron or muon, and not a \wtl.

Most \wtls produced within \lhcb decay before reaching the \velo which
is described in \sec{Exp:Trk}, and so the experimental selection of
${\z \to \ditau}$ events must use \wtl decay products rather than the
\wtls themselves, contrary to $\z \to \dimu$ and $\z \to \die$
analyses where lepton pairs can be directly selected. Because the \wtl
is heavier than the lightest mesons and all other leptons, it can
decay into a variety of final states with multiplicities of up to
seven particles, as detailed in \sec{Tau:Dec} and summarised in
\tab{Tau:Decays}. However, the experimental reconstruction of high
multiplicity \wtl final states can have large backgrounds from \qcd
induced processes at hadron colliders which introduce undesirable
systematic uncertainties. Consequently, this analysis only utilises
leptonic \wtl decays into a \wtl neutrino with a lepton and lepton
neutrino, or semi-leptonic \wtl decays into a \wtl neutrino and single
charged hadron; only a single muon, electron, or charged hadron is
selected as a \wtl decay product, as neutrinos are not
reconstructed. The matrix elements for these decays are given by
\equs{Tau:Meson} and \ref{equ:Tau:TwoLeptons} of \sec{Tau:Dec}.

\begin{subfigures}{2}{The signal ${\z \to \ditau}$ final states
    considered with \subfig{Z2TauTau2LL}~two leptonic \wtl decays and
    \subfig{Z2TauTau2LJet}~one leptonic decay and one semi-leptonic
    decay. The experimentally selected final state is highlighted in
    red.\labelfig{Signals}}
  \fmpbeg
  \fmp{Z2TauTau2LL} & \fmp{Z2TauTau2LJet} \fmpend
\end{subfigures}

\begin{subfigures}[p]{2}{Examples of the \subfig{Z2LL}~${\z \to
      \dilep}$, \subfig[W2LJet]{Z2LJet}~\ewk, \subfig{GG2JetJet}~\qcd,
    \subfig{WW2LL}~\ww, and \subfig{TT2LL}~\ttbar backgrounds are
    given, with the experimentally selected final state highlighted in
    red.\labelfig{Backgrounds}}
  \fmpbeg
  \fmp{Z2LL}   & \fmp{W2LJet}    \fmpsep
  \fmp{Z2LJet} & \fmp{GG2JetJet} \fmpsep
  \fmp{WW2LL}  & \fmp{TT2LL}     \fmpend
\end{subfigures}

A lepton is required to experimentally trigger ${\z \to \ditau}$
events, and so only final states where both \wtl decays are leptonic,
or where one \wtl decay is leptonic and one \wtl decay is
semi-leptonic, are selected. Feynman diagrams for these two final
states are given in \fig{Zed:Signals}. A variety of processes, with
examples diagrammed in \fig{Zed:Backgrounds}, contribute experimental
backgrounds to the ${\z \to \ditau}$ final states of
\fig{Zed:Signals}. For this analysis the \sm \whb provides a
negligible contribution, as demonstrated in \chp{Hig}, and
consequently is not considered as a background. In both
\fig{Zed:Signals} and \fig{Zed:Backgrounds} the possibly selected \wtl
decay product candidates are highlighted in red.

A ${\z \to \dilep}$ background is shown in \fig{Zed:Z2LL} where an
electron or muon pair is produced from a \wzb. This acts as a
background to ${\z \to \ditau}$ signal events consisting of an
electron or muon pair where both \wtls decay leptonically. If one of
the leptons from the event type of \fig{Zed:Z2LL} is mis-identified as
a hadron, then these events can also act as a background to the ${\z
  \to \ditau}$ signal events of \fig{Zed:Z2TauTau2LJet} where only one
of the \wtls decays leptonically and the other decays
semi-leptonically. For signal events where one \wtl decays
leptonically into an electron and the other into a muon, the ${\z \to
  \dilep}$ background is not considered, as the probability of
mis-identifying a muon as an electron, or conversely an electron as a
muon, is negligible.

In \figs{Zed:Z2LJet} and \ref{fig:Zed:W2LJet} a single charged lepton
from a \z or \wwb decay falls within \lhcb acceptance and an
associated jet provides either an additional charged lepton or
hadron. Alternatively, this jet could be produced from jet activity
not associated with the vector boson. These two backgrounds are
grouped together as an electroweak background, \ewk, and are
considered for both ${\z \to \ditau}$ signal event types. Backgrounds
consisting of particles produced from jet activity,
\fig{Zed:GG2JetJet}, are grouped into a general \qcd background where
a hard lepton is typically produced from the decay of a heavy flavour
meson and either an additional lepton or a charged hadron is produced
within the event. This background, like the \ewk background, is
considered for both ${\z \to \ditau}$ signal event types.

Pairs of \wwbs can be produced within \lhcb, resulting in the
background shown in \fig{Zed:WW2LL}. Here, both the \wwbs decay
leptonically, providing a background to signal events where both \wtls
decay leptonically. One of the \wwbs can also decay into quarks,
providing a background to signal events with a leptonic and
semi-leptonic \wtl decay. In \fig{Zed:TT2LL} a background to signal
events where both \wtls decay leptonically is produced from a pair of
\wtqs decaying leptonically. One of the \wtqs can also decay
hadronically, producing a background to signal events with a leptonic
and semi-leptonic \wtl decay.

\newsubsection{Particle Identification}{Rec}

In order to select ${\z \to \ditau}$ signal events, the muons,
electrons, and charged hadrons within an event must first be
reconstructed and identified. All reconstructed particles considered
as \wtl decay product candidates are long tracks, described in
\sec{Exp:RecTrk}, which are required to have a $\chi^2$ probability
greater than $0.1\%$ and a pseudo-rapidity within the range $2.0 \leq
\eta \leq 4.5$. Hits associated with the long tracks from the muon
system are used to identify muons, while associated calorimeter
energy, described in \sec{Exp:RecCal}, from the \prs, \ecal, and \hcal
is used to differentiate between electrons and charged
hadrons. Additionally, the momenta for electron candidates is
corrected for bremsstrahlung radiation losses using the method of
\sec{Exp:RecCal}.

Four variables associated with long tracks are used for particle
identification: the number of muon stations with hits, the \prs
energy, the \ecal energy over the momentum of the track, and the \hcal
energy over the momentum of the track. Distributions of these
variables, normalised to an integral of one, for muons, electrons, and
hadrons are shown in \fig{Zed:Rec}. The muon distributions are taken
from data events consistent with a ${\z \to \dimu}$ signal, which are
selected by requiring an isolated muon and an isolated opposite-sign
track. Both the muon and track must have ${\pt >20~\gev}$, a combined
invariant mass within the range ${60 \leq \m \leq 120~\gev}$, and an
azimuthal angle separation of greater than $2.7$ radians. The
distributions are then produced from the variables of the isolated
track. The same method is used to obtain the electron distributions,
but a ${\z \to \die}$ signal is selected by requiring isolated
electrons rather than muons. The charged hadron distributions are
obtained from long tracks in minimum bias data with ${\pt > 5~\gev}$.

The distributions of \fig{Zed:Rec} are intended to illustrate the
particle identification requirements and will be similar to, but not
the same, as the muon, electron, and charged hadron distributions from
${\z \to \ditau}$ signal events which are unavailable directly from
data. Primarily, the \pt spectrum of the electrons and muons used to
produce these distributions will be harder than the spectrum for
electrons and muons from ${\z \to \ditau}$ events. The vertical black
lines of \fig{Zed:Rec} indicate the particle identification
requirements used for selecting \wtl decay product candidates and are
summarised in \tab{Zed:Rec}. A more detailed motivation of the
selection criteria for each particle type is given in the remainder of
this section.

\begin{subfigures}{2}{The \subfig{Rec.Muon}~number of muon stations
    with associated hits, \subfig{Rec.Prs} \prs energy,
    \subfig{Rec.Ecal} fractional \ecal energy over track momentum, and
    \subfig{Rec.Hcal} fractional \hcal energy over track momentum
    distributions for muons (red), electrons (blue), and pions (green)
    from data. The particle identification requirements, as described
    in the text, are indicated by the black lines.\labelfig{Rec}}
  \svgbeg
  \svg{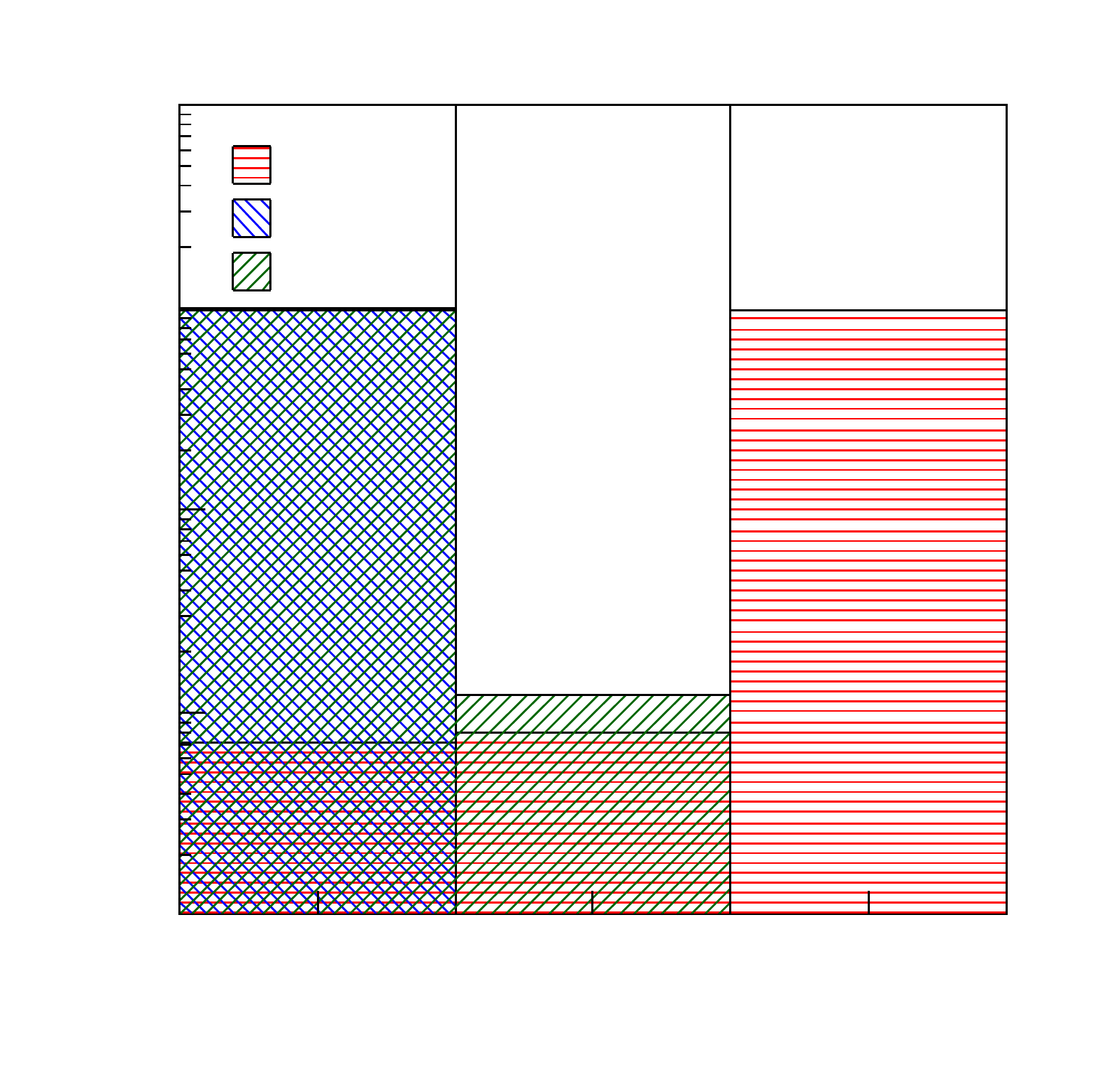} & \svg{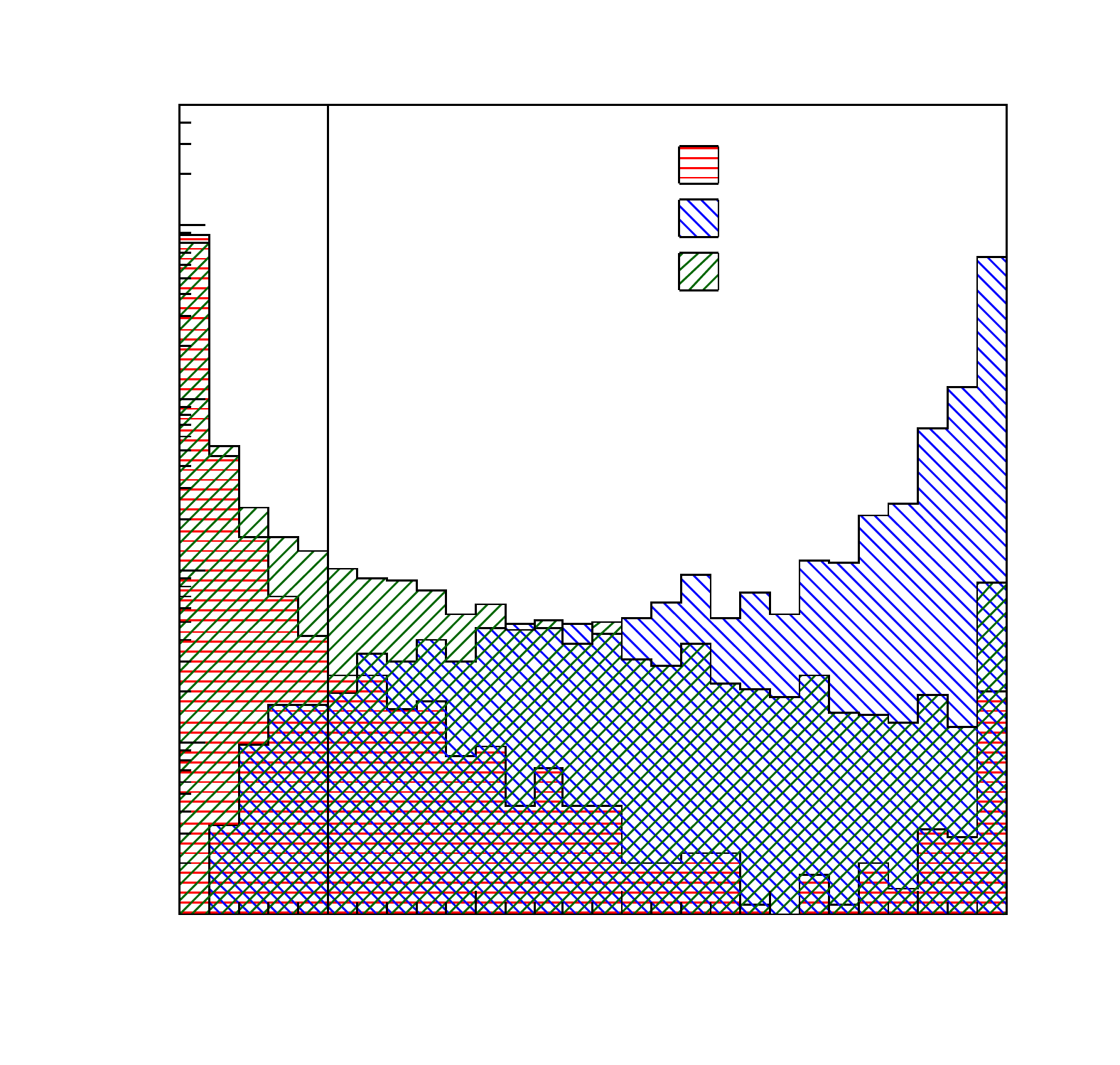}  \svgsep
  \svg{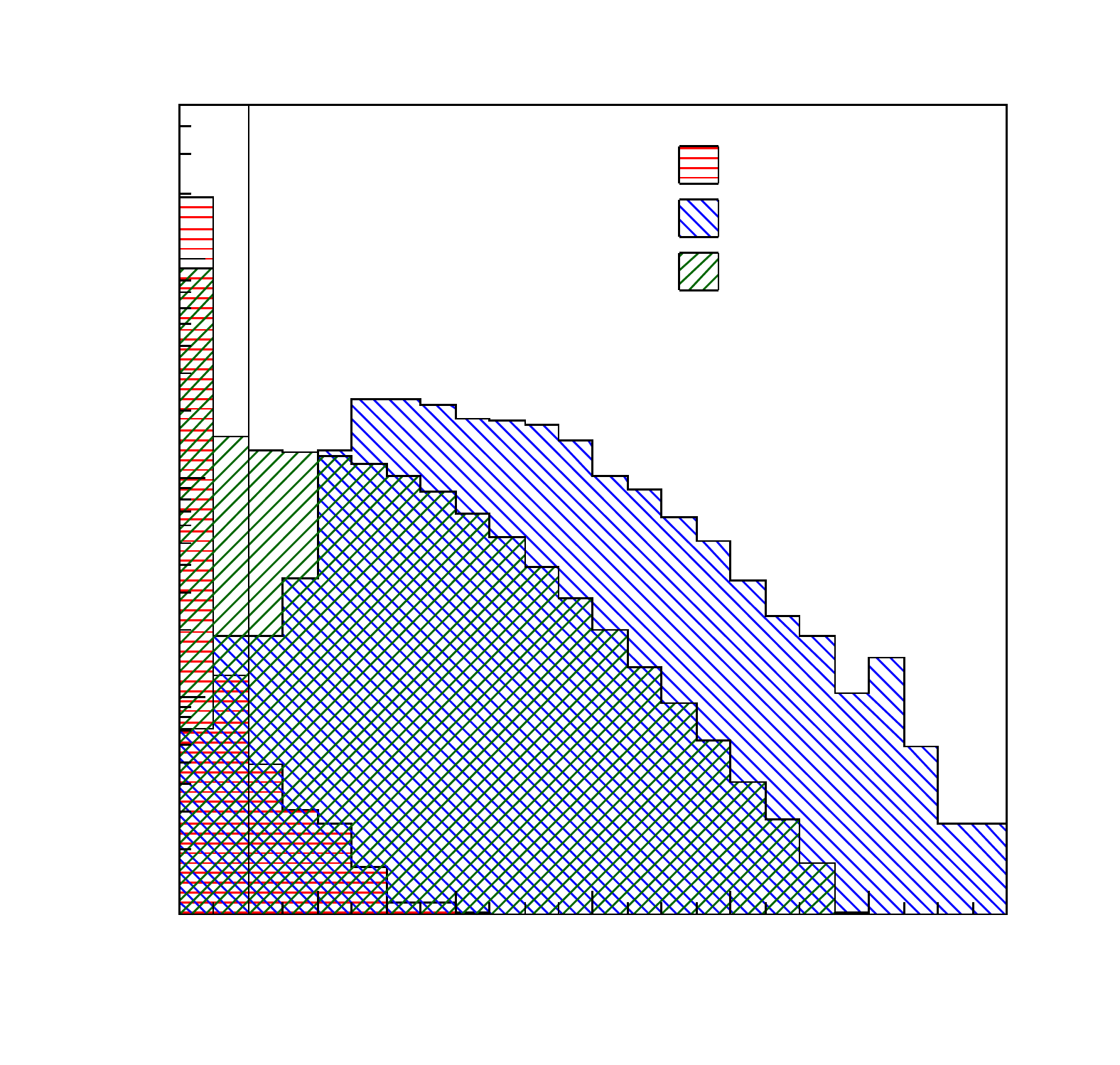} & \svg{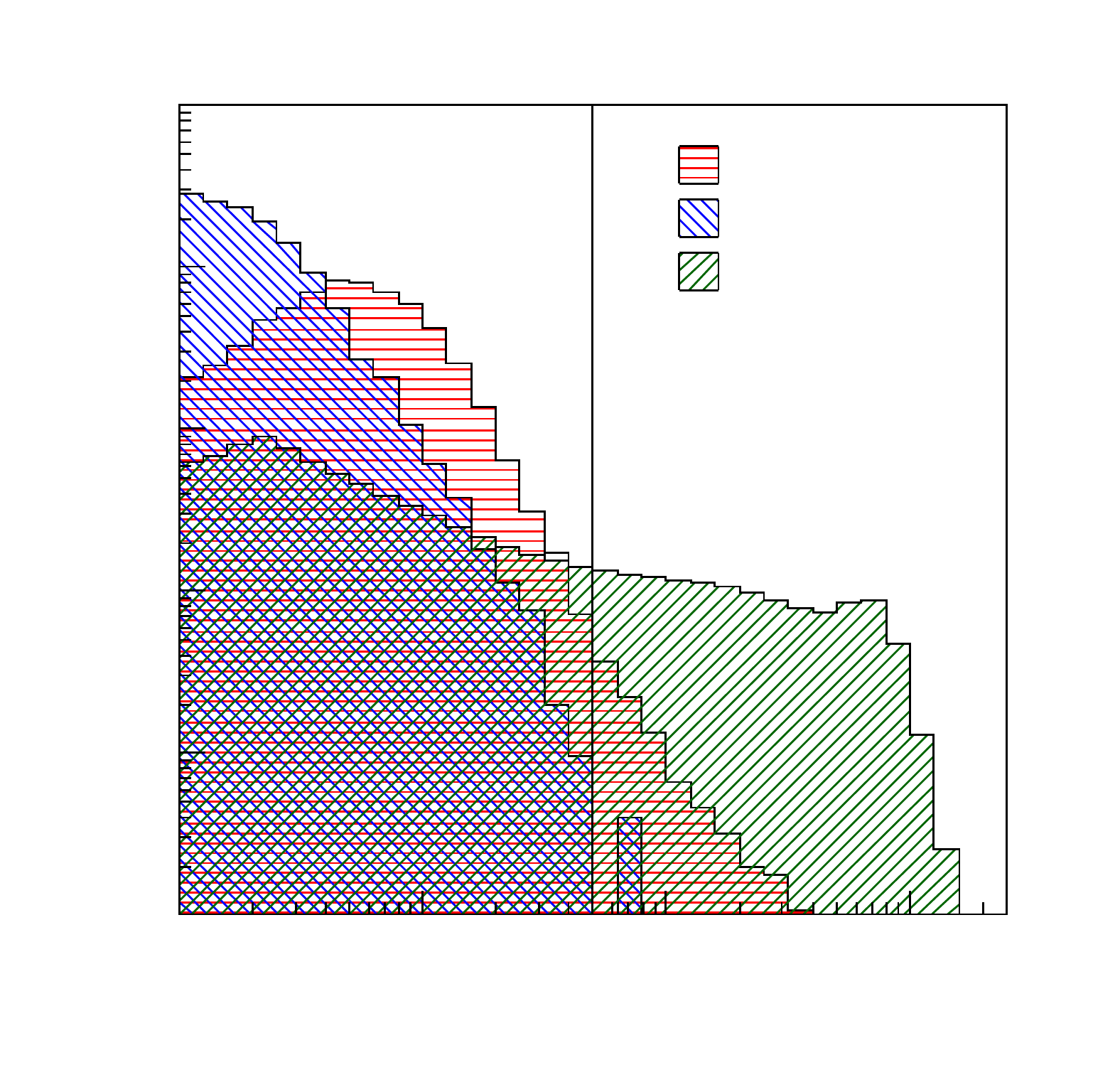} \fmpend
\end{subfigures}

\begin{table}\centering
  \captionabove{A summary of the particle identification requirements for
    muons, electrons, and charged hadrons.\labeltab{Rec}}
  \begin{tabular}{l|c|c|c}
    \toprule
    & \multicolumn{1}{c|}{muons} & \multicolumn{1}{c|}{electrons}
    & \multicolumn{1}{c}{hadrons} \\
    \midrule
    track $P(\chi^2)~[\%]$  & $> 0.01$ & $> 0.01$ & $> 0.01$ \\
    track $\eta$ & $2.0 \leq \eta \leq 4.5$ & $2.0 \leq \eta \leq 4.5$ 
    &$2.25 \leq \eta \leq 3.75$ \\
    muon stations & $4$ & \makebox[\widthof{$> 0.01$}][l]{$< 3$} 
    & \makebox[\widthof{$> 0.01$}][l]{$< 3$} \\
    $E_\prs~[\gev]$ & \multicolumn{1}{c|}{$-$} & $> 0.05$ &
    \multicolumn{1}{c}{$-$} \\
    $E_\ecal / p$ & \multicolumn{1}{c|}{$-$} 
    & \makebox[\widthof{$> 0.01$}][l]{$> 0.1$}
    & \multicolumn{1}{c}{$-$} \\
    $E_\hcal / p$ & \multicolumn{1}{c|}{$-$} & $< 0.05$ & $> 0.05$ \\
    \bottomrule
  \end{tabular}
\end{table}

\newsubsubsection{Muon Identification}{}

Muons are reconstructed as described in \sec{Exp:RecTrk}, where tracks
are extrapolated to the four muon stations downstream of the
calorimeters and matched to compatible hits. Muon candidates must have
an associated hit in each of the four muon stations, requiring the
candidate to traverse approximately twenty hadronic interaction
lengths of material. The high efficiency of this requirement can be
seen in the distribution of muon stations with associated hits for
muons in \fig{Zed:Rec.Muon}, where nearly all muons have associated
hits in all four muon stations. Charged hadrons and electrons
typically have less than three stations with associated hits. Because
the muons from \wtl decays have a lower \pt spectrum than the muons
from the ${\z \to \dimu}$ events of \fig{Zed:Rec.Muon}, the muon
identification efficiency for ${\z \to \ditau}$ events is slightly
reduced.

\newsubsubsection{Electron Identification}{}

Electrons are reconstructed by extrapolating tracks into the \ecal,
matching \ecal clusters with the track, and then performing
bremsstrahlung recovery, as described in \sec{Exp:RecCal}. The \ecal
cells saturate at transverse energies above $10~\gev$, also described
in \sec{Exp:RecCal}, and so bremsstrahlung recovery for high \pt
electrons from \wzbs is incomplete. The standard \lhcb electron
identification requirements were designed for low \pt electrons from
$B$-hadron decays, relying upon \rich{} information, and so different
requirements suited for high momentum electrons are used instead.

High momentum electrons begin showering within the \prs, but deposit
most of their energy within the \ecal. Conversely, most hadrons do not
shower until the \hcal, and muons typically do not shower. The
$E_\prs$ distributions for muons and electrons with ${\pt > 20~\gev}$
and charged hadrons with ${\pt > 5~\gev}$ are shown in
\fig{Zed:Rec.Prs}. An associated \prs energy, $E_\prs$, greater than
$0.05~\gev$ is required for electron identification, as most electrons
from \wzb decays will have energies well above this requirement.

The \ecal and \hcal energies are dependent upon the energy and
momentum of the electron and so the fractional \ecal and \hcal energy
with respect to the electron momentum are used as electron
identification variables. The \ecal energy over momentum distribution
is shown in \fig{Zed:Rec.Ecal} and the \hcal energy over momentum
distribution is shown in \fig{Zed:Rec.Hcal} for muons, electrons, and
pions. Electrons are expected to deposit a large fraction of their
momentum within the \ecal, as can be seen in \fig{Zed:Rec.Ecal}, and
so electron candidates must satisfy $E_\ecal/p > 0.1$. Conversely,
electrons are not expected to reach the \hcal and are required to have
$E_\hcal/p < 0.05$, as shown in \fig{Zed:Rec.Hcal}. By requiring less
than three muon stations with associated hits, electron candidates by
definition are mutually exclusive to muon candidates.

\newsubsubsection{Charged Hadron Identification}{}

Charged hadrons are identified using a mutually exclusive selection to
both muons and electrons. They are expected to shower within the
\hcal, as described in \sec{Exp:Pid}, and so all charged hadron
candidates must satisfy $E_\hcal/p > 0.05$, in addition to the
requirement of less than three muon stations with associated hits. To
ensure coverage of the \hcal, charged hadron candidates must also fall
within the reduced pseudo-rapidity range of $2.25 \leq \eta \leq
3.75$.

\newsubsection{Event Selection}{Sel}

Once reconstructed particles are identified using the criteria of
\sec{Zed:Rec}, ${\z \to \ditau}$ events from data can be selected. The
data used have been collected via a single muon trigger requiring $\pt
> 10~\gev$ and a single electron trigger requiring $\pt > 15~\gev$.
Further details on the trigger can be found in \sec{Exp:RecTrg}. The
selection of ${\z \to \ditau}$ events is divided into five mutually
exclusive categories: \mumu, \mue, \emu, \muh, \eh. The first \wtl
decay product candidate is labelled by the first \wtl subscript, while
the second candidate is labelled by the second subscript; here $\mu$,
$e$, and $\had$ indicate muons, electrons, and charged hadrons
respectively. The first three categories select signal events where
both \wtls decay leptonically, while the last two categories select
signal events where the first \wtl decays leptonically and the second
\wtl decays semi-leptonically into a \wtl neutrino and single charged
hadron. No di-electron category is considered due to the \ecal
saturation described in \sec{Exp:Rec} which results in poor separation
of this signal from large backgrounds.

To eliminate large \qcd backgrounds with soft \pt spectra, the first
\wtl decay product candidate is required to have $\pt > 20~\gev$ while
the second candidate must have $\pt > 5~\gev$. The following
additional trigger and particle identification requirements are
applied for each category.
\begin{itemize}
  \settowidth{\itemindent}{\mumu}
\item[{\makebox[\itemindent][l]{\mumu:}}] Requires two
  oppositely-charged muons, where the muon with the larger \pt is
  considered the first \wtl decay product candidate. Either or both
  muons can trigger the event.
\item[{\makebox[\itemindent][l]{\mue:}}] Requires a muon and an
  oppositely-charged electron. Only the muon can trigger the event.
\item[{\makebox[\itemindent][l]{\emu:}}] Requires an electron and an
  oppositely-charged muon with $\pt < 20~\gev$. Either or both the
  electron and muon can trigger the event.
\item[{\makebox[\itemindent][l]{\muh:}}] Requires a muon and an
  oppositely-charged hadron. Only the muon can trigger the event.
\item[{\makebox[\itemindent][l]{\eh:}}] Requires an electron and an
  oppositely-charged hadron. Only the electron can trigger the event.
\end{itemize}

Additional selection requirements, dependent upon the category, are
also applied to further separate the ${\z \to \ditau}$ signal from its
backgrounds. These variables, as well as the requirements placed on
them, are described in the remainder of this section. The selection
requirements are not optimised using multivariate techniques, but are
manually selected to adequately separate signal and background without
severely limiting signal statistics. For each variable, its
distributions for ${\z \to \dimu}$ events from simulation and data are
compared to validate simulation. The ${\z \to \dimu}$ data events are
selected by requiring two opposite-sign muons with ${\pt > 20~\gev}$
and a combined invariant mass within the range ${80 \leq \m \leq
  100~\gev}$. If the simulation does not match well with data, it is
corrected to data and the distribution for this calibrated simulation
is also provided.

Distributions for the ${\z \to \ditau}$ signal and its backgrounds,
described in the introduction of \sec{Zed:Ana}, are also given for
each variable. The ${\z \to \ditau}$, \ewk, $\ww$, and $\ttbar$
distributions are taken from calibrated simulation, while the \qcd and
${\z \to \dilep}$ background distributions are estimated from data
using the methods described in \sec{Zed:Bkg}. All distributions are
normalised to an integral of one for comparison purposes and are
provided for only the \mumu event category.

\newsubsubsection{Invariant Mass}{}

The invariant mass of the two \wtl decay product candidates is defined
as,
\begin{equation}
  \m \equiv \sqrt{(E_1 + E_2)^2 - (\vec{p}_1 + \vec{p}_2)^2} 
\end{equation}
where $E_1$ and $E_2$ are the energies of the two \wtl decay product
candidates and $\vec{p}_1$ and $\vec{p}_2$ are their
three-momenta. The momentum resolution observed in data is
underestimated in simulation, resulting in narrower mass distributions
from simulation. This effect can be seen by comparing the invariant
mass distribution from ${\z \to \dimu}$ data to the distribution from
simulation, as shown in \fig{Zed:Sel.Mass.Compare}. The momentum
components, $p_i$, for a reconstructed particle in simulation are
calibrated using,
\begin{equation}
  p_i = \fshift\left({p_i}_\gen + \fwidth({p_i}_\rec -
    {p_i}_\gen)\right)
  \labelequ{Momentum}
\end{equation}
where $p_\gen$ is the generated momentum of the particle and $p_\rec$
is the reconstructed momentum of the particle. The parameter \fshift
shifts the distribution, while \fwidth adjusts the width of the
distribution.  Values of ${\fshift = 0.998 \pm 0.001}$ and ${\fwidth =
  2.0 \pm 0.1}$ were obtained by fitting the simulated invariant mass
distribution to the data distribution, resulting in the calibrated
distribution of \fig{Zed:Sel.Mass.Compare}. This calibration is used
for all mass distributions taken from simulation and is assumed to
remain constant for the lower momentum range of the $\z \to \ditau$
signals and backgrounds.

\begin{subfigures}{2}{\subfig{Sel.Mass.Compare}~A comparison of the
    invariant mass distributions between data (points), simulation
    (red), and calibrated simulation (blue) for ${\z \to \dimu}$
    events. Invariant mass distributions for \subfig{Sel.Mass.A}~${\z
      \to \ditau}$ (red), \qcd (blue), and \ewk (green) events and
    \subfig{Sel.Mass.B}~\ttbar (orange), \ww (magenta), and ${\z \to
      \dilep}$ (cyan) events. The distributions from simulation are
    corrected with the momentum calibration of \equ{Zed:Momentum} and the
    mass ranges excluded by the \mumu selection requirement are shaded
    in grey. \labelfig{Sel.Mass}}
  \svgbeg \svg{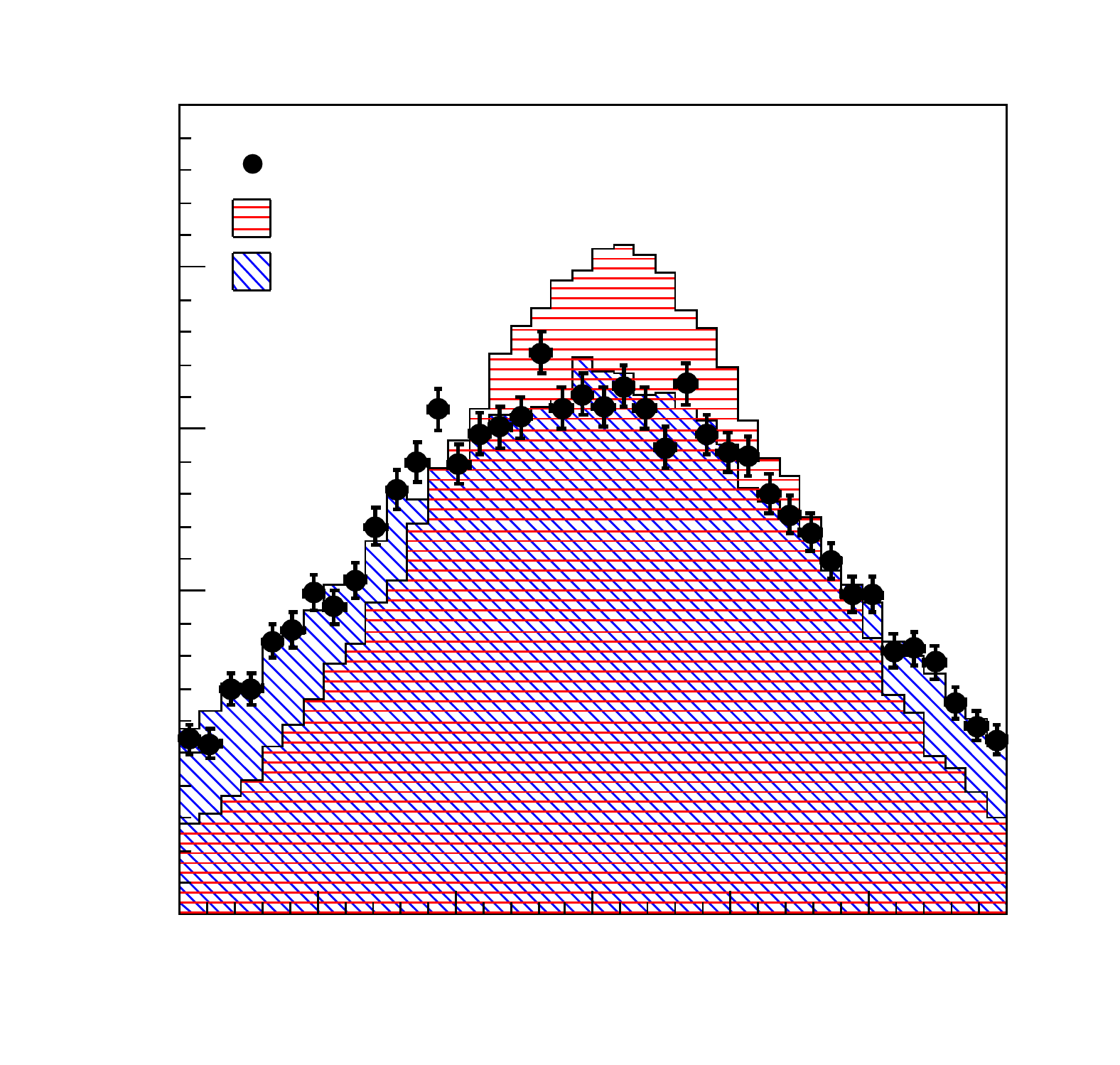} & \sidecaption \svgsep
  \svg{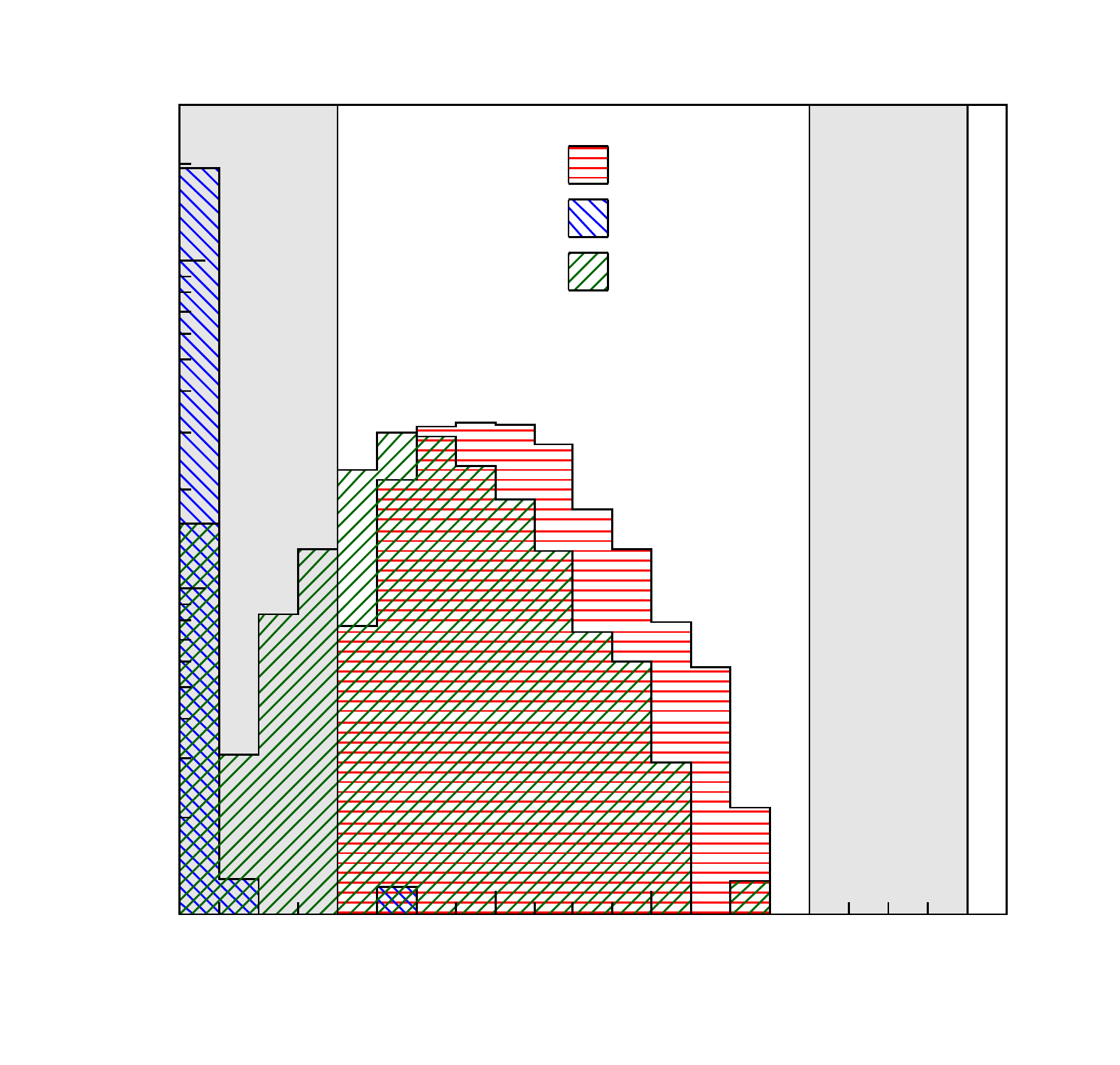} & \svg{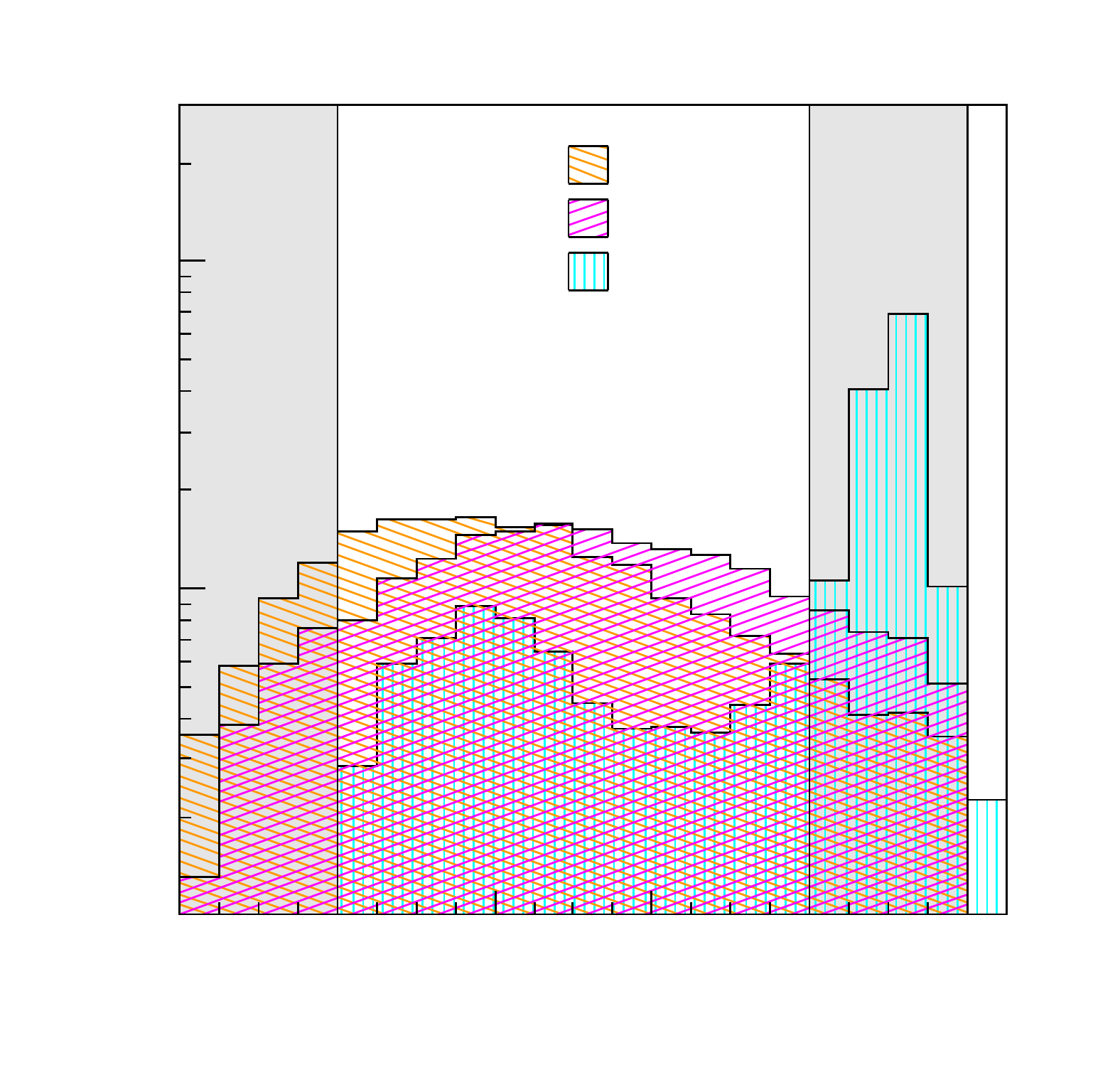} \svgend
\end{subfigures}

The invariant mass for a \wzb decay into a back-to-back \wtl pair in
the transverse plane is,
\begin{equation}
  \m^2 \geq \left(\sqrt{\pt_1^2 + \m_1^2} + \sqrt{\pt_2^2 + \m_2^2}\right)^2 -
  (\pt_1 - \pt_2)^2 \geq 4\pt_1\pt_2
  \labelequ{Mass}
\end{equation}
where $\pt_1$ and $\pt_2$ are the transverse momenta of the two \wtls,
and $\m_1$ and $\m_2$ are their masses. As $\m_\z \gg \m_\tau$, the
single charged visible decay products of the \wtls are approximately
collinear with their parent \wtls. Consequently, the inequality of
\equ{Zed:Mass} also holds for the invariant mass of the decay products
from the \wtls.

Since \wtl pairs produced from \wzbs within \lhcb fulfil the
assumptions of \equ{Zed:Mass} and $\pt_1 > 20~\gev$ and $\pt_2 >
5~\gev$ are required for the \wtl decay product candidates, the
invariant mass of the two \wtl decay products must be greater than
$20~\gev$. \Fig{Zed:Sel.Mass.A} shows the ${\z \to \ditau}$ invariant
mass distribution for the \mumu category of events.

Of the five backgrounds, only the ${\z \to \dilep}$ background fulfils
the assumptions of \equ{Zed:Mass} and consequently has no events below
$20~\gev$. The ${\z \to \dimu}$ background distribution is shown in
\fig{Zed:Sel.Mass.B} and has a sharp peak at $90~\gev$, corresponding
to the on-shell mass of the \wzb. The ${\z \to \dimu}$ background
dominates the signal for the \mumu category, and so the mass window of
${80 < \m < 100~\gev}$ is excluded for the \mumu category only.

The invariant mass distribution for the \qcd background is given in
\fig{Zed:Sel.Mass.A} and shows that \qcd background events have a
significantly lower invariant mass than signal ${\z \to \ditau}$
events. For \ewk events, a single lepton with hard \pt from a \w or
\wzb is combined with a candidate from the underlying event to produce
an invariant mass distribution harder than the \qcd background, yet
softer than the signal, see \fig{Zed:Sel.Mass.A}. In
\fig{Zed:Sel.Mass.B} the \mumu invariant mass distributions are given
for the \ttbar and \ww backgrounds. The \wtl decay product candidates
from these events are produced from massive parents and produce
distributions with higher mass tails than either the \qcd or \ewk
backgrounds.

\newsubsubsection{Track Isolation}{}

The track isolation associated to a candidate track, \iso, is defined as,
\begin{equation}
  \iso \equiv \sum_i^\mathrm{tracks} {p_x}_i \oplus
  \sum_i^\mathrm{tracks} {p_y}_i
  \labelequ{Iso}
\end{equation}
where the $x$ and $y$-momenta of all tracks within a cone of $\Delta R
\equiv \Delta \phi \oplus \Delta \eta < 0.5$ around the candidate
track are summed and then added in quadrature. Here, $\Delta \phi$ and
$\Delta \eta$ are the differences in $\phi$ and $\eta$, defined in
\sec{Exp:Det}, between the candidate and the track. The track of the
candidate itself is excluded from the sum. Physically, \iso is the \pt
of the vectorial sum of all tracks in a cone surrounding the candidate
and quantifies the charged isolation of the candidate. Note that a
large \iso corresponds to a poorly isolated track.

In \fig{Zed:Sel.Iso.Compare}, the distributions of the maximum \iso of
the two muons from ${\z \to \dimu}$ data and simulation are
compared. The isolation of the two muon candidates is dependent
primarily upon the underlying event activity, which is underestimated
in simulation. Consequently, the \iso distribution produced from
simulation is softer than that from data. The soft underlying event
cannot be calculated perturbatively, and must be modelled
phenomenologically in simulation, as described in \chp{Thr}. As there
is no simple relation between data and simulation, calibration of the
\iso distribution from simulation is not possible. However, the
efficiency of the \iso selection is calculated from data, as described
in \sec{Zed:SelEff}, and so the cross-section determination of
\sec{Zed:Sig} does not depend upon the accuracy of the simulated \iso
distribution.

\begin{subfigures}{2}{\subfig{Sel.Iso.Compare}~A comparison of the
    maximum track isolation distributions between data (points) and
    simulation (red) for ${\z \to \dimu}$ events. Distributions of the
    maximum track isolation between the two \wtl decay product
    candidates for \subfig{Sel.Iso.A} ${\z \to \ditau}$ (red), \qcd
    (blue), and \ewk (green) events and \subfig{Sel.Iso.B} \ttbar
    (orange), \ww (magenta), and ${\z \to \dilep}$ (cyan) events. The
    isolation range excluded by the \mumu selection requirement is
    shaded in grey.\labelfig{Sel.Iso}}
  \svgbeg
  \svg{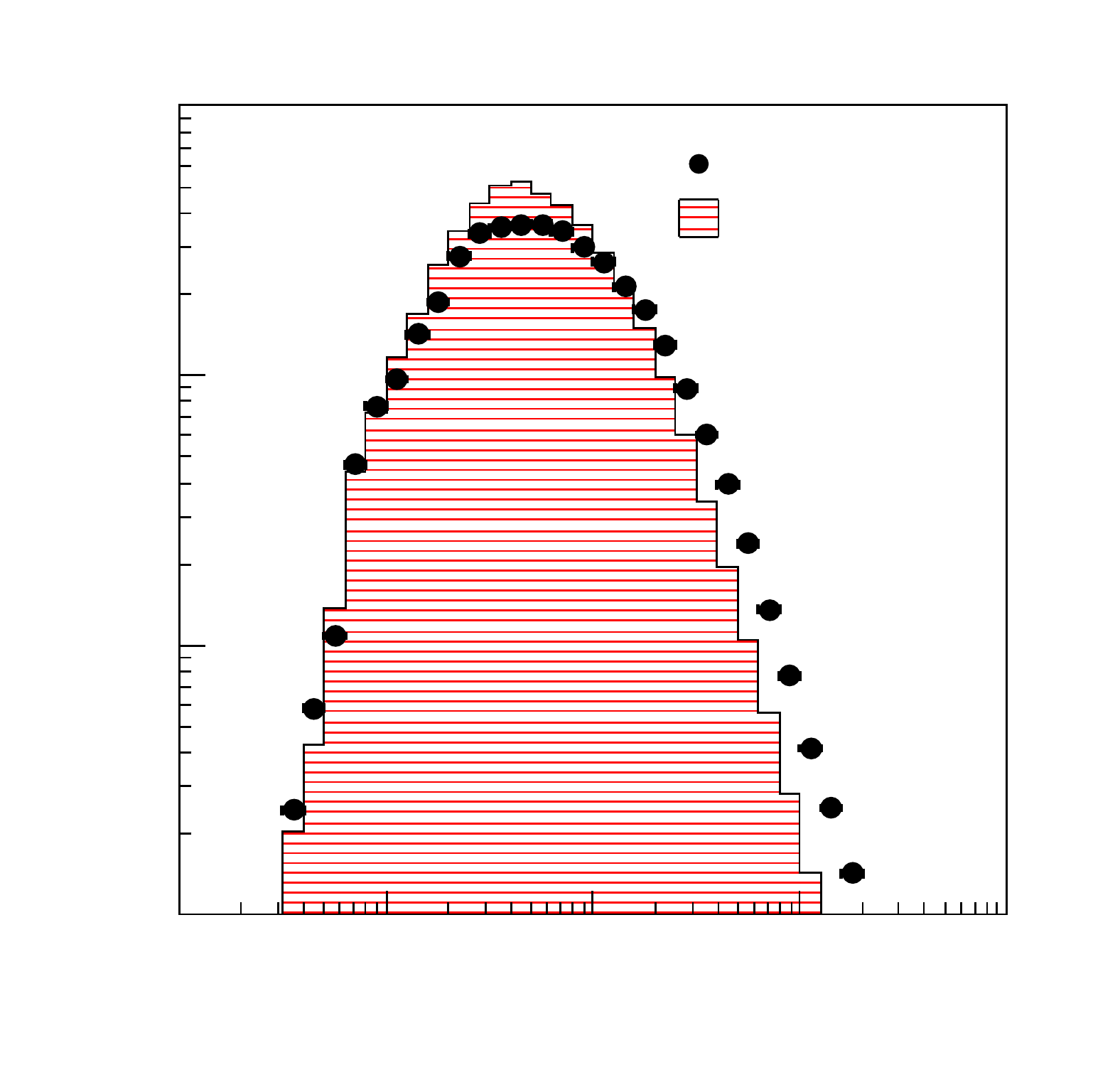}  & \sidecaption \svgsep
  \svg{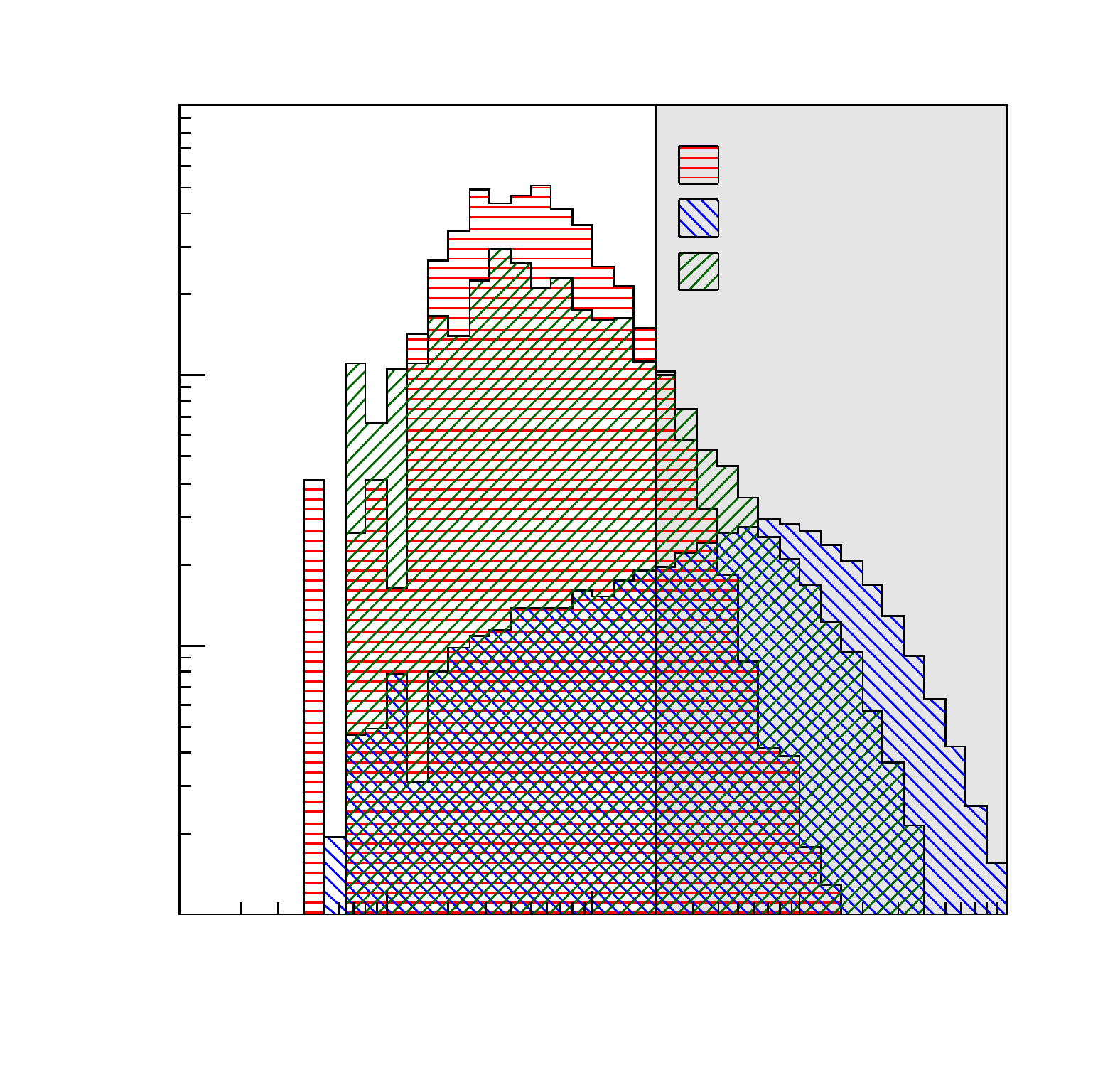} & \svg{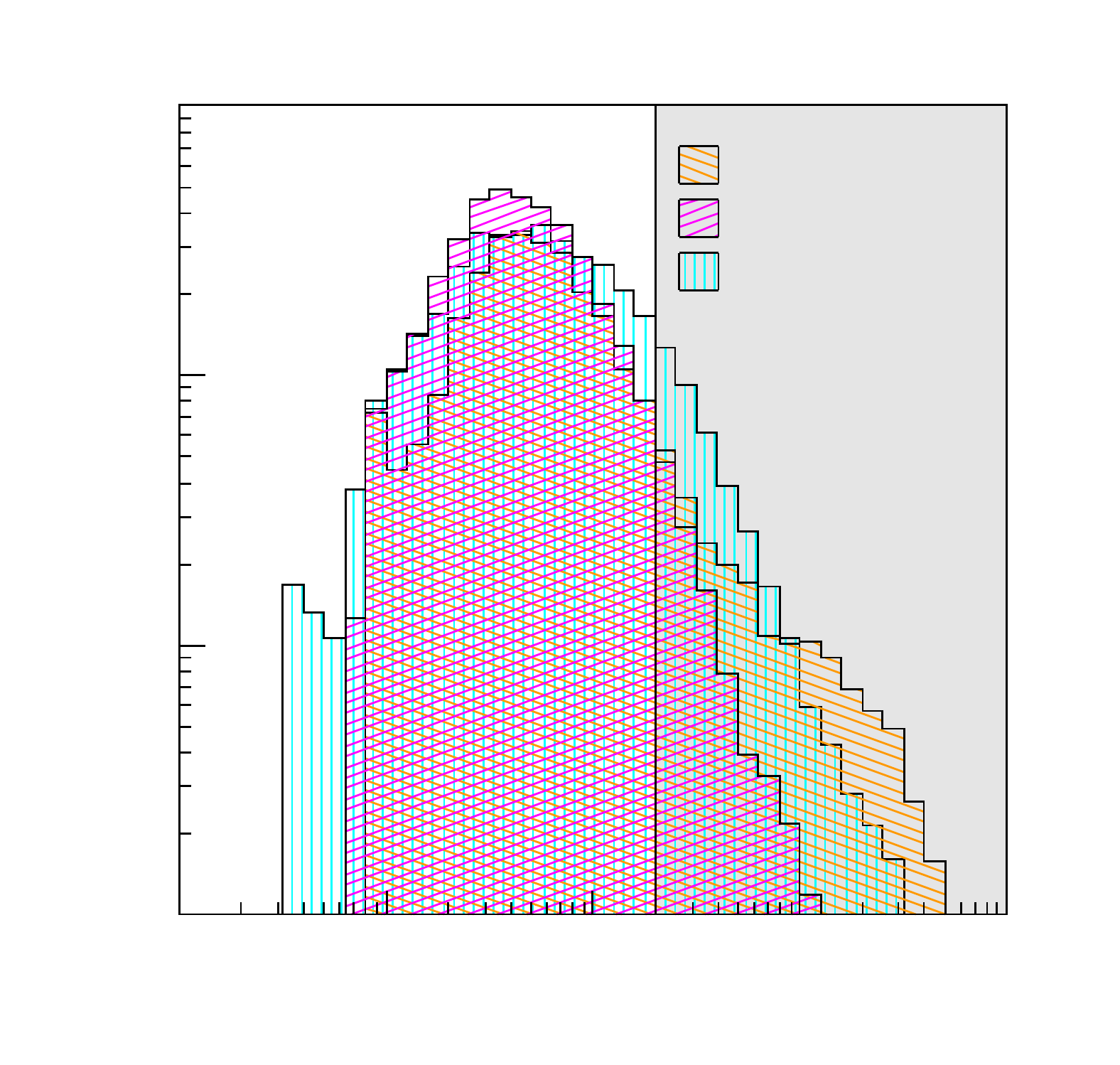} \svgend
\end{subfigures}

\Figs{Zed:Sel.Iso.A} and \ref{fig:Zed:Sel.Iso.B} show the distribution
of the maximum \iso of the two \wtl decay product candidates for
signal and background events. The \wtl decay product candidates
produced from ${\z \to \ditau}$, \ww, and ${\z \to \dilep}$ events are
expected to be relatively isolated, as the candidates are produced
from the decays of massive electroweak bosons, and the jets from the
underlying event will be uncorrelated with the candidate
direction. For \ttbar events, the associated $b$-jets from the decay
of the \wtq slightly contaminate the isolation of the \wtl decay
product candidates.

For the \qcd backgrounds the candidates are produced from jet
activity, resulting in tracks that are not isolated, as can be seen
\fig{Zed:Sel.Iso.A}. Similarly, for the \ewk backgrounds, one of the
\wtl decay product candidates typically is produced from a jet and is
not isolated. The \qcd and \ewk backgrounds are separated from \mumu,
\mue, and \emu signal events by requiring the \iso for both \wtl decay
product candidates to be less than $2~\gev$. For the \muh and \eh
signals a harsher criteria of $\iso < 1~\gev$ is necessary as the
initial \qcd and \ewk backgrounds are larger.

\newsubsubsection{Azimuthal Separation}{}

For \wzbs produced at the \lhc, their \pt is typically small. In the
subsequent decay of the \wzb, transverse momentum must be conserved,
and so the two \wzb decay products are approximately back-to-back in
the transverse plane. For ${\z \to \ditau}$ decays ${\m_\tau \ll
  \m_\z}$, resulting in the decay products of the \wtls produced in a
collinear direction with their parent \wtl, and so the decay products
of the two \wtls will also be back-to-back.

The azimuthal separation of the observed \wtl decay product
candidates, \dphi, is defined as,
\begin{equation}
  \dphi \equiv \begin{cases}
    \abs{\phi_1 - \phi_2}
    & \mathrm{if}~\abs{\phi_1 - \phi_2} \leq \pi \\
    2\pi - \abs{\phi_1 - \phi_2} & \mathrm{else} \\
  \end{cases}
  \labelequ{Phi}
\end{equation}
where values near $\pi~\mathrm{radians}$ indicate events where the two
\wtl decay product candidates are back-to-back. Here $\phi_1$ and
$\phi_2$ are the azimuthal angles of the first and second \wtl decay
product candidates. Given the definition of \equ{Zed:Phi}, \dphi must
range between $0$ and $\pi~\mathrm{radians}$.

A comparison between the ${\z \to \dimu}$ \dphi distributions from
data and simulation is given in \fig{Zed:Sel.Phi.Compare}. The
distribution is described by the decay kinematics of the \wzb which
are well modelled in simulation, and so there is good agreement
between data and simulation.

\begin{subfigures}[t]{2}{\subfig{Sel.Phi.Compare}~A comparison of the
    azimuthal separation distributions between data (points) and
    simulation (red) for ${\z \to \dimu}$ events. Distributions of the
    azimuthal separation between the two \wtl decay product candidates
    for \subfig{Sel.Phi.A}~${\z \to \ditau}$ (red), \qcd (blue), and
    \ewk (green) events and \subfig{Sel.Phi.B}~\ttbar (orange), \ww
    (magenta), and ${\z \to \dilep}$ (cyan) events. The range excluded
    by the azimuthal separation selection requirement is shaded in
    grey.\labelfig{Sel.Phi}}
  \svgbeg
  \svg{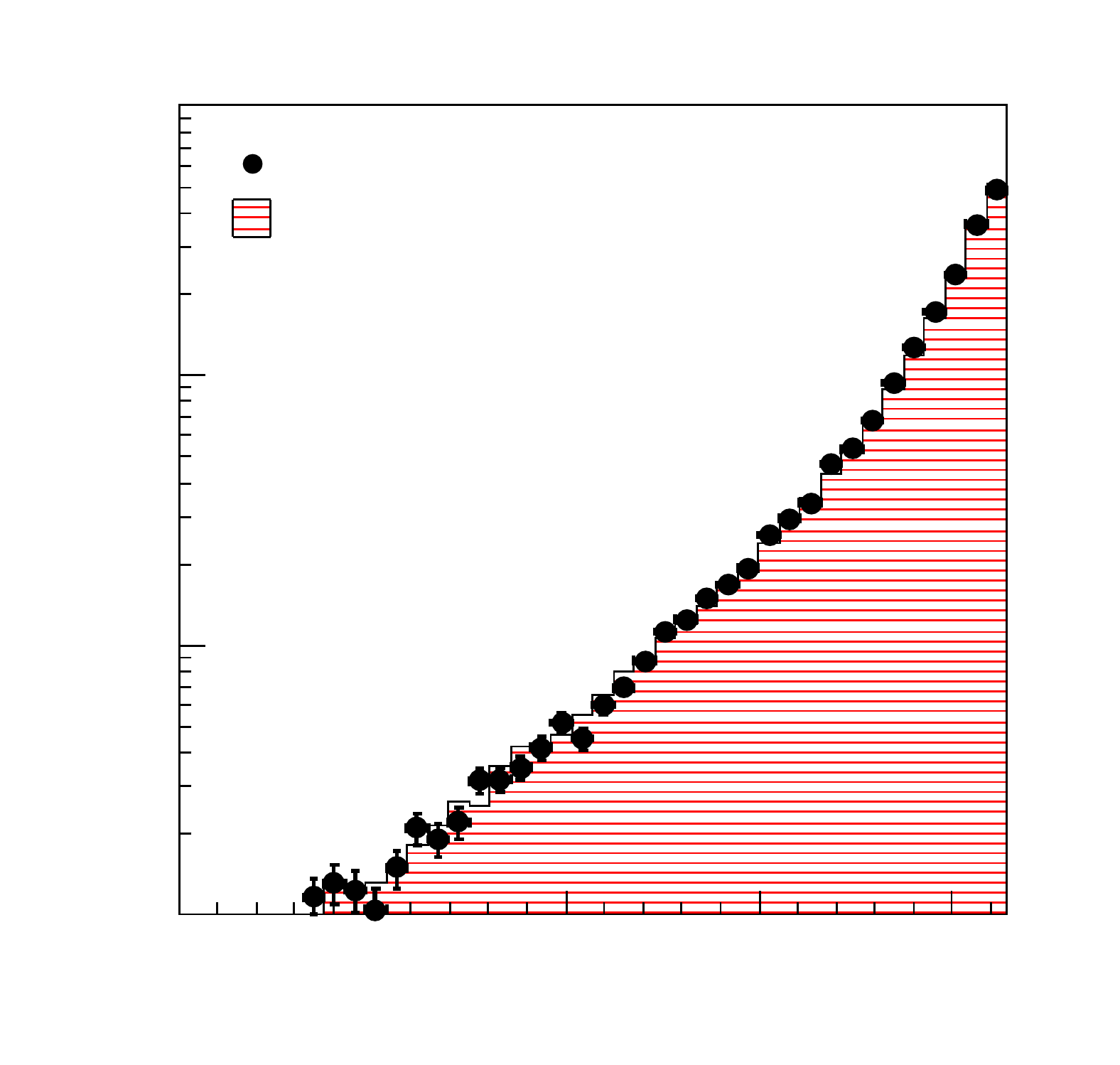} & \sidecaption    \svgsep
  \svg{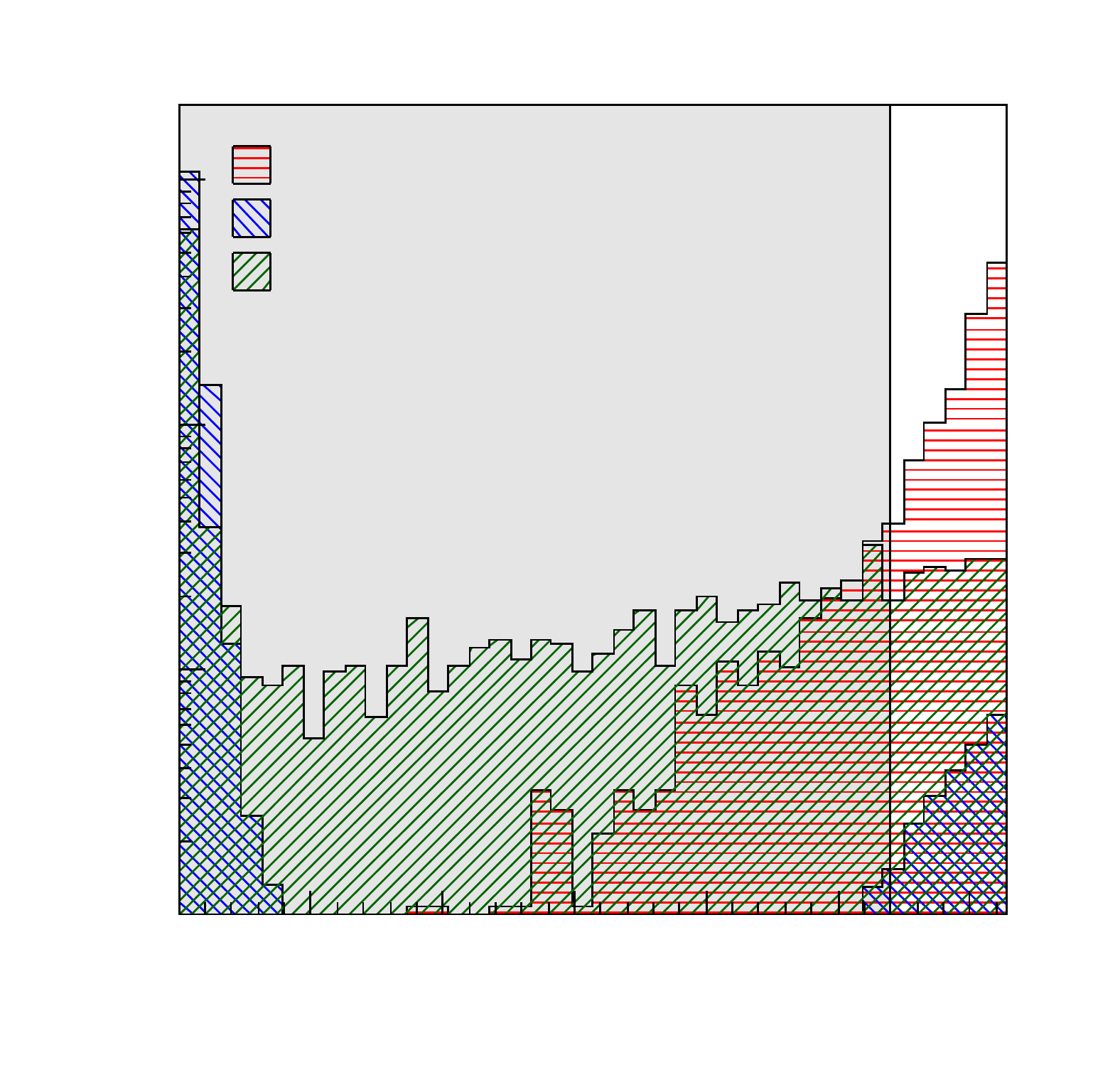}       & \svg{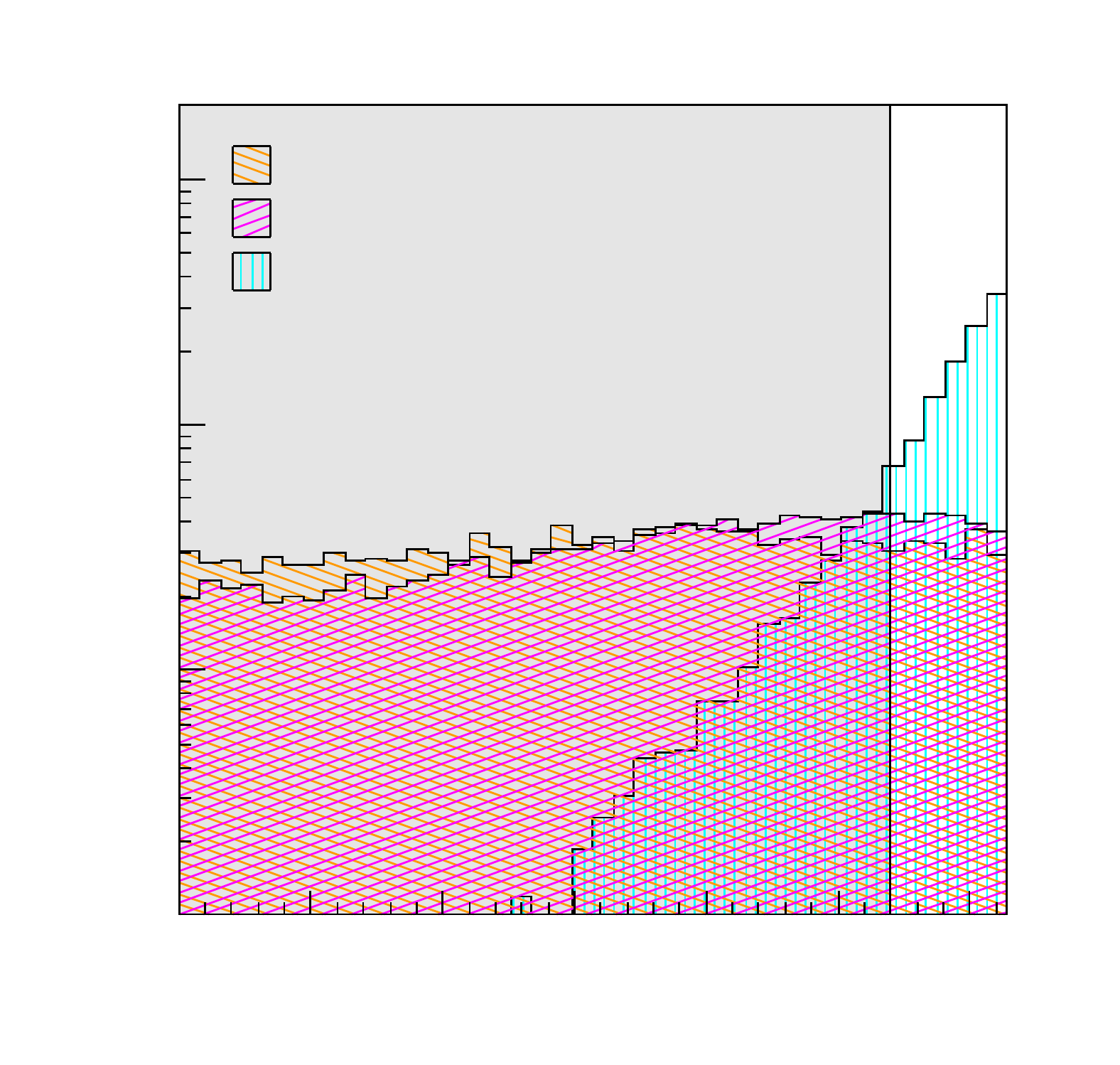} \svgend
\end{subfigures}

The distributions of \dphi are shown in \figs{Zed:Sel.Phi.A} and
\ref{fig:Zed:Sel.Phi.B} for \mumu signal and background events. The
back-to-back nature of ${\z \to \ditau}$ and ${\z \to \dilep}$ events
is apparent and these distributions are nearly identical, validating
the collinear approximation of the \wtl decays. The \wtl decay product
candidates from the \qcd background are primarily produced from the
same jet, and travel approximately in the same direction, producing
events with small \dphi. However, the candidates can also be produced
from di-jet events, contributing to the high \dphi tail of the \qcd
distribution. The \dphi distributions of the \ewk, \ttbar, and \ww
backgrounds are relatively flat, as expected due to the small
correlations between the production of the two candidates. For all
five event categories, $\dphi > 2.7~\mathrm{radians}$ is required.

\newsubsubsection{Impact Parameter Significance}{}

The mean lifetime of the \wtl is experimentally known to be ${0.2906
  \pm 0.0010~\mathrm{ps}}$~\cite{pdg.12.1} and so \wtls produced from
\wzb decays are expected to travel on the order of a centimetre within
LHCb before decaying. While the decay vertex position for a \wtl
decaying into a single visible track cannot be directly measured
within \lhcb, the impact parameter of the \wtl decay product candidate
can be measured. Here, the impact parameter, $\vec{r}$, is defined as
the vector of closest approach between the \wtl decay product
candidate track and the associated primary vertex. The associated
primary vertex is refitted, as described in \sec{Exp:RecTrk}, without
including the candidate track in the fit.

The uncertainty on the impact parameter resolution is dependent upon
the uncertainty of not only the track fit but also the primary vertex
location. However, by summing the signed impact parameters of the two
\wtl decay product candidates, the associated primary vertex
uncertainties can be effectively cancelled \cite{aleph.92.1,
  ferrante.95.1}, yielding a variable more sensitive to the lifetime
of the \wtl. The signed impact parameter, $r$, is defined as,
\begin{equation}
  r \equiv \abs{\vec{r}} \left(\frac{\left(\vec{r} \times \vec{p}
    \right)\cdot \hat{k}}{\abs{\left(\vec{r} \times \vec{p}\right)
      \cdot \hat{k}}}\right)
  \labelequ{Sip}
\end{equation}
where $\vec{p}$ is the momentum of the track and $\hat{k}$ is the
$z$-direction unit vector. A diagram of the signed impact parameter is
shown in \fig{Zed:Sel.Sip}.

\begin{subfigures}[t]{2}{Schematic of the components used to build the
    signed impact parameter, $r$, of \equ{Zed:Sip}. The diagram on the
    left-hand side is in the plane of the particle momentum and impact
    parameter. The diagram on the right-hand side is in the plane of
    ${\vec{r} \times \vec{p}}$ and the $z$-axis.}
  \svgbeg 
  \svg[1]{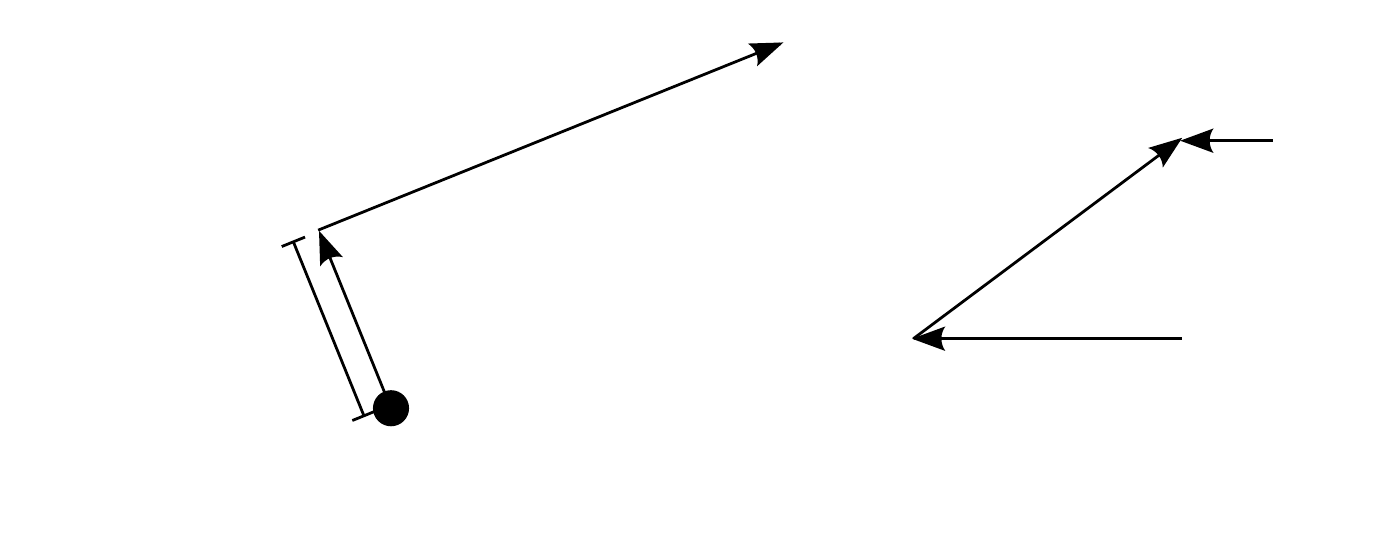} & \sidecaption \svgend
\end{subfigures}

The summed impact parameter significance, \ips, can then be written
as,
\begin{equation}
  \ips \equiv \frac{\abs{r_1 + r_2}}{\delta_1 \oplus \delta_2}
  \labelequ{Ips}
\end{equation}
where $r_1$ and $r_2$ are the signed impact parameters for the first
and second \wtl decay product candidates and $\delta_1$ and $\delta_2$
are their associated uncertainties calculated from the track fit
covariance matrices.

For accurate simulation of the \ips, correct modelling of the track
fit and its associated uncertainty is critical. The \ips is
underestimated in simulation as demonstrated in
\fig{Zed:Sel.Ips.Compare} which plots the \ips distributions for ${\z
  \to \dimu}$ events from data and simulation. The \ips distribution
from simulation can be corrected to match the distribution from data
by multiplying the \ips for each simulated event by a factor
\fscale. An $\fscale$ of ${1.12 \pm 0.01}$ is found to minimise the
$\chi^2$ between the data distribution and the corrected simulated
distribution, and is used to calibrate all simulated samples. This
calibration is assumed to propagate from the lower \ips values of ${\z
  \to \dimu}$ events to the higher \ips values of ${\z \to \ditau}$
events.

\begin{subfigures}[t]{2}{\subfig{Sel.Phi.Compare}~A comparison of the
    impact parameter significance distributions between data (points),
    uncalibrated simulation (red), and calibrated (blue) simulation
    for ${\z \to \dimu}$ events. Distributions of the impact parameter
    significance for \subfig{Sel.Phi.A}~${\z \to \ditau}$ (red), \qcd
    (blue), and \ewk (green) events and \subfig{Sel.Phi.B}~\ttbar
    (orange), \ww (magenta), and ${\z \to \dilep}$ (cyan) events. The
    distributions from simulation are corrected with the \fscale
    factor described in the text and the range excluded by the impact
    parameter significance selection requirement is shaded in
    grey.\labelfig{Sel.Ips}}
  \svgbeg 
  \svg{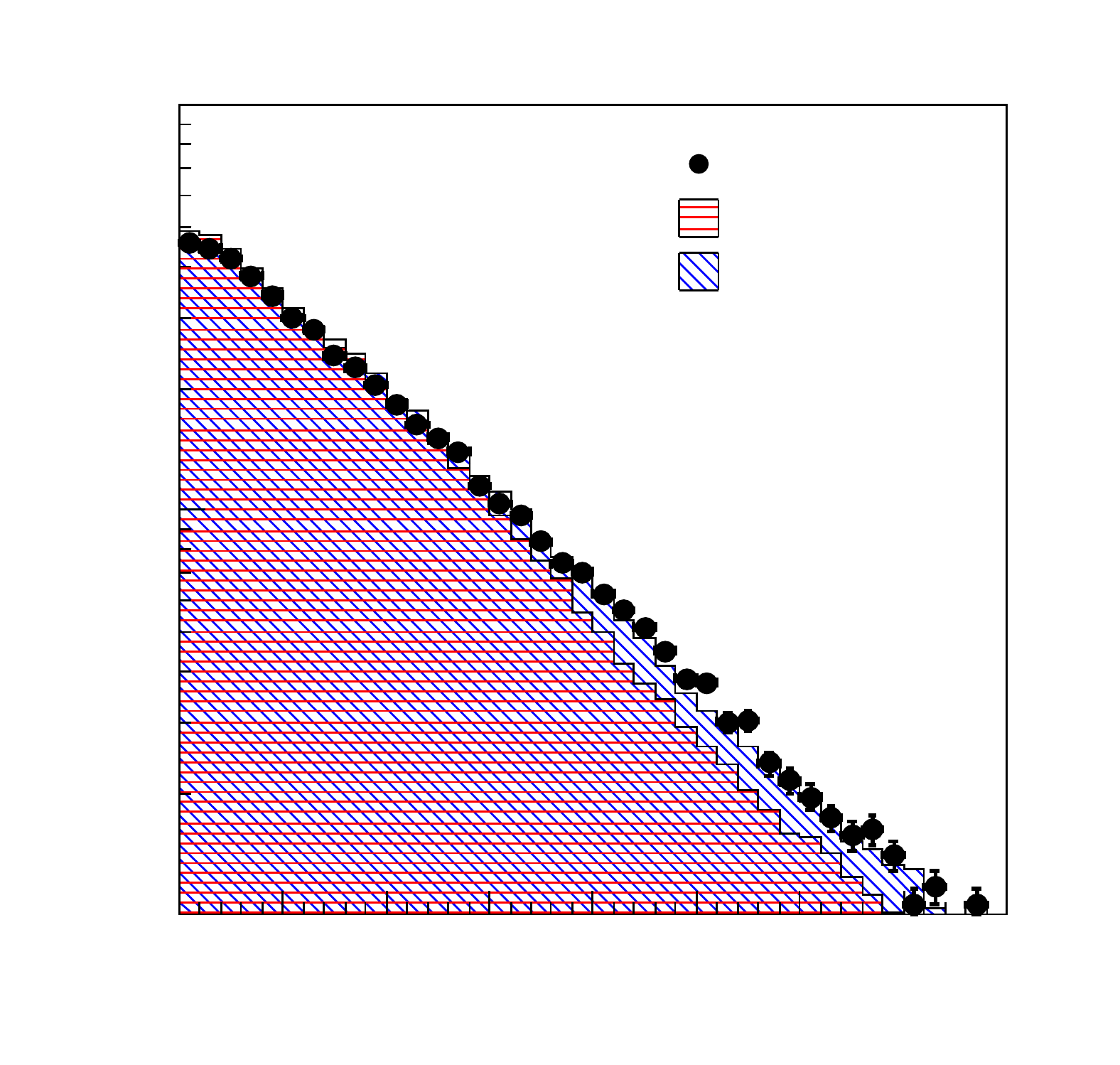} & \sidecaption    \svgsep
  \svg{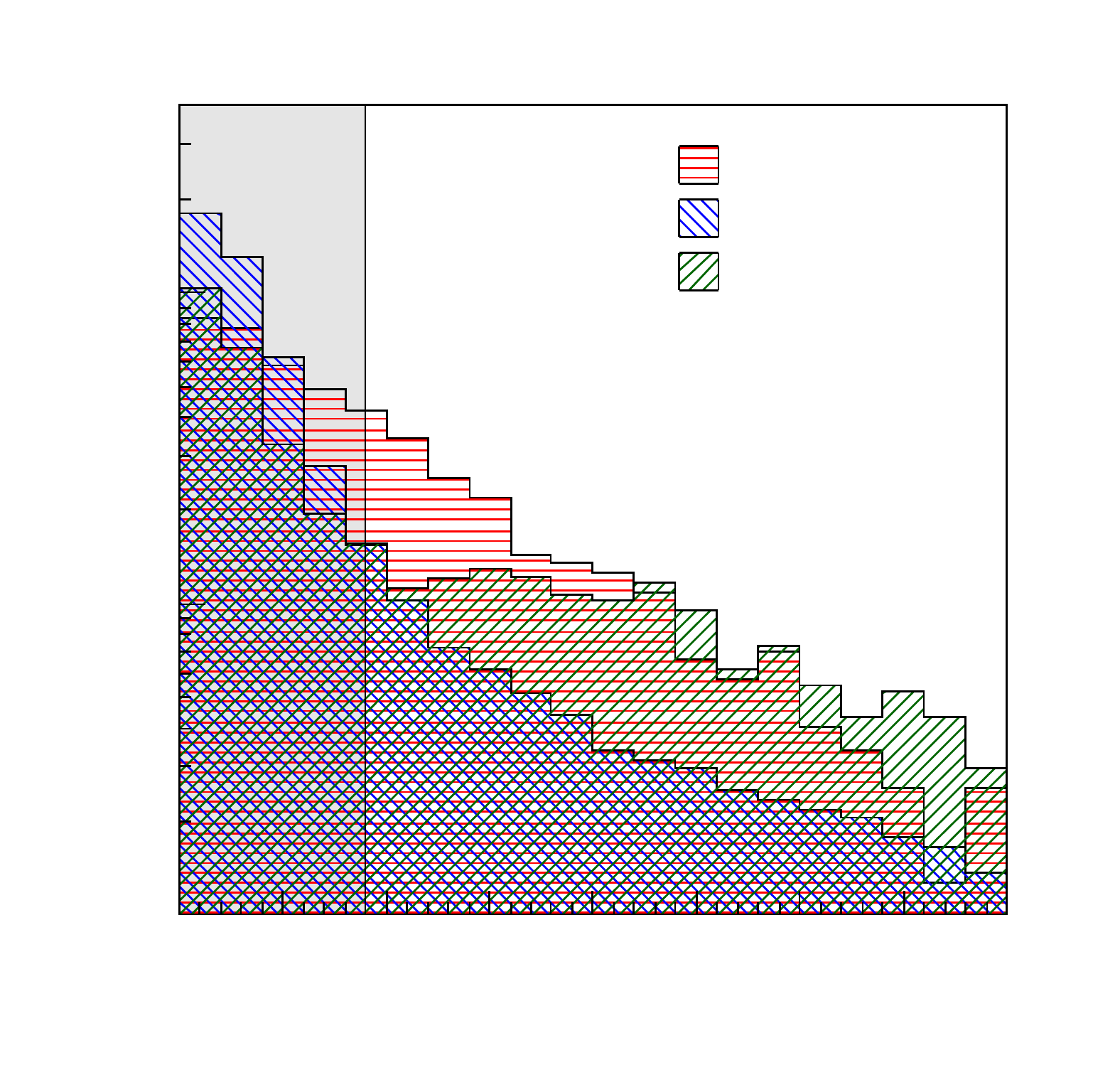}       & \svg{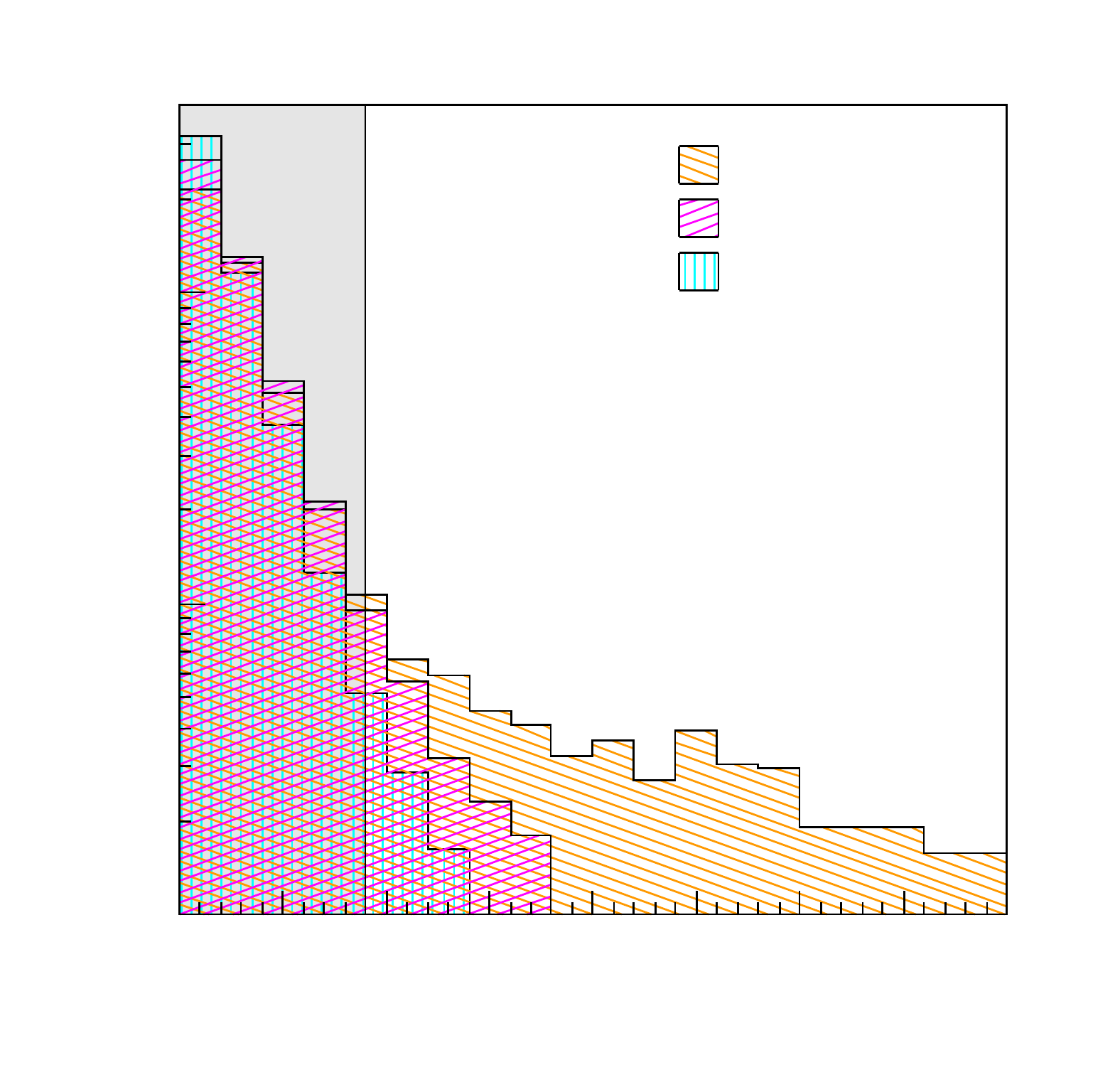} \svgend
\end{subfigures}

In \Fig{Zed:Sel.Ips.A} the \ips distribution for ${\z \to \ditau}$
signal events of the \mumu category is shown, as well as the
distributions for the \qcd and \ewk backgrounds. The signal
distribution has a longer tail than the \qcd distribution as most
particles produced within a jet are either stable or have a shorter
lifetime than the \wtl. However, heavy-flavour mesons can have longer
lifetimes than the \wtl and produce the harder tail of the \qcd
distribution. The \ewk background has a much harder \ips spectrum than
either the signal or \qcd distributions, which could be a result of
the two candidates being produced from a separate electroweak boson
and jet in the event.

The \ips distributions for the \ttbar, \ww, and ${\z \to \dimu}$
backgrounds are provided in \fig{Zed:Sel.Ips.B}. The \ttbar background
distribution has a much harder tail than both the signal distribution
and all other background distributions. This is most probably caused
by one of the candidates being produced by a heavy-flavour meson decay
from one of the $b$-jets. The \ww and ${\z \to \dimu}$ distributions,
however, are both softer than the ${\z \to \ditau}$ signal
distribution. The ${\z \to \dimu}$ distribution is softer because both
\wtl decay product candidates are produced directly from the same
\wzb, while the \ww distribution is softer because the lifetime of
\wwbs is approximately twelve orders of magnitude smaller than the
\wtl.

For the \mumu, \muh, and \eh categories a requirement of $\ips > 9$ is
applied, as indicated by the shaded grey areas of \figs{Zed:Sel.Ips.A}
and \ref{fig:Zed:Sel.Ips.B}. For the \mumu category this reduces the
dominant ${\z \to \dilep}$ background, while for the \muh and \eh
categories this requirement reduces the \qcd backgrounds. The \ips
requirement is not necessary for the cleaner \mue and \emu categories.

\newsubsubsection{Transverse Momentum Asymmetry}{}

While the selection requirements detailed above are sufficient for
separating signal from background for the \mue, \emu, \muh, and \eh
event categories, these requirements are inadequate in eliminating the
${\z \to \dimu}$ background from \mumu signal events. The transverse momentum
asymmetry is defined as,
\begin{equation}
  \apt \equiv \frac{\abs{\pt_1 - \pt_2}}{\pt_1 + \pt_2}
\end{equation}
where $\pt_1$ and $\pt_2$ are the transverse momenta of the first and
second \wtl decay product candidates respectively. The muons from
background ${\z \to \dimu}$ events will have balanced \pt, resulting
in a low \apt while ${\z \to \ditau}$ signal events have missing
neutrinos, oftentimes resulting in a larger \apt.

\begin{subfigures}{2}{\subfig{Sel.Phi.Compare}~A comparison of the \pt
    asymmetry distributions between data (points) and simulation (red)
    for ${\z \to \dimu}$ events. Distributions of the \pt asymmetry
    between the two \wtl decay product candidates for
    \subfig{Sel.Phi.A}~${\z \to \ditau}$ (red), \ewk (blue), and \qcd
    (green) events and \subfig{Sel.Phi.B}~\ttbar (orange), \ww
    (magenta), and ${\z \to \dilep}$ (cyan) events. The range excluded
    by the \pt asymmetry selection requirement is shaded in
    grey.\labelfig{Sel.Apt}}
  \svgbeg
  \svg{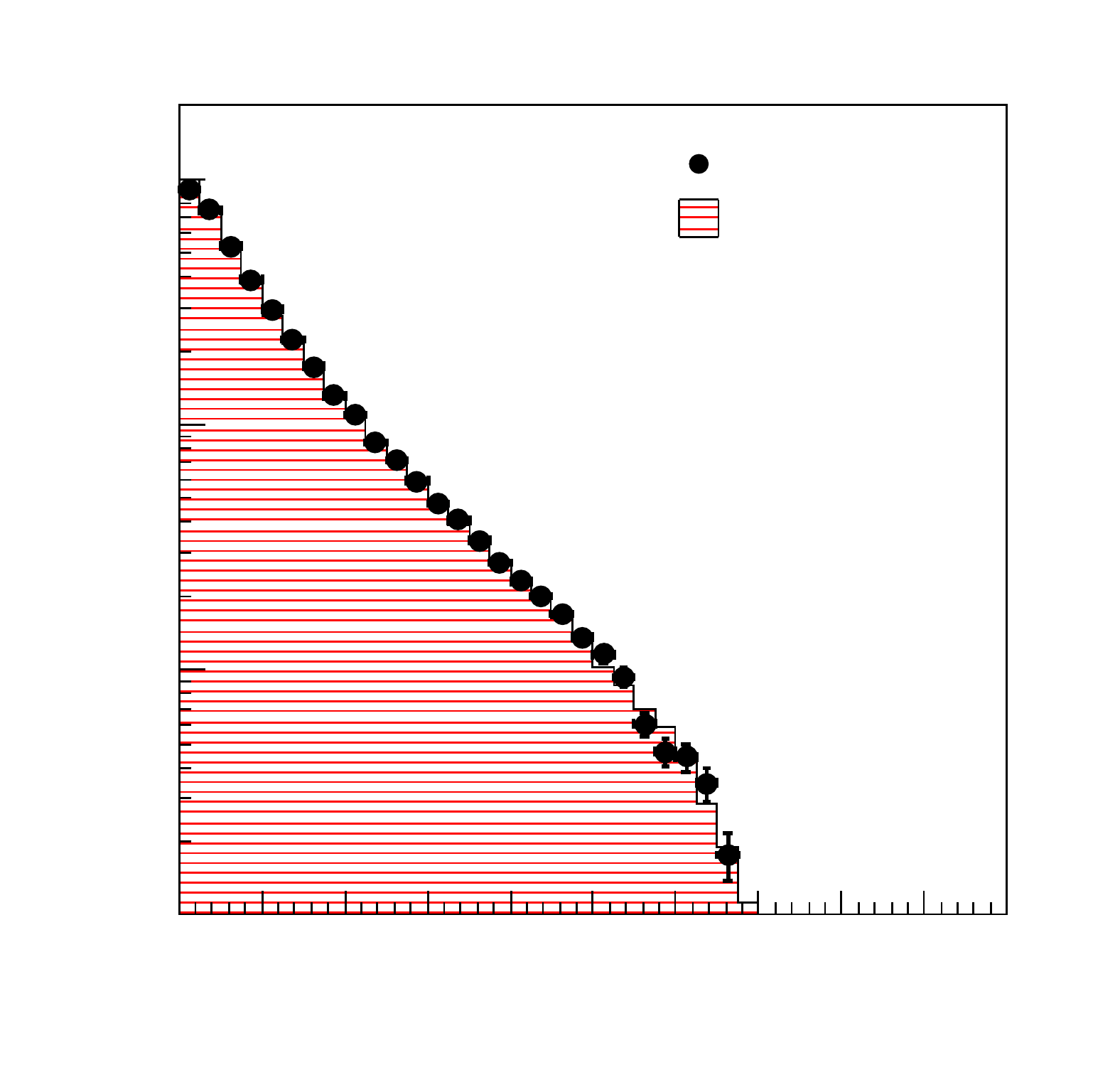} & \sidecaption    \svgsep
  \svg{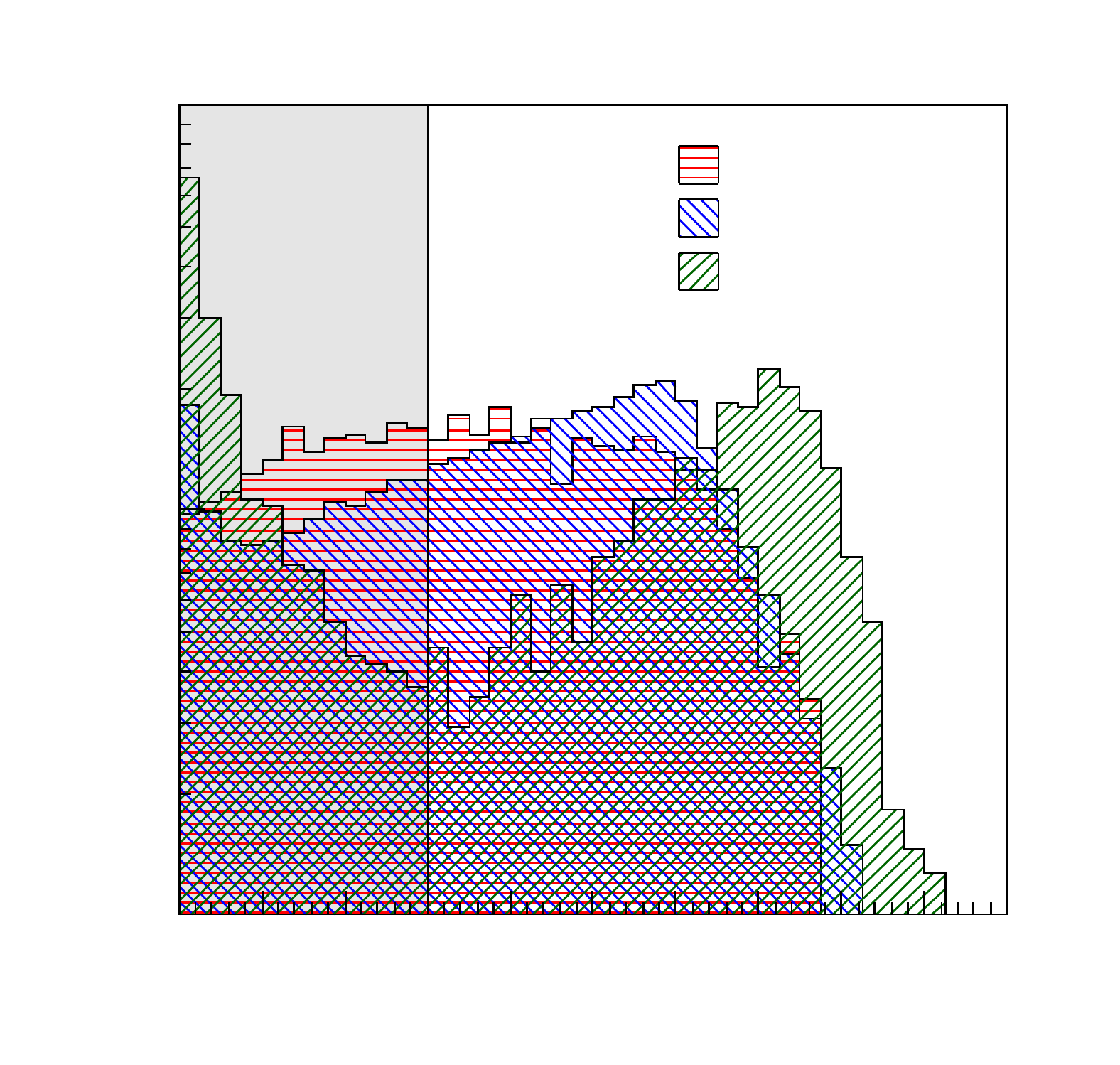}       & \svg{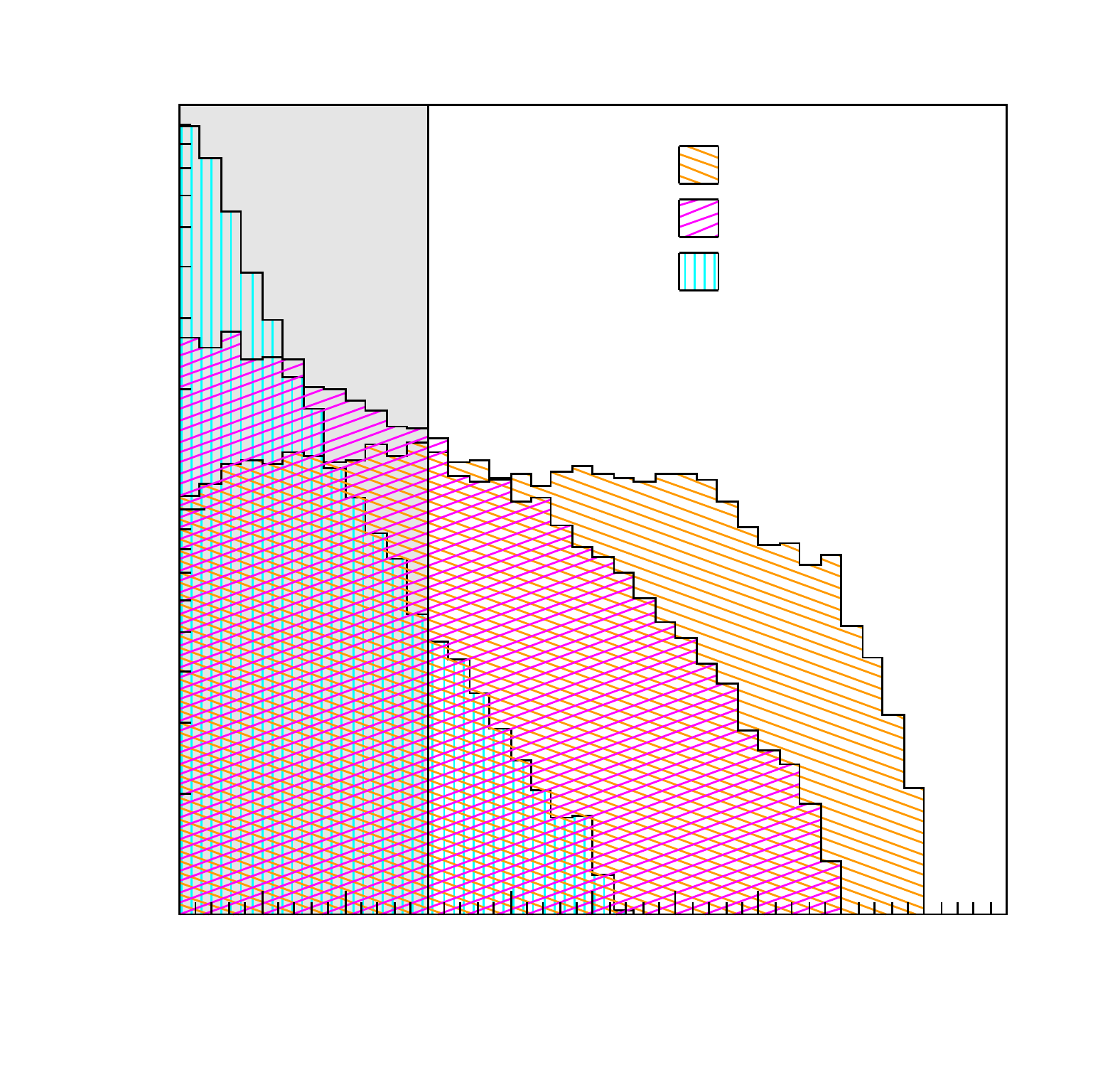} \svgend
\end{subfigures}

The \apt is primarily dependent upon the decays of the parent
particle, and for the case of ${\z \to \ditau}$, the decays of the
\wtls. These decays are well modelled in simulation, and so no
calibration of the simulation is necessary, as can be seen in the \apt
distributions for ${\z \to \dimu}$ events from data and simulation of
\fig{Zed:Sel.Apt.Compare}. The \apt distributions for ${\z \to
  \ditau}$, \qcd, and \ewk events for the \mumu category are shown in
\fig{Zed:Sel.Apt.A}, while the \apt distributions for the \ttbar, \ww,
and ${\z \to \dimu}$ backgrounds are shown in \fig{Zed:Sel.Apt.B}. As
can be seen, the \apt for ${\z \to \ditau}$ events can be large, while
the \apt for ${\z \to \dimu}$ background events is much smaller. For
the \mumu event category, the criteria $\apt > 0.3$ is required, as
shown by the grey exclusions. No \apt minimum is required for the
\mue, \emu, \muh, and \eh event categories.

\begin{table}\centering
  \captionabove{A summary of the event selection requirements: invariant
    mass (\m), charged track isolation (\iso), azimuthal separation
    (\dphi), impact parameter significance (\ips), and \pt asymmetry
    (\apt). The requirements applied to each of the five event
    categories is given.\labeltab{Sel}}
  \begin{tabular}{l|l|l|l|l|l}
    \toprule
    & \multicolumn{1}{c|}{\mumu} & \multicolumn{1}{c|}{\mue} 
    & \multicolumn{1}{c|}{\emu} & \multicolumn{1}{c|}{\muh} 
    & \multicolumn{1}{c}{\eh} \\
    \midrule
    \m $[\gev]$ & $> 100$ or $20 < \m < 80$  & $> 20$ & $> 20$ 
    & $> 20$ & $> 20$  \\
    \iso $[\gev]$ & $< 2$ & $< 2$ & $< 2$ & $< 1$ & $< 1$ \\
    \dphi $[\rad]$ & $> 2.7$ & $> 2.7$ & $> 2.7$ & $> 2.7$ & $> 2.7$ \\
    \ips & $> 9$ & \multicolumn{1}{c|}{$-$} &
    \multicolumn{1}{c|}{$-$} 
    & $> 9$ & $> 9$ \\
    \apt & $> 0.3$ & \multicolumn{1}{c|}{$-$} & \multicolumn{1}{c|}{$-$} 
    & \multicolumn{1}{c|}{$-$} & \multicolumn{1}{c}{$-$} \\
    \bottomrule
  \end{tabular}
\end{table}

A summary of the event selection requirements placed on the five
variables of this section for each event category is given in
\tab{Zed:Sel}. Only the \mumu category places requirements on all five
variables, with the \ips and \apt requirements designed to remove ${\z
  \to \dimu}$ background. The \ips requirement is also kept for the
\muh and \eh categories to help reduce \qcd background. The \mue and
\emu categories have much less \qcd background than the \muh and \eh
categories and do not utilise the \ips requirement.

\newsubsection{Background Estimation}{Bkg}

The selection criteria developed in \sec{Zed:Sel} are applied to data
to select ${\z \to \ditau}$ signal events. In order to calculate the
${\dip \to \z \to \ditau}$ cross-section in \sec{Zed:Sig} the number
of background events in the selected data must be estimated. The
methods used to estimate the number of events from the backgrounds
categorised in the beginning of \sec{Zed:Ana} and shown in
\fig{Zed:Backgrounds} are now described.

The invariant mass distributions for ${\z \to \ditau}$ candidates from
data for all five event categories, together with the estimated
backgrounds which are described in the remainder of this section, are
given in \fig{Zed:Mass}. No events were observed in data above an
invariant mass of $120~\gev$. The simulated signal is normalised to
the difference between the number of observed and estimated background
events. A summary of these values is given in
\tab{Zed:Events}. Further validation plots are provided in
\figs{Zvr:Eta} through \ref{fig:Zvr:Eta2} of \app{Zvr} for the \pt and
$\eta$ distributions of the combined \wtl decay product candidates, as
well as the \pt and $\eta$ distributions of the individual candidates
and the number of primary vertices for the event.

\begin{table}\centering
  \captionabove{Estimated number of events for each background component
    and their sum, together with the observed number of candidates for
    each of the five event categories.\labeltab{Events}}
  \settowidth{\backspace}{$1$}
  \begin{tabular}{>{$}l<{$}|E|E|E|E|E}
    \toprule
    & \multicolumn{2}{c|}{\mumu} & \multicolumn{2}{c|}{\mue} &
    \multicolumn{2}{c|}{\emu} & \multicolumn{2}{c|}{\muh} &
    \multicolumn{2}{c}{\eh} \\
    \midrule
    \qcd 
    & 11.7&3.4 & 72.4 &2.2  &\p54.0&3.0  & 41.9 &0.5 & 24.5&0.6 \\
    \ewk
    & 0.0 &3.5 & 40.3 &4.3  & 0.0  &1.3  & 10.8 &0.5 & 9.3 &0.5 \\
    \ttbar
    & <0.1&0.1 & 3.6  &0.4  & 1.0  &0.1  & <0.1 &0.1 & 0.7 &0.4 \\
    \ww 
    & <0.1&0.1 & 13.3 &1.2  & 1.6  &0.2  & 0.2  &0.1 & <0.1&0.1 \\
    {\z \to \dilep}
    & 29.8&7.0 & \multicolumn{2}{c|}{$-$} & \multicolumn{2}{c|}{$-$}
    & 0.4 &0.1 & 2.0&0.2 \\
    \midrule
    {\rm Background}
    & 41.6&8.5 & 129.7&4.9  & 56.6 &3.3  & 53.3 &0.8 & 36.6&0.9 \\
    \midrule
    {\rm Observed}
    & \multicolumn{2}{l|}{124} & \multicolumn{2}{l|}{421} 
    & \multicolumn{2}{l|}{155} & \multicolumn{2}{l|}{189}
    & \multicolumn{2}{l}{101} \\
    \bottomrule
  \end{tabular}
\end{table}

\begin{subfigures}[p]{2}{The invariant mass distributions of the two \wtl
    decay product candidates from data (points) for the
    \subfig{Z2TauTau2MuMu.Mass}~\mumu,
    \subfig{Z2TauTau2MuE.Mass}~\mue,
    \subfig{Z2TauTau2EMu.Mass}~\emu,
    \subfig{Z2TauTau2MuPi.Mass}~\muh, and
    \subfig{Z2TauTau2EPi.Mass}~\eh categories. The simulated
    signal (red) is normalised to the number of signal events, while
    the \qcd (blue), \ewk (green), \ttbar (orange), \ww (magenta), and
    ${\z \to \dilep}$ (cyan) backgrounds are estimated as described in
    the text.\labelfig{Mass}}
  \svgbeg
  \svg{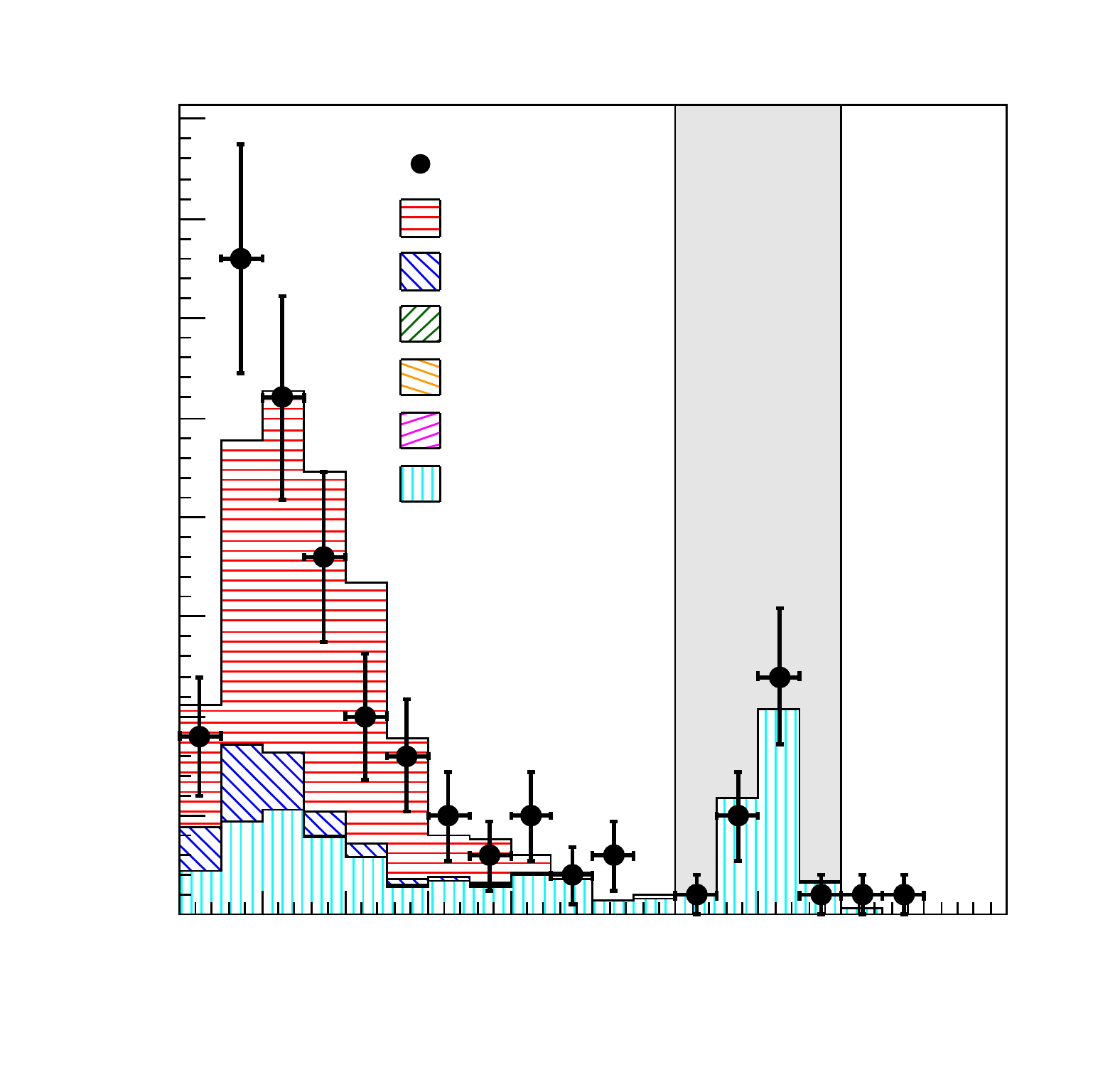} & \svg{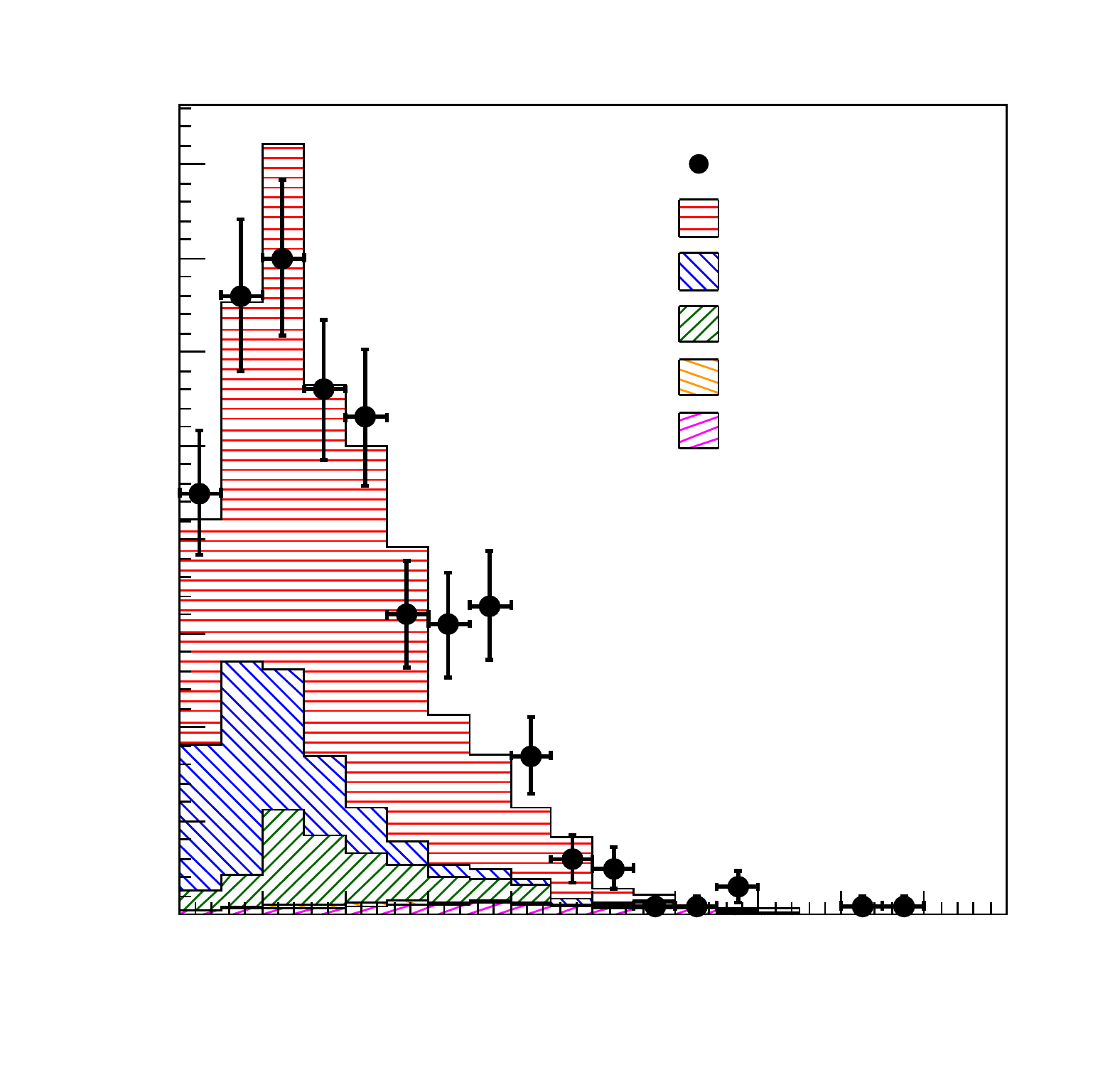}  \svgsep
  \svg{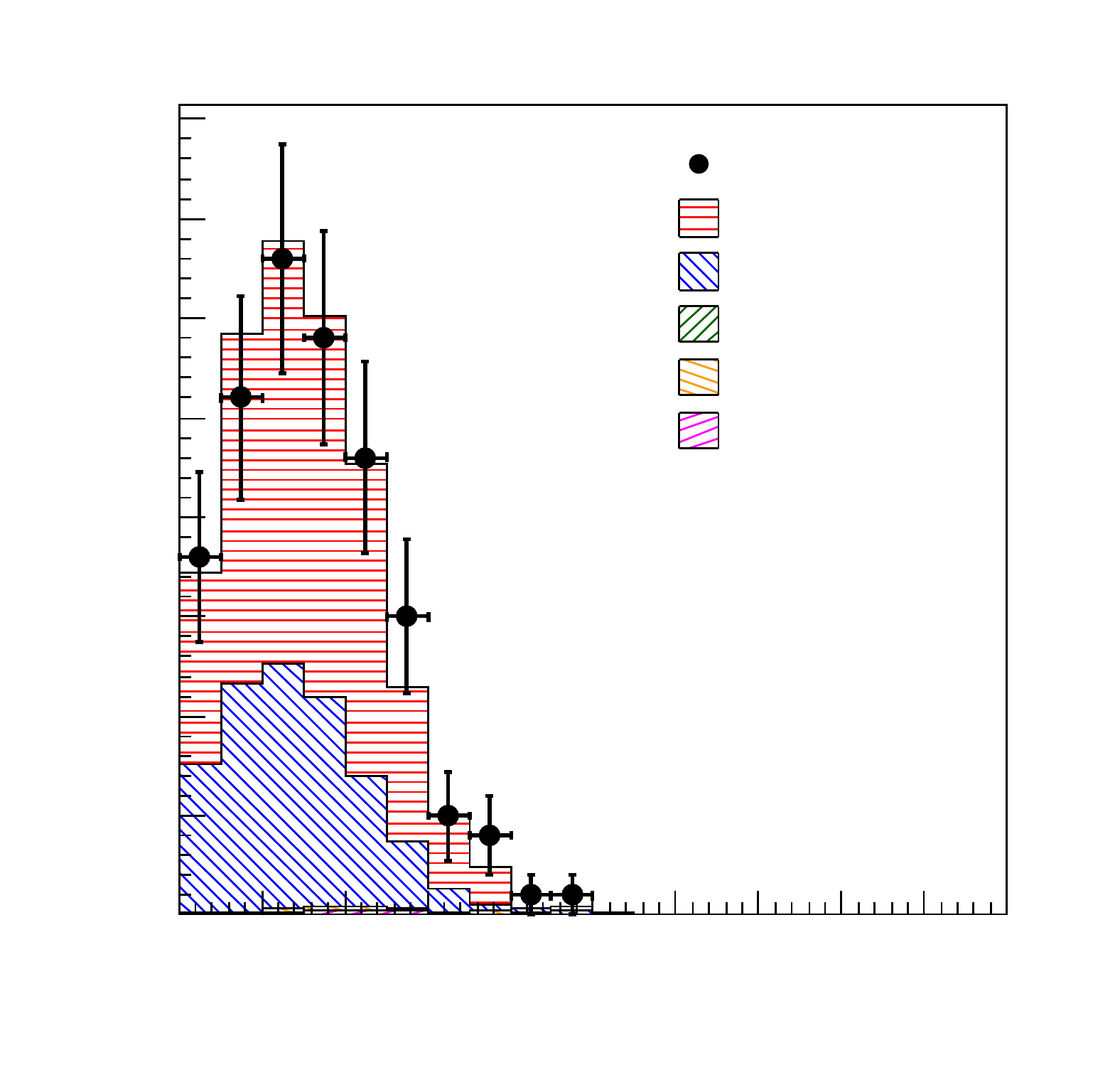}  & \svg{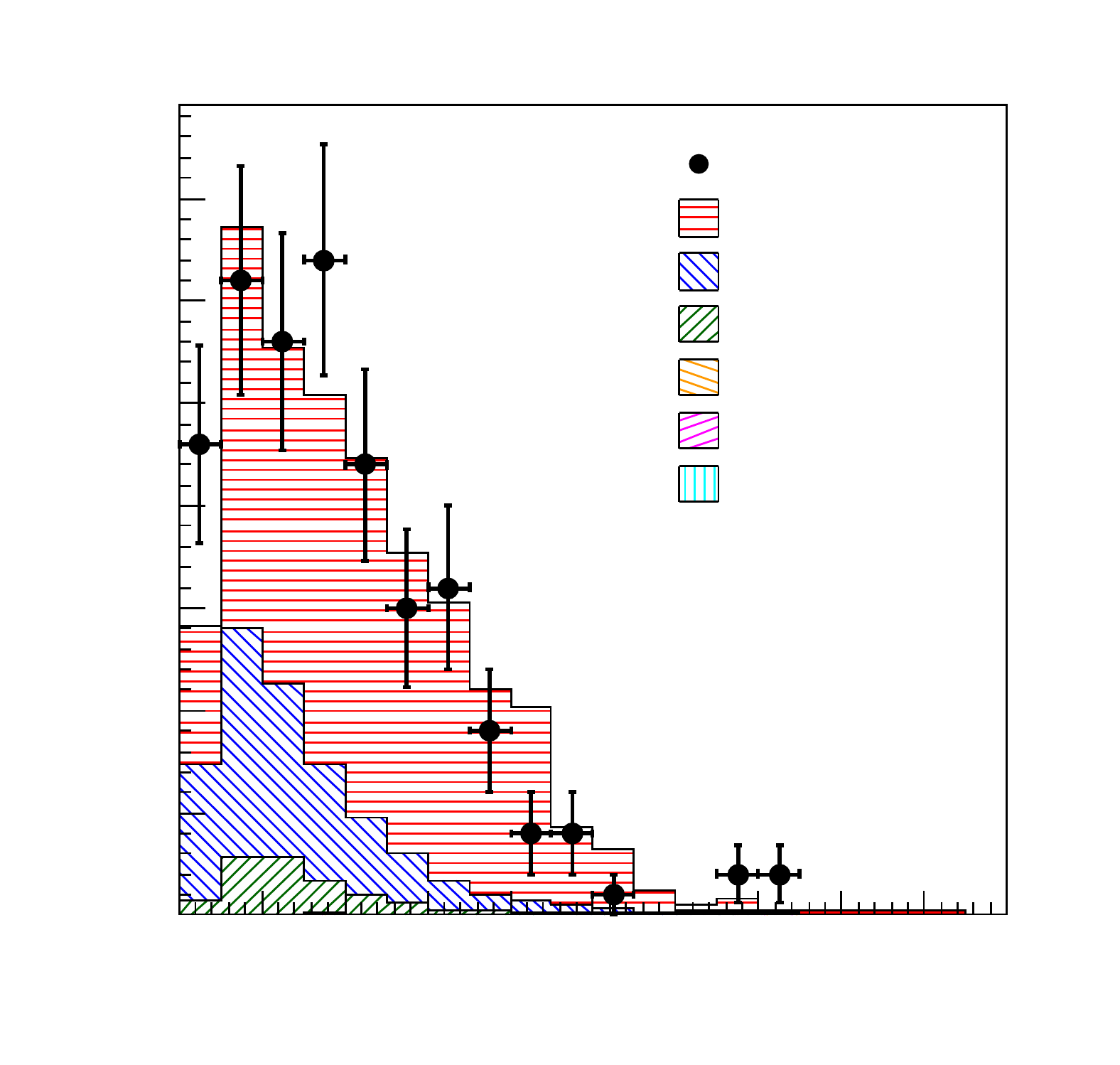} \svgsep
  \svg{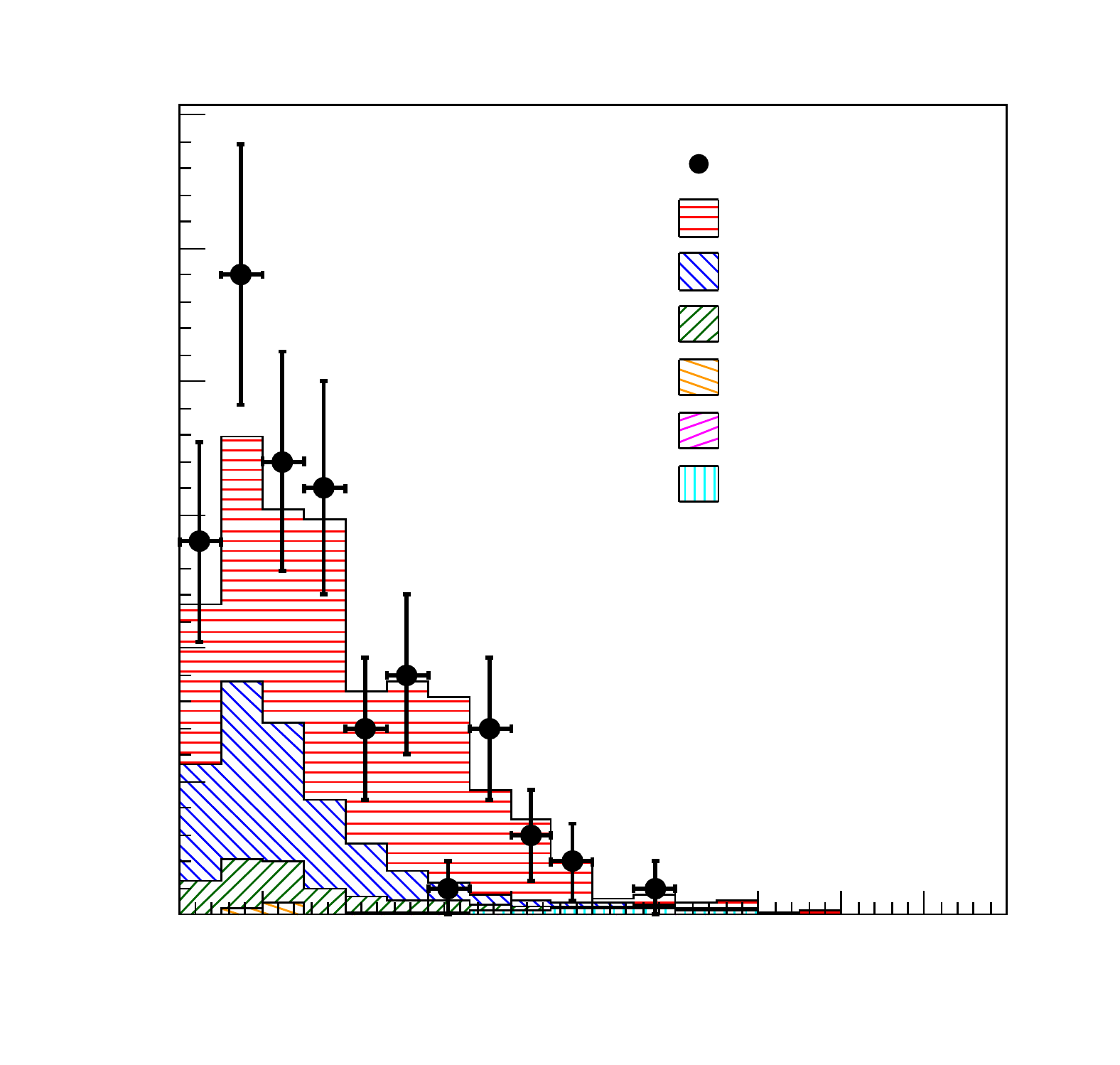}  & \sidecaption             \svgend
\end{subfigures}

\newsubsubsection{QCD and EWK Backgrounds}{}

The number of \qcd and \ewk background events given in
\tab{Zed:Events} are estimated from data. A signal-depleted data
sample is produced by applying the selection criteria of \sec{Zed:Sel}
to data, but requiring the two \wtl decay product candidates have the
same-sign charge. The distributions from the same-sign sample of the
\pt difference between the first and second \wtl decay product
candidates, ${\pt_1 - \pt_2}$, are fitted with \qcd
and \ewk templates to determine the number of same-sign charge \qcd
and \ewk events, \sqcd and \sewk. The same-sign \qcd template is taken
from data fulfilling the selection requirements of \sec{Zed:Sel}, but
requiring ${\iso > 10~\gev}$ for both \wtl decay product
candidates. The \ewk template is taken from simulation without the
\iso selection applied.

The ${\pt_1 - \pt_2}$ distributions from same-sign data, with their
respective \qcd and \ewk template fits, are given in
\fig{Zed:QcdEwkFit} for the five event categories. In \qcd events both
\wtl decay product candidates are typically produced from the same
jet, resulting in a similar \pt for the two candidates and a small \pt
difference. For \ewk events the first candidate produced is generally
from an electroweak boson and will have a hard \pt, while the second
candidate will have a softer \pt from jet activity, and so the \pt
difference distribution is shifted upwards to larger \pt differences
than the \qcd distribution.

\begin{subfigures}[p]{2}{The distributions of the \pt difference between
    the two \wtl decay product candidates for opposite-sign data
    (points) and fitted \qcd (blue) and \ewk (green)
    templates. These fits are used to determine the \qcd and \ewk
    backgrounds, as described in the text, and are given for the
    \subfig{Z2TauTau2MuMu.QcdEwkFit}~\mumu,
    \subfig{Z2TauTau2MuE.QcdEwkFit}~\mue,
    \subfig{Z2TauTau2EMu.QcdEwkFit}~\emu,
    \subfig{Z2TauTau2MuPi.QcdEwkFit}~\muh, and
    \subfig{Z2TauTau2EPi.QcdEwkFit}~\eh
    categories. \labelfig{QcdEwkFit}}
  \svgbeg
  \svg{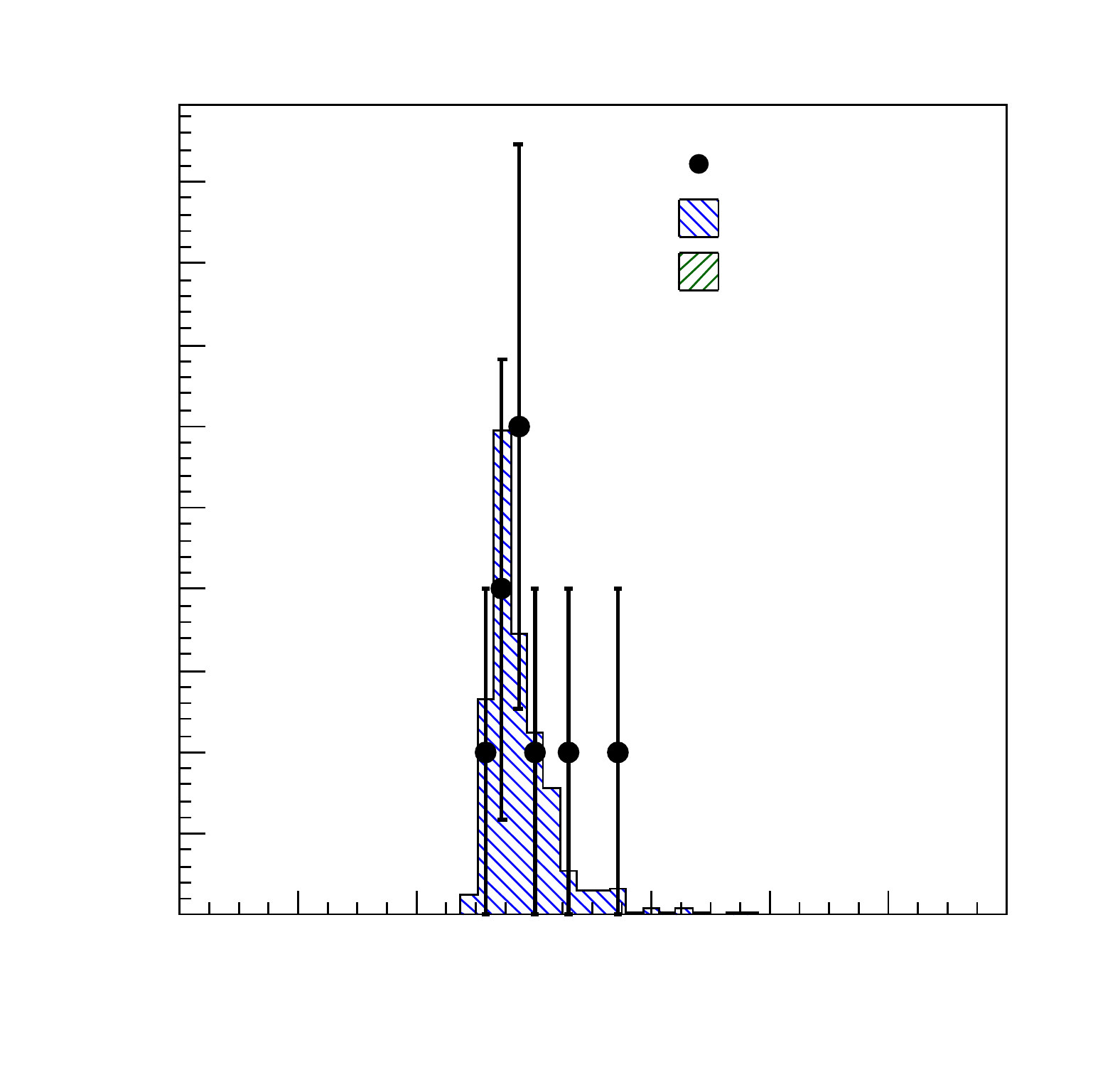} & \svg{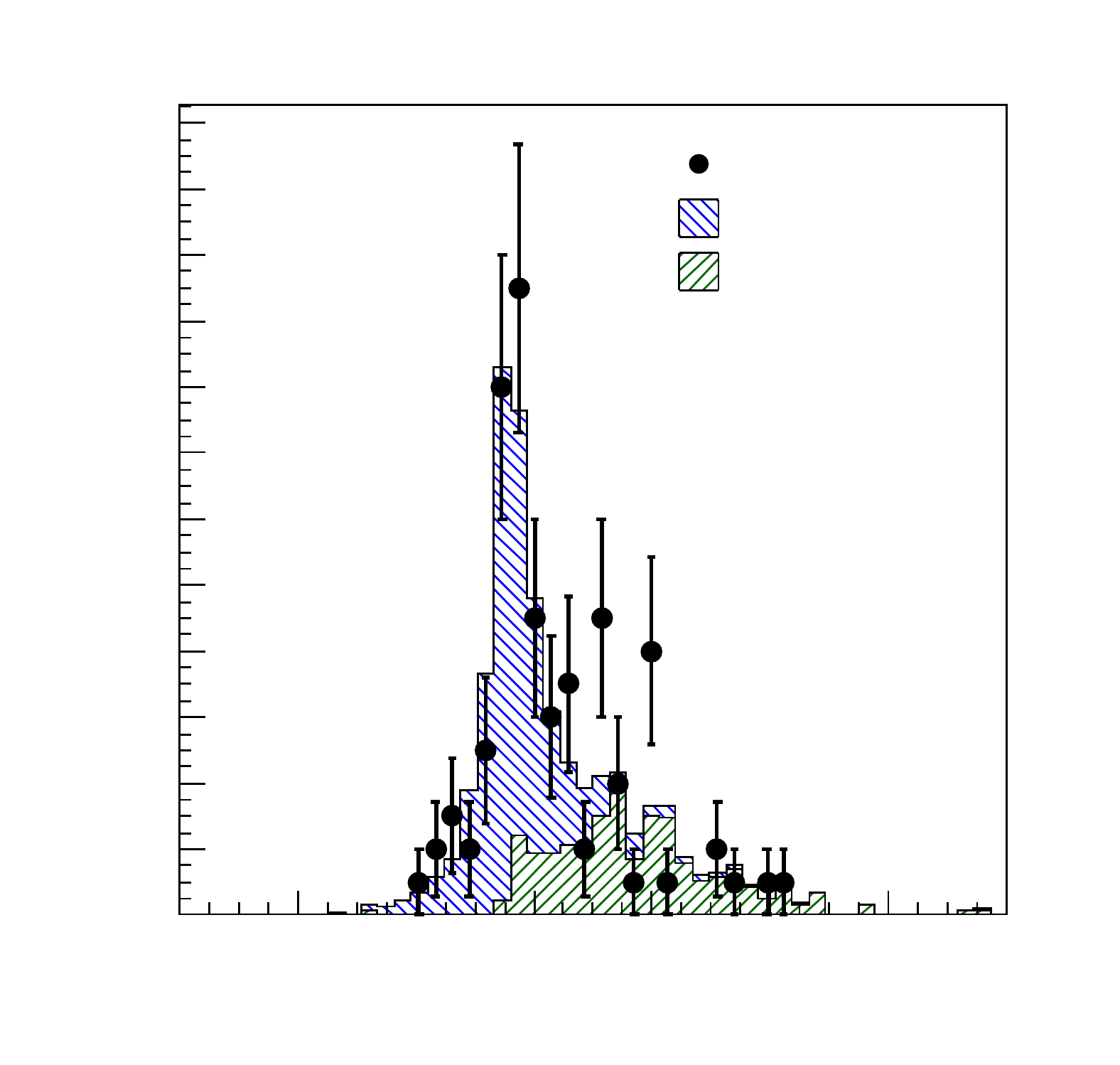}  \svgsep
  \svg{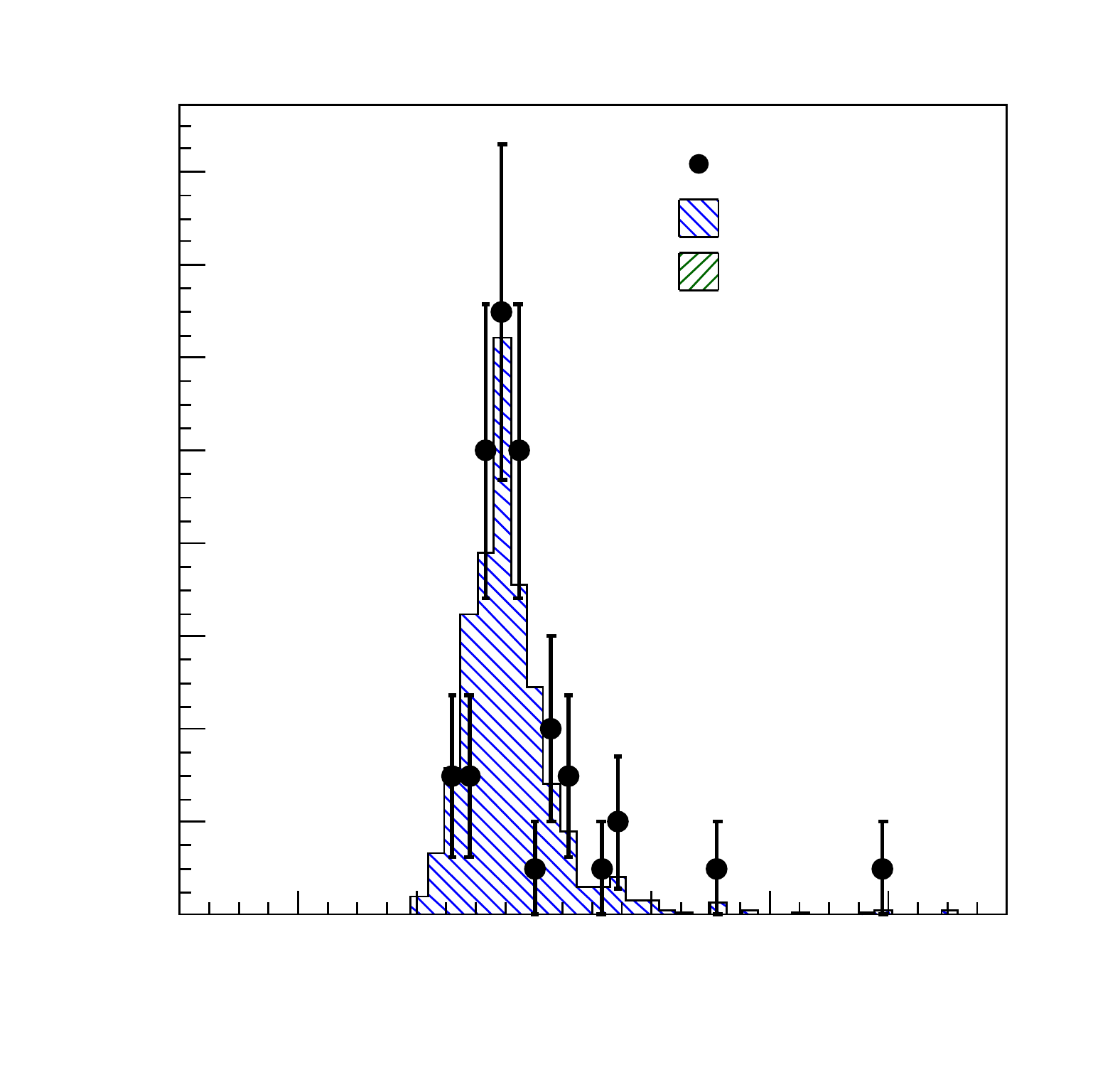}  & \svg{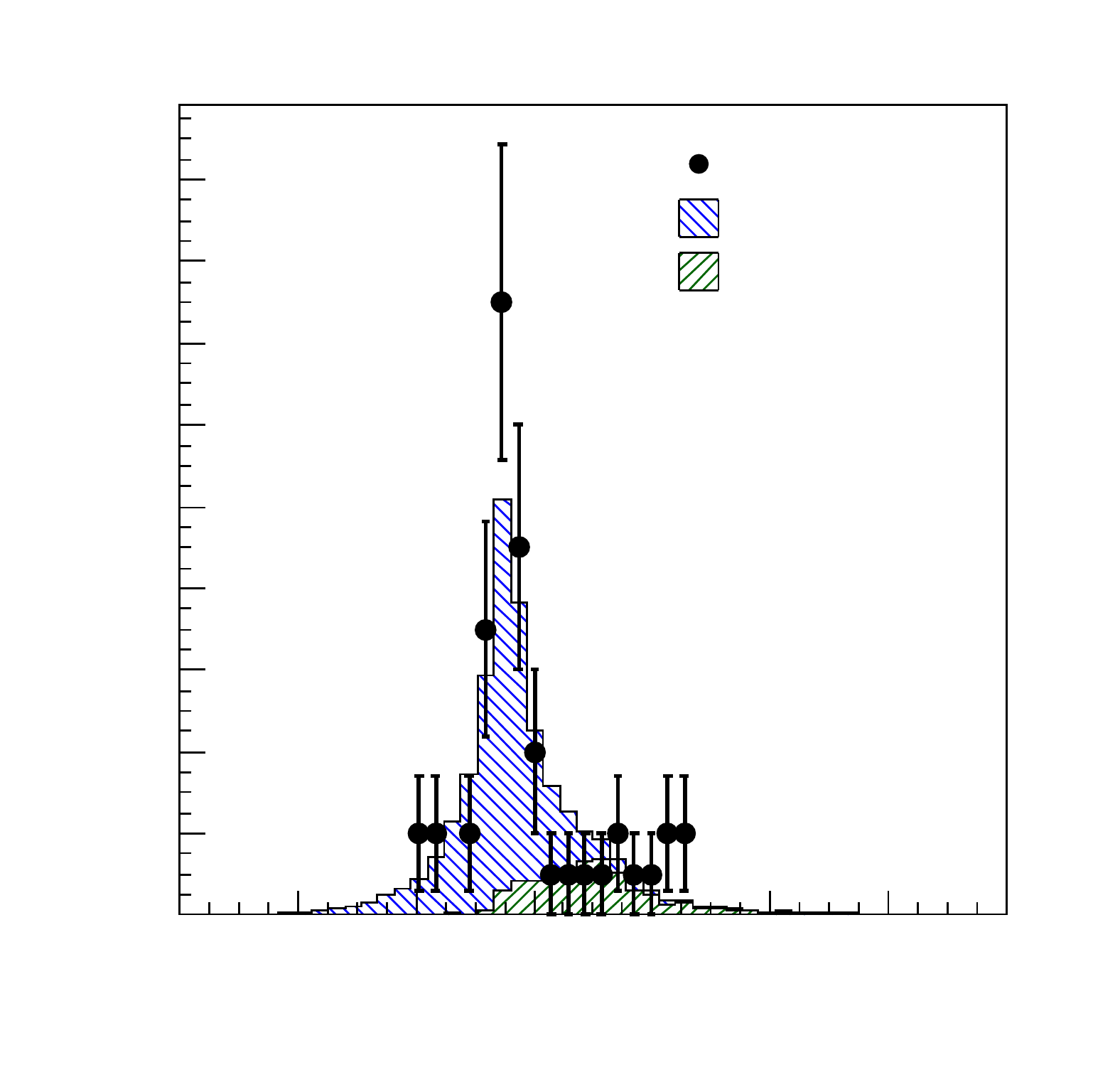} \svgsep
  \svg{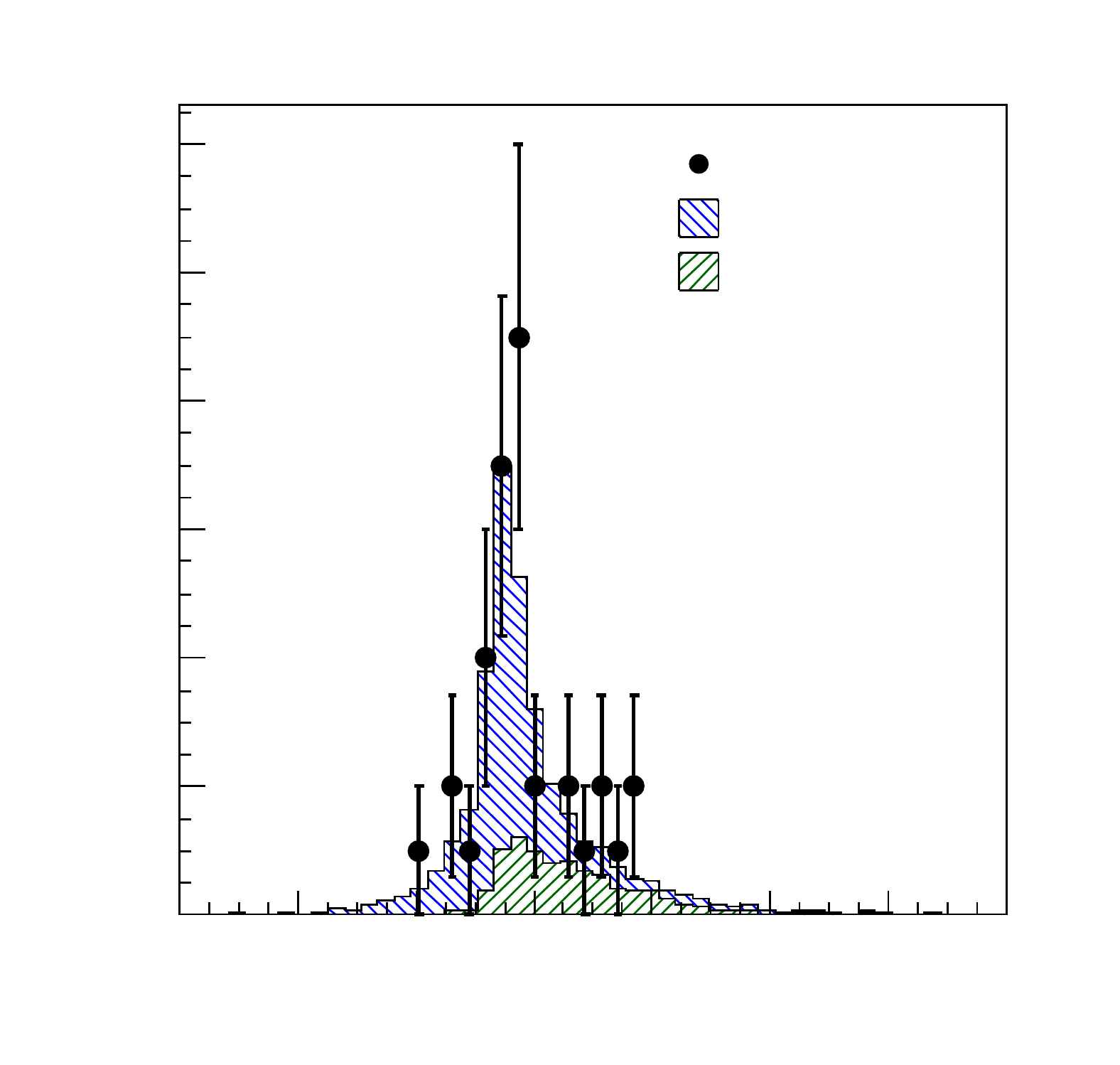}  & \sidecaption                  \svgend
\end{subfigures}

The number of same-sign background events \sqcd and \sewk are
extrapolated to the number of opposite-sign events, $N_\qcd$ and
$N_\ewk$, using,
\begin{equation}
  N_\qcd = \rqcd \sqcd \equcomma N_\ewk = \rewk \sewk
  \labelequ{QcdEwk}
\end{equation}
where $r$ is the ratio of opposite-sign to same-sign events evaluated
from data for the \qcd background and from simulation for the \ewk
background. These $r$ are calculated from the ratio of opposite-sign
to same-sign events satisfying the \qcd and \ewk sample
requirements. The value of $r_\qcd$ was found to remain consistent,
within statistical uncertainty, when varying the \qcd sample \iso
requirement between $5$ and $20~\gev$, and consequently is assumed to
remain valid for the ${\z \to \ditau}$ \iso selection requirements. A
cross-check with data was made for $r_\ewk$ by selecting an \ewk
sample from data using ${\iso < 1~\gev}$ for the first \wtl decay
product candidate and ${\iso > 10~\gev}$ for the second candidate.

The uncertainties on the \qcd and \ewk backgrounds are taken from the
uncertainties of the same-sign template fit, \rqcd, and \rewk. The
uncertainties on \rqcd and \rewk are statistical and uncorrelated
between the two backgrounds. The same-sign template fit requires the
sum of the \qcd and \ewk events, ${\sqcd + \sewk}$, to equal the total
number of same-sign data events, $S$, and so the fit uncertainty is
fully correlated between the two backgrounds. The combined number of
\qcd and \ewk background events, using \equ{Zed:QcdEwk} is,
\begin{equation}
  N_{\qcd+\ewk} = \rqcd \sqcd + \rewk \sewk = \rqcd \ffit S +
  \rewk (1 - \ffit) S
\end{equation}
where \ffit is defined as $\sqcd / S$. Assuming the uncertainties on
\ffit, \rqcd, and \rewk are uncorrelated and normally distributed, the
uncertainty on $N_{\qcd + \ewk}$ is,
\begin{equation}
  \delta_{N_{\qcd + \ewk}}^2 = \left(\rqcd S - \rewk S\right)^2
  \delta_\ffit^2 + \left(\ffit S\right)^2 \delta_\rqcd^2 + \left(S -
    fS\right)^2 \delta_\rewk^2
  \labelequ{DeltaQcdEwk}
\end{equation}
where $\delta_\ffit$, $\delta_\rqcd$, and $\delta_\rewk$ are the
uncertainties on \ffit, \rqcd, and \rewk respectively.

In \tab{Zed:Events}, the number of \qcd and \ewk backgrounds, with
associated uncertainty, have been presented separately for
clarity. Here, the uncertainty for the two backgrounds from
\equ{Zed:DeltaQcdEwk} is split between the two backgrounds using the
arbitrary convention,
\begin{subequations}
  \begin{align}
    \delta_{N_\qcd}^2 &= \frac{1}{2}\left(\rqcd S - \rewk S\right)^2
    \delta_\ffit^2 + \left(\ffit S\right)^2
    \delta_\rqcd^2 \labelequ{DeltaQcd} \\
    \delta_{N_\ewk}^2 &= \frac{1}{2}\left(\rqcd S - \rewk S\right)^2
    \delta_\ffit^2 + \left(S - fS\right)^2 \delta_\rewk^2
    \labelequ{DeltaEwk}
  \end{align}
\end{subequations}
where the template fit uncertainty term is evenly divided between the
\qcd and \ewk background uncertainties.

\newsubsubsection{$\bm{Z \to \ell\ell}$ Background}{}

The number of ${\z \to \dilep}$ background events, given in
\tab{Zed:Events} are evaluated only for the \mumu, \muh, and \eh
categories and not for the \mue and \emu categories, as the background
for these categories is negligible. For the \mumu and \muh categories
the ${\z \to \dilep}$ background consists of ${\z \to \dimu}$ events,
while for the \mue category the ${\z \to \dilep}$ background is from
${\z \to \die}$ events.

In the \mumu invariant mass distribution from data of
\fig{Zed:Z2TauTau2MuMu.Mass}, the ${\z \to \dimu}$ events from an
on-shell \wzb are clearly visible in the excluded mass range of ${80 <
  m < 100~\gev}$. The shape for the ${\z \to \dimu}$ background is
obtained from data by applying the \mumu selection of \sec{Zed:Sel},
but requiring $\ips < 1$ to eliminate ${\z \to \ditau}$ events. This
shape is then normalised to the number of \mumu events within the
excluded mass range ${80 < \m < 100~\gev}$. The uncertainty on the
number of ${\z \to \dimu}$ background events is estimated from the
statistical uncertainty on the normalisation of the background sample
and is the primary systematic uncertainty for the \mumu category.

Events from the ${\z \to \dimu}$ process contribute a small background
to the \muh category when one of the muons is mis-identified as a
hadron. The data sample for this background is found by applying the
\muh requirements of \sec{Zed:Sel}, but requiring that the second \wtl
decay product candidate fulfil the muon identification criteria of
\sec{Zed:Rec}. The sample is scaled by the probability of
mis-identifying a muon as a hadron. The muon mis-identification
probability is determined from data where ${\z \to \dimu}$ events are
selected by requiring a single well defined muon and a second isolated
track with a combined invariant mass within the range ${80 < \m <
  100~\gev}$. Only \prc{0.06 \pm 0.01} of the isolated tracks pass the
hadron identification requirement. The uncertainty on the \muh ${\z
  \to \dilep}$ background is estimated from the statistical
uncertainty of the background sample and the muon mis-identification.

For \eh events, a small ${\z \to \die}$ background can contribute when
one of the two electrons is mis-identified as a hadron. The background
sample is found by applying the \eh selection of \sec{Zed:Sel} to data,
but requiring the second \wtl decay product candidate to fulfil the
electron identification criteria of \sec{Zed:Rec}. The background
sample is scaled by the probability for an electron to be
misidentified as a hadron, which is determined from simulation to be
\prc{0.63 \pm 0.02}. The uncertainty on the background is estimated as
the statistical uncertainty of the background sample and the electron
mis-identification.

\newsubsubsection{$\bm{WW}$ and $\bm{t\bar{t}}$ Backgrounds}{}

Both the \ww and \ttbar samples are estimated to be small for all
event categories as shown in \tab{Zed:Events}. These backgrounds are
estimated from simulation which has been calibrated for the variables
described in \sec{Zed:Sel} and normalised to theoretical
cross-sections. Additionally, the simulation samples are corrected on
an event-by-event basis for the differences observed in the
reconstruction efficiencies between data and simulation. Details on
the reconstruction efficiencies are provided in \sec{Zed:RecEff}.

\newsection{Cross-Section}{Sig}

For consistency with the ${\dip \to \z \to \dimu}$ and ${\dip \to \z
  \to \die}$ cross-section measurements published by the \lhcb \collab
in \rfrs{lhcb.12.1} and \cite{lhcb.13.2}, the ${\dip \to \z \to
  \ditau}$ cross-section is evaluated in the kinematic region ${60 <
  \m_\ditau < 120~\gev}$, ${\pt_\tau > 20~\gev}$, and ${2.0 \leq
  \eta_\tau \leq 4.5}$, where $\tau$ indicates the \wtl before
decaying. The cross-section is calculated with,
\begin{equation}
  \sigma_{pp \to \z \to \ditau} =
  \frac{\displaystyle\sum\limits{_i^{N_\obs}} \left(
    \eff[_\rec^{-1}]_i \right) - \sum\limits{_j} \left({N_\bkg}_j
  \langle \eff[_\rec^{-1}]\rangle_j \right)}
  {\displaystyle\eff[_\sel] \lum 
  \acc[_{\tau_1\tau_2}] \br[_{\tau_1\tau_2}]}
  \labelequ{Xs}
\end{equation}
where $N_\obs$ is the number of candidate events observed in data,
${N_\bkg}_j$ is the number of estimated background events for each
background source $j$, and $\langle \eff[_\rec^{-1}]_j \rangle$ is the
average $\eff[_\rec^{-1}]$ for each background source $j$. The
reconstruction efficiency, \eff[_\rec], is calculated from data or
simulation and is dependent upon the momentum and pseudo-rapidity of
the \wtl decay product candidates for each event, while the event
selection efficiency, \eff[_\sel], is an average efficiency for all
events. The integrated luminosity is given by \lum, while
\acc[_{\tau_1\tau_2}] is an acceptance and final state radiation
factor and \br[_{\tau_1\tau_2}] is the branching fraction for the
event category.

The first summation with index $i$ over all observed events, corrects
each event by the reconstruction efficiency, $\eff[_\rec]_i$,
evaluated for the \wtl decay product candidates of that event. The
second summation with index $j$ over all background sources, is the
addition of the estimated number of events for each background source,
weighted by the average event reconstruction efficiency for the data
or simulation sample used to evaluate that background.

In \sec{Zed:RecEff}, the methods used to calculate the reconstruction
efficiency are described, while in \sec{Zed:SelEff} the selection
efficiency is calculated and in \sec{Zed:Acc}, the acceptance and
branching fractions are determined. The reconstruction efficiencies
are tabulated in \tab{Zed:RecEff} of \sec{Zed:RecEff}, while the
selection efficiencies are provided in \tab{Zed:SelEff} of
\sec{Zed:SelEff}. The acceptance and branching fractions are given in
\tab{Zed:Acc} of \sec{Zed:Acc}.

The integrated luminosity was determined using the Van de Meer
scan~\cite{meer.68.1} and beam-gas imaging~\cite{luzzi.05.1} methods
described in \sec{Exp:RecLum}. These methods provide similar results
and so the integrated luminosity is taken as the average of the two
with an estimated uncertainty of $3.5\%$~\cite{lhcb.12.4}. For the
\mumu, \mue, and \muh categories the integrated luminosity is ${1028
  \pm 28~\ipb}$, while the integrated luminosity for the \emu and \eh
channels is ${955 \pm 33~\ipb}$. The reduction in integrated
luminosity for the \emu and \eh categories is due to a change in the
electron triggers during the $2011$ data-taking period.

\newsubsection{Reconstruction Efficiency}{RecEff}

The reconstruction efficiency, \eff[_\rec], used in the cross-section
determination of \equ{Zed:Xs}, is defined as,
\begin{equation}
  \eff[_\rec] \equiv \eff[_\gec] ~ \eff[_\trg] ~ \eff[_\trk]_1 
  ~ \eff[_\trk]_2 ~ \eff[_\id]_1 ~ \eff[_\id]_2
  \labelequ{RecEff}
\end{equation}
where \eff[_\gec] is the global event cut (\gec) efficiency,
\eff[_\trg] the trigger efficiency, \eff[_\trk] the track finding
efficiency, and \eff[_\id] the particle identification efficiency.
The numerical subscripts indicate whether the efficiency is evaluated
for the first or second \wtl decay product candidate. A summary of
these component reconstruction efficiencies for muons, electrons, and
charged hadrons is given in \tab{Zed:RecEff}. The component
efficiencies are calculated in the order indicated in
\equ{Zed:RecEff}, {\it e.g.} the muon identification efficiency is
determined for muons with reconstructed tracks from events passing the
\gec and single-muon trigger requirement.

\begin{table}\centering
  \captionabove{The individual reconstruction efficiencies for muons,
    electrons, and charged hadrons. The track and identification
    requirements are given in \sec{Zed:Rec} and the trigger
    requirements for each category in
    \sec{Zed:Sel}.\labeltab{RecEff}}
  \begin{tabular}{>{$}l<{$}|E|E|E}
    \toprule
    & \multicolumn{2}{c|}{muons} & \multicolumn{2}{c|}{electrons} &
    \multicolumn{2}{c}{hadrons} \\
    \midrule
    \eff[_\gec]
    & 0.955&0.001 & 0.951&0.001 & \multicolumn{2}{c}{$-$} \\
    \eff[_\trg]
    & \multicolumn{2}{c|}{$0.76 - 0.79$} & \multicolumn{2}{c|}{$0.65 -
      0.72$} & \multicolumn{2}{c}{$-$} \\
    \eff[_\trk]
    & \multicolumn{2}{c|}{$0.87 - 0.93$} & 0.83&0.03\p 
    & \multicolumn{2}{c}{$0.73 - 0.79$} \\
    \eff[_\id]
    & \multicolumn{2}{c|}{$0.93 - 0.99$} & \multicolumn{2}{c|}{$0.79 -
      0.93$} & \multicolumn{2}{c}{$0.92 - 0.96$} \\
    \bottomrule
  \end{tabular}
\end{table}

The trigger efficiencies are evaluated individually for muons and
electrons and combined to determine the trigger efficiency for the
given event category. For the \mumu and \emu categories either the
first or second \wtl decay product candidate can trigger the event,
and so \eff[_\trg] is,
\begin{equation}
  \eff[_\trg] = \eff[_\trg]_1 + \eff[_\trg]_2 - \eff[_\trg]_1 ~ \eff[_\trg]_2
  \labelequ{Trg}
\end{equation}
where the numerical subscripts indicate the first or second
candidate. For the \mue, \muh, and \eh categories the first \wtl decay
product is required to trigger the event, and so the corresponding
trigger efficiency for the candidate type is used.

The lepton trigger, track finding, and identification efficiencies are
evaluated using tag-and-probe methods on ${\z \to \dilep}$ events from
data. The ${\z \to \dilep}$ events are selected by requiring a tag
lepton passing the full trigger, track, and identification
reconstruction requirements and a probe lepton passing looser
reconstruction requirements, where the requirement being tested is
omitted; oftentimes further event requirements are necessary to ensure
a pure ${\z \to \dilep}$ sample. The efficiency is then calculated as
the percentage of probes passing the test requirement. The event
topologies for ${\z \to \dilep}$ and ${\z \to \ditau}$ are nearly
identical except for the momenta of the final state candidates, due to
the decays of the \wtls, and so the lepton reconstruction efficiencies
from ${\z \to \dilep}$ samples are evaluated only as a function of
lepton momentum, when practicable.

In the remainder of this section the methods used to determine the
reconstruction efficiencies are detailed. Plots of the trigger, track
finding, and identification efficiencies for muons, electrons, and
charged hadrons are also provided. In these plots the efficiency
determined from data is compared to the biased and unbiased
efficiencies from simulation. The biased efficiency is found by
applying the tag-and-probe method to the reconstructed level of
simulated events, while the unbiased efficiency is found directly from
the generator level of simulated events.

\newsubsubsection{Global Event Cut}{}

The global event cut is applied at the L$0$ trigger, described in
\sec{Exp:RecTrg}, to eliminate high multiplicity events which require
significant processing time, and the \gec efficiency, \eff[_\gec], is
the probability for an event to pass the \gec requirement. For the
single-muon and single-electron triggers, the \spd multiplicity for an event
is required to be less than $600$ hits, while for the di-muon
trigger, the \spd multiplicity must be less than $900$ hits.

In \fig{Zed:RecEff.MuGec.Effs} the \spd distribution for ${\z \to
  \dimu}$ events from data requiring a di-muon trigger is given. The
${\z \to \dimu}$ data is selected by requiring two opposite-sign muons
with ${\pt > 20~\gev}$ and an invariant mass within the range ${60
  \leq \m \leq 120~\gev}$. The distribution for ${\z \to \dimu}$
events from data requiring a single-muon trigger is also given in
\fig{Zed:RecEff.MuGec.Effs}, but normalised so the integral of the
distribution equals the integral for the di-muon trigger distribution
below $600$ hits.

\begin{subfigures}{2}{\subfig{RecEff.MuGec.Effs} The \spd hit
    distributions from ${\z \to \dimu}$ data events passing the
    di-muon (points) and single-muon (cyan) triggers, where the
    single-muon distribution is normalised to the di-muon distribution
    below $600$ \spd hits. Between $600$ and $900$ \spd hits the
    di-muon distribution is fit with the $\Gamma$-function of
    \equ{Zed:Gamma} (line). \subfig{RecEff.EGec.Effs} The \spd hit
    distribution from ${\z \to \die}$ events passing the
    single-electron trigger (cyan) normalised to the ${\z \to \dimu}$
    di-muon distribution (points), where the shift upwards of $20$
    \spd hits is apparent in the ${\z \to \die}$
    distribution.\labelfig{RecEff.Gec}}
  \svgbeg
  \svg{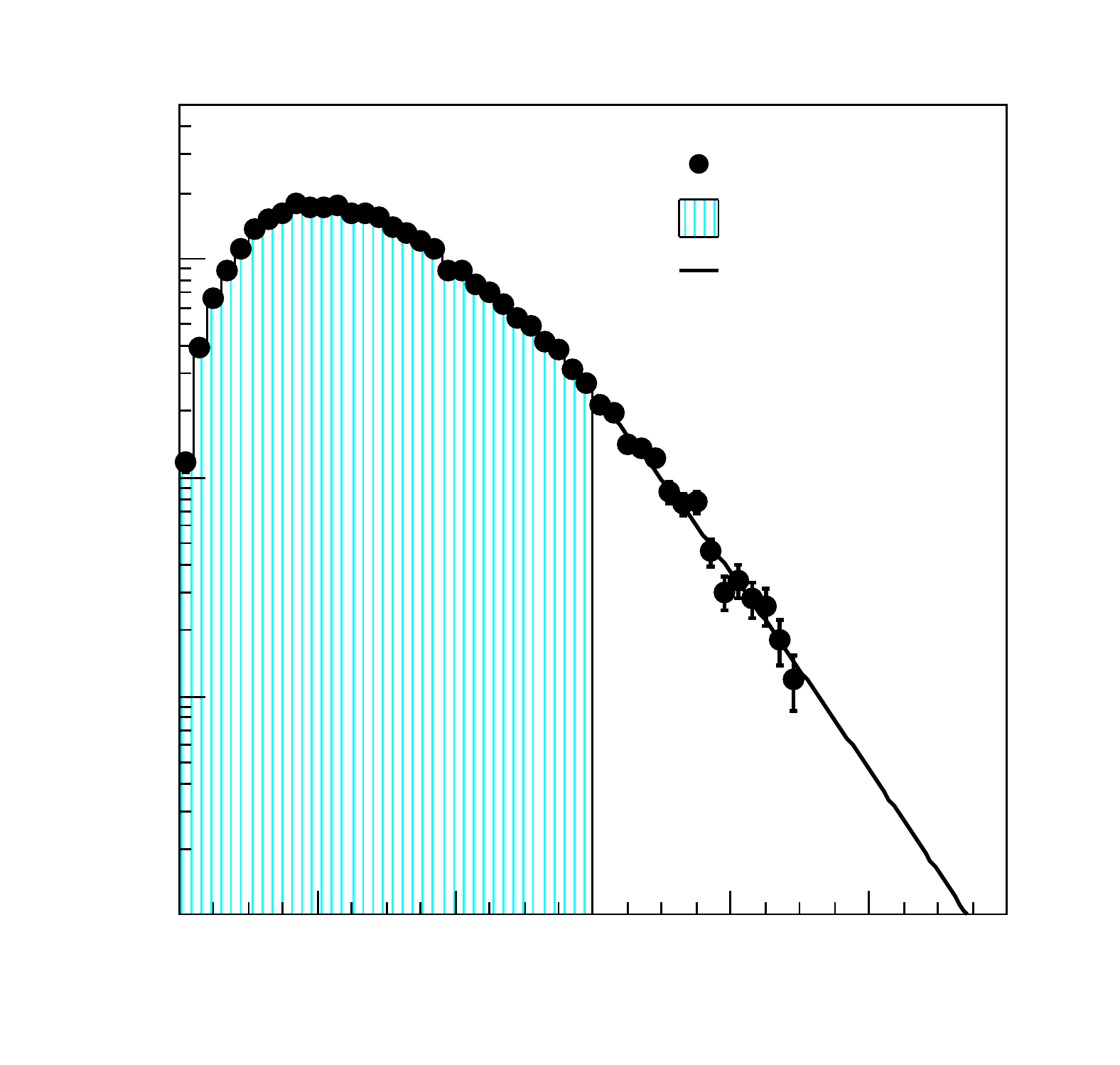} & \svg{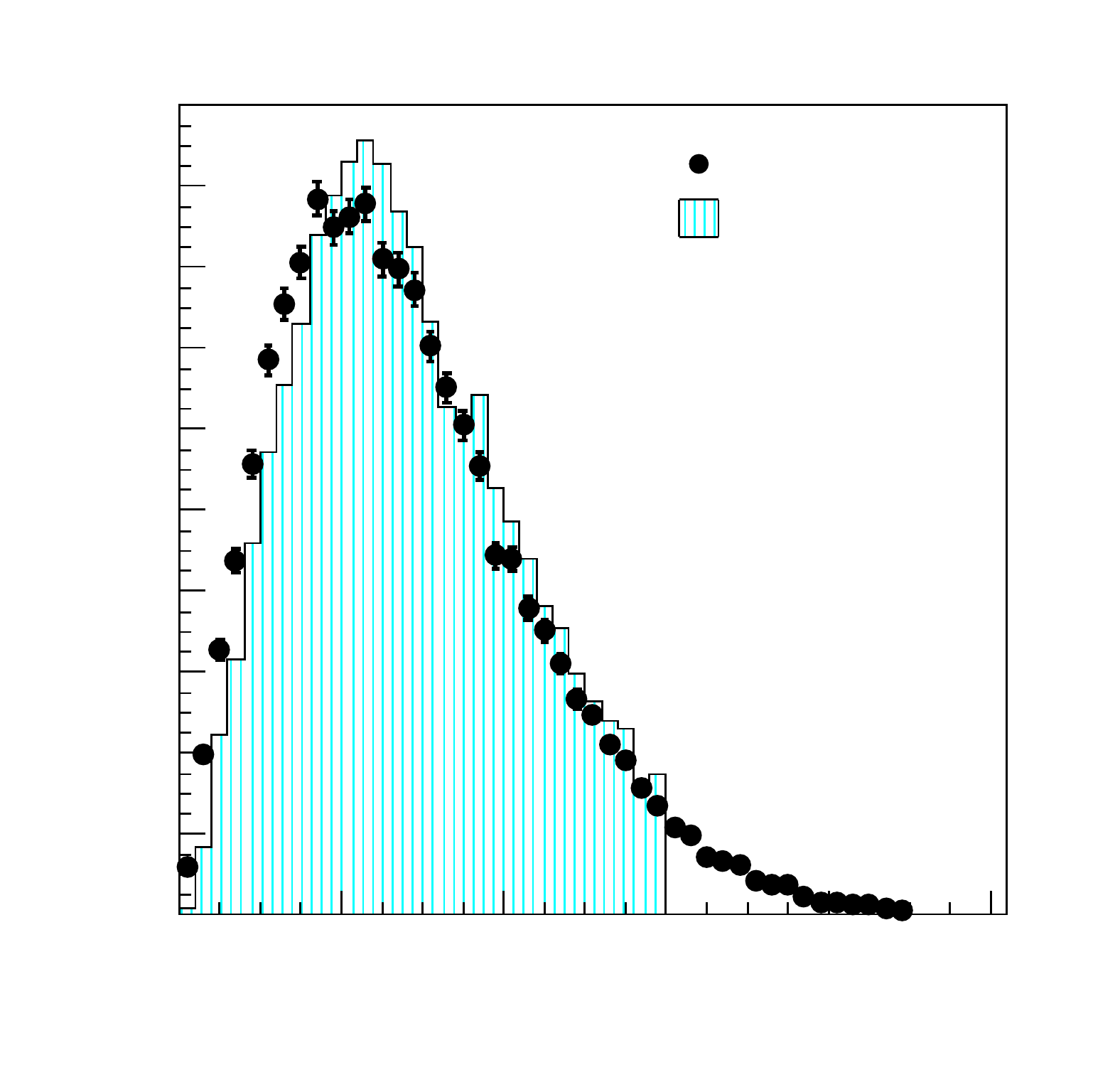} \svgend
\end{subfigures}

The tail of the di-muon trigger distribution of
\fig{Zed:RecEff.MuGec.Effs} is fit over the range $600$ to $900$
hits with a $\Gamma$-function,
\begin{equation}
  \Gamma(x) \equiv P_0 x^{P_1} e^{-\frac{x}{P_2}}
  \labelequ{Gamma}
\end{equation}
where $P_0$, $P_1$, and $P_2$ are free parameters. This function is
chosen as the \spd multiplicity should be roughly Poisson in
shape. The single-muon trigger \gec efficiency, \eff[_\gec], is then
evaluated as the number of events in the di-muon trigger distribution
with less than $600$ hits over the total number of events in the
distribution plus the integrated tail of \equ{Zed:Gamma}. The
uncertainty on the efficiency is determined from the statistical
uncertainty of the di-muon distribution and the uncertainty on the
integral of the tail from the fit of the $\Gamma$-function.

The \spd multiplicity distribution for ${\z \to \die}$ events from
data requiring a single-electron trigger is given in
\fig{Zed:RecEff.EGec.Effs}, where the distribution has been normalised
to the integral below $600$ hits of the di-muon trigger distribution
of \fig{Zed:RecEff.MuGec.Effs}. The ${\z \to \die}$ events from data
are selected by requiring opposite-sign electrons with ${\pt >
  20~\gev}$ and an invariant mass within the range ${60 \leq \m \leq
  120~\gev}$. The di-muon trigger distribution from ${\z \to \dimu}$
events of \fig{Zed:RecEff.MuGec.Effs} is also plotted in
\fig{Zed:RecEff.EGec.Effs} to provide a comparison.

As can be seen in \fig{Zed:RecEff.EGec.Effs}, the ${\z \to \die}$
single-eletron trigger distribution is shifted upwards by $20$ \spd
hist with respect to the di-muon trigger distribution, due to
additional \spd activity in the event from early showering of the
electrons. The single-electron trigger \gec efficiency is evaluated
with the same method used for the single-muon \gec efficiency, but
with the di-muon distribution shifted upwards by $20$ hits. The
\eff[_\gec] for both muons and electrons is found to be approximately
$95\%$. The muon \eff[_\gec] is used for the \mumu, \mue, and \muh
categories, while the electron \eff[_\gec] is used for the \emu and
\eh categories.

\newsubsubsection{Muon and Electron Trigger}{}

The muon trigger efficiency, $\eff[_\trg]_\mu$, is evaluated using a
tag-and-probe method on ${\z \to \dimu}$ events from data and is the
probability for a muon to pass the triggers of \sec{Exp:RecTrg}. The
tag is a muon passing trigger, track, and identification requirements,
while the probe is a muon passing only the track and identification
requirements. The tag and probe are required to have ${\pt >
  10~\gev}$, opposite charge, and a combined invariant mass within the
range ${60 < \m < 120~\gev}$. The efficiency is calculated as the
number of probes passing the muon trigger requirement over the total
number of probes, and is evaluated as a function of the probe muon
momentum in bins of $50~\gev$ within the range $0$ to $500~\gev$.

In \fig{Zed:RecEff.MuTrg.Effs} the efficiency is plotted for data,
biased simulation, and unbiased simulation, and ranges from $87\%$ to
$93\%$ for the data determined efficiency. The uncertainty on the
efficiencies is determined from the statistical uncertainty on the
number of events used to determine the efficiency for each bin. The
biased and unbiased efficiencies match within uncertainty, indicating
no bias is introduced via the event requirements.

\begin{subfigures}{2}{\subfig{RecEff.MuTrg.Effs} The muon trigger
    efficiency as a function of muon momentum from
    ${\z \to \dimu}$ data (points), biased simulation (red), and
    unbiased simulation (blue). The $\eff[_\trg]_\mu$ is taken from
    the data-driven
    efficiency. \subfig{RecEff.ETrg.Effs} The electron trigger
    efficiency as a function of electron momentum from ${\z \to \die}$
    data (points), biased simulation (red), and unbiased simulation
    (blue). The $\eff[_\trg]_e$ is taken from the data-driven
    efficiency.\labelfig{RecEff.Trg}}
  \svgbeg \svg{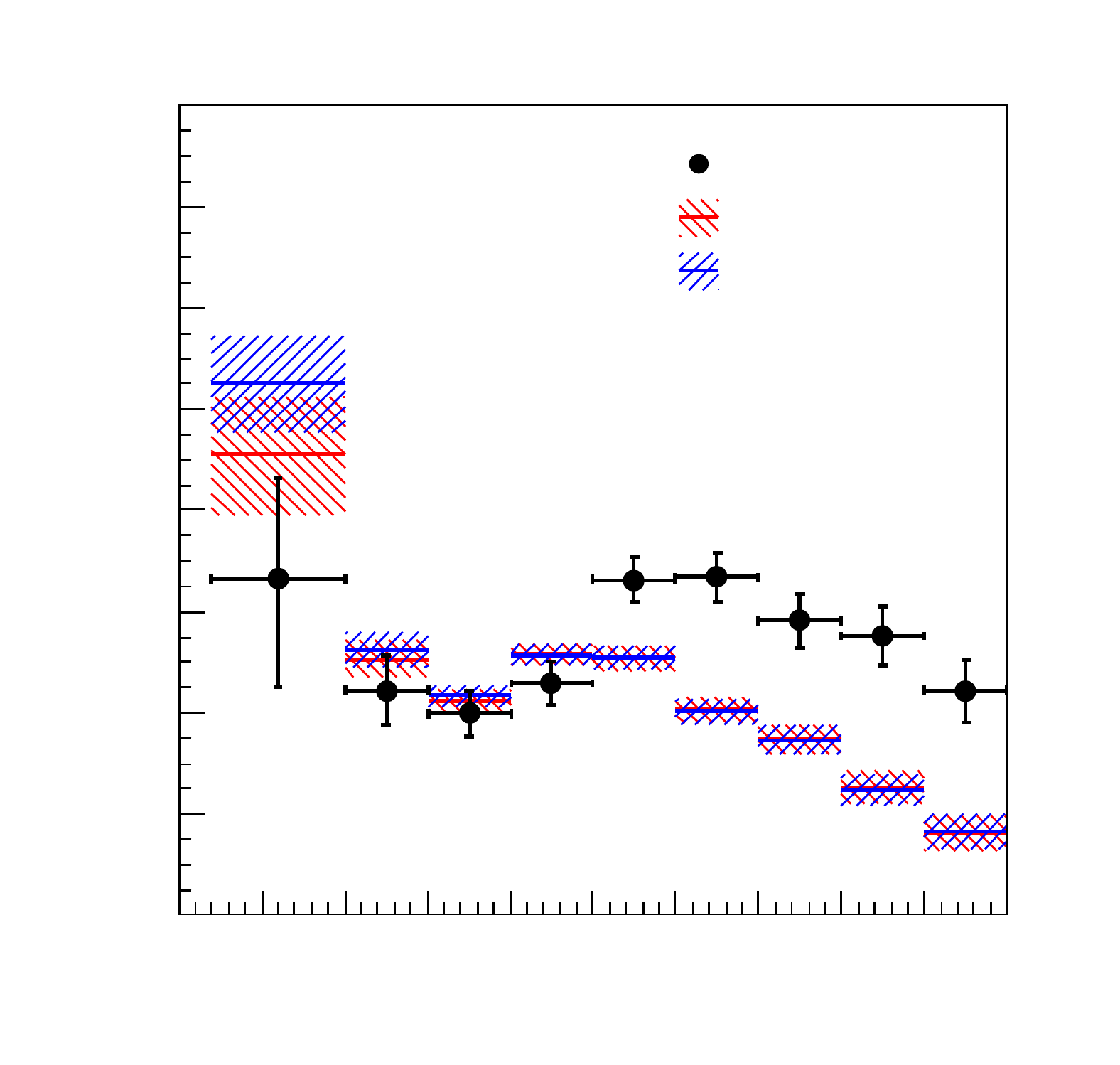} & \svg{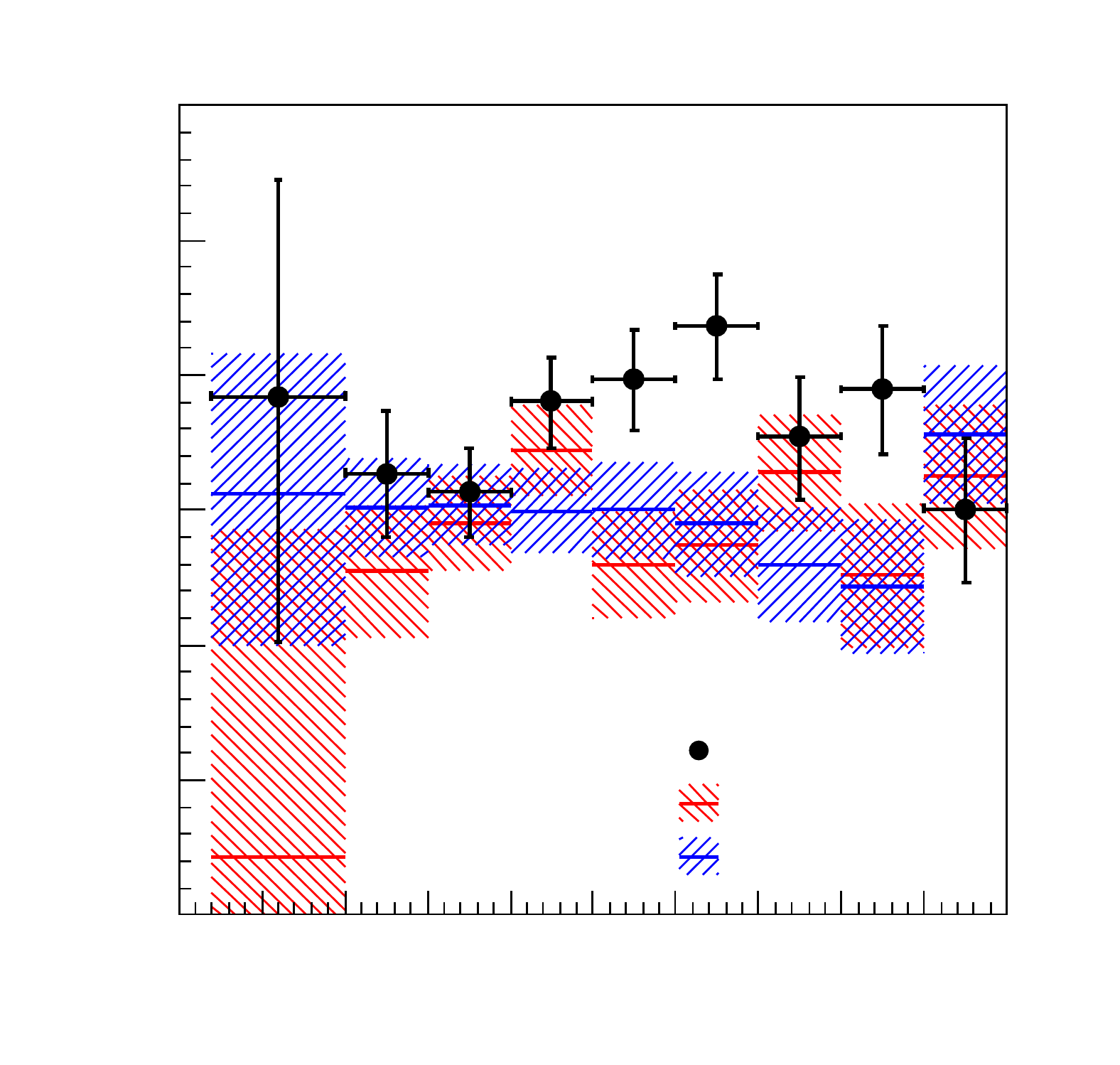} \svgend
\end{subfigures}

The electron trigger efficiency, $\eff[_\trg]_e$, is the probability
for an electron to pass the triggers of \sec{Exp:RecTrg} and is
evaluated using a tag-and-probe method similar to the method for
muons, but now on ${\z \to \die}$ events from data. The ${\z \to
  \die}$ sample, if selected using the same requirements as the ${\z
  \to \dimu}$ trigger efficiency sample, is contaminated by a
background on the order of $5\%$. This lowers $\eff[_\trg]_e$ on the
percent level and so stricter requirements on the the tag and probe of
${\pt > 20~\gev}$ and a combined invariant mass within the range ${70
  < \m < 120~\gev}$ are required. Additionally, the tag electrons must
be isolated with ${\iso < 2~\gev}$.

Again, the efficiency is evaluated as a function of probe momentum in
bins of $50~\gev$ within the range $0$ to $500~\gev$ and is given in
\fig{Zed:RecEff.ETrg.Effs}. The efficiency is found to vary from
$65\%$ to $72\%$. No bias is observed between the simulated
samples. The uncertainty for each bin is evaluated from the
statistical uncertainties of the data sample and is the primary
systematic uncertainty for the \eh category cross-section measurement.
 
\newsubsubsection{Muon Track Finding}{}

The muon track finding efficiency, $\eff[_\trk]_\mu$, is the
probability for a muon to have a reconstructed track fulfilling the
requirements of \sec{Zed:Rec}. The efficiency is determined from ${\z
  \to \dimu}$ data using the tag-and-probe method diagrammed in
\fig{Zed:RecEff.MuTrk.TagProbe}, first proposed in
\rfr{lhcb.10.1}. The tag is a muon passing the trigger, track
finding, and identification requirements. The probe is a track
reconstructed from hits within the \ttt and muon system, as neither
set of hits is used in the initial reconstruction of the muon
track. The tag and probe are required to have ${\pt > 20~\gev}$ and
opposite charge with a separation of ${\dphi > 1~\mathrm{radians}}$ to
ensure the tag and probe are not produced from the same muon. The tag
and probe are also required to be produced from the same vertex with a
$\chi^2$ less than $5$.

\begin{subfigures}{2}{\subfig{RecEff.MuId.Effs} The muon track finding
    efficiency as a function of muon momentum from ${\z \to \dimu}$
    data (points), ${\jpsi \to \dimu}$ data (grey points with
    uncertainty given by fill), biased simulation (red), and unbiased
    simulation (blue). The first three \jpsi bins and the remaining \z
    bins are used for $\eff[_\trk]_\mu$. \subfig{RecEff.MuId.TagProbe}
    A schematic, in the bending $xz$-plane of the detector, for the
    tag-and-probe method used to determine the $\eff[_\trk]_\mu$ from
    data.\labelfig{RecEff.MuTrk}}
  \svgbeg
  \svg{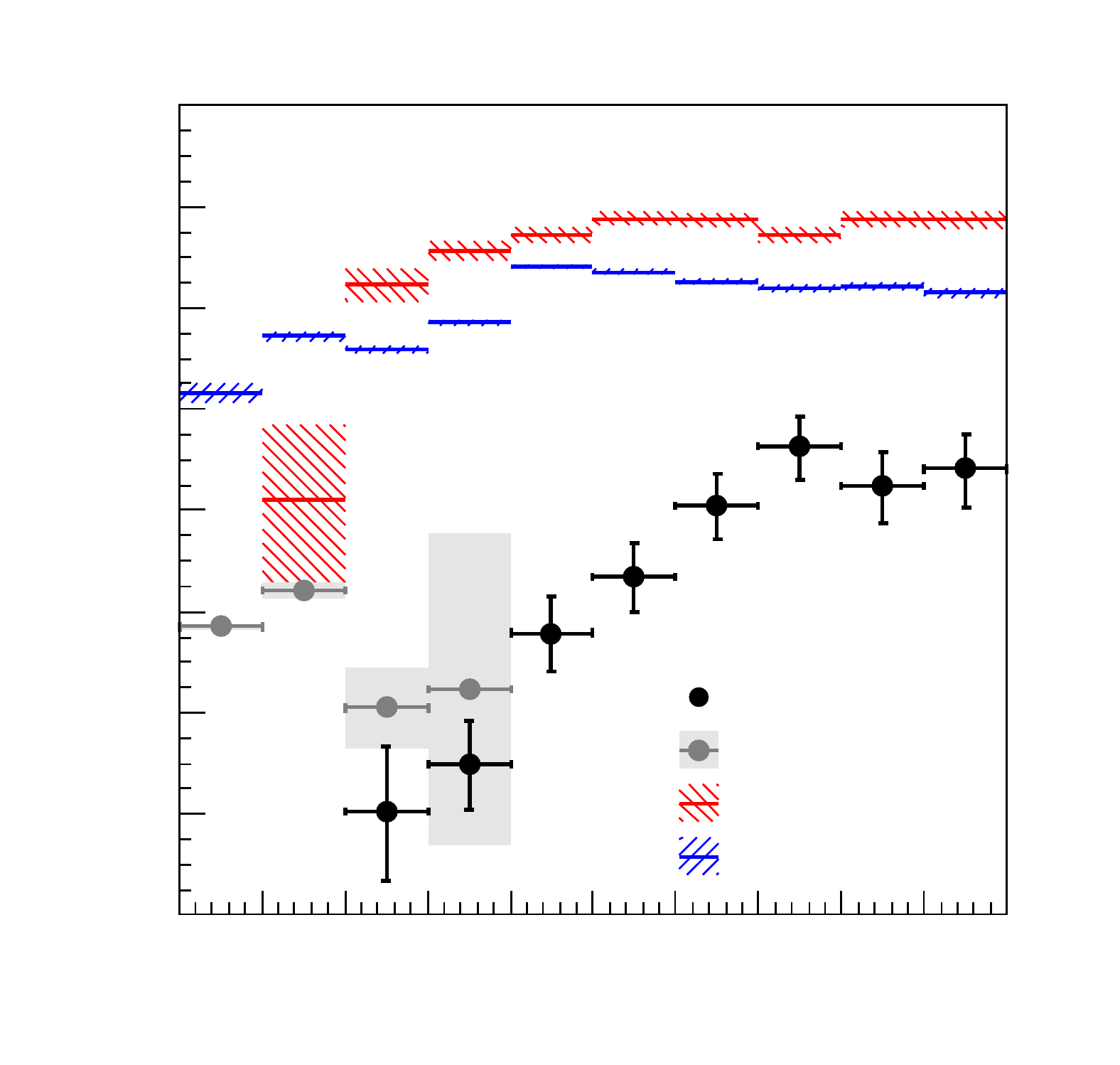} & \svg{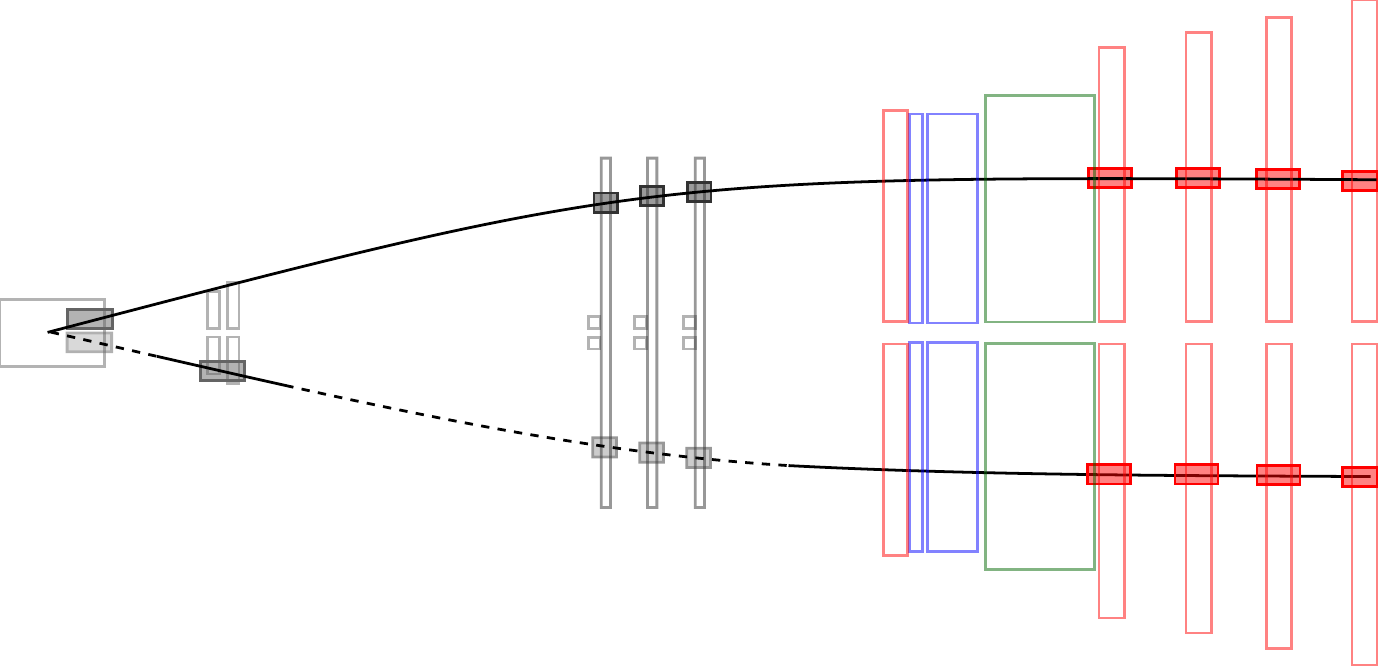} \svgend
\end{subfigures}

The probe is matched with tracks fulfilling the requirements of
\sec{Zed:Rec}. If more than $40\%$ of the \ttt and muon system hits
from the probe match the hits from the track, the probe is considered
to have an associated reconstructed track. The tracking efficiency is
evaluated as the number of probes with a reconstructed track over the
total number of probes, and is given as a function of probe momentum
in \fig{Zed:RecEff.MuTrk.Effs} with bins of $50~\gev$ from $0$ to
$500~\gev$. The large difference between the efficiencies from data
and simulation is primarily due to poor modeling of the probability
$\chi^2$ variable in simulation. A slight bias between the biased and
unbiased efficiencies from simulation is introduced by the selection
used to obtain the tag-and-probe sample, but is less than the
uncertainty on the efficiency evaluated from data. The uncertainty is
estimated as the statistical uncertainty on the sample used to
calculate each bin.

The $\pt$ requirement eliminates low momentum probes, and no
efficiencies from ${\z \to \dimu}$ data events are available for the
probe momentum range of $0$ to $100~\gev$ in
\fig{Zed:RecEff.MuTrk.Effs}. To access this low momentum range, the
same tag-and-probe method is applied to low mass ${\jpsi \to \dimu}$
events. Now, the tag is required to have ${\pt > 1.3~\gev}$ and the
combined tag and probe are required to have ${\pt > 1~\gev}$ with an
invariant mass within the range ${3.0 < \m < 3.2~\gev}$ and vertex
${\chi^2 < 5}$. The ${\jpsi \to \dimu}$ sample is not pure, and so the
\jpsi mass peak must be fit to determine the number of signal
events. Two different fit methods are used, and the difference between
the tracking efficiencies determined from the two fits is estimated as
the systematic uncertainty. Further details can be found in
\sap{Zvr:RecEff}.

The pseudo-rapidity distribution of ${\jpsi \to \dimu}$ events differs
from ${\z \to \dimu}$ events due to the mass difference between the
on-shell \wzb and \jpsi masses. Consequently, the muon tracking
efficiency from ${\jpsi \to \dimu}$ events is evaluated as a function
of both probe $\eta$ and $p$. The efficiency is reduced to only a
function of $p$ by taking the weighted average of the $\eta$ bins for
a given $p$, where the weight for each $\eta$ bin is taken from the
${\z \to \dimu}$ pseudo-rapidity distribution. The resultant
efficiency from ${\jpsi \to \dimu}$ events is given in
\fig{Zed:RecEff.MuTrk.Effs} and matches the overlapping ${\z \to
  \dimu}$ efficiencies, within uncertainty. The uncertainty on this
efficiency is the combination of the systematic uncertainty from
fitting the \jpsi peak and the statistical uncertainty of the
sample. This uncertainty is indicated by the grey error bands of
\fig{Zed:RecEff.MuTrk.Effs}.

The $\eff[_\trk]_\mu$ used in \equ{Zed:RecEff} and given in
\tab{Zed:RecEff} is evaluated from the ${\jpsi \to \dimu}$ efficiency
for muon momenta between $0$ to $150~\gev$ and from the ${\z \to
  \dimu}$ efficiency for momenta between $150$ and $500~\gev$. The
efficiency is found to range from $87\%$ to $93\%$.

\newsubsubsection{Electron Track Finding}{}

The electron track finding efficiency, $\eff[_\trk]_e$, is the
probability that an electron produces a reconstructable track, and is
determined using the tag-and-probe method diagrammed in
\fig{Zed:RecEff.ETrk.TagProbe} on ${\z \to \die}$ events from
data. The tag is an electron passing the trigger, track finding, and
identification requirements of \sec{Zed:Rec} with a ${\pt > 20~\gev}$,
while the probe is an \ecal cluster with an ${\et > 5~\gev}$. The tag
is required to be isolated with ${\iso < 2~\gev}$ and the \hcal energy
in a cone of ${\Delta R < 5}$ about the probe must be less than $50\%$
of the \ecal energy in the same cone. The tag and probe are required
to be back-to-back in the azimuthal plane with ${\dphi >
  3~\mathrm{radians}}$ and the difference in \ecal \et between the
probe and tag must be less than $20\%$ of the \ecal \et of the probe.

\begin{subfigures}{2}{Template fit of the electron tag \pt
    distribution from electron tracking tag-and-probe data (points)
    with ${\z \to \die}$ (cyan) and \qcd events (blue) from data
    \subfig{RecEff.ETrk.Ttl} before requiring a test track and
    \subfig{RecEff.ETrk.Pss} after. \subfig{RecEff.ETrk.TagProbe}~A
    schematic, in the bending $xz$-plane of the detector, for the
    tag-and-probe method used to determine $\eff[_\trk]_e$ from
    data.\labelfig{RecEff.ETrk}}
  \svgbeg
  \svg{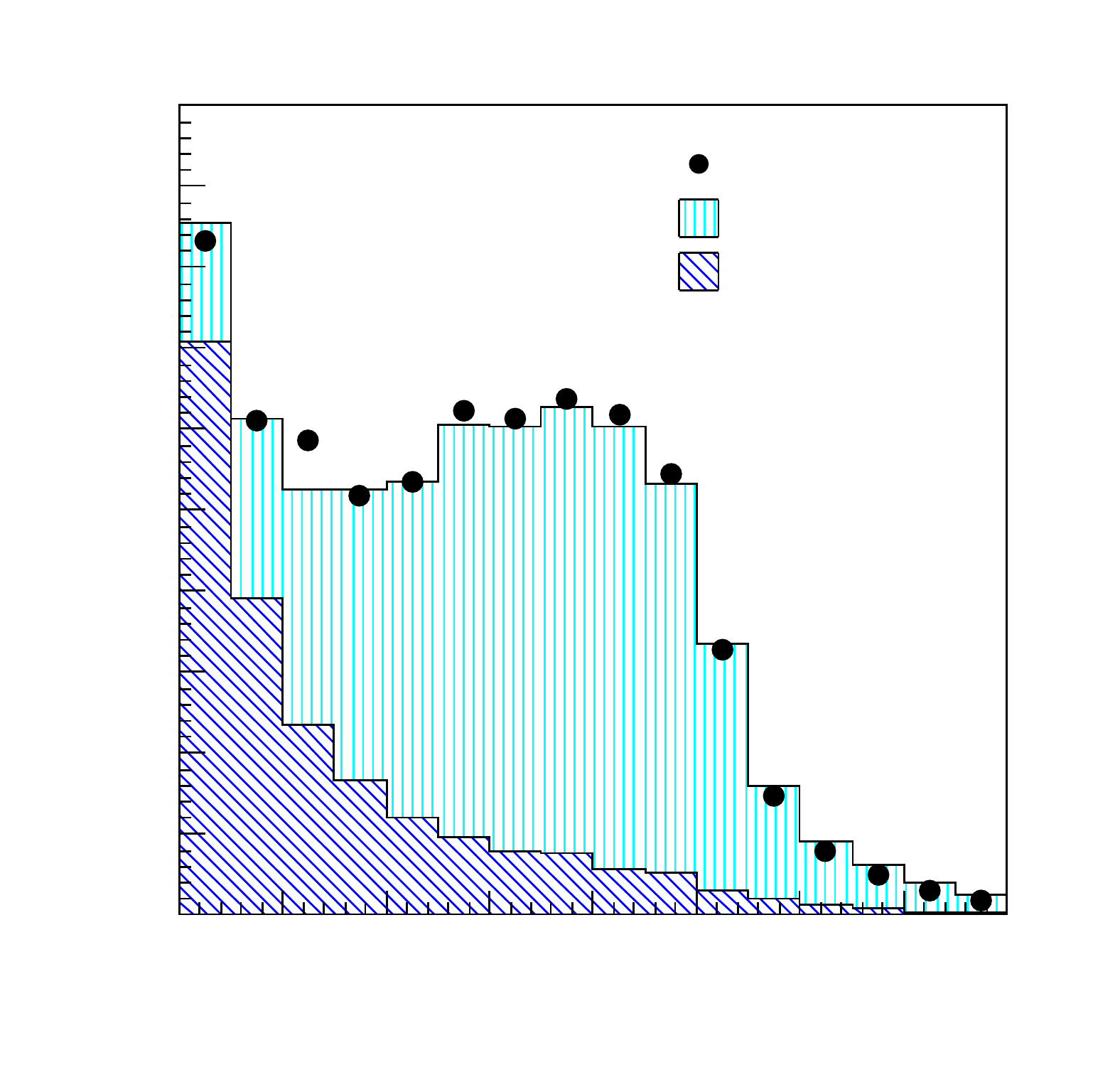} & \svg{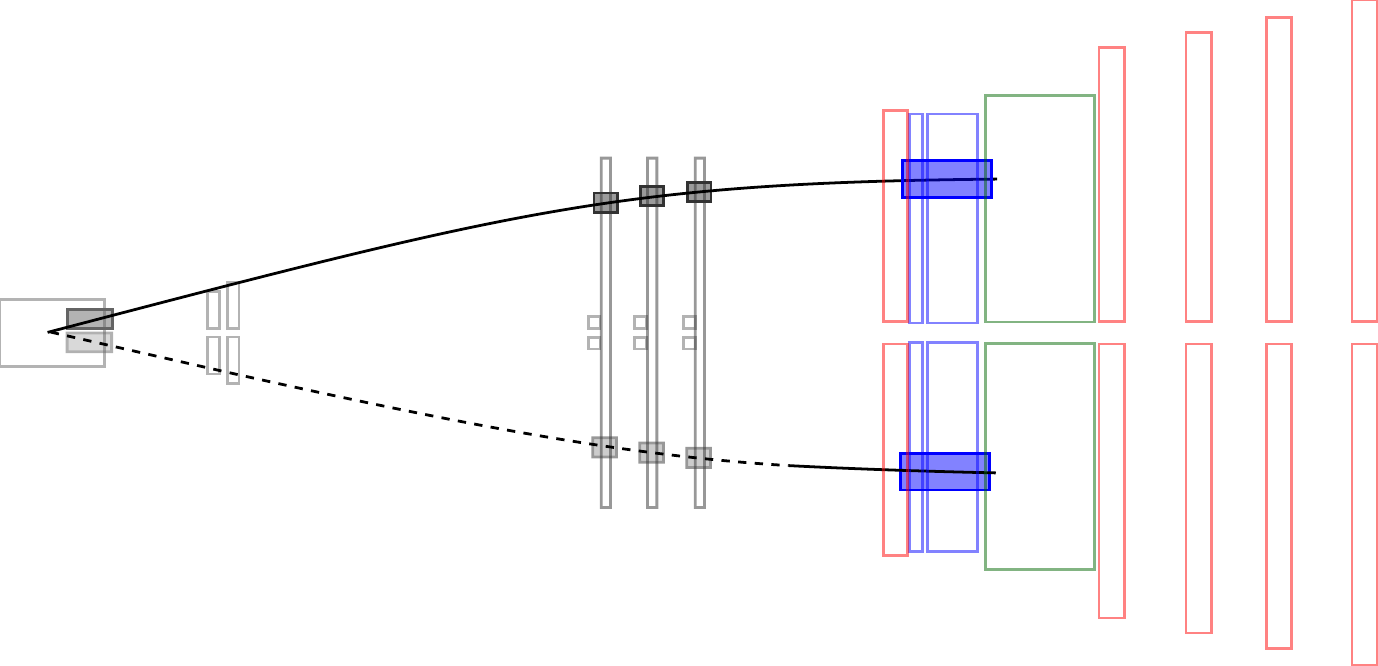}  \svgsep
  \svg{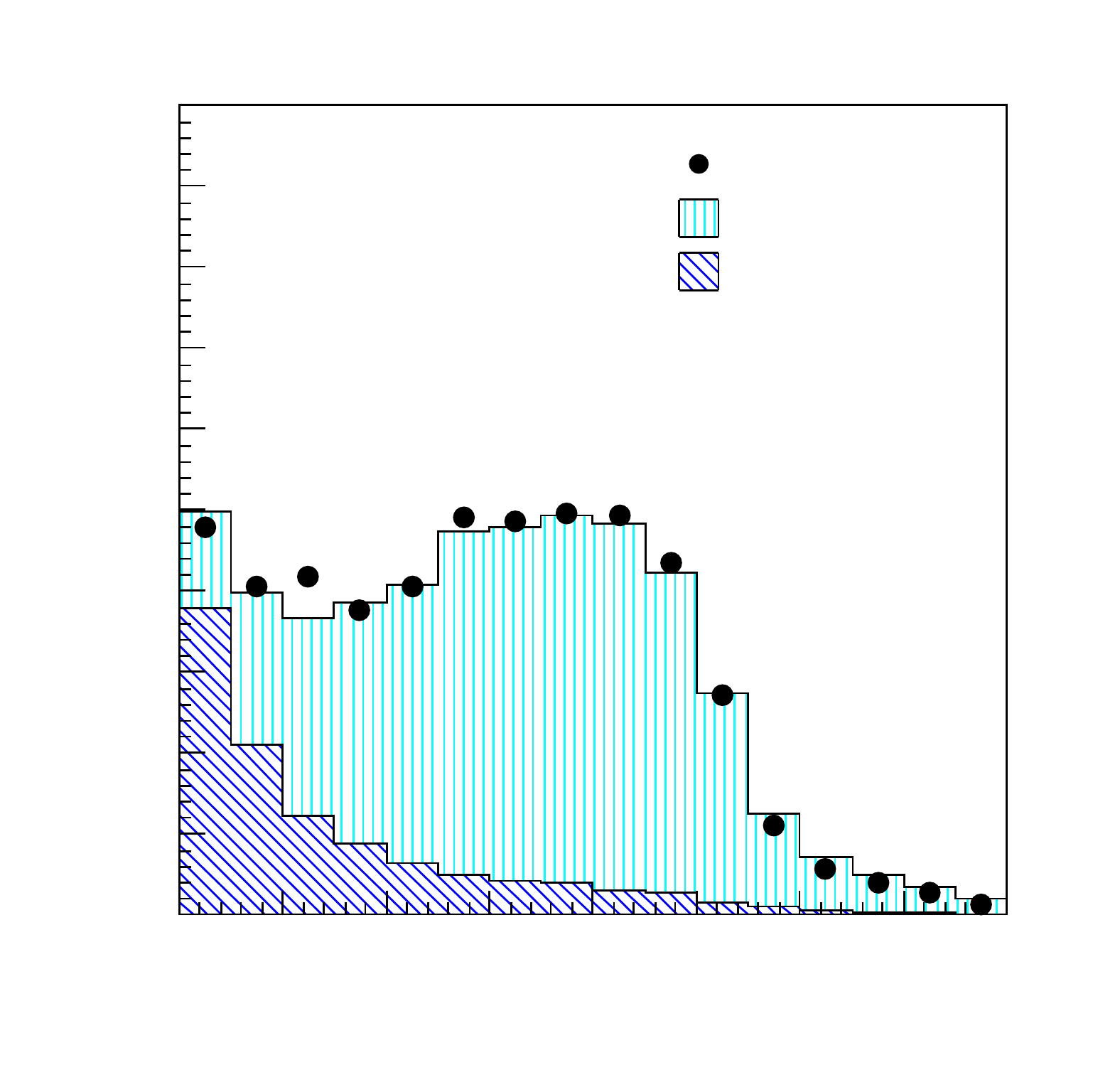} & \sidecaption      \svgend
\end{subfigures}

This selection results in a sample of ${\z \to \die}$ events
contaminated with a small \qcd background. The purity of the sample is
determined by fitting the \pt distribution of the tag with ${\z \to
  \die}$ and \qcd templates. For ${\z \to \die}$ events a peak is
expected in the tag \pt distribution at half the on-shell mass of the
\wzb, while a much softer distribution is expected from \qcd
events. The ${\z \to \die}$ template is constructed by selecting
events from data with two identified electrons with opposite charge
and a combined invariant mass in the range ${60 < \m < 120~\gev}$. The
\qcd template is taken from data using the same requirements as the
tag-and-probe sample, but omitting the tag isolation, azimuthal
separation, and \et balance requirements. Additionally, the
sub-leading \pt track in the event must have the same charge as the
leading \pt tag.

The fit of the ${\z \to \die}$ tag-and-probe \pt distribution with the
${\z \to \die}$ and \qcd templates, is given in
\fig{Zed:RecEff.ETrk.Ttl}, resulting in a signal purity of
approximately $70\%$. The tag-and-probe sample is tested for electron
track finding by requiring a track, fulfilling the requirements of
\sec{Zed:Rec}, with opposite charge and an azimuthal separation of
${\dphi > 2.5~\mathrm{radians}}$ with the tag. The fit of this
distribution is given in \fig{Zed:RecEff.ETrk.Pss} and is estimated to
have a signal purity of approximately $80\%$. As expected, both the
total number of events and the \qcd background in this distribution
are reduced.

The electron track finding efficiency is calculated as the number of
${\z \to \die}$ signal events from \fig{Zed:RecEff.ETrk.Pss}, after
requiring a track, over the number of signal events from
\fig{Zed:RecEff.ETrk.Ttl}, prior to requiring a track, and is found to
be \prc{83.0 \pm 0.3}. The uncertainty is estimated from the fit
uncertainty combined in quadrature with the statistical uncertainty of
the efficiency. While no momentum information is available, the
electron track finding efficiencies from simulated ${\z \to \die}$ and
${\z \to \ditau}$ events are found to be in statistical agreement with
values of \prc{85.2 \pm 0.2} and \prc{85.7 \pm 4.9},
respectively. These two efficiencies demonstrate that while the
electron tracking efficiency is estimated from generator level
simulation to have a momentum dependence, the net effect is small. No
additional systematic uncertainty is included in the $\eff[_\rec]_e$
uncertainty for this effect. The $\eff[_\rec]_e$ uncertainty is the
dominant systematic uncertainty for the cross-section measurements of
the \mue and \emu categories, as well as a large systematic
uncertainty for the \eh category cross-section measurement.

\newsubsubsection{Charged Hadron Track Finding}{}

The charged hadron track finding efficiency, $\eff[_\trk]_\had$, is
the probability for a charged hadron to produce a reconstructable
track. Prior to the final tracking station, particles pass through
approximately $20\%$ of a hadronic interaction length of material,
resulting in the early showering of charged hadrons caused by their
nuclear interactions with the detector material. Consequently, the
muon track finding efficiency is used as the charged hadron track
finding efficiency, but with a correction for additional material
interactions which is estimated from ${\z \to \ditau}$ simulation to
be \prc{84.3 \pm 1.5}. The uncertainty on this correction factor is
from the uncertainty of the \lhcb material budget of
$10\%$~\cite{lhcb.12.3}.

The grey points in \fig{Zed:RecEff.HadTrk.Effs} are the muon track
finding efficiency from \fig{Zed:RecEff.MuTrk.Effs} as a function of
momentum, corrected for material interactions, and are used as the
hadron track finding efficiency. The grey bands indicate the
uncertainty on the efficiency which is the combination of the muon
track finding efficiency uncertainty and the correction
uncertainty. The uncorrected muon track finding efficiency from data
is also plotted in \fig{Zed:RecEff.HadTrk.Effs}, as well as the
unbiased muon and charged hadron track finding efficiencies from
simulation. The hadron track finding efficiency ranges from values of
$73\%$ to $79\%$.

\begin{subfigures}{2}{The charged hadron track finding efficiency
    (points with fill) determined from the ${\z \to \dimu}$
    data-driven muon track finding efficiency (points) corrected for
    hadronic nuclear interactions. The associated uncertainty is the
    \lhcb material budget uncertainty combined in quadrature with the
    muon track finding uncertainty. The unbiased muon (red) and
    charged hadron (blue) track finding efficiencies from simulation
    are shown for comparison.}
  \svgbeg
  \svg[1]{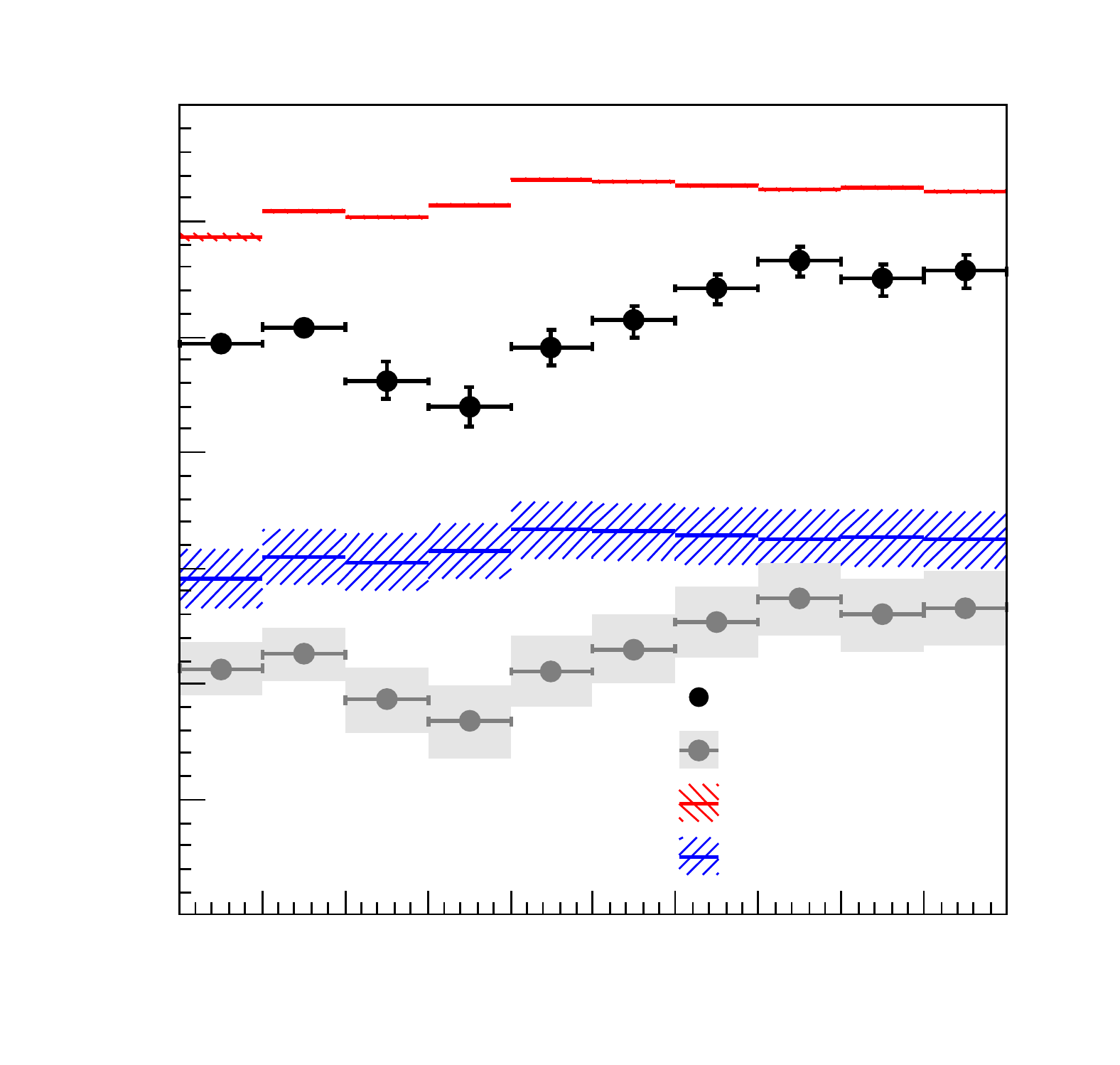} & \sidecaption \svgend
\end{subfigures}

\newsubsubsection{Muon Identification}{}

The muon identification efficiency is the probability for a muon with
a track to pass the muon identification requirements of
\sec{Zed:Rec}. Here, the muon is already required to have a
reconstructed track from hits within the \velo and \ttt stations and a
search for hits from the muon system is performed, in contrast to the
muon track finding efficiency where a track is not required {\it a
  priori} and hits from the muon system are not required. The muon
identification efficiency is evaluated from data using a tag-and-probe
method on ${\z \to \dimu}$ data, diagrammed in
\fig{Zed:RecEff.MuId.TagProbe}, where the tag is a muon passing the
trigger, track, and identification requirements and the probe is a
muon passing the track requirements. Additionally, the tag and probe
are required to have ${\pt >20~\gev}$, an isolation of ${\iso <
  2~\gev}$, opposite charge, and a combined invariant mass within the
range ${60 < m < 120~\gev}$. The efficiency is then taken as the
number of probes with associated hits in each of the four outermost
muon stations over the total number of probes.

\begin{subfigures}[t]{2}{\subfig{RecEff.MuId.Effs} The muon
    identification efficiency as a function of muon momentum from ${\z
      \to \dimu}$ data (points), ${\jpsi \to \dimu}$ data (points with
    fill), biased simulation (red), and unbiased
    simulation (blue). The first three \jpsi bins and the remaining \z
    bins are used for
    $\eff[_\id]_\mu$. \subfig{RecEff.MuId.TagProbe} A schematic,
    in the bending $xz$-plane of the detector, for the tag-and-probe
    method used to determine $\eff[_\id]_\mu$ from
    data.\labelfig{RecEff.MuId}}
  \svgbeg
  \svg{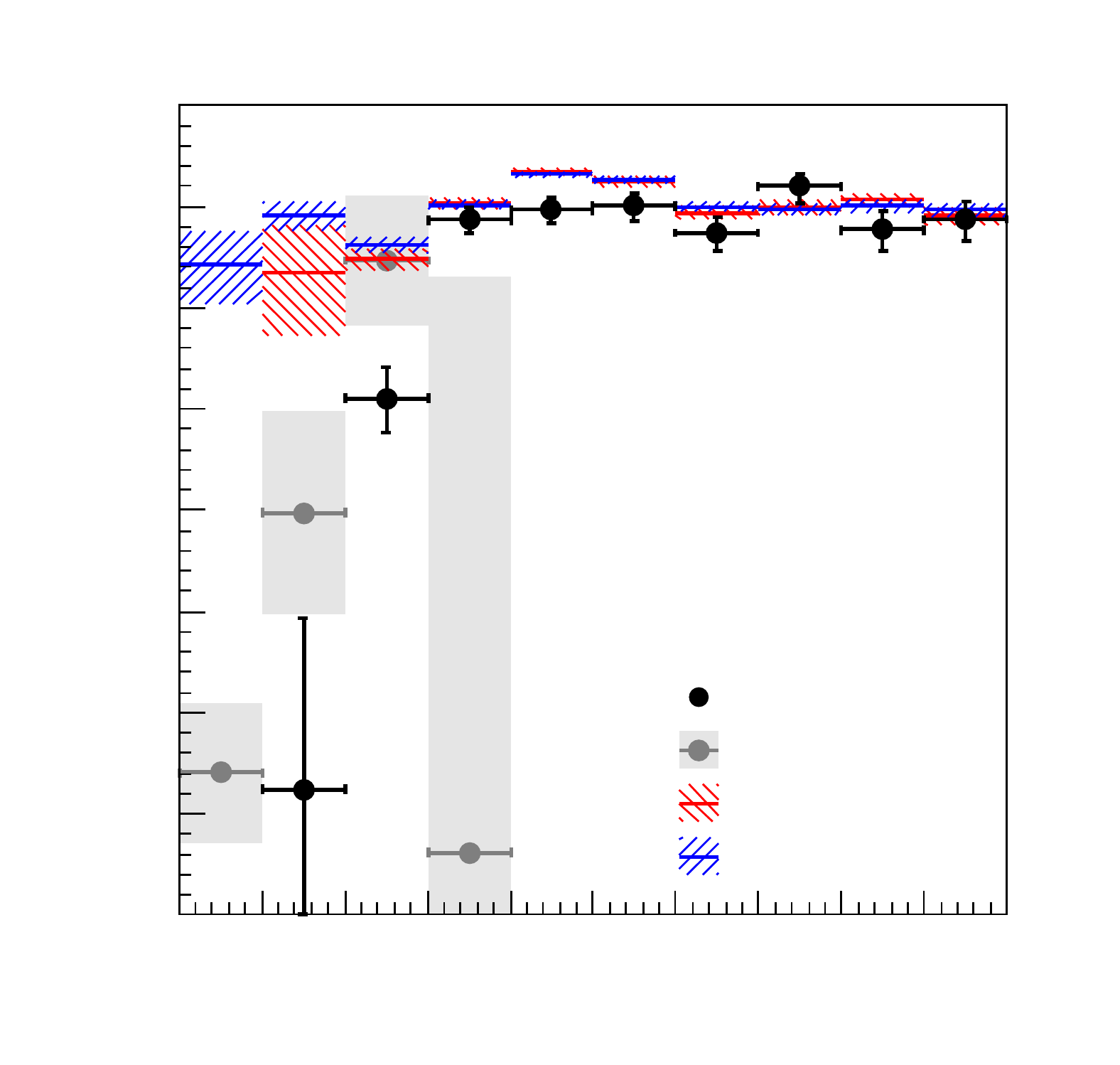} & \svg{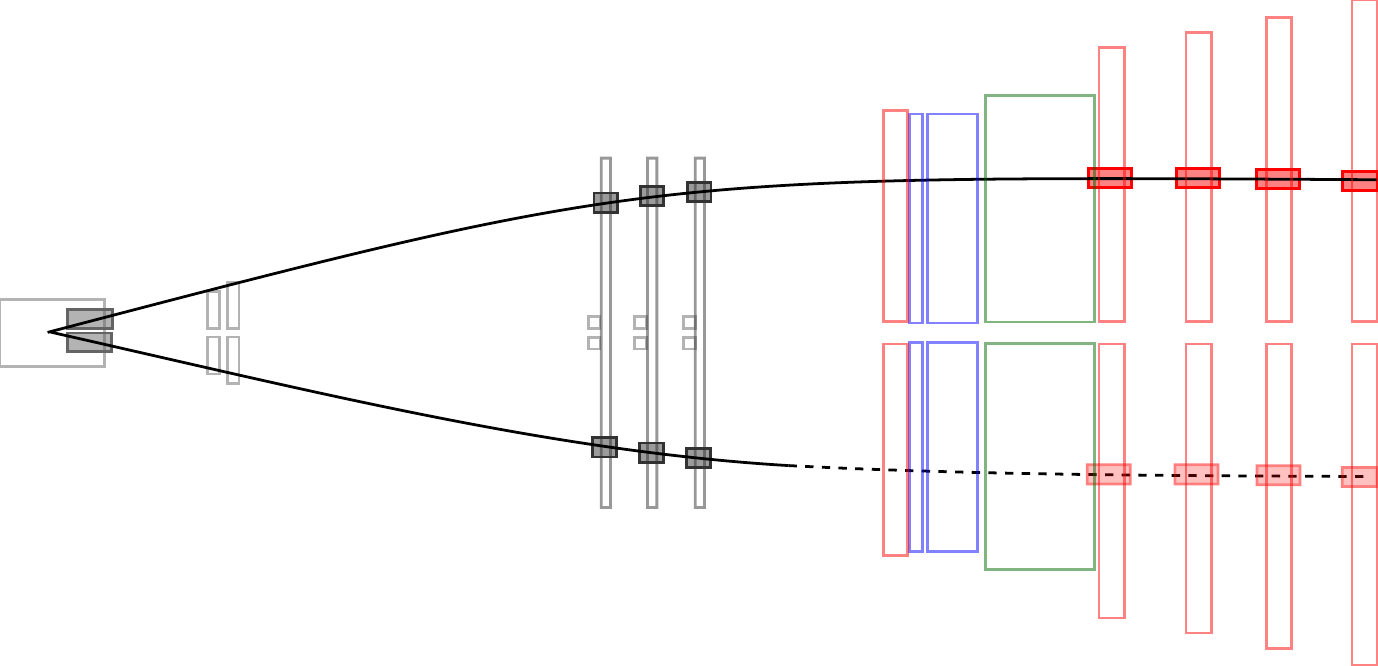} \svgend
\end{subfigures}

The muon identification efficiency from ${\z \to \dimu}$ events is
determined with respect to the probe momentum, and is plotted in
\fig{Zed:RecEff.MuId.Effs}.  Due to the \pt requirements on the tag
and probe, the low momementum bins between $0$ and $150~\gev$ must be
suplemented using efficiencies from ${\jpsi \to \dimu}$ events. For
the ${\jpsi \to \dimu}$ sample, the probe must have ${\pt > 1.5~\gev}$
and ${p > 6~\gev}$ while the tag must have ${\pt > 0.8~\gev}$ and ${p
  > 3~\gev}$. The combined invariant mass of the ${\jpsi \to \dimu}$
tag and probe is required to be within the range $3.0 < \m < 3.2$ and
the vertex $\chi^2$ for the event must be less than $8$.

Just as for the muon track finding efficiency, the pseudo-rapidity
distributions between the ${\jpsi \to \dimu}$ and ${\z \to \dimu}$
samples differ. Consequently, the muon identification efficiency from
${\jpsi \to \dimu}$ events is evaluated as a function of probe $\eta$
and $p$. This efficiency is reduced to a function of only $p$ using
the same weighted average method as the muon track finding efficiency
and is given in \fig{Zed:RecEff.MuId.Effs}. The uncertainty on this
efficiency is the combination of the statistical uncertainty for the
sample and the uncertainty from the fit of the \jpsi peak.

The two overlapping bins between the ${\jpsi \to \dimu}$ and ${\z \to
  \dimu}$ tag-and-probe efficiencies match within uncertainty,
although the uncertainty for the final ${\jpsi \to \dimu}$ bin is
large due to poor fits from low statistics. Additionally, the decrease
in efficiency at low momentum for the ${\z \to \dimu}$ tag-and-probe
efficiency arises from minor background contamination in the data
sample. There is good agreemant between the biased and unbiased
efficiencies from simulation, verifying no bias has been introduced by
the ${\z \to \dimu}$ selection requirements. The biased efficiency is
determined from applying the ${\z \to \dimu}$ tag-and-probe method to
simulation, while the unbiased efficiency is taken directly from the
generator level information in simulation. The muon identification
efficiency, used in \equ{Zed:RecEff} and given in \tab{Zed:RecEff}, is
taken from the ${\jpsi \to \dimu}$ data for ${0 < p < 150~\gev}$ and
from the ${\z \to \dimu}$ data for ${150 < p < 500~\gev}$, with values
ranging from $93\%$ to $99\%$.

\newsubsubsection{Electron Identification}{}

The probability for an electron with a track fulfulling the
requirements of \sec{Zed:Rec} to pass the electron identification
requirements is given by the electron identification efficiency,
$\eff[_\id]_e$. This efficiency is determined using the tag-and-probe
method of \fig{Zed:RecEff.EId.TagProbe} on ${\z \to \die}$ events from
data. The tag is an electron passing the trigger, track, and
identification requirements with ${\pt > 20~\gev}$, while the probe is
a track, also with ${\pt > 20~\gev}$. Both the tag and probe are
required to be isolated with ${\iso < 2~\gev}$, have a combined
invariant mass within the range ${60 < \m < 120~\gev}$, and an
azimuthal separation of ${\dphi > 3.0~\mathrm{radians}}$.

\begin{subfigures}{2}{\subfig{RecEff.EId.Effs} The electron
    identification efficiency as a function of electron momentum from
    ${\z \to \die}$ data (points), biased simulation (red), and
    unbiased simulation (blue). The $\eff[_\id]_e$ is taken as the
    unbiased efficiency with the total efficiency difference between
    data and biased simulation estimated as the associated uncertainty
    (grey fill). \subfig{RecEff.MuId.TagProbe} A schematic, in the
    bending $xz$-plane of the detector, for the tag-and-probe method
    used to determine $\eff[_\id]_e$ from data.\labelfig{RecEff.EId}}
  \svgbeg
  \svg{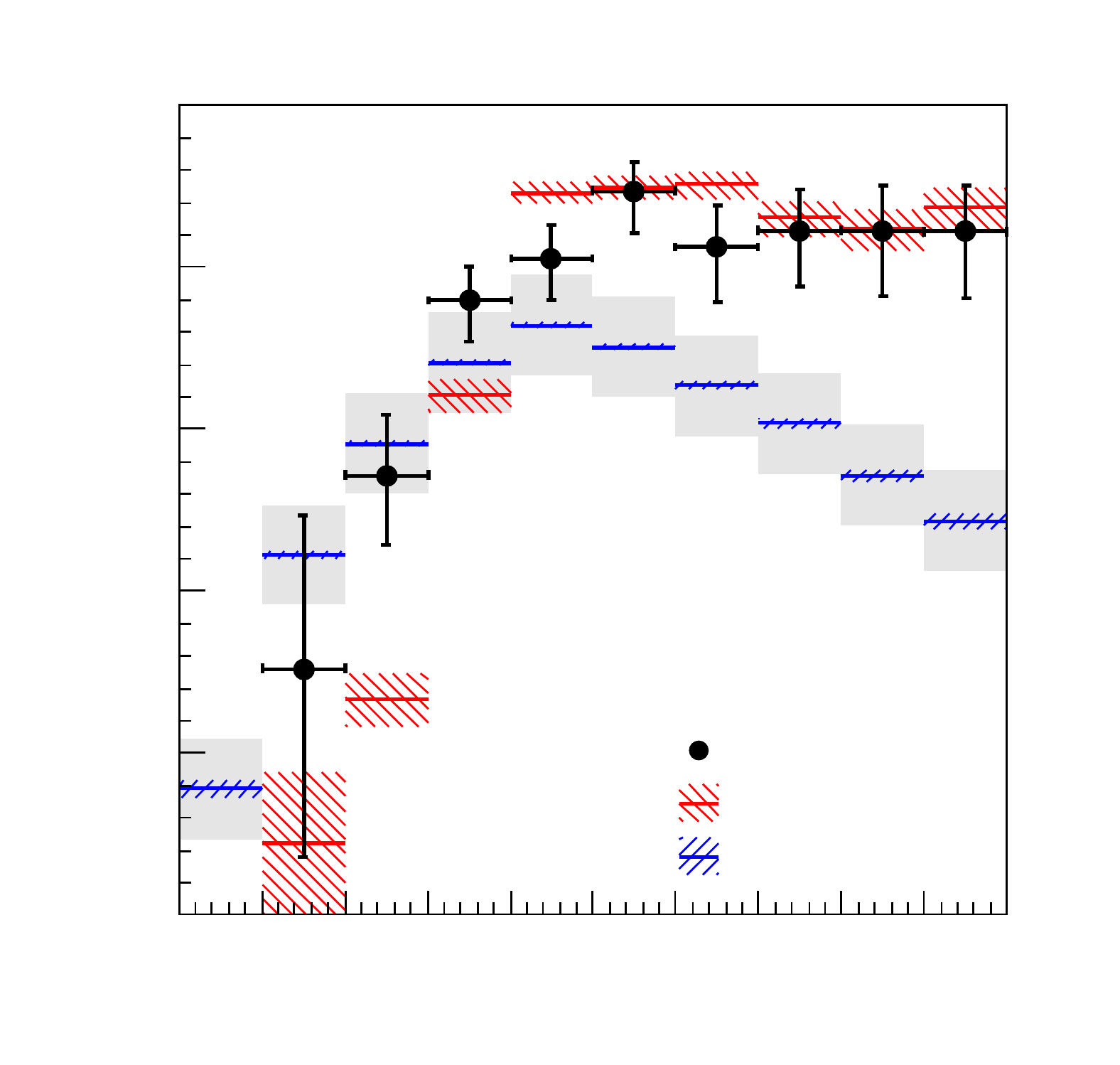} & \svg{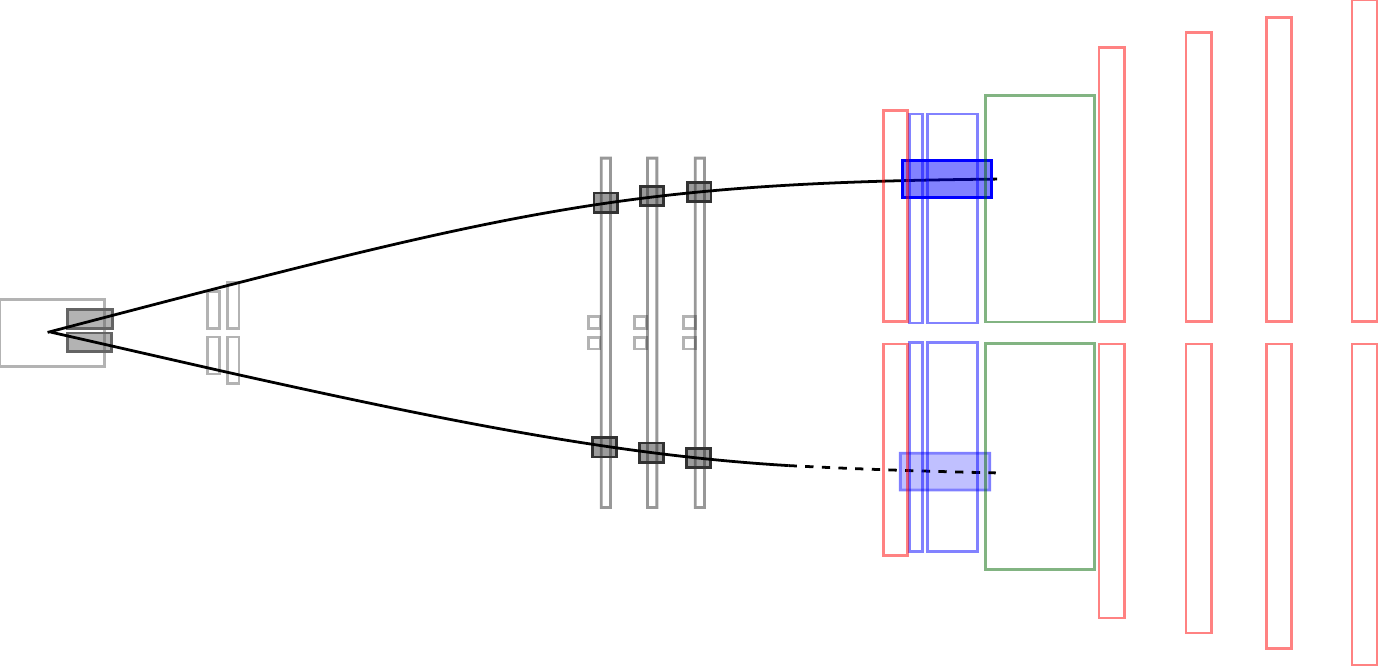} \svgend
\end{subfigures}

The efficiency is then calculated as the number of probes passing the
electron identification requirements over the total number of probes,
and is plotted as a function of probe momemntum in
\fig{Zed:RecEff.EId.Effs}. A large bias is introduced by the
tag-and-probe selection requirements, as can be seen by the
disagreemant between the biased and unbiased efficiencies from
simulation. Relaxing the selection requirements reduces the bias, but
introduces background contamination to the ${\z \to \die}$ signal.

However, the efficiency from data matches the efficiency from biased
simulation, and so the unbiased efficiency from simulation is taken to
describe the inaccessible unbiased efficiency from data. Consequently,
the electron identification efficiency used in \equ{Zed:RecEff} and
given in \tab{Zed:RecEff} is evaluated as the unbiased efficiency from
simulation, and ranges between values of $79\%$ to $93\%$. The
uncertainty is estimated as the difference between the total biased
efficiencies from simulation and data, combined in quadrature with the
statistical uncertainty from simulation. This uncertainty is shown in
\fig{Zed:RecEff.EId.Effs} as the grey band about the unbiased
efficiency.

\begin{subfigures}{2}{Hadron identification efficiency as a function
    of hadron pseudo-rapidity from minimum bias data (points), biased
    simulation (red), and unbiased simulation (blue). The drop-off in
    efficiency at low and high pseudo-rapidity is due to \hcal
    geometrical acceptance.\labelfig{RecEff.HadId}}
  \svgbeg
  \svg[1]{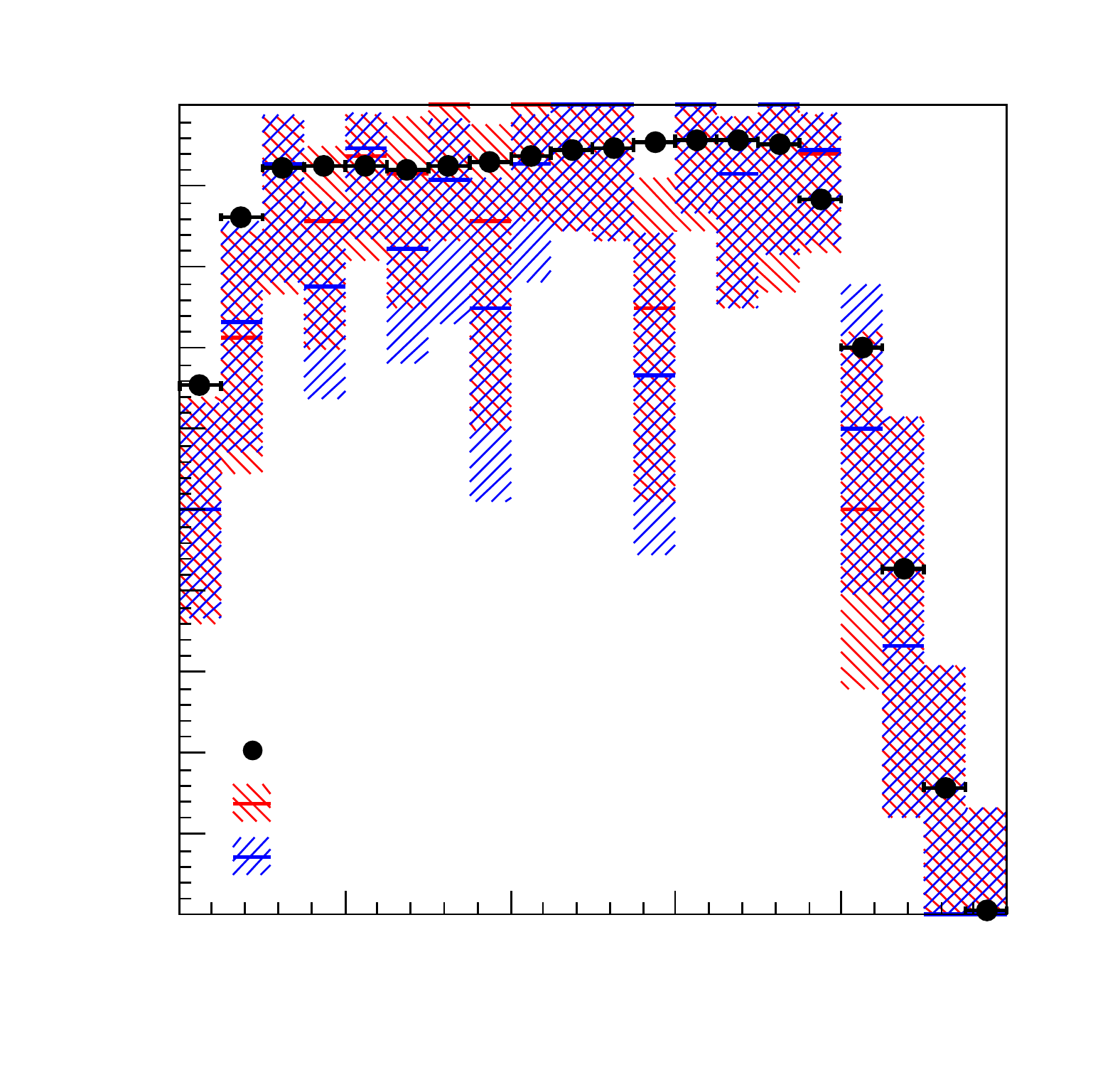} & \sidecaption \svgend
\end{subfigures}

\newsubsubsection{Hadron Identification}{}

The charged hadron identification, $\eff[_\id]_\had$, is the
probability for a hadron with a reconstructed track fulfilling the
requirements of \sec{Zed:Rec} to pass the charged hadron particle
identification requirements. Events selected using a minimum bias
trigger from data are assumed to consist primarily of charged hadrons,
validated with simulation, and so this efficiency is taken as the
percentage of events where the highest \pt track in the event, with a
minimum $\pt$ of ${5~\gev}$ passes the charged hadron identification
requirements. The efficiency as a function of hadron momentum is flat,
but is highly dependent upon $\eta$ due to the acceptance of the
\hcal, as can be seen in \fig{Zed:RecEff.HadId}. Consequently,
$\eff[_\id]_\had$ is evaluated from minimum bias data in
pseudo-rapidity bins of $0.125$ and found to range from $92\%$ to
$96\%$.

\newsubsection{Selection Efficiency}{SelEff}

The event selection efficiency, \eff[_\sel], is the probability for an
event with two reconstructed \wtl decay product candidates to pass the
selection requirements of \sec{Zed:Sel} summarised in
\tab{Zed:Sel}. The event selection efficiency is defined as,
\begin{equation}
  \eff[_\sel] \equiv \eff[_\kin] ~ \eff[_\iso] ~ \eff[_\dphi] ~ \eff[_\ips]
  ~ \eff[_\apt]
  \labelequ{RecSel}
\end{equation}
where each component is the efficiency for a reconstructed event to
pass the corresponding selection of \tab{Zed:Sel}. The invariant mass
selection efficiency is excluded, as this efficiency is by definition
one. However, a kinematic efficiency, \eff[_\kin], is included which
is the probability for a true reconstructable event passing the
kinematic requirements of \sec{Zed:Sel} to have its reconstruction
also fulfil the same kinematic requirements. Each component efficiency
from \equ{Zed:RecSel} is determined from either data or simulation
calibrated to data, with the component values for each event category
tabulated in \tab{Zed:SelEff}. More details on the determination for
each component efficiency are provided in the remainder of this
section.

\begin{table}\centering
  \captionabove{The total selection efficiency and component selection
    efficiencies corresponding to the selection requirements of
    \sec{Zed:Sel}.\labeltab{SelEff}}
  \begin{tabular}{>{$}l<{$}|E|E|E|E|E}
    \toprule
    & \multicolumn{2}{c|}{\mumu} & \multicolumn{2}{c|}{\mue} &
    \multicolumn{2}{c|}{\emu} & \multicolumn{2}{c|}{\muh} &
    \multicolumn{2}{c}{\eh} \\
    \midrule
    \eff[_\sel]
    & 0.138&0.006 & 0.517&0.012 & 0.344&0.016 & 0.135&0.004 & 0.082&0.004 \\
    \midrule
    \eff[_\kin]
    & \multicolumn{2}{c|}{$-$} & 0.993&0.010 & 0.668&0.019 
    & \multicolumn{2}{c|}{$-$} & 0.670&0.013 \\
    \eff[_\iso]
    & 0.660&0.012 & 0.613&0.012 & 0.623&0.020 & 0.413&0.007 & 0.386&0.011 \\
    \eff[_\dphi]
    & 0.848&0.009 & 0.850&0.009 & 0.827&0.015 & 0.845&0.005 & 0.838&0.008 \\
    \eff[_\ips]
    & 0.414&0.011 & \multicolumn{2}{c|}{$-$} & \multicolumn{2}{c|}{$-$}
    & 0.387&0.007 & 0.379&0.011 \\
    \eff[_\apt]
    & 0.597&0.012 & \multicolumn{2}{c|}{$-$} & \multicolumn{2}{c|}{$-$} 
    & \multicolumn{2}{c|}{$-$} & \multicolumn{2}{c}{$-$} \\
    \bottomrule
  \end{tabular}
\end{table}

\newsubsubsection{Kinematic}{}

The kinematic efficiency is found by applying the pseudo-rapidity, \pt,
and invariant mass requirements of \sec{Zed:Sel} to simulated ${\z \to
  \ditau}$ events. The efficiency is defined as the number of events
fulfilling these requirements at the reconstructed level of simulation
divided by the number of events fulfilling the requirements at the
generated level of simulation. For the \mumu and \muh categories
\eff[_\rec] is found to be consistant with unity, as these variables
are well reconstructed for both muons and hadrons.  Because of \ecal
saturation, described in \sec{Exp:RecCal}, the brehmsstralung recovery
for high \pt electrons is incomplete, and their reconstructed momenta
is lower than their generated momenta. This leads to low kinematic
efficiencies for the \emu and \eh categories, on the order of $70\%$,
and an efficiency near one for the \mue category, as given in
\tab{Zed:SelEff}.

\begin{subfigures}{2}{The \pt distribution of ${\z \to \die}$ events
    from data (points), uncalibrated simulation (red), and calibrated
    simulation (blue). The calibrated distribution of the electron \pt
    is scaled by a factor of ${1.02 \pm 0.01}$ and is applied when
    evaluating \eff[_\kin] for the \mue, \emu, and \eh categories.}
  \svgbeg
  \svg[1]{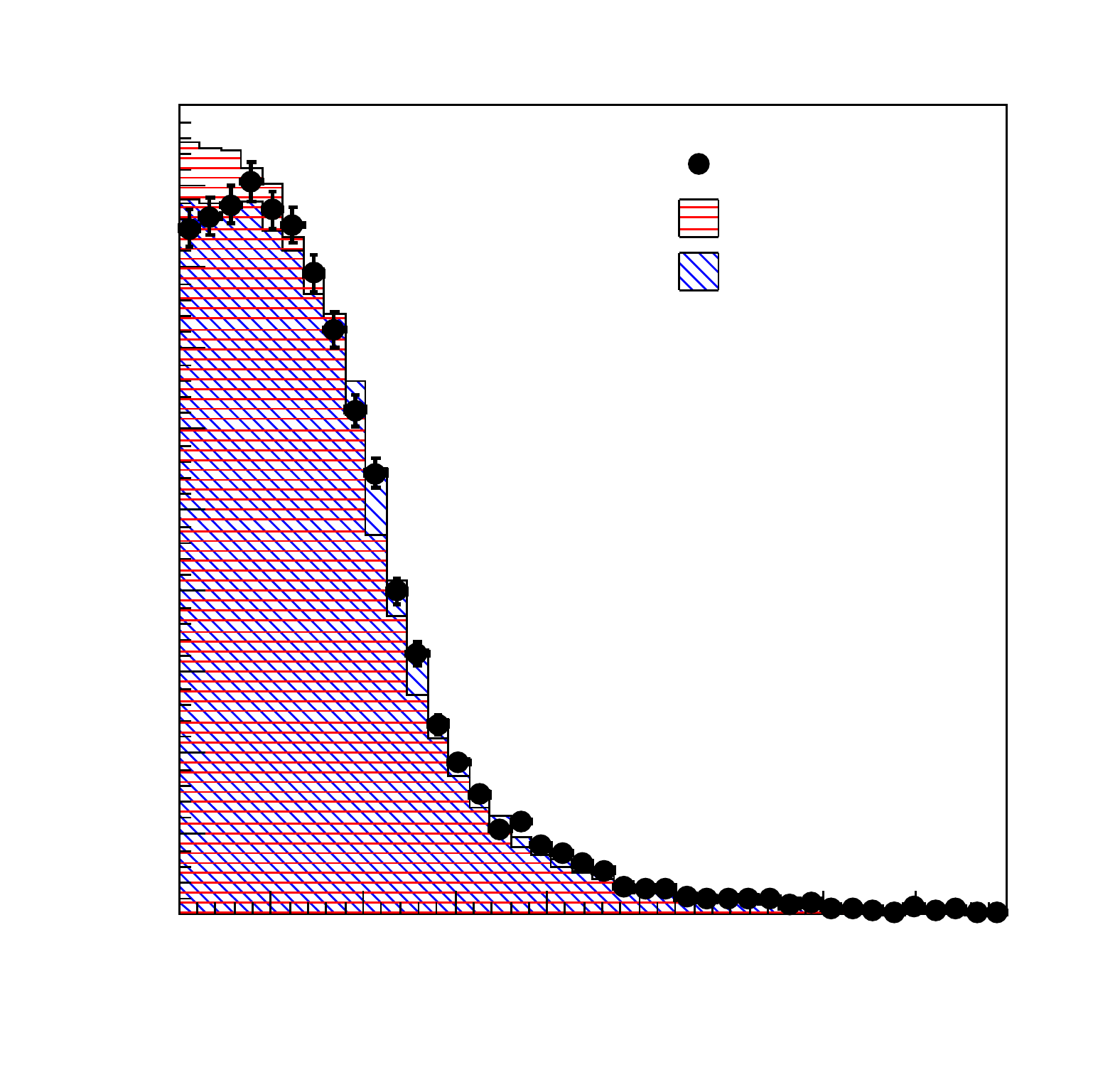} & \sidecaption \svgend
\end{subfigures}

When calculating \eff[_\kin] for the categories containing an
electron, the \pt distribution of the electrons is scaled by a factor
of $1.02 \pm 0.01$. The scale is determined from a fit between the
electron \pt distributions of ${\z \to \die}$ events from simulation
and data, where the procedure for the fit is similar to the invariant
mass fit of \sec{Zed:Sel}; the result of the fit is given in
\fig{Zed:SelEff.Kin}. The associated uncertainties for the \eff[_\kin]
values are estimated as the propogated electron \pt scale uncertainty
combined in quadrature with the statistical uncertainty from
simulation.

\newsubsubsection{Track Isolation}{}
 
The track isolation variable of \equ{Zed:Iso}, \iso, is not well
described by simulation, as previously shown in
\fig{Zed:Sel.Iso.Compare}, due to an underestimation of the underlying
event. However, the \iso distributions for ${\z \to \dimu}$ and ${\z
  \to \mumu}$ events from simulation match, as the underlying event
topologies are identical. Consequently, the track isolation selection
efficiency, \eff[_\iso], for each category is calculated from ${\z \to
  \ditau}$ simulation and calibrated to data by the ratio of the \iso
selection efficiencies from ${\z \to \dimu}$ data to ${\z \to \ditau}$
simulation. The uncertainty for \eff[_\iso] is estimated as the
difference between the \iso selection efficiencies evaluated using
events from ${\z \to \dimu}$ and ${\z \to \mumu}$ simulation.

The \iso selection efficiencies for the \mumu, \mue, and \emu
categories are given in \tab{Zed:SelEff} and range from approximately
\prc{60-70}. For the \mumu and \emu categories, both with an electron
in the final state, the efficiency is slightly lower than the \mumu
category due to contamination of the electron candidate isolation by
pair production from brehmstrahlung photons. The harsher isolation
requirement for the semi-leptonic \muh and \eh categories, ${\iso <
  1~\gev}$, results in an efficiency of approximately $40\%$ which is
slighly lower for the \eh category, again as a result of
brehmstrahlung radiation from the electron.

\newsubsubsection{Azimuthal Angle and Transverse Momentum Asymmetry}{}

Both the variables of azimuthal angle separation and transverse
momentum asymmetry are well described by simulation, as previously
shown in \figs{Zed:Sel.Phi.Compare} and
\ref{fig:Zed:Sel.Apt.Compare}. The two selection efficiencies,
\eff[_\dphi] and \eff[_\apt] are determined from simulation and are
given for each category in \tab{Zed:Sel}. The \eff[_\dphi] is found to
be approximately $85\%$ for all categories, while the \eff[_\apt] is
found to be approximately $60\%$ for \mumu events. The associated
uncertainty for each category is estimated as the difference between
these efficiencies evaluated in ${\z \to \dimu}$ data and simulation,
combined in quadrature with the statistical uncertainty on the
efficiency from ${\z \to \ditau}$ simulation.

\newsubsubsection{Impact Parameter Significance}{}

The impact parameter signficance requirement efficiency, \eff[_\ips],
is calculated from ${\z \to \ditau}$ simulation using the \ips
calibration of \sec{Zed:Sel}, with the agreemant between data and
calibrated simulation previously shown in
\fig{Zed:Sel.Ips.Compare}. The uncertainty on \eff[_\ips] is
determined by re-calculating the efficiency in simulated ${\z \to
  \ditau}$ events, where the calibration factor has been varied within
uncertainty. This uncertainty is the dominant systematic uncertainty
for the cross-section measurement of the \muh category. For the \mumu,
\muh, and \eh categories the \eff[_\ips] is on the order of $40\%$, as
given in \tab{Zed:SelEff}.

\newsubsection{Acceptance and Branching Fractions}{Acc}

Without an acceptance factor, \acc[_{\tau_1\tau_2}], the cross-section
calculation of \equ{Zed:Xs} would yield a measurement dependent upon
the the kinematic requirements of \sec{Zed:Sel} on the
pseudo-rapidities, transverse momenta, and combined invariant mass of
the \wtl decay products for each category. Consequently, to allow
comparison between these cross-section measurements and the ${\dip \to
  \z \to \dimu}$ and ${\dip \to \z \to \die}$ cross-sections of
\rfrs{lhcb.12.1} and \cite{lhcb.13.2}, the acceptance factor corrects
the kinematics for each category to ${60 < \m_\ditau < 120~\gev}$,
${\pt_\tau > 20~\gev}$, and ${2.0 \leq \eta_\tau \leq 4.5}$ with no
\wtl final state radiation. The acceptance factor is taken from ${\z
  \to \ditau}$ simulation and is defined for each category as the
number of events, after electroweak final state radiation, passing the
$\eta$, \pt, and \m requirements of \sec{Zed:Sel} over the number of
events, before electroweak final state radiation, passing the
requirements ${60 < \m_\ditau < 120~\gev}$, ${\pt_\tau > 20~\gev}$,
and ${2.0 \leq \eta_\tau \leq 4.5}$.

Simulation samples are generated for each category using \pythia{8}
\cite{\citepythiaeight} at leading order, \herwig{++}
\cite{\citeherwigpp} at leading order, and \herwig{++} at
next-to-leading order using the \powheg method
\cite{frixione.07.1}. The \wtl decays in \pythia{8} are simulated
using the methods of \chp{Tau} with full spin correlations, while the
\wtl decays in \herwig{++} are also decayed with full spin
correlations and decay models as outlined in
\rfr{grellscheid.07.1}. The \cteq leading-order \PDF set
\cite{lai.99.1} was used with \pythia{8} while the \mstw \PDF sets
\cite{martin.09.1} were used with \herwig{++}. A sufficient number of
events were generated to ensure the statistical uncertainties for each
sammple are much less than the associated systematic uncertainties.

\begin{table}\centering
  \captionabove{The acceptances
    \acc[_{\tau_1\tau_2}], and branching fractions
    \br[_{\tau_1\tau_2}], as a percent, for each
    of the five categories.\labeltab{Acc}}
  \begin{tabular}{>{$}l<{$}|E|E|E|E|E}
    \toprule
    & \multicolumn{2}{c|}{\mumu} & \multicolumn{2}{c|}{\mue} &
    \multicolumn{2}{c|}{\emu} & \multicolumn{2}{c|}{\muh} &
    \multicolumn{2}{c}{\eh} \\
    \midrule
    \acc[_{\tau_1\tau_2}]
    & 0.405&0.006 & 0.248&0.004 & 0.152&0.002 & 0.182&0.002 & 0.180&0.002 \\
    \br[_{\tau_1\tau_2}]
    & 3.031&0.014 & 6.208&0.020 & 6.208&0.020 &16.933&0.056 &17.341&0.057 \\
    \bottomrule
  \end{tabular}
\end{table}

The acceptance factors for each category are given in \tab{Zed:Acc}
and are calculated as the mean of the maximum and minimum acceptances
from the three samples. The uncertainty is taken as half the
difference between the maximum and minimum values. Because the three
samples encompass different \PDF sets, \wtl decay and correlation
mechanisms, hard matrix elements, intial state radiation, and final
state radiation, this uncertainty determination is expected to provide
a conservative estimate.

The \mumu acceptance of \tab{Zed:Acc} is the largest of the five
categories, as both muons can fulfill either of the two \pt
requirements. The remaining acceptances are smaller than the \mumu
acceptance as the two \wtl decay product candidates are not the same
particle type. The \emu acceptance is smaller than the \mue acceptance
because of the additional ${\pt_\mu < 20~\gev}$ requirement. The \muh
and \eh acceptances are also smaller than the \mue acceptance,
primarily from the additional ${2.25 \leq \eta_\had \leq 3.75}$
requirement.

The branching fractions for each category, given in \tab{Zed:Acc}, are
calculated using the world averaged \wtl decay branching fractions of
\rfr{pdg.12.1},
\begin{equation}
  \begin{array}{ll}
    \br[_\mumu] = \br[_{\tauto \mu^-\bar{\nu}_\mu}]\br[_{\tauto
      \mu^-\bar{\nu}_\mu}], &
    \br[_{\mue,\emu}] = 2\br[_{\tauto \mu^-\bar{\nu}_\mu}]\br[_{\tauto
      e^-\bar{\nu}_e}] \equsep
    \br[_{\muh}] = 2\br[_{\tauto \mu^-\bar{\nu}_\mu}]\br[_{\tauto
      \had^-\geq0\had^0}], &
    \br[_{\eh}] = 2\br[_{\tauto e^-\bar{\nu}_e}]\br[_{\tauto
      \had^-\geq0\had^0}] \\
  \end{array}
\end{equation}
where $\br[_{\tauto \had^-\geq0\had^0}]$ is the branching fraction of
a \wtl to a single charged hadron with zero or more neutral
hadrons. The uncertainty on the branching fraction for each category
is propogated from the uncertainties on the \wtl branching fractions,
assuming the uncertainty for each unique channel is uncorrelated and
normally distributed.

\newsection{Results}{Res}

The cross-sections for each category are determined using \equ{Zed:Xs}
and the values presented in \tabs{Zed:Events}, \ref{tab:Zed:RecEff},
\ref{tab:Zed:SelEff}, and \ref{tab:Zed:Acc}. These cross-sections are
calculated for the production of \wzbs, with photon interference, from
proton-proton collisions at ${\sqrt{s} = 7~\tev}$ where the \wzb mass
is between $60$ and $120~\gev$ and the \wzb decays into a \wtl pair,
both within the pseudo-rapidity range ${2.0 \leq \eta \leq 4.5}$ and
with transverse momenta greater than $20~\gev$. A summary of the
systematic uncertainties for these values propogated to percentage
uncertainties on the cross-section is given in
\tab{Zed:Uncertainty}. The results for each event category are,
\begin{equation}
  \begin{aligned}
    \sigma_{\dip \to \z \to \ditau} ~ (\mumu) ~ 
    &=~ 77.4 \pm           10.4 \pm 8.6 \pm 2.7 ~\pb \\
    \sigma_{\dip \to \z \to \ditau} ~ (\mue ) ~ 
    &=~ 75.2 \pm \phantom{1}5.4 \pm 4.1 \pm 2.6 ~\pb \\
    \sigma_{\dip \to \z \to \ditau} ~ (\emu ) ~ 
    &=~ 64.2 \pm \phantom{1}8.2 \pm 4.9 \pm 2.2 ~\pb \\
    \sigma_{\dip \to \z \to \ditau} ~ (\muh ) ~ 
    &=~ 68.3 \pm \phantom{1}7.0 \pm 2.6 \pm 2.4 ~\pb \\
    \sigma_{\dip \to \z \to \ditau} ~ (\eh  ) ~ 
    &=~ 77.9 \pm           12.2 \pm 6.1 \pm 2.7 ~\pb \\
  \end{aligned}
  \labelequ{Categories}
\end{equation}
where the first uncertainty is statistical, the second is systematic,
and the third is due to the uncertainty on the integrated luminosity.

\begin{table}[p]\centering
  \captionabove{Systematic uncertainties expressed as a percentage
    of the cross-section for each $\z \to \ditau$ category. The
    acceptance \acc[_{\tau_1\tau_2}], branching fractions
    \br[_{\tau_1\tau_2}], number of background
    events $N_\bkg$, reconstruction efficiency \eff[_\rec], and
    selection efficiency \eff[_\sel] contributions are
    listed, where the numerical subscripts indicate the first or
    second \wtl decay product candidate. The percentage uncertainties
    on the cross-section for $N_\bkg$ are given for each background
    and the total background. A similar splitting of the efficiency
    uncertainties is also provided.\labeltab{Uncertainty}}
  \begin{tabular}{ll|rrrrr}
    \toprule
    && \multicolumn{5}{c}{$\delta\sigma_{\dip \to \z \to \ditau}~[\%]$} \\
    & & \mumu & \mue & \emu & \muh & \eh \\
    \midrule
    \multicolumn{2}{l|}{\acc[_{\tau_1\tau_2}]}
    & $1.48$  & $1.61$ & $1.32$  & $1.10$ & $1.11$  \\
    \midrule
    \multicolumn{2}{l|}{\br[_{\tau_1\tau_2}]}
    & $0.46$  & $0.32$ & $0.32$  & $0.32$ & $0.33$  \\
    \midrule
    \multicolumn{1}{l|}{\multirow{5}{*}{$N_\bkg$}}
    & \qcd
    & $4.33$  & $0.80$ & $3.08$  & $0.40$ & $0.92$  \\
    \multicolumn{1}{l|}{} & \ewk
    & $4.22$  & $1.54$ & $1.52$  & $0.40$ & $0.72$  \\
    \multicolumn{1}{l|}{} & \ttbar
    & $0.02$  & $0.08$ & $0.12$  & $0.00$ & $0.58$  \\
    \multicolumn{1}{l|}{} & \ww
    & $0.02$  & $0.14$ & $0.13$  & $0.09$ & $0.08$  \\
    \multicolumn{1}{l|}{} & $\z \to \dilep$
    & $8.00$  & $-$    &  $-$    & $0.22$ & $0.23$  \\
    \midrule
    \multicolumn{2}{l|}{Total $N_\bkg$}
    & $10.03$ & $1.75$ & $3.44$  & $0.61$ & $1.32$  \\
    \midrule
    \multicolumn{1}{l|}{\multirow{6}{*}{\eff[_\rec]}}
    & \gec
    & $0.10$  & $0.10$ & $0.10$  & $0.10$ & $0.10$  \\
    \multicolumn{1}{l|}{} & \trg
    & $0.88$  & $0.71$ & $2.29$  & $0.72$ & $4.30$  \\
    \multicolumn{1}{l|}{} & $\trk_1$
    & $0.71$  & $0.74$ & $3.67$  & $0.79$ & $3.67$  \\
    \multicolumn{1}{l|}{} & $\trk_2$
    & $0.34$  & $3.67$ & $0.61$  & $1.76$ & $1.68$  \\
    \multicolumn{1}{l|}{} & $\id_1$
    & $0.38$  & $0.28$ & $1.72$  & $0.29$ & $1.73$  \\
    \multicolumn{1}{l|}{} & $\id_2$
    & $0.78$  & $0.18$ & $0.56$  & $0.03$ & $0.09$  \\
    \midrule
    \multicolumn{2}{l|}{Total \eff[_\rec]}
    & $1.47$  & $4.21$ & $4.73$  & $2.08$ & $6.15$  \\
    \midrule
    \multicolumn{1}{l|}{\multirow{5}{*}{\eff[_\sel]}} 
    & \kin
    & $-$     & $1.04$ & $2.89$  & $-$    & $1.91$  \\
    \multicolumn{1}{l|}{} & \iso
    & $1.79$  & $1.91$ & $3.19$  & $1.65$ & $2.75$  \\
    \multicolumn{1}{l|}{} & \dphi
    & $1.08$  & $1.03$ & $1.86$  & $0.60$ & $0.97$  \\
    \multicolumn{1}{l|}{} & \ips
    & $2.70$  & $-$    &  $-$    & $1.92$ & $2.85$  \\
    \multicolumn{1}{l|}{} & \apt
    & $2.03$  & $-$    &  $-$    & $-$    & $-$     \\
    \midrule
    \multicolumn{2}{l|}{Total \eff[_\sel]}
    & $3.97$  & $2.41$ & $4.69$  & $2.60$ & $4.50$  \\
    \midrule
    \multicolumn{2}{l|}{Total systematic}
    & $11.13$ & $5.41$ & $7.56$  & $3.88$ & $7.88$  \\
    \bottomrule
  \end{tabular}
\end{table}

% See Code/Zed/macro/CombinedFit.m for details.
A fit of the five cross-sections is performed using the method of the
best linear unbiased estimator \cite{lyons.88.1} with further details
given in \sap{Zvr:Fit}. A combined result of,
\begin{equation}
  \sigma_{\dip \to \z \to \ditau} = 72.3 \pm 3.5 \pm 2.9 \pm 2.5
  \labelequ{Combined}
\end{equation}
with a $\chi^2$ per degrees of freedom of $0.40$ is obtained. Each
category is a mutually exclusive dataset, and so the statistical
uncertainties were assumed to be uncorrelated, while the luminosity
and shared reconstruction and selection efficiencies are assumed to be
fully correlated. The theoretical cross-section for $\sigma_{\dip \to
  \z \to \ditau}$ is ${74.3_{-2.1}^{+1.9}~\pb}$, and was calculated at
next-to-next-to-leading order using \dynnlo \cite{\citedynnlo} with
the \mstw NNLO \PDF set \cite{martin.09.1}.

\begin{subfigures}[p]{1}{A comparison between the \lhcb ${\dip \to \z
      \to \dimu}$, ${\dip \to \z \to \die}$, and ${\dip \to \z \to
      \ditau}$ cross-section measurements divided by their expected
    \sm theoretical values. The equivalent ${\dip \to \z \to \ditau}$
    measurements from \atlas and \cms are also provided. The red
    points indicate the muon, electron, or combined \wtl decay
    channels of the \wzb while the black points represent the
    individual \wtl categories. The blue bar provides the theoretical
    uncertainty, centred about unity. The dark inner error bars
    correspond to the statistical uncertainty for the individual
    measurements, while the light outer error bars are the combined
    systematic and luminosity uncertainties. \labelfig{Results}}
  \svgbeg
  \includesvg[width=\columnwidth]{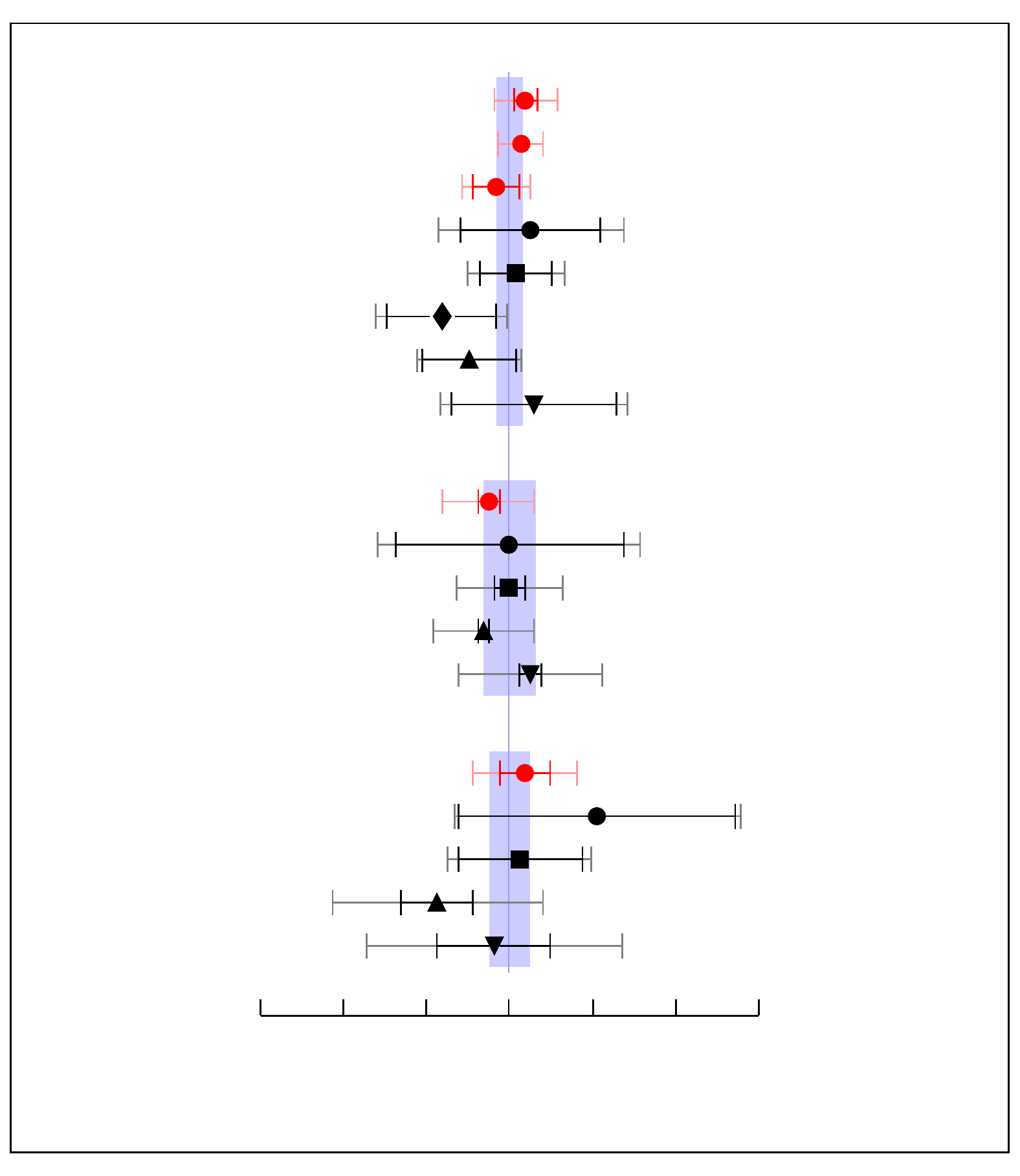} \svgend
\end{subfigures}

A graphical comparison between the combined result of
\equ{Zed:Combined} and the individual results of \equ{Zed:Categories}
is given in \fig{Zed:Results} where the values are expressed as the
ratio of experiment to theory. The theoretical uncertainty is given by
the blue band centered about unity, while the statistical uncertainty
is given by the dark inner error bars and the systematic and
luminosity uncertainty are given by the light outer error bars for
each point. Each decay channel of the \wzb is highlighted in red,
while the \wtl decay categories for the ${\z \to \ditau}$ channels are
given in black. Comparisons to the ${\dip \to \z \to \dimu}$
\cite{lhcb.12.1} and ${\dip \to \z \to \die}$ \cite{lhcb.13.2}
cross-sections from \lhcb are made, as well as the ${\dip \to \z \to
  \ditau}$ cross-section measurements from \atlas \cite{atlas.11.1,
  atlas.12.1} and \cms \cite{cms.11.1}.

All measurements, within uncertainty, are consistant with their
corresponding \sm theory predictions. Of the three combined ${\dip \to
  \z \to \ditau}$ measurements, the \lhcb measurement of this chapter
is the most precise, primarily due to a much lower systematic
uncertainty than either the \atlas or \cms measurements, specifically
in the semi-hadronic final states. Both \atlas and \cms reconstruct
hadronic final states of the \wtl decay using jet reconstruction
algorithms, resulting in large uncertainties on the jet energy scale
and identification efficiency. The combined cross-section uncertainty
for the \lhcb measurement is $7.2\%$ while the \atlas uncertainty is
$9.5\%$ and the \cms uncertainty is $10.2\%$.

The efficacy of the \velo subdetector in separating ${\z \to \ditau}$
events in the \mumu category from ${\z \to \dimu}$ events is clear, in
comparison to the \mumu results from \atlas and \cms. In this
category, \lhcb achieves a signal purity of approximately $70\%$ while
\atlas attains a purity of $50\%$ and \cms a purity of $60\%$. The
purity of the \lhcb \mue category is slightly reduced to the \atlas
and \cms measurements as \met cannot be reconstructed within \lhcb and
used to further reduce the \qcd and \ewk backgrounds. The purities for
the semi-leptonic \muh and \eh categories are approximately equivalent
between the three experiments.

The lepton universality test of \equ{Zed:Lep} can be performed,
\begin{equation}
  \frac{\sigma_{\dip \to \z \to \dimu}}{\sigma_{\dip \to \z \to \die}}
  = 1.01 \pm 0.08 \equcomma
  \frac{\sigma_{\dip \to \z \to \ditau}}{\sigma_{\dip \to \z \to \die}}
  = 0.95 \pm 0.07
  \labelequ{RatioEE}
\end{equation}
using the combined result of this chapter and the ${\dip \to \z \to
  \dimu}$ and ${\dip \to \z \to \die}$ cross-section measurements of
\lhcb. Here, the luminosity uncertainty is assumed to be fully
correlated as the ${\z \to \ditau}$ and ${\z \to \die}$ analyses were
both performed using the same luminosity measurement. Additionally the ratio,
\begin{equation}
  \frac{\sigma_{\dip \to \z \to \ditau}}{\sigma_{\dip \to \z \to \dimu}}
  = 0.94 \pm 0.09
  \labelequ{RatioMuMu}
\end{equation}
can be calculated where all uncertainties between the two
cross-sections are assumed to be uncorrelated as the ${\z \to \dimu}$
\lhcb analysis uses the $2010$ dataset and not the $2011$ dataset of
the ${\z \to \ditau}$ analysis. All three ratios from
\equs{Zed:RatioEE} and \ref{equ:Zed:RatioMuMu} verify lepton
universality under the unique conditions observed by \lhcb.

\newchapter{Higgs Boson Limits}{Hig}

The upper limits on the production of neutral \whbs decaying into \wtl
pairs using the data from the analysis of \chp{Zed} are presented
within this chapter. The \whb phenomenology needed to determine the
event model is introduced in \sec{Hig:Phe} while the event model
itself is described in \sec{Hig:Mod}, the statistical methods used to
calculate the limits are outlined in \sec{Hig:Sta}, and the limits are
presented in \sec{Hig:Res}.

Further investigation of the boson with a mass of approximately
$125~\gev$, discovered by \atlas \cite{atlas.12.2} and \cms
\cite{cms.13.1} is required to determine whether the properties of the
boson match those of a \whb from the standard model (\sm),
supersymmetry (\susy), or other models beyond the \sm.  Within both
the \sm and the minimal supersymmetric model (\mssm), the \whb is
predicted to couple to fermions with a strength proportional to the
mass of the fermion, as previously shown by the vertices of
\fig{Thr:H02FF}, and \figs{Thr:H12FuFu} through
\ref{fig:Thr:H4m2FF}. Consequently, the neutral \whbs, whether \sm or
\mssm, are expected to decay into \wtl pairs over two orders of
magnitude more often than into muon pairs, and seven orders of
magnitude more often than into electron pairs.

Within this chapter the neutral \sm \whb is denoted by \hH, while for
the \mssm \whbs the light \cp-even \whb is denoted by \hhz, the heavy
\cp-even \whb by \hHz, the \cp-odd \whb by \hAz, and the charged \whbs
by \hHpm. Any neutral \whb, whether \sm, \mssm, or otherwise, is
indicated by \h, while any neutral \mssm \whb is indicated by
\hz. Model independent upper limits on the production for a neutral
\whb decaying into a \wtl pair, $\sigma_{\dip \to \h \to \ditau}$,
within the \lhcb acceptance, ${2.0 \leq \eta_\tau \leq 4.5}$, as a
function of the mass of the \whb are set using both the individual
event categories of \chp{Zed} and their combination. The combined
limit is compared to the expected cross-section from the \sm
\whb. Model dependent limits on \tanb are set for the three neutral
\mssm \whbs decaying into \wtl pairs as a function of the mass of the
\cp-odd \whb. These limits are also set using the individual
categories and their combination.

The \mssm limits are set using the \mhmax scenario of
\rfr{carena.03.1} where the parameter space is selected to maximise
the mass of the light \cp-even \whb. This benchmark provides the most
conservative limits on \tanb for a given mass of the \cp-odd \whb and
is commonly used amongst experiments when reporting \mssm limits. This
allows the limits of this chapter to be compared to results from
\atlas, \cms, and \dlep. The \sm parameters for this scenario are set
at $172.5~\gev$ for the \wtq mass $\m_t$, $4.213~\gev$ for the bottom
quark mass $\m_b$ using the \msbar scheme, and $0.119$ for the strong
coupling $\aS(\m_\z)$. The \susy parameters are set at $1~\tev$ for
the soft \susy-breaking mass $\m_\susy$, $2~\tev$ for the stop mixing
parameter $X_t$, $200~\gev$ for the \su[2] gaugino mass parameter
$\m_2$, $200~\gev$ for the Higgs mixing parameter $\mu$, and
$800~\gev$ for the gluino mass parameter $\m_3$.

\newsection{Higgs Phenomenology}{Phe}

To set limits on the production of \whbs, both the production and
decay of the \whbs must be known. Within this section the branching
fractions are given in \sec{Hig:Br} and the cross-sections in
\sec{Hig:Xs} for both the \sm \whb and the neutral \mssm \whbs. The
branching fractions and cross-sections depend upon the mass of the
\whb, and for the \mssm, also depend upon \tanb. However, the masses
of the light and heavy \cp-even \whbs of the \mssm can be written in
terms of the \cp-odd \whb mass, $\m_\hAz$, and \tanb. Consequently,
the branching fractions and cross-sections for the \sm are given as a
function of only $\m_\hH$, while the \mssm branching fractions and
cross-sections are given as a function of $\m_\hAz$ and \tanb.

In \fig{Hig:Mssm.Mass} the light and heavy \cp-even \whb masses are
plotted as a function of the \cp-odd \whb mass and \tanb. The masses
are calculated using the program \feynhiggs~\cite{\citefeynhiggs}
which performs the calculations up to the order $\aE\aS$. The features
of these mass functions can be understood at tree-level using the
relations,
\begin{align*}\labelali{Mssm.Mass}
  \m_\hhz^2 &= \frac{1}{2} \left(\m_\hAz^2 + \m_\z^2 - \left(
      (m_\hAz^2 - m_\z^2)^2 +
      4\m_\z^2\m_\hAz^2\sinb[^22]\right)^{\frac{1}{2}} \right) \\
  \m_\hHz^2 &= \frac{1}{2} \left(\m_\hAz^2 + \m_\z^2 + \left(
      (m_\hAz^2 - m_\z^2)^2 +
      4\m_\z^2\m_\hAz^2\sinb[^22]\right)^{\frac{1}{2}} \right) \\
  \m_\hHpm^2 &= \m_\hAz^2 + \m_\w^2
\end{align*}
which are determined from the \mssm \whb mass eigenstate matrix of
\equ{Thr:HiggsMass}. Here, $\mu$ is the Higgs mixing parameter as
described in \sec{Thr:Alt}.

\begin{subfigures}{2}{Masses of the \subfig{Mssm.M.H1}~light \cp-even
    and \subfig{Mssm.M.H2}~heavy \cp-even \mssm \whbs as a function of
    the \cp-odd \whb mass and \tanb, calculated using
    \feynhiggs~\cite{\citefeynhiggs}.\labelfig{Mssm.Mass}}
  \svgbeg
  \svg{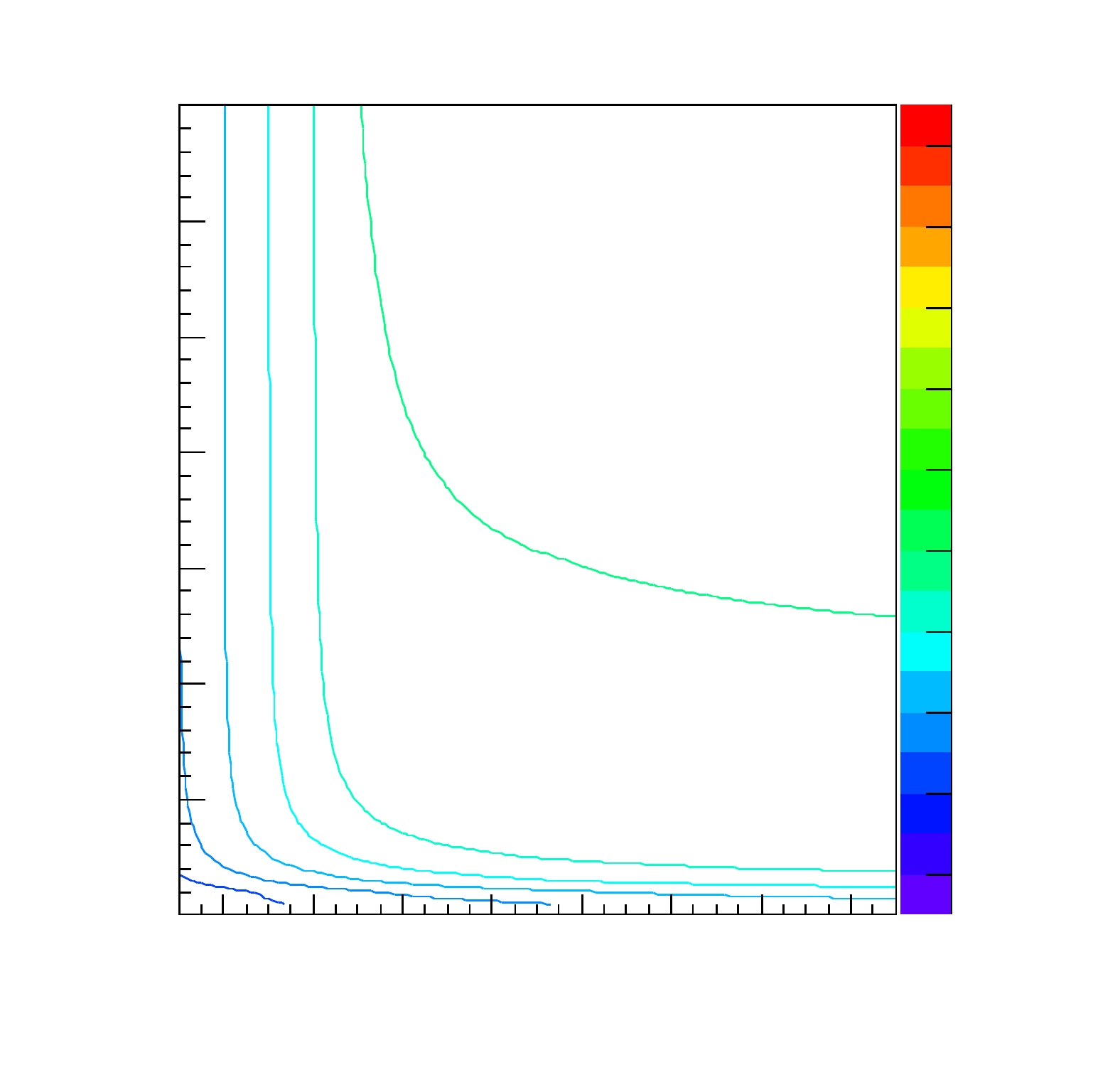} & \svg{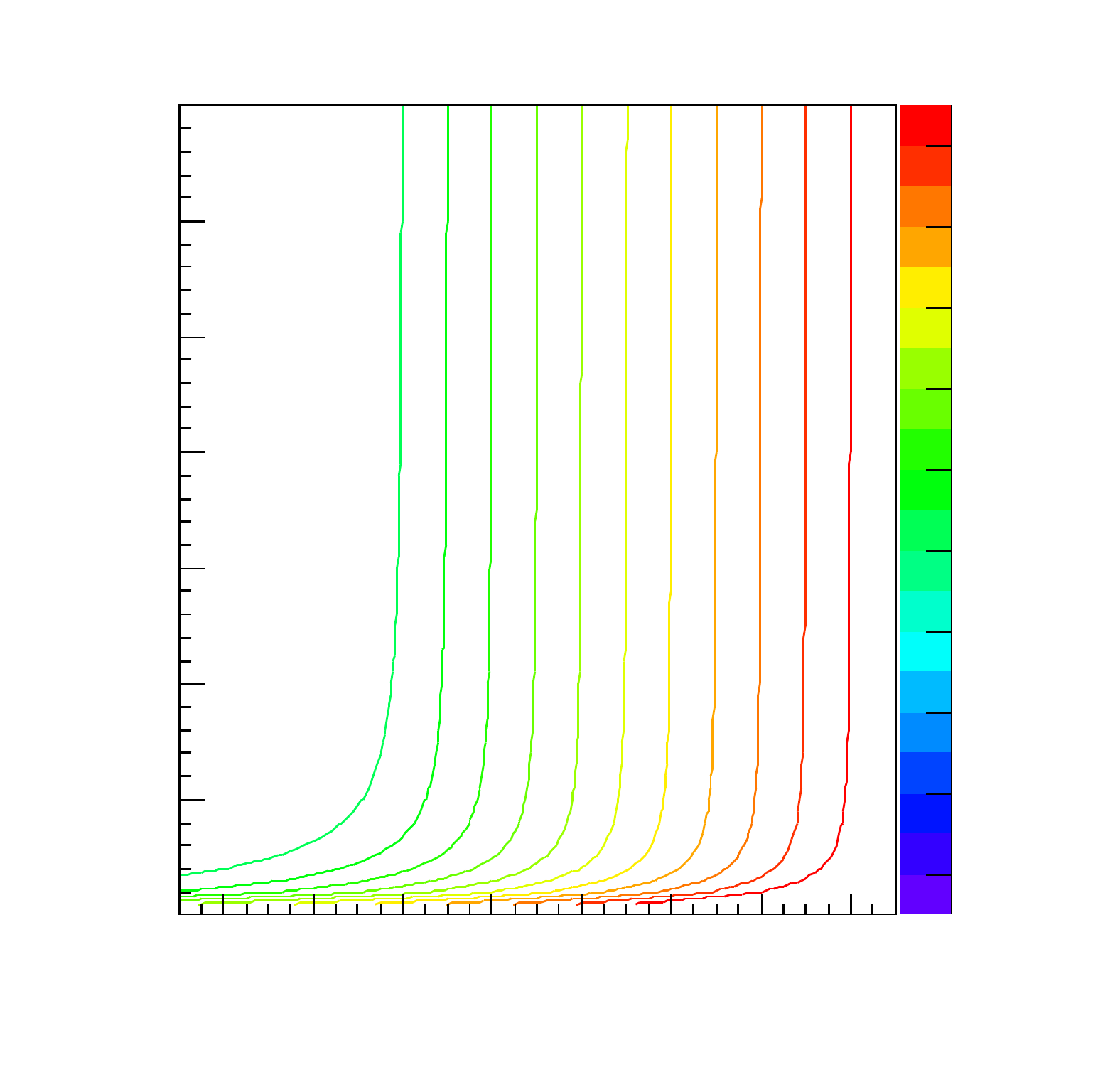} \svgend
\end{subfigures}

The relations of \equ{Hig:Mssm.Mass} result in a light \cp-even \whb
mass that must be less than or equal to $\m_\z\abs{\cosb[2]}$, a heavy
\cp-even \whb mass that must be greater than or equal to $\m_\z$, and
a charged \whb mass that must be greater than $\m_\w$. The mass of the
\cp-odd \whb is bounded by the light and heavy \cp-even masses,
$\m_\hhz \leq \m_\hAz \leq \m_\hHz$, and for large $\m_\hAz$ the mass
of the heavy \cp-even \whb approaches that of the \cp-odd \whb.

In \fig{Hig:Mssm.Mass} the light \cp-even \whb mass plateaus at
$130~\gev$ for larger $\m_\hAz$ rather than $\m_\z$, as heavy quark
and squark loops provide sizable corrections at the one-loop
level. The heavy \cp-even \whb mass approaches the maximum $\m_\hhz$
at low $\m_\hAz$ as expected, and for $\m_\hAz > 140~\gev$ is nearly
degenerate with the \cp-odd \whb at \tanb values greater than
$20$. However, for a given $\m_\hAz$, the heavy \cp-even \whb mass
increases asymptotically as \tanb approaches zero.

\newsubsection{Branching Fractions}{Br}

The decay width for the \sm or \mssm \whbs is calculated as the
sum of the partial decay widths,
\begin{equation}
    \Gamma_\h = \sum_{i}\Gamma_{\h \to f_i\bar{f}_i} + \Gamma_{\h \to gg}
    + \Gamma_{\h \to \gamma\gamma} + \Gamma_{\h \to \z\gamma} +
    \Gamma_{\h \to \z\z} + \Gamma_{\h \to \w\w}
    \labelequ{Sm.Width.Total}
\end{equation}
where the summation is over muon, \wtl, \wsq, \wcq, \wbq, and \wtq
pairs, and the remaining terms are the possible gauge boson
combinations. The branching fraction for a given channel is then,
\begin{equation}
  \br[_{\h \to X}] = \frac{\Gamma_{\h \to X}}{\Gamma_\h}
\end{equation}
where $\Gamma_{\h \to X}$ is the partial decay width for the
channel. First the \sm \whb branching fractions are described,
followed by a description of the \mssm \whbs branching fractions.

\newsubsubsection{SM Branching Fractions}{}

Over the \sm \whb mass range considered in this chapter, ${90 < \m_\h
  < 250~\gev}$, the \wtl, \wbq, gluon, \wwb, and \wzb pair partial
widths dominate the total width of the \sm \whb. The branching
fractions for these channels as a function of the \sm \whb mass are
plotted in \fig{Hig:Sm.Br}. These branching fractions, with
uncertainties, are calculated following the prescription of
\rfr{denner.11.1} which uses the programs \hdecay~\cite{\citehdecay}
and \prophecy~\cite{\citeprophecy}. Further details on these
calculations are given in \sap{Hvr:BrSm}, as well as tabulated values
for the $\hH \to \ditau$ branching fraction. The features of the
branching fractions plotted in \fig{Hig:Sm.Br} can be understood from
the simpler tree-level calculations of the partial widths which can be
determined using the vertices of \fig{Thr:Vertices.Sm.Higgs},
\equ{Thr:DecayWidth}, and the methods outlined in \chp{Thr}.

\begin{subfigures}[t]{2}{Branching fractions, as a percentage, for the
    five leading decay channels of the \sm \whb in the relevant mass
    range: $\tau\tau$ (red), $b\bar{b}$ (blue), $gg$ (green), $\ww$
    (orange), and $\z\z$ (magenta). The branching fractions and
    uncertainties, indicated by the coloured bands, are calculated using
    the results of \rfr{denner.11.1} and plotted as a function of the
    \whb mass.}
  \svgbeg
  \svg[1]{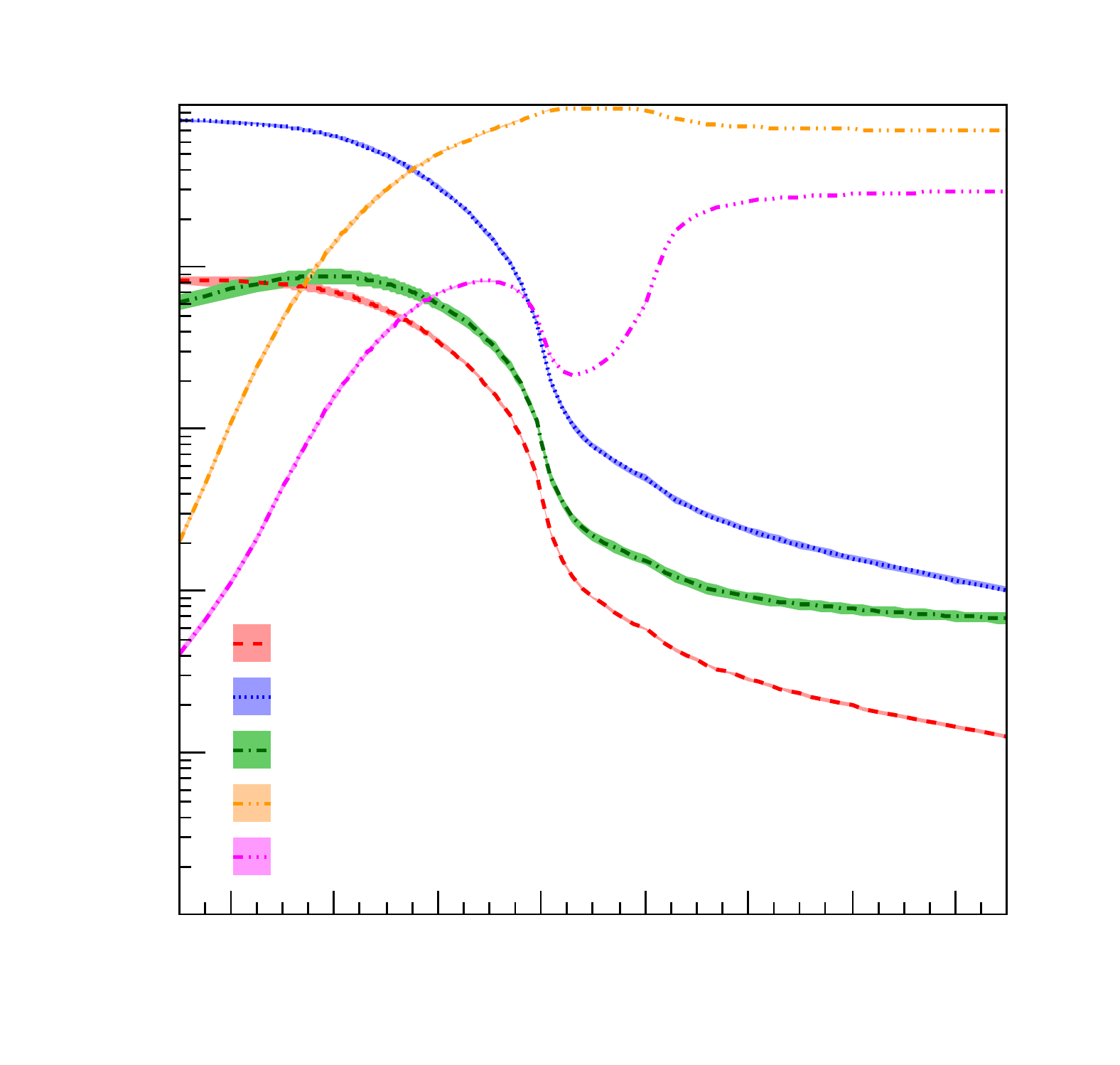} & \sidecaption \svgend
\end{subfigures}

For fermions above the threshold ${\m_\hH > 2\m_f}$ the relevant
vertex is \fig{Thr:H02FF} and the tree-level decay width from
\rfr{resnick.73.1} is,
\begin{equation}
  \Gamma_{\hH \to f\bar{f}} = \frac{N_c \gW[^2] \m_f^2}{32 \pi \m_\w^2}
  \left(1 -  \frac{4\m_f^2}{\m_\hH^2}\right)^{\frac{3}{2}}\m_\hH
  \labelequ{Sm.Width.FF}
\end{equation}
where $N_c$ is the number of colours, $1$ for leptons and $3$ for
quarks, and $\gW$ is the \su[2] gauge coupling. For the \whb masses
considered, the \wtl and \wbq channels dominate the fermion partial
width, and the \wbq pair partial width is expected to be a factor of
${3\m_b^2/\m_\tau^2 \approx 15}$ times larger than the \wtl pair width
as can be seen in \fig{Hig:Sm.Br}. Electron, \wuq, and \wdq pairs are
not considered in the calculation of the total decay width, as their
partial widths are less than six orders of magnitude smaller than the
${\hH \to b\bar{b}}$ width resulting in a maximum branching fraction
of less than $10^{-4}\%$.

For on-shell vector bosons, \w and \z, the corresponding vertices are
\figs{Thr:H02WW} and \ref{fig:Thr:H02ZZ} and the tree-level decay
width from \rfr{lee.77.1} is,
\begin{equation}
  \Gamma_{\hH \to VV} = \frac{N_m \gW[^2]}{64 \pi \m_\w^2}
  \left(1 - \frac{4\m_\vb^2}{\m_\hH^2}\right)^{\frac{1}{2}}
  \left(1 - \frac{4\m_\vb^2}{\m_\hH^2} +
    \frac{3}{4}\left(\frac{4\m_\vb^2}{\m_\hH^2}\right)^2 \right) \m_\hH^3
  \labelequ{Sm.Width.VV}
\end{equation}
for a \whb mass above threshold, ${\m_\hH > 2\m_\vb}$, where $N_m$ is a
multiplicity factor of $1/2$ for the \wzb and $1$ for the \wwb. This
factor results in a \wwb pair partial width approximately double that
of the \wzb pair width for ${\m_\hH > 2\m_\z}$, as can be seen in
\fig{Hig:Sm.Br}. Additionally, the $\m_\hH^3$ term ensures that the
partial widths of \equ{Hig:Sm.Width.VV} are on the order of $\m_\hH^2$
times larger than the fermionic partial widths for large $\m_\hH$.

However, the on-shell decay width of \equ{Hig:Sm.Width.VV} is not
sufficient to describe the full \w and \wzb branching fractions. It is
also necessary to include decays where one or both of the vector bosons
are off-shell. From \rfr{keung.84.1}, this partial width can be
written as,
\begin{equation}
  \newcommand{\x}[1][]{x^{#1}}
  \begin{aligned}
    \Gamma_{\hH \to \vb\vb^*}  = \frac{F_\vb\gW[^4]}{\pi^3}
    \Bigg(&\frac{3\left(1 - 8\x[2] +
          20\x[4]\right)} {\sqrt{4\x[2] - 1}}
      \cos^{-1}\left(\frac{3\x[2] - 1}{2\x[3]}\right) \\
      & - 3\left(1 - 6\x[2] + 4\x[4]\right)\ln\left(\x\right) \\
      &- \left(1 - \x[2]\right) \left(\frac{47}{2}\x[2] - \frac{13}{2}
        + \x[-2]\right)
    \Bigg) \m_\hH \\
  \end{aligned}
  \labelequ{Sm.Width.VVstar}
\end{equation}
where $x$ is ${\m_\vb / \m_\hH}$ and the pre-factors $F_\vb$ are,
\begin{equation}
  F_\w = \frac{3}{512}\equcomma F_\z = \frac{7 - \frac{40}{20}\sintw[^2]
    + \frac{160}{9}\sintw[^4]}{2048\,\costw[^4]}
\end{equation}
for the \wwb and \wzb respectively; here, \tw is the weak-mixing
angle. Over the mass range $2\m_\w < \m_\hH < 2\m_\z$ the partial
width for the \wwb pair channel is given by \equ{Hig:Sm.Width.VV} while
the width for the \wzb pair channel is given by \equ{Hig:Sm.Width.VVstar}
resulting in a dip in the $\hH \to \z\z$ branching fraction, which is
clearly visible in \fig{Hig:Sm.Br}. Over this range the \wwb pair
partial width is anywhere between ten to fifty times larger than the
\wzb pair width. For \whb masses below the mass of the \w and \wzbs,
the double off-shell width is necessary, which is not discussed here,
but can be found in \rfr{spira.97.1}.

The decay widths for the additional gauge boson combinations, gluon
pairs, photon pairs, and \wzbs with photons, do not have tree-level
diagrams due to the massless gluon and photon. Of these three
channels, only the gluon pair channel provides a considerable
contribution to the total decay width for the \whb masses considered
here. From \rfr{gunion.90.1}, the width for \whb decays into gluon
pairs can be written as,
\begin{equation}
  \Gamma_{\hH \to gg} = \frac{\gS[^4]\gW[^2]}{2048\pi^5\m_\w^2} \left(
    \sum_i \big(x_i + (x_i - x_i^2)F(x_i)\big)\right) \left(
    \sum_i \big(x_i + (x_i - x_i^2)F(x_i)\big)\right)^\dagger \m_\hH^3
  \labelequ{Sm.Width.GG}
\end{equation}
where \gS is the \su[3] gauge coupling and $x_i$ is $4\m_i^2/\m_\hH^2$
for fermion $i$. The summations are over all contributing fermion
loops where,
\begin{equation}
  F(x) = \begin{cases}
    \left(\sin^{-1} x^{-\frac{1}{2}}\right)^2 & \mathrm{if}~x \geq 1 \\
    \frac{1}{4}\left(\pi + i\ln\left(\frac{2+2\sqrt{1-x} -
          x}{x}\right)\right)^2 &
    \mathrm{else} \\
  \end{cases}
\end{equation}
and only massive quarks are considered, with the \wtq loop dominating
the decay width. From the ratio of \equ{Hig:Sm.Width.GG} to
\equ{Hig:Sm.Width.FF} for both \wtl and \wbq pairs, one can see that
for the mass range considered in \fig{Hig:Sm.Br}, the gluon pair width
will remain above the \wtl pair width, but below the \wbq pair width.

\newsubsubsection{MSSM Branching Fractions}{}

The branching fractions for the \hhz, \hHz, and \hAz \whbs decaying
into a \wtl pair as a function of the \cp-odd \whb mass, $\m_\hAz$,
and \tanb are plotted in \fig{Hig:Mssm.Br}. These branching fractions
are calculated using the programs \hdecay and \prophecy, and dressed
with \mssm couplings from \feynhiggs following the recommendations of
\rfr{denner.11.1}. Details on the calculation, as well as plots with
numerical values of these branching fractions, are provided in
\sap{Hvr:BrMssm}. Just as for the \sm \whb, the features of
\fig{Hig:Mssm.Br} can be understood using tree-level calculations of
the partial widths.

\begin{subfigures}[t]{2}{Branching fractions, as a percentage, for
    \subfig{Mssm.Br.H1}~light \cp-even, \subfig{Mssm.Br.H2}~heavy
    \cp-even, and \subfig{Mssm.Br.H3}~\cp-odd \mssm \whbs decaying
    into \wtl pairs as a function of the \cp-odd \whb mass and
    \tanb. The \mssm couplings are calculated with
    \feynhiggs.\labelfig{Mssm.Br}}
  \svgbeg
  \svg{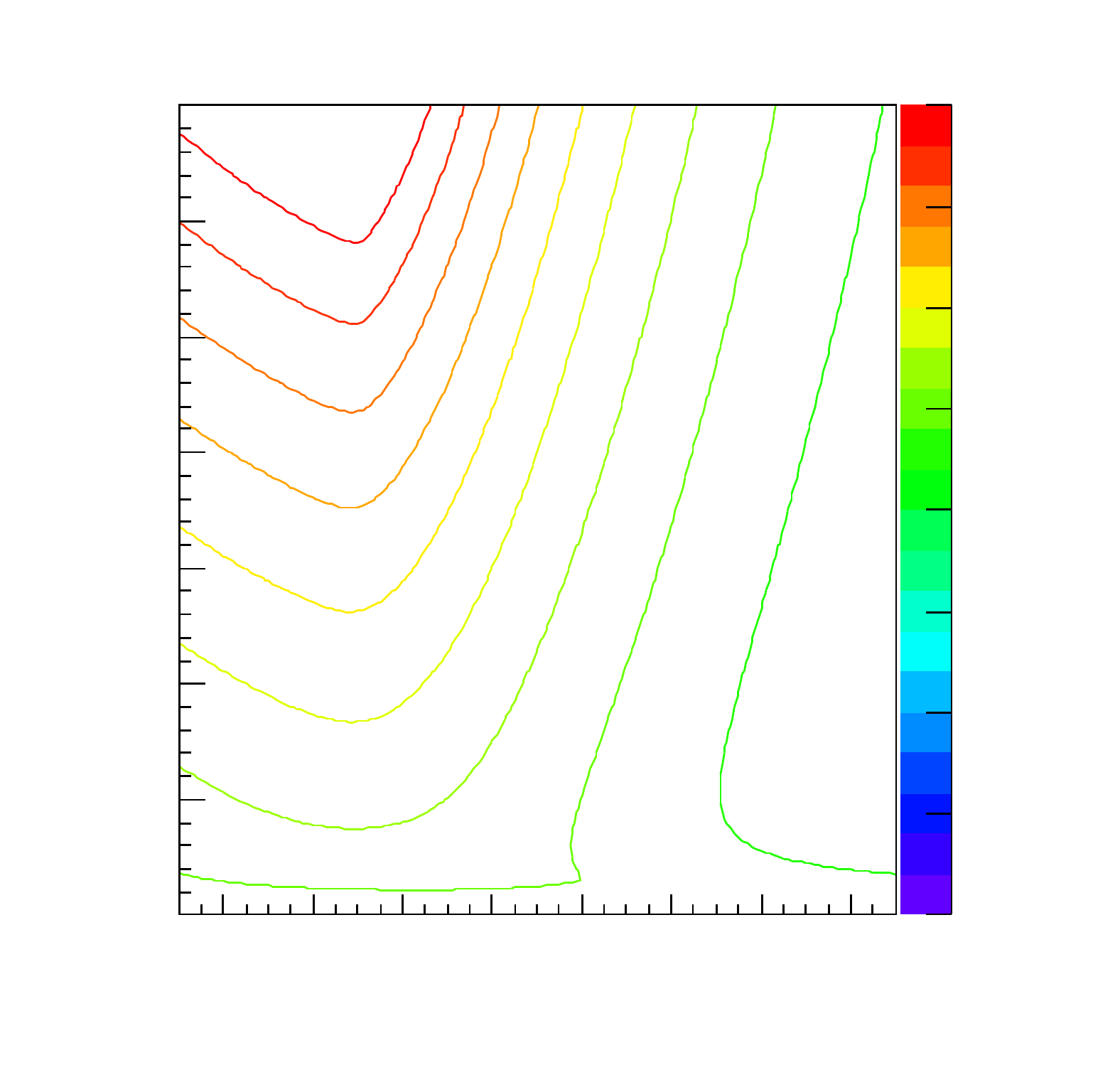} & \svg{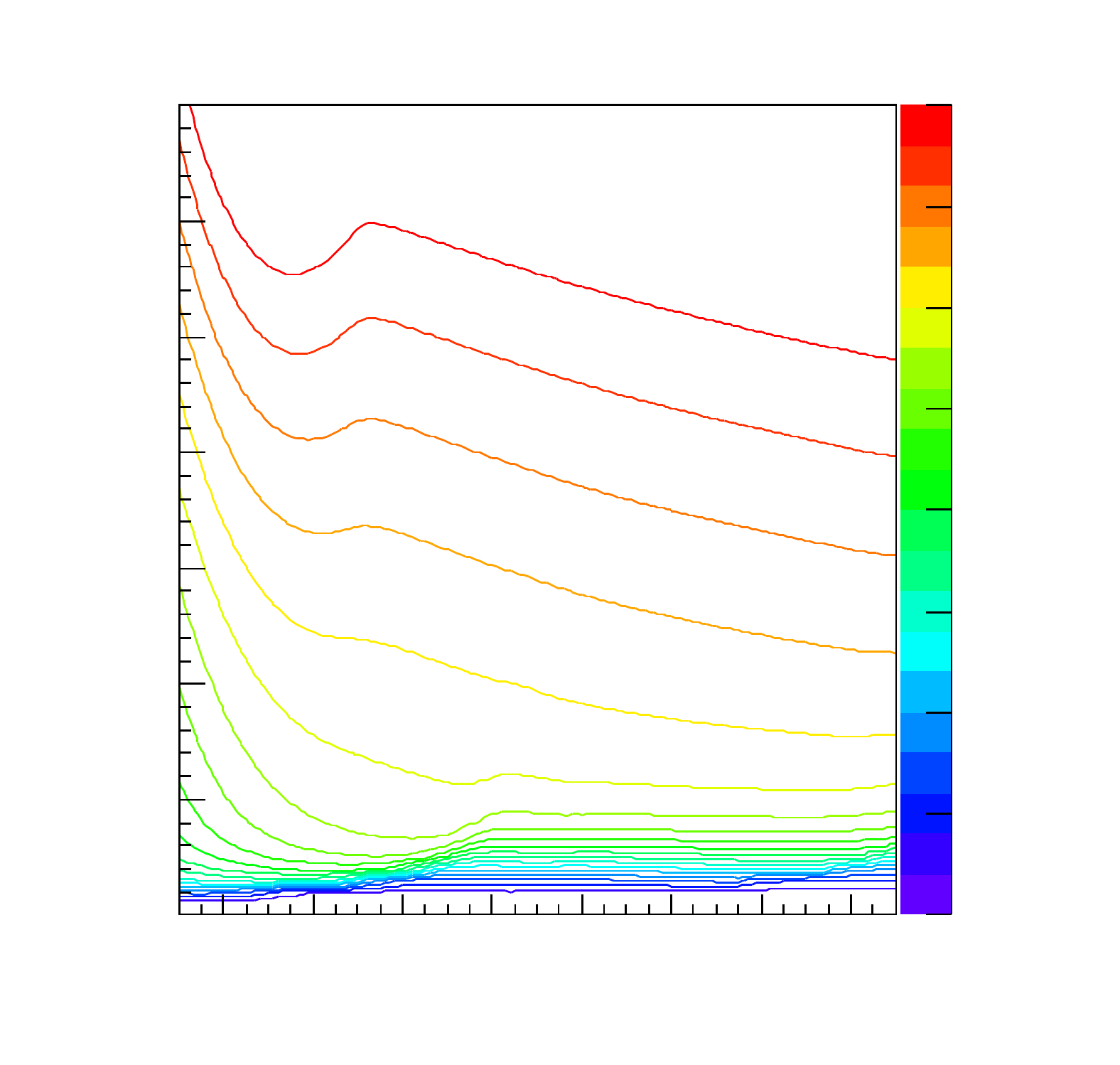} \svgsep
  \sidecaption     & \svg{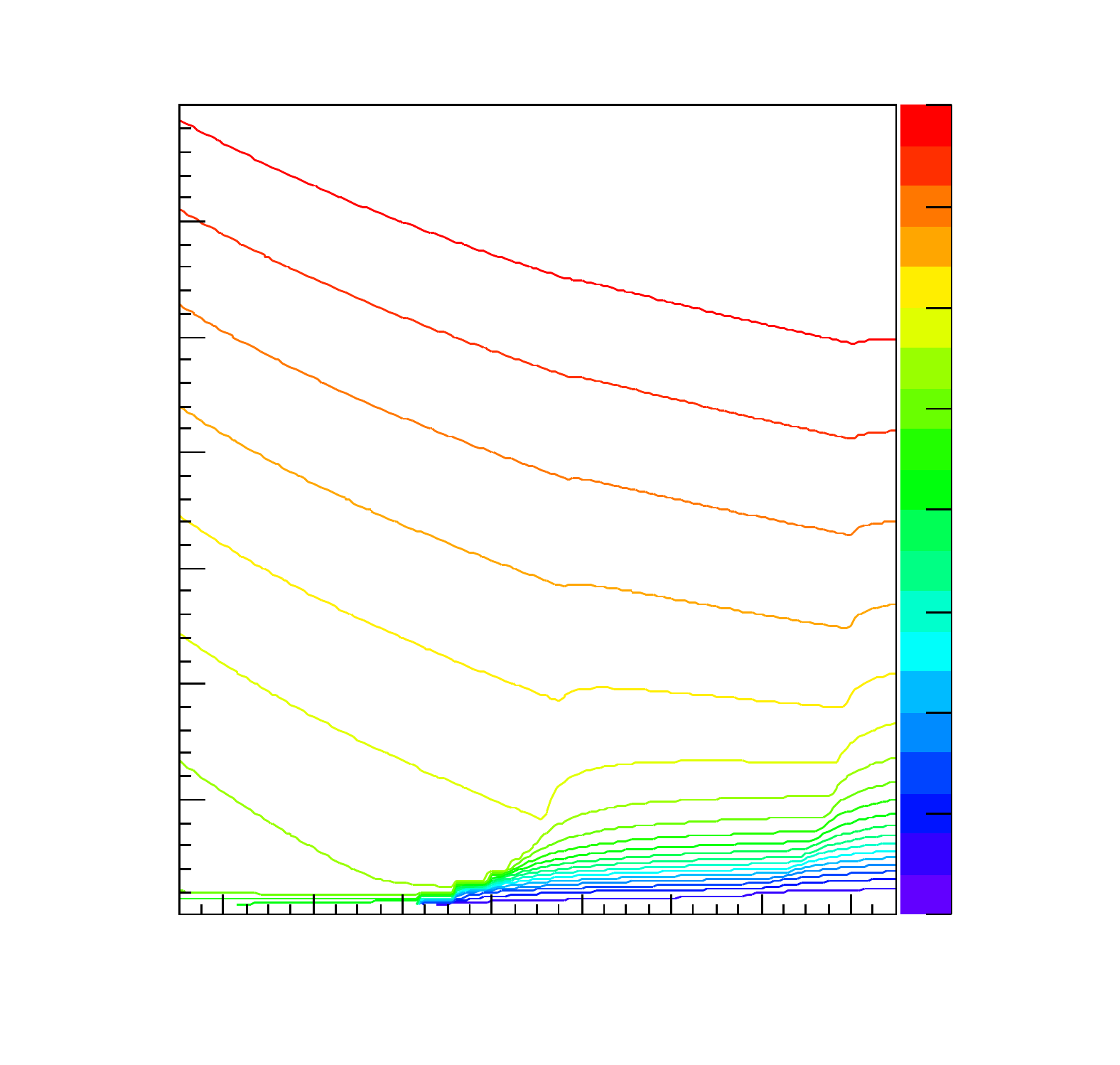} \svgend
\end{subfigures}

Comparing the couplings of the neutral \mssm \whbs
with fermions from the vertex factors of \figs{Thr:H12FuFu} through
\ref{fig:Thr:H32FdFd} with the \sm \whb coupling given in
\fig{Thr:H02FF}, the partial decay width for all three neutral \whbs
can be written as,
\begin{equation}
  \Gamma_{\hz \to f\bar{f}} = F_\hz\Gamma_{\hH \to
    f\bar{f}}
  \labelequ{Mssm.Width.FF}
\end{equation}
where $\Gamma_{\hz \to f\bar{f}}$ is given by \equ{Hig:Sm.Width.FF}. For
the \cp-odd \whb the exponent of $3/2$ in \equ{Hig:Sm.Width.FF} is
reduced to an exponent of $1/2$ due to the additional $\gm{}{5}$
factor in the couplings of \figs{Thr:H32FuFu} and
\ref{fig:Thr:H32FdFd}. The pre-factors $F_\hz$ are given by,
\begin{equation}
  \begin{aligned}
    F_\hhz &= \begin{cases}
      \frac{\cosa[^2]}{\sinb[^2]} & \textrm{for}~\nu,u \\
      \frac{\sina[^2]}{\cosb[^2]} & \textrm{for}~\lep,d \\
    \end{cases}\equcomma
    F_\hHz = \begin{cases}
      \frac{\sina[^2]}{\sinb[^2]} & \textrm{for}~\nu,u \\
      \frac{\cosa[^2]}{\cosb[^2]}  & \textrm{for}~\lep,d \\
    \end{cases}, \\
    F_\hAz &= \begin{cases}
      \tanb[^-2] & \textrm{for}~\nu,u \\
      \tanb[^2] & \textrm{for}~\lep,d \\
    \end{cases} \\
  \end{aligned}
  \labelequ{Mssm.Width.FF.Prefactors}
\end{equation}
for the three neutral \whbs where $u$ is a $u$-type quark, $\nu$ is a
neutrino, $d$ is a $d$-type quark, and \lep is a charged lepton. The
parameter $\alpha$ is the \whb mass mixing angle.

A similar procedure for calculating the vector boson pair partial
width is possible by comparing the \sm \whb vertices from
\figs{Thr:H02WW} and \ref{fig:Thr:H02ZZ} with the \mssm \whbs vertices
from \figs{Thr:H12WW} through \ref{fig:Thr:H22ZZ}. Notice that here,
the \cp-odd \whb does not couple with \w or \wzbs. The vector boson
partial decay width can then be written as,
\begin{equation}
  \Gamma_{\hz \to VV} = F_\hz\Gamma_{\hH \to VV}
  \labelequ{Mssm.Width.VV}
\end{equation}
where the pre-factors are given by,
\begin{equation}
  F_\hhz = \sinbma[^2] \equcomma F_\hHz = \cosbma[^2] \equcomma F_\hAz
  = 0
  \labelequ{Mssm.Width.VV.Prefactors}
\end{equation}
for the light and heavy \cp-even \whbs and the \cp-odd \whb. The
factor \cosbma[^2] can be rewritten in terms of,
\begin{equation}
  \cosbma[^2] = \frac{1}{2} - \frac{\m_\hAz^2 - \m_\z^2\cosb[4]}{2\left(m_\hAz^4 +
      \m_\z^4 - 2\m_\hAz^2\m_\z^2\cosb[4]
      \right)^{\frac{1}{2}}}
\end{equation}
for a given $\m_\hAz$ and \tanb. In the limit of large $\m_\hAz$ the
term \cosbma[^2] approaches zero and so the vector boson pair partial
widths are suppressed at large $\m_\hAz$ for the heavy \cp-even \whb,
while they approach the \sm widths for the light \cp-even \whb.

The branching fractions for the light \cp-even \whb, given in
\fig{Hig:Mssm.Br.H1}, range from $8\%$ at large $\m_\hAz$ up to values
of $16\%$ for large \tanb and low $\m_\hAz$. The upper limit on the
branching fraction is governed by the ratio of the \wtl and \wbq pair
partial widths from \equ{Hig:Mssm.Width.FF}, which are enhanced by a
factor of ${1+\tanb[^2]}$. However, for large $\m_\hAz$ the vector
boson pair widths increase, just as in the \sm, and so the \wtl
branching fraction is reduced.

For the heavy \cp-even \whb, the vector boson pair partial widths of
\equ{Hig:Mssm.Width.VV} are suppressed by a factor of \cosbma[^2], and
so the \wtl and \wbq pair partial widths dominate the total width at
large $\m_\hAz$, as can be seen in \fig{Hig:Mssm.Br.H2}, resulting in
the maximum \wtl pair branching fraction of $16\%$. Similar behaviour
can be seen in \fig{Hig:Mssm.Br.H3} at large $\m_\hAz$ for the \cp-odd
\whb due to the lack of couplings with vector bosons. The \wtl and
\wbq partial widths for the heavy \cp-even \whb are enhanced by a
factor of ${1+\tanb[^2]}$, while the \cp-odd widths are enhanced by a
factor of $\tanb[^2]$.

\newsubsection{Cross-Sections}{Xs}

The cross-section for incoming particles ${p_1}$ and ${p_2}$ producing a \whb
can be related to the decay widths of \sec{Hig:Br} using a result of
\rfr{bijnens.06.1},
\begin{equation}
  \sigma_{{p_1}{p_2} \to \h}(s) \approx \frac{16\pi}{(2N_{s_{p_1}}+1)(2N_{s_{p_2}} +
    1)N_{c_{p_1}}N_{c_{p_2}}} \frac{\Gamma_{\h \to {p_1}{p_2}}
    \Gamma_\h}{(s-\m_\z^2)^2 + \m_\h^2\Gamma_\h}
  \labelequ{{p_1}s}
\end{equation}
where $s$ is the centre-of-mass energy, and $N_c$ and $N_s$ are the
colour and spin multiplicities for ${p_1}$ and ${p_2}$. At the \lhc,
${p_1}$ and ${p_2}$ are two partons from the colliding protons, and so
the observable cross-section $\sigma_{\dip \to \h}$ is calculated
using the partonic cross-section $\sigma_{{p_1}{p_2} \to \h}$ and the
factorisation theorem of \equ{Thr:Factorisation}. Thus, the
cross-section for \whb production at the \lhc depends upon the
knowledge of the proton \PDF, which introduces an uncertainty within
the range of $\prc{2-14}$.

\begin{subfigures}{2}{Example diagrams of \sm \whb production at the
    \lhc from \subfig{GG2H0}~\ggf, \subfig{QQ2VV2H0}~\vbf,
    \subfig{QQ2H0V}~\avp, and
    \subfig{GG2H0QQ}~\aqp.\labelfig{Production}}
  \fmpbeg
  \fmp{GG2H0}  & \fmp{QQ2VV2H0} \fmpsep
  \fmp{QQ2H0V} & \fmp{GG2H0QQ}  \fmpend
\end{subfigures}

Production of \whbs at the \lhc in proton-proton collisions occurs
primarily through four partonic processes, outlined in the example
diagrams of \fig{Hig:Production}. \Ggf is shown in \fig{Hig:GG2H0},
\vbf in \fig{Hig:QQ2VV2H0}, \avp in \fig{Hig:QQ2H0V}, and \aqp in
\fig{Hig:GG2H0QQ}. The cross-sections for each of these mechanisms is
dependent upon the mass and type of the \whb, as well as \tanb for the
\mssm \whbs. The cross-sections for \sm and \mssm \whbs produced in
proton-proton collisions at a centre-of-mass energy of $7~\tev$
through these production mechanisms are given in this section.

\newsubsubsection{SM Cross-Sections}{}

The \sm \whb cross-sections, as a function of mass, are shown in
\fig{Hig:Sm.Xs}. These cross-sections are calculated using the
programs \higlu~\cite{\citehiglu}, \dfg~\cite{\citedfg}, and
\vbfh~\cite{\citevbfh} following the methods of
\rfr{hxswg.11.1}. Further details on these calculations can be found
in \sap{Hvr:XsSm}.

\begin{subfigures}{2}{Cross-sections for ${\dip \to \hH}$ production
    at the \lhc with a centre-of-mass energy of $7~\tev$. The
    inclusive cross-section (black) is the sum of \ggf (red), \vbf
    (blue), \awp (green), \azp (orange), and \aqp (magenta). The
    coloured bands provide the linearly combined \qcd scale, \aS, and
    \PDF uncertainties for the cross-sections.}
  \svgbeg
  \svg[1]{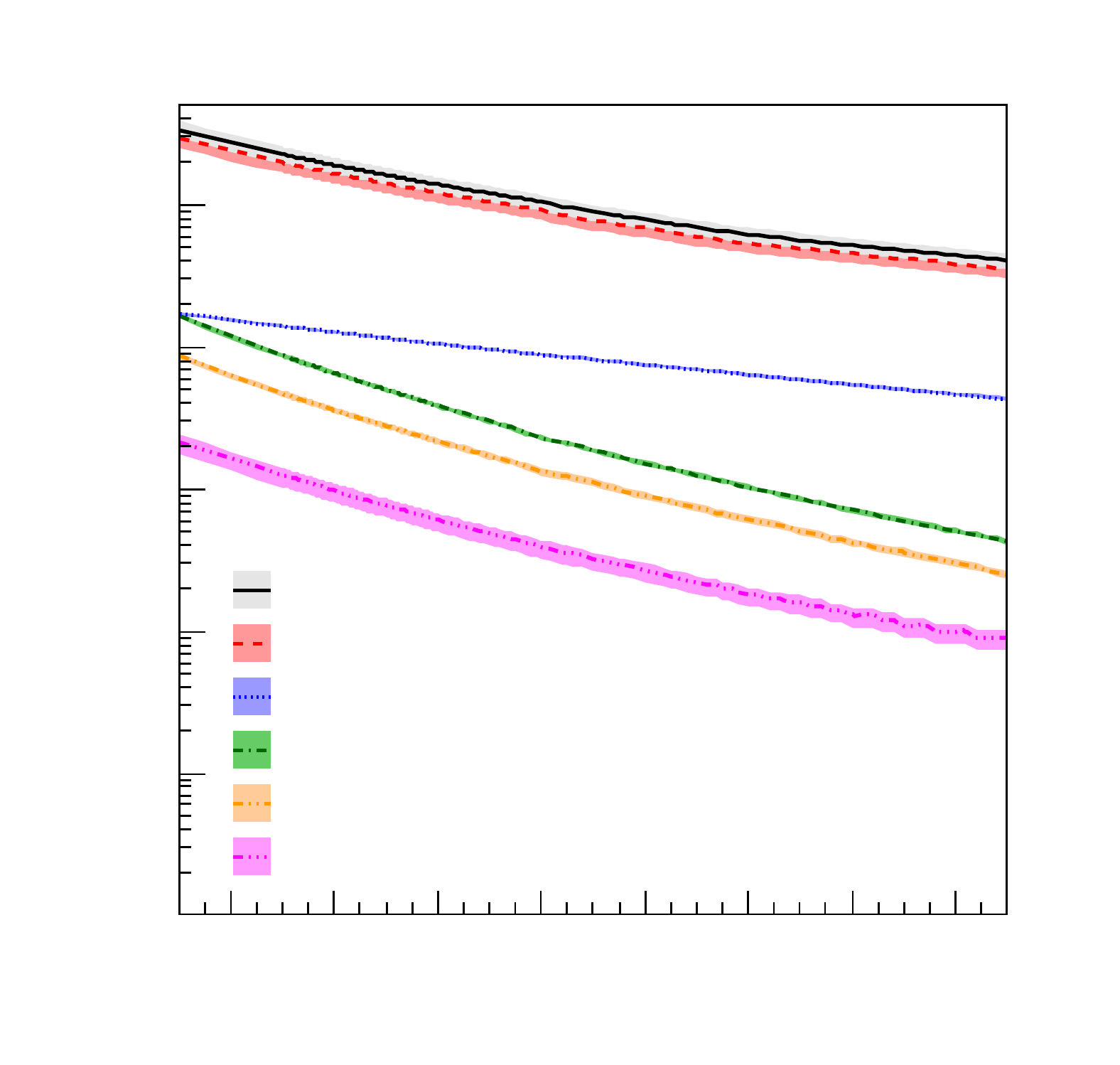} & \sidecaption \svgend
\end{subfigures}

The leading \sm \whb production mechanism is \ggf which, just like the
\whb decay into a gluon pair of \equ{Hig:Sm.Width.GG}, must proceed
through a fermion loop, as the massless gluon does not couple directly
with the \whb. At lower masses this loop is dominated by \wbqs, but
for $\m_\hH > 2\m_t$ the \wtq loop contributes. While the gluon pair
partial decay width of \equ{Hig:Sm.Width.GG} is not the largest width
for the \whb, the gluon contribution to the proton at low momentum
transfer is much larger than the other partons, as shown in
\fig{Thr:Pdf.All}, resulting in a large cross-section from \ggf.

The \vbf contribution to the inclusive \sm \whb cross-section is
nearly an order of magnitude smaller than the \ggf contribution, as
the partons for this process are quarks and not gluons. The
cross-section for \avp is even smaller with respect to \ggf, but at
lower \whb masses is comparable to the \vbf contribution, as can be
seen in \fig{Hig:Sm.Xs}. The \aqp cross-section is calculated for
associated \wtqs and is nearly two orders of magnitude smaller than
the \ggf cross-section.

\newsubsubsection{MSSM Cross-Sections}{}

Two cross-sections are included in the inclusive cross-sections for
the \mssm \whbs, \ggf and \abp, and are given for the three neutral
\whbs as a function of the \cp-odd \whb mass and \tanb in
\fig{Hig:Mssm.Xs}. These cross-sections are calculated following the
recommendations of \rfr{hxswg.11.1} using the programs \higlu and
\ggh~\cite{\citeggh} for the \ggf cross-section and the program
\bbh~\cite{\citebbh} for the \abp cross-section. Both of these
calculations are modified with \mssm couplings from
\feynhiggs. Further details on the calculations, including additional
plots, can be found in \sap{Hvr:XsMssm}.

\begin{subfigures}{2}{Inclusive cross-sections for the production of
    the \mssm \subfig{Mssm.Xs.All.H1}~light \cp-even,
    \subfig{Mssm.Xs.All.H2}~heavy \cp-even, and
    \subfig{Mssm.Xs.All.H3}~\cp-odd \mssm \whbs as a function of the
    \cp-odd \whb mass and \tanb. The cross-sections are calculated
    using \higlu, \ggh, and \bbh with the \mssm couplings calculated
    using \feynhiggs.\labelfig{Mssm.Xs}}
  \svgbeg
  \svg{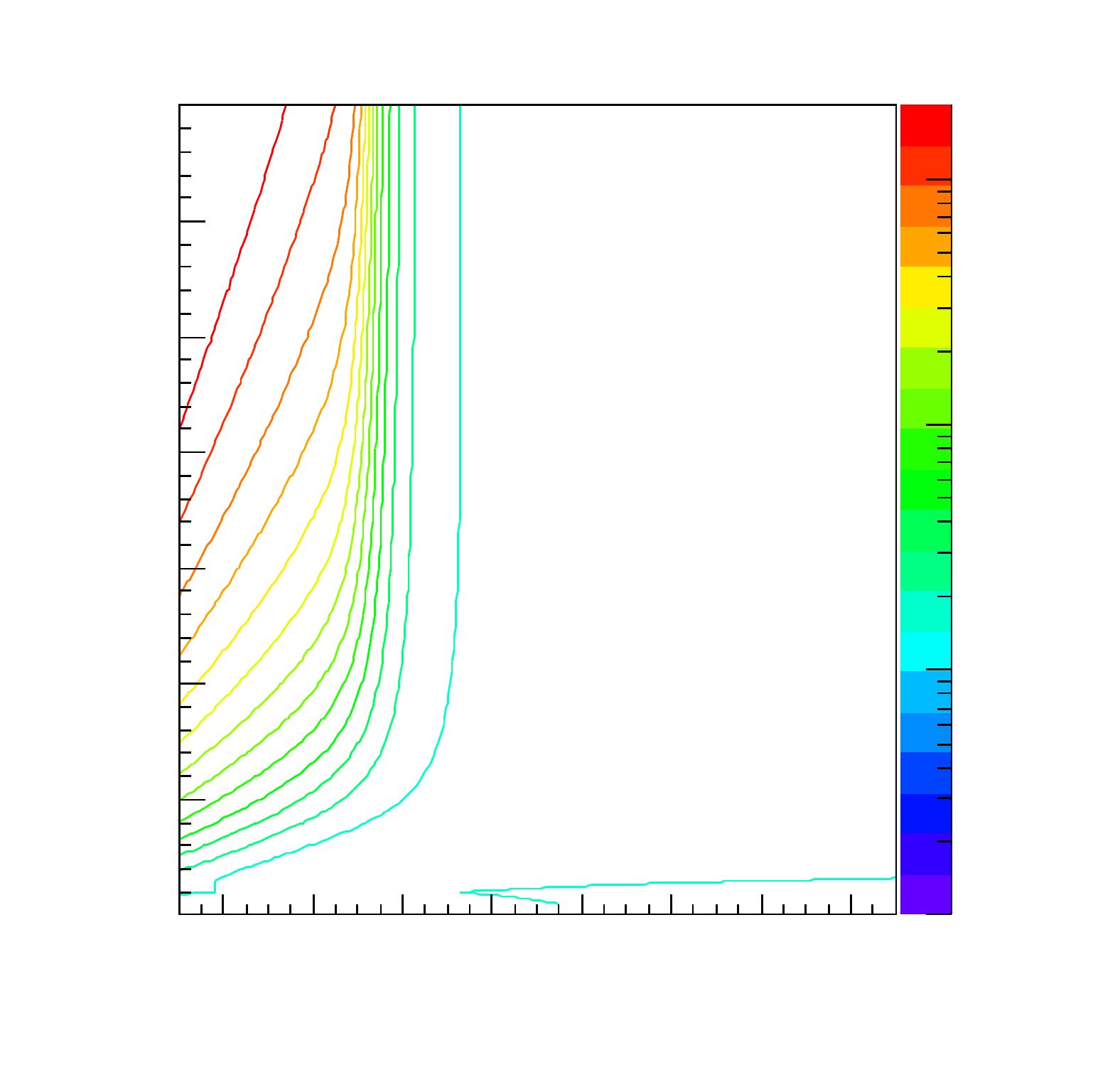} & \svg{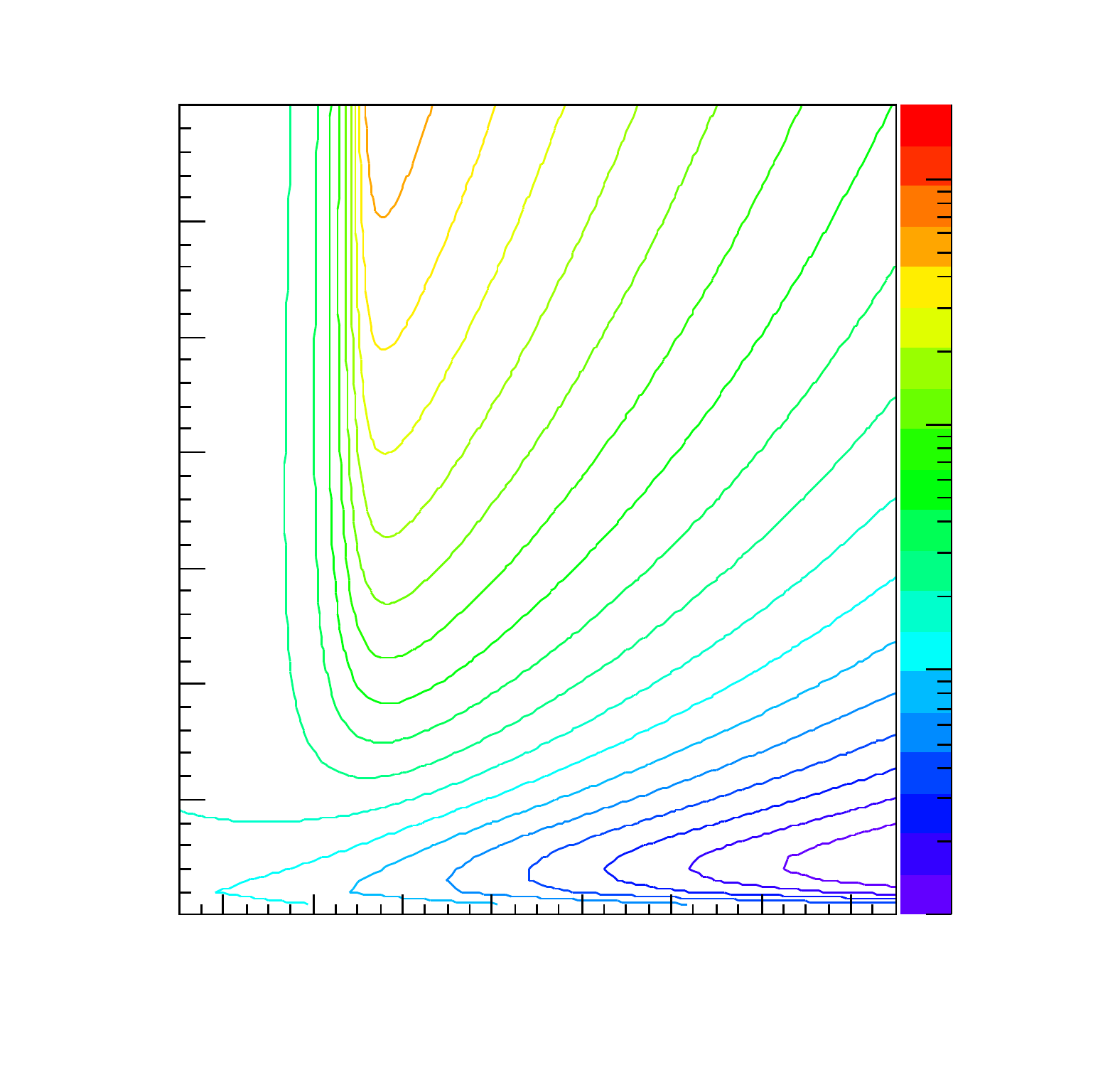} \svgsep
  \sidecaption         & \svg{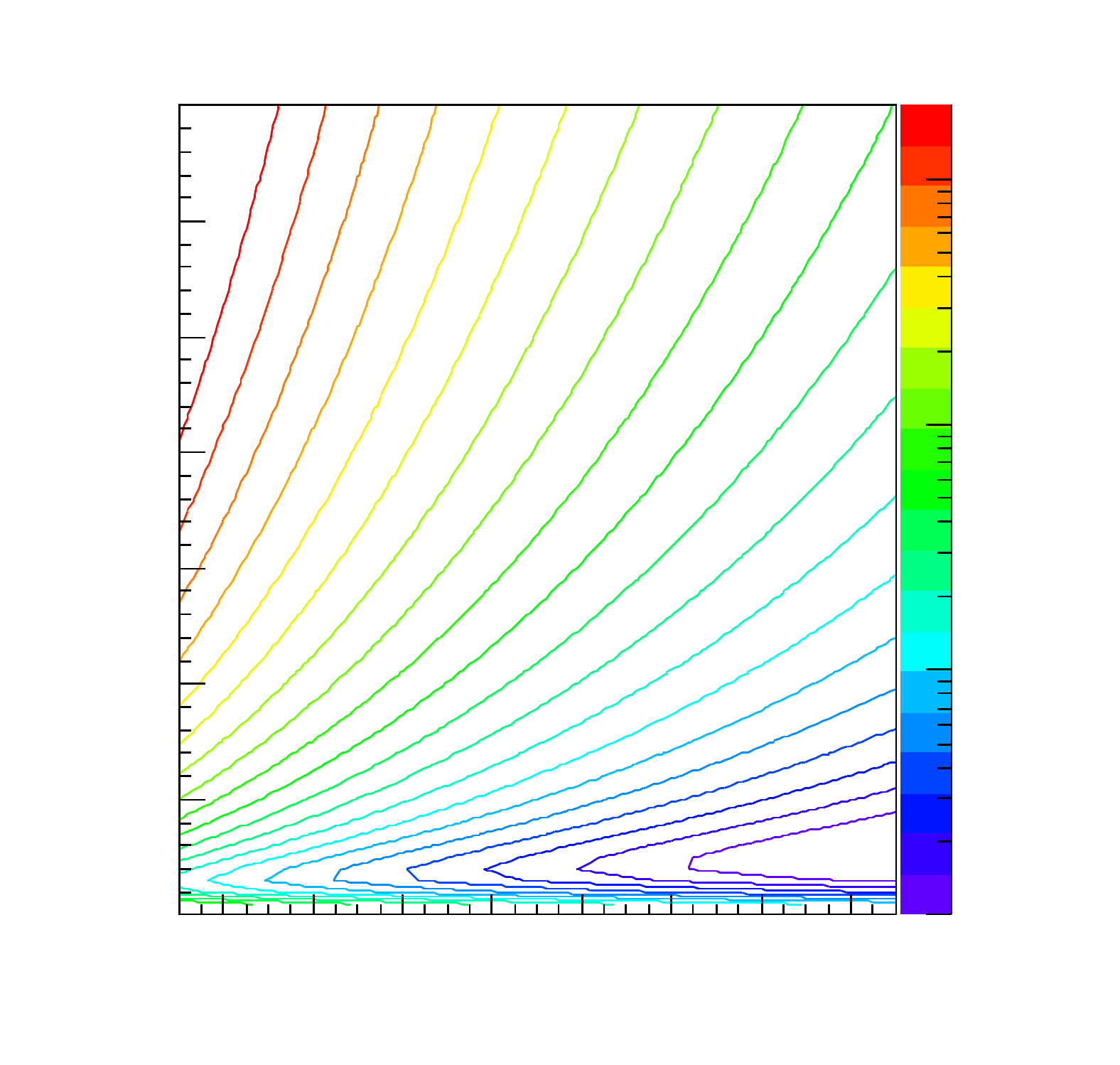} \svgend
\end{subfigures}

At large \tanb the inclusive \mssm \whb cross-sections can be enhanced
by nearly two orders of magnitude with respect to the \sm \whb
cross-section, due to the additional \tanb dependence in the couplings
of \fig{Thr:Vertices.Mssm.Higgs}. The light \cp-even \whb approaches a
maximum mass of $130~\gev$ for large $\m_\hAz$ resulting in an
inclusive cross-section similar to the \sm \whb for all \tanb. As can
be seen in \fig{Hig:Mssm.Xs.All.H1}, this maximum inclusive
cross-section is on the order of $10~\pb$ which is comparable to the
inclusive cross-section for the \sm \whb given in \fig{Hig:Sm.Xs}.

\newsection{Event Model}{Mod}

The decay of a neutral \whb into a \wtl pair produces an experimental
signature similar to ${\z \to \ditau}$ events, and so the analysis of
\chp{Zed} can also be used to determine the cross-section for neutral
\whb production. However, the ${\dip \to \z \to \ditau}$
cross-sections measured in \chp{Zed} match well with the theoretical
prediction, and so if $\h \to \ditau$ events are present within the
data, their contribution is too small to measure a
cross-section. Instead, this data can be used to set upper limits on
neutral \whb production.

To set these limits, the expected number of background and \whb signal
events is required, as well as the observed number of events. Here,
the ${\z \to \ditau}$ signal of \chp{Zed} is now considered a
background. In \sec{Hig:Sim} the simulation samples used to determine
the number of expected \whb signal events is described, and in
\sec{Hig:Nrm} both the expected number of background and signal events
are estimated. However, using just the number of expected and observed
events to set upper limits does not utilise differences between the
signal ${\h \to \ditau}$ events and the dominant ${\z \to \ditau}$
background events. The primary difference between these two event
types is the invariant mass of the \wtl decay products, and so the
invariant mass distributions of \fig{Zed:Mass}, normalised to the
number of expected events, can be used to increase the information
used to set the limits. In \sec{Hig:Mass} the signal and background
mass distributions are given, using both the simulation samples of
\sec{Hig:Sim} and the normalisation of \sec{Hig:Nrm}.

\newsubsection{Simulation}{Sim}

The ${\h \to \ditau}$ simulation samples are used to determine the
efficiency and acceptance corrections necessary to calculate the
number of expected signal events in \sec{Hig:Nrm}, as well as
determine the invariant mass distributions of \sec{Hig:Mass}. The
samples are generated, simulated, digitised, and reconstructed
following the process described in \sec{Exp:Rec}. All samples are
generated with \pythia{6}~\cite{\citepythiasix} and simulated with
\gauss using the \lhcb simulation configuration \MC[11a]. Over the
\whb mass range considered in this chapter the leading order treatment
of \pythia{6} is sufficient, as effects from higher order corrections,
off-shell effects, and signal and background interference are
small~\cite{hxswg.12.1}.

The \whb signals are generated for seventeen mass steps between
$90~\gev$ and $250~\gev$ in steps of $10~\gev$. For each mass step
five samples are generated, one for each event category of
\sec{Zed:Sel}. Only events with generator level \wtl decay products
matching the $\eta$, \pt, and particle type requirements of
\sec{Zed:Sel} are selected for full simulation and reconstruction. A
total of $10^4$ events are fully reconstructed and simulated for each
mass step and event category.

The primary production mechanism for the \sm \whb, as described in
\sec{Hig:Xs}, is \ggf and so the ${\hH \to \ditau}$ samples are
generated for \ggf production. In the \mssm, the dominant \whb
production mechanisms are both \ggf and \abp. However, the expected
number of signal events and invariant mass distribution when
determined from samples produced with either mechanism are found to be
consistent within uncertainty. Additionally, the \cp of the \whb
affects neither the expected number of signal events nor the mass
distribution, within uncertainty, and so the same simulation samples
used for the \sm \whb are also used for all three neutral \mssm \whbs.

\newsubsection{Event Yields}{Nrm}

The expected number of background events for the \qcd, \ewk, \ttbar,
\ww, and ${\z \to \dilep}$ backgrounds have already been estimated in
\sec{Zed:Bkg}, and so only the expected number of ${\z \to \ditau}$
background events and ${\h \to \ditau}$ signal events need to be
estimated. In this section these expected number of events are
calculated using the results of \sec{Zed:Sig}, the theoretical
branching fractions and cross-sections of \sec{Hig:Phe}, and the
simulation of \sec{Hig:Sim}.

\newsubsubsection{Expected $\bm{Z \to \tau\tau}$ Background}{}

Given a simulated ${\z \to \ditau}$ sample, the expected number of
${\z \to \ditau}$ events in data is given by,
\begin{equation}
  N_{\z \to \ditau} = \frac{\sigma_{\dip \to
      \z \to \ditau}  \eff[_\sel] \lum
    \acc[_{\tau_1\tau_2}] \br[_{\tau_1\tau_2}]}
  {\frac{1}{N_\simu} \displaystyle \sum \limits{_i^{N_\simu}}
    \left(\eff[_\rec^{-1}]_i \right)}
  \labelequ{Events.Z2TauTau}
\end{equation}
where $\sigma_{\dip \to \z \to \ditau}$ is ${74.3_{-2.1}^{+1.9}~\pb}$,
$N_\simu$ is the number of events in the simulation sample, and all
remaining variables are the same as for the cross-section formula of
\equ{Zed:Xs}. In the summation, $\eff[_\rec]_i$ is the reconstruction
efficiency evaluated for event $i$ in the simulated ${\z \to \ditau}$
sample. The theoretical cross-section for $\sigma_{\dip \to \z \to
  \ditau}$ from \sec{Zed:Res} is used as this cross-section has a
higher precision than either of the experimental ${\dip \to \z \to
  \dilep}$ cross-sections, $\sigma_{\dip \to \z \to \ditau}$ and
$\sigma_{\dip \to \z \to \ditau}$, measured with \lhcb.

In \tab{Hig:Events} the number of expected ${\z \to \ditau}$
background events in data, calculated using \equ{Hig:Events.Z2TauTau},
is given. The remaining backgrounds from \tab{Zed:Events} have been
summed and are also provided, as well as the sum of these backgrounds
and the ${\z \to \ditau}$ background. The number of events observed in
data from \tab{Zed:Events} is given for comparison. As can be seen,
the expected number of total background events and the number of
events observed in data are compatible within uncertainty.

\begin{table}\centering
  \captionabove{Expected number of events for the ${\z \to \ditau}$
    background, remaining backgrounds summed from \tab{Zed:Events},
    and total background for each 
    event category. The number of events observed in data, from
    \tab{Zed:Events}, as well as the expected number of \sm \whb
    events multiplied by a factor of $100$ for $\m_\h = 125$,
    are also given.\labeltab{Events}}
  \settowidth{\backspace}{$1$}
  \begin{tabular}{>{$}l<{$}|E|E|E|E|E}
    \toprule
    & \multicolumn{2}{c|}{\mumu} & \multicolumn{2}{c|}{\mue} &
    \multicolumn{2}{c|}{\emu} & \multicolumn{2}{c|}{\muh} &
    \multicolumn{2}{c}{\eh} \\
    \midrule
    {\z \to \ditau}
    & 79.8&5.6 & 288.2&26.2 & 115.8&12.7 & 146.1&9.7 &\p62.1&8.0 \\
    {\rm other~bkg.}
    & 41.6&8.5 & 129.7&4.9  & 56.6 &3.3  & 53.3 &0.8 & 36.6&0.9 \\
    {\rm total~bkg.}
    &121.4&10.2& 417.9&26.7 & 172.4&13.1 & 199.3&9.7 & 98.7&8.0 \\
    \midrule
    {\rm observed}
    & \multicolumn{2}{l|}{124} & \multicolumn{2}{l|}{421} 
    & \multicolumn{2}{l|}{155} & \multicolumn{2}{l|}{189}
    & \multicolumn{2}{l}{101} \\
    \midrule
    {\hH \to \ditau \times 100}
    & 3.9 &0.5 & 11.9 &1.6  &  3.8 &0.5  &  9.7 &1.3 &  4.2&0.6 \\
    \bottomrule
  \end{tabular}
\end{table}

\newsubsubsection{Expected Higgs Boson Signal}{}

For the \whb signal, the expected number of events is,
\begin{equation}
  N_{\h \to \ditau}(\m_\h) = \frac{\sigma_{\dip \to
      \h}\left(\m_\h\right) \br[_{\h \to \ditau}]\left(\m_\h\right)
    \eff[_\sel](\m_\h) \lum
    \acc[_{\tau_1\tau_2}]\left(\m_\h\right) \br[_{\tau_1\tau_2}]}
  {\frac{1}{N_\simu} \displaystyle \sum \limits{_i^{N_\simu}}
    \left(\eff[_\rec^{-1}]_i \right)}
  \labelequ{Events.H2TauTau}
\end{equation}
where $\m_\h$ indicates the term is dependent upon the \whb mass and
$N_\simu$ is the number of events in the simulated ${\h \to \ditau}$
sample. The term $\br[_{\h \to \ditau}]$ is the branching fraction of
the \whb into a \wtl pair and $\sigma_{\dip \to \h}$ is the \whb
inclusive cross-section. Both terms are mass dependent and are
provided by the calculations of \secs{Hig:Br} and \ref{sec:Hig:Xs} for
\sm and \mssm \whbs. For the \mssm these terms also depend upon
\tanb.

The cross-section $\sigma_{\dip \to \h}$ is calculated with no
kinematic requirements, unlike $\sigma_{\dip \to \z \to \ditau}$ of
\equ{Hig:Events.Z2TauTau}, and so the mass dependent acceptance is
defined as the fraction of generator level events from ${\h \to
  \ditau}$ simulation passing the \pt and $\eta$ requirements of
\sec{Zed:Sel}. Because the same simulation samples are used for all
\whb types and production mechanisms, the acceptance can be plotted as
a function of the \whb mass, as is done for the five event categories
in \fig{Hig:Info.Acc}. As $\m_\h$ increases, the longitudinal boost of
the \whb is reduced, resulting in less events where both \wtls from
the \whb fall within \lhcb. Consequently, the acceptance decreases
for increasing $\m_\h$. Numerical values for these acceptances are
provided in \fig{Hvr:Acceptance} of \sap{Hvr:Acc}.

\begin{subfigures}[t]{2}{\subfig{Info.Acc}~Acceptances and
    \subfig{Info.Eff}~efficiencies for the \whb signal as a function
    of the \whb mass for the five event categories. The uncertainties
    for the \mumu (red), \mue (blue), \emu (green), \muh (orange), and
    \eh (magenta) event categories are provided by their corresponding
    coloured bands.}
  \svgbeg
  \svg{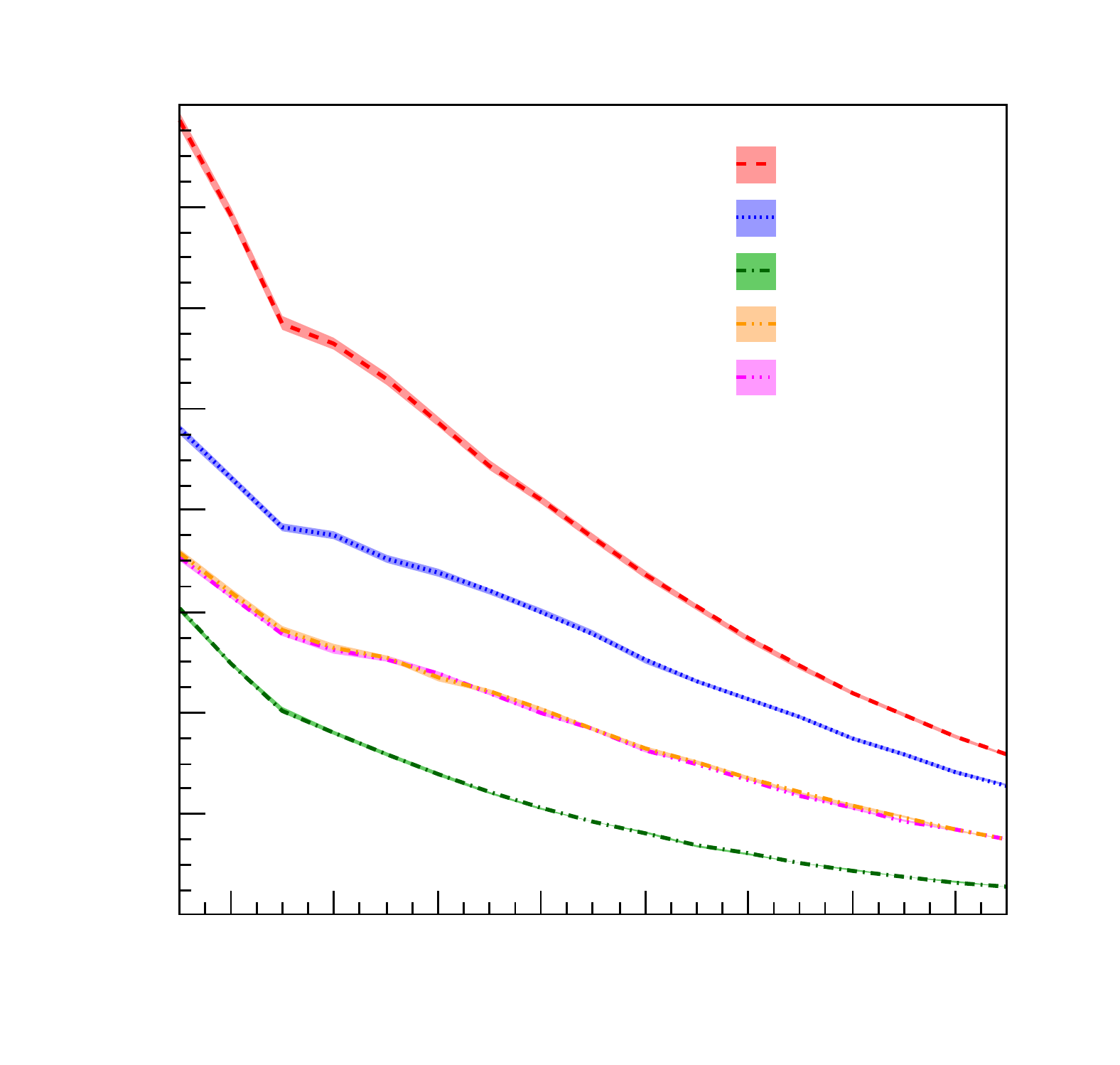} & \svg{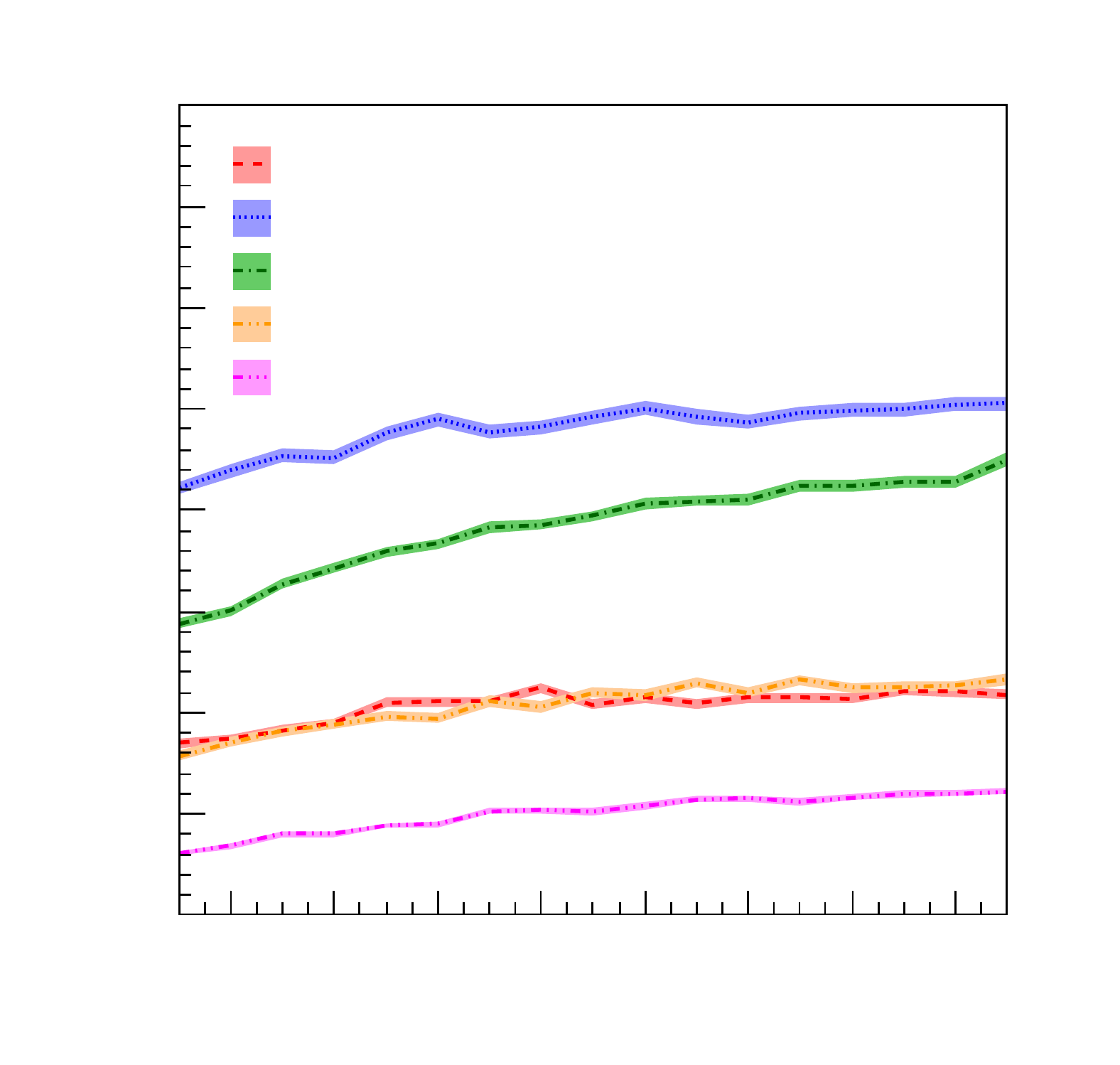} \svgend
\end{subfigures}

The selection efficiency \eff[_\sel] is mass dependent and is
calculated using the same methods as \sec{Zed:SelEff} but with ${\h
  \to \ditau}$ simulation samples. The combined selection and
reconstruction efficiency correction from \equ{Hig:Events.H2TauTau} is,
\begin{equation}
  \eff[_{\tau_1\tau_2}](\m_\h) \equiv
  \frac{\eff[_\sel](\m_\h)}{\frac{1}{N_\simu} \displaystyle \sum
    \limits{_i^{N_\simu}} \left(\eff[_\rec^{-1}]_i \right)}
  \labelequ{Efficiency}
\end{equation}
and is plotted in \fig{Hig:Info.Eff} for all five event
categories. Again, the same \whb simulation samples are used for all
\whb types, and so $\eff[_{\tau_1\tau_2}]$ is only a function of the
\whb mass. As $\m_\h$ increases, the boosts of the \wtls from the \whb
increase, producing more collinear \wtl decays and subsequent decay
products with larger momenta. Consequently, both the \dphi selection
efficiency is expected to increase, as well as the reconstruction
efficiencies, resulting in $\eff[_{\tau_1\tau_2}]$ rising for
increasing $\m_\h$. Numerical values of the efficiency for each event
category are given in \fig{Hvr:Efficiency} of \sap{Hvr:Acc}.

In \tab{Hig:Events}, the expected number of \sm \whb events multiplied
by a factor of $100$ and assuming ${\m_\hH = 125~\gev}$ is given for
each event category. The number of expected \sm \whb events, without
uncertainty included, is tabulated as a function of $\m_\h$ in
\fig{Hig:Events.Sm} for all five event categories, as well as the sum
of all five categories. The expected number of neutral \mssm \whb
events for each category is tabulated in \fig{Hig:Events.Mssm} and for
the sum of all categories in \fig{Hig:Events.Mssm.All} as a function
of $\m_\hAz$ and \tanb.  As can be seen, the expected number of \sm
\whb signal events is less than $1$ for all event categories, while
for the \mssm as many as $120$ events are expected.

\begin{subfigures}{2}{Values of the
    expected number of signal events from the \sm \whb for the
    \subfig{H2TauTau2MuMu.Sm.Signal.Text}~\mumu,
    \subfig{H2TauTau2MuE.Sm.Signal.Text}~\mue,
    \subfig{H2TauTau2EMu.Sm.Signal.Text}~\emu,
    \subfig{H2TauTau2MuPi.Sm.Signal.Text}~\muh,
    \subfig{H2TauTau2EPi.Sm.Signal.Text}~\eh categories as a function
    of mass. \subfig{H2TauTau.Sm.Signal.Text}~The sum of all five
    event categories. Uncertainty is excluded for
    clarity.\labelfig{Events.Sm}}
  \multicolumn{2}{c}{\svg[1]{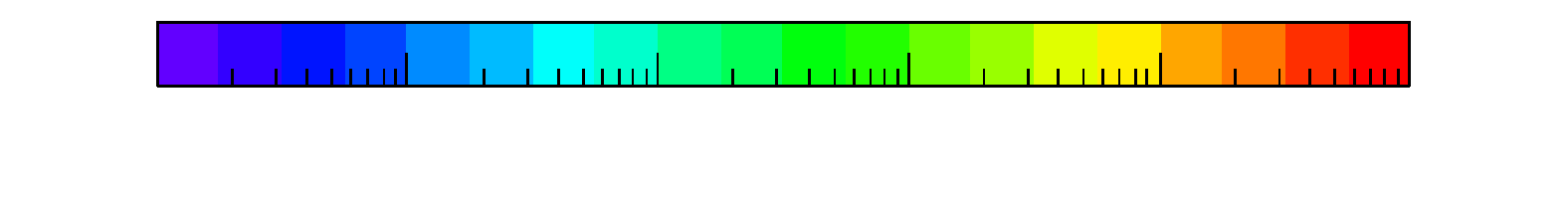}} \\
  \svg{H2TauTau2MuMu.Sm.Signal.Text} 
  & \svg{H2TauTau2MuE.Sm.Signal.Text} \barsep
  \svg{H2TauTau2EMu.Sm.Signal.Text} 
  & \svg{H2TauTau2MuPi.Sm.Signal.Text} \barsep
  \svg{H2TauTau2EPi.Sm.Signal.Text}
  & \svg{H2TauTau.Sm.Signal.Text} \barend
\end{subfigures}

\begin{subfigures}[p]{2}{Values of the
    expected number of signal events from \mssm \whbs for the
    \subfig{H2TauTau2MuMu.Mssm.Signal.Text}~\mumu,
    \subfig{H2TauTau2MuE.Mssm.Signal.Text}~\mue,
    \subfig{H2TauTau2EMu.Mssm.Signal.Text}~\emu,
    \subfig{H2TauTau2MuPi.Mssm.Signal.Text}~\muh, and
    \subfig{H2TauTau2EPi.Mssm.Signal.Text}~\eh categories as a function
    of the \cp-odd \whb mass and \tanb. Uncertainty is excluded for
    clarity.\labelfig{Events.Mssm}}
  \svgbeg
  \svg{H2TauTau2MuMu.Mssm.Signal.Text} 
  & \svg{H2TauTau2MuE.Mssm.Signal.Text} \svgsep
  \svg{H2TauTau2EMu.Mssm.Signal.Text} 
  & \svg{H2TauTau2MuPi.Mssm.Signal.Text} \svgsep
  \sidecaption
  & \svg{H2TauTau2EPi.Mssm.Signal.Text} \svgend
\end{subfigures}

\begin{subfigures}{2}{Values of the
    expected number of signal events from \mssm \whbs, summed for all
    event categories, as a function
    of the \cp-odd \whb mass and \tanb. Uncertainty is excluded for
    clarity.\labelfig{Events.Mssm.All}}
  \svgbeg
  \sidecaption
  & \svg{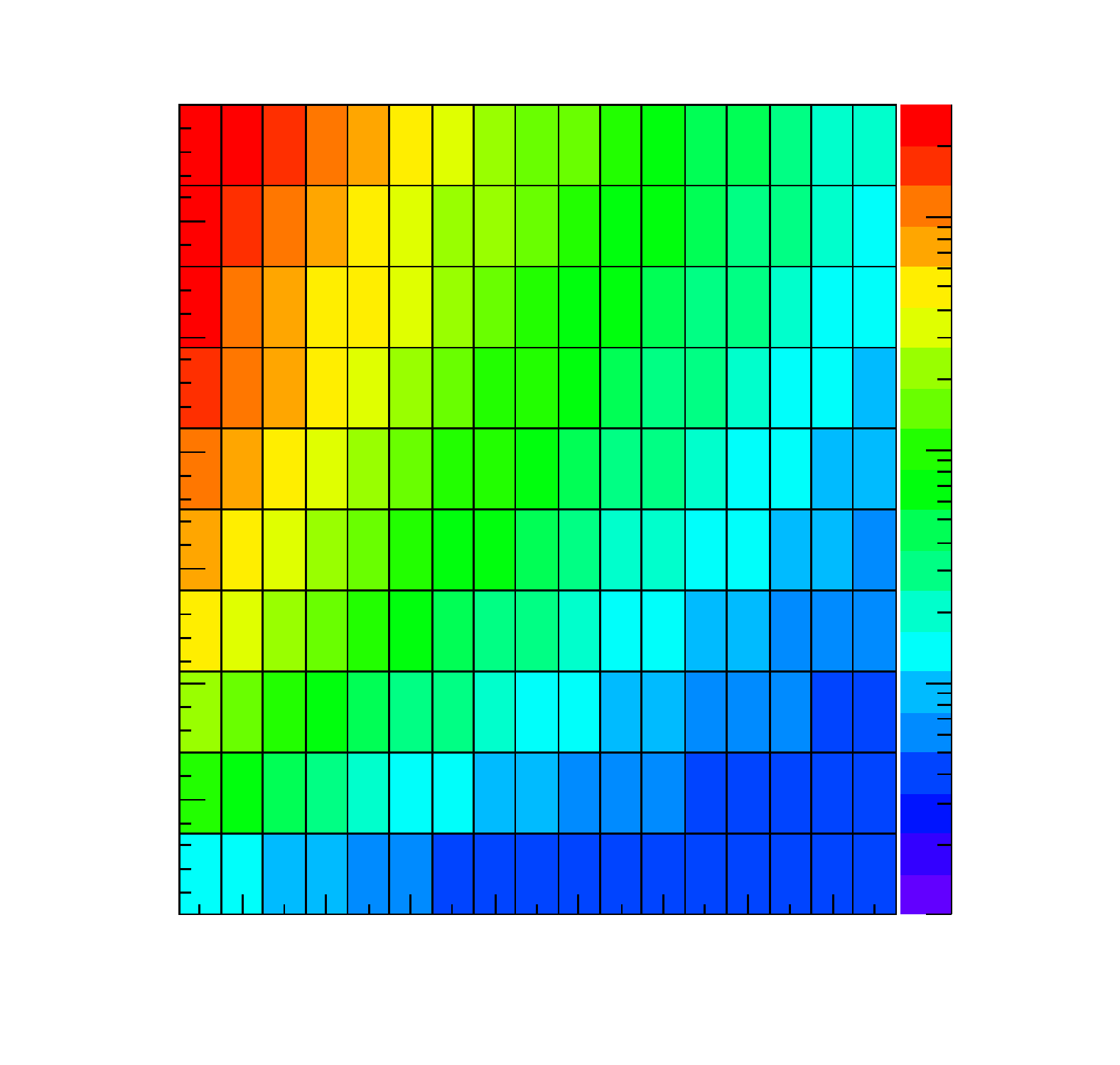} \svgend
\end{subfigures}

\newsubsection{Mass Distributions}{Mass}

The invariant mass distributions for the \qcd and ${\z \to \dilep}$
backgrounds are determined from data as described in \sec{Zed:Bkg} The
distributions for the \ewk, \ttbar, \ww, and ${\z \to \ditau}$
backgrounds as well as the ${\h \to \ditau}$ signal, are taken from
simulation which has been calibrated as described in
\sec{Zed:Sel}. Additionally, the events used to produce the
distributions are weighted on an event-by-event basis for the
differences between the efficiencies evaluated from simulation and
data as determined in \secs{Zed:RecEff} and \ref{sec:Zed:SelEff}. This
correction is negligible in comparison to the momentum resolution
calibration.

\begin{subfigures}[t]{2}{\subfig{H2TauTau.Mssm.Shape.Mass125.Tanb60}~Invariant
    mass distributions for the five combined event categories of the
    backgrounds and an example expected \mssm \whb signal (grey) with
    ${\m_\hAz = 125~\gev}$ and ${\tanb = 60}$. The backgrounds from
    the five event categories are grouped into ${\z \to \ditau}$
    (red), \qcd (blue), \ewk (green), \ttbar (orange), \ww (magenta),
    and ${\z \to \dilep}$ (cyan). The ${\z \to \ditau}$ and \whb
    signal distributions are normalised following the prescriptions of
    \equs{Hig:Events.Z2TauTau} and
    \ref{equ:Hig:Events.H2TauTau}. \subfig{H2TauTau.Sm.Shape.Mass250}~The
    same invariant mass distribution, but with an example model
    independent neutral \whb signal (grey) with ${\m_\h = 250~\gev}$
    which can be excluded is shown.}
  \svgbeg
  \svg{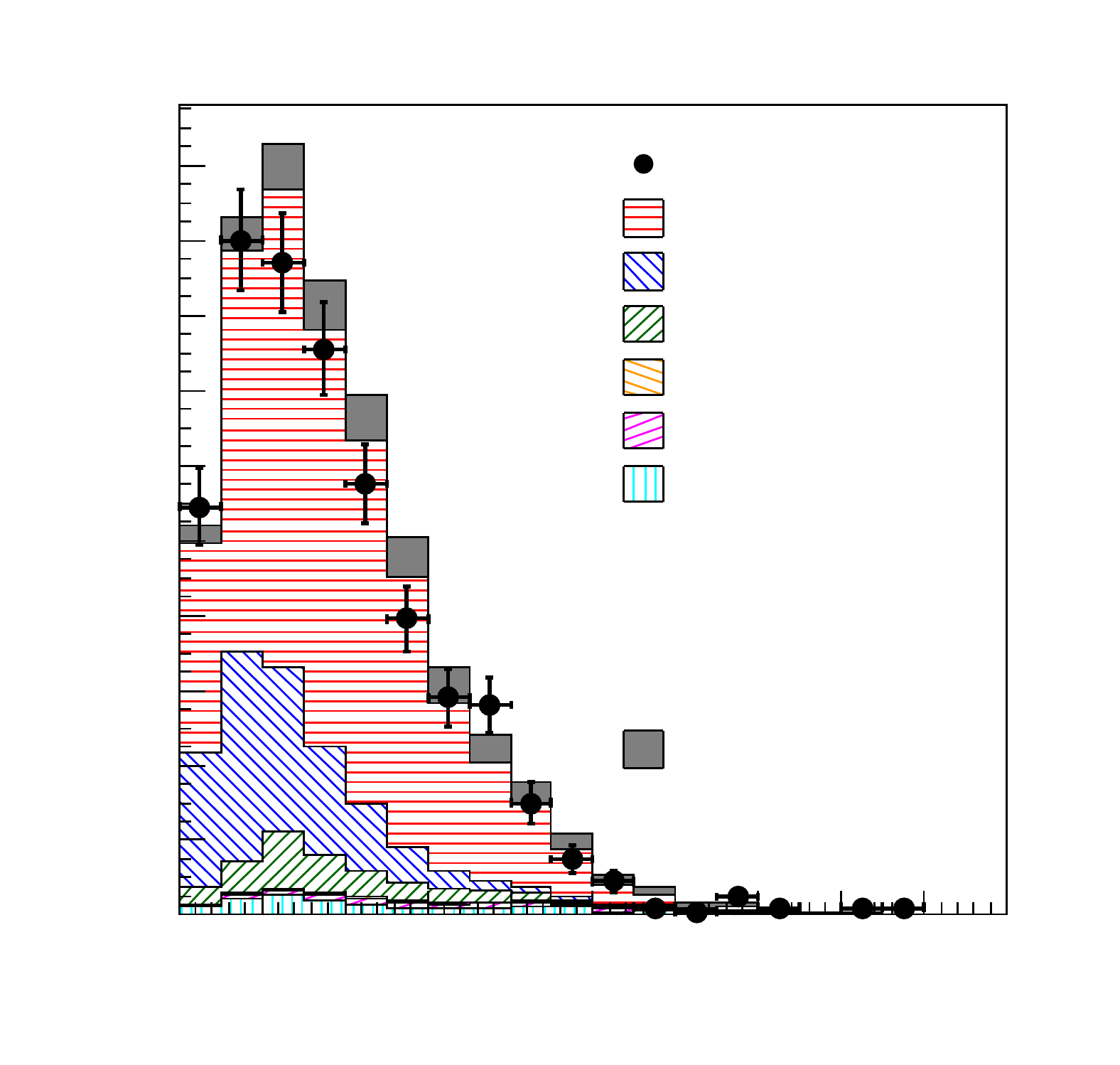} & 
  \svg{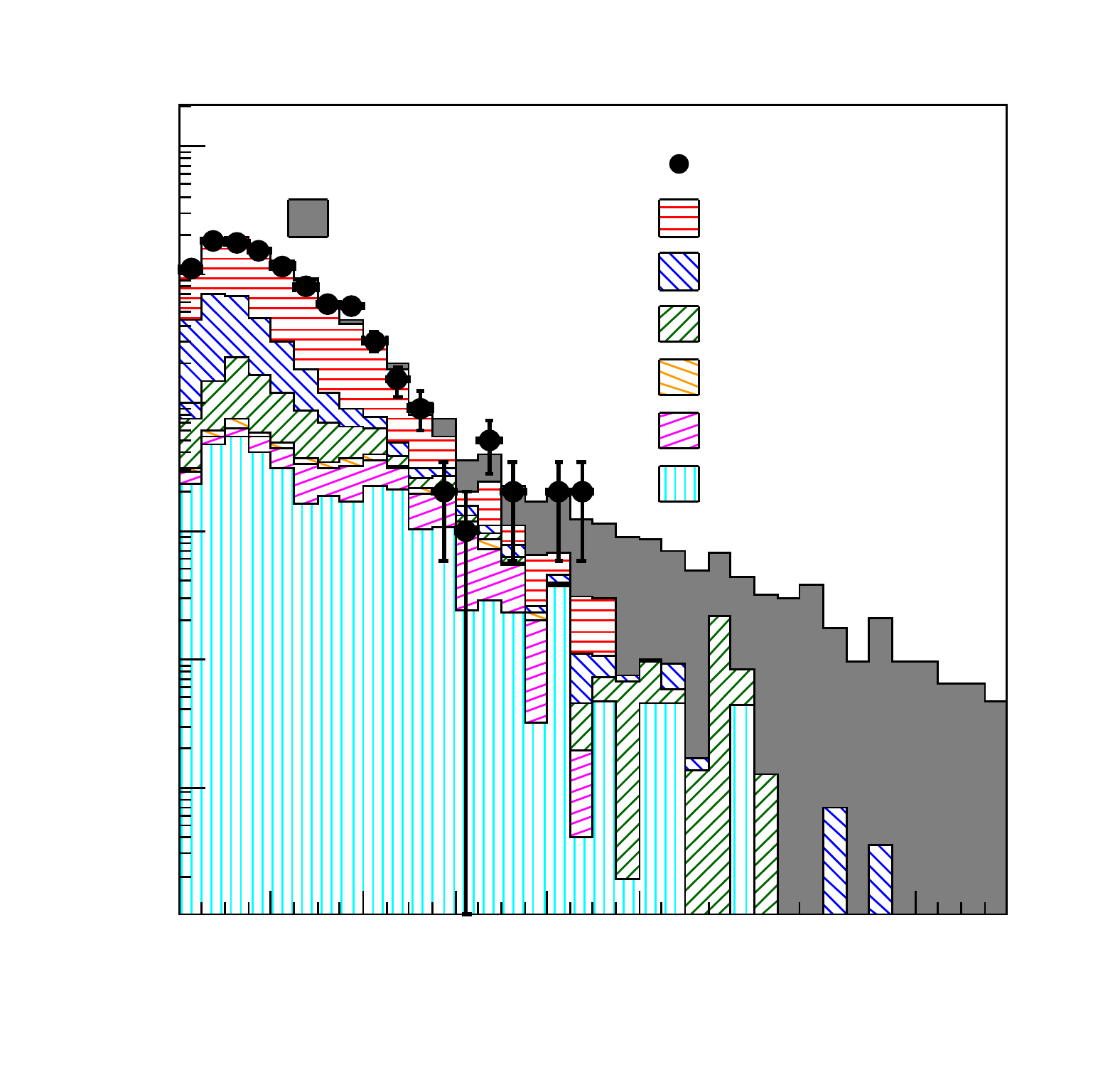}
  \svgend
\end{subfigures}

\begin{subfigures}[p]{2}{Invariant mass distributions of the
    backgrounds and an example expected \mssm \whb signal (grey) with
    ${\m_\hAz = 125~\gev}$ and ${\tanb = 60}$ for the
    \subfig{H2TauTau2MuMu.Mssm.Shape.Mass125.Tanb60}~\mumu,
    \subfig{H2TauTau2MuE.Mssm.Shape.Mass125.Tanb60}~\mue,
    \subfig{H2TauTau2EMu.Mssm.Shape.Mass125.Tanb60}~\emu,
    \subfig{H2TauTau2MuPi.Mssm.Shape.Mass125.Tanb60}~\muh, and
    \subfig{H2TauTau2EPi.Mssm.Shape.Mass125.Tanb60}~\eh event
    categories. The \qcd (blue), \ewk (green), \ttbar (orange), \ww
    (magenta), and ${\z \to \dilep}$ (cyan) backgrounds are determined
    from \sec{Zed:Bkg}, while the ${\z \to \ditau}$ background (red) is
    normalised via \equ{Hig:Events.Z2TauTau} and the signal is
    normalised with \equ{Hig:Events.H2TauTau}.\labelfig{Mass}}
  \svgbeg
  \svg{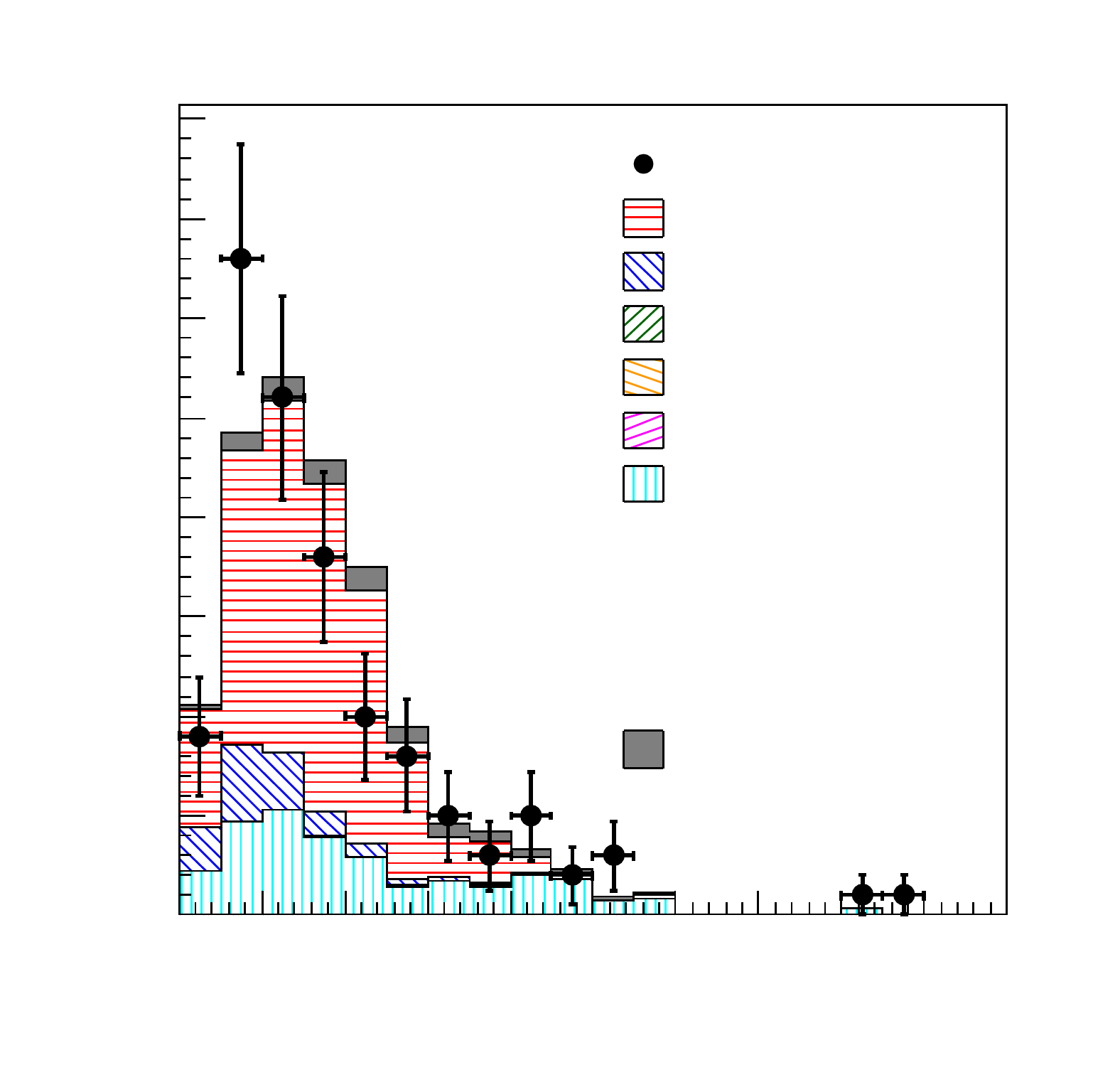}
  & \svg{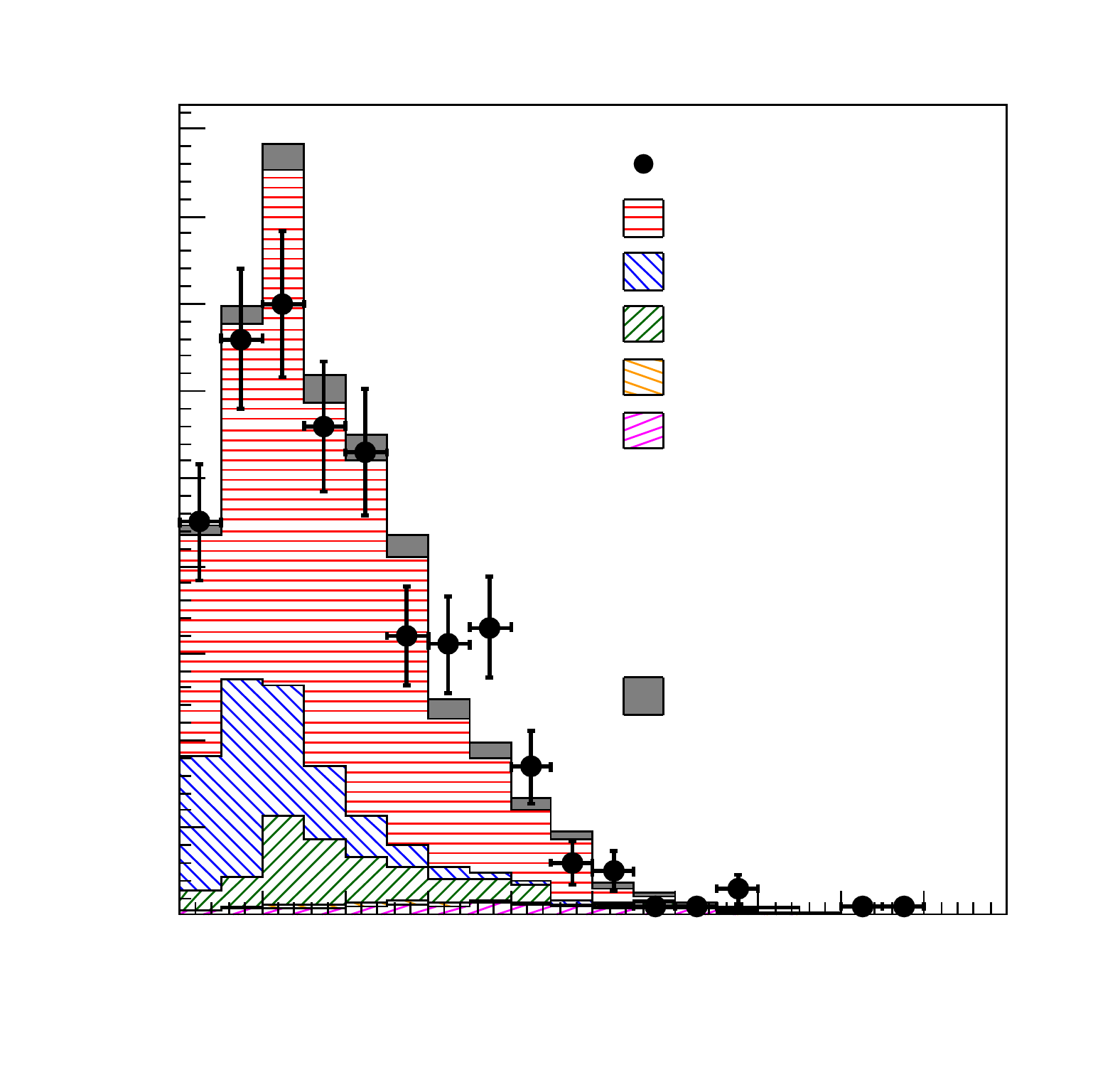}  \svgsep
  \svg{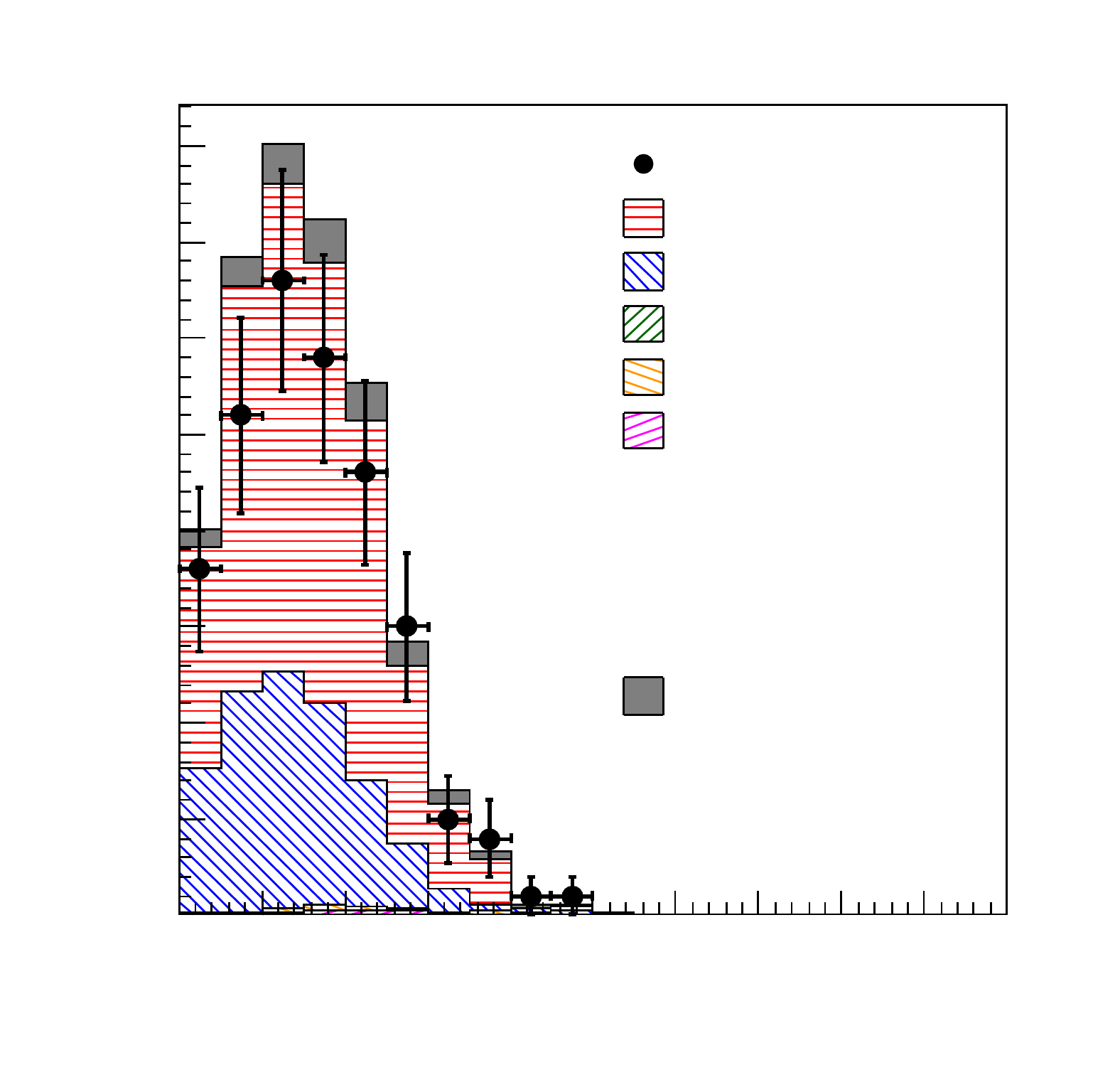}
  & \svg{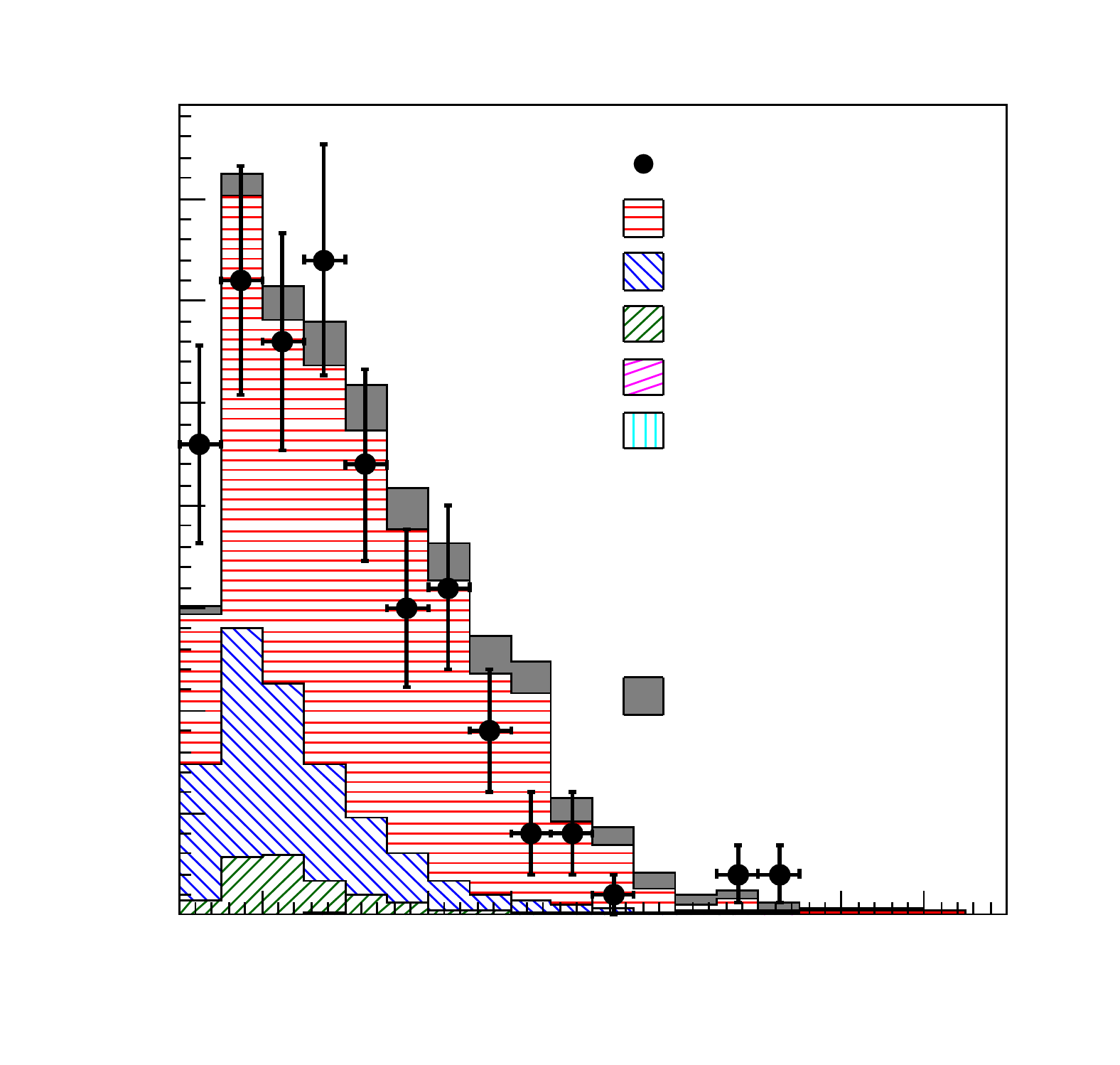} \svgsep
  \svg{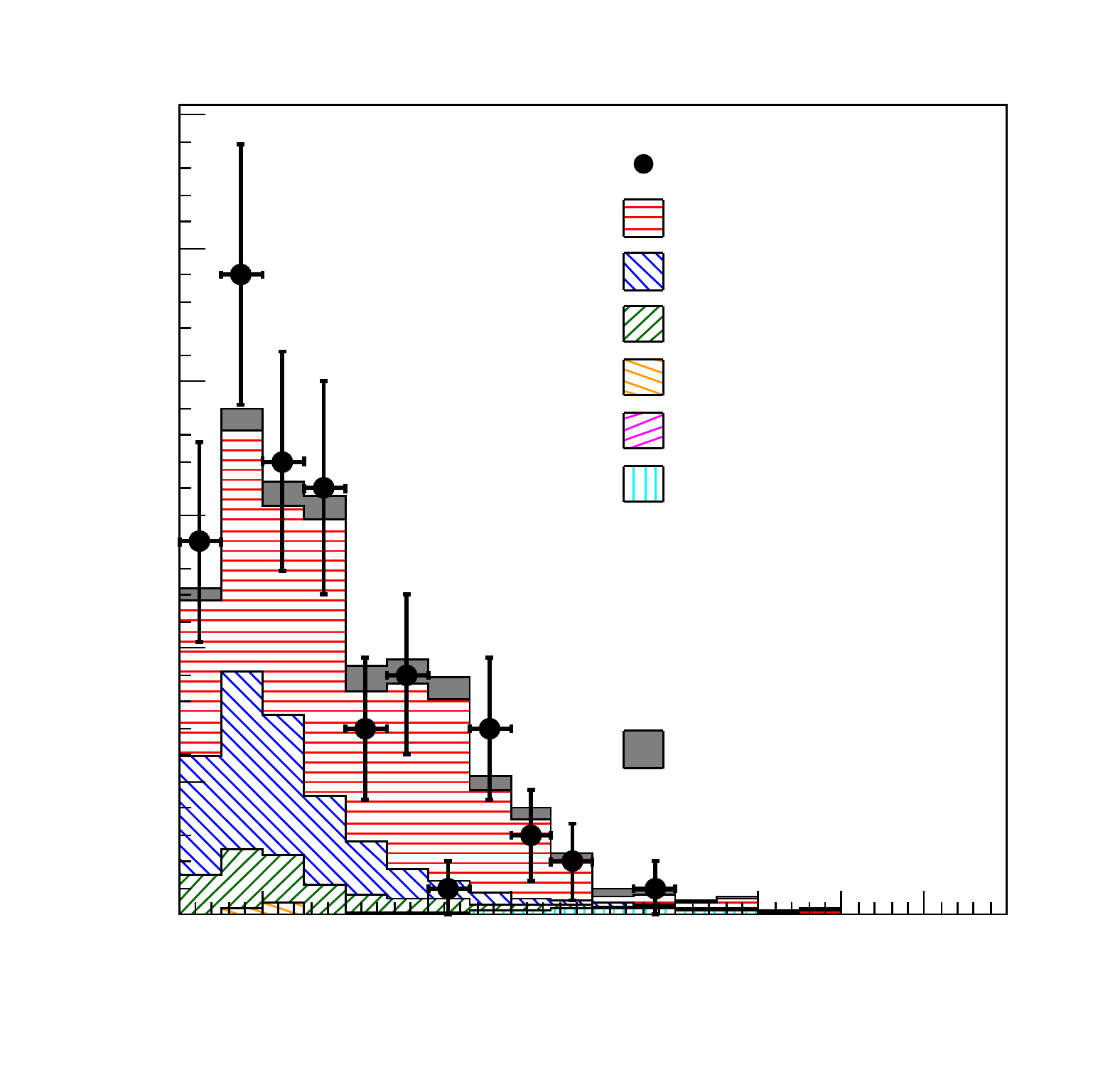}  & \sidecaption \svgend
\end{subfigures}

The invariant mass distributions for the expected backgrounds and an
example \mssm signal, with ${\m_\hAz = 125~\gev}$ and ${\tanb = 60}$,
is given for the combination of the five event categories in
\fig{Hig:H2TauTau.Mssm.Shape.Mass125.Tanb60} and separated into the
five individual categories in \fig{Hig:Mass}. The individual
background distributions are normalised to the expected number of
events from \tabs{Hig:Events} for the ${\z \to \ditau}$ backgrounds
and \tab{Zed:Events} for all other backgrounds. The signal
distribution is normalised to the number of events calculated with
\equ{Hig:Events.H2TauTau}. Additionally, an example model independent
neutral \whb signal which can be excluded, using the invariant mass
distribution information, is provided in
\fig{Hig:H2TauTau.Sm.Shape.Mass250} for the combination of the five
event categories.

\newsection{Statistical Methods}{Sta}

The tool-set used to set model independent and dependent limits on
\whb production within the forward region of \lhcb is now
presented. The primary background can be found in the general
statistics textbooks of \rfrs{kendall.45.1, *kendall.46.1} and
\cite{brandt.99.1}, and the more particle physics oriented textbooks
of \rfrs{cowan.98.1} and \cite{barlow.89.1}, as well as the Particle
Data Group review of statistics in \rfr{pdg.12.1}. This analysis also
relies heavily upon asymptotic approximations of test statistics, for
which a comprehensive guide is given in \rfr{cowan.10.1}.

Two hypothesis are considered: a background only hypothesis,
\hypo[_0], and a signal plus background hypothesis, \hypo[_1]. The
hypotheses can be parametrised as \hypo[_\mu] by a fractional signal
strength factor $\mu$, where ${\mu = 0}$ is the background only
hypothesis and ${\mu = 1}$ is the signal plus background
hypothesis. If a hypothesis is fully specified with no unknown
parameters, the hypothesis is simple, while if the hypothesis depends
upon one or more unknown parameters, {\it e.g} reconstruction
efficiencies, the hypothesis is complex.

A set of random variables \vars observed in data is compared with the
hypotheses. The agreement between the set of random variables and the
proposed hypothesis can be expressed via a test statistic,
$\vec{\stat}(\vec{x})$. Ideally the test statistic can be expressed as
a single random variable, $\stat(\vec{x})$, without a loss of
discrimination between the hypotheses.

\newsubsection{Hypothesis Testing}{Hyp}

The probability density function for a test statistic is dependent
upon the hypothesis considered, ${\pdf(\stat|\mu)}$, where $\mu$ is
the fractional signal strength parameter of the hypothesis
\hypo[_\mu]. In \fig{Hig:Statistics.Pdfs} an example of the \pdf{s}
for a test statistic is given. The dashed red curve gives the test
statistic probability density function for the background only
hypothesis, ${\pdf(\stat|0)}$, while the dotted blue curve gives
${\pdf(\stat|1)}$ for the signal plus background hypothesis. An
example experimental measurement is made with a result of \vars[_\obs]
and the corresponding test statistic is $\stat_\obs$, given by the
vertical black line of \fig{Hig:Statistics.Pdfs}.

\begin{subfigures}{2}{Probability density functions for the test
    statistic \stat, using the test statistic of
    \equ{Hig:Statistics.Npr} with the likelihood function of
    \equ{Hig:Statistics.Lhs.Simple}, assuming the background only
    hypothesis (red) and the signal plus background hypothesis
    (blue). The \wpvs are given by the fills for \pval[_0]
    (red) and \pval[_1] (blue).}
  \svgbeg \svg[1]{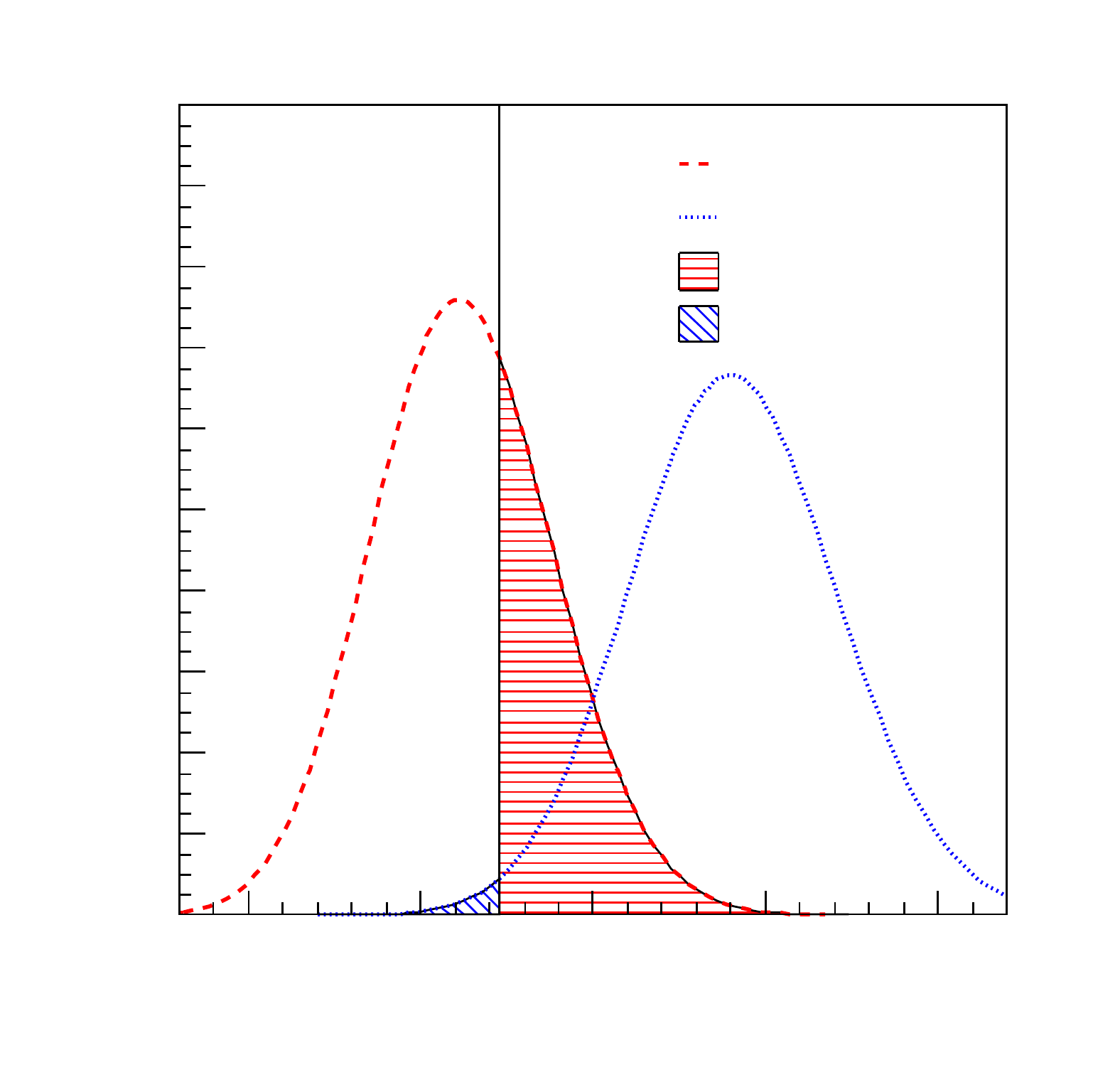} & \sidecaption \svgend
\end{subfigures}

A hypothesis \hypo[_\mu] is always tested via rejection, not
acceptance, by defining a critical region, \crit, within the space of
possible test statistics, such that the probability of observing
$\vec{\stat}$ in \crit, assuming \hypo[_\mu], is \pval. An
experimental measurement is then made, and if $\vec{\stat}_\obs$ falls
within the critical region the hypothesis is rejected with a
significance level \pval. For a test statistic of dimension one like
that of \fig{Hig:Statistics.Pdfs}, the critical region can be defined
by,
\begin{equation}
  \pval \equiv \int_{\lcrit_1}^{\lcrit_2} \pdf(\stat | \mu) \sdif{\stat}
\end{equation}
where $\lcrit_1$ and $\lcrit_2$ are the limits of the critical
region. The critical region is determined by which hypothesis is being
tested and at what confidence level the test is being performed.

\newsubsubsection{Confidence Levels}{}

If looking for evidence of a new signal, the background only
hypothesis must be rejected. For an observed test statistic
$\stat_\obs$, the background only hypothesis can be
rejected at a maximum significance of,
\begin{equation}
  \pval[_0] \equiv \int_{\stat[_\obs]}^{\infty} \pdf(\stat|0) \sdif{\stat}
  \labelequ{Statistics.P0}
\end{equation}
where \pval[_0] is the background only \wpv, given by the red fill in
the example of \fig{Hig:Statistics.Pdfs}. The background only \wpv is
then the probability of observing a \stat larger than \stat[_\obs] for
an ensemble of repeated experiments, assuming
\hypo[_0]. Alternatively, the background only hypothesis confidence
level is oftentimes defined as,
\begin{equation}
  \clb \equiv 1 - \pval[_0]
  \labelequ{Statistics.Clb}
\end{equation}
for which a large value indicates a low level of confidence in the
hypothesis. Typically to claim the discovery of a new signal in
particle physics, the background only hypothesis must be rejected with
a $p_0$ of $\prc{2.9 \times 10^{-5}}$ or less. Further discussion of
this convention can be found in \rfr{feldman.06.1}.

When testing the signal plus background hypothesis \hypo[_1] the \wpv,
\begin{equation}
  \pval[_1] \equiv \int_{-\infty}^{\stat[_\obs]} \pdf(\stat|1) \sdif{\stat}
  \labelequ{Statistics.P1}
\end{equation}
is used, which in the example of \fig{Hig:Statistics.Pdfs} is given by
the blue fill. Assuming \hypo[_1] is true, \pval[_1] is the
probability of observing a \stat lower than \stat[_\obs] for an
ensemble of repeated experiments. The signal plus background
confidence level is defined as,
\begin{equation}
  \clsb \equiv \pval[_1]
  \labelequ{Statistics.Clsb}
\end{equation}
where now a small value indicates a low level of confidence in the
hypothesis. In particle physics, a signal plus background hypothesis
is oftentimes considered to be excluded at a $95\%$ confidence level
if \clsb is found to be less than $5\%$. Here, the
confidence level is defined as ${1 - \clsb}$. At this confidence
level, assuming the signal plus background hypothesis, $95\%$ of
repeated experiments will produce a \stat greater than \stat[_\obs].

\begin{subfigures}{2}{The test statistic \pdf{s} for an example
    experiment resulting in a mis-leading exclusion at
    $95\%~\clsb$. The black line indicates the observed test statistic
    $\stat_\obs$.}
  \svgbeg
  \svg[1]{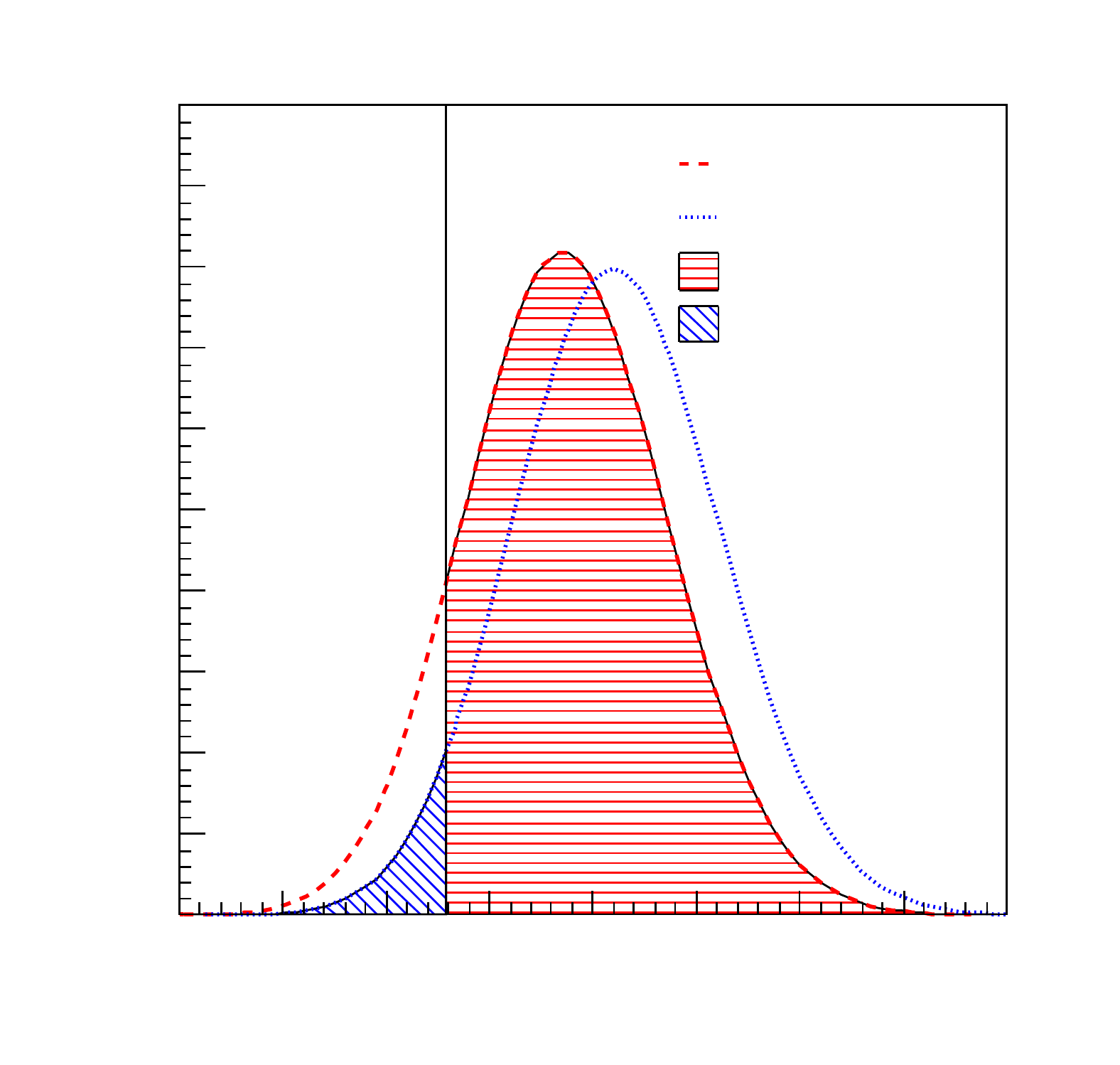} & \sidecaption \svgend
\end{subfigures}

Using \clsb for an experiment with a small signal and large background
can result in mis-leading exclusions of the signal plus background
hypothesis when there is a large downward fluctuation in the observed
number of events. The \pdf{s} for the test statistic assuming
\hypo[_0] and \hypo[_1] for such an example are given in
\fig{Hig:Statistics.Cls.Pdfs}. Here, the signal plus background
hypothesis can be rejected at a confidence level of $95\%$ using the
\clsb method, despite poor agreement of the observed test statistic
with the background only hypothesis. One method to combat mis-leading
exclusions like this is to report both $p_0$ and $p_1$ so the
agreement with not only the signal plus background hypothesis but also
the background hypothesis is known.

However, this method does not allow for a simple comparison between
different experimental results. In particle physics
the \cls method,
\begin{equation}
  \cls \equiv \frac{\clsb}{\clb} = \frac{\pval[_1]}{1 - \pval[_0]}
  \labelequ{Statistics.Cls}
\end{equation}
from \rfrs{junk.99.1} and \cite{read.02.1} is oftentimes used instead
of the \clsb method. Using this method for the example of
\fig{Hig:Statistics.Cls.Pdfs} does not result in a mis-leading exclusion
at the $95\%$ confidence level, because while $p_1$ is less than
$5\%$, $p_0$ is also large, resulting in a larger \clsb.

Particle physics lends itself towards a frequentist interpretation of
probability, where probability is defined as the relative frequency
for an event to occur. The \cls method, like the \clsb method, also
has a frequentist interpretation as shown in \rfr{birnbaum.62.1}, and
when the expected number of signal events is large with respect to the
expected background the limits produced from the \clsb and \cls
methods converge. Alternatives to \cls have been proposed, {\it e.g.}
\rfr{cowan.11.1}, but for comparison purposes with other experiments
the \cls method is used in this analysis.

\newsubsubsection{Test Statistics}{}

The above discussion relies upon the test statistic \stat, which is
defined here. From the result of Neyman and Pearson~\cite{neyman.32.1}
the most powerful test statistic, assuming simple hypotheses, is
\begin{equation}
  t \equiv \llh(\vars|1) - \llh(\vars|0) = \ln
  \left(\frac{\pdf(\vars|1)}{\pdf(\vars|0)}\right)
  \labelequ{Statistics.Npr}
\end{equation}
when using the background only hypothesis test of
\equ{Hig:Statistics.P0} and the signal plus background hypothesis test
of \equ{Hig:Statistics.P1}. Here, $\llh(\vars|\mu)$ is the natural
logarithm of the likelihood function,
\begin{equation}
  \lh(\vars|\mu) \equiv \prod_i \pdf(\var_i|\mu)
  \labelequ{Statistics.Lh}
\end{equation}
given the random variables $\var_i$ of \vars are independent. The
log-likelihood function is used for numerical stability.

The \pdf of \stat can be determined analytically when the \pdf{s} of
$\llh(\stat|0)$ and $\llh(\stat|1)$ are known and
\equ{Hig:Statistics.Npr} is an invertible function. However, this is
oftentimes not the case, and so the \pdf of \stat must be built using
a Monte Carlo technique: a large number of pseudo-experiments are
generated, the test statistic for each experiment is built, and the
\pdf is taken as the normalised distribution of \stat. This process
can be computationally expensive, particularly for complicated
likelihood functions, and so a numerically simpler alternative which
provides a similar separation power is preferable to \stat of
\equ{Hig:Statistics.Npr}.

Consider maximising $\llh(\vars|\nu)$ with respect to a signal
strength parameter $\nu$ such that $\llh(\vars|\hat{\nu})$ is the
maximum log-likelihood for a given \vars and $\hat{\nu}$ is the
maximum likelihood estimator. The first derivative of
$\llh(\vars|\nu)$ can be expanded about the point $\hat{\nu}$,
\begin{equation}
  \frac{\partial \llh(\vars|\nu)}{\partial\nu} = \left.\frac{\partial
      \llh(\vars|\nu)}{\partial\nu}\right|_{\hat{\nu}} + (\nu - \hat{\nu})
  \left.\frac{\partial^2\llh(\vars|\nu)}{\partial\nu^2}\right|_{\hat{\nu}}
  + \ldots
  \labelequ{Statistics.Plr.Expand}
\end{equation}
where the first term vanishes since $\llh(\vars|\nu)$ is at a maximum
for $\hat{\nu}$. In the limit for a large number of repeated
experiments $N$, the second derivative approaches the expectation
value of the set of experiments and can be written as $-1/\sigma^2$
where $\sigma^2$ is the variance. Additionally, the maximum likelihood
estimator $\hat{\nu}$ becomes normally distributed,
\begin{equation}
  \pdf(\hat{\nu}|\mu) = \frac{1}{\sigma\sqrt{2\pi}}e^{-\frac{(\hat{\nu} -
      \mu)^2}{2\sigma^2}}
  \labelequ{Statistics.Plr.Mle}
\end{equation}
with the same variance $\sigma^2$ and a mean of $\mu$; see
\rfr{brandt.99.1} for further details.

\Equ{Hig:Statistics.Plr.Expand} can then be written as,
\begin{equation}
  \frac{\partial \llh(\vars|\nu)}{\partial\nu} = -\frac{\nu -
    \hat{\nu}}{\sigma^2}
\end{equation}
where the higher orders terms have been neglected. Integrating yields,
\begin{equation}
  \llh(\vars|\nu) = -\frac{(\nu - \hat{\nu})^2}{2\sigma^2} +
  \llh(\vars|\hat{\nu})
  \labelequ{Statistics.Plr.Integral}
\end{equation}
where the initial condition $\nu = \hat{\nu}$ is used to determine the
constant of integration $\llh(\vars|\hat{\nu})$. Using
\equ{Hig:Statistics.Plr.Integral} the profile likelihood ratio test
statistic,
\begin{equation}
  \stat[_\nu] \equiv -2\left(\llh(\vars|\nu) -
    \llh(\vars|\hat{\nu})\right) = \frac{(\nu - \hat{\nu})^2}{\sigma^2}
  \labelequ{Statistics.Plr}
\end{equation}
is defined. This provides a test statistic that can be written without
likelihood functions but also provides a discrimination close to the
most powerful test statistic of \equ{Hig:Statistics.Npr}. The only random
variable of \stat[_\nu] is $\hat{\nu}$ and so the \pdf of \stat[_\nu]
can be found by,
\begin{equation}
  \pdf(\stat[_\nu]|\mu) = \pdf(\hat{\nu}(\stat[_\nu])|\mu) \abs{\frac{\partial
      \hat{\nu}(\stat[_\nu])}{\partial \stat[_\nu]}}
  \labelequ{Statistics.Plr.Transform}
\end{equation}
where the \pdf of $\hat{\nu}$ is transformed to the variable
\stat[_\nu]. The inverse of $\stat[_\nu]$, $\hat{\nu}(\stat[_\nu]),$
is ${\nu \pm \sigma\sqrt{\stat[_\nu]}}$ and so
\equ{Hig:Statistics.Plr.Transform} becomes,
\begin{equation}
  \pdf(\stat[_\nu]|\mu) = \frac{1}{\sqrt{8\pi\stat[_\nu]}}
  e^{-\frac{1}{2}(\frac{\nu - \mu}{\sigma} + \sqrt{\stat[_\nu]})^2} +
  \frac{1}{\sqrt{8\pi\stat[_\nu]}}
  e^{-\frac{1}{2}(\frac{\nu - \mu}{\sigma} - \sqrt{\stat[_\nu]})^2}
  \labelequ{Statistics.Plr.Pdf}
\end{equation}
where $\pdf(\hat{\nu}|\mu)$ is given by \equ{Hig:Statistics.Plr.Mle},
the first term is from the ${\nu + \sigma\sqrt{\stat[_\nu]}}$
solution, and the second term is from the ${\nu -
  \sigma\sqrt{\stat[_\nu]}}$ solution of $\hat{v}(\stat[_\nu])$. This
\pdf, further discussed in \rfr{cowan.10.1}, is a non-central
chi-squared distribution for one degree of freedom. From a result of
\rfr{wald.43.1} the remaining terms of
\equ{Hig:Statistics.Plr.Expand}, which have not been explicitly
written, can be shown to contribute to \stat[_\nu] on the order of
$1/\sqrt{n}$ where $n$ is the size of \vars.

\begin{subfigures}{2}{Probability density functions from $10^6$
    pseudo-experiments for the two-sided test statistic \stat[_0],
    using the test statistic of \equ{Hig:Statistics.Plr} with the
    likelihood function of \equ{Hig:Statistics.Lhs.Simple}, assuming
    the background hypothesis (red) and the signal plus background
    hypothesis (blue). The two \pdf{s} are compared to
    \equ{Hig:Statistics.Plr.Pdf}, a chi-squared distribution with one
    degree of freedom for \hypo[_0] (solid black) and a fitted
    non-central chi-squared distribution with one degree of freedom
    for \hypo[_1] (dashed grey).}
  \svgbeg \svg[1]{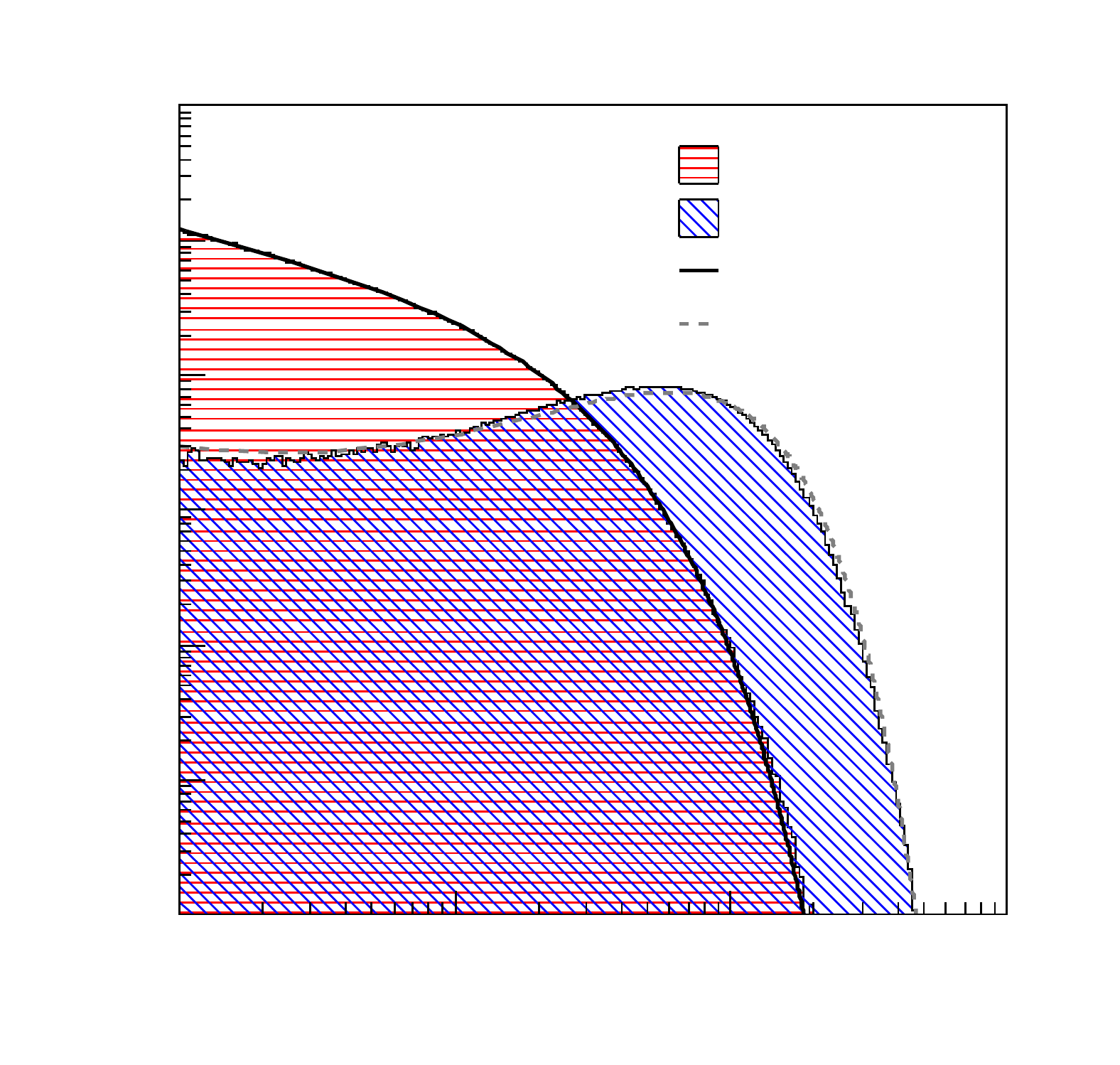} & \sidecaption \svgend
\end{subfigures}

For the case when $\nu = \mu$, \equ{Hig:Statistics.Plr.Pdf} is a
chi-squared distribution of one degree of freedom, a result first
shown by Wilks in \rfr{wilks.38.1}. The distributions necessary for
background and signal plus background hypothesis testing,
${\pdf(t_0|0)}$ and ${\pdf(t_1|1)}$, are then fully defined by
\equ{Hig:Statistics.Plr.Pdf} as $\nu$ and $\mu$ cancel and the terms
with $\sigma$ are zero.

In \fig{Hig:Statistics.Wilks} the same example of
\fig{Hig:Statistics.Pdfs} is used and the distributions for
${\pdf(\stat[_0]|0)}$ and ${\pdf(\stat[_0]|1)}$, the two-sided test
statistic of \equ{Hig:Statistics.Plr}, are generated using the Monte
Carlo technique from $10^6$ pseudo-experiments. These distributions
are shown in \equ{Hig:Statistics.Plr.Pdf}, where ${\pdf(\stat[_0]|0)}$
approaches a chi-squared distribution of one degree of freedom and
${\pdf(\stat[_0]|0)}$ approaches a non-central chi-squared
distribution. The two \pdf{s} are also compared to the chi-squared
functions of \equ{Hig:Statistics.Plr.Pdf} which they approach.

In \rfr{cowan.10.1} a test statistic for performing an upper limit test
on the signal plus background hypothesis is proposed,
\begin{equation}
  q_\nu \equiv \begin{cases}
    -2\left(\llh(\vars|\nu) - \llh(\vars|\hat{\nu})\right)
    & \mathrm{if}~ \hat{\nu} \leq \nu \\
    0 & \mathrm{else}\\
  \end{cases}
  \labelequ{Statistics.Ulr}
\end{equation}
which, following the same transformation procedure of
\equ{Hig:Statistics.Plr.Transform}, results in the \pdf,
\begin{equation}
  \pdf(q_\nu|\mu) = \Phi\left(-\frac{\nu - \mu}{\sigma}\right)\delta(q_\nu) + 
  \frac{1}{\sqrt{8\pi q_\nu}}
  e^{-\frac{1}{2}(\frac{\nu - \mu}{\sigma} - \sqrt{q_\nu})^2}
  \labelequ{Statistics.Ulr.Pdf}
\end{equation}
for the limit of large $N$ where $\Phi$ is the standard normal
cumulative distribution function. The cumulative distribution function
is then given by,
\begin{equation}
  \cdf(q_\nu|\mu) = \int_{-\infty}^{q_\nu} \pdf(q_\nu'|\mu) \sdif
  q_\nu' = \Phi\left(\sqrt{q_\nu} - \frac{\nu - \mu}{\sigma}\right)
  \labelequ{Statistics.Ulr.Cdf}
\end{equation}
and so for the special case of ${\nu = \mu}$ the \cdf is given by
$\Phi(\sqrt{q_\nu})$. Consequently, \pval[_0] and \pval[_1] can be
calculated quickly by,
\begin{equation}
  \pval[_0] = \Phi({q_0}_\obs) \equcomma \pval[_1] = 1 - \Phi({q_1}_\obs)
  \labelequ{Statistics.Ulr.Pval}
\end{equation}
without the need to employ Monte Carlo techniques. In the example of
\figs{Hig:Statistics.Pdfs} and \ref{fig:Hig:Statistics.Wilks} the
\wpvs, assuming the signal plus background hypothesis, are found to be
${\pval[_1] = 1.07 \times 10^{-2}}$ using \stat of
\equ{Hig:Statistics.Npr}, and ${\pval[_1] = 0.91 \times 10^{-2}}$ using
$q_\nu$ of \equ{Hig:Statistics.Ulr}.

\newsubsection{Likelihoods, Medians, and Uncertainties}{Lik}

The test statistics of \sec{Hig:Hyp} are built from likelihood
functions, which have not yet been defined. In this analysis two
likelihood functions are used, a simple likelihood function for
consistency checks and an extended likelihood function which utilises
the mass distributions of \sec{Hig:Mass} for producing the final upper
limits of \sec{Hig:Res}. In this section these likelihoods are
described. Additionally, a method for building the median test
statistic with these likelihood functions as well as how uncertainty
can be incorporated into a likelihood function is outlined.

\newsubsubsection{Likelihood Functions}{}

The simple likelihood function is a Poisson \pdf,
\begin{equation}
  \lh_s(\var|\mu) = \pdf_p(x|N_\bkg + \mu N_\sig) =
  \frac{(N_\bkg + \mu N_\sig)^{\var} e^{-N_\bkg - \mu N_\sig}}{\var!}
  \labelequ{Statistics.Lhs.Simple}
\end{equation}
where \var is the number of observed events, $N_\bkg$ is the expected
number of background events, $N_\sig$ is the number of expected signal
events, and $\mu$ is the signal strength parameter of
\sec{Hig:Hyp}. In the test statistics of \equs{Hig:Statistics.Npr},
\ref{equ:Hig:Statistics.Plr}, and \ref{equ:Hig:Statistics.Ulr} the
ratio of two likelihoods is always taken for a given \var and so the
discrete \var factorial terms cancel. Consequently, while the observed
\var from an experiment must be an integer, the defined test
statistics can still be calculated for any real-valued
\var. Additionally, the $e^{N_\bkg}$ terms cancel and can be omitted
when calculating the test statistics with this likelihood function.

The simple likelihood only utilises the number of expected and observed
events, without considering any additional information from the
events. The extended likelihood function utilises not only the number
of events, but also any other observables which are measured for each
event. The extended likelihood function is defined as,
\begin{equation}
  \lh_e(\vars|\mu) = e^{-N_\bkg - \mu N_\sig} \prod_i \big((N_\bkg +
  \mu N_\sig) \pdf(\vec{o}_i|\mu) \big)
  \labelequ{Statistics.Lhs.Extended}
\end{equation}
where \vars consists of a set of observables $\vec{o}_i$ for each
event $i$, and the product is over all events. The probability density
function for the set of observables $\vec{o}_i$, assuming a signal
strength $\mu$, is given by $\pdf(\vec{o}_i|\mu)$. For this analysis,
only the invariant mass of the \wtl decay products is measured per
event, and so $\vec{o}_i$ is just the invariant mass for event
$i$. The extended likelihood function of
\equ{Hig:Statistics.Lhs.Extended} is the limit of the binned
likelihood function, which is the product of the simple likelihood
function for every bin, as the widths of the bins approach zero. For
further details on the properties of the extended likelihood function,
refer to \rfr{barlow.90.1}.

\begin{subfigures}{2}{\subfig{Statistics.Lhs.Pdfs}~The example of
    \fig{Hig:Statistics.Pdfs} but now using the extended likelihood
    function of \equ{Hig:Statistics.Lhs.Extended} rather than the
    simple likelihood function to calculate the test
    statistic. \subfig{Statistics.Lhs.Shape} Invariant mass
    distribution for pseudo-data (points), expected background (red),
    and expected signal (grey) used when calculating the extended
    likelihood for this example.\labelfig{Statistics.Lhs}}
  \svgbeg
  \svg{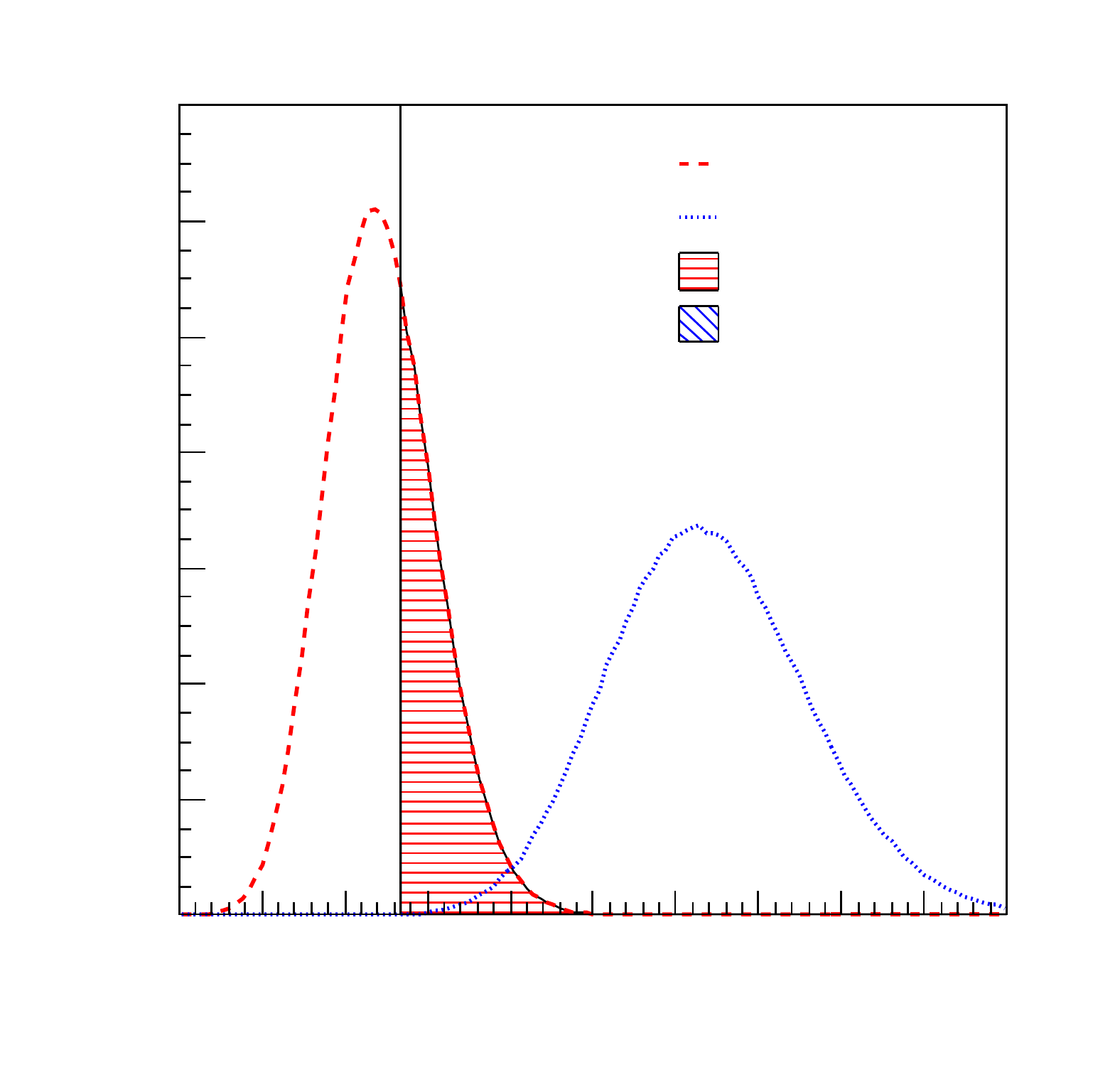} & \svg{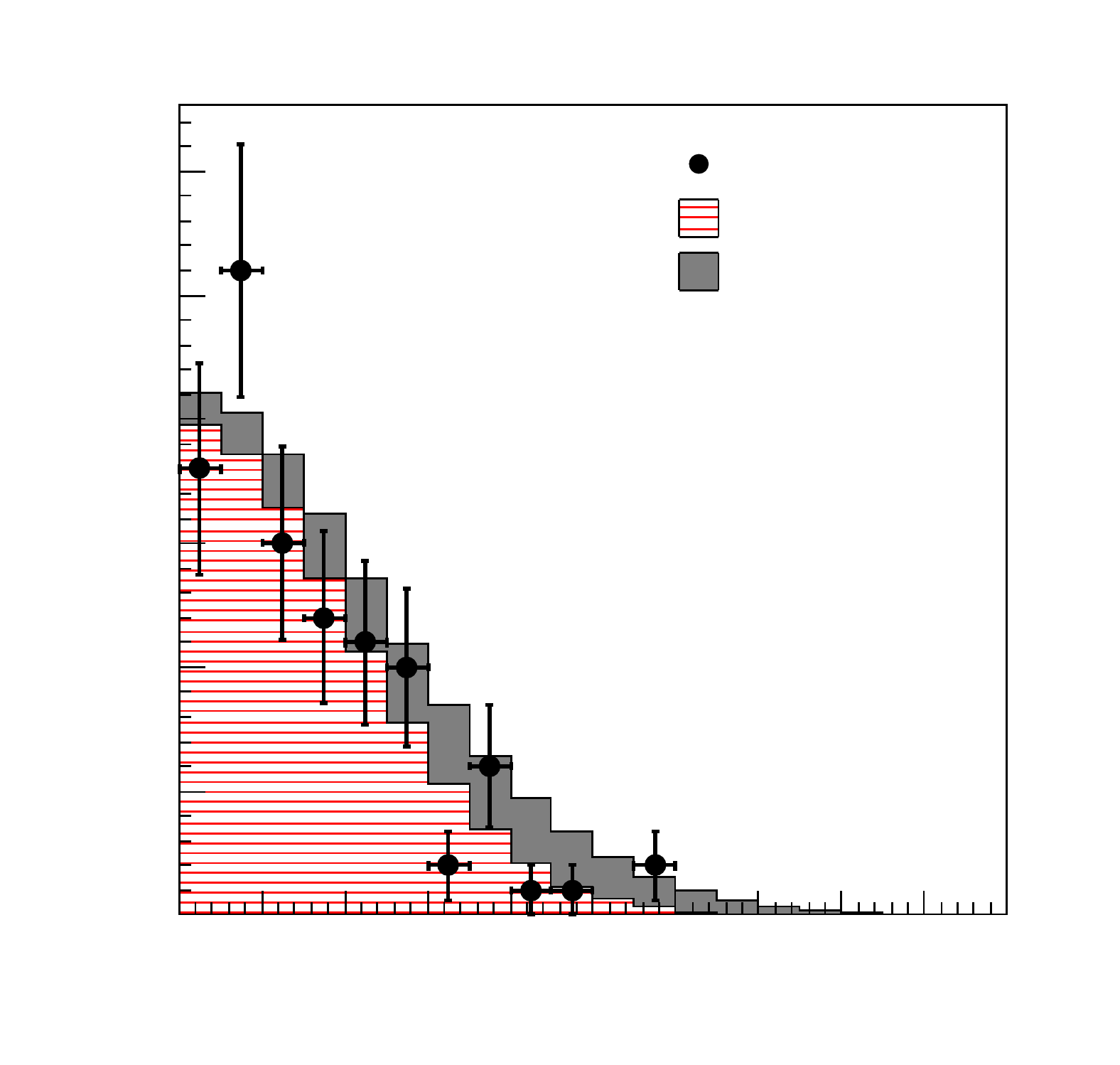} \svgend
\end{subfigures}

An example demonstrating the additional separation power between the
background only and signal plus background hypotheses when using the
extended likelihood function is given in
\fig{Hig:Statistics.Lhs}. This is the same example as
\fig{Hig:Statistics.Pdfs}, but now the extended likelihood is used to
calculate the test statistic of \equ{Hig:Statistics.Npr} rather than
the simple likelihood function. The invariant mass distributions used
in this example are given in \fig{Hig:Statistics.Lhs.Shape}. As can be
seen, the additional information from the mass for each event produces
a more significant exclusion of the signal plus background hypothesis.

\newsubsubsection{Median Values}{}

The sensitivity of an experiment in rejecting a given hypothesis is
quantified by the \wpvs for the median test statistics when testing
the null hypothesis, assuming the alternative hypothesis. When
rejecting \hypo[_0], \pval[_0] for the median test statistic assuming
\hypo[_1], \mval{\stat|1}, characterises the experimental
sensitivity. Conversely, when rejecting \hypo[_1], \pval[_1] for
\mval{\stat|0} characterises the sensitivity. The \wpvs from the
median test statistics can also be used to perform consistency checks.

The median test statistic is determined from,
\begin{equation}
  \cdf(\mval{\stat|\mu}|\mu) = \int_{-\infty}^{\mval{\stat|\mu}}
  \pdf(\stat'|\mu) \sdif{\stat'} = \frac{1}{2}
  \labelequ{Statistics.Median}
\end{equation}
where the test statistics for half of all repeated experiments are
expected to fall below \mval{\stat|\mu} and half above, assuming the
signal strength parameter $\mu$ is the true signal strength.

For the Neyman-Pearson ratio test statistic of
\equ{Hig:Statistics.Npr}, the median test statistic can be determined
by integrating the test statistic \pdf{s}. When using the simple
likelihood function of \equ{Hig:Statistics.Lhs.Simple} the \pdf of the
test statistic is,
\begin{equation}
  \pdf(t|\mu) = \frac{\pdf_p\left(\left. \frac{t+N_\sig}{\ln(N_\bkg + N_\sig) -
        \ln(N_\bkg)} \right| N_\bkg + \mu N_\sig\right)}{\ln(N_\bkg + N_\sig)
    - \ln(N_\bkg)}
\end{equation}
which is a modified Poisson distribution. The mean of this
distribution is \stat evaluated for an \var of ${N_\bkg + \mu
  N_\sig}$. If the expected background plus signal is large, the
${\pdf(t|\mu)}$ approaches a normal distribution and,
\begin{equation}
  \mval{t|\mu} = t(N_\sig + \mu N_\bkg)
\end{equation}
since the mean of the test statistic \pdf approaches the median.

For the test statistic $q_\nu$ of \equ{Hig:Statistics.Ulr}, the median
test statistic can be found using the Asimov dataset proposed
in \rfr{cowan.10.1} and inspired by \rfr{asimov.55.1}. The extended
log-likelihood function evaluated with the Asimov data set is,
\begin{equation}
  \labelequ{Statistics.Asimov}
  \llh_e(\asis(\mu)|\nu) = \begin{aligned}[t] &\int \ln\Big((N_\bkg + \nu
    N_\sig)\pdf(\vec{o}|\nu)\Big) (N_\bkg +  \mu N_\sig)
    \pdf(\vec{o}|\mu)\sdif{\vec{o}}\\
    &-N_\bkg - \nu N_\sig \\
  \end{aligned}
\end{equation}
as shown in \sap{Hvr:Asi}. Here, the integral is over the joint \pdf
of the observables $\vec{o}$, where the first instance of the \pdf is
evaluated with signal strength $\nu$ of the test statistic $q_\nu$ and
the second instance of the \pdf is evaluated with signal strength
$\mu$, the signal strength of the Asimov dataset. In this analysis,
the integral is only over the invariant mass \pdf. As an example, the
median value for the $q_1$ test statistic of
\equ{Hig:Statistics.Ulr.Pdf} can be evaluated as,
\begin{equation}
  \mval{q_1|0} = \llh_e(\asis(0)|1) - \llh_e(\asis(0)|0)
\end{equation}
when using the extended likelihood function and assuming the
background only hypothesis is true.

\newsubsubsection{Uncertainties}{}

The test statistics of \equs{Hig:Statistics.Npr},
\ref{equ:Hig:Statistics.Plr}, and \ref{equ:Hig:Statistics.Ulr} as well
as the likelihood functions of \equs{Hig:Statistics.Lhs.Simple} and
\ref{equ:Hig:Statistics.Lhs.Extended} do not incorporate any
experimental systematic uncertainties. While there is no standardised
method for incorporating uncertainties, a variety of methods can be
used.  For the frequentist hypothesis testing of \sec{Hig:Hyp} the
systematic uncertainties are introduced into the likelihood function
and the method of maximum likelihood from
\equ{Hig:Statistics.Plr.Expand} is used. The alternative methods of
marginalisation and hybrid marginalisation are described in
\sap{Hvr:Mar}, but are not used here.

If the uncertainties factorise, the likelihood function
becomes,
\begin{equation}
  \lh(\vars|\mu,\uncs) = \prod_i \pdf(\var[_i]|\mu, \uncs) \prod_j
  \pdf(\unc[_j])
\end{equation}
where the first product is over all \var[_i] of \vars and the second
product is over all \unc[_j] of \uncs. The probability density
function for each nuisance parameter is given by $\pdf(\unc_j)$. This
likelihood function is then introduced into the profiled likelihood
ratios which become,
\begin{equation}
  \stat[_\nu] = -2(\llh(\vars|\nu,\hat{\hat{\uncs}}) -
  \llh(\vars,\hat{\nu},\hat{\uncs}))
  \labelequ{Statistics.Plr.Unc}
\end{equation}
and,
\begin{equation}
  q_\nu = \begin{cases}
    -2\left(\llh(\vars|\nu,\hat{\hat{\uncs}}) -
      \llh(\vars|\hat{\nu},\hat{\uncs})\right)
    & \mathrm{if}~ \hat{\nu} \leq \nu \\
    0 & \mathrm{else}\\
  \end{cases}
  \labelequ{Statistics.Ulr.Unc}
\end{equation}
where $\hat{\hat{\uncs}}$ is the maximum likelihood estimator for a
given \vars and $\nu$, and $\hat{\uncs}$ is the maximum likelihood
estimator for a given \vars and $\hat{\nu}$. The \pdf{s} for
\stat[_\nu] and $q_\nu$ of \equs{Hig:Statistics.Plr.Pdf} and
\ref{equ:Hig:Statistics.Ulr.Pdf} still hold, as does the method of
obtaining the median test statistic using the Asimov dataset.

In this analysis, each systematic uncertainty is introduced as a
normally distributed nuisance parameter with a mean of $\epsilon$ and
a deviation of $\delta$. The likelihood functions of
\equs{Hig:Statistics.Lhs.Simple} and
\ref{equ:Hig:Statistics.Lhs.Extended} then become,
\begin{equation}
  \lh(\vars|\mu,\uncs) = \lh(\vars|\mu) \prod_i
  \left(\frac{1}{\delta_i\sqrt{2\pi}} e^{-\frac{(\unc[_i] -
        \epsilon_i)^2}{2\delta_i^2}} \right)
  \labelequ{Statistics.Lhs.Unc}
\end{equation}
where $\lh(\vars|\mu)$ is determined using $\pdf(\vars|\mu,\unc)$
evaluated at \uncs rather than the central values $\vec{\epsilon}$,
and the product is over all nuisance parameters \unc[_i].

\newsection{Results}{Res}

Because no excess is seen in the number of observed events shown in
\fig{Hig:Mass} and tabulated in \tab{Hig:Events} when compared to the
background only hypothesis, upper limits are set on neutral \whb
production. Limits on model independent production of neutral \whbs
decaying into \wtl pairs, $\sigma_{\dip \to \hz \to \ditau}$, within
the \lhcb acceptance $2.0 \leq \eta_\tau \leq 4.5$, using the \whb
phenomenology of \sec{Hig:Phe}, the event model of \sec{Hig:Mod}, and
the statistical methods of \sec{Hig:Sta}, are calculated for the
individual event categories. The limits are set at a $95\%$ confidence
level, using \cls with the profiled likelihood method and are
determined as a function of the neutral \whb mass, $\m_\h$. Similarly,
upper limits on \tanb from the production of the three neutral \mssm
\whbs decaying into \wtl pairs, assuming the \mhmax scenario, are
calculated for the five event categories as a function of the \cp-odd
\whb mass, $\m_\hAz$.

The observed limits are calculated using the test statistic $q_\nu$ of
\equ{Hig:Statistics.Ulr.Unc} and the extended likelihood function of
\equ{Hig:Statistics.Lhs.Extended}. The likelihood function is
determined using the invariant mass distributions of
\fig{Hig:Mass}. The systematic uncertainty on each background
component, given in \tab{Hig:Events} for the ${\z \to \ditau}$
background and \tab{Zed:Events} for the remaining backgrounds, is
introduced into the extended likelihood function as a nuisance
parameter using \equ{Hig:Statistics.Lhs.Unc}. Each uncertainty affects
the shape of the expected invariant mass distribution,
$\pdf(m|\mu,\uncs)$, by scaling the individual background distribution
component associated with the uncertainty. The expected number of
background events is also affected by each nuisance parameter such
that ${{N_\bkg}_i = \theta_i}$. The uncertainty on the simulated mass
shape, determined in \sec{Zed:Sel} using the calibration of
\fig{Zed:Sel.Mass.Compare}, is also introduced as a nuisance
parameter, where the parameter only affects the component mass
distributions from simulation, {\it e.g.} the ${\z \to \ditau}$
background mass shape.

These limits are given as the solid black lines in
\figs{Hig:Sm.Limits.Validate} through \ref{fig:Hig:Mssm.Limits.Final}
for each of the five event categories, and in \fig{Hig:Limits.Compare}
for the combination of the five categories. In \sec{Hig:Val}, the
observed limits for each event category are checked for consistency
against additional validation limits in \fig{Hig:Sm.Limits.Validate}
for the \sm and \fig{Hig:Mssm.Limits.Validate} for the \mssm. In
\sec{Hig:Lim} the final observed limits are compared with the expected
limits in \fig{Hig:Sm.Limits.Final} for the \sm and
\fig{Hig:Mssm.Limits.Final} for the \mssm, as well as limits from
previous results in \fig{Hig:Limits.Compare}.

\newsubsection{Validation}{Val}

The final observed limit is compared against three validation limits
for each of the five event categories in \fig{Hig:Sm.Limits.Validate}
for the model independent limits and \fig{Hig:Mssm.Limits.Validate}
for the \mssm limits. The dashed green line is the observed limit,
using the same statistical methods, but without uncertainties
introduced. The dash-dotted orange line is the observed limit using
the $q_\nu$ test statistic with a simple likelihood and with
uncertainties. For this limit, because no mass distribution
information is utilised, the mass shape uncertainty is not
included. Finally, the the dash-dot-dotted magenta line is the
observed limit using the $q_\nu$ test statistic with a simple
likelihood and no uncertainties. These validation limits can be
compared using either \fig{Hig:Sm.Limits.Validate} or
\fig{Hig:Mssm.Limits.Validate}, but the results are more easily
interpreted for the model independent limits, as the \mssm limits are
presented in $\m_\hAz$ and \tanb space. Consequently, only the model
independent limits are discussed here.

\begin{subfigures}[p]{2}{Upper limits on neutral \whbs production as
    function of the \whb mass for the
    \subfig{H2TauTau2MuMu.Sm.Limits.Validate}~\mumu,
    \subfig{H2TauTau2MuE.Sm.Limits.Validate}~\mue,
    \subfig{H2TauTau2EMu.Sm.Limits.Validate}~\emu,
    \subfig{H2TauTau2MuPi.Sm.Limits.Validate}~\muh, and
    \subfig{H2TauTau2EPi.Sm.Limits.Validate}~\eh categories. The
    observed limit using $q_\nu$ and $\lh_e$ (black) is compared to
    the observed limit without uncertainty (green) as well as the
    limits using $q_\nu$ and $\lh_s$ with uncertainty (orange) and
    without uncertainty (magenta).\labelfig{Sm.Limits.Validate}}
  \svgbeg 
  \svg{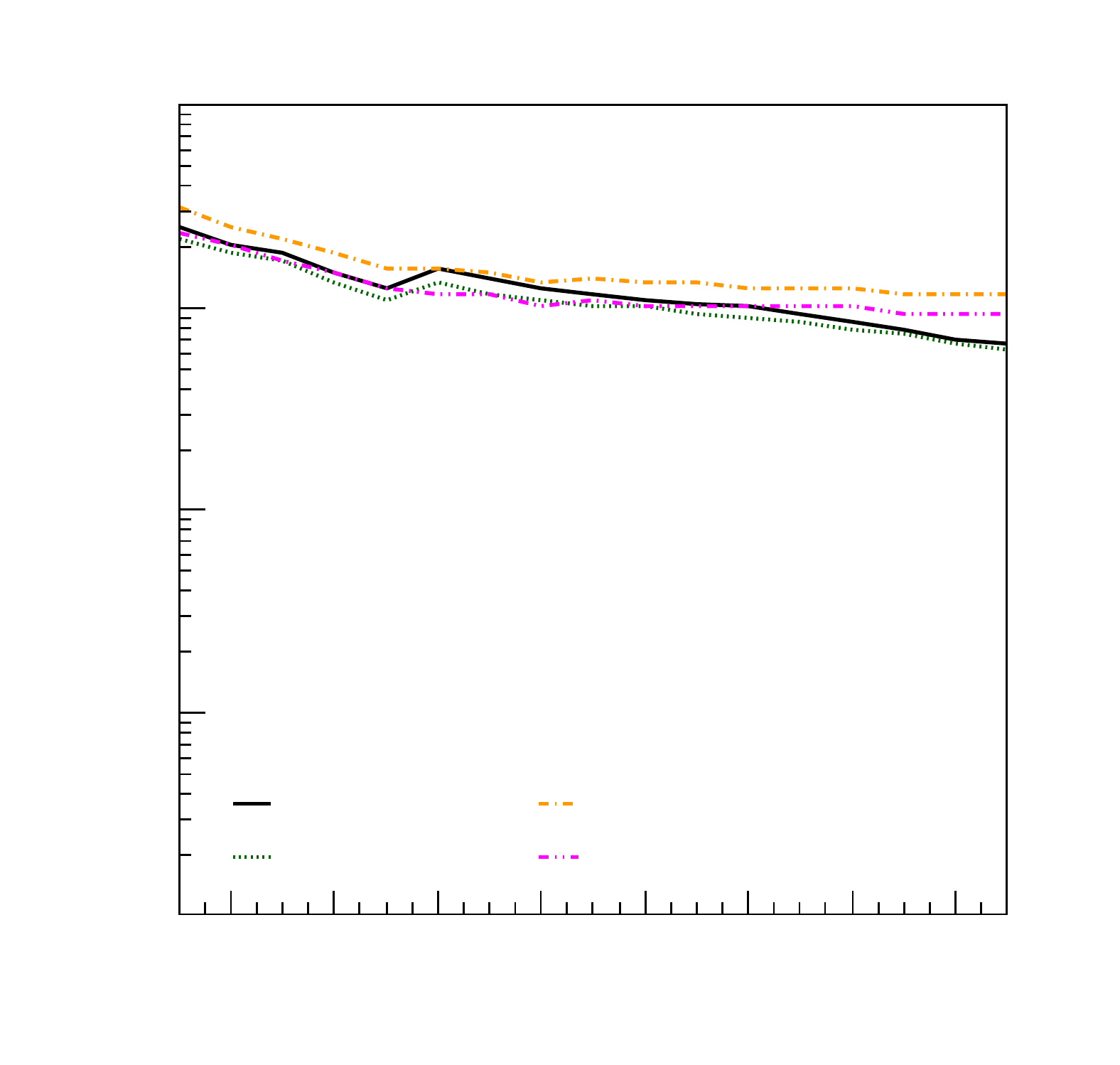} 
  & \svg{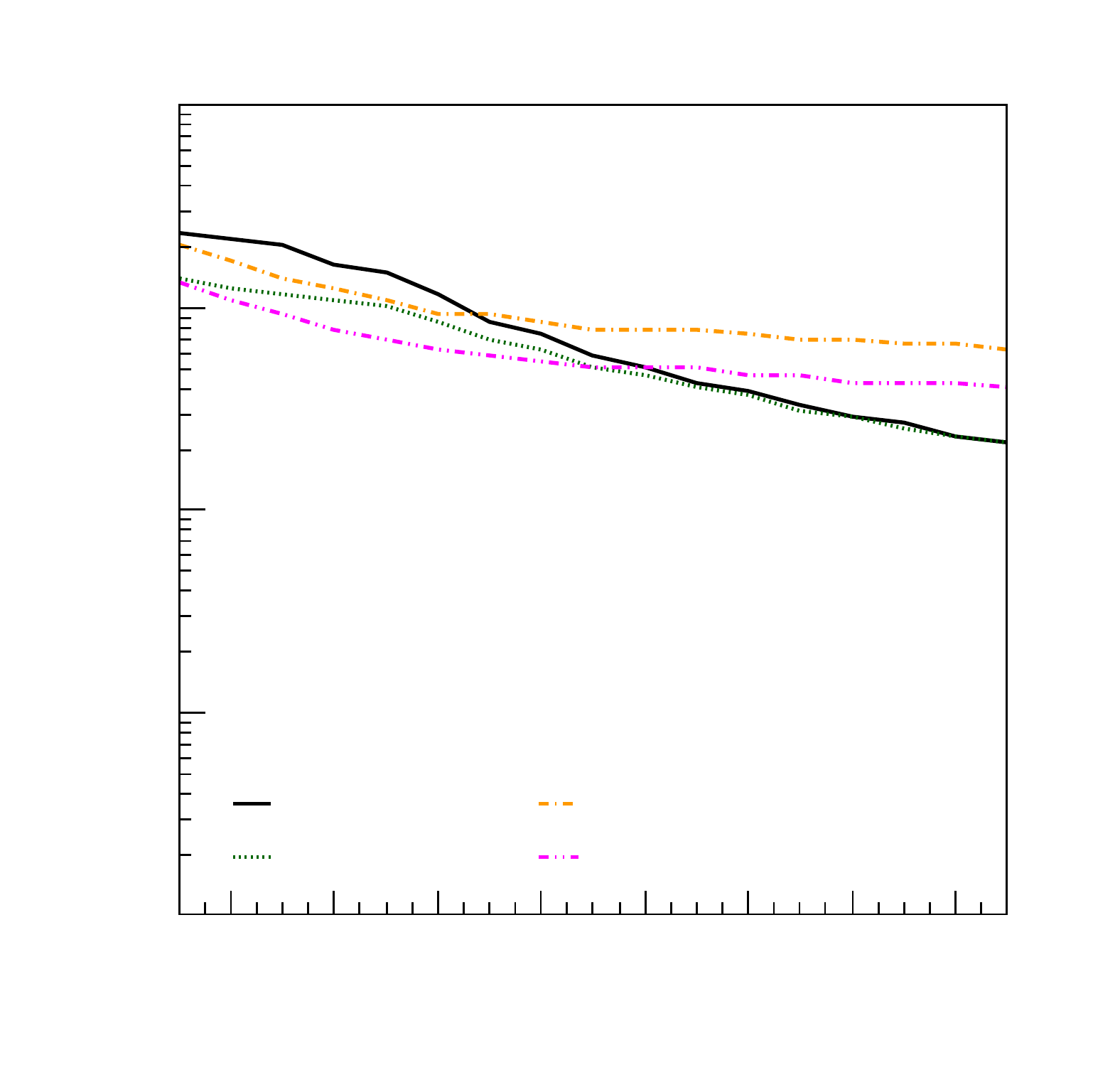} \svgsep
  \svg{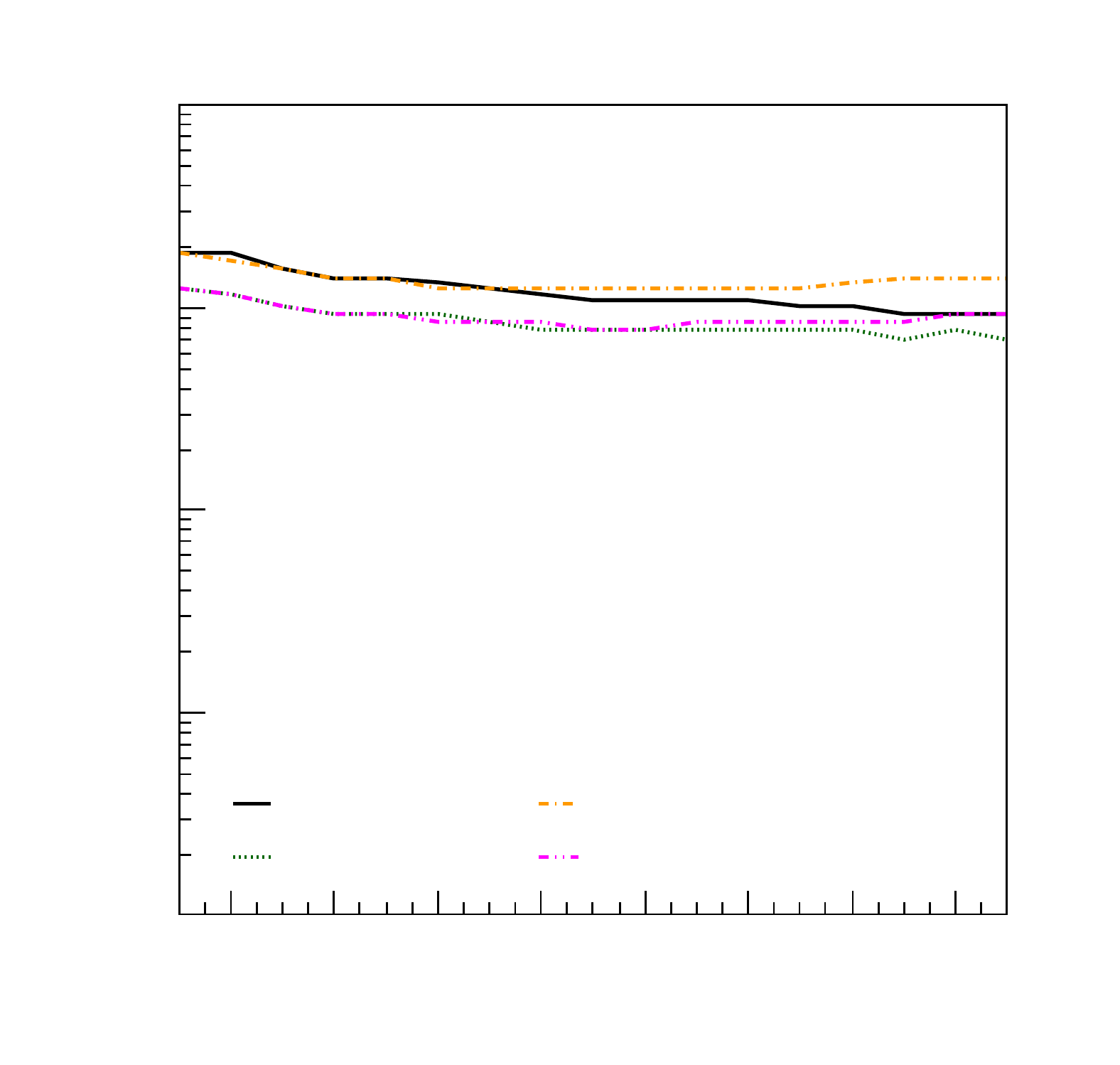} 
  & \svg{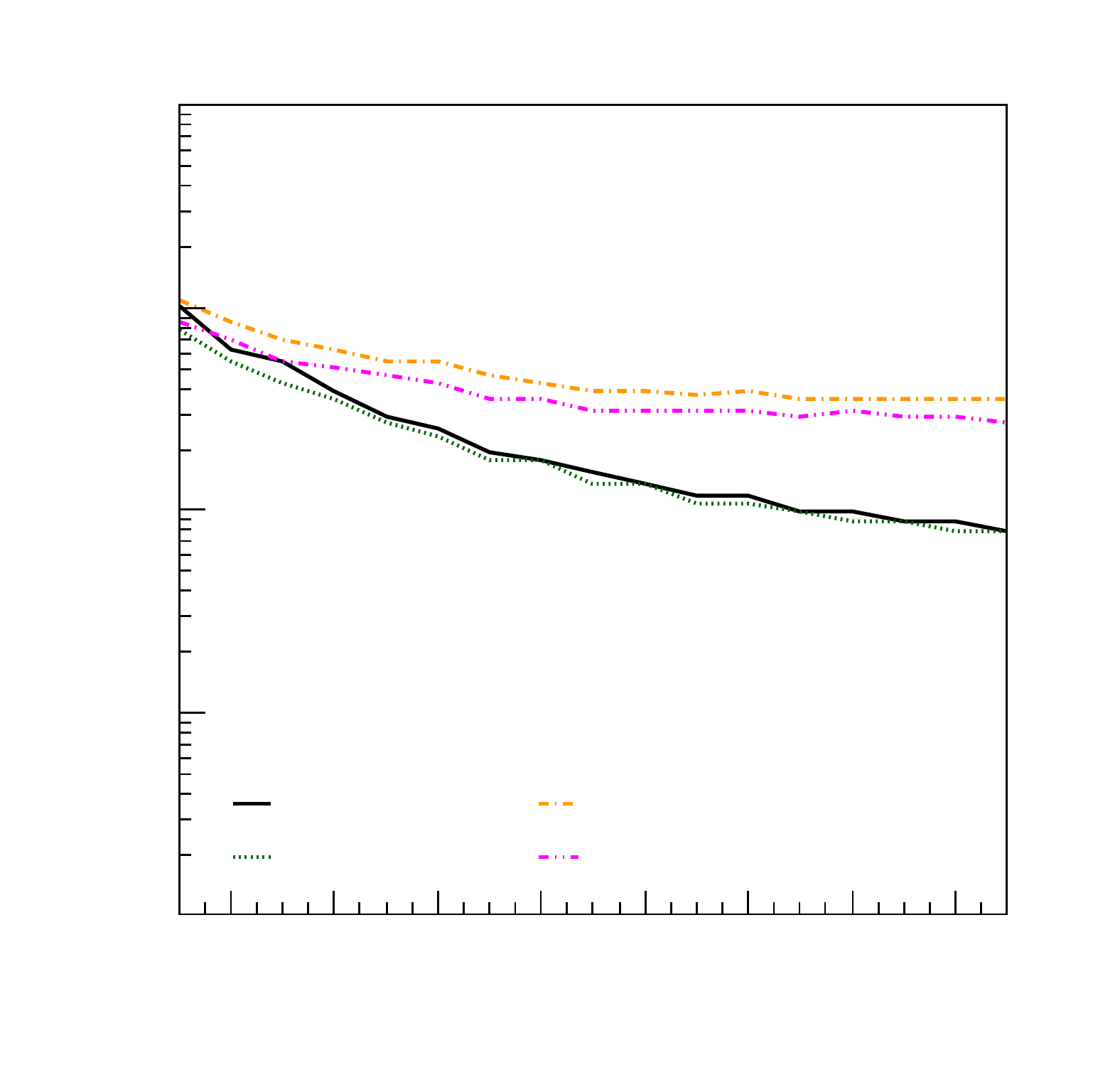} \svgsep
  \svg{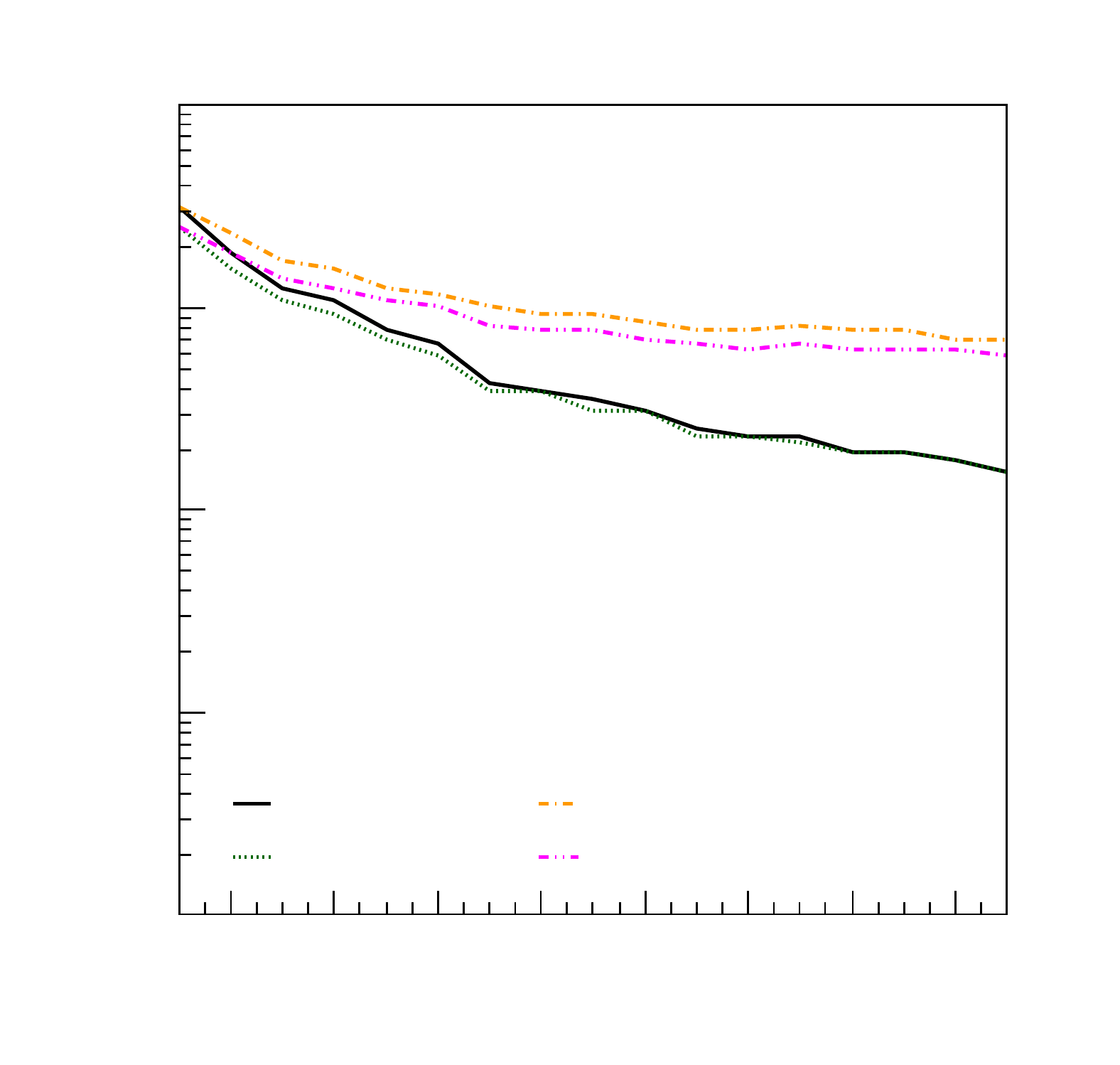} & \sidecaption \svgend
\end{subfigures}

\begin{subfigures}[p]{2}{Upper limits on \tanb as a function of the
    \cp-odd \whb mass for the
    \subfig{H2TauTau2MuMu.Sm.Limits.Validate}~\mumu,
    \subfig{H2TauTau2MuE.Sm.Limits.Validate}~\mue,
    \subfig{H2TauTau2EMu.Sm.Limits.Validate}~\emu,
    \subfig{H2TauTau2MuPi.Sm.Limits.Validate}~\muh, and
    \subfig{H2TauTau2EPi.Sm.Limits.Validate}~\eh categories. The
    observed limit using $q_\nu$ and $\lh_e$ (black) is compared to
    the observed limit without uncertainty (green) as well as the
    limits using $q_\nu$ and $\lh_s$ with uncertainty (orange) and
    without uncertainty (magenta).\labelfig{Mssm.Limits.Validate}}
  \svgbeg
  \svg{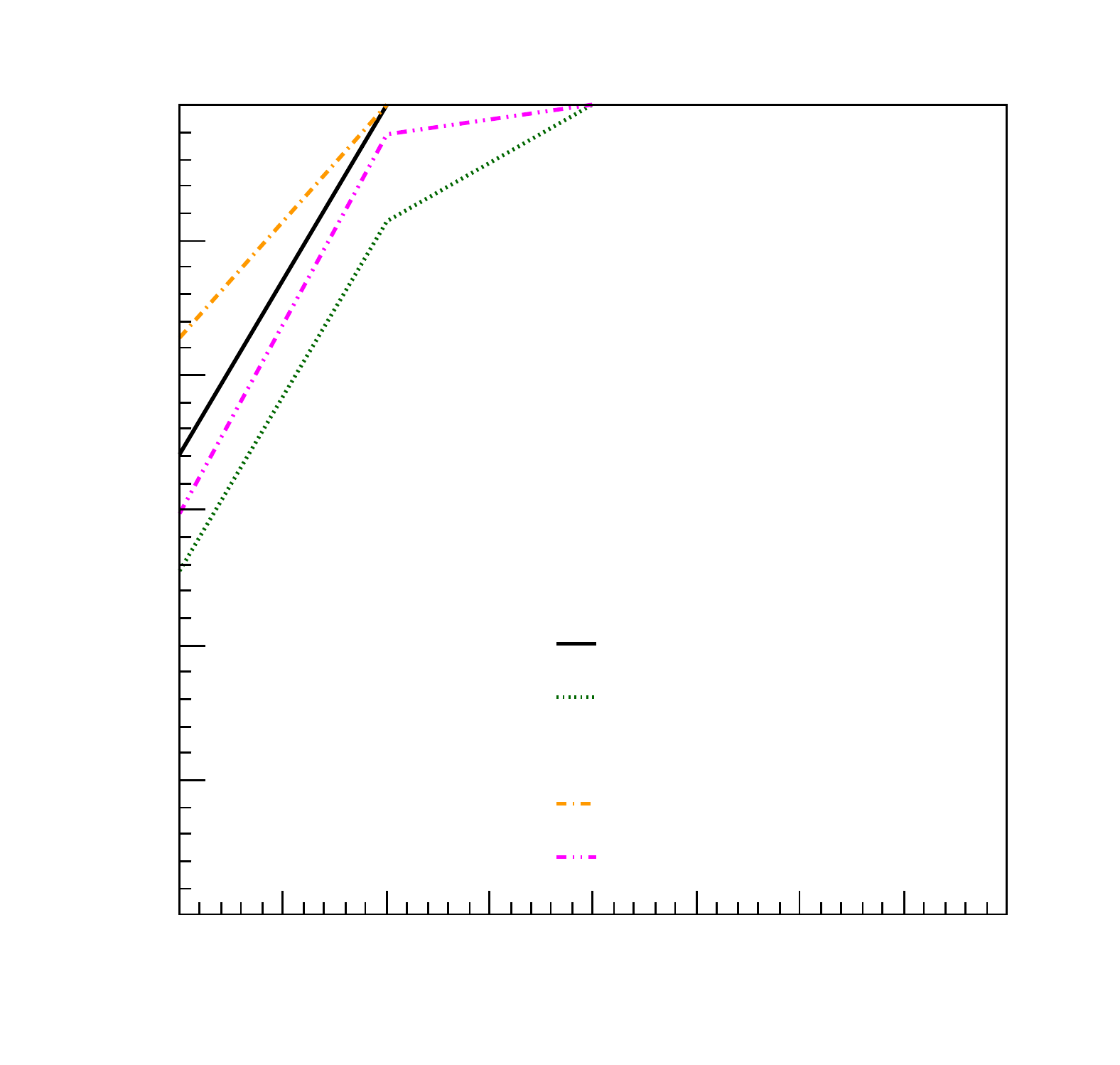} 
  & \svg{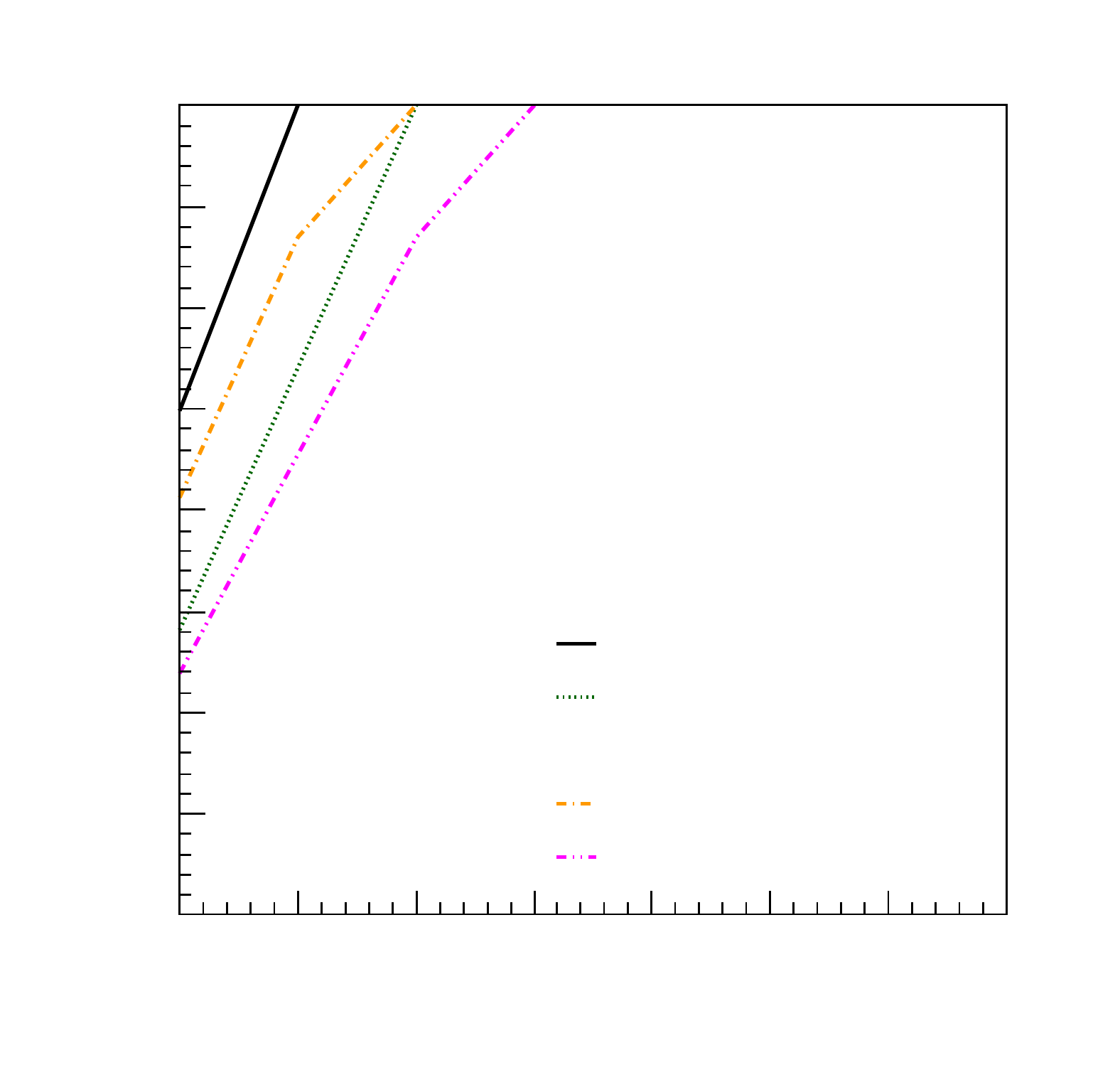} \svgsep
  \svg{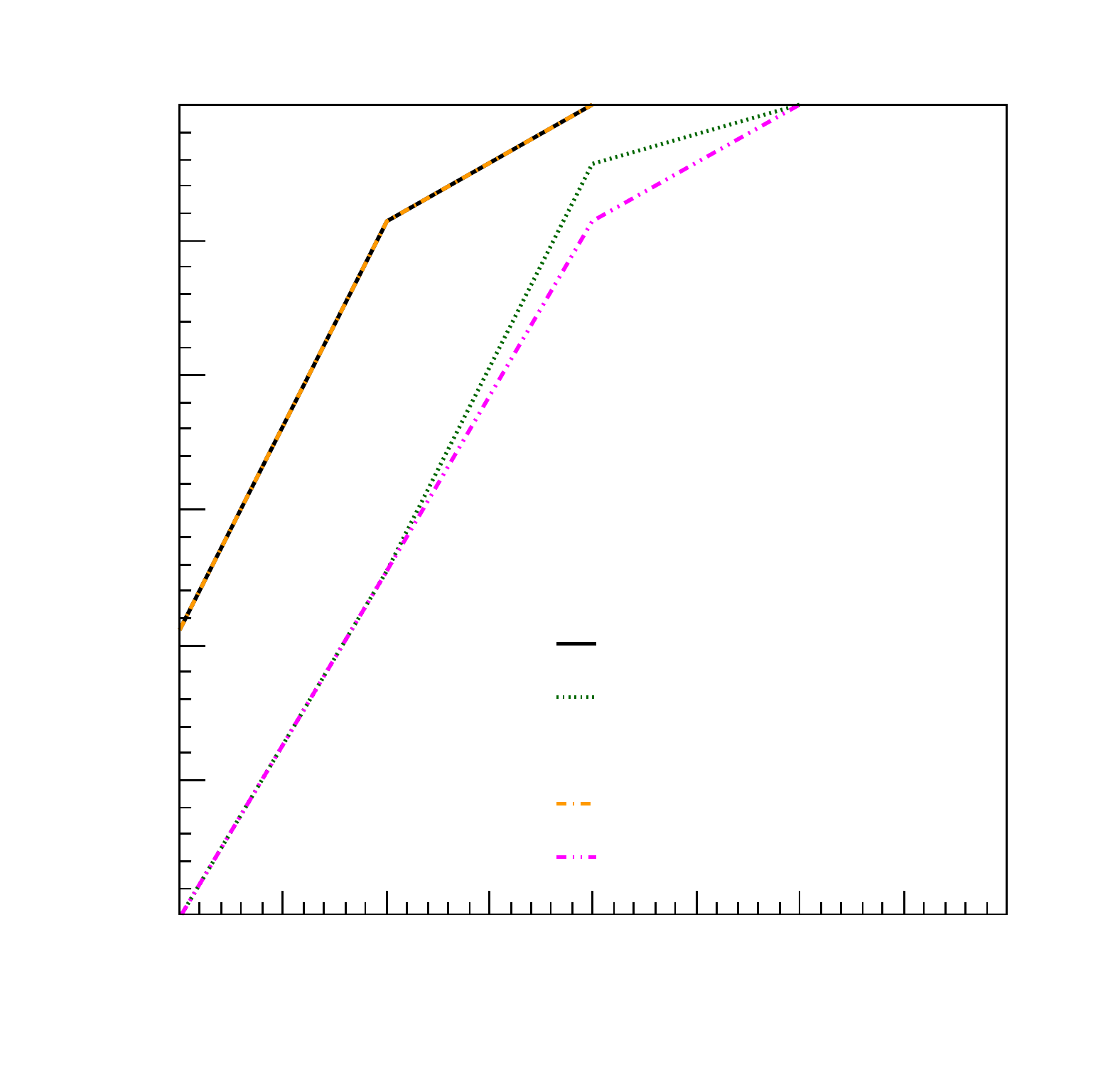}  
  & \svg{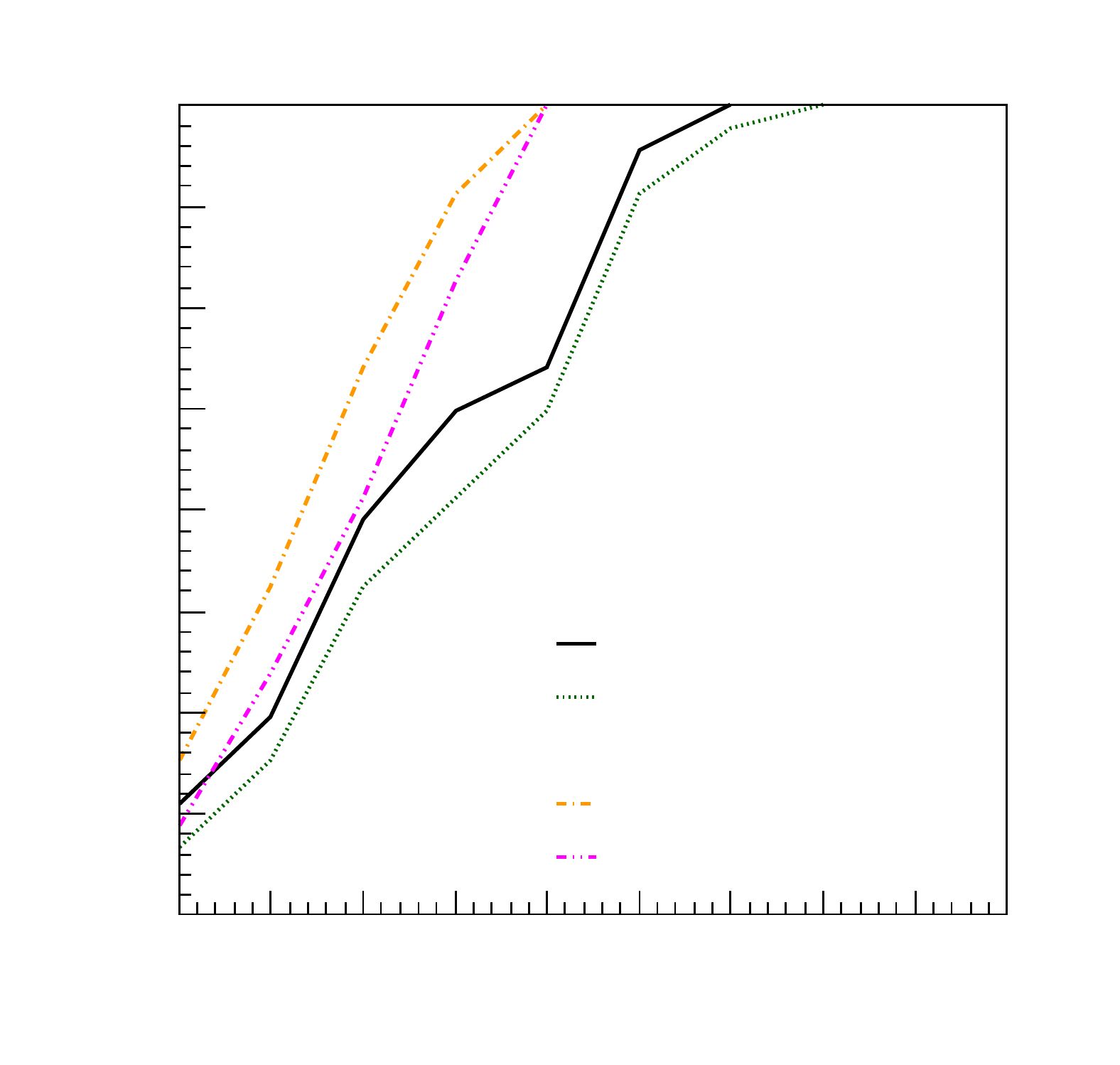} \svgsep
  \svg{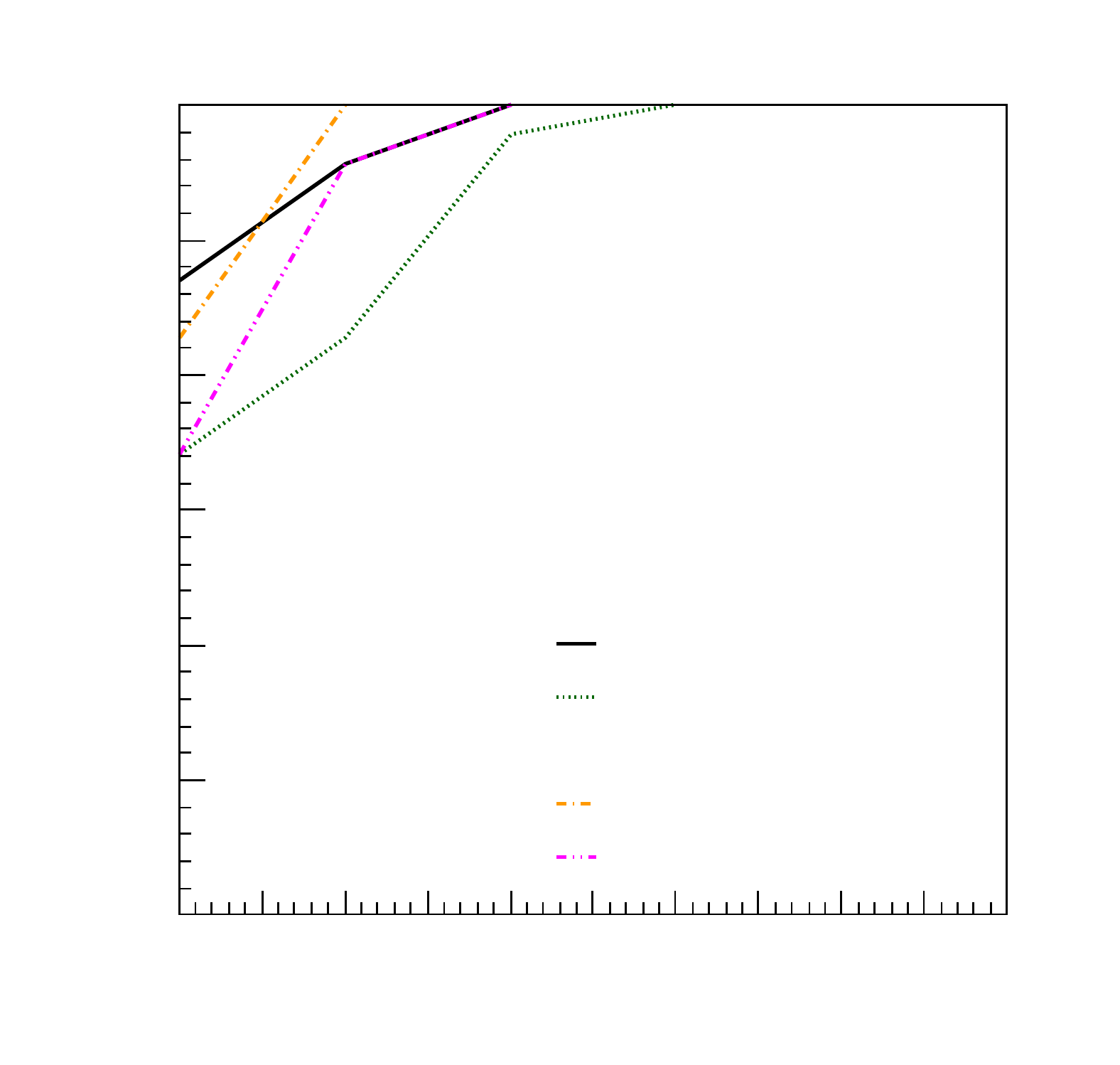}  
  & \sidecaption \svgend
\end{subfigures}

The effect of including systematic uncertainties in the limits is seen
for the extended likelihood function by comparing the green and black
lines of \fig{Hig:Sm.Limits.Validate}. As expected, the limit
calculated without uncertainties is lower than the limit with
uncertainties. Additionally, for larger $\m_\h$ the difference between
the two limits decreases, as the limit becomes more influenced by the
shape of the mass distribution which has a smaller uncertainty than
the relative scaling of the background components. The introduction of
the systematic uncertainties can also be seen in the magenta and
orange lines, calculated with the simple likelihood function, but now
the difference between the two limits is not dependent upon
$\m_\h$. The pulls of the nuisance parameters, $(\theta - \epsilon) /
\delta$, where $\theta$ is the nuisance parameter, $\epsilon$ is its
mean, and $\delta$ is its deviation, are given in
\fig{Hig:Nuisance.Mu2} for ${\m_\h = 90~\gev}$ and $\mu = \hat{\mu}$
when calculated using the extended likelihood function. Here the
nuisance parameters for each of the five event categories as well as
the combined result are given.

\begin{subfigures}{2}{Pulls of the nuisance parameters for ${\m_h =
      90~\gev}$ and ${\mu = \hat{\mu}}$ calculated using the extended
    likelihood for the five event categories \mumu (circles), \mue
    (square), \emu (diamond), \muh (up triangle), and \eh (down
    triangle), as well as the combined results. The colours of the
    points correspond to the colour of the associated background as
    given in \fig{Hig:Mass} for which the nuisance parameter is
    assigned. The black points are for mass shape uncertainty nuisance
    parameters.}
    \svgbeg 
    \svg[1]{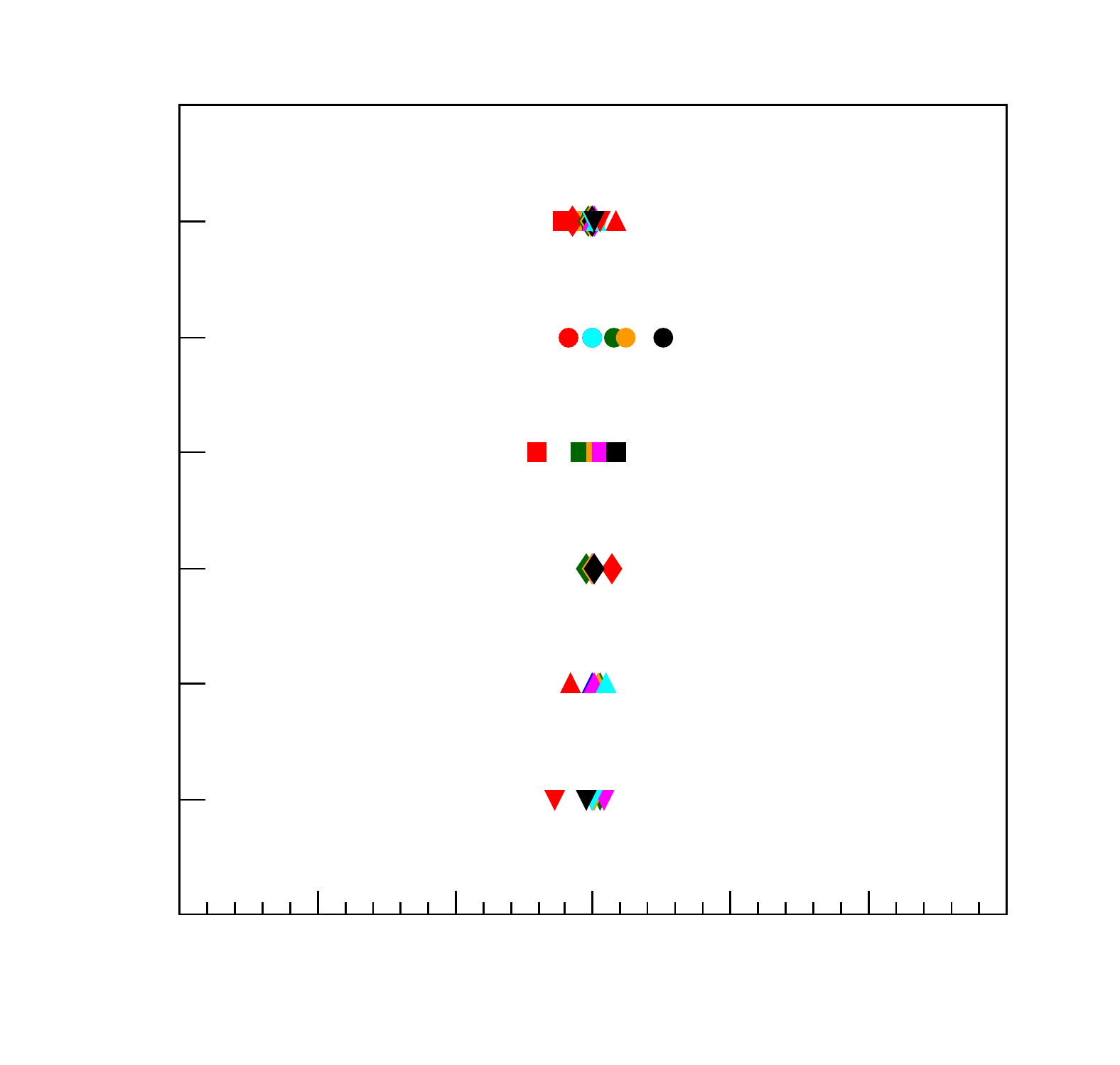} & \sidecaption \svgend
\end{subfigures}

The effect of using the extended likelihood function, rather than the
simple likelihood function can be seen by comparing the black line
with the orange line or green line with the magenta line in
\fig{Hig:Sm.Limits.Validate}. The limits calculated using the extended
likelihood function are consistently lower than the limits calculated
using the simple likelihood function, demonstrating the additional
exclusion power gained from utilising the mass shape information. The
difference between these two sets of limits is particularly pronounced
at large $m_\h$ where the mass distribution expected for the signal
differs significantly from the background mass distribution.

% This check is done using the modified flag in Exclude.cpp.
However, in \fig{Hig:H2TauTau2MuE.Sm.Limits.Validate}, the extended
likelihood function limits are consistently worse than the simple
likelihood function limits for the \whb mass range of ${90 \leq \m_\h
  \leq 145~\gev}$. This behaviour can be attributed to an observed
invariant mass shape, given in
\fig{Hig:H2TauTau2MuE.Mssm.Shape.Mass125.Tanb60}, that is more
consistent with the signal than with the background within this
$\m_\h$ range. To check this, the data has been modified to be more
consistent with the expected background in
\fig{Hig:H2TauTau2MuE.Mssm.Shape.Mass125.Tanb60.Modified} which
results in the limits of
\fig{Hig:H2TauTau2MuE.Sm.Limits.Validate.Modified}, where the limits
calculated using the extended likelihood function are now found to
outperform the limits calculated using the simple likelihood function.

\begin{subfigures}[t]{2}{
    \subfig{H2TauTau2MuE.Mssm.Shape.Mass125.Tanb60.Modified}~Invariant
    mass distribution of
    \fig{Hig:H2TauTau2MuE.Mssm.Shape.Mass125.Tanb60} modified to
    produce an observed distribution more consistent with the
    background only
    hypothesis. \subfig{H2TauTau2MuE.Sm.Limits.Validate.Modified}~Validation
    limits for this modified data, equivalent to
    \fig{Hig:H2TauTau2MuE.Sm.Limits.Validate.Modified}, where the
    limits calculated using the extended likelihood function now
    outperform the limits calculated using the simple likelihood
    function.}
  \svgbeg
  \svg{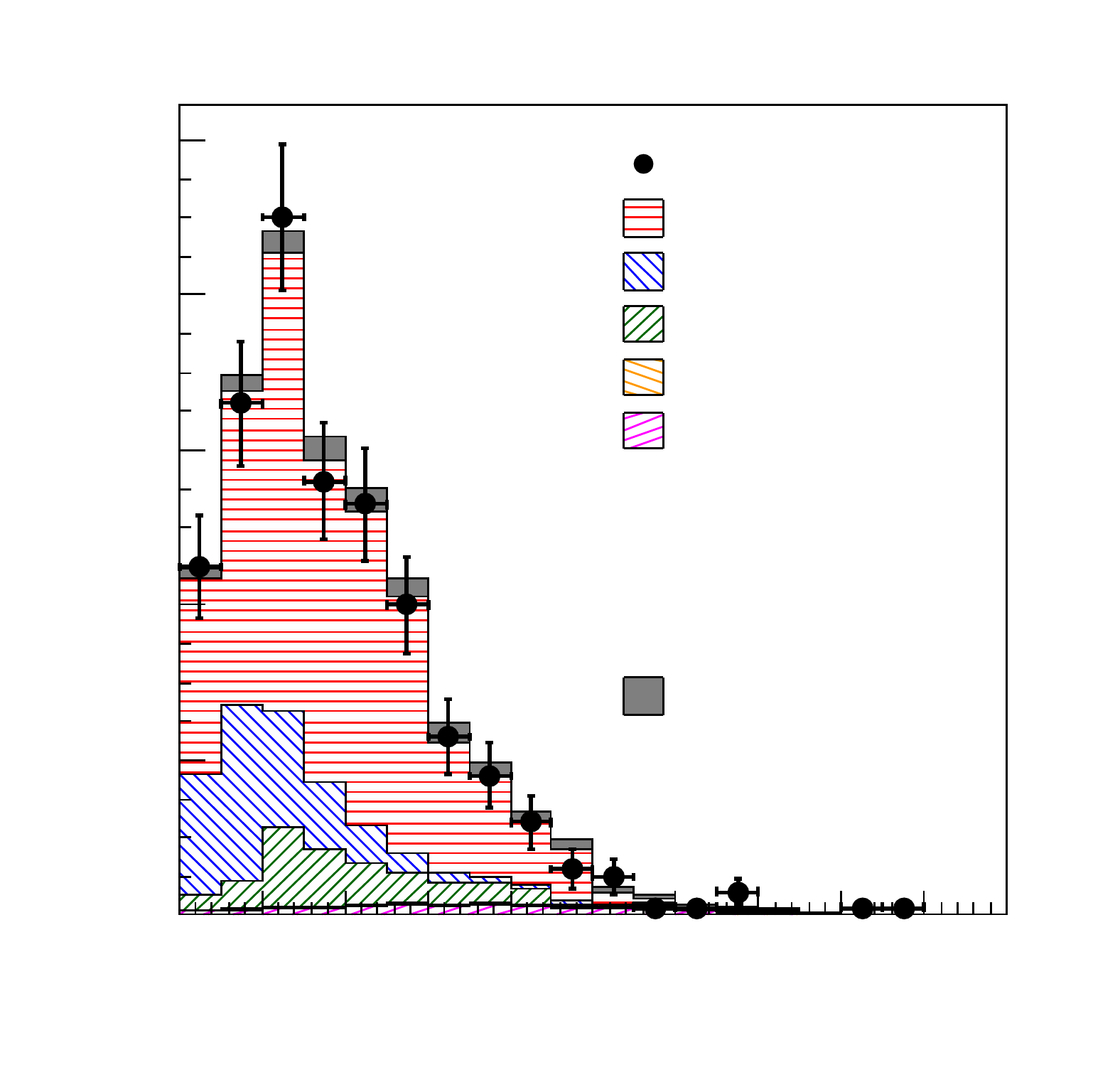} 
  & \svg{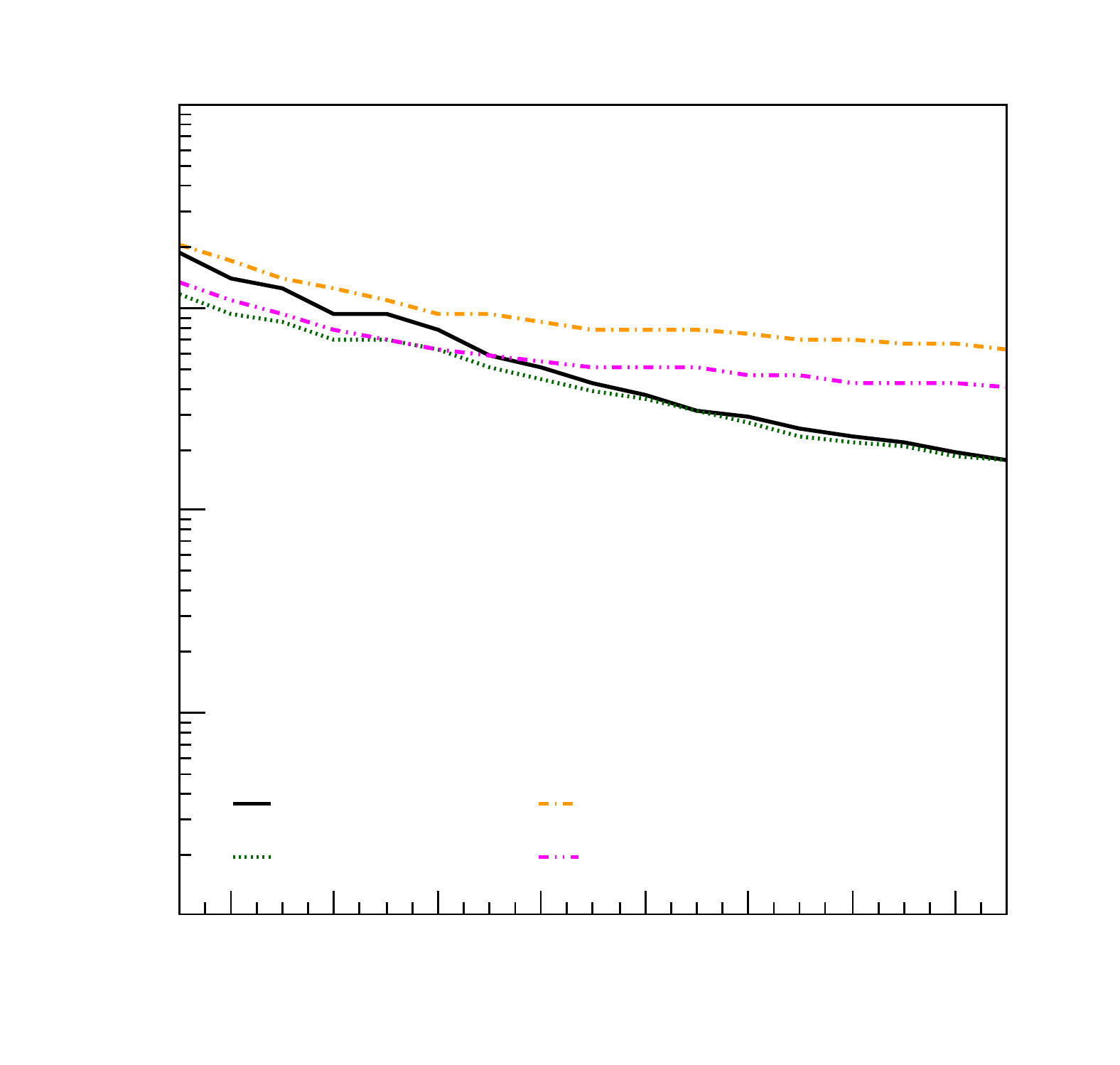} \svgend
\end{subfigures}

The asymptotic approximation made in calculating $q_\nu$ can be
checked by using the Monte Carlo technique, where a large number of
pseudo-experiments are randomly generated and a \pdf for $q_\nu$ is
built. Because this process is time consuming, checks were made for
only a few limit points. The limits calculated at these points using
the Monte Carlo technique are consistent with the limits calculated
assuming the asymptotic approximation.

\newsubsection{Final Limits}{Lim}

In \figs{Hig:Sm.Limits.Final} and \ref{fig:Hig:Mssm.Limits.Final}, the
final limit is plotted with the expected limits for a median
experiment, assuming the background only hypothesis. The dashed red
line provides the central expected limit. The dark blue band is the
$\pm1\sigma$ range for the median experiment, {\it i.e.} $68\%$ of
experiments, assuming the background only hypothesis, are expected to
produce a limit within this band. Similarly, the light blue band
provides the $\pm2\sigma$ range about the limits expected from a
median experiment. The expected limit is calculated using $q_\nu$
evaluated with the Asimov dataset of \sec{Hig:Lik}, which for the
extended likelihood function is given by \equ{Hig:Statistics.Asimov}.

\begin{subfigures}[p]{2}{Upper limits on neutral
    \whbs production as function of the \whb mass for
    the \subfig{H2TauTau2MuMu.Sm.Limits.Final}~\mumu,
    \subfig{H2TauTau2MuE.Sm.Limits.Final}~\mue,
    \subfig{H2TauTau2EMu.Sm.Limits.Final}~\emu,
    \subfig{H2TauTau2MuPi.Sm.Limits.Final}~\muh, and
    \subfig{H2TauTau2EPi.Sm.Limits.Final}~\eh categories. The
    observed limit using $q_\nu$ and $\lh_e$ (black) is compared
    to the expected limit (red, blue
    bands).\labelfig{Sm.Limits.Final}}
  \svgbeg 
  \svg{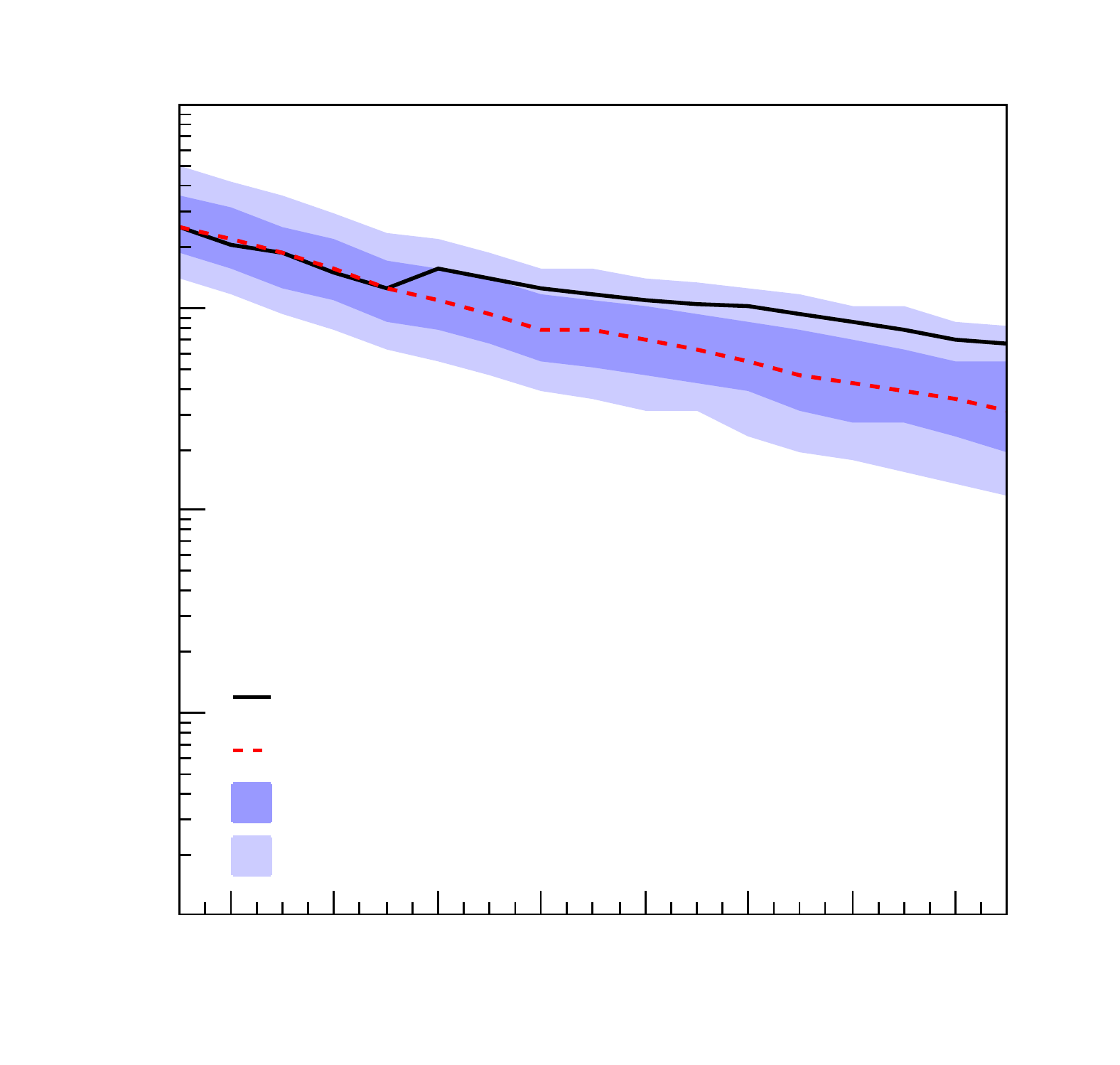} 
  & \svg{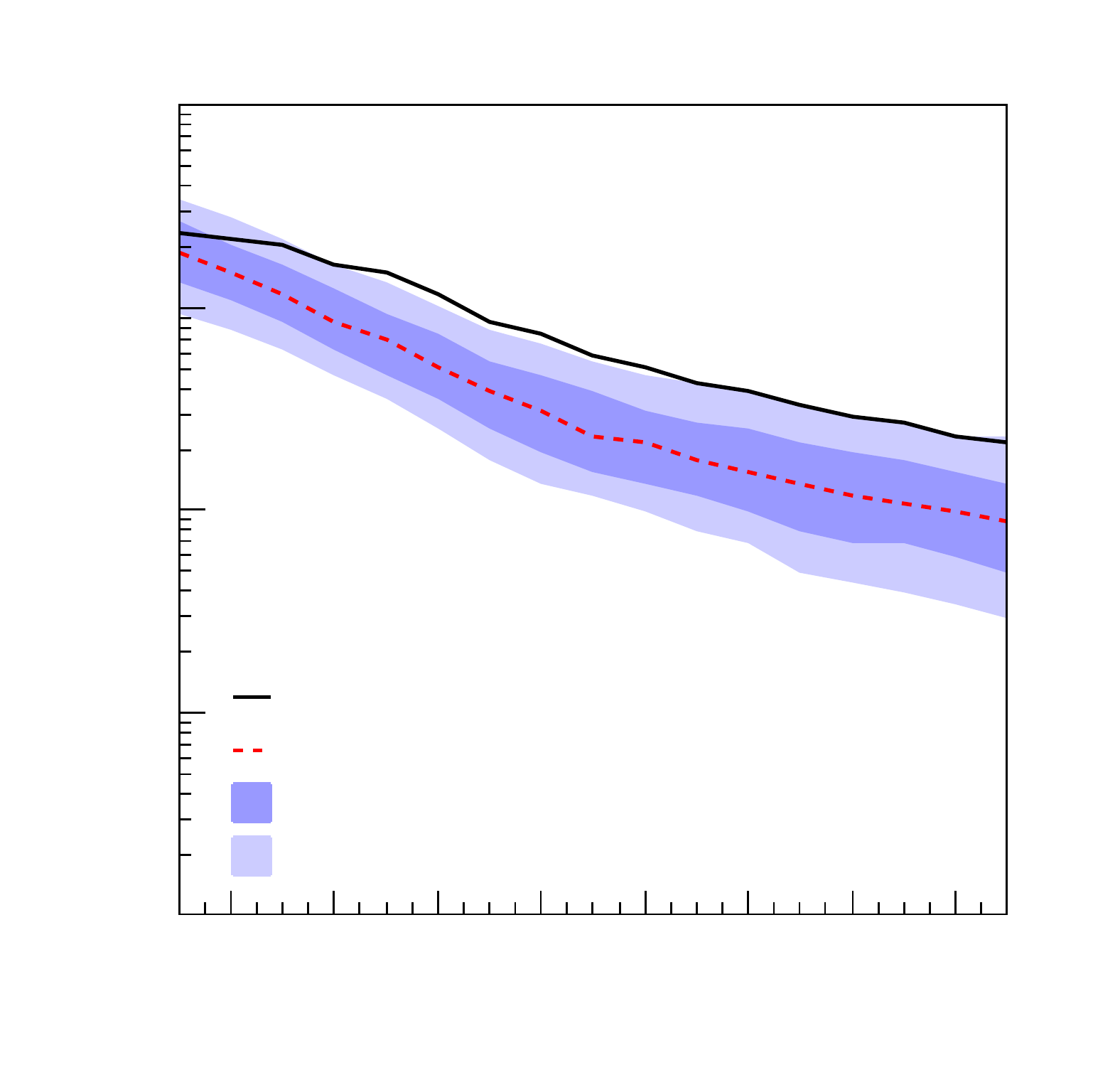} \svgsep
  \svg{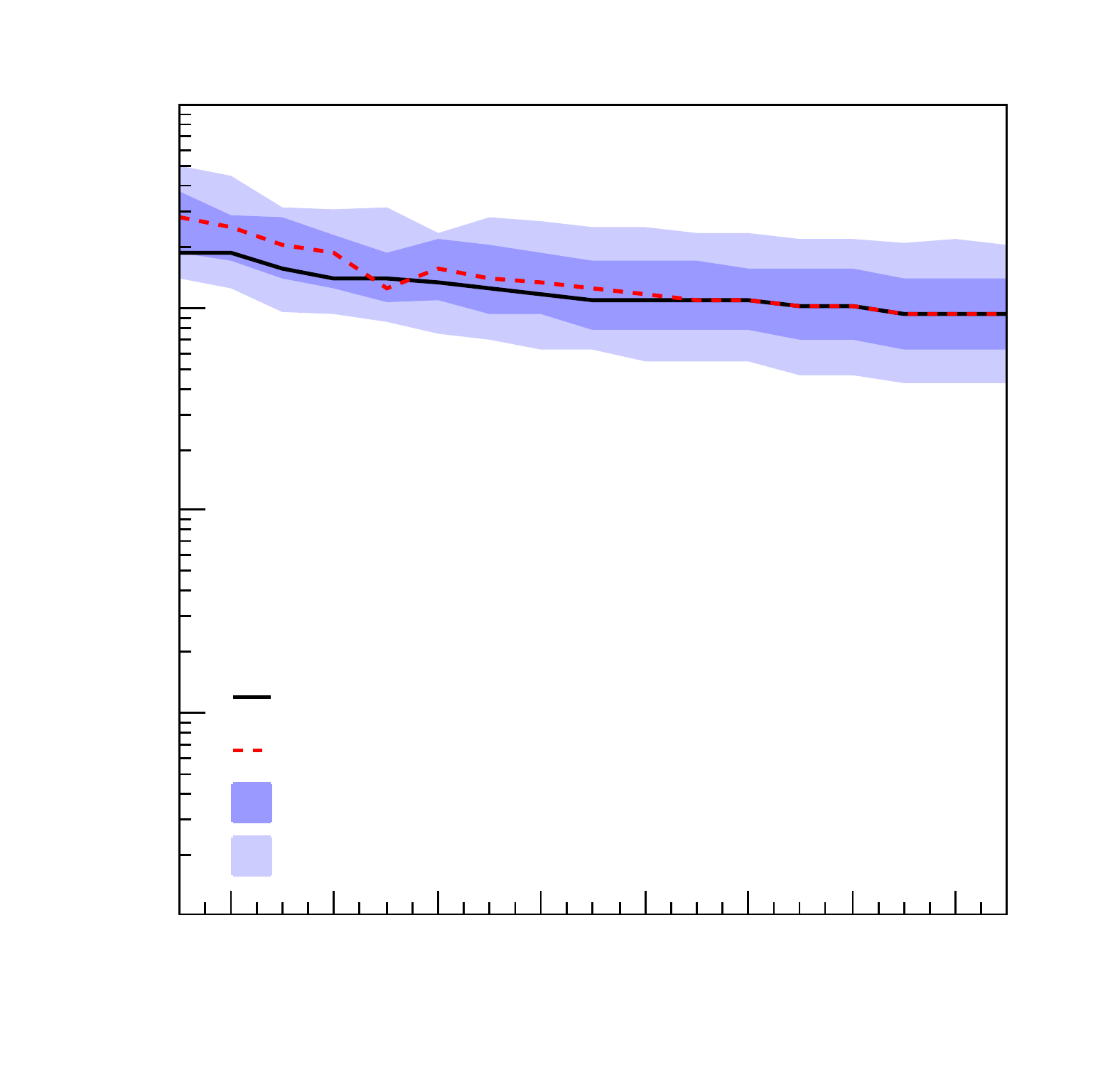} 
  & \svg{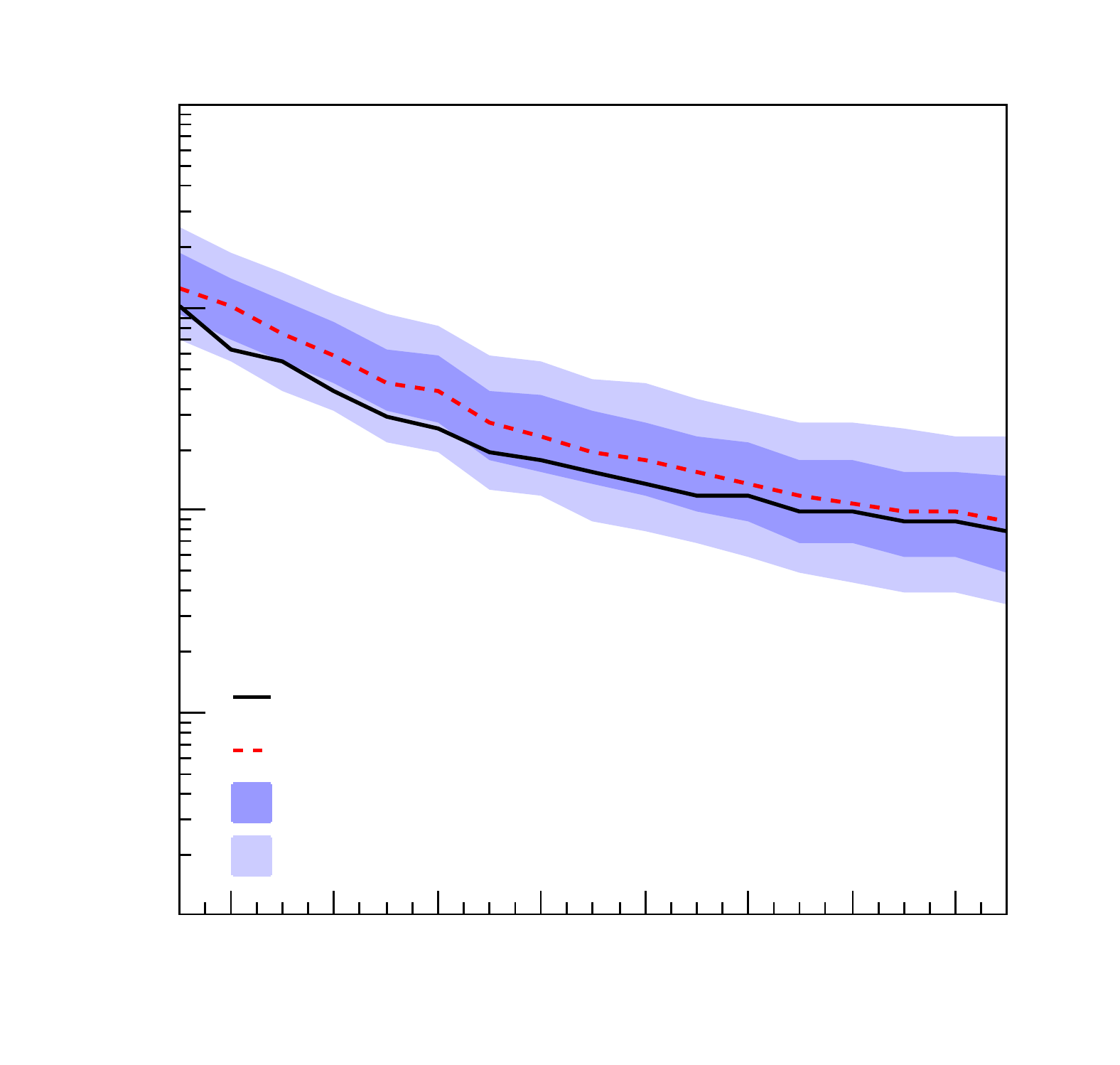} \svgsep
  \svg{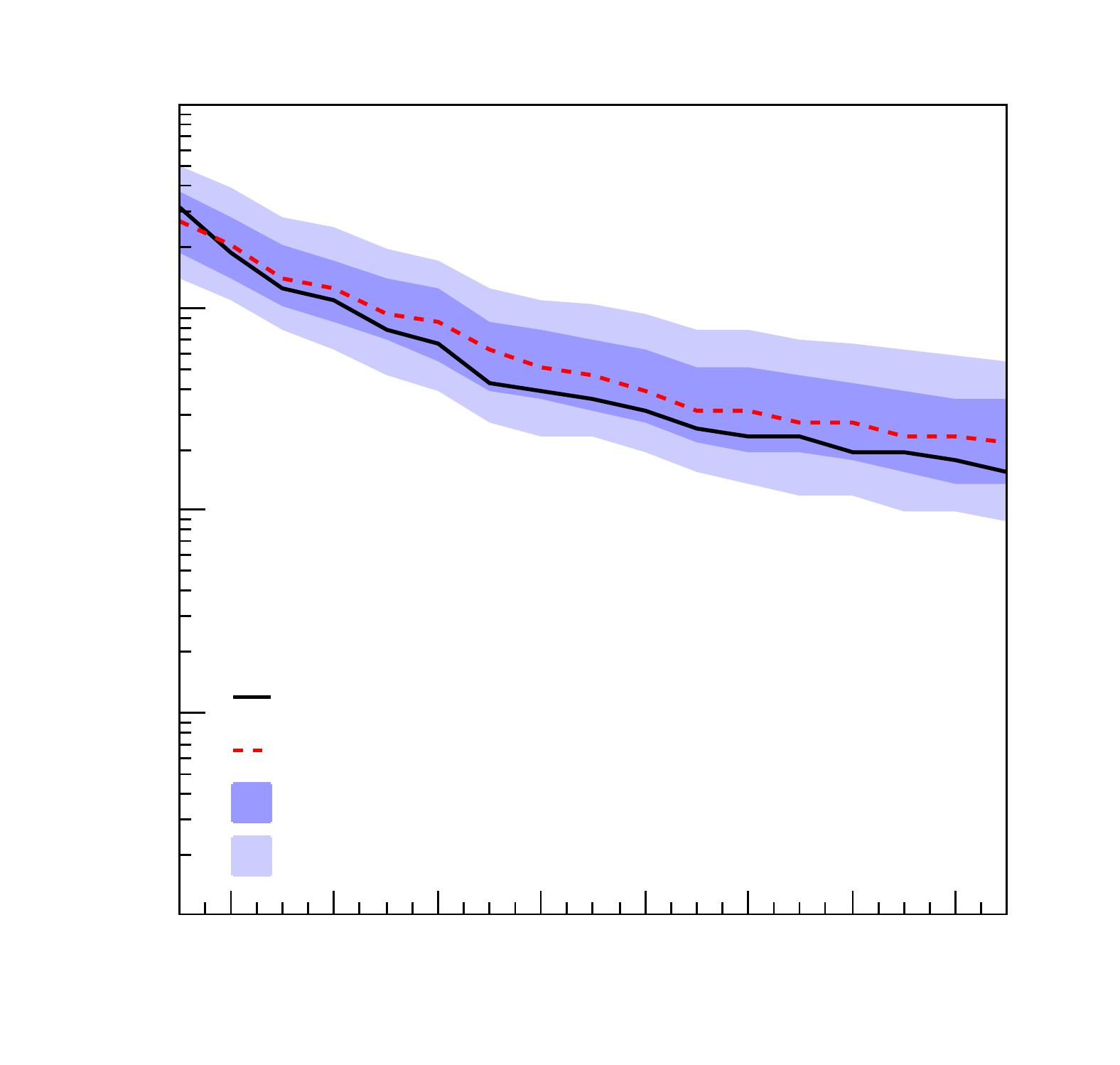} & \sidecaption \svgend
\end{subfigures}

\begin{subfigures}[p]{2}{Upper limits on \tanb as a function of the
    \cp-odd \whb mass for the
    \subfig{H2TauTau2MuMu.Sm.Limits.Final}~\mumu,
    \subfig{H2TauTau2MuE.Sm.Limits.Final}~\mue,
    \subfig{H2TauTau2EMu.Sm.Limits.Final}~\emu,
    \subfig{H2TauTau2MuPi.Sm.Limits.Final}~\muh, and
    \subfig{H2TauTau2EPi.Sm.Limits.Final}~\eh categories. The observed
    limit using $q_\nu$ and $\lh_e$ (black) is compared to the
    expected limit (red, blue bands).\labelfig{Mssm.Limits.Final}}
  \svgbeg
  \svg{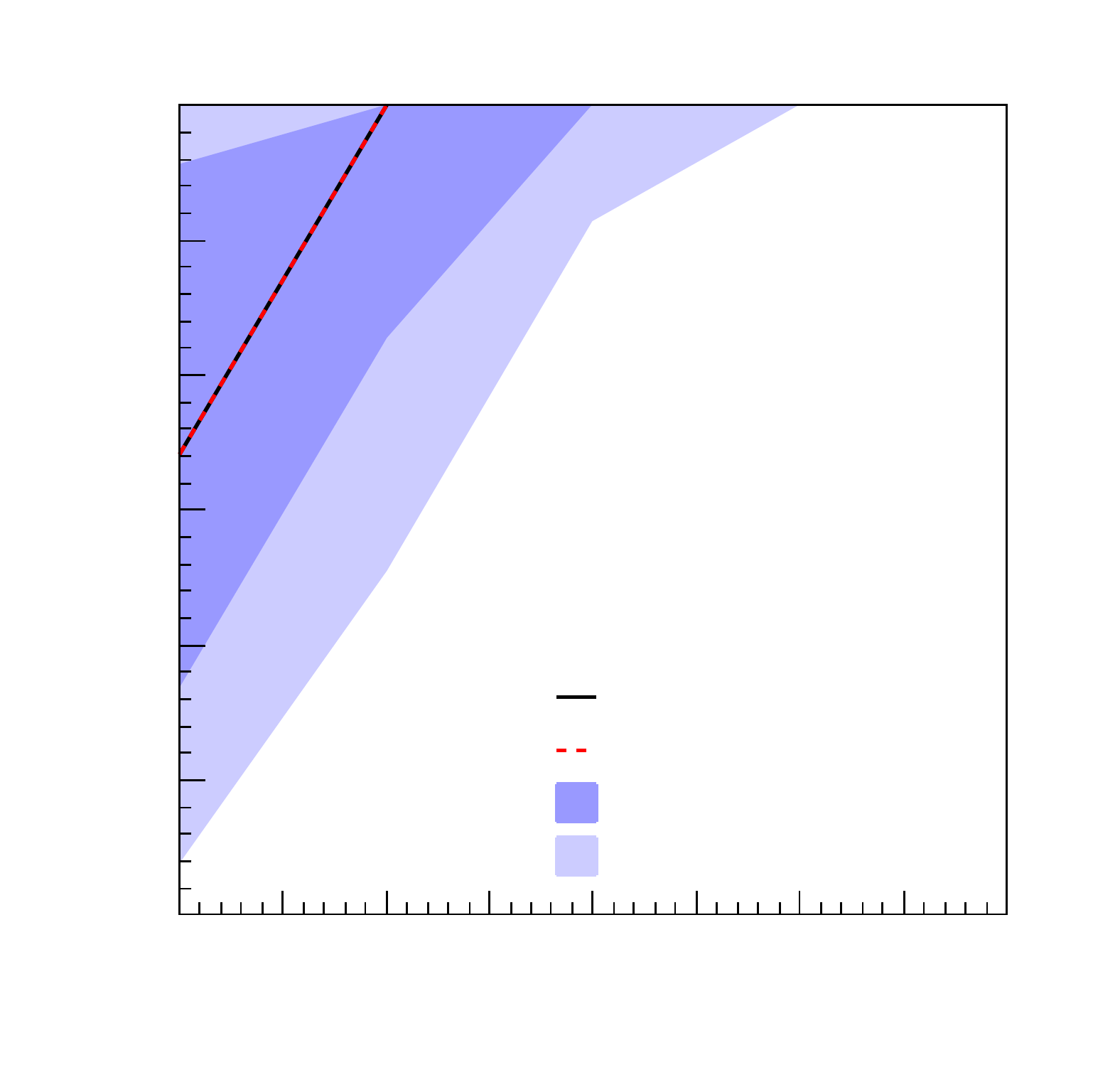} 
  & \svg{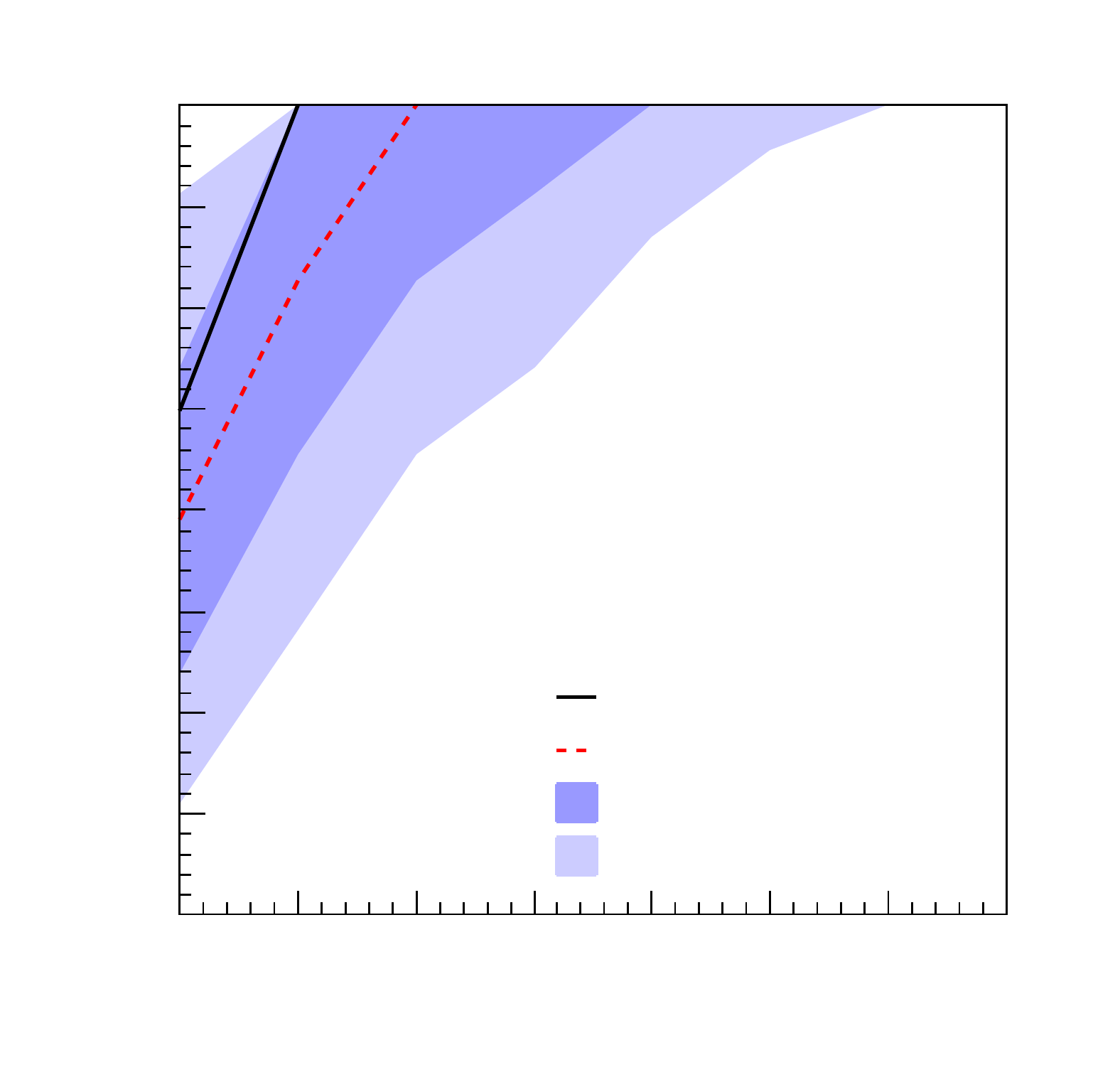} \svgsep
  \svg{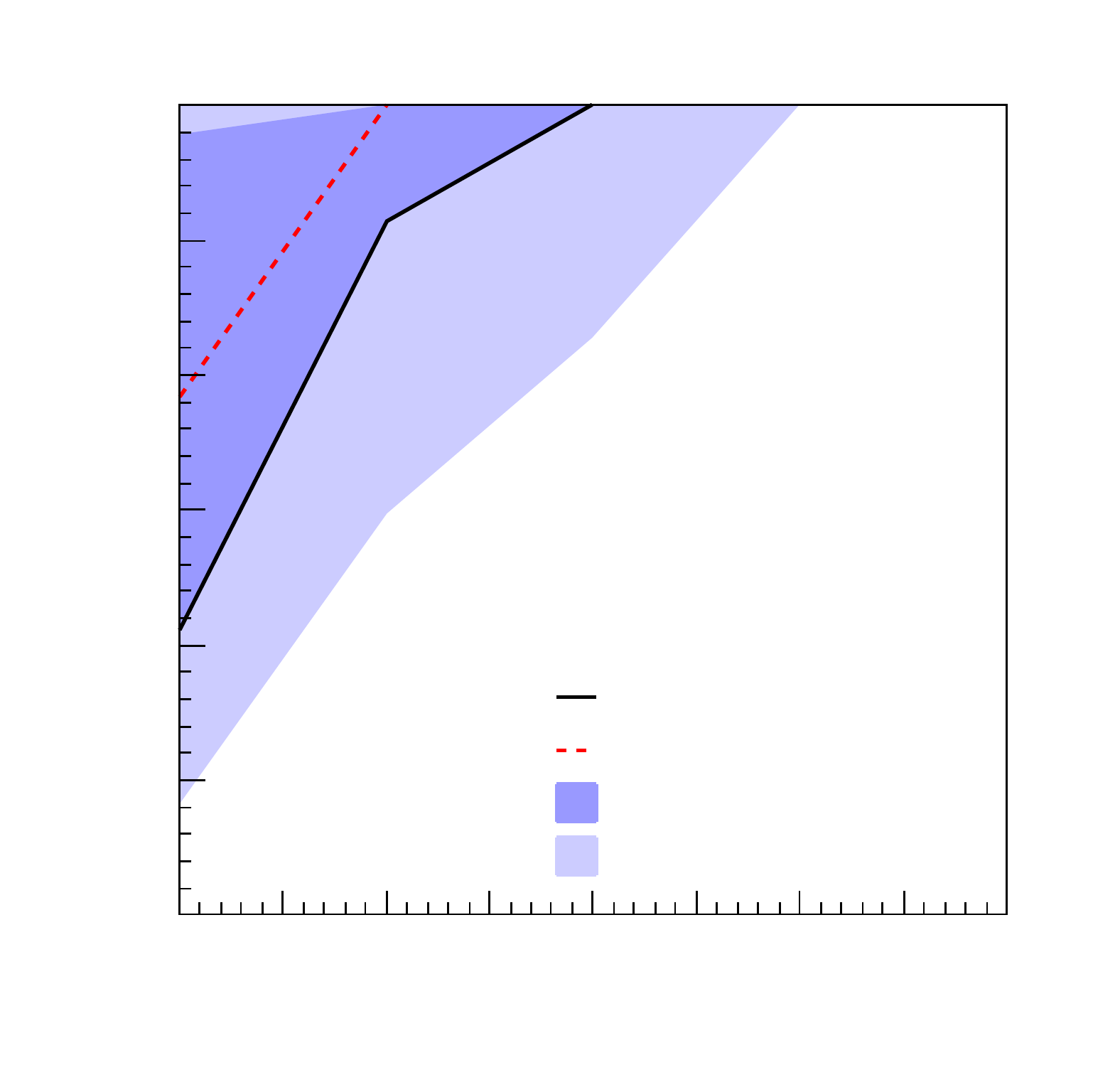}  
  & \svg{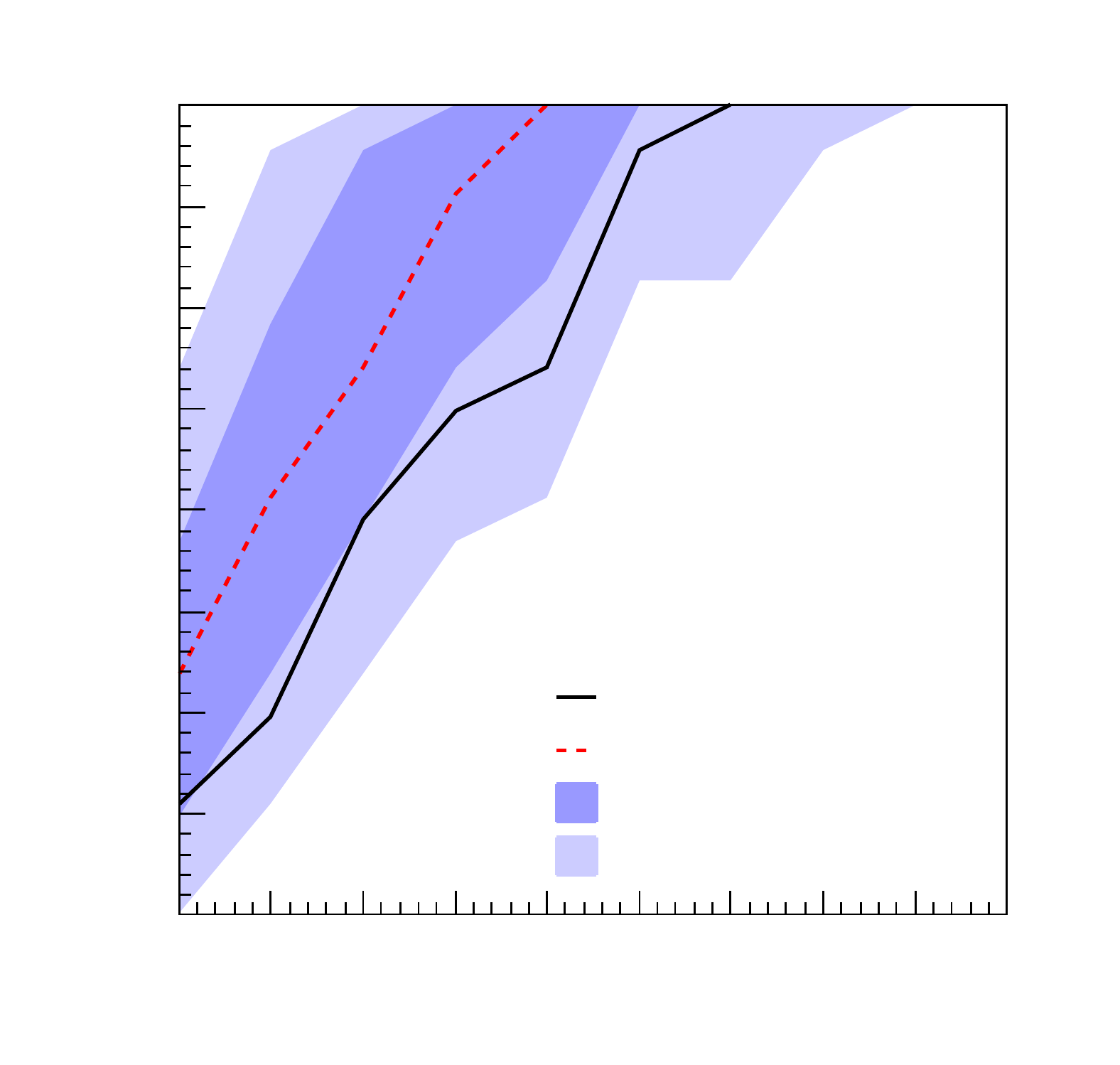} \svgsep
  \svg{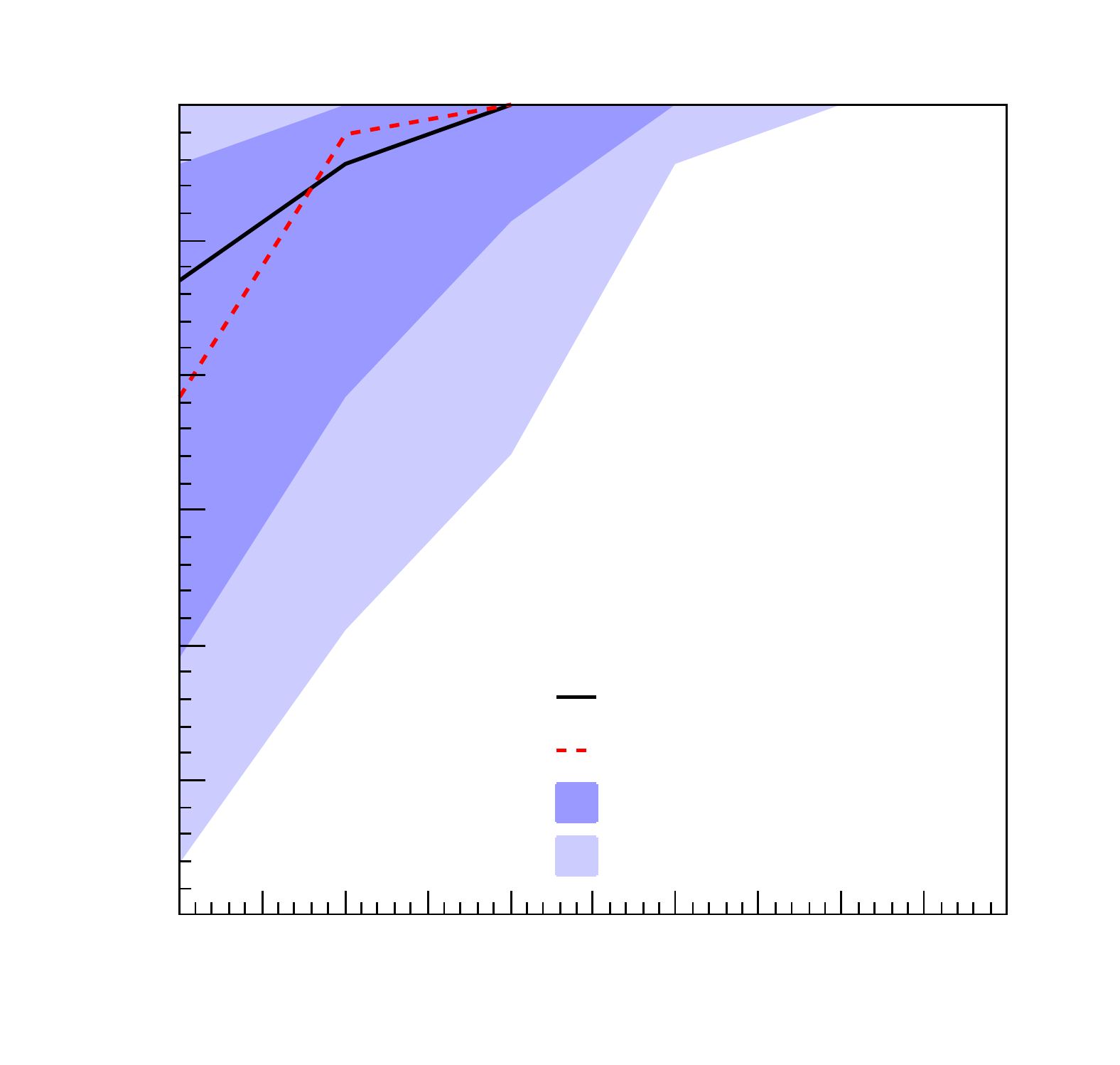}  
  & \sidecaption \svgend
\end{subfigures}

The expected limits can be compared to the observed limits in either
\fig{Hig:Sm.Limits.Final} or \fig{Hig:Mssm.Limits.Final}, but just as
for the validation limits, the results are most easily interpreted for
the model independent limits. However, for both
\figs{Hig:Sm.Limits.Final} and \ref{fig:Hig:Mssm.Limits.Final} the
observed limits are consistent with the expected limits for the
background only hypothesis and fall within the expected $\pm2\sigma$
band for most of the five categories.

In the \mue category, given by \fig{Hig:H2TauTau2MuE.Sm.Limits.Final},
a slight excess of events with respect to the background only
hypothesis is observed, resulting in a higher than expected upper
limit, slightly above the $\pm2\sigma$ band at large $\m_\h$. The
excess of events is particularly visible in the ${55-60~\gev}$
invariant mass bin of
\fig{Hig:H2TauTau2MuE.Mssm.Shape.Mass125.Tanb60}, resulting in the
difference between the expected limits and the observed limits
increasing for larger values of $\m_\h$, as the mass distribution for
large $\m_\h$ becomes more consistent with this excess.

The \mumu category of \fig{Hig:H2TauTau2MuMu.Sm.Limits.Final} also has
an observed number of events larger than the background only
hypothesis. This excess is more evenly spread across the mass
distribution of \fig{Hig:H2TauTau2MuMu.Mssm.Shape.Mass125.Tanb60} than
for the \mue category, although a small excess is observed in the two
invariant mass bins from ${100-110~\gev}$ resulting in the difference
between the observed and expected limits increasing slightly for large
$\m_\h$. The \eh category has an overall excess of observed events,
but a slight deficiency in events for large invariant masses,
resulting in an observed limit above the expected limit for small
$\m_\h$ and below the expected limit for large $\m_\h$. The \emu and
\muh categories both have a deficiency in the observed number of
events with respect to the background only hypothesis, yielding
observed limits slightly lower than the expected limits.

The five event categories are combined to produce the model
independent limits and the \mssm limits of
\fig{Hig:Limits.Compare}. The limits are calculated using the same
methods described for the individual limits of
\figs{Hig:Sm.Limits.Final} and \ref{fig:Hig:Mssm.Limits.Final}. In
\fig{Hig:H2TauTau.Sm.Limits.Compare} the expected cross-section
$\sigma_{\dip \to \hH \to \ditau}$ from the \sm \whb is given by the
dotted black line, with the theoretical uncertainty given by the grey
band. A neutral \whb mass of $125~\gev$ is indicated by the vertical
black line. The ratio of this limit to the \sm expectation is given in
\fig{Hig:H2TauTau.Sm.Limits.Ratio} for further comparison. As can be
seen, the limit from data is nearly two orders of magnitude larger
than the expected \sm cross-section at lower \whb masses. In
\fig{Hig:H2TauTau.Mssm.Limits.Compare} the \lhcb limit from this
analysis is compared with limits from \atlas, \cms, and \dlep. The
$36~\ipb$ and $4.7~\ifb$ \atlas limits from \rfrs{atlas.11.2} and
\cite{atlas.13.1} correspond to analyses performed using the $2010$
and $2011$ \atlas datasets respectively. Similarly the $36~\ipb$ and
$4.6~\ifb$ \cms limits from \rfrs{cms.11.2} and \cite{cms.12.1}
correspond to the $2010$ and $2011$ \cms datasets. The \dlep limit is
a combined lower limit for all \dlep data from \rfr{lep.06.2}. The
\lhcb limit of this analysis is competitive with the \atlas and \cms
limits using $2010$ datasets, but not with the \atlas and \cms limits
using $2011$ datasets. While \atlas and \cms maintain or even gain
sensitivity for large $\m_\hAz$, the results of this chapter lose
sensitivity due to the decreased acceptance of ${\hAz/\hHz \to
  \ditau}$ events where both \wtls are produced within \lhcb.

\begin{subfigures}{2}{Upper limits at $95\%$ \cls using the five
    combined event categories on \subfig{H2TauTau.Sm.Limits.Compare}
    neutral \whb production into \wtl pairs,
    \subfig{H2TauTau.Sm.Limits.Ratio} the ratio of this limit with SM
    expectation, and \subfig{H2TauTau.Mssm.Limits.Compare} \tanb for
    the \mssm as a function of the \cp-odd \whb mass. The observed
    limit (solid black) is compared to the expected limit (dashed red,
    blue bands). The model independent limit is compared to the
    expected \sm cross-section (dotted black), where a neutral \whb
    mass of $125~\gev$ is indicated by the vertical black line, and
    the \mssm limit is compared to limits from \atlas, \cms, and
    \dlep.\labelfig{Limits.Compare}}
  \svgbeg
  \svg{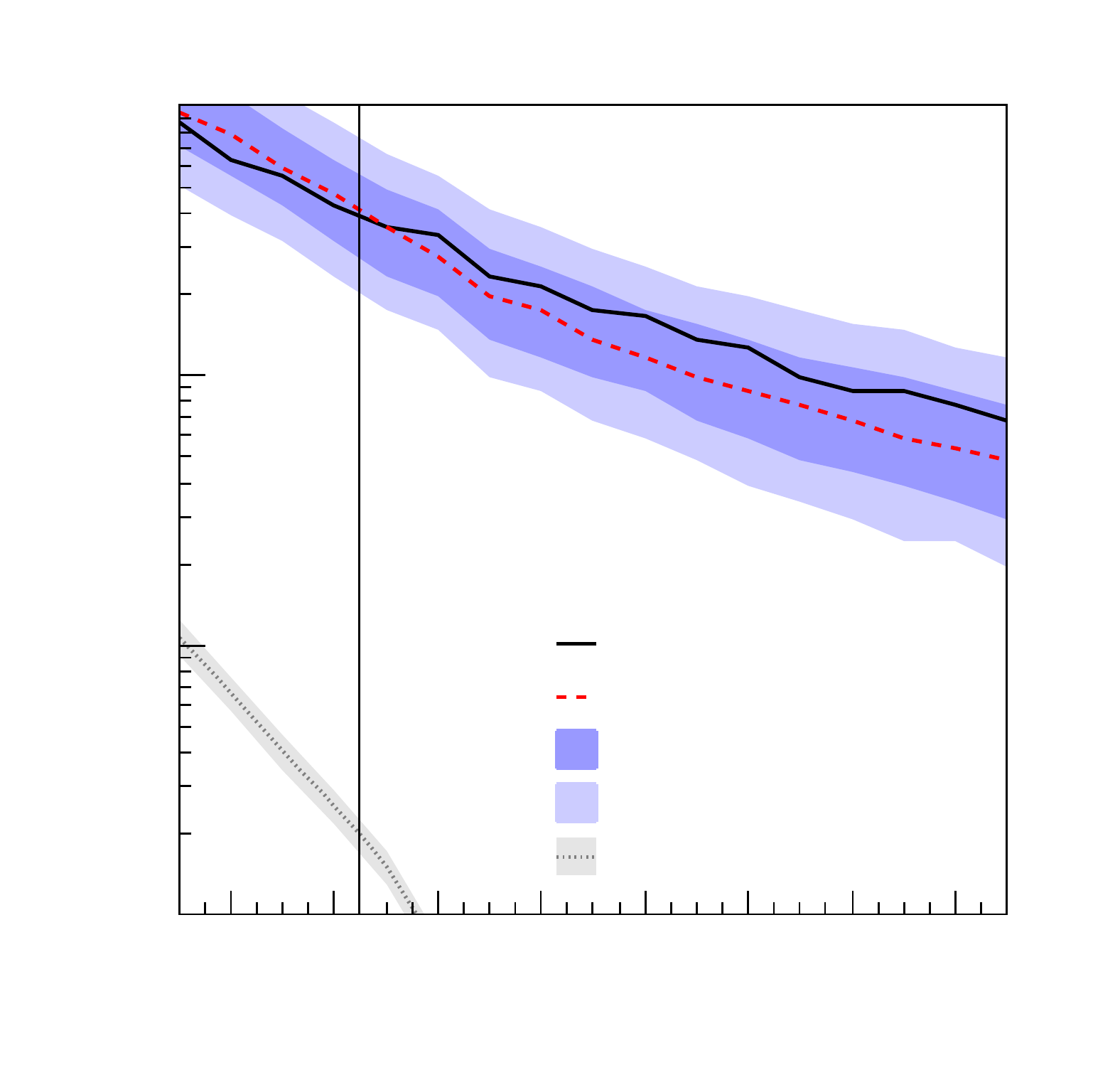}   & \svg{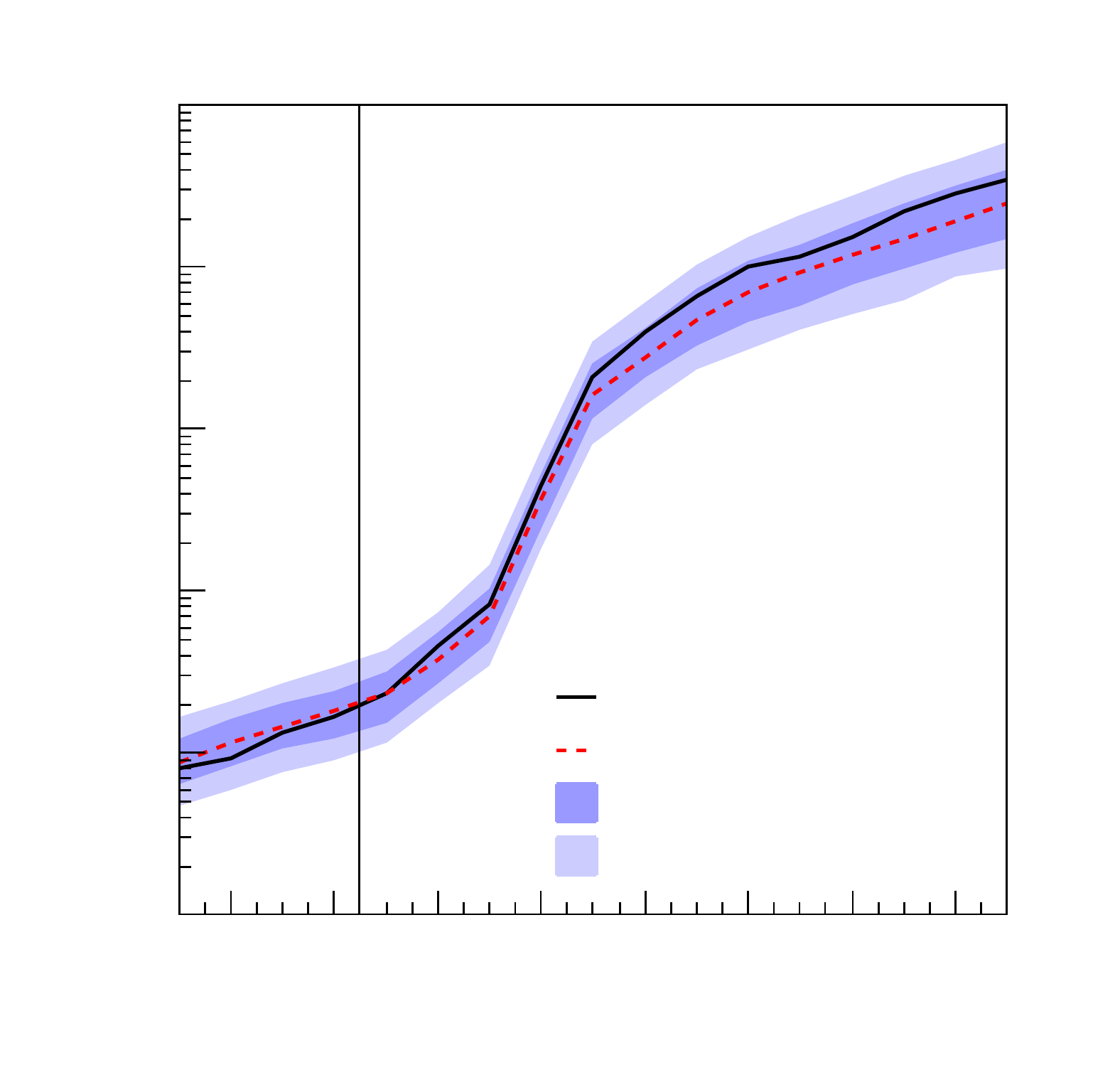} \svgsep
  \svg{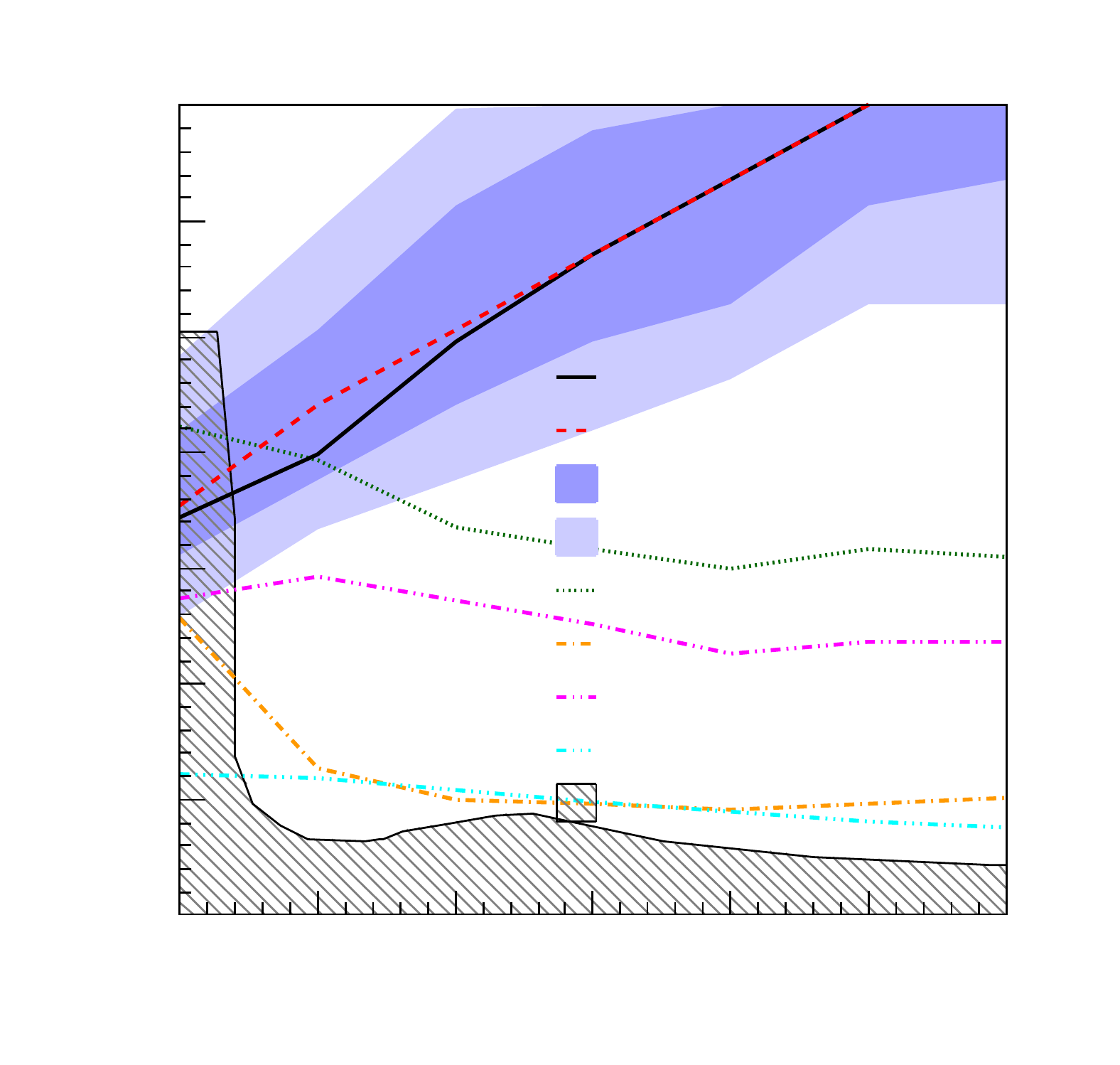} & \sidecaption \svgend
\end{subfigures}

\newchapter{Conclusion}{Con}

Within this thesis a full analysis, from theory to experimental result,
has been presented. The results of \chp{Tau} demonstrate how \wtl
decays may be used to differentiate between their production
mechanisms. Future measurements of the \whb, in conjunction with \wtl
decays from \pythia{8}, could verify both the spin and \cp of the
\whb. A similar analysis can be applied to ${\w \to \nu_\tau \tau}$
measurements to search for charged \whbs, supersymmetric particle
decays, and other new physics which produce unique helicity
correlations in the decays of \wtls. The results of \chp{Tau} can also
be used in lower energy particle physics. The models implemented in
\pythia{8} can be tuned to data from \wtl factories such as BaBar and
provide insights into \qcd at energies where perturbative calculations
fail. These tunes could also be used to improve meson resonance decay
models.

The work of \chp{Tau} can be expanded with the addition of a variety
of features to \pythia{8}. Currently the decay widths for the \wtl
decay channels are not calculated, as constants of proportionality
have been omitted from the implemented hadronic currents. Including
these constants of proportionality, as well as an extension of the
code to perform the decay width integration, can be implemented so
that \pythia{8} could be used to calculate \wtl decay widths in a
fashion similar to its cross-section calculations. Additionally, the
parameters for the hadronic currents should be migrated to user
settings so they can easily be modified by the user without the need
to recompile source code. With added decay width calculations and
modifiable parameters, the \pythia{8} \wtl decay machinery could more
easily be used to perform phenomenological tunes of \wtl decays
similar to the three pion CLEO fit. Common preset fits could also be
included in the \pythia{8} user settings, allowing users to quickly
and simply switch between parameter sets for a \wtl decay channel.

The forward cross-section for the production of \wzbs decaying into
\wtls was measured to a precision of $7\%$ in \chp{Zed}. This
measurement, as well as the individual measurements for each final
state category, are consistent with the expected cross-section from
the \sm. The final cross-section can be used in \PDF fits to constrain
the proton \PDF at both high and low momentum fractions. The ratios of
the ${\z \to \ditau}$ cross-section with the ${\z \to \die}$ and ${\z
  \to \dimu}$ cross-sections from \lhcb, test lepton universality in
the forward region to a precision of $7\%$ and $9\%$ respectively. No
excess was observed in either of these ratios, testing the absence of
additional heavy particles decaying into \wtls, as predicted by the
\sm.

The cross-section measurement of \chp{Zed} should be extended to both
$2012$ and future data from \lhcb. In the process, improvements can be
made to the analysis. One of the more confusing aspects of the
analysis is the difference between the \mue and \emu categories. The
reason for this differentiation is purely historical, and in an
updated version of this analysis, the two categories should be
simplified into a single \mue category. A similar change could also be
made in the handling of the efficiencies. Here, the efficiencies are
split into a reconstruction efficiency and a selection efficiency but
this division is somewhat artificial, and is not necessary.

Currently, the ${\z \to \ditau}$ cross-section measurement is
dominated by statistical uncertainties, but in the future the
systematic uncertainties will need to be reduced. One of the largest
sources of uncertainty for the event categories containing an electron
is due to the electron track finding efficiency which is determined
from data using a tag-and-probe method. Currently, the probe is a high
energy \ecal cluster. However, a two part approach could be taken
where a probe of a \velo track and \ecal cluster is used, followed by
a probe of a downstream track and an \ecal cluster. These two probes
could then be used to calculate the \velo and downstream electron
track finding efficiencies separately, potentially resulting in a
reduced systematic uncertainty on the electron track finding
efficiency.

The selection requirements for the ${\z \to \ditau}$ cross-section
measurement of \chp{Zed} were optimised manually, rather than with
multivariate techniques. Further gains in reducing the background with
respect to the signal could be accomplished by optimising the
selection requirements using multivariate techniques such as boosted
decision trees or artificial neural nets. Additionally, final state
signals with multi-pronged \wtl decays, rather than just the
single-pronged \wtl decays of this analysis, could be included.

The limits of \chp{Hig} exclude the production of neutral \whbs
decaying into \wtls within the forward region with cross-sections
larger than $8.6$ to $0.7~\pb$ over the \whb mass range $90$ to
$250~\gev$. This result is two orders of magnitude larger than the
expected cross-section from the \sm, and so no conclusions can be
drawn about the \sm \whb. The limits for the \mssm exclude a \tanb
greater than $34$ for a \cp-odd \whb mass of $90~\gev$ to a \tanb
greater than $70$ for a \cp-odd \whb mass of $140~\gev$. These limits,
in conjunction with the limits from \atlas and \cms, are rapidly
reducing the remaining parameter space for the \mssm which has not yet
been excluded.

Unique forward measurements of the \whb in the future, using the \lhcb
detector, will need to combine a variety of \whb decay channels and
cannot rely solely upon \wtl pair decays of the \whb. Within the
analysis of \chp{Hig} the simulated \whb signal samples were generated
using \pythia{6}. The simulation framework of \lhcb, \gauss, needs to
be expanded to allow the simulation of samples generated with
higher-order hard matrix elements, such as those produced using the
\powheg method. Particularly, this simulation will be critical for
associated vector boson analyses where well modelled \w and \wzb
simulation samples with associated hard jets will be needed.

The statistical analysis of \chp{Hig} could also be improved with a
refinement of the nuisance parameters, specifically with the addition
of correlated systematic uncertainties. Currently there is no
dedicated statistics group for \whb analyses within the \lhcb
collaboration. Upcoming \whb analyses could benefit from such a group,
which would provide a consistent statistical treatment for all \whb
analyses produced by \lhcb.

Future analyses at \lhcb, based on results from this thesis, should
provide complementary measurements of the properties for the
Higgs-like boson discovered by \atlas and \cms. The forward
capabilities of \lhcb will allow models such as the diffractive
production of \whbs from intrinsic heavy flavours within the
proton~\cite{brodsky.06.1} to be tested. The selection for \wtls
developed in this thesis could also be used to select \wwbs decaying
into \wtls and search for heavy charged particles such as the charged
\mssm \whb.

\phantomsection\addcontentsline{toc}{chapter}{Bibliography}
\newboolean{articletitles}
\setboolean{articletitles}{true}
\bibliographystyle{lhcb}
\bibliography{Latex/bib}
\appendix
\phantomsection\addcontentsline{toc}{chapter}{Appendices}
\newappendix{Theory}{Tvr}

In this appendix, additional material for the theoretical review of
\chp{Thr} is provided. In \sap{Tvr:Not} the notational conventions
used are outlined. In \sap{Tvr:Ewk} the electroweak Lagrangian is
given and in \sap{Tvr:Mhm} the minimal Higgs mechanism Lagrangian is
given.

\newsection{Notation}{Not}

The momentum four-vector is given by ${q = (E,q_x,q_y,q_z)}$ and the
spatial component by $\vec{q} = (q_x,q_y,q_z)$. The Einstein notation
$q^\mu$ and $q_\mu$ indicate the four-vectors,
\begin{equation}
  q^\mu = \begin{pmatrix} E \\ q_x \\ q_y \\ q_z \\ \end{pmatrix}\equcomma
  q_\mu = \begin{pmatrix} E, & -q_x, & -q_y, & -q_z \\ \end{pmatrix}\equcomma
\end{equation}
where the metric,
\begin{equation}
  g_{\mu\nu} = g^{\mu\nu} = \begin{pmatrix} 
    1 & 0  & 0  & 0  \\
    0 & -1 & 0  & 0  \\
    0 & 0  & -1 & 0  \\
    0 & 0  & 0  & -1 \\
  \end{pmatrix}
\end{equation}
is used to raise and lower indices, {\it i.e.} ${q_\mu = g_{\mu\nu}
  q^\nu}$. Furthermore, Greek indices indicate summation over the four
components and Latin indices indicate summation over the three spatial
components where the conventions,
\begin{alignat*}{2}\labelali{}
  q^2 &= qq = q_\mu q^\mu = E^2 - q_x^2 - q_y^2 - q_z^2 \equcomma
  &\vec{q}^2 &= q_i q^i = q_x^2 + q_y^2 + q_z^2, \\
  \abs{\vec{q}} &= \sqrt{q_x^2 + q_y^2 + q_z^2},
  &q_T &= \sqrt{q_x^2 + q_y^2}
\end{alignat*}
are used. The notation $q_a$ where $a \neq x,~y,~z$, indicates the
momentum four-vector for particle $a$ and subsequently the notation
${q_x}_a$ indicates the $x$-component of the momentum for particle
$a$.

The units throughout this thesis follow the convention of
\rfr{peskin.95.1} where $\hbar$, the reduced Planck constant, and $c$,
the speed of light, are taken as unity. However, the value of ${c =
  3.0 \times 10^8}~\mathrm{m/s}$ is sometimes needed, particularly in
\chp{Exp}. The explicit use of $c$ in an equation indicates this value
of $c$ should be used rather than unity.

\newsection{Electroweak Lagrangian}{Ewk}

Following the discussion of \sec{Thr:Lag} the Lagrangian for unified
electroweak theory can written as,
\begin{align*}\labelali{EwkL} 
  \ld_\ewk = 
  &
  % Free fermions.
  \termlabel{i\bar{\psi}_f \gamma^\mu \partial_\mu \psi_f}
  {\equs{Thr:Spin1P}, \ref{equ:Thr:Spin1E}}
  % Free photon.
  \termlabel{-\frac{1}{2}\left(\partial^\mu A^\nu - \partial^\nu
      A^\mu\right)\left(\partial_\mu A_\nu - \partial_\nu
      A_\mu\right)}
  {\equs{Thr:Spin2MP}, \ref{equ:Thr:Spin2E}}
  \lgdsep &
  % Free Z.
  \termlabel{-\frac{1}{2}\left(\partial^\mu Z^\nu - \partial^\nu
      Z^\mu\right)\left(\partial_\mu Z_\nu - \partial_\nu
      Z_\mu\right)}
  {\equs{Thr:Spin2P}, \ref{equ:Thr:Spin2E}}
  \lgdsep &
  % Free W.
  \termlabel{-\left(\partial^\mu W^{+\nu} - \partial^\nu
      W^{+\mu}\right)\left(\partial_\mu W_\nu^- - \partial_\nu
      W_\mu^-\right)}
  {\equs{Thr:Spin2P}, \ref{equ:Thr:Spin2E}}
  \lgdsep &
  % Gm2FF interaction.
  \termlabel{+i Q_f \gE A_\mu \bar{\psi}_f \gamma^\mu \psi_f}
  {\fig{Thr:Gm2FF}}
  % Z2FF interaction.
  \termlabel{-\frac{i\gW}{2\costw} Z_\mu \bar{\psi}_f \gamma^\mu
    (\vc[_f] - \ac[_f]\gamma^5)\psi_f}
  {\fig{Thr:Z2FF}}
  \lgdsep &
  % W2LL interaction.
  \termlabel{-\frac{i\gW}{2\sqrt{2}} \left(W_\mu^-\bar{\lep}_f
      \gamma^\mu(1-\gamma^5) \nu_f + W_\mu^+\bar{\nu}_f
      \gamma^\mu(1-\gamma^5) \lep_f\right)}
  {\fig{Thr:W2LL}}
  \lgdsep &
  % W2QQ interaction.
  \termlabel{-\frac{i\gW V_{ij}}{2\sqrt{2}} \left(W_\mu^-\bar{q}_f^j
      \gamma^\mu(1-\gamma^5) q_f^i + W_\mu^+\bar{q}_f^i
      \gamma^\mu(1-\gamma^5) q_f^j\right)}
  {\fig{Thr:W2QQ}}
  \lgdsep &
  % Gm2WW interaction.
  \termlabel[8mm]{\begin{aligned}-i\gE\big(&\partial_\nu A_\mu (W_\mu^+
      W_\nu^- - W_\nu^+ W_\mu^-) - A_\nu(W_\mu^+ \partial_\nu
      W_\mu^- - W_\mu^- \partial_\nu W_\mu^+) \\ &+ A_\mu(W_\nu^+ \partial_\nu
      W_\mu^- - W_\nu^- \partial_\nu W_\mu^+)\big)\end{aligned}}
  {\fig{Thr:Gm2WW}}
  \lgdsep &
  % Z2WW interaction.
  \termlabel[8mm]{\begin{aligned}-i\gW\costw\big(&\partial_\nu Z_\mu (W_\mu^+
      W_\nu^- - W_\nu^+ W_\mu^-) - Z_\nu(W_\mu^+ \partial_\nu
      W_\mu^- - W_\mu^- \partial_\nu W_\mu^+) \\ &+ Z_\mu(W_\nu^+ \partial_\nu
      W_\mu^- - W_\nu^- \partial_\nu W_\mu^+)\big)\end{aligned}}
  {\fig{Thr:Z2WW}}
  \lgdsep &
  % WW2WW interaction.
  \termlabel{-i\gW[^2] \left(W_\mu^+W_\mu^-W_\nu^+W_\nu^- -
      W_\mu^+W_\nu^-W_\mu^+W_\nu^-\right)}
  {\fig{Thr:WW2WW}}
  \lgdsep &
  % WW2ZZ interaction.
  \termlabel{+i\gW[^2]\costw[^2] \left(Z_\mu W_\mu^+Z_\nu W_\nu^- -
      Z_\mu Z_\mu W_\nu^+W_\nu^-\right)}
  {\fig{Thr:WW2ZZ}}
  \lgdsep &
  % WW2GmGm interaction.
  \termlabel{+i\gE[^2] \left(A_\mu W_\mu^+A_\nu W_\nu^- -
      A_\mu A_\mu W_\nu^+W_\nu^-\right)}
  {\fig{Thr:WW2GmGm}}
  \lgdsep &
  % WW2GmZ interaction.
  \termlabel{+i\gE\gW\costw \left(A_\mu Z_\nu(W_\mu^+W_\nu^- -
      W_\nu^+W_\nu^-) - 2A_\mu Z_\mu W_\nu^+W_\nu^-\right)}
  {\fig{Thr:WW2GmZ}}
\end{align*}
where the covariant derivatives of \equ{Thr:U2D} have been introduced
into the free Lagrangians for $y_f$ of \equ{Thr:Su2F} and $t_f$ of
\equ{Thr:U1F}, as well as free Lagrangians, using \equ{Thr:Spin2L},
for the photon, \wwbs, and \wzb. Here, $\psi_f$ indicates any fermion
field of flavour $f$. Note that mass terms for the fermions and bosons
have not been introduced, as these terms are introduced via the Higgs
mechanism and are included in $\ld_{\mathrm{MHM}}$ of \equ{Tvr:MhmL}.

\newsection{Higgs Lagrangian}{Mhm}

The Lagrangian for the minimal Higgs mechanism of the \sm can be
written as,
\begin{align*}\labelali{MhmL}
  \ld_\mathrm{MHM} = 
  &
  % Fermion mass terms.
  \termlabel{-\m_f^2\bar{\psi}_f\psi_f}
  {$\m_f$}
  % W mass term.
  \termlabel{-\m_\w^2W_\mu^{+}W^{-\mu}}
  {$\m_\w$}
  % Z mass term.
  \termlabel{-\frac{\m_\z^2}{2}Z_\mu Z^\mu}
  {$\m_\z$}
  % Free Higgs boson.
  \termlabel{-\frac{1}{2}\partial_\mu H \partial^\mu H -
    \frac{\m_H^2}{2}HH}
  {\equ{Thr:Spin0P}} 
   \lgdsep &
  % H02H0H0 interaction.
  \termlabel{-\frac{3i\gW\m_H^2}{2\m_\w}HHH}
  {\fig{Thr:H02H0H0}}
  % H02WW interaction.
  \termlabel{+i\gW\m_WW_\mu^{+}W^{-\mu}H}
  {\fig{Thr:H02WW}}
  % H02ZZ interaction.
  \termlabel{+\frac{i\gW\m_z}{\costw} Z_\mu Z^\mu H}
  {\fig{Thr:H02ZZ}}
  \lgdsep &
  % H0H02H0H0 interaction.
  \termlabel{-\frac{3i\gW[^2]\m_H^2}{4\m_W^2}HHHH}
  {\fig{Thr:H0H02H0H0}}
  % H0H02WW interaction.
  \termlabel{+\frac{i\gW[^2]}{4} W_\mu^+W^{-\mu}HH}
  {\fig{Thr:H0H02WW}}
  % H0H02ZZ interaction.
  \termlabel{+\frac{i\gW[^2]}{8\costw[^2]} Z_\mu Z^\mu HH}
  {\fig{Thr:H0H02ZZ}}
  \lgdsep &
  % H02FF interaction.
  \termlabel{-\frac{i\gW\m_f}{2\m_W}H\bar{\psi}_f\psi_f}
  {\fig{Thr:H02FF}}
\end{align*}
where the terms follow the conventions of \sec{Thr:Lag}.
\newappendix{Tau Leptons}{Dvr}

Distributions of $\m_{23}$, the invariant mass of the second and third
\wtl decay products, for the general three meson current of
\sec{Tau:DecFour} by Decker {\it et al.}~\cite{decker.93.1} are given
for the channels which do not use this model by default. In
\fig{Dvr:ThreeMesonsGeneric.OnePion} the distributions for \wtl decays
with one pion and two kaons are given, while in
\fig{Dvr:ThreeMesonsGeneric.TwoPions} the distributions for decays
with two or more pions are given.

\begin{subfigures}[b!]{2}{Distributions of $\m_{23}$ for the
    \subfig{q_qp.W.16_211_321_321.me_1543.m_211_321n}~$K^- \pi^- K^+$,
    \subfig{q_qp.W.16_211_311_311.me_1543.m_211_311}~$K^0 \pi^- \bar{K}^0$, and
    \subfig{q_qp.W.16_111_311_321.me_1543.m_111_321}~$K^- \pi^0 K^0$
    decay channels using the general three meson model by
    Decker {\it et
      al.}~\cite{decker.93.1}.\labelfig{ThreeMesonsGeneric.OnePion}}
  \svgbeg
  \svg{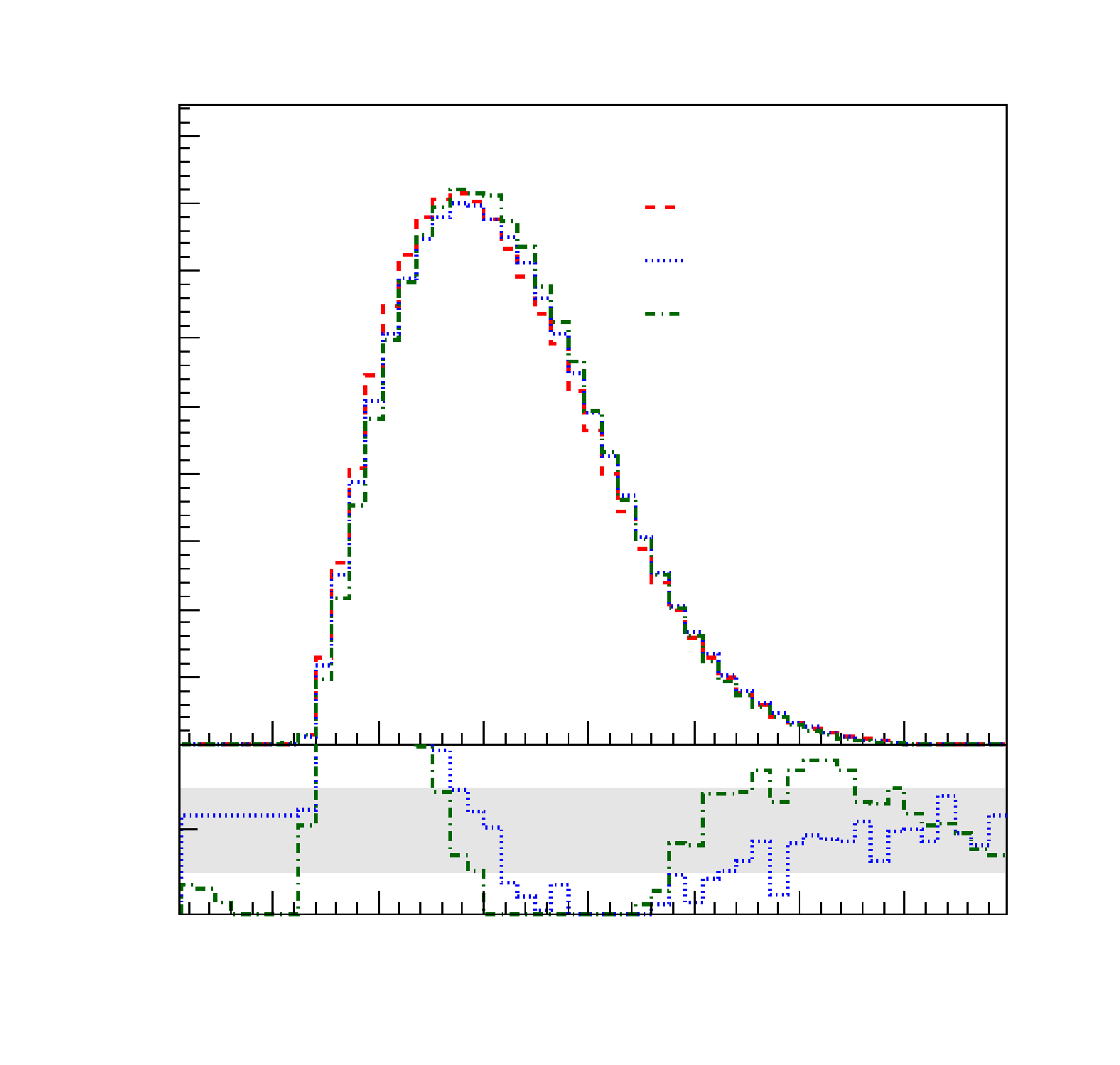} &
  \svg{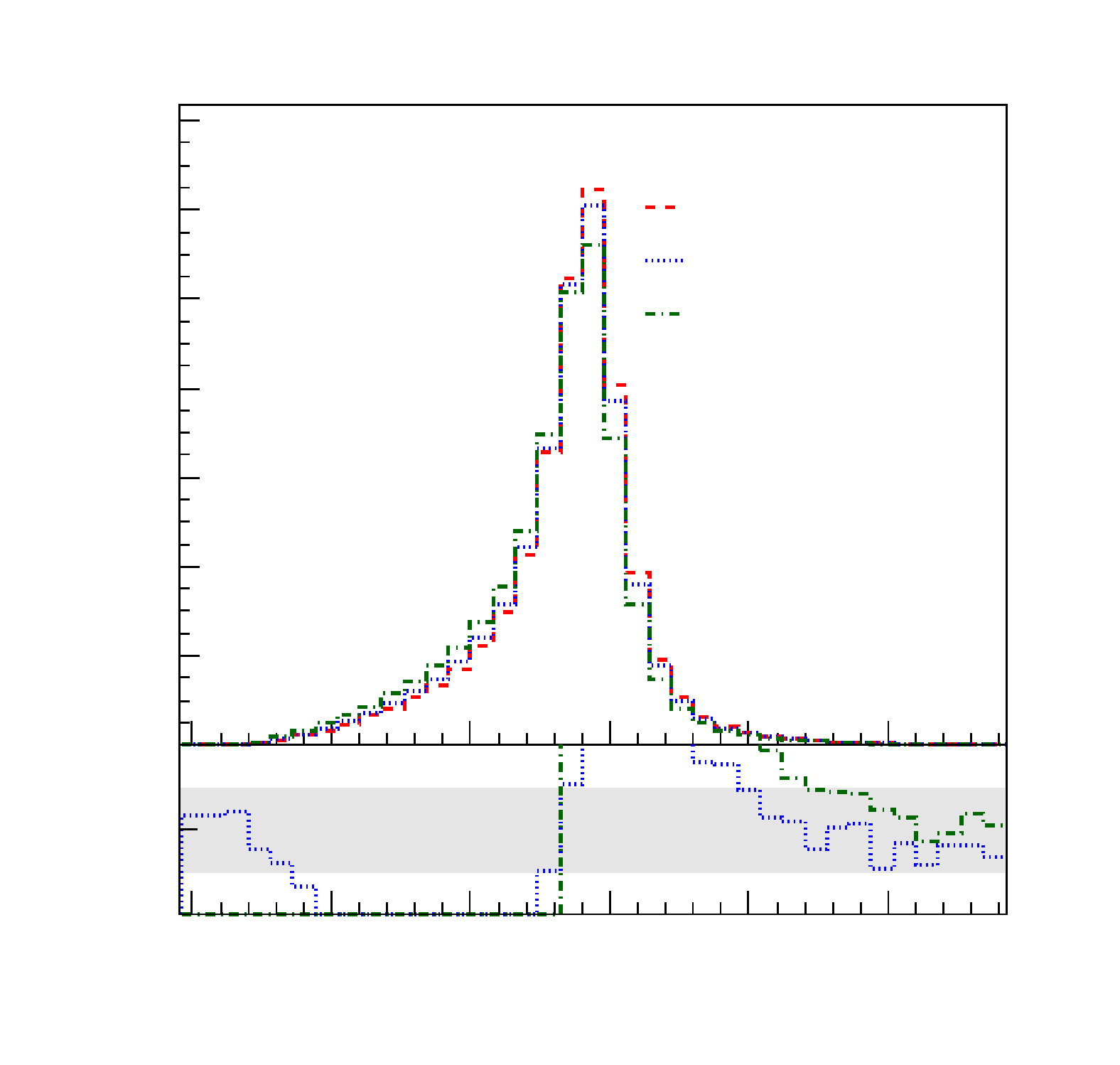} \svgsep
  \svg{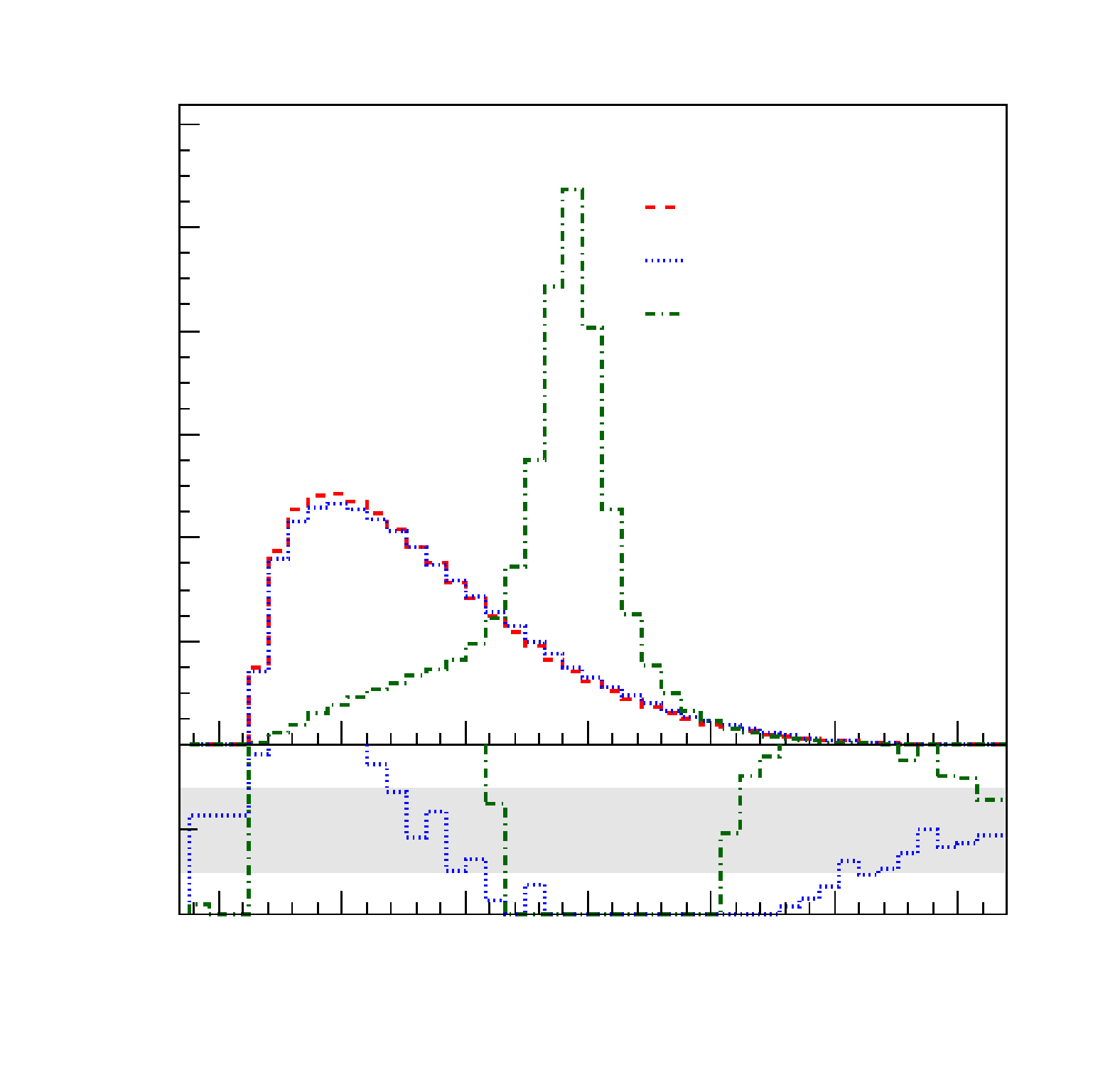} & \sidecaption \svgend
\end{subfigures}

\begin{subfigures}[p]{2}{Distributions of $\m_{23}$ for the
    \subfig{q_qp.W.16_111_111_211.me_1543.m_111_211}~$\pi^0 \pi^0 \pi^-$,
    \subfig{q_qp.W.16_211_211_211.me_1543.m_211n_211p}~$\pi^- \pi^- \pi^+$,
    \subfig{q_qp.W.16_111_111_321.me_1543.m_111_321}~$\pi^0 \pi^0 K^-$,
    \subfig{q_qp.W.16_211_211_321.me_1543.m_211n_321}~$K^- \pi^- \pi^+$, and
    \subfig{q_qp.W.16_111_211_311.me_1543.m_211_311}~$\pi^- \bar{K}^0
    \pi^0$ decay channels using the general three meson model by
    Decker {\it et
      al.}~\cite{decker.93.1}.\labelfig{ThreeMesonsGeneric.TwoPions}}
  \svgbeg
  \svg{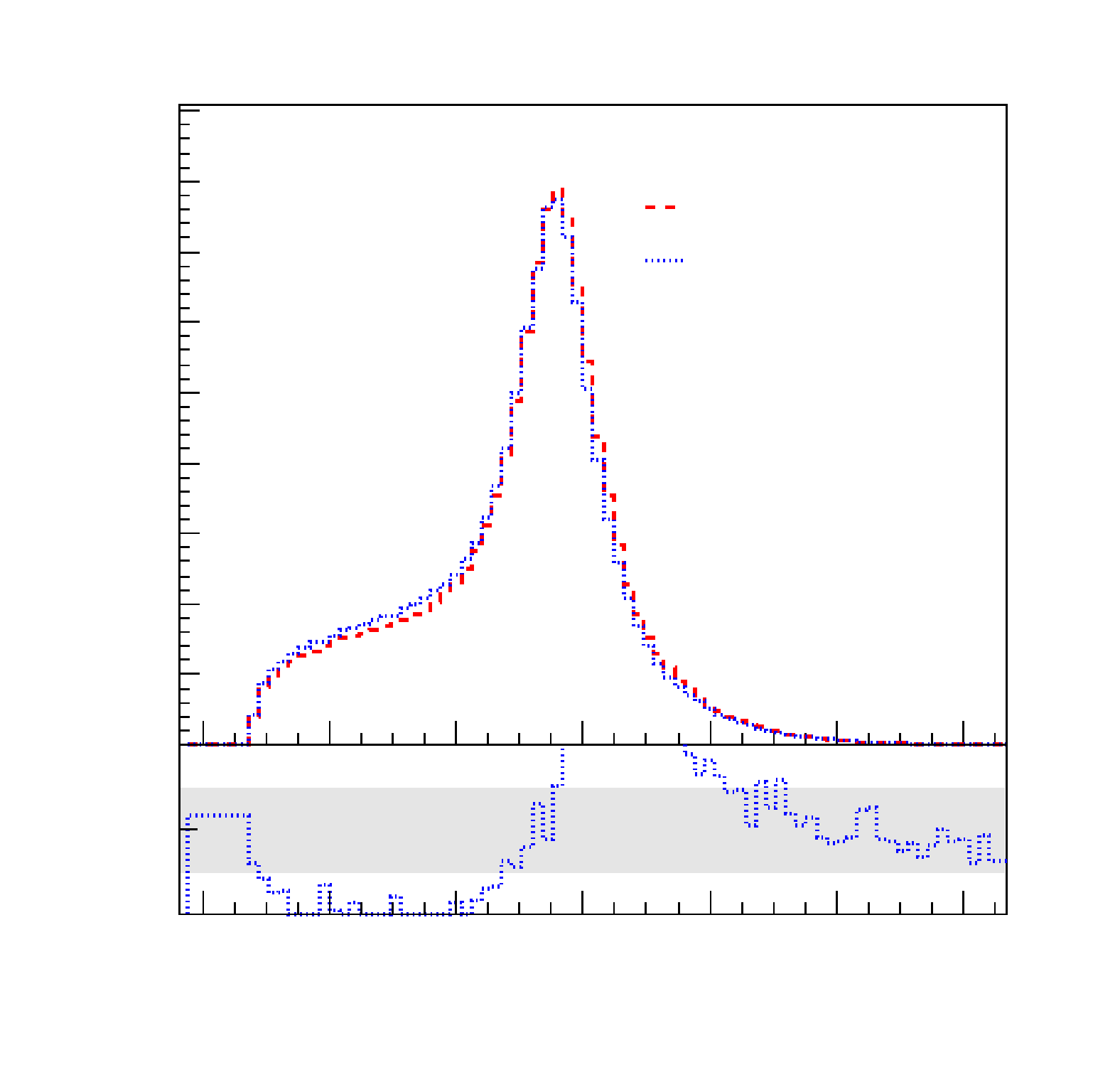} &
  \svg{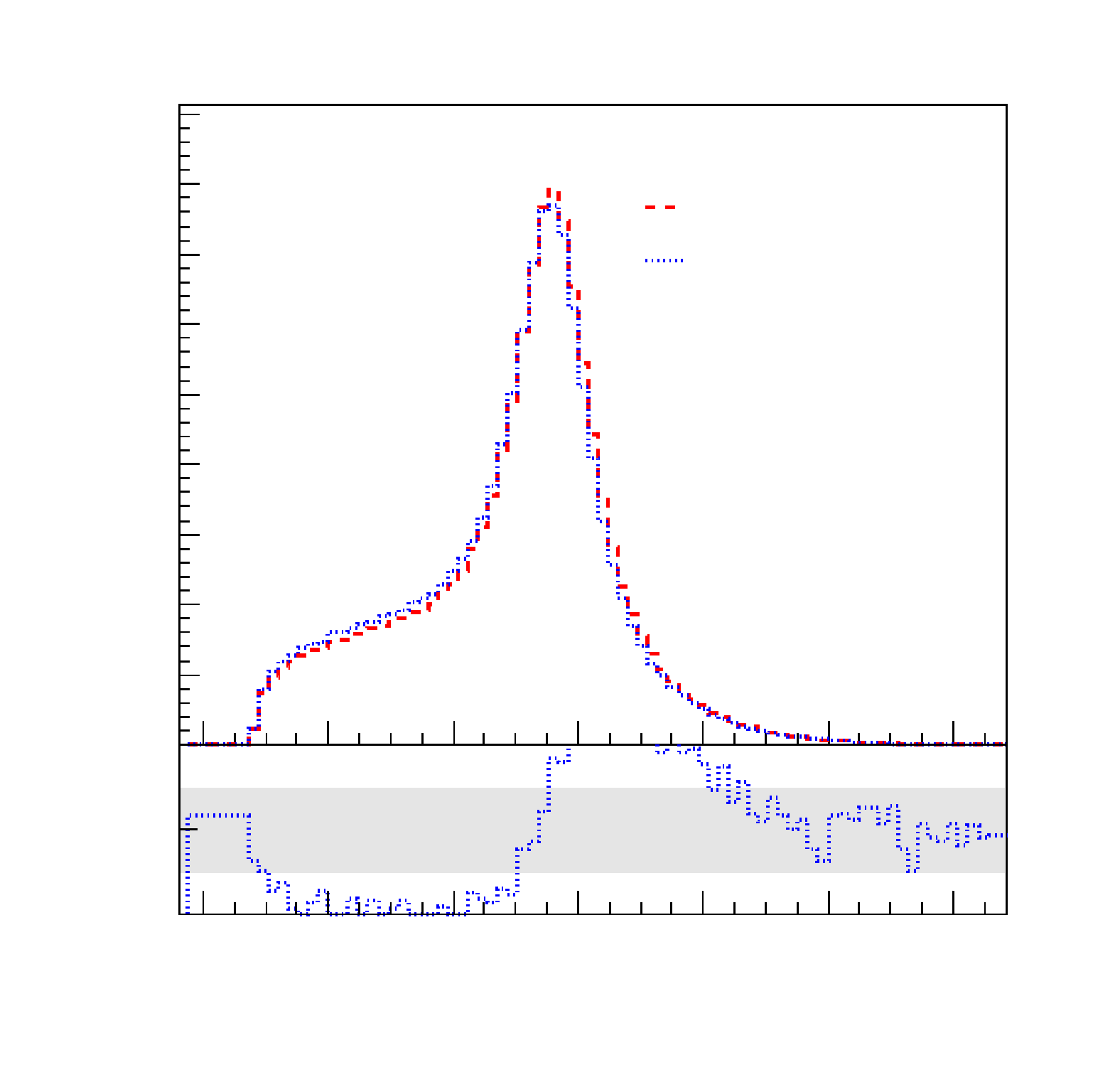} \svgsep
  \svg{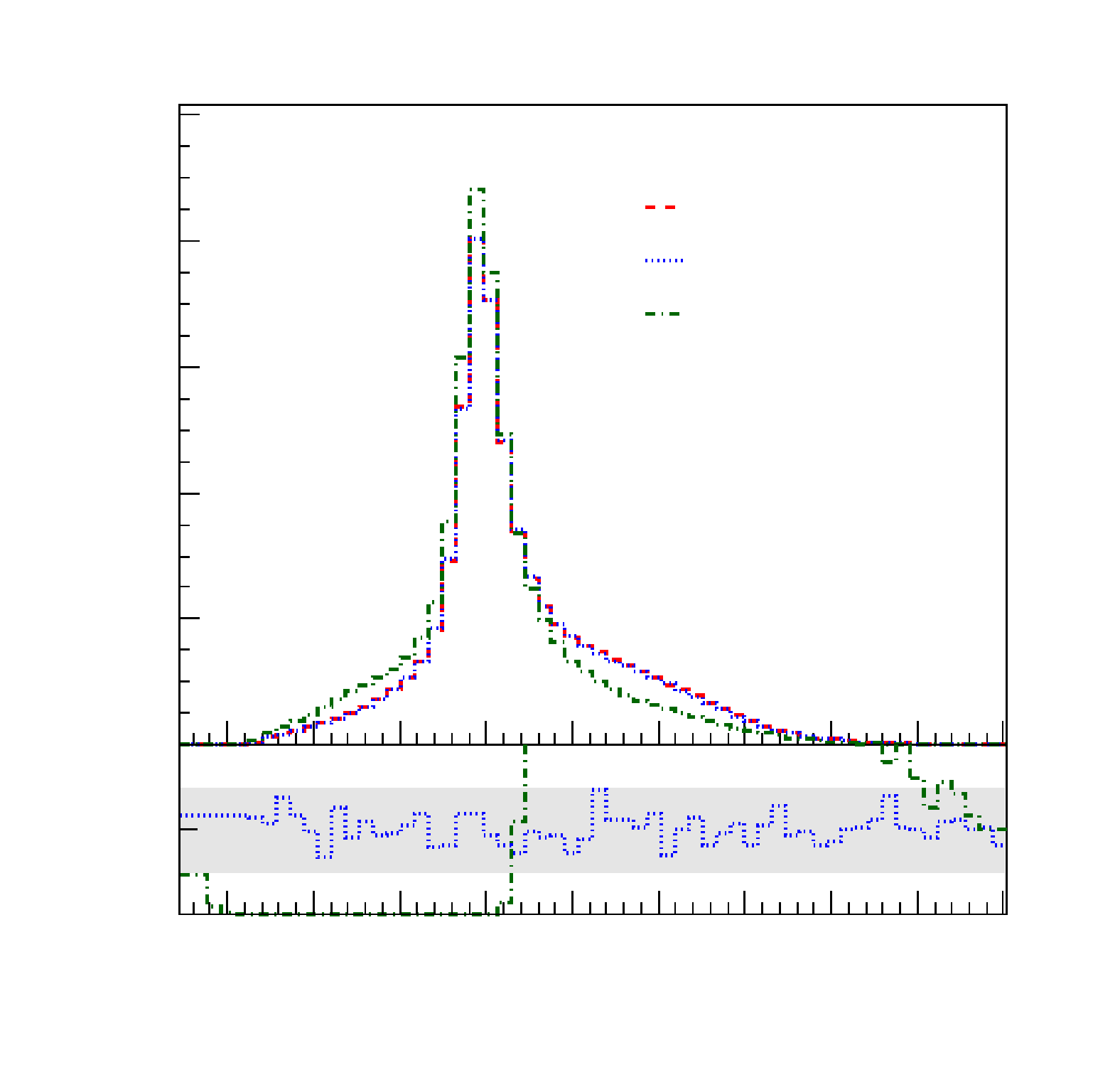} &
  \svg{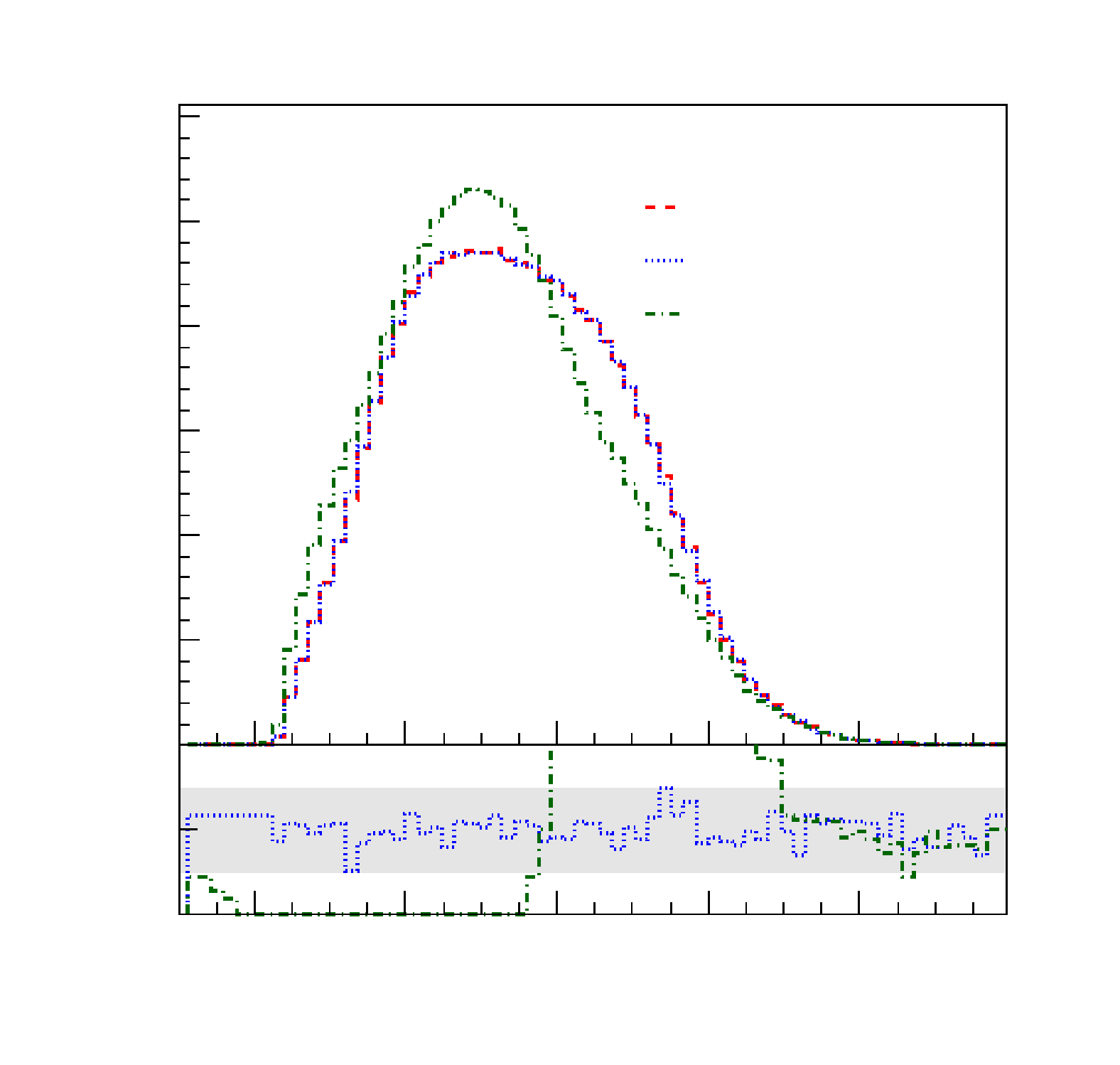} \svgsep
  \svg{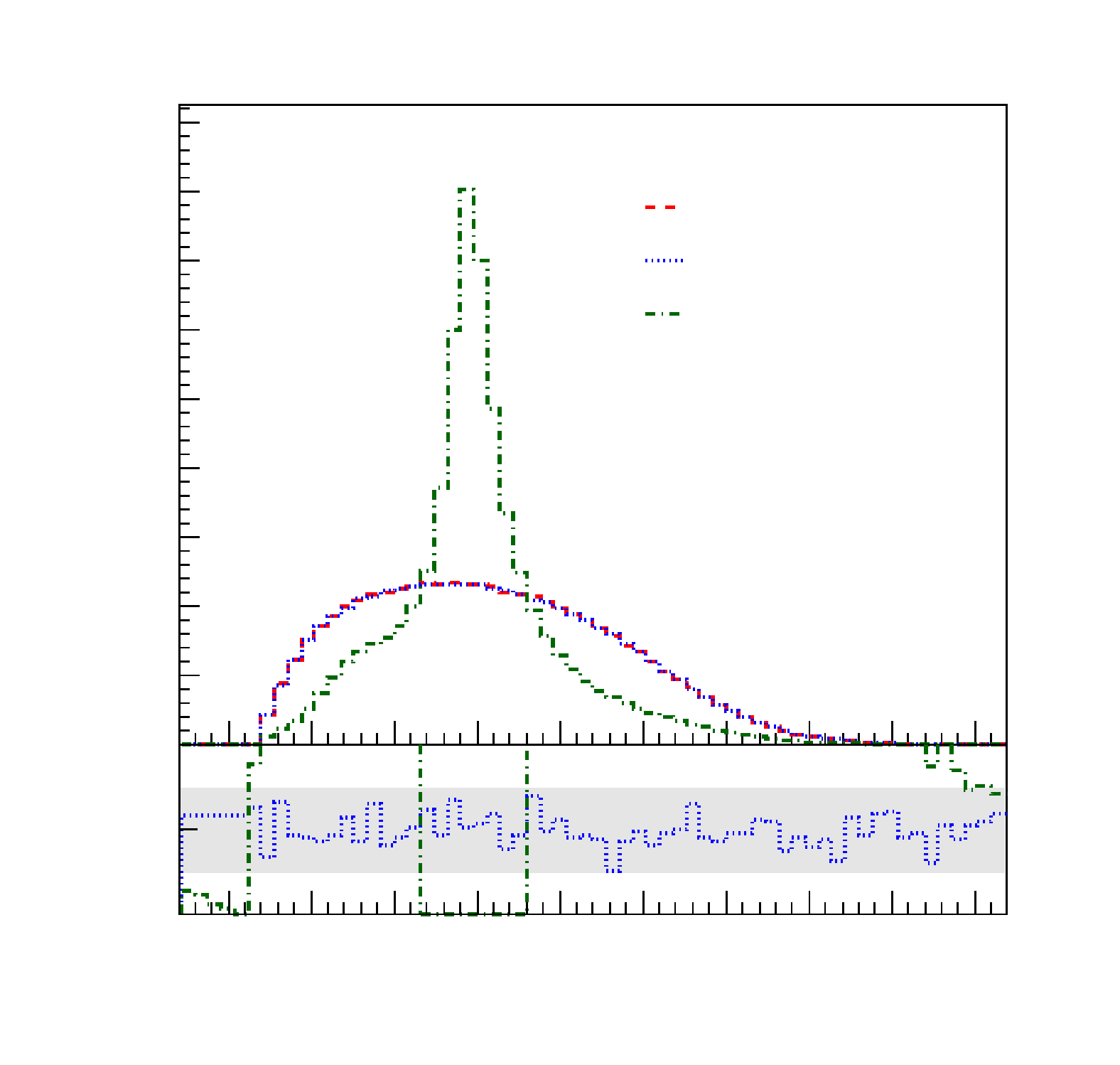} & \sidecaption \svgend
\end{subfigures}
\newappendix{\textit{Z} Boson}{Zvr}

Additional information for the ${\z \to \ditau}$ cross-section
measurement of \chp{Zed} is given in this appendix. Additional
distributions for the data, estimated signal, and estimated
backgrounds are given in \sap{Zvr:Dis}. The method for determining low
momentum muon track finding and identification efficiencies using the
tag-and-probe method on ${\jpsi \to \dimu}$ events from data is
outlined in \sap{Zvr:RecEff}. Finally, the combined cross-section fit
of the cross-sections from each event category is detailed in
\sap{Zvr:Fit}.

\newsection{Distributions}{Dis}

The pseudo-rapidity distributions for the combination of the two \wtl
decay product candidates are given in \fig{Zvr:Eta} for each event
category, while the distributions for the number of primary vertices
in the event are given in \fig{Zvr:Pv}. The transverse momentum
distributions for the first and second \wtl decay products are given
in \figs{Zvr:Pt1} and \ref{fig:Zvr:Pt2} respectively. Similarly, their
pseudo-rapidity distributions are given in \figs{Zvr:Eta1} and
\ref{fig:Zvr:Eta2}. All these distributions were determined using the
selection of \sec{Zed:Sel} and the background estimation of
\sec{Zed:Bkg}.

\begin{subfigures}[p]{2}{The pseudo-rapidity, $\eta$, distributions of the
    two \wtl decay product candidates from data (points) for the
    \subfig{Z2TauTau2MuMu.Eta}~\mumu,
    \subfig{Z2TauTau2MuE.Eta}~\mue,
    \subfig{Z2TauTau2EMu.Eta}~\emu,
    \subfig{Z2TauTau2MuPi.Eta}~\muh, and
    \subfig{Z2TauTau2EPi.Eta}~\eh categories. The simulated
    signal (red) is normalised to the number of signal events, while
    the \qcd (green), \ewk (blue), \ttbar (orange), \ww (magenta), and
    ${\z \to \dilep}$ (cyan) backgrounds are estimated as described in
    \sec{Zed:Bkg}.\labelfig{Eta}}
  \svgbeg
  \svg{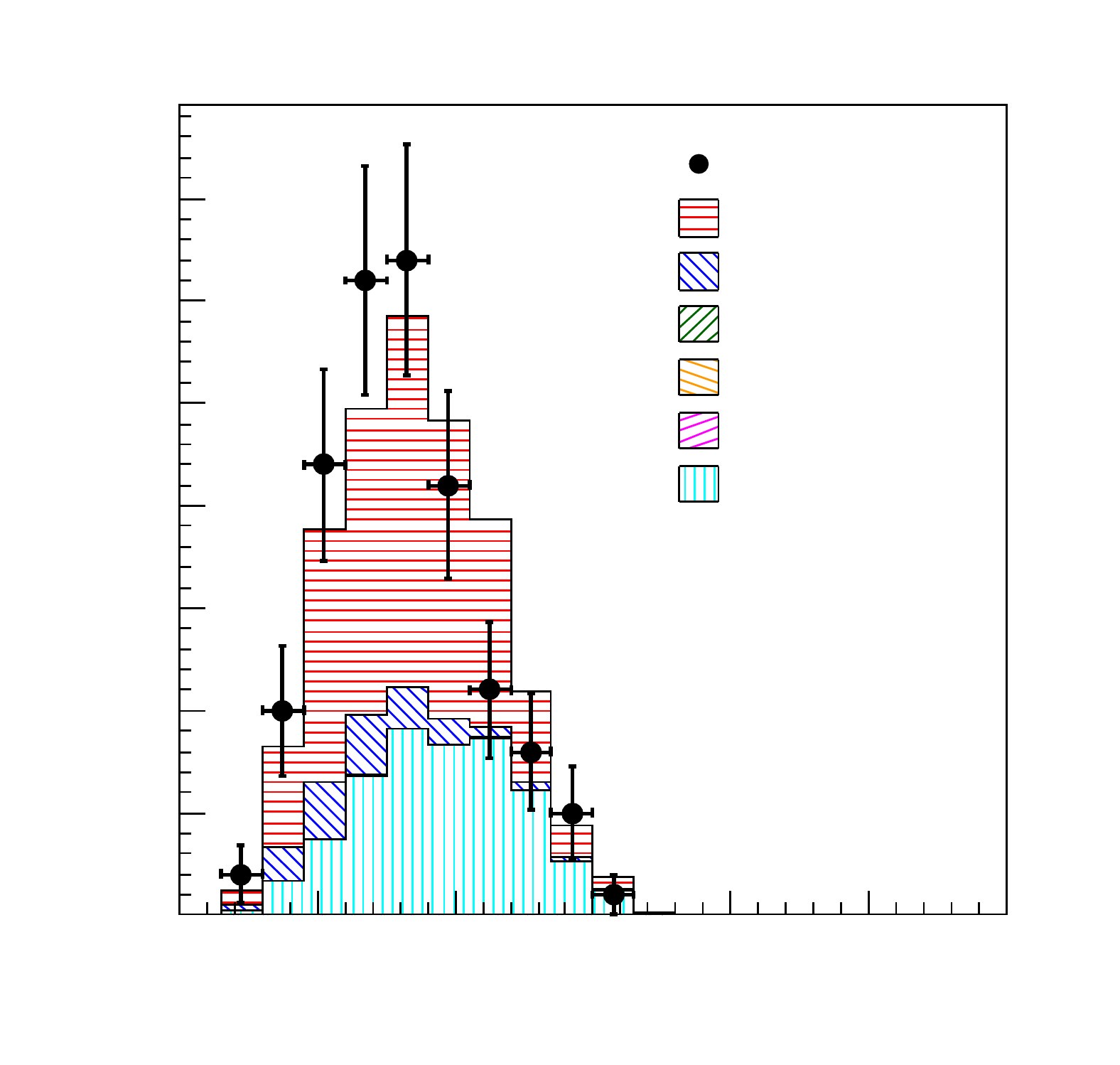} & \svg{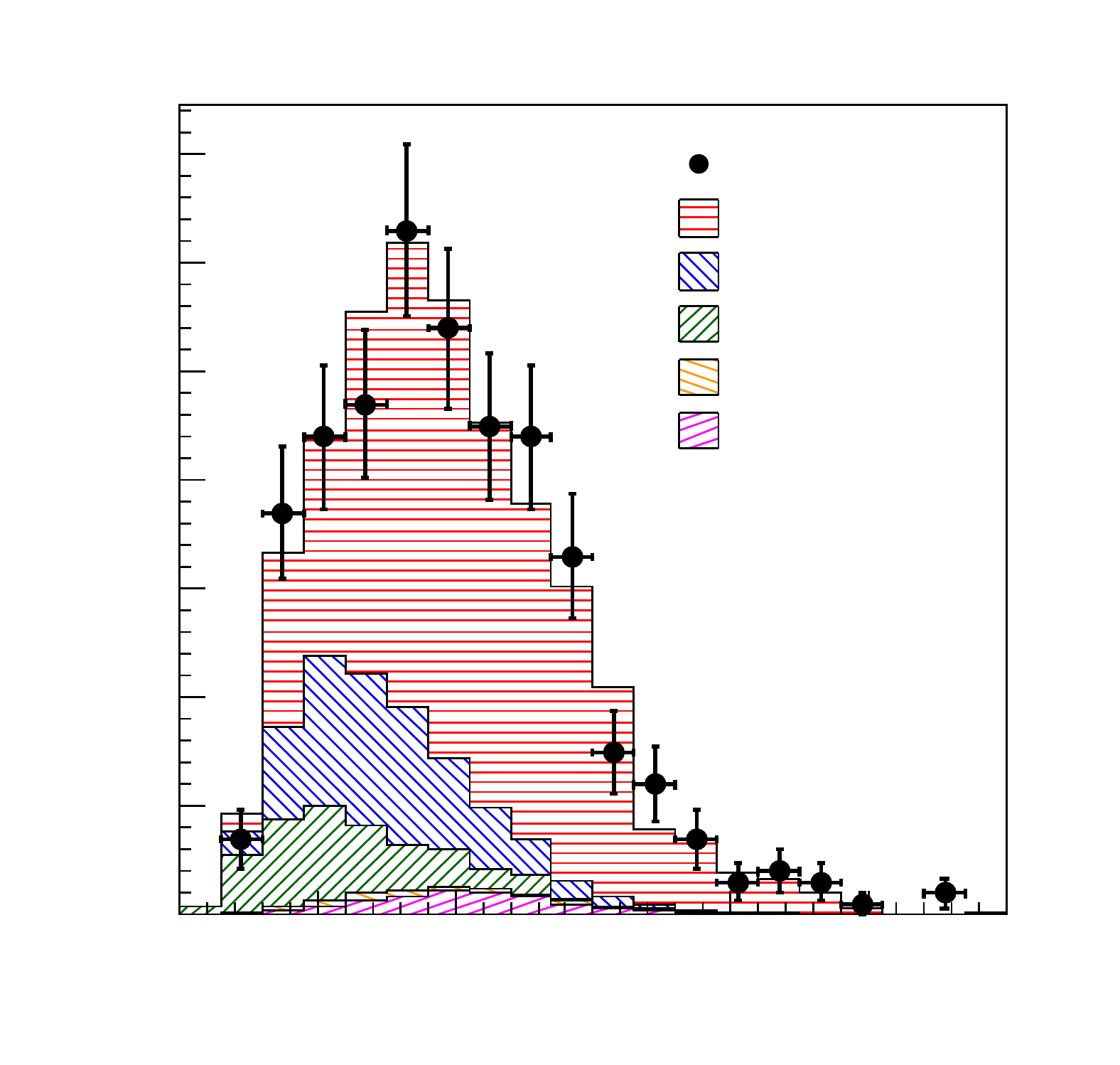}  \svgsep
  \svg{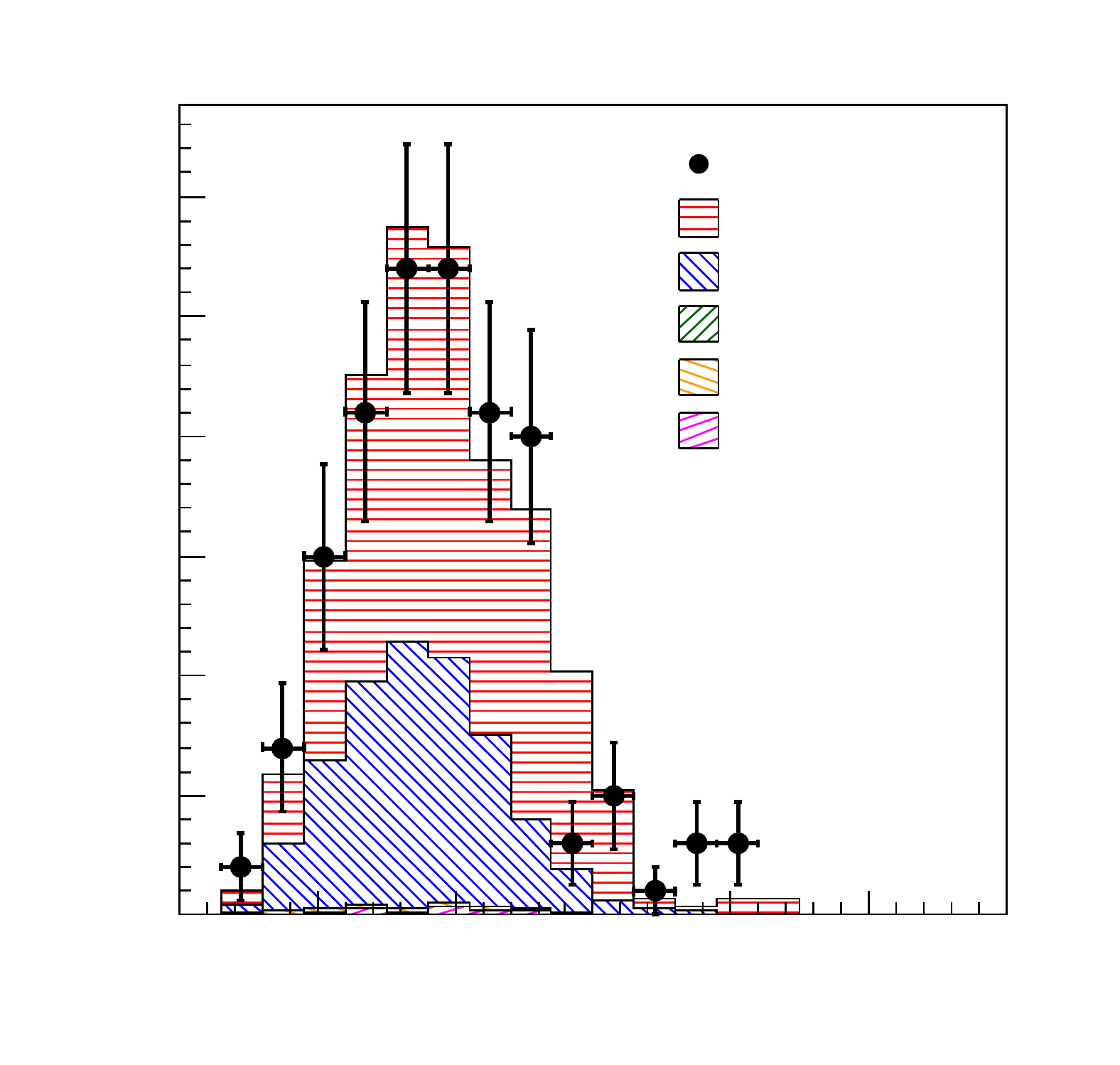}  & \svg{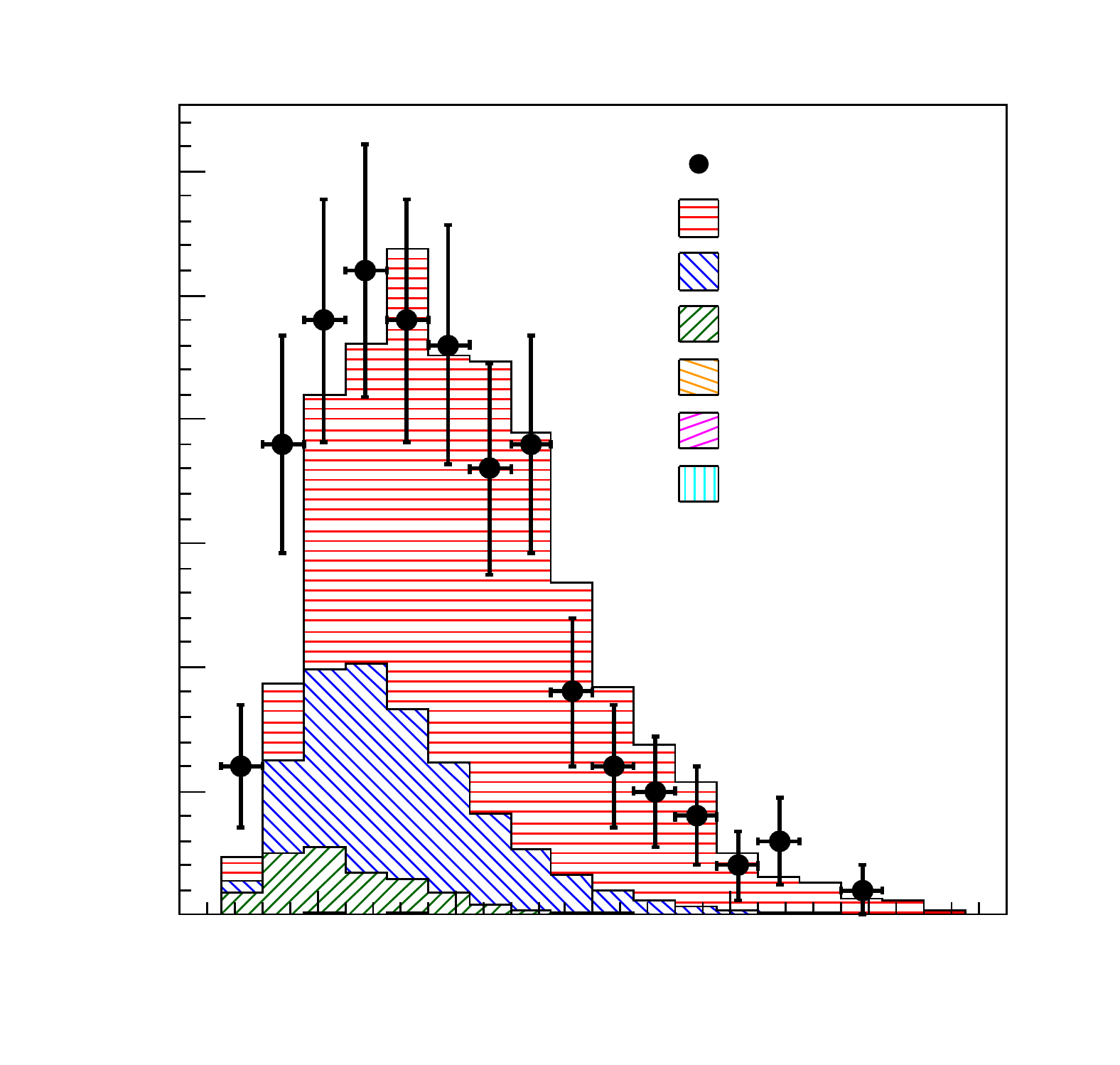} \svgsep
  \svg{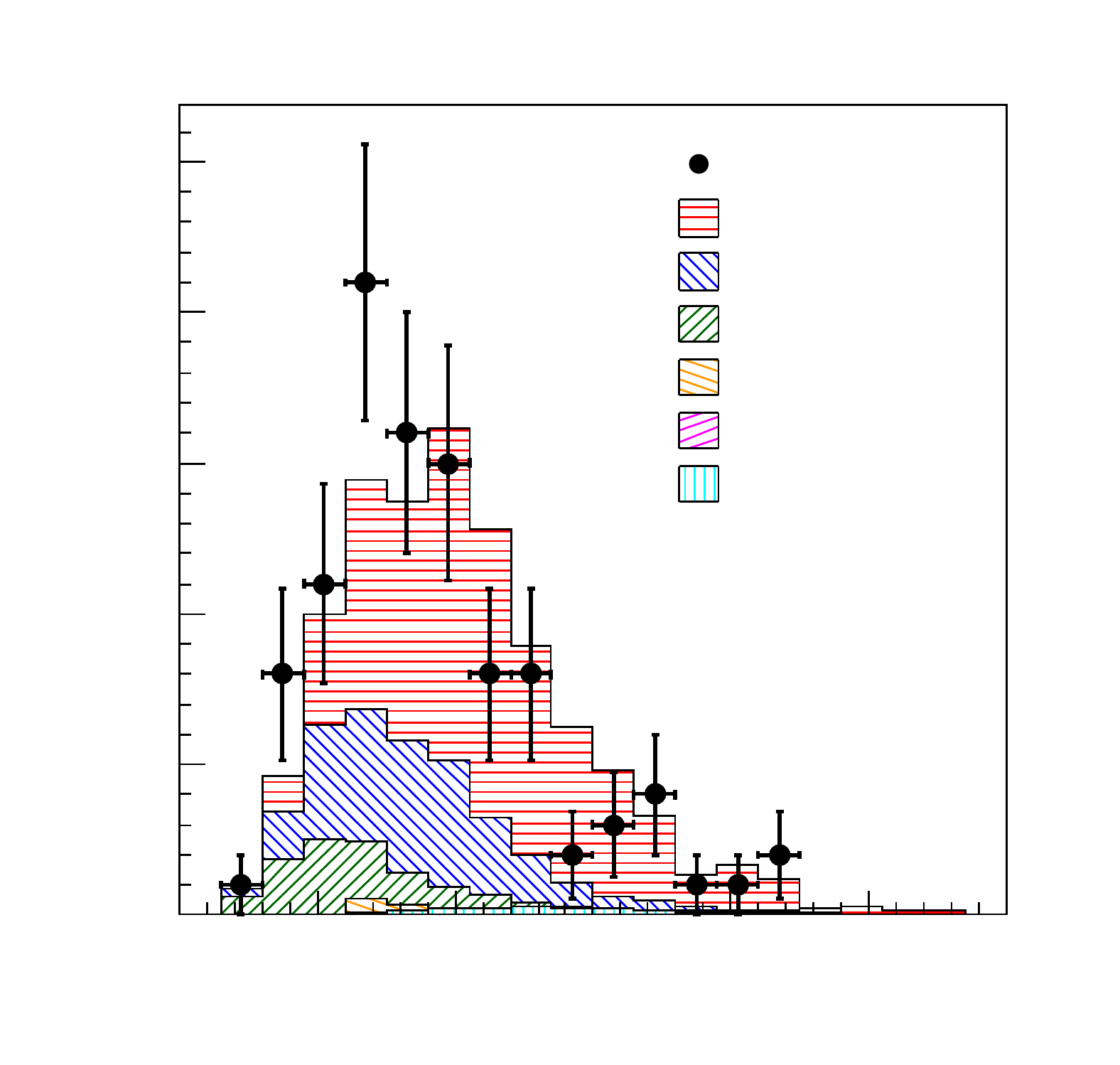}  & \sidecaption            \svgend
\end{subfigures}

\begin{subfigures}[p]{2}{The number of primary vertices from data
    (points) for the \subfig{Z2TauTau2MuMu.Pv}~\mumu,
    \subfig{Z2TauTau2MuE.Pv}~\mue,
    \subfig{Z2TauTau2EMu.Pv}~\emu,
    \subfig{Z2TauTau2MuPi.Pv}~\muh, and
    \subfig{Z2TauTau2EPi.Pv}~\eh categories. The simulated
    signal (red) is normalised to the number of signal events, while
    the \qcd (green), \ewk (blue), \ttbar (orange), \ww (magenta), and
    ${\z \to \dilep}$ (cyan) backgrounds are estimated as described in
    \sec{Zed:Bkg}.\labelfig{Pv}}
  \svgbeg
  \svg{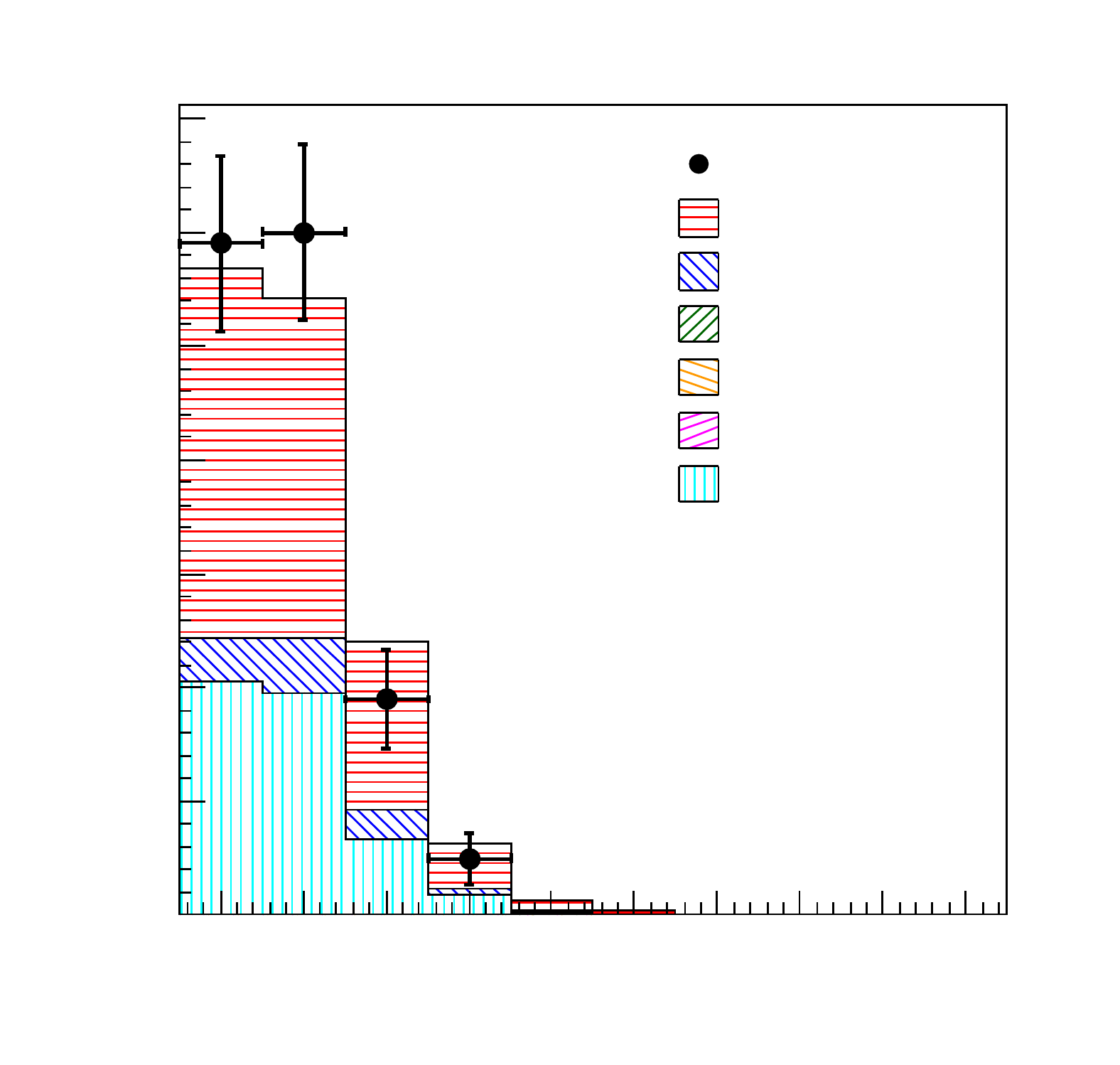} & \svg{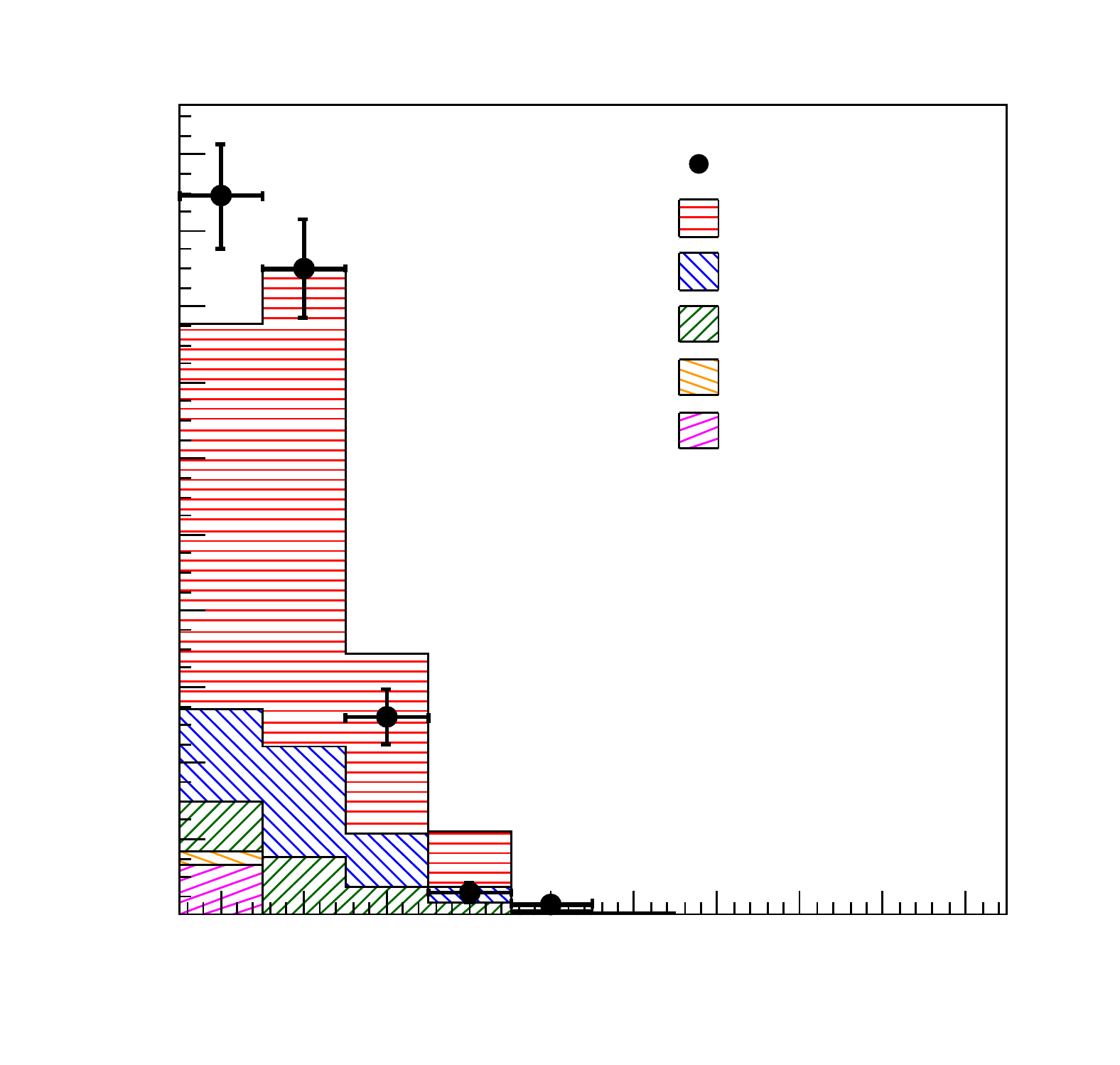}  \svgsep
  \svg{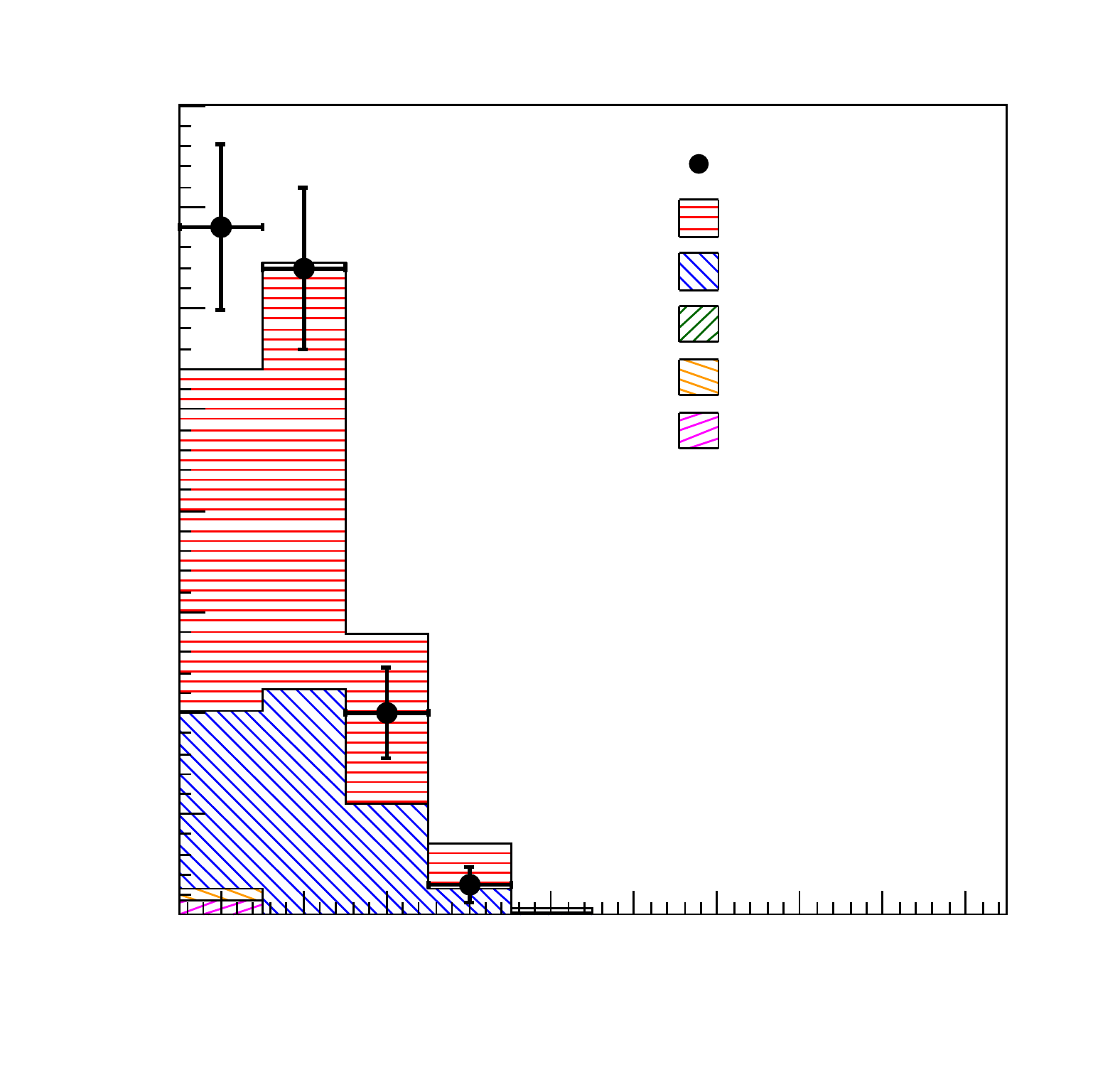}  & \svg{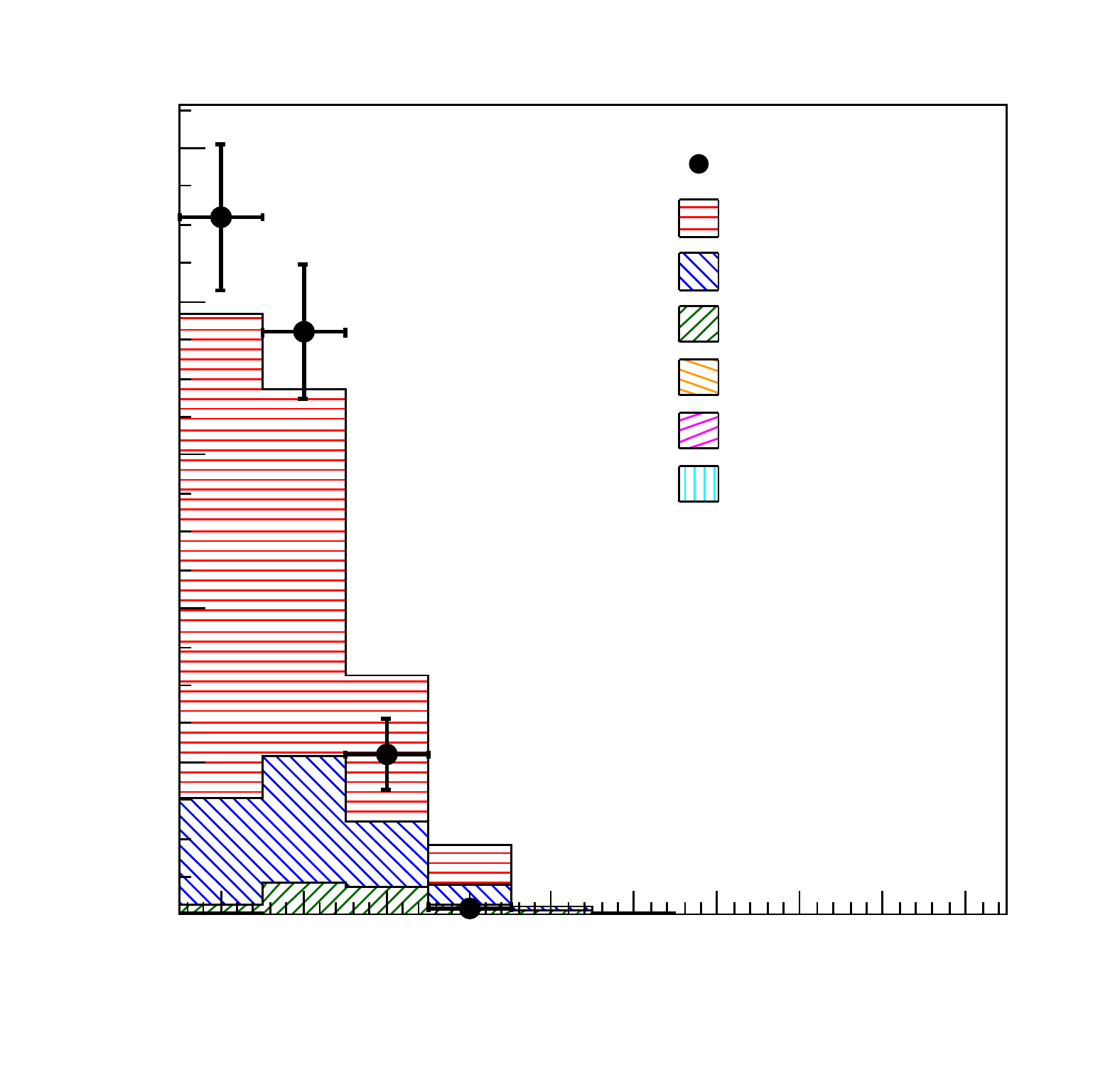} \svgsep
  \svg{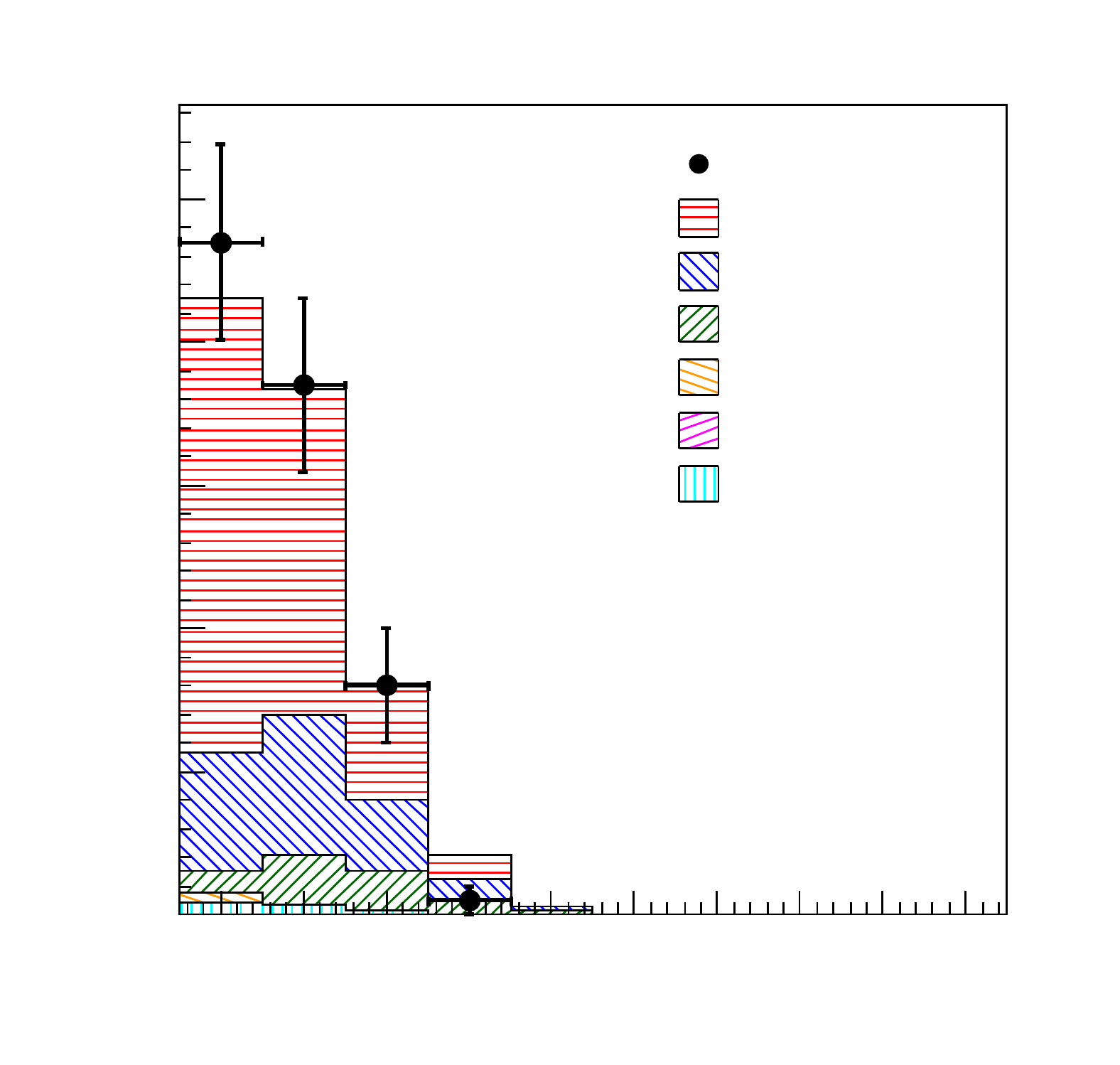}  & \sidecaption           \svgend
\end{subfigures}

\begin{subfigures}[p]{2}{The \pt distributions for the first \wtl decay
    product candidate from data (points) for the
    \subfig{Z2TauTau2MuMu.Pt1}~\mumu,
    \subfig{Z2TauTau2MuE.Pt1}~\mue,
    \subfig{Z2TauTau2EMu.Pt1}~\emu,
    \subfig{Z2TauTau2MuPi.Pt1}~\muh, and
    \subfig{Z2TauTau2EPi.Pt1}~\eh categories. The simulated
    signal (red) is normalised to the number of signal events, while
    the \qcd (green), \ewk (blue), \ttbar (orange), \ww (magenta), and
    ${\z \to \dilep}$ (cyan) backgrounds are estimated as described in
    \sec{Zed:Bkg}.\labelfig{Pt1}}
  \svgbeg
  \svg{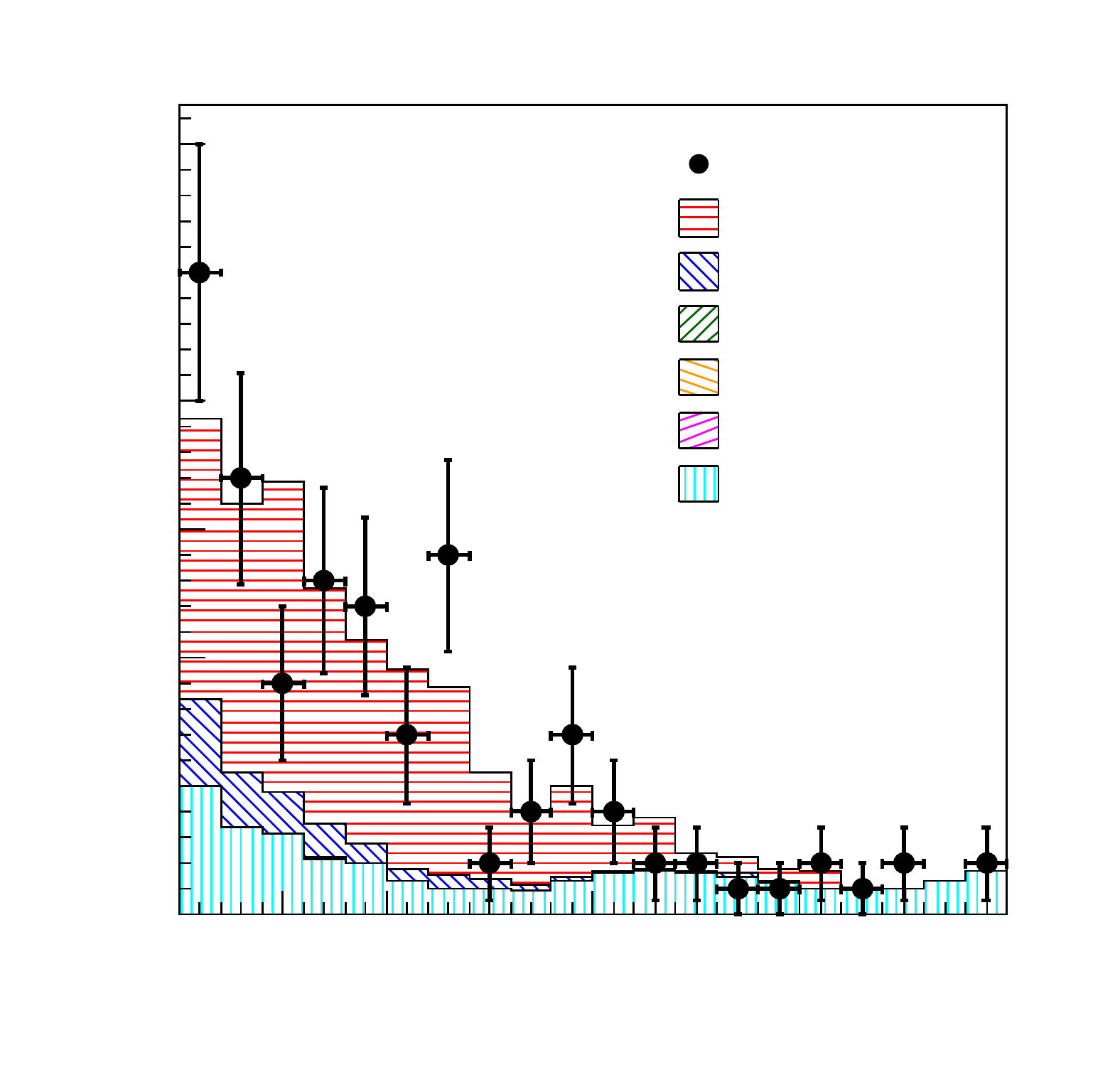} & \svg{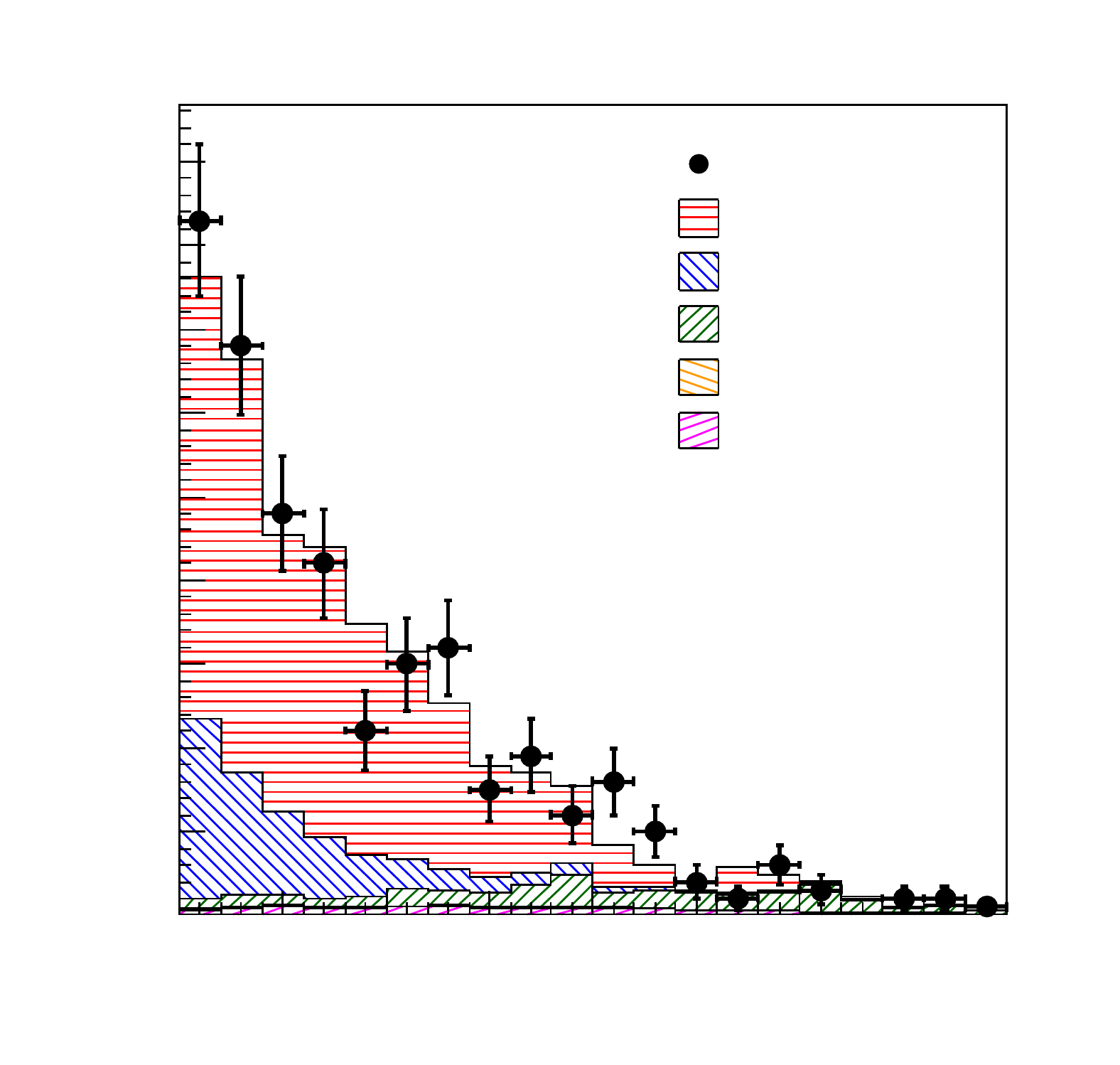}  \svgsep
  \svg{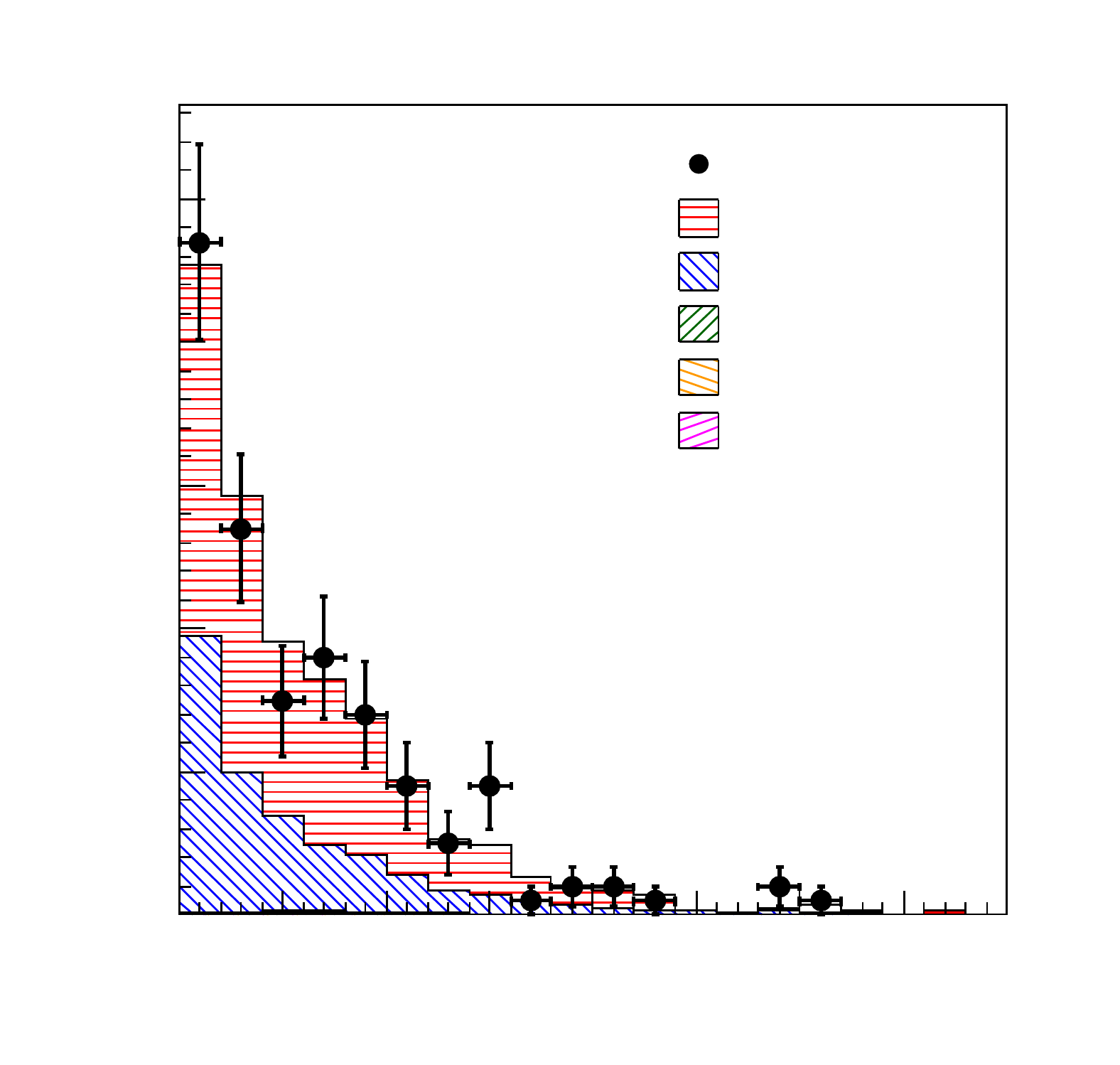}  & \svg{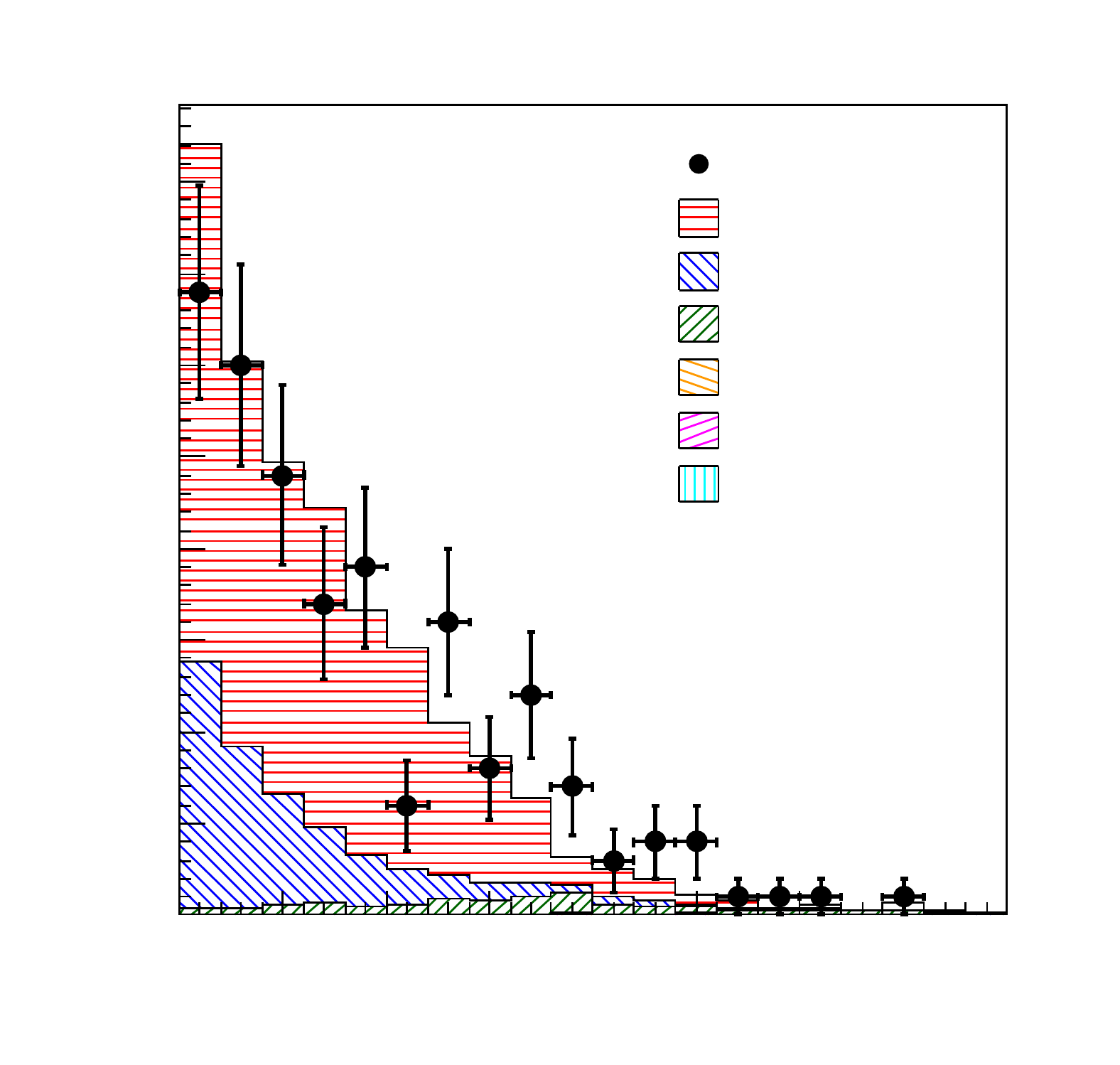} \svgsep
  \svg{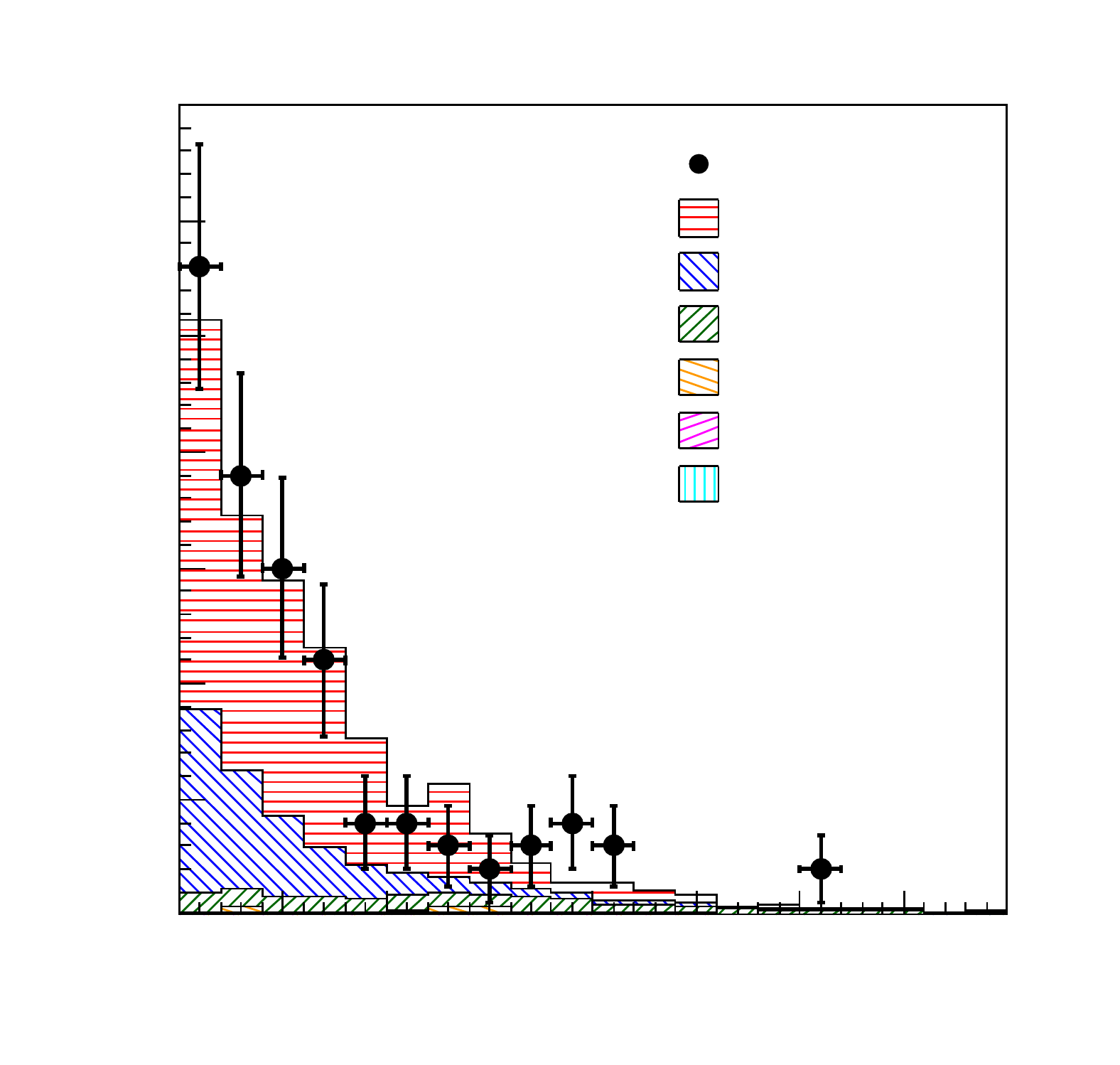}  & \sidecaption            \svgend
\end{subfigures}

\begin{subfigures}[p]{2}{The \pt distributions for the second \wtl decay
    product candidate from data (points) for the
    \subfig{Z2TauTau2MuMu.Pt2}~\mumu,
    \subfig{Z2TauTau2MuE.Pt2}~\mue,
    \subfig{Z2TauTau2EMu.Pt2}~\emu,
    \subfig{Z2TauTau2MuPi.Pt2}~\muh, and
    \subfig{Z2TauTau2EPi.Pt2}~\eh categories. The simulated
    signal (red) is normalised to the number of signal events, while
    the \qcd (green), \ewk (blue), \ttbar (orange), \ww (magenta), and
    ${\z \to \dilep}$ (cyan) backgrounds are estimated as described in
    \sec{Zed:Bkg}.\labelfig{Pt2}}
  \svgbeg
  \svg{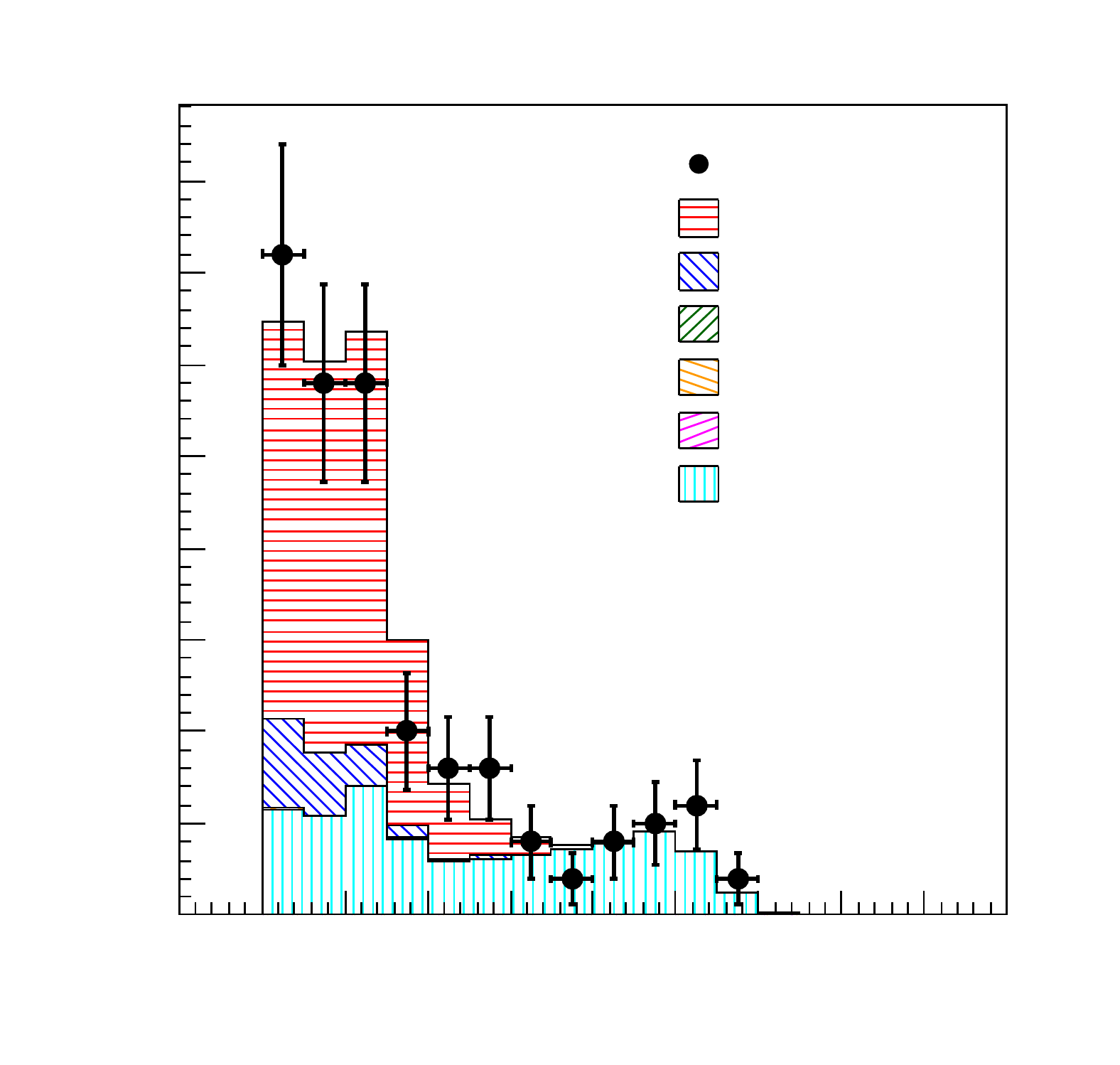} & \svg{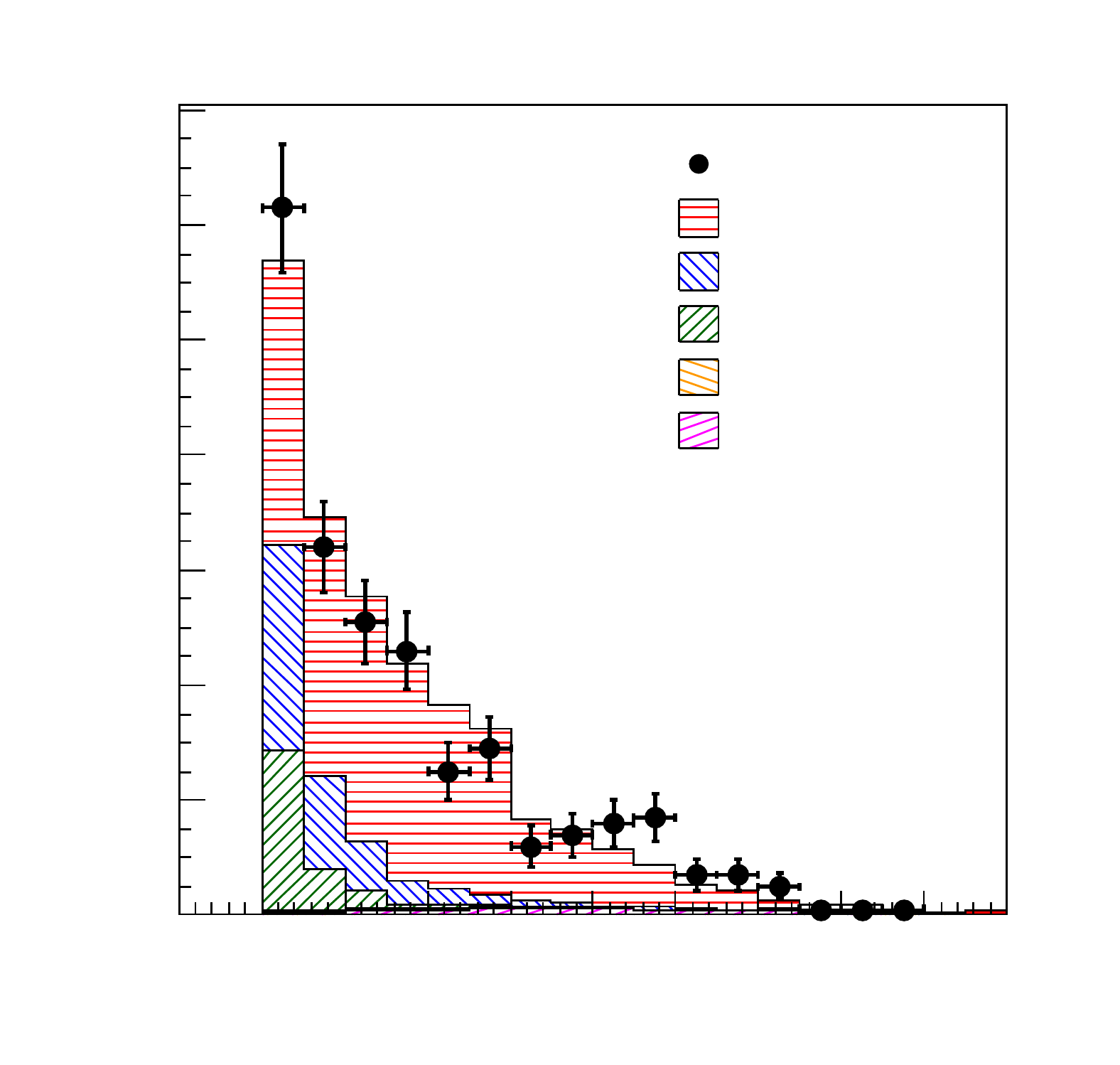}  \svgsep
  \svg{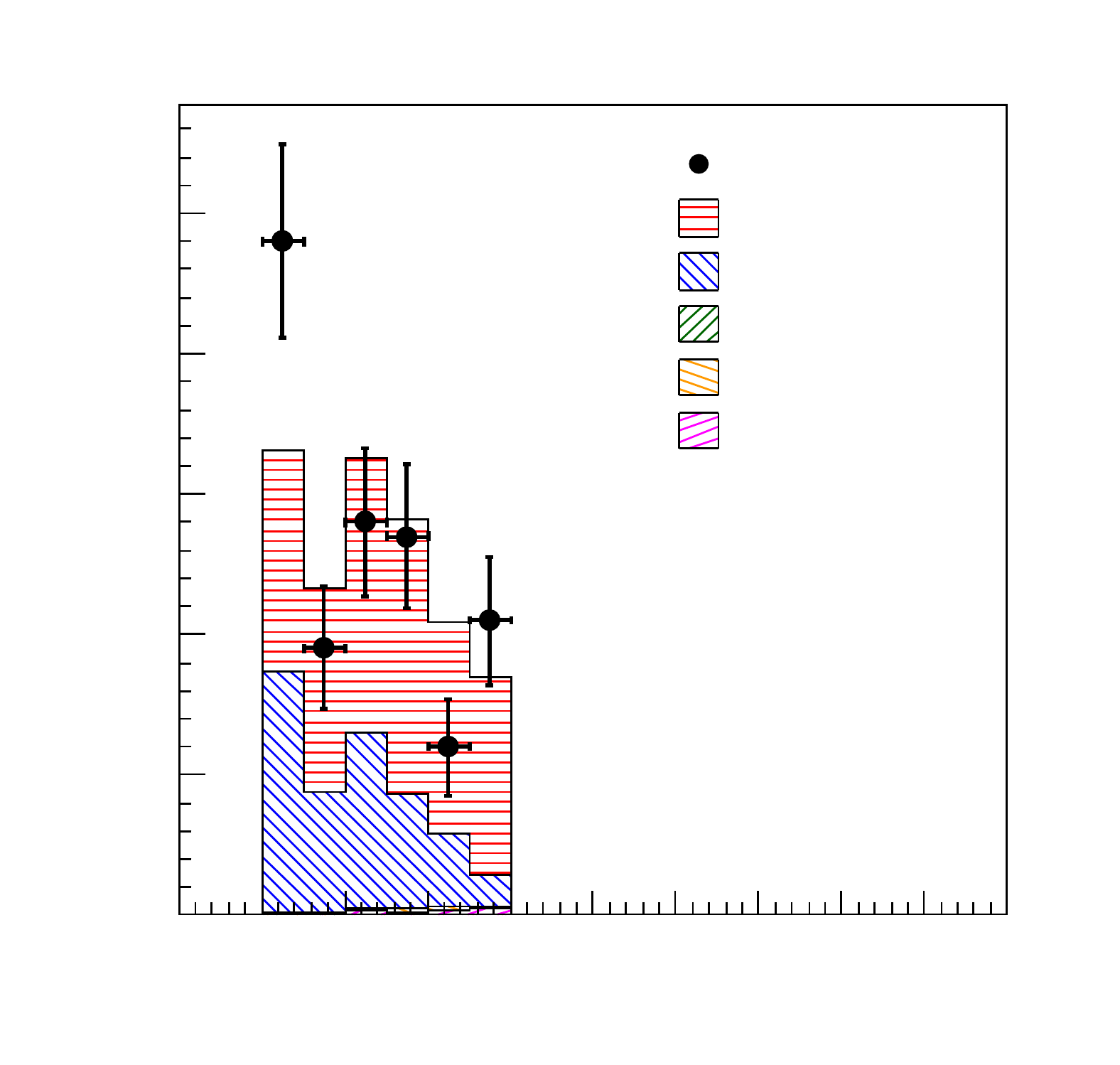}  & \svg{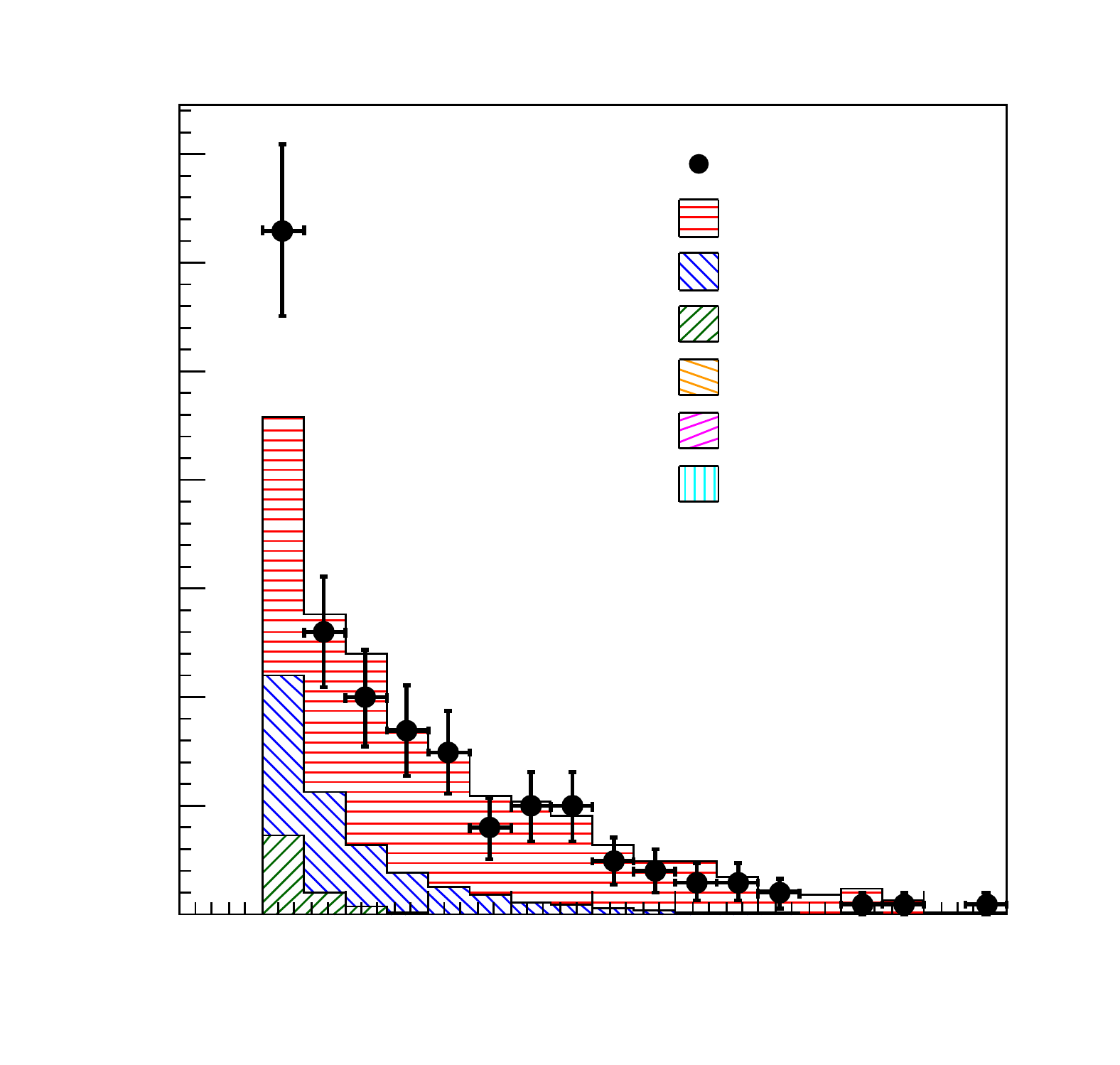} \svgsep
  \svg{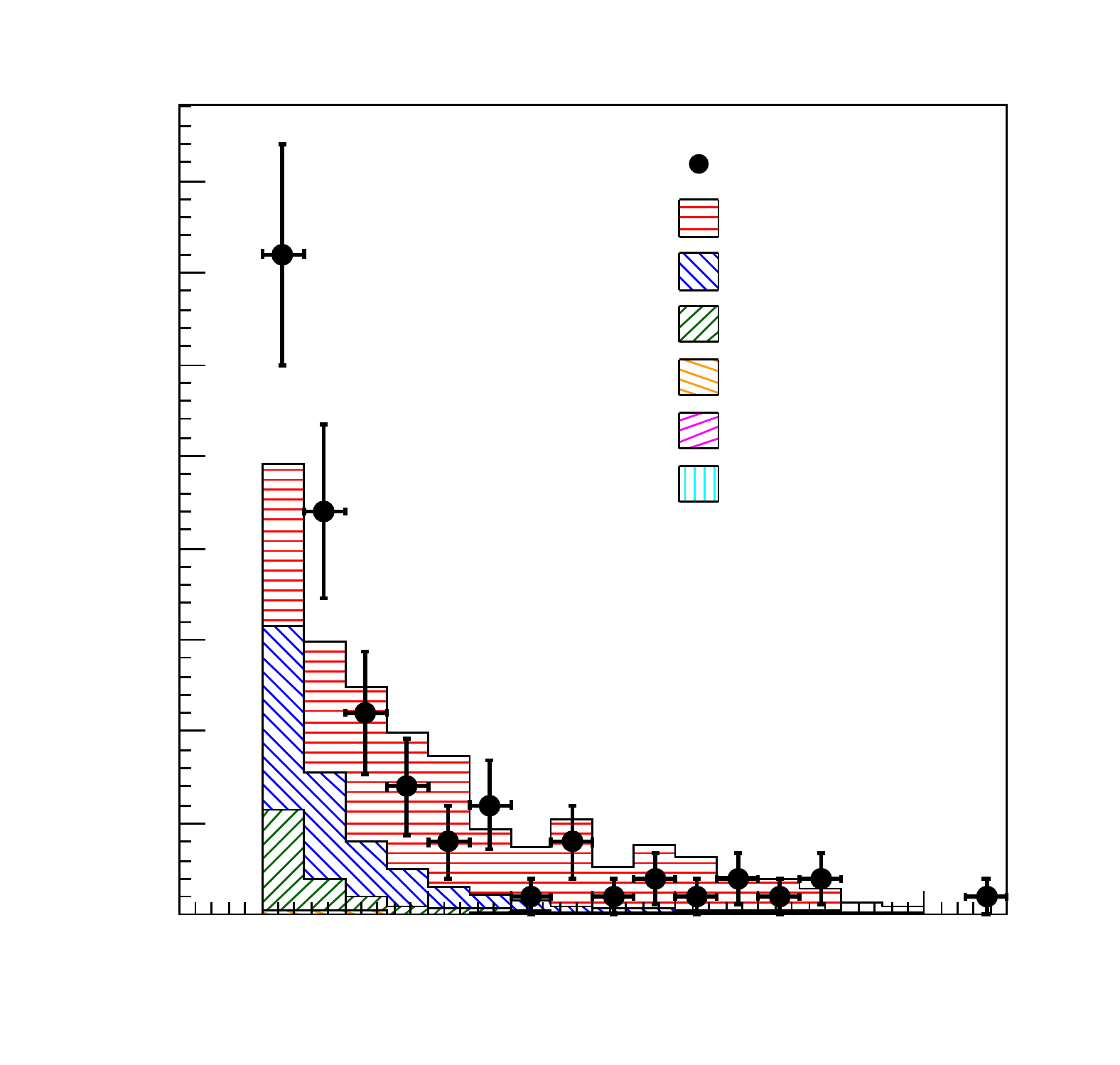}  & \sidecaption            \svgend
\end{subfigures}

\begin{subfigures}[p]{2}{The pseudorapidity, $\eta$, distributions for
    the first \wtl decay product candidate from data (points) for the
    \subfig{Z2TauTau2MuMu.Eta1}~\mumu,
    \subfig{Z2TauTau2MuE.Eta1}~\mue,
    \subfig{Z2TauTau2EMu.Eta1}~\emu,
    \subfig{Z2TauTau2MuPi.Eta1}~\muh, and
    \subfig{Z2TauTau2EPi.Eta1}~\eh categories. The simulated
    signal (red) is normalised to the number of signal events, while
    the \qcd (green), \ewk (blue), \ttbar (orange), \ww (magenta), and
    ${\z \to \dilep}$ (cyan) backgrounds are estimated as described in
    \sec{Zed:Bkg}.\labelfig{Eta1}}
  \svgbeg
  \svg{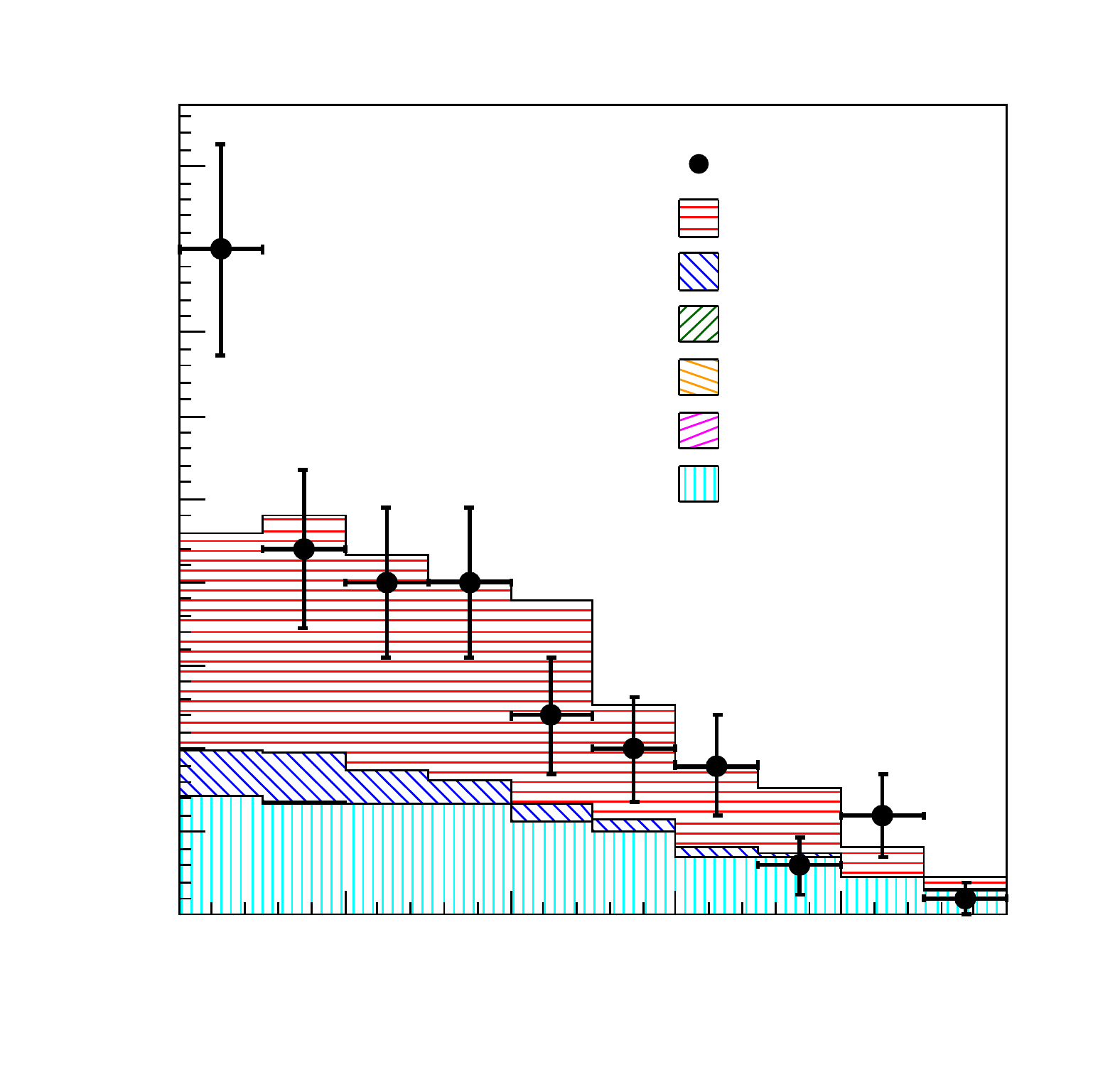} & \svg{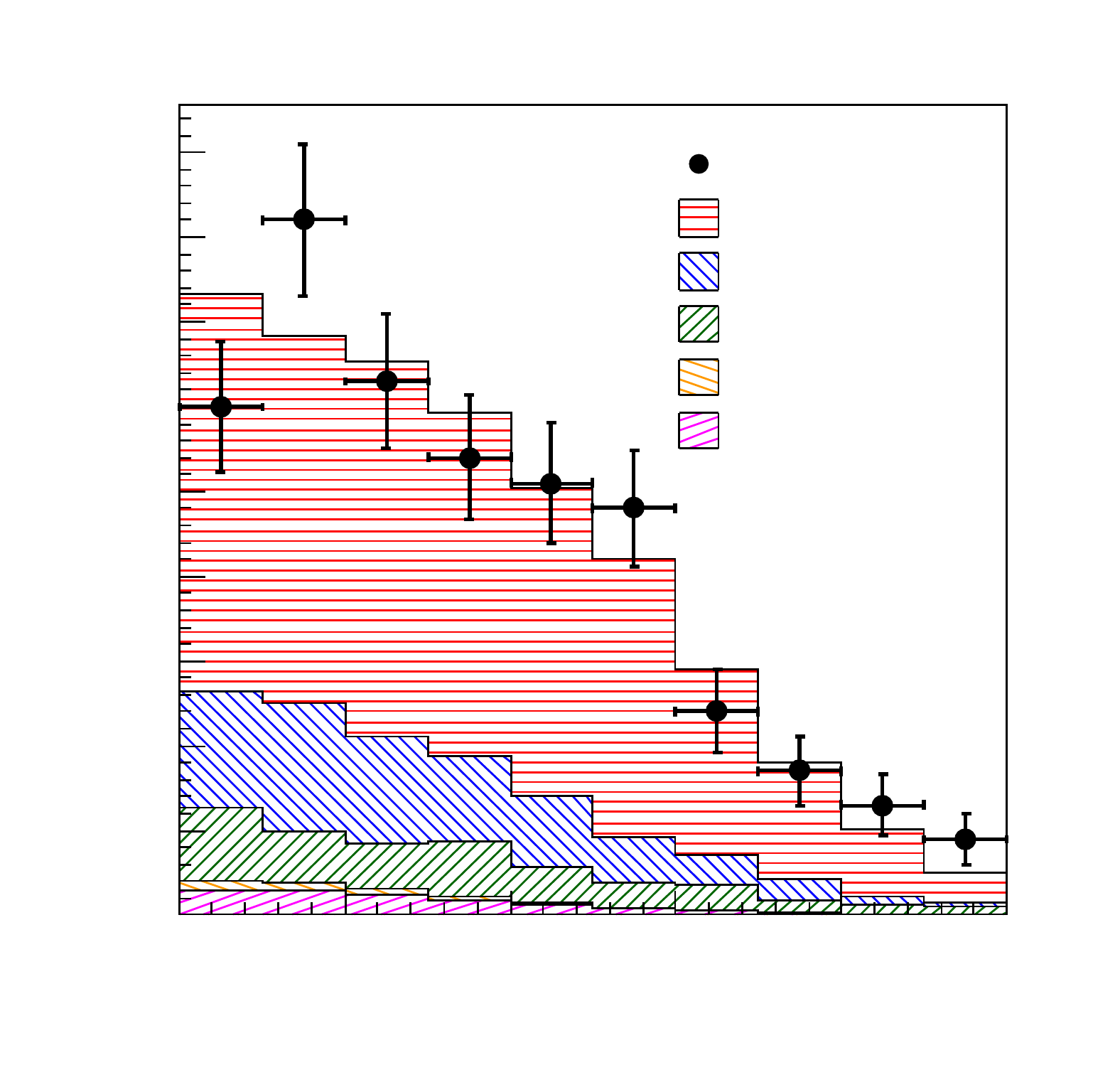}  \svgsep
  \svg{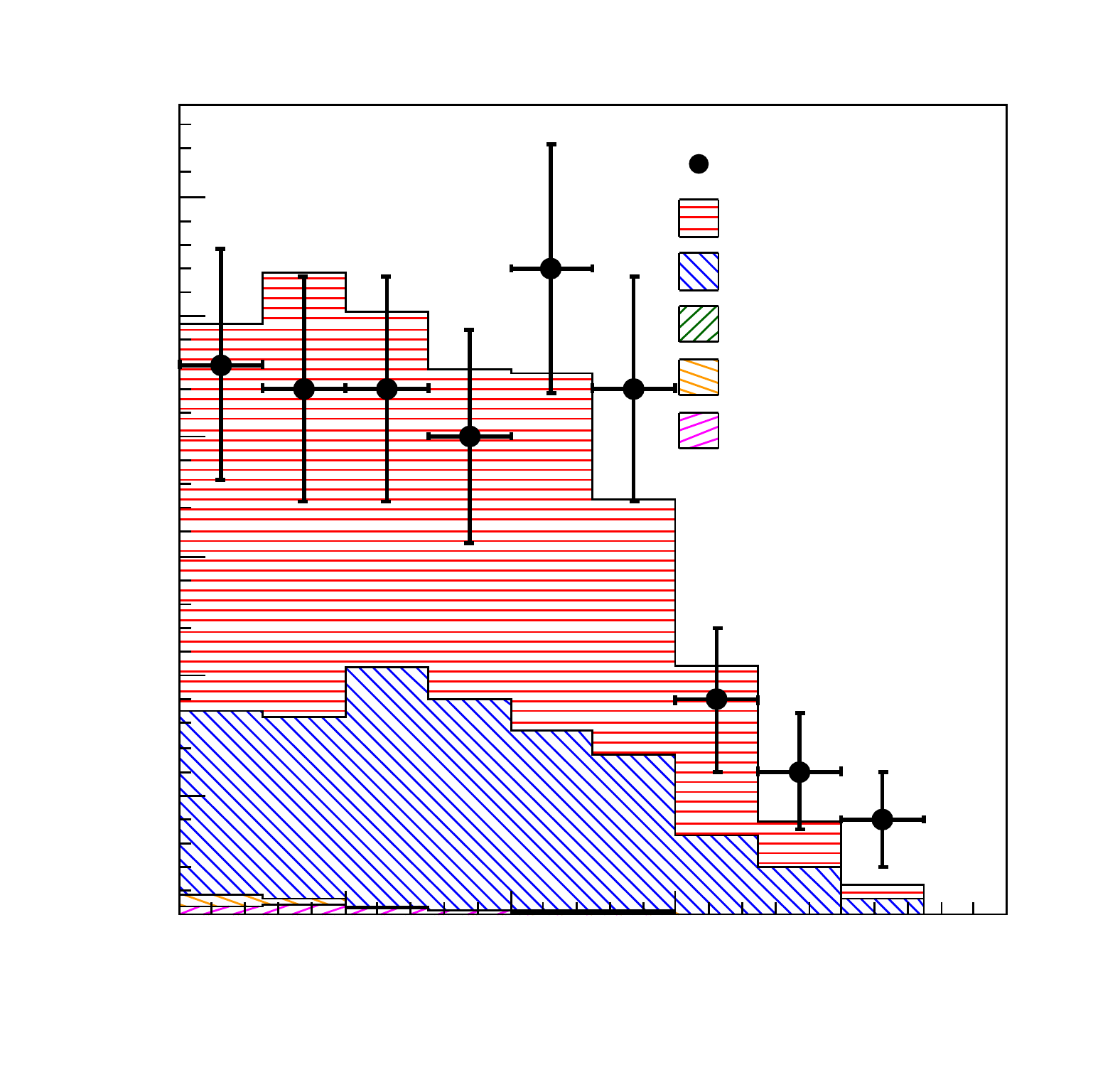}  & \svg{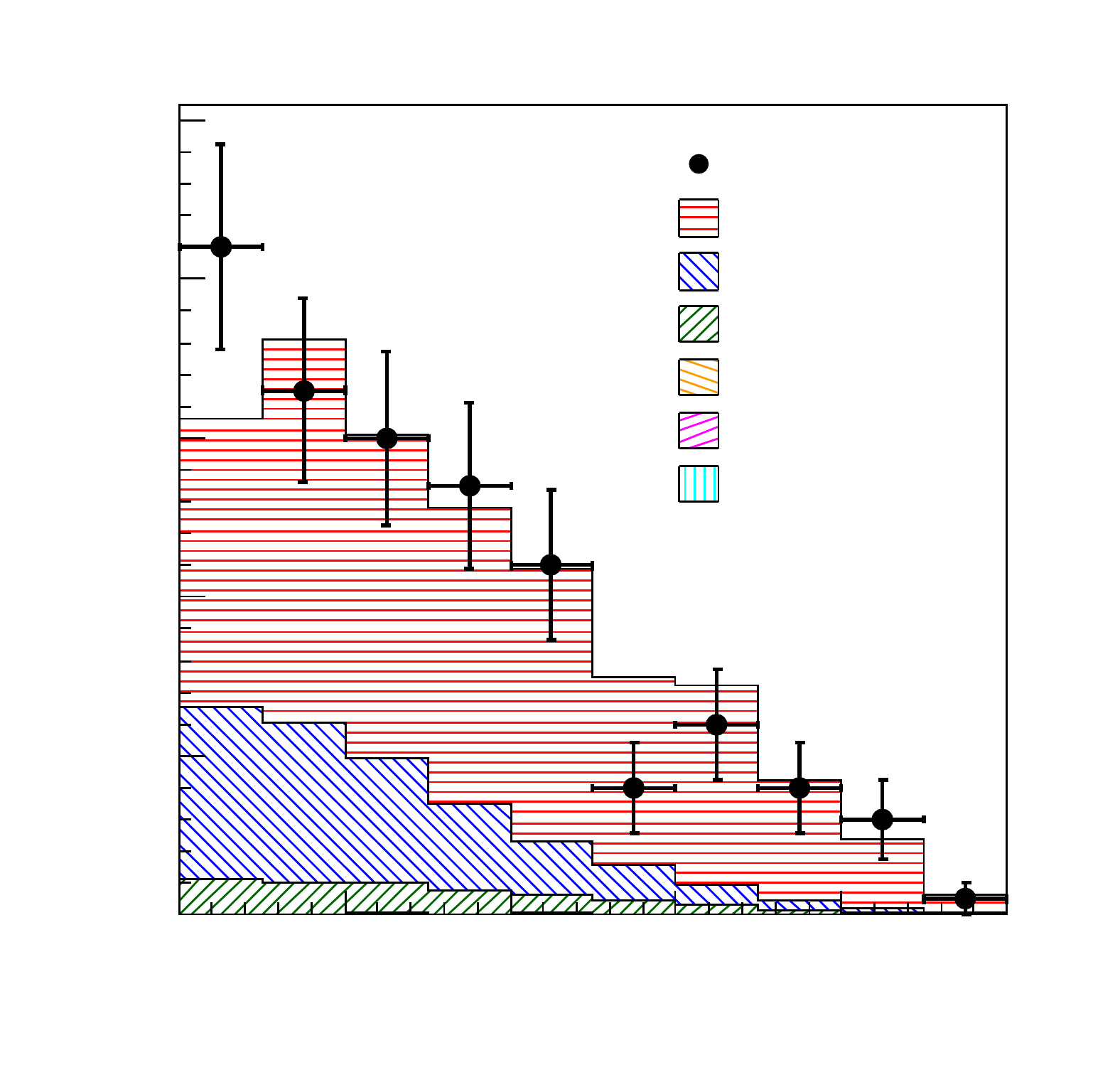} \svgsep
  \svg{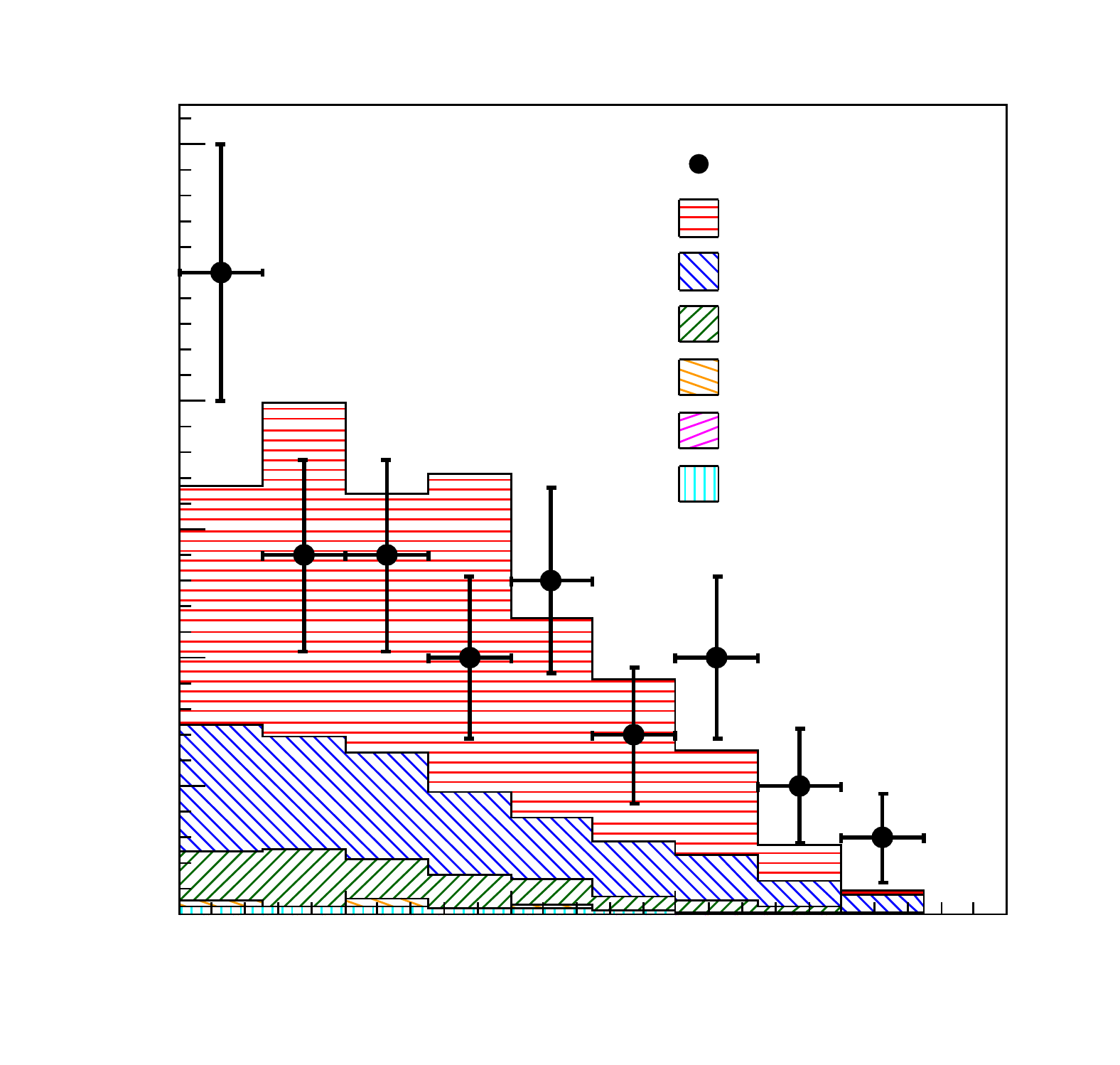}  & \sidecaption             \svgend
\end{subfigures}

\begin{subfigures}[p]{2}{The pseudorapidity, $\eta$, distributions for
    the second \wtl decay product candidate from data (points) for the
    \subfig{Z2TauTau2MuMu.Eta2}~\mumu,
    \subfig{Z2TauTau2MuE.Eta2}~\mue,
    \subfig{Z2TauTau2EMu.Eta2}~\emu,
    \subfig{Z2TauTau2MuPi.Eta2}~\muh, and
    \subfig{Z2TauTau2EPi.Eta2}~\eh categories. The simulated
    signal (red) is normalised to the number of signal events, while
    the \qcd (green), \ewk (blue), \ttbar (orange), \ww (magenta), and
    ${\z \to \dilep}$ (cyan) backgrounds are estimated as described in
    \sec{Zed:Bkg}.\labelfig{Eta2}}
  \svgbeg
  \svg{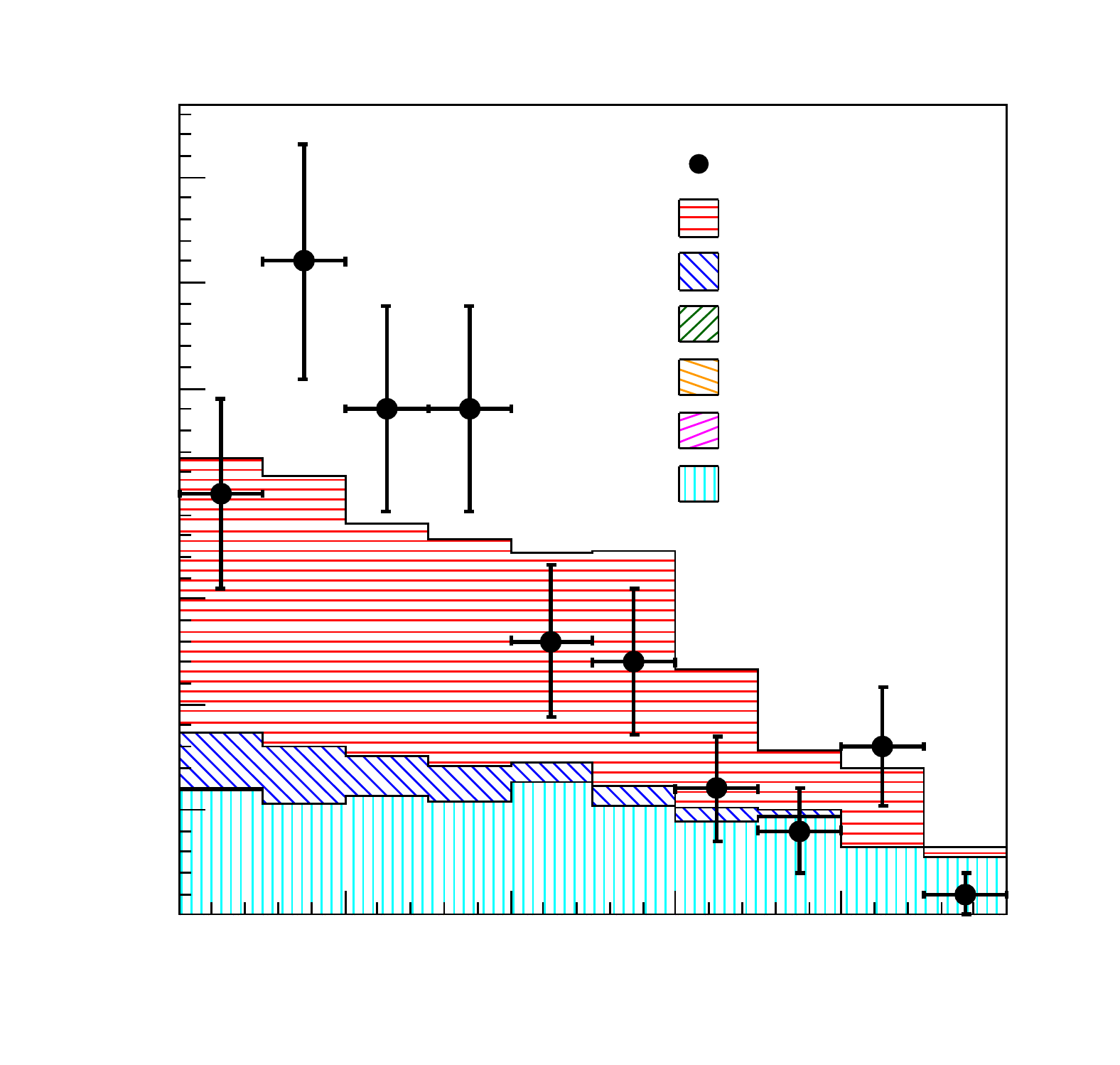} & \svg{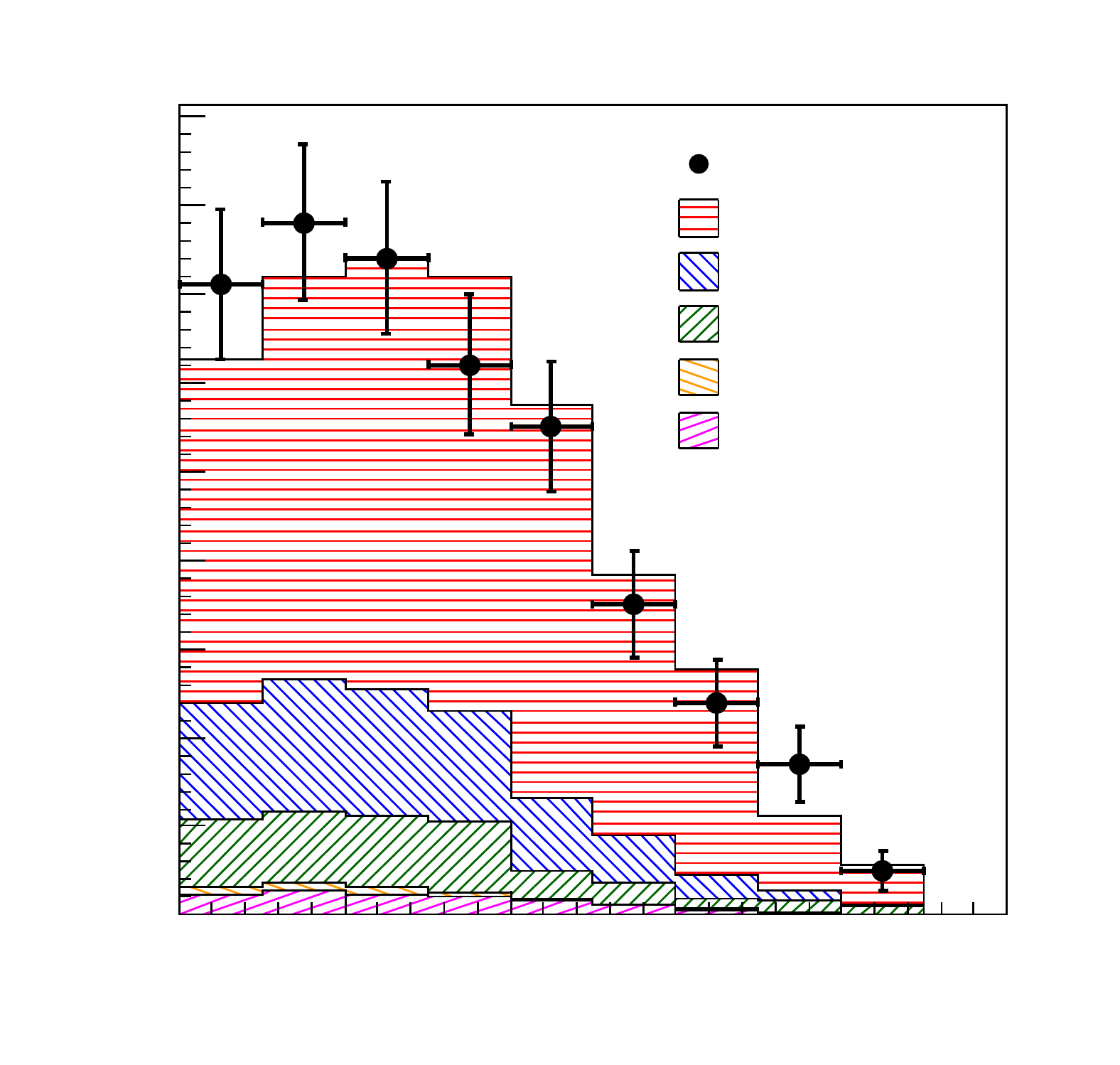}  \svgsep
  \svg{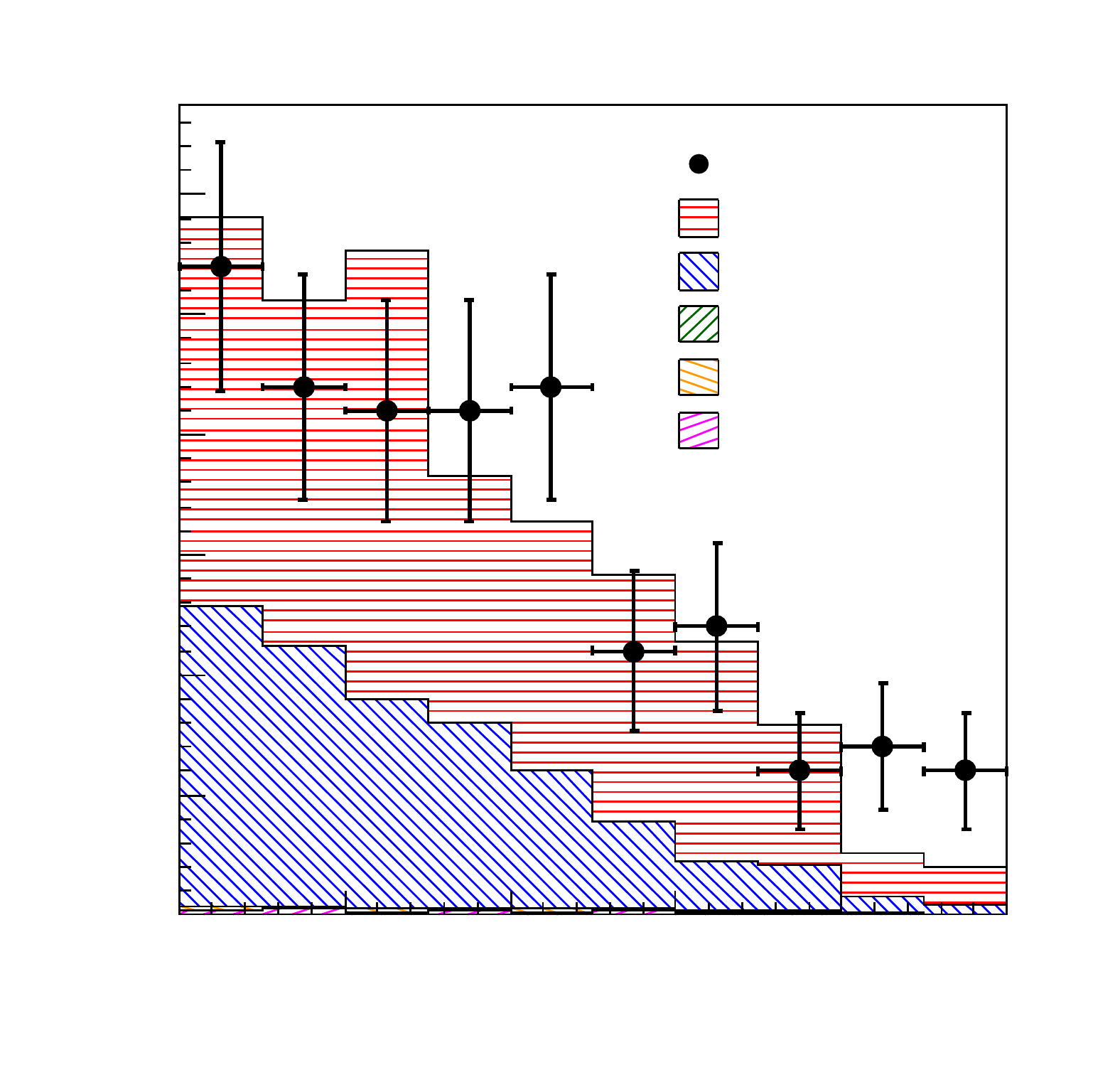}  & \svg{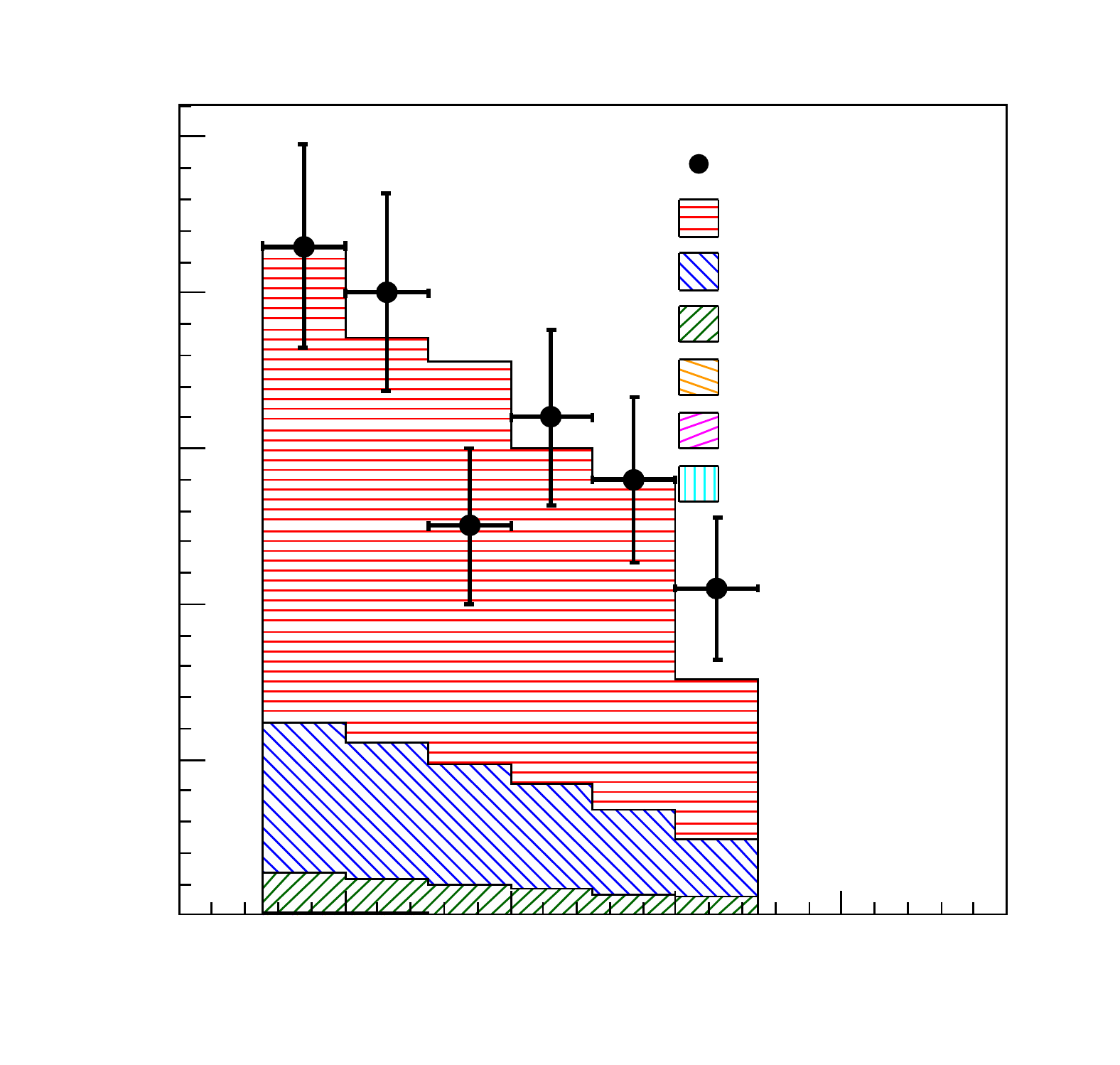} \svgsep
  \svg{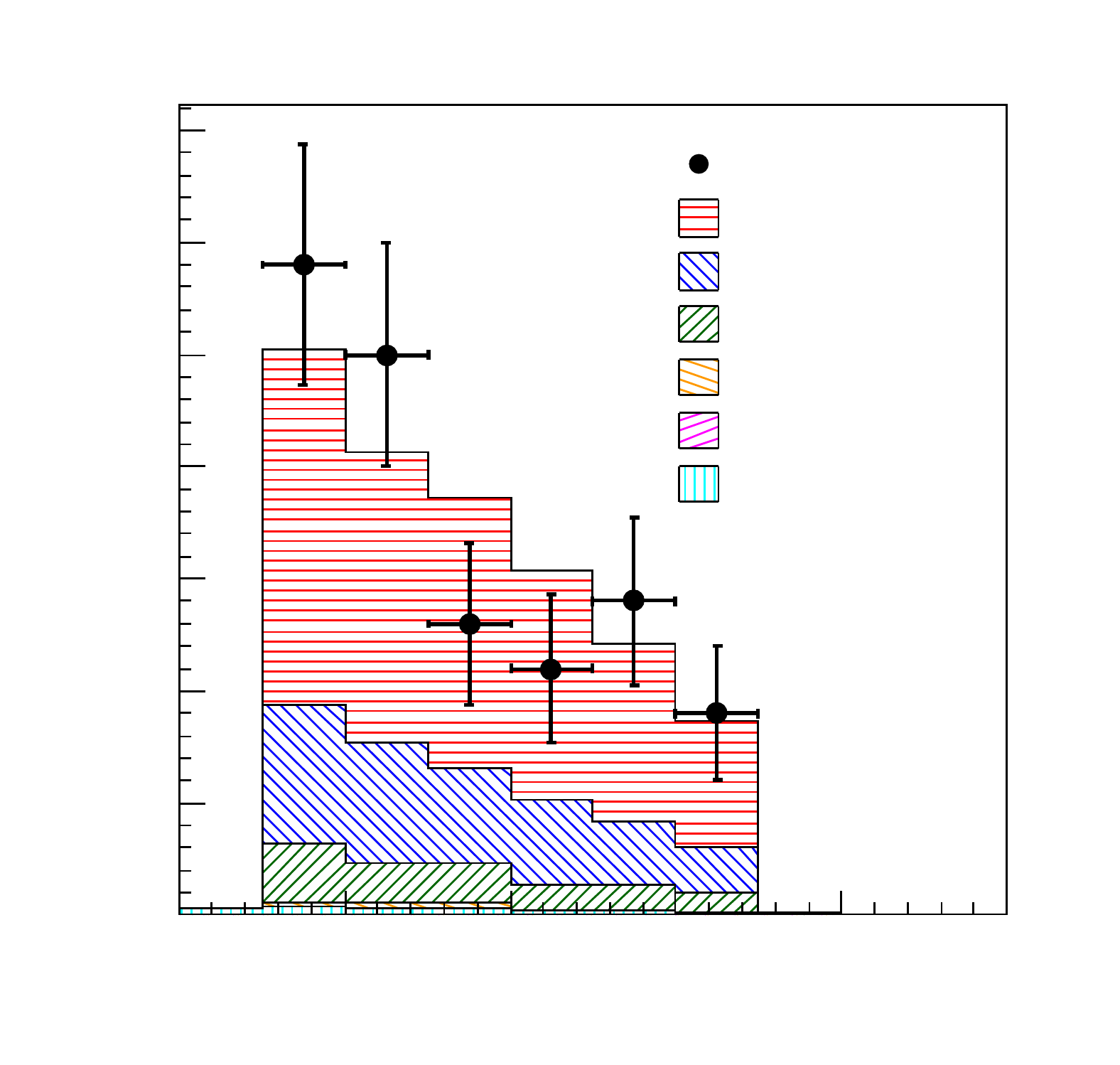}  & \sidecaption             \svgend
\end{subfigures}

\newsection{Efficiencies}{RecEff}

The low momentum, $0$ to $150~\gev$, muon track finding and
identification efficiencies are determined using a tag-and-probe
method on ${\jpsi \to \dimu}$ events selected from data, as described
in \sec{Zed:RecEff}.These events, unlike the ${\z \to \dimu}$ events
used for the high momentum muon efficiencies, are contaminated with a
significant background contribution. Additionally, because these
events do not have the same pseudo-rapidity distribution as ${\z \to
  \ditau}$ signal events, the efficiencies from ${\jpsi \to \dimu}$
events must be determined as a function of both the momentum and
pseudo-rapidity of the probe muon candidate, $\eta_\mu$ and $p_\mu$.

The ${\jpsi \to \dimu}$ tag-and-probe events are separated into twelve
bins, three bins in pseudo-rapidity for each of the four bins in
momentum. The pseudo-rapidity bin widths are $5/6$ of $\eta$ and range
from $2.0$ to $4.5$, while the momentum bin widths are $50~\gev$ and
range from $0$ to $200~\gev$. The last momentum bin, ${150 < p_\mu <
  200~\gev}$, is only used as a consistency check and is not used for
final efficiencies. The distributions of the di-muon invariant mass
for each of the twelve bins are shown in \fig{Zvr:RecEff.MuTrk.Total}
for the muon track finding efficiency before the track selection
criteria is applied to the probe. The \jpsi resonance is clearly
visible, with its width increasing for larger momenta due to a
decrease in momentum resolution. This is particularly visible in the
final pseudo-rapidity column, ${3.67 < \eta_\mu < 4.50}$.

The same distributions are given in \fig{Zvr:RecEff.MuTrk.Pass}, but
now for events where the probe muon candidate passes the track
selection requirement. As expected, there are fewer events in these
distributions than the distributions of
\fig{Zvr:RecEff.MuTrk.Total}. The same plots, but for muon
identification, are given in \fig{Zvr:RecEff.MuId.Total} before
requiring the probe muon candidate pass the muon identification of
\sec{Zed:Rec}, and after, in \fig{Zvr:RecEff.MuId.Pass}. The \jpsi
resonances in these distributions are much narrower as the probe track
has a much better resolution than the probe muon-track for the track
finding efficiency. The widening of the \jpsi resonance due to a
decrease in momentum resolution can also be seen in both of these
plots.

\begin{subfigures}[p]{4}{Crystal ball fits (dashed red) with a linear
    background (dotted blue) of the ${\jpsi \to \dimu}$ data (points)
    used to determine the muon track finding efficiency, before
    requiring a reconstructed track. The fits are given in three bins
    of pseudo-rapidity between $2.0$ and $4.5$ and four bins of
    momentum between $0$ and $200~\gev$.\labelfig{RecEff.MuTrk.Total}}
  \effbeg
  && $\quad\quad2.00 < \eta_\mu < 2.83$ 
  & $\quad\quad2.83 < \eta_\mu < 3.67$ 
  & $\quad\quad3.67 < \eta_\mu < 4.50$ \\
  \midrule
  \svg[$0 < p_\mu < 50~\gev$]{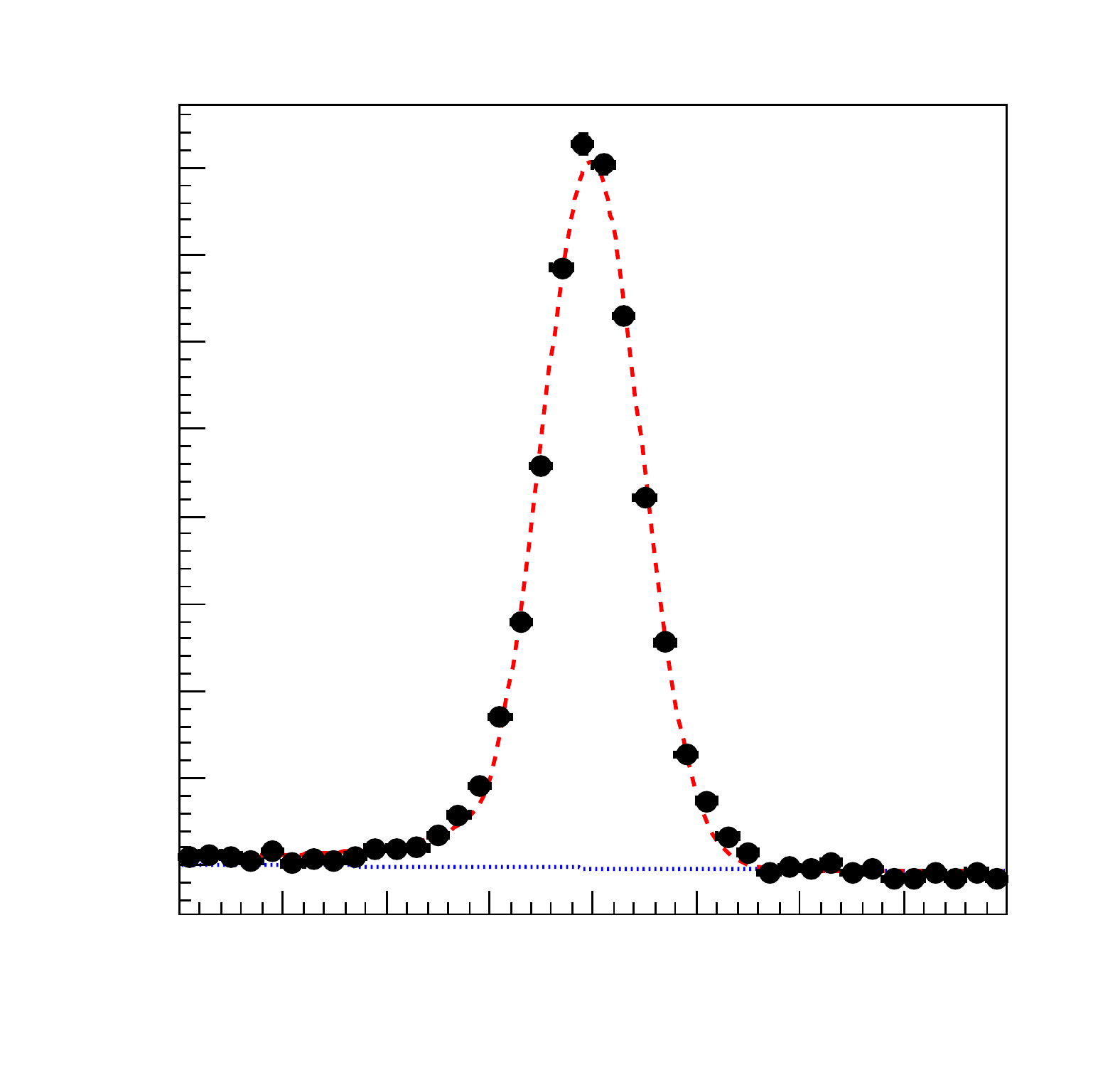}
  & \svg[2]{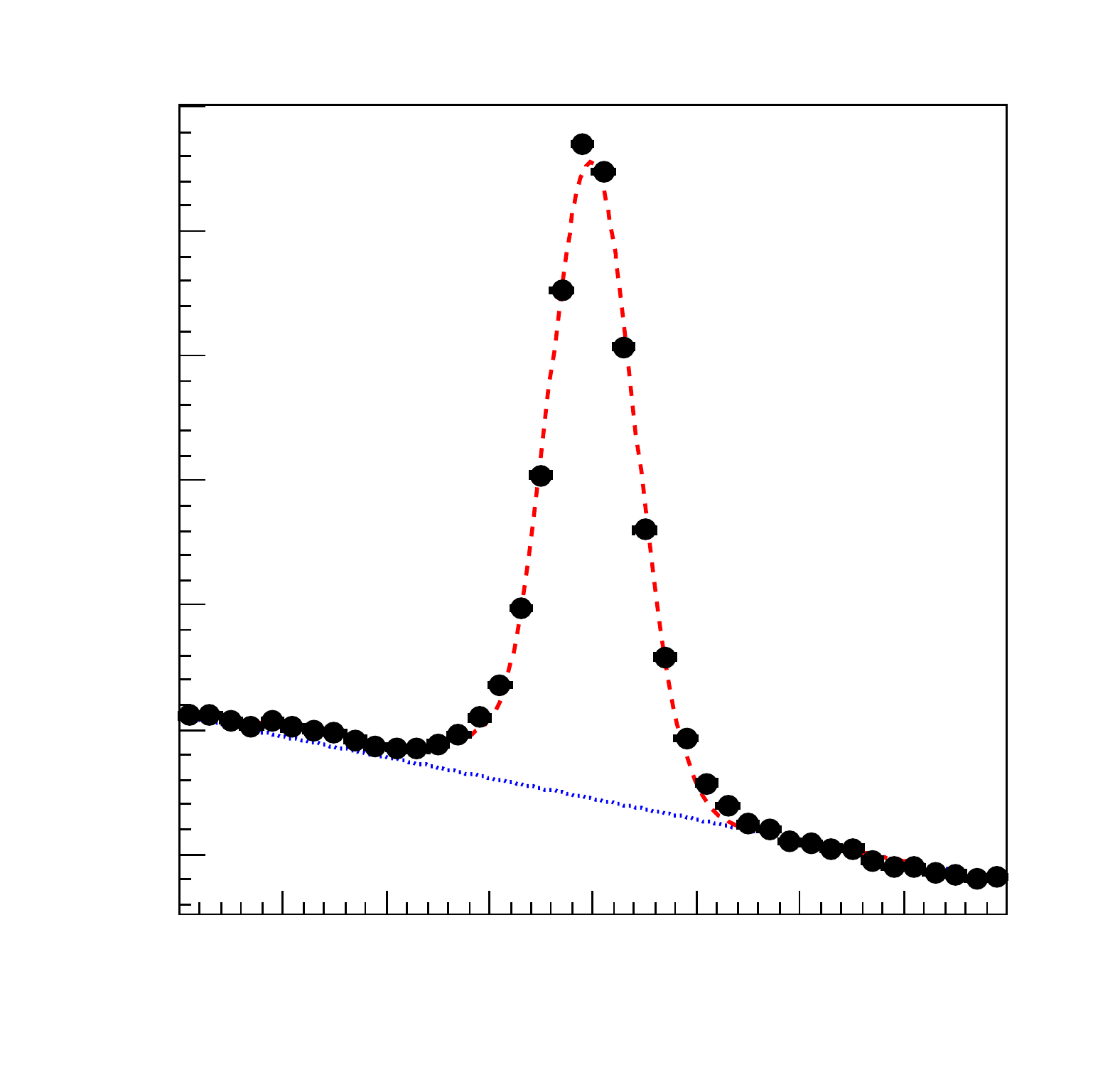}
  & \svg[2]{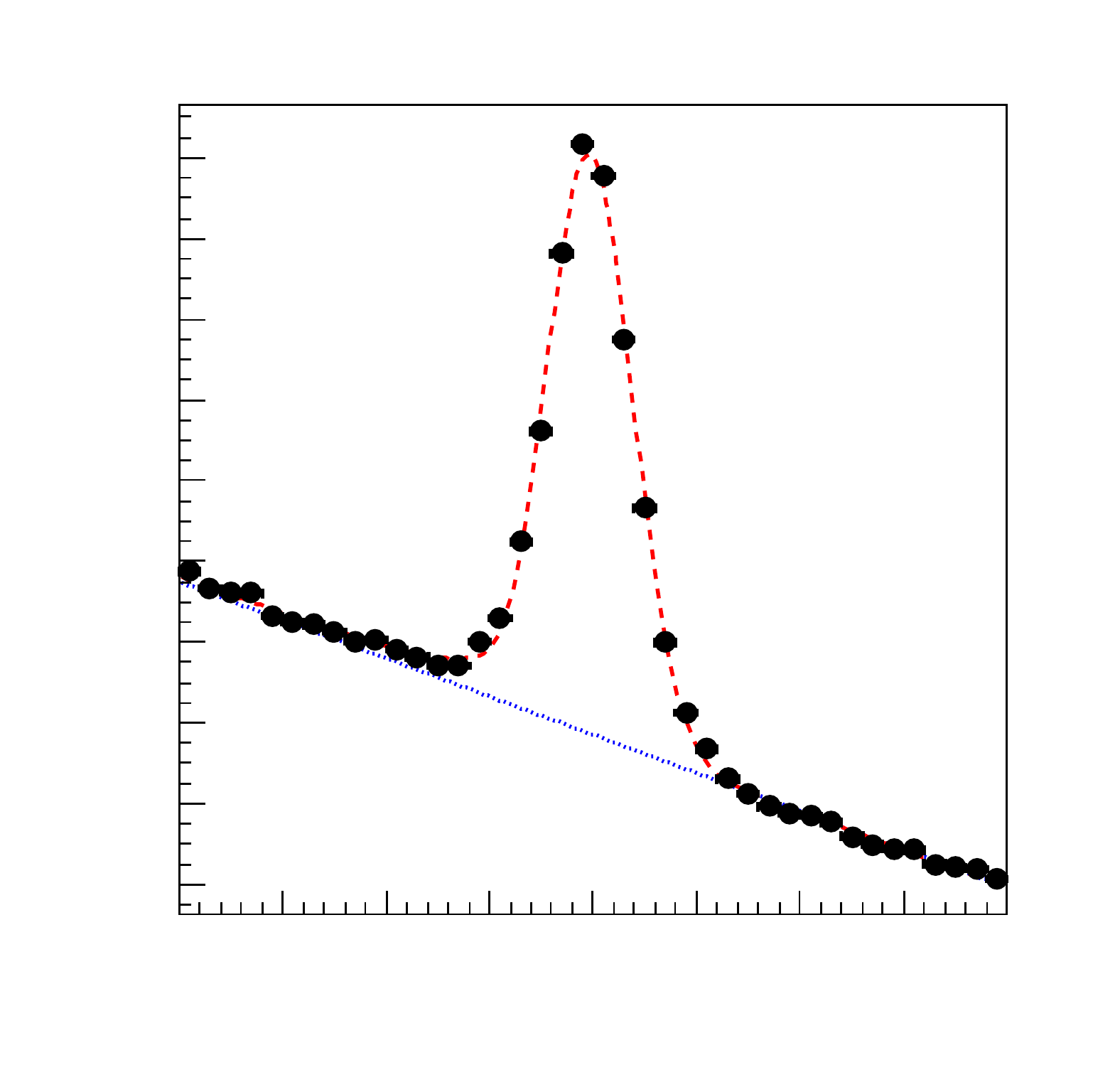} \effsep
  \svg[$50 < p_\mu < 100~\gev$]{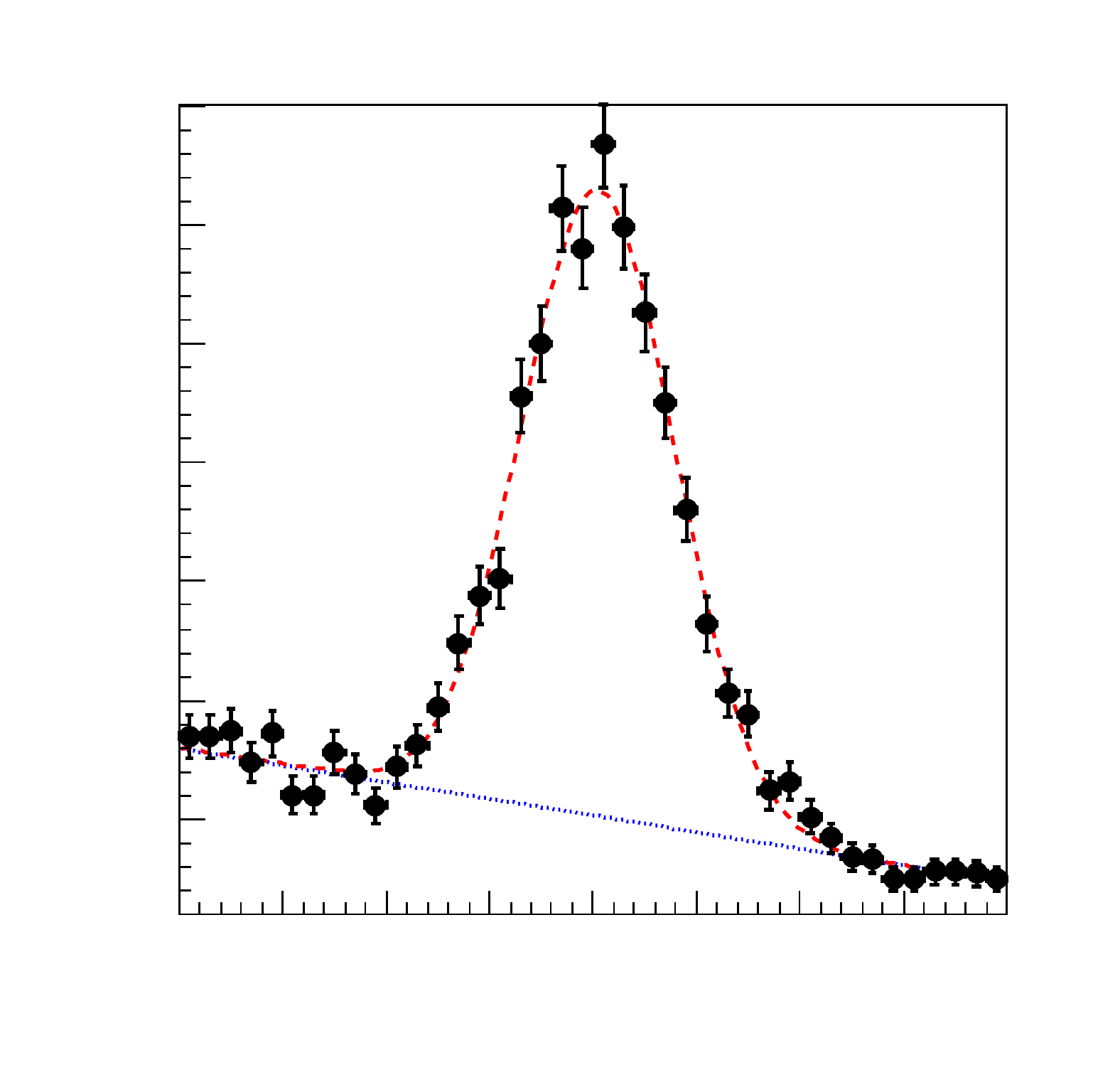}
  & \svg[2]{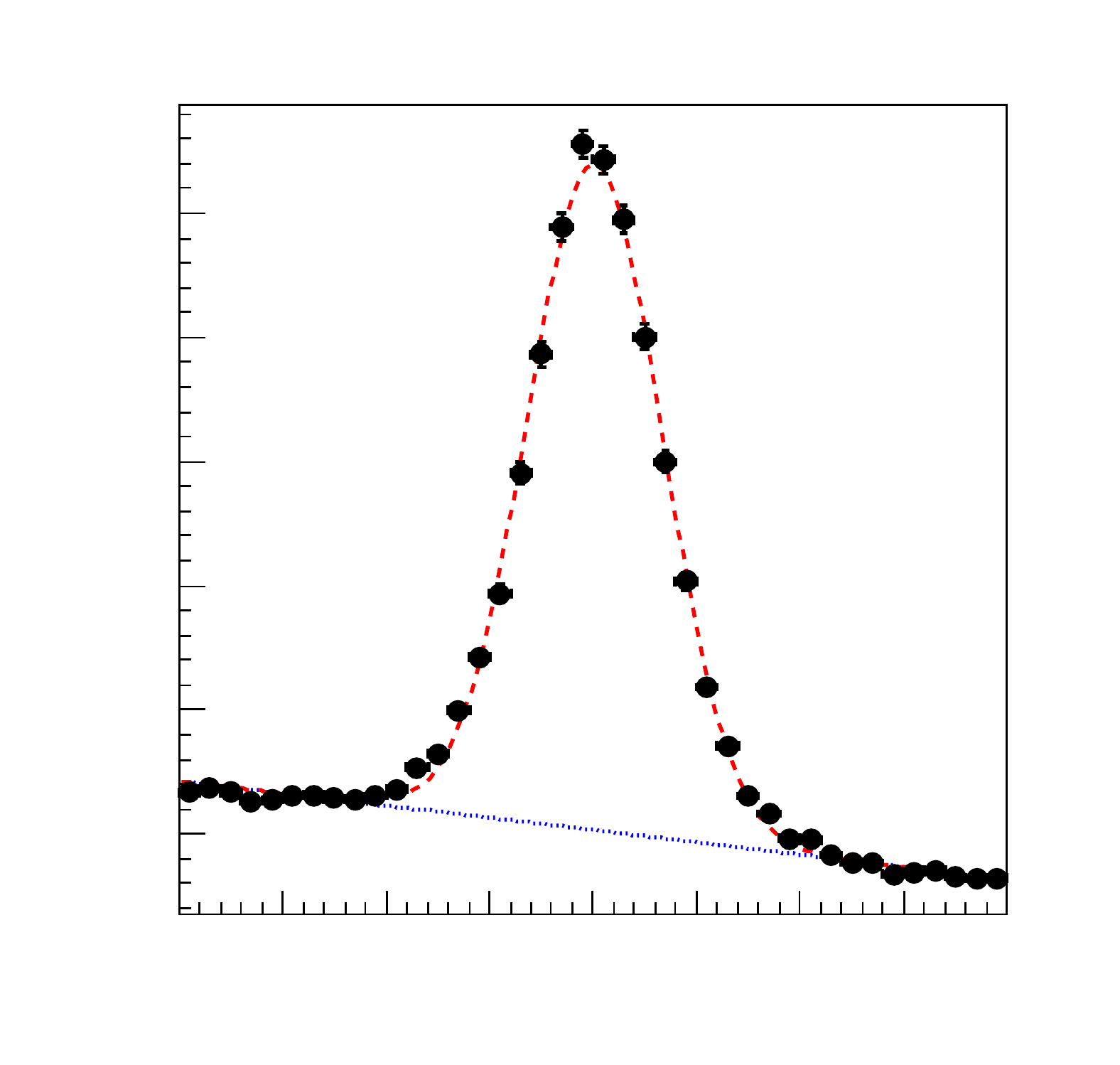}
  & \svg[2]{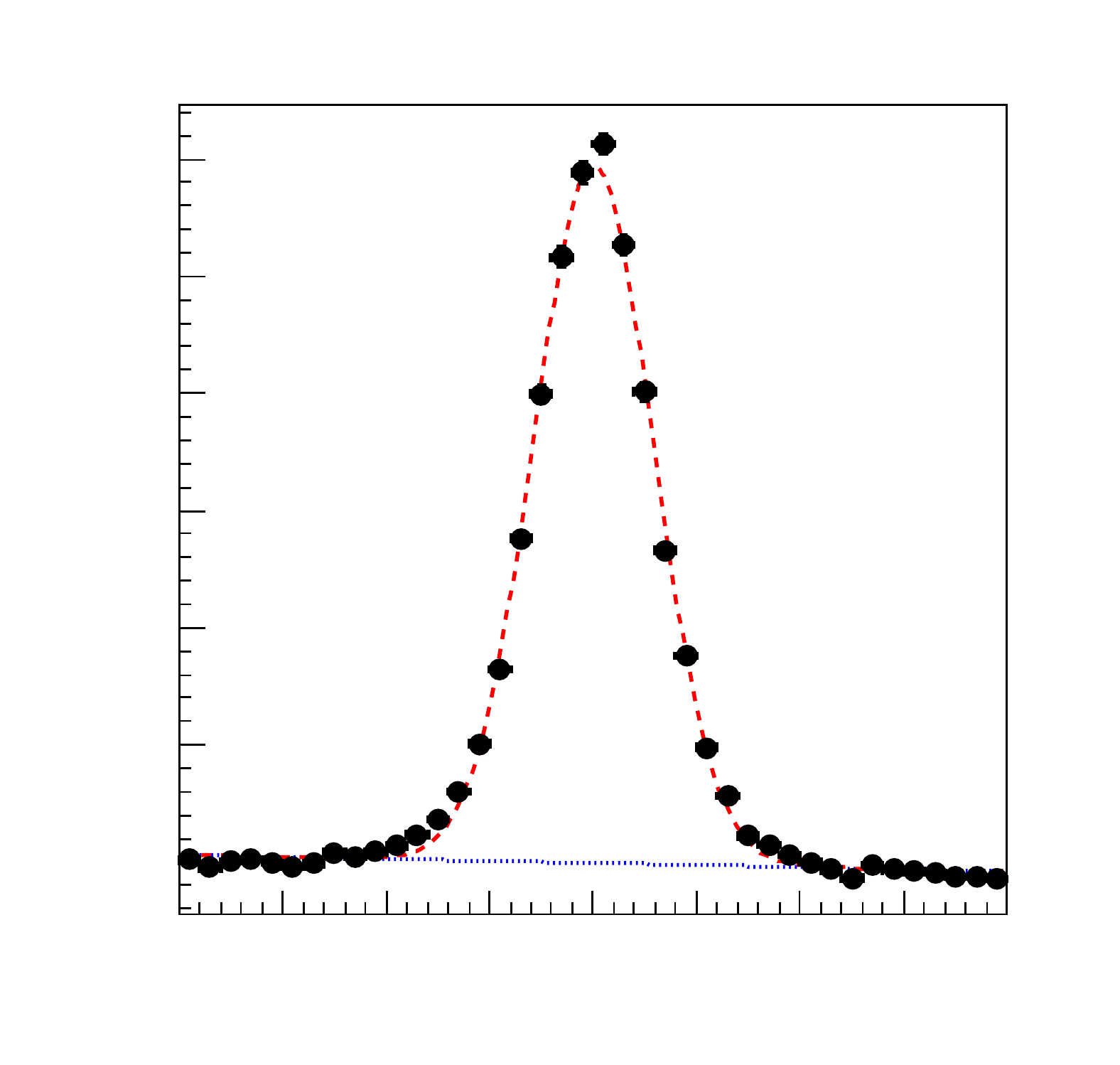} \effsep
  \svg[$100 < p_\mu < 150~\gev$]{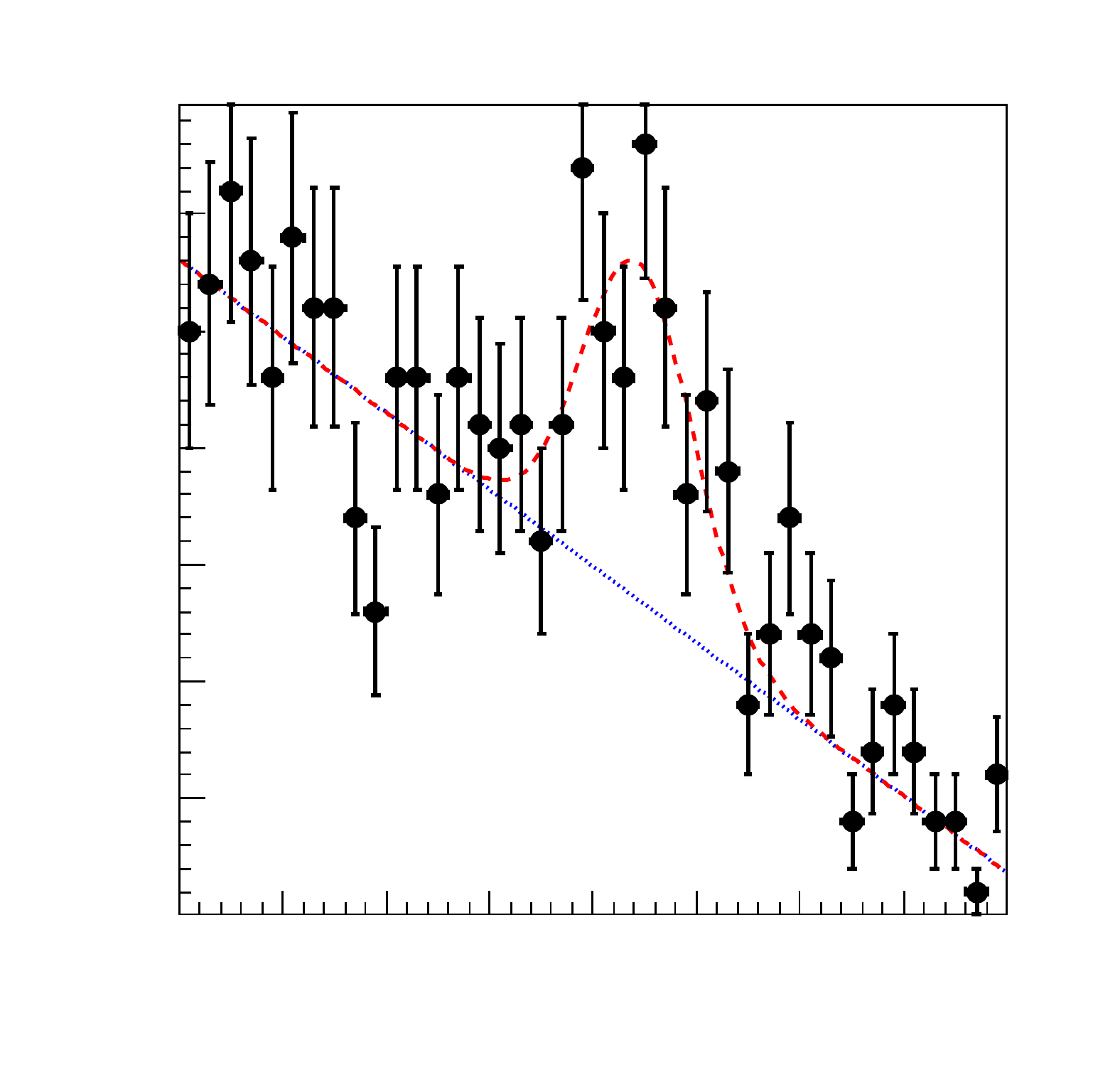}
  & \svg[2]{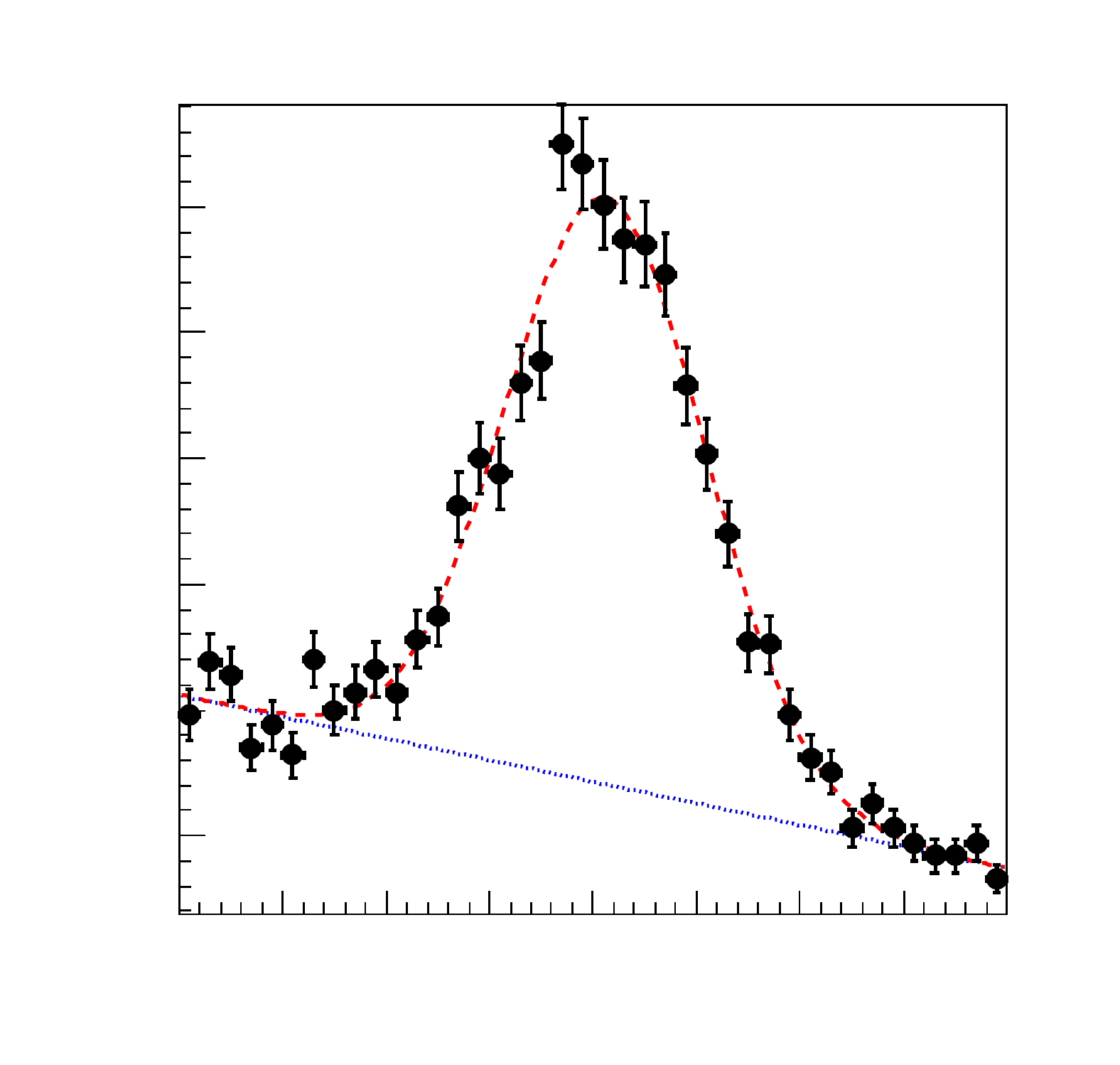}
  & \svg[2]{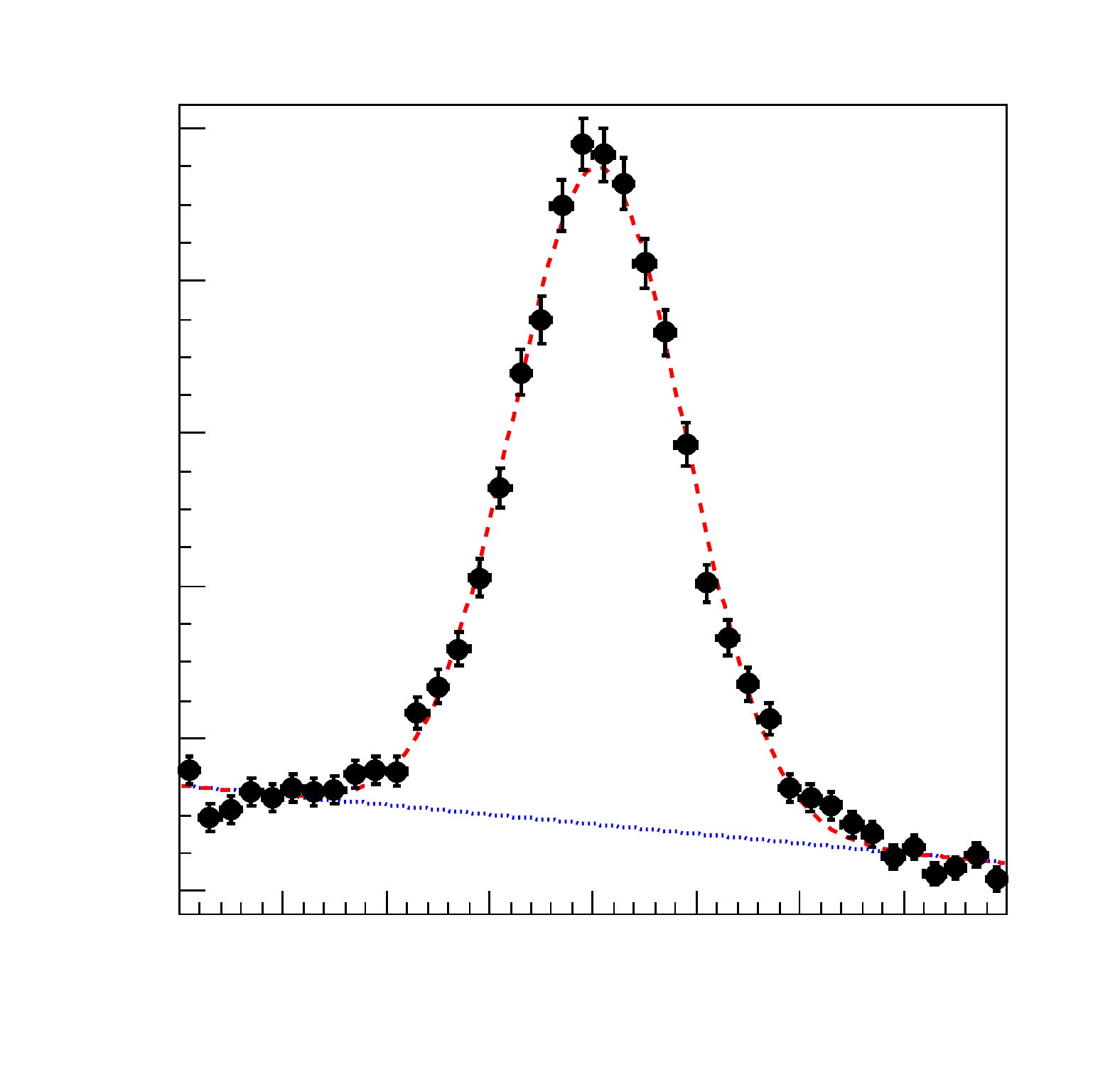} \effsep
  \svg[$150 < p_\mu < 200~\gev$]{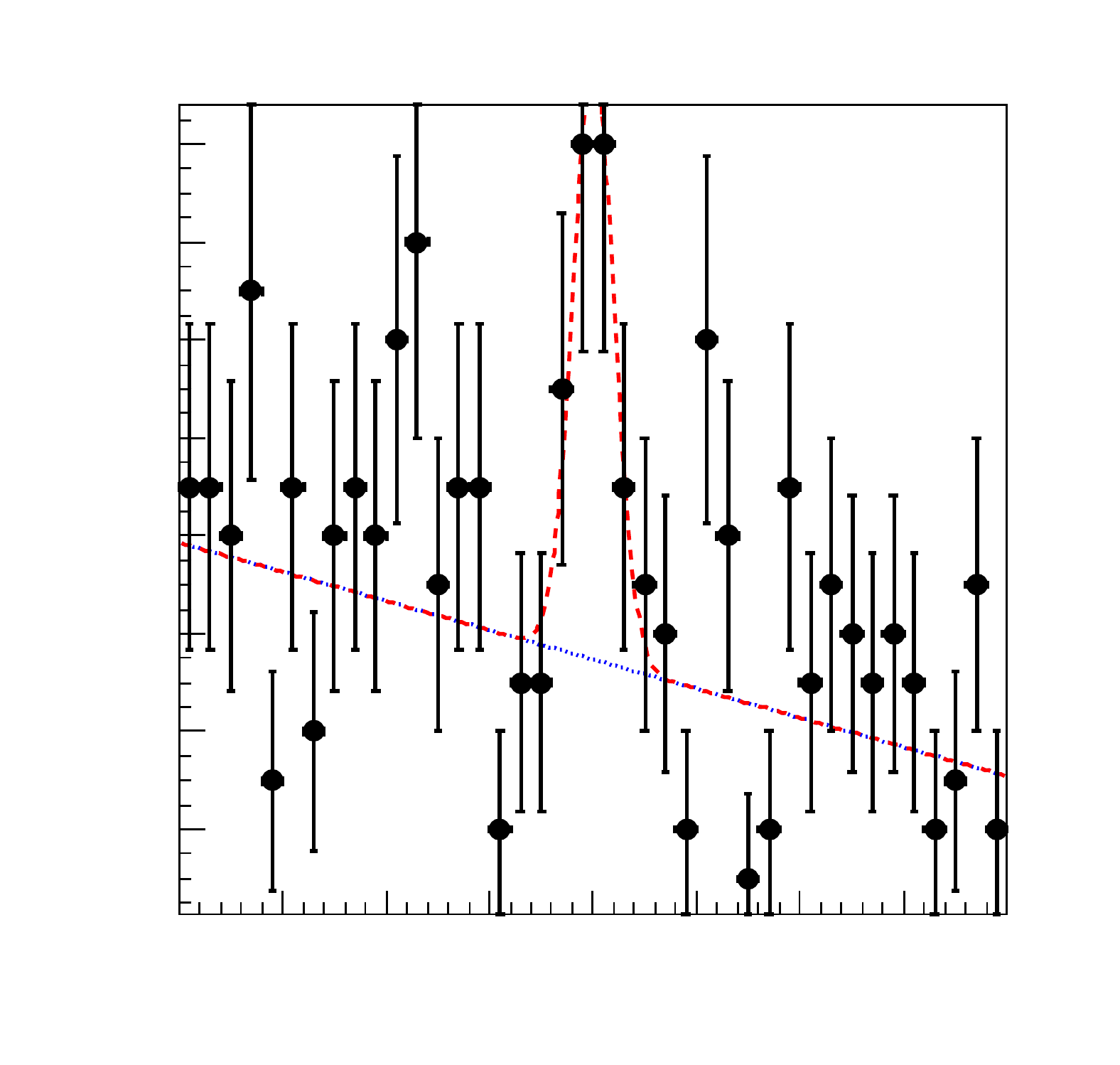}
  & \svg[2]{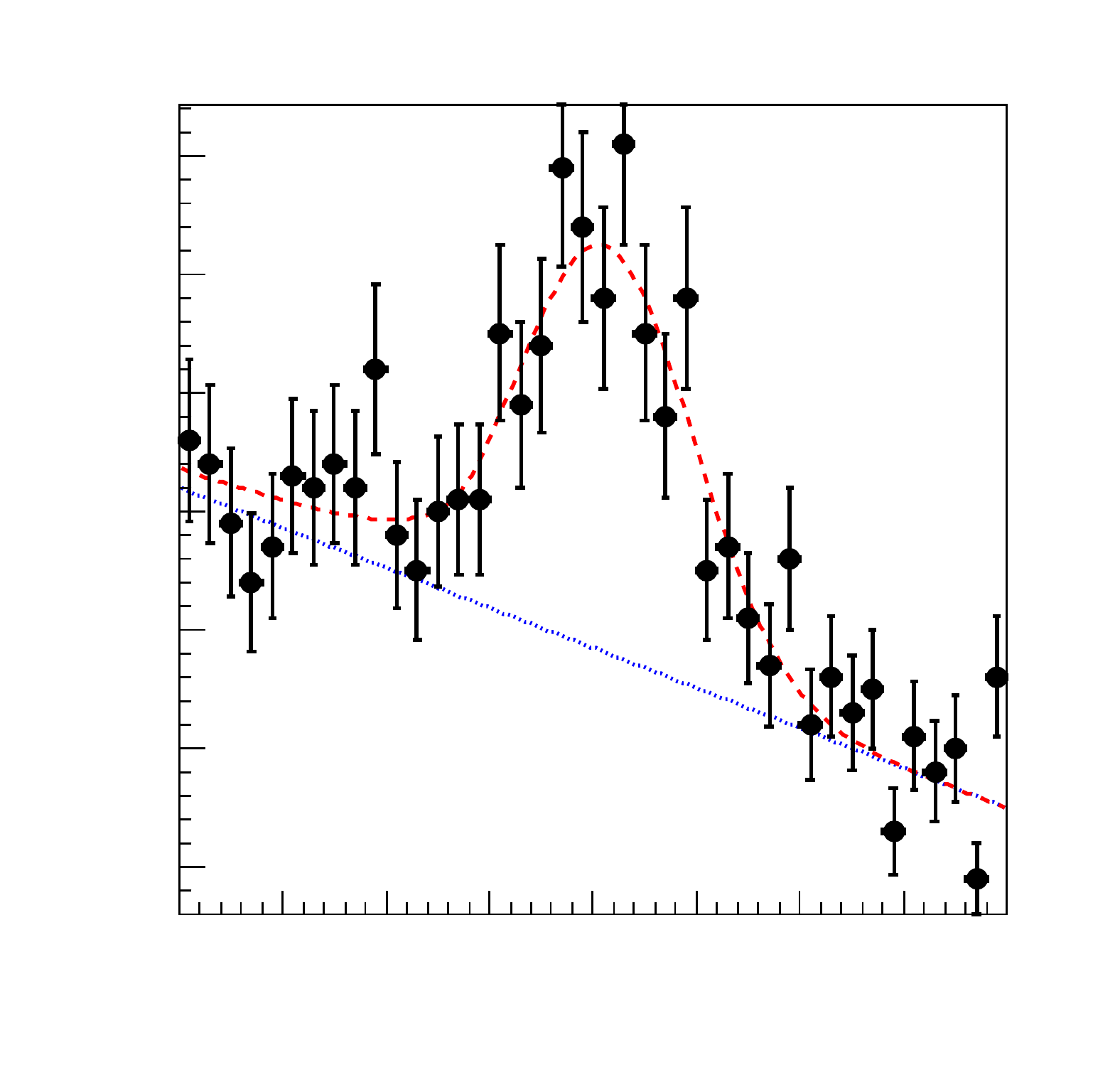}
  & \svg[2]{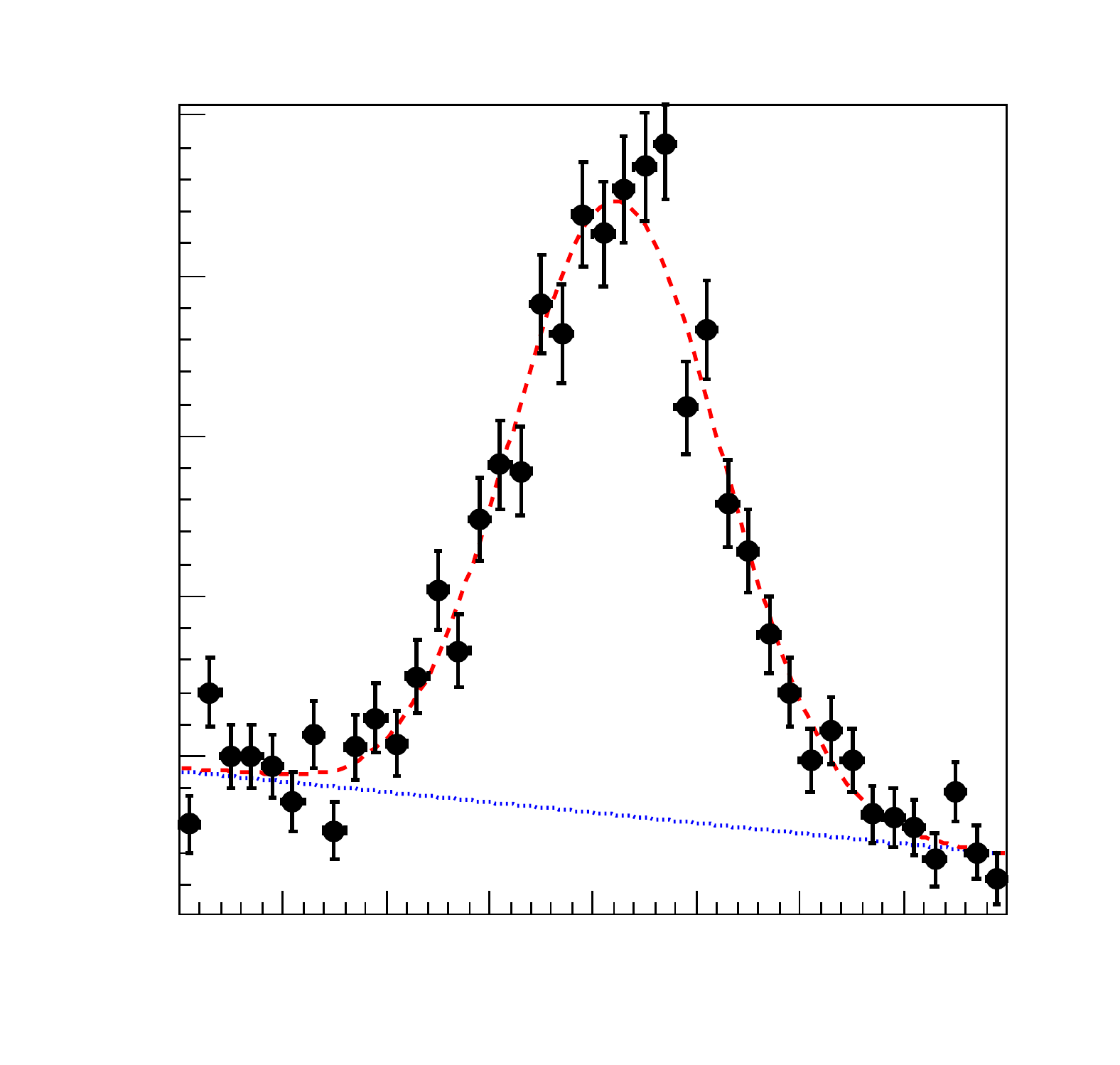} \effend
\end{subfigures}

\begin{subfigures}[p]{4}{Crystal ball fits (dashed red) with a linear
    background (dotted blue) of the ${\jpsi \to \dimu}$ data (points)
    used to determine the muon track finding efficiency, after
    requiring a reconstructed track. The fits are given in three bins
    of pseudo-rapidity between $2.0$ and $4.5$ and four bins of
    momentum between $0$ and $200~\gev$.\labelfig{RecEff.MuTrk.Pass}}
  \effbeg
  && $\quad\quad2.00 < \eta_\mu < 2.83$ 
  & $\quad\quad2.83 < \eta_\mu < 3.67$ 
  & $\quad\quad3.67 < \eta_\mu < 4.50$ \\
  \midrule
  \svg[$0 < p_\mu < 50~\gev$]{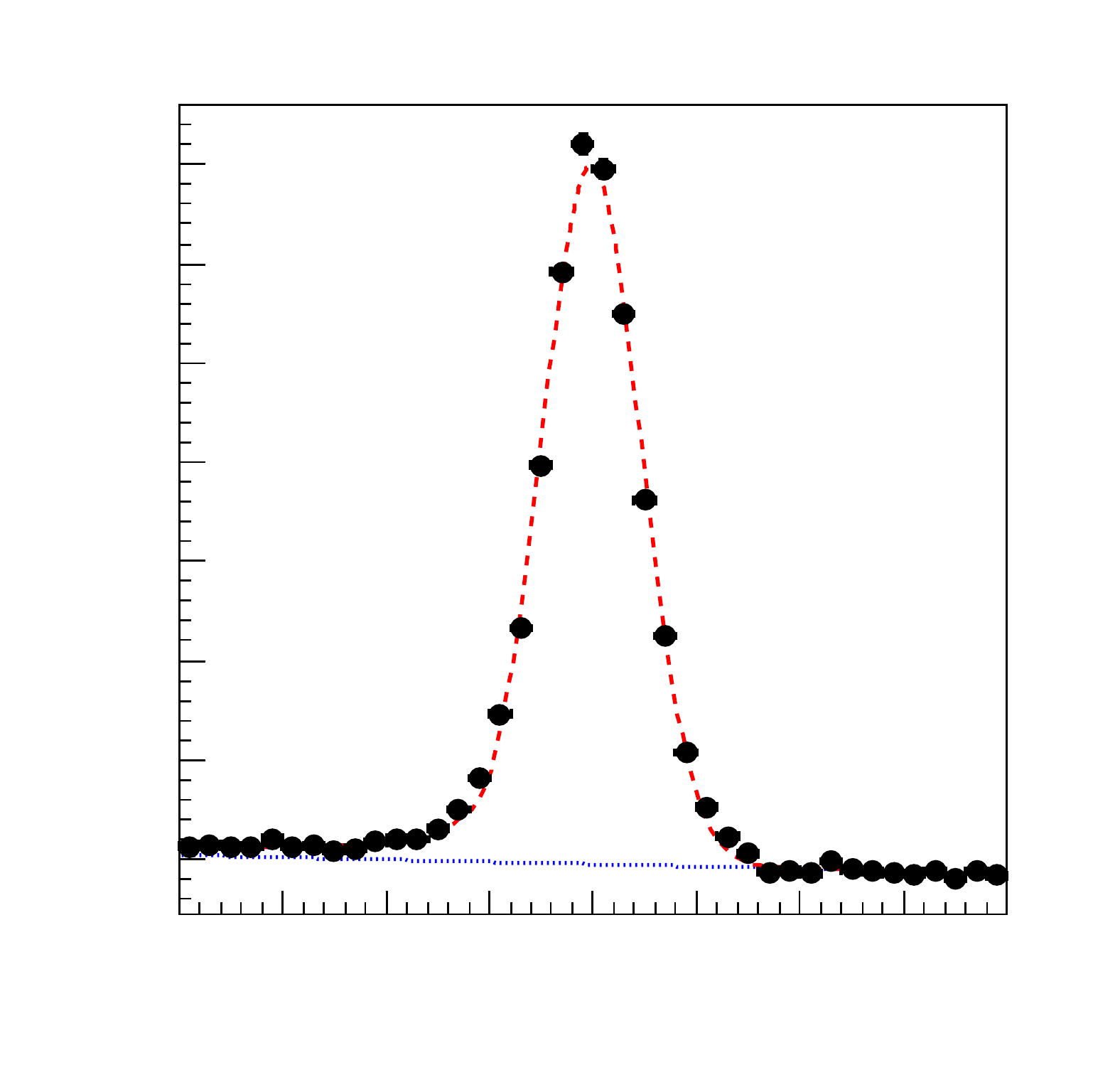}
  & \svg[2]{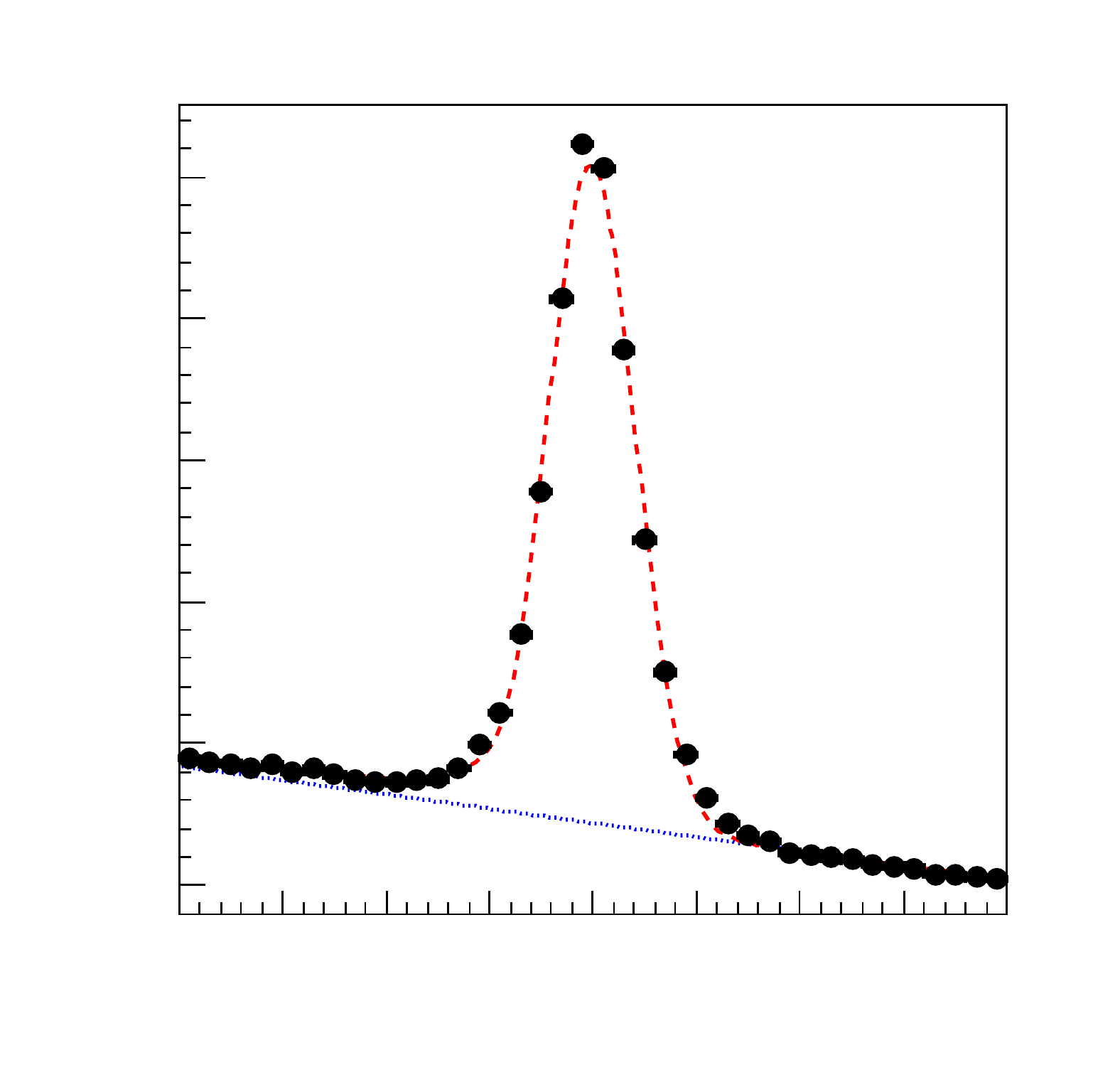}
  & \svg[2]{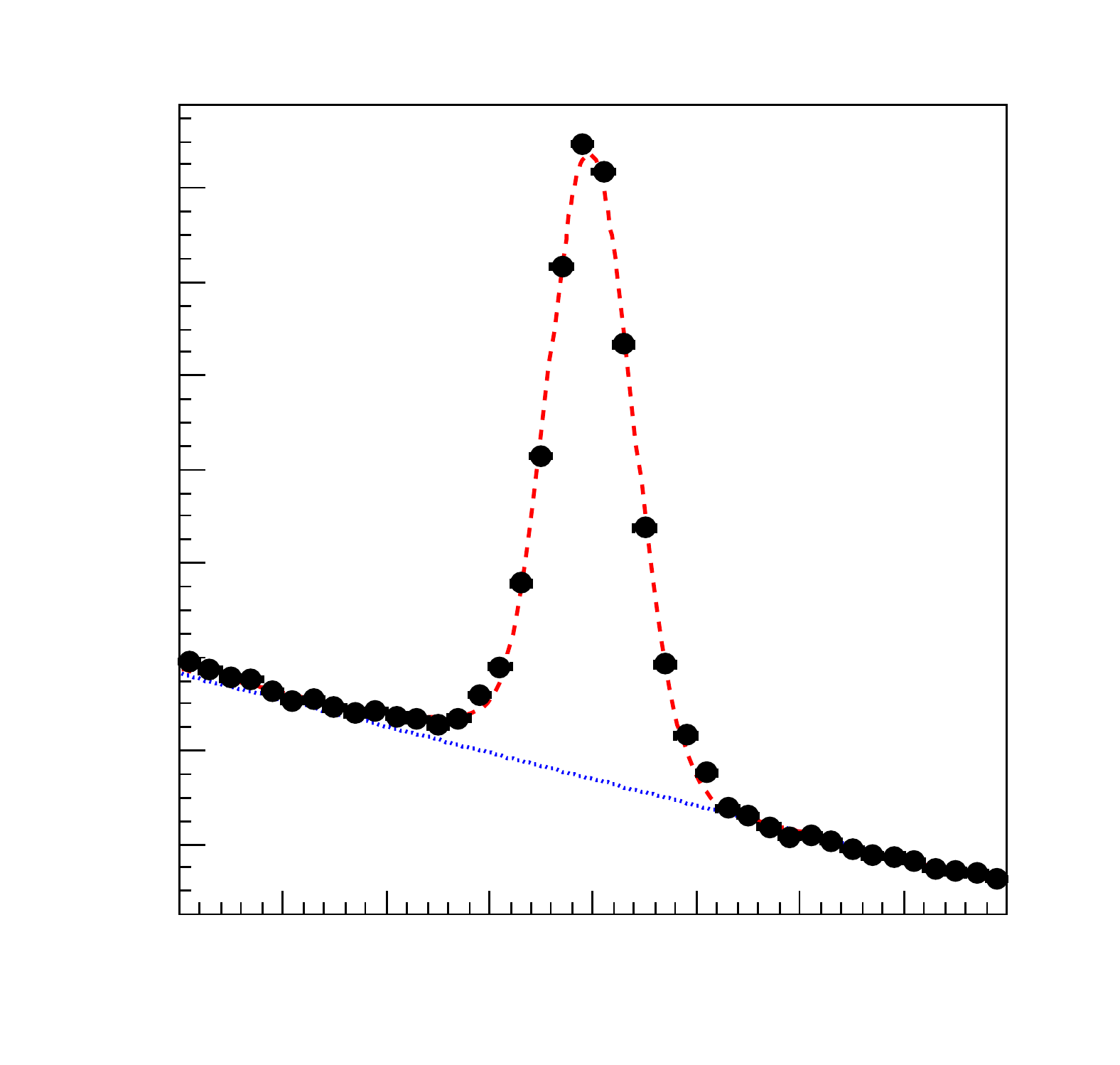} \effsep
  \svg[$50 < p_\mu < 100~\gev$]{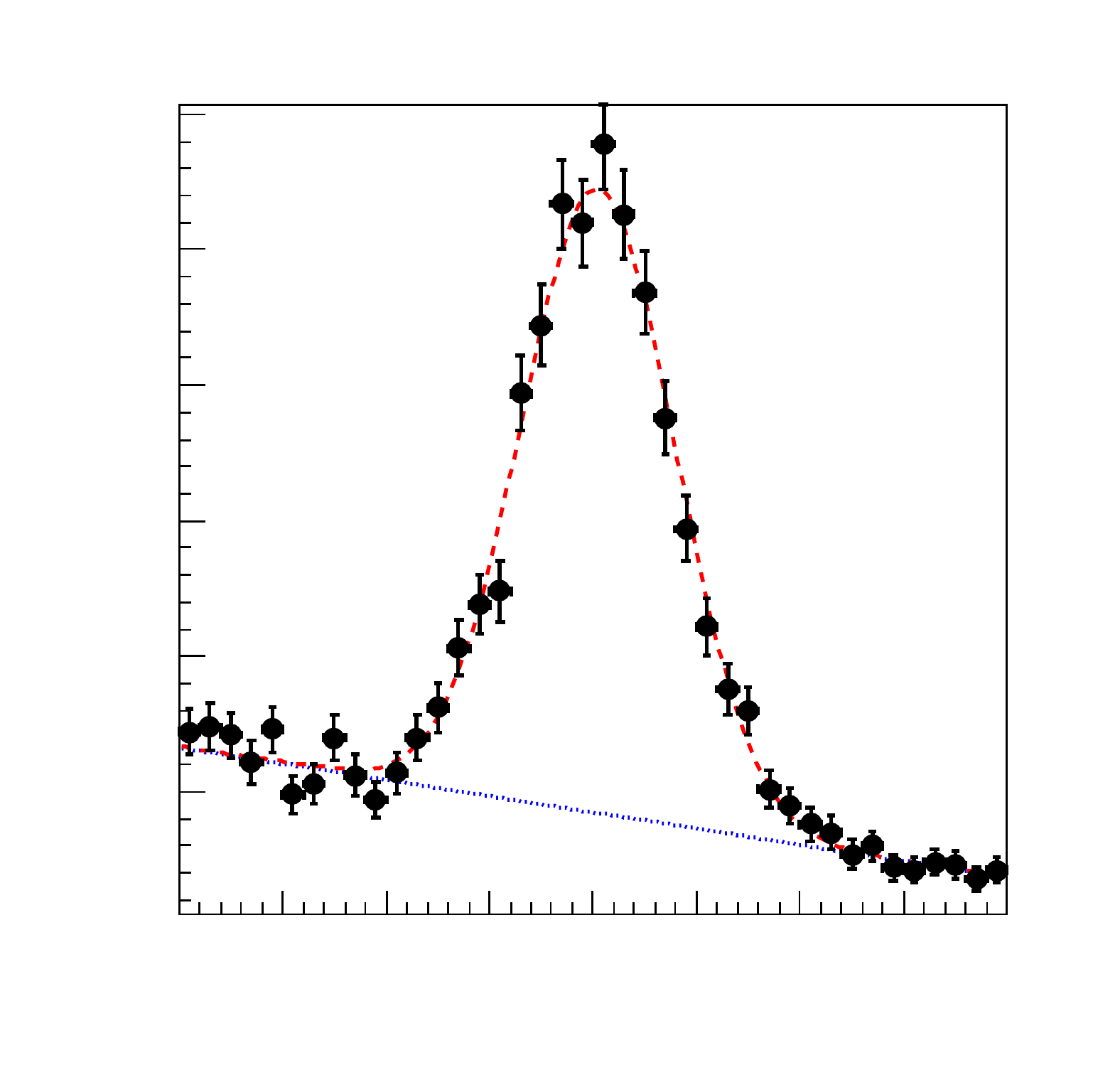}
  & \svg[2]{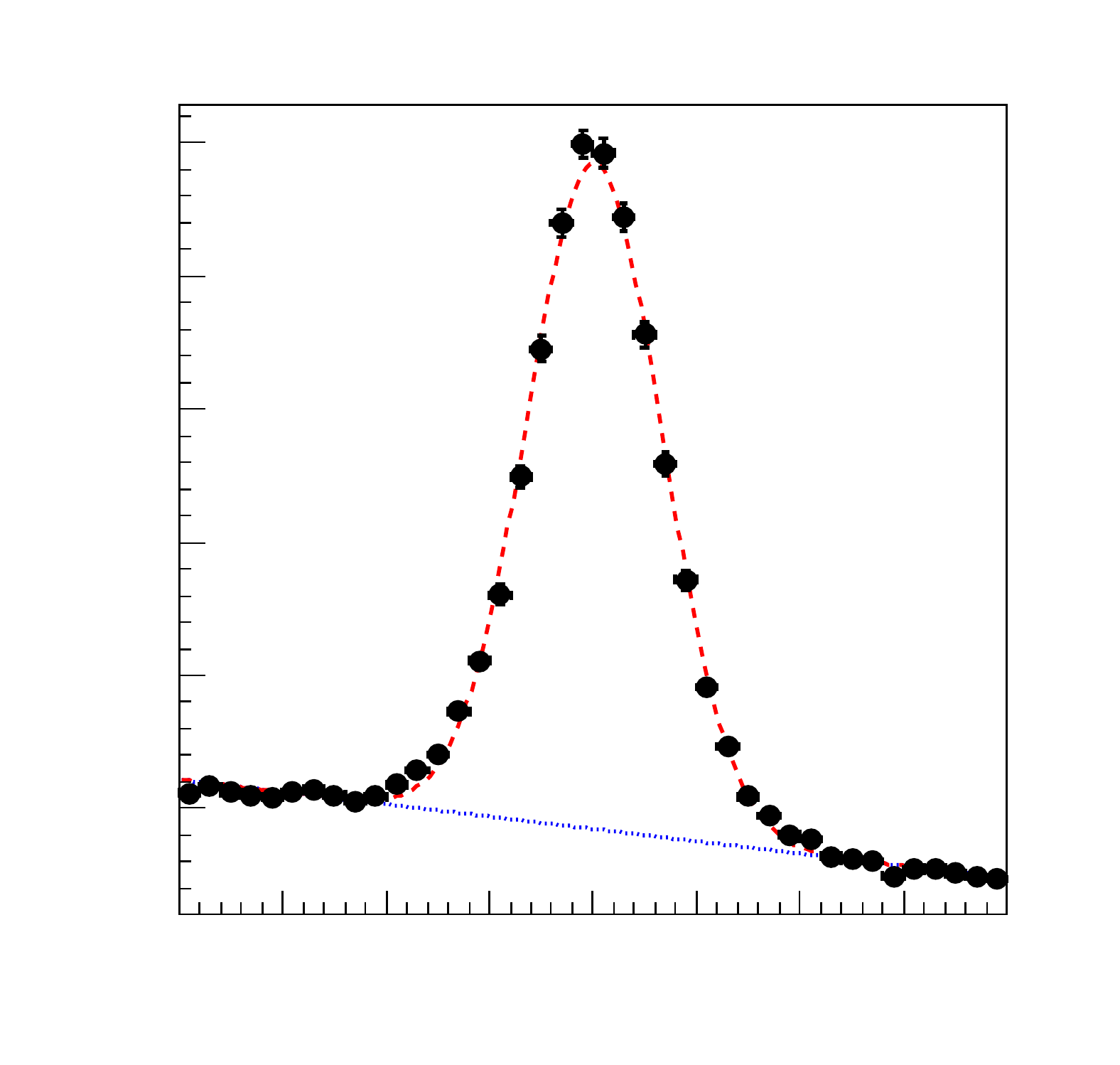}
  & \svg[2]{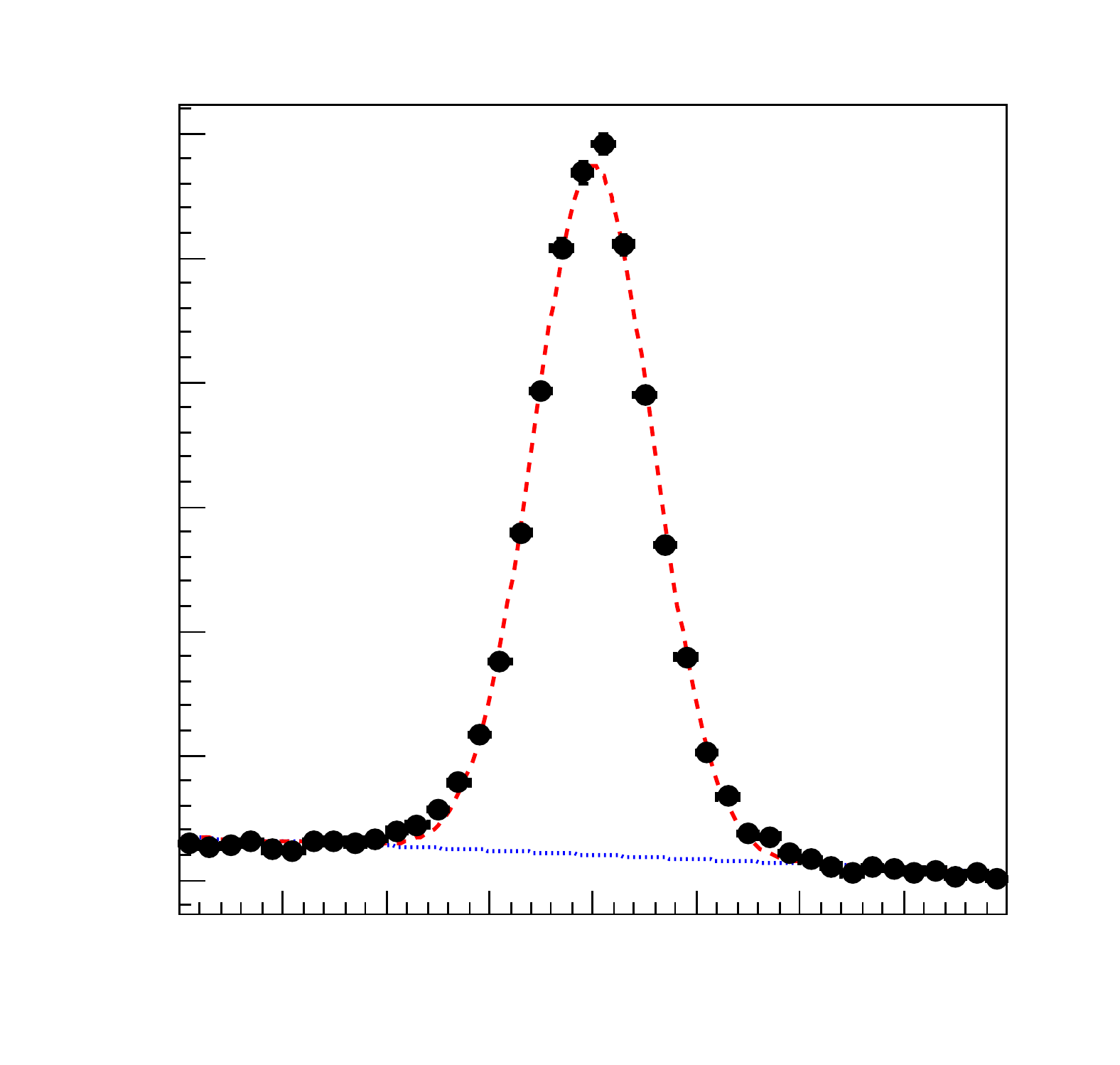} \effsep
  \svg[$100 < p_\mu < 150~\gev$]{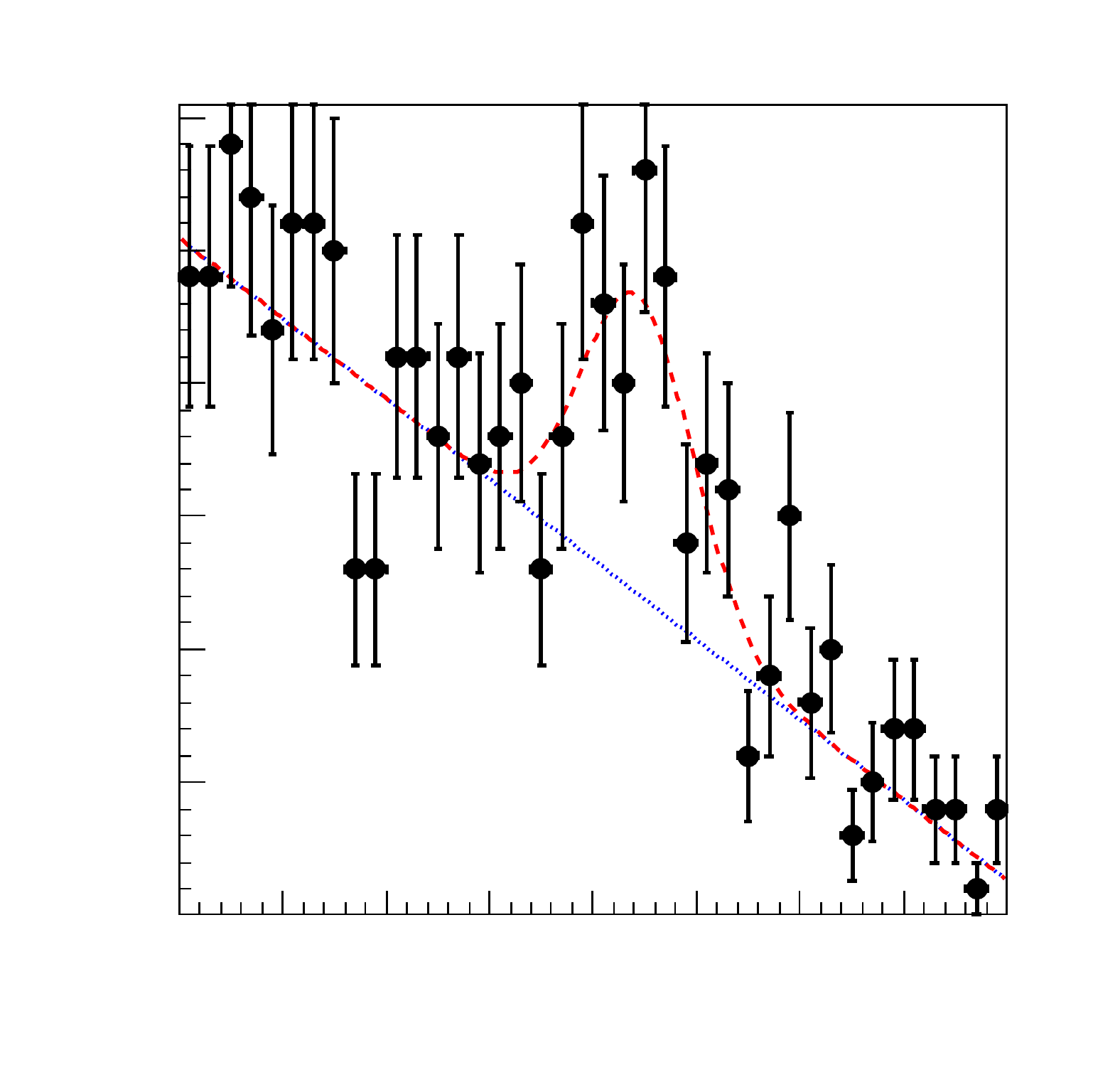}
  & \svg[2]{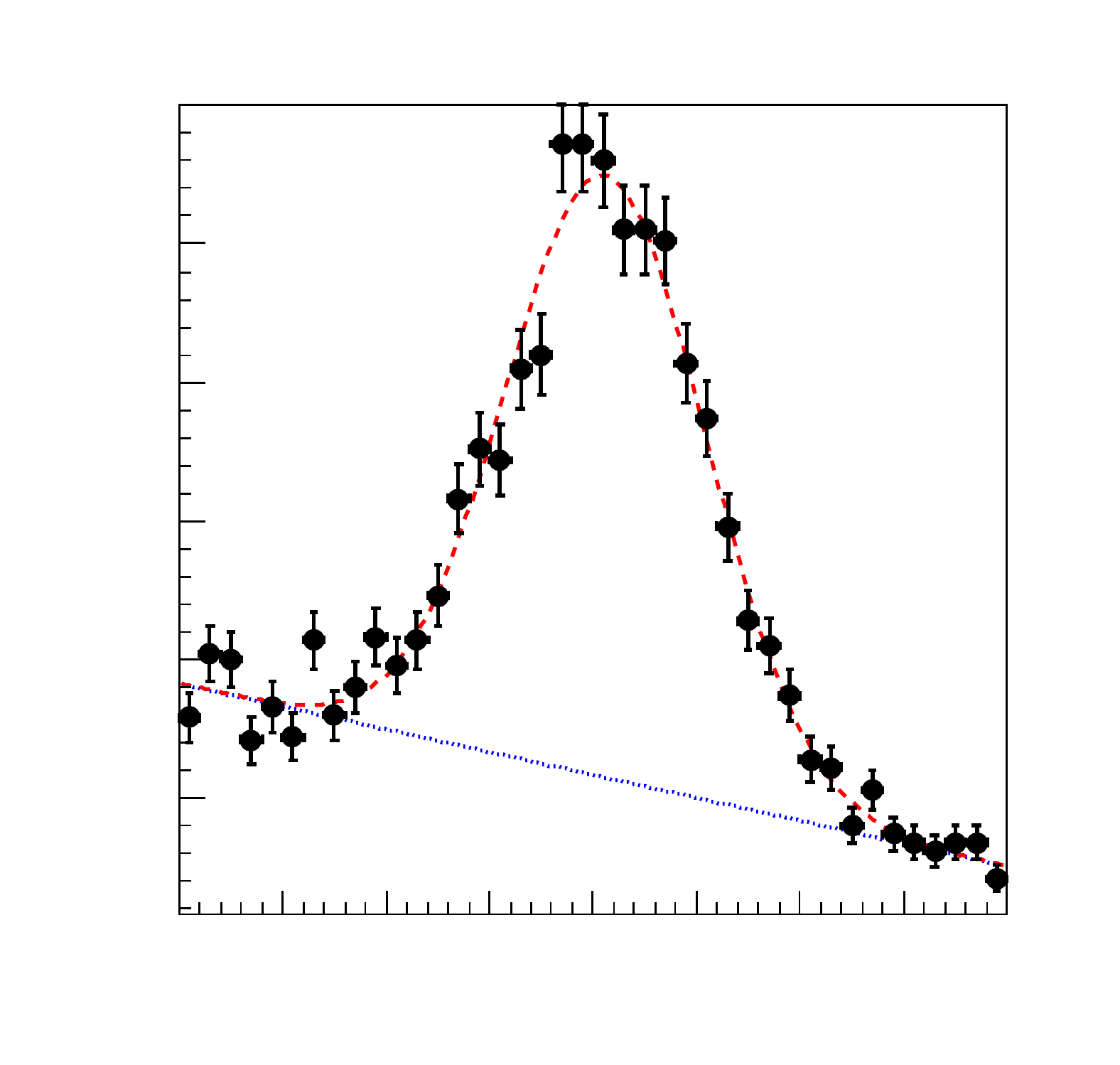}
  & \svg[2]{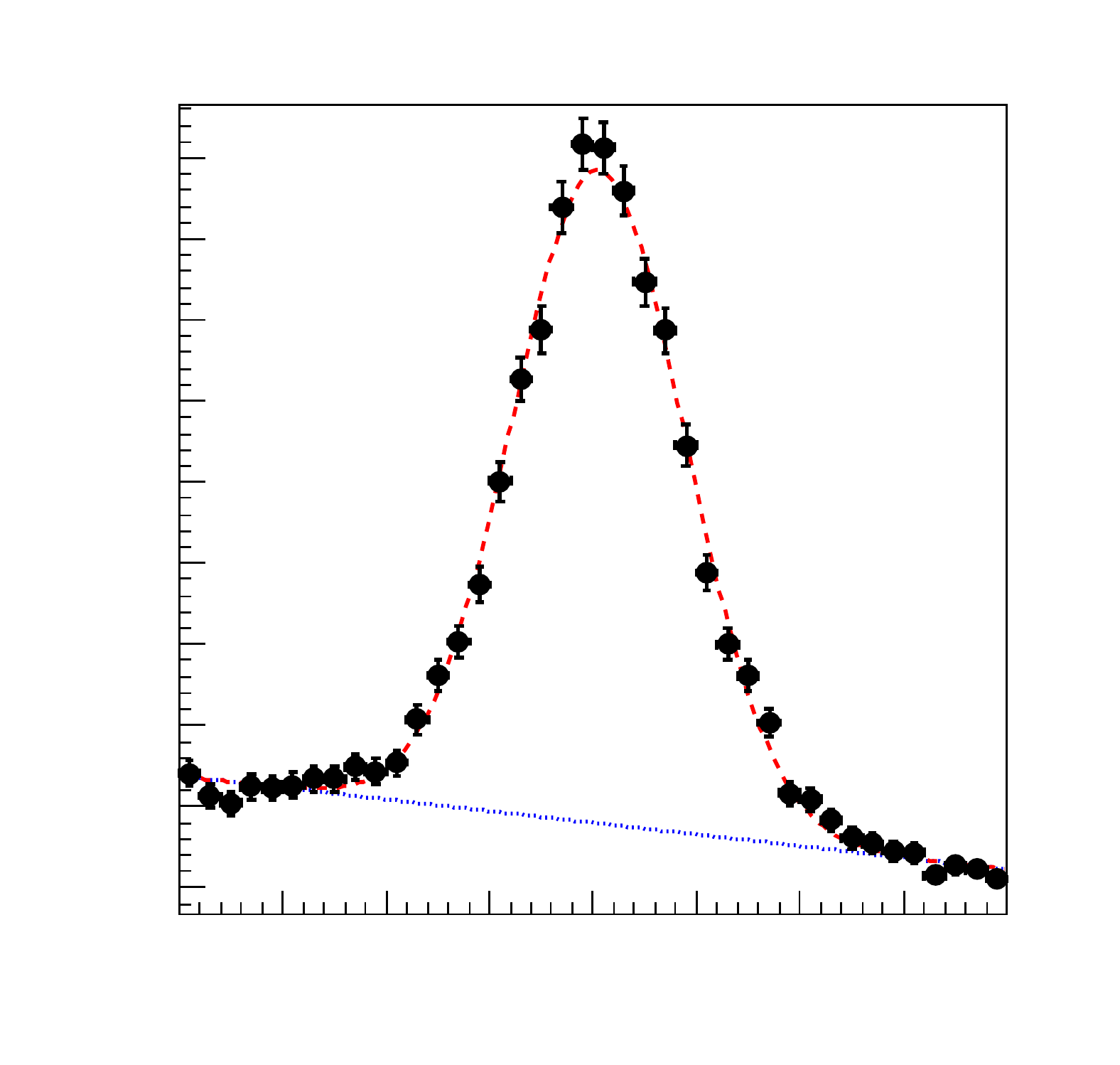} \effsep
  \svg[$150 < p_\mu < 200~\gev$]{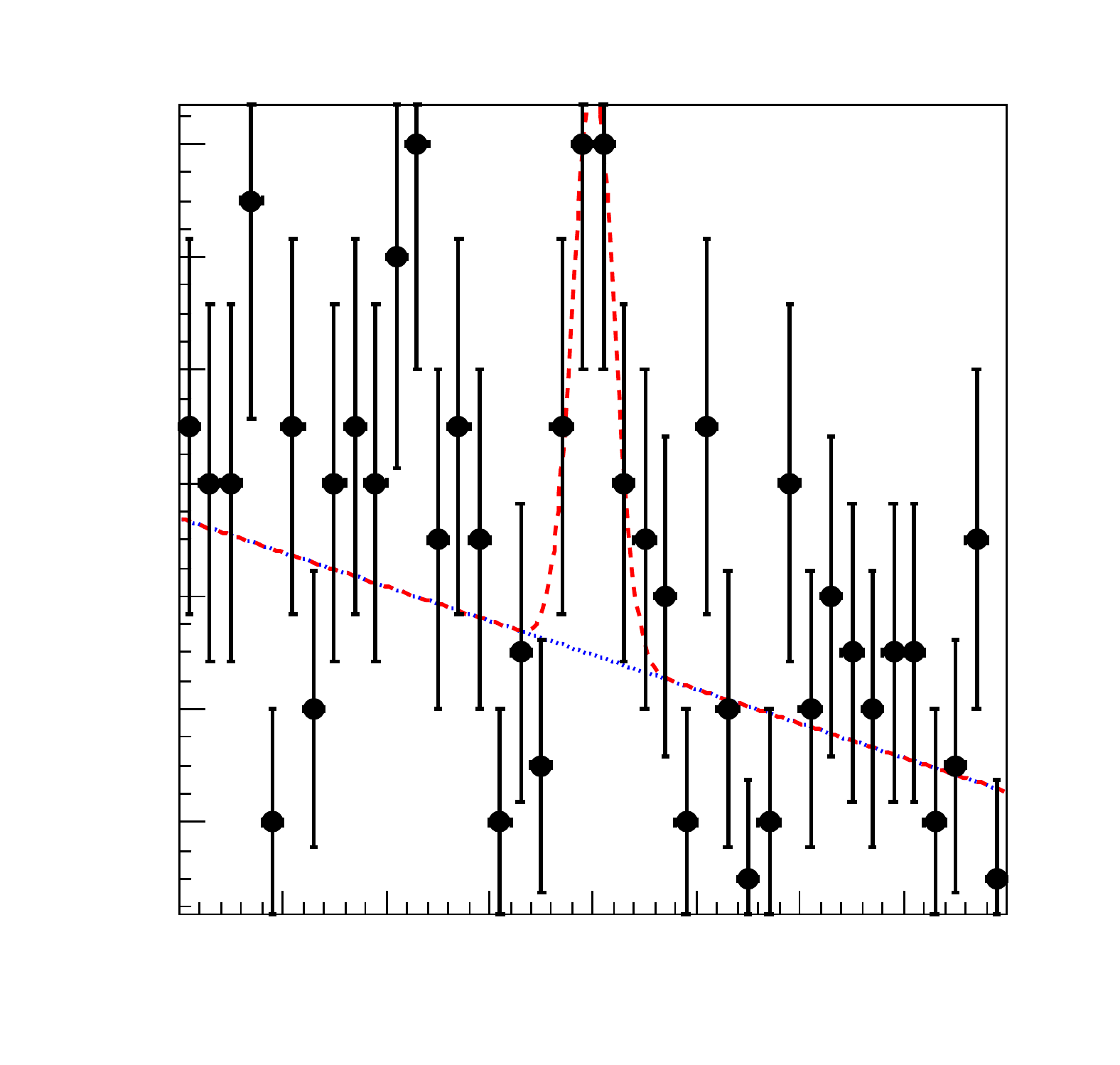}
  & \svg[2]{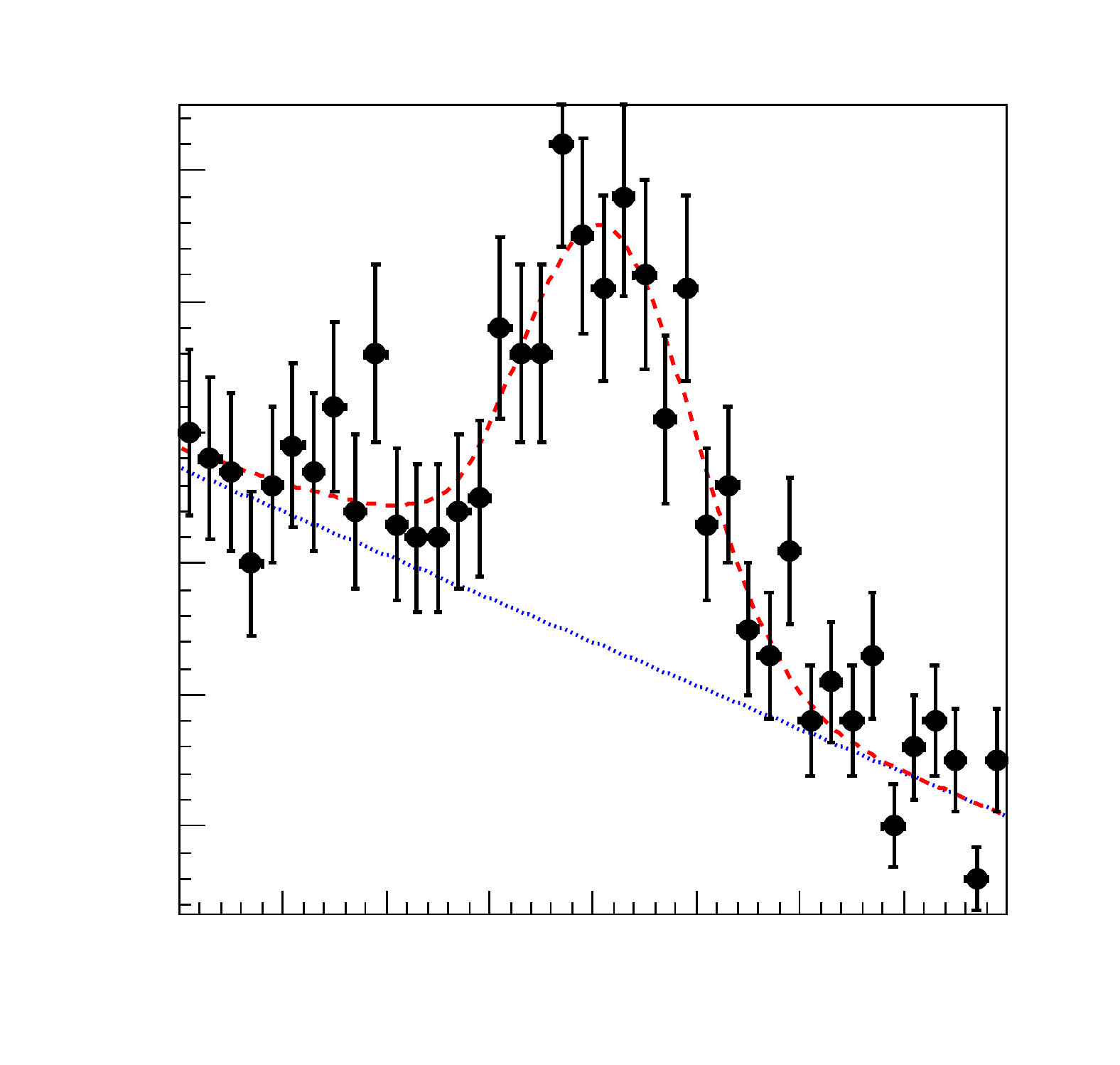}
  & \svg[2]{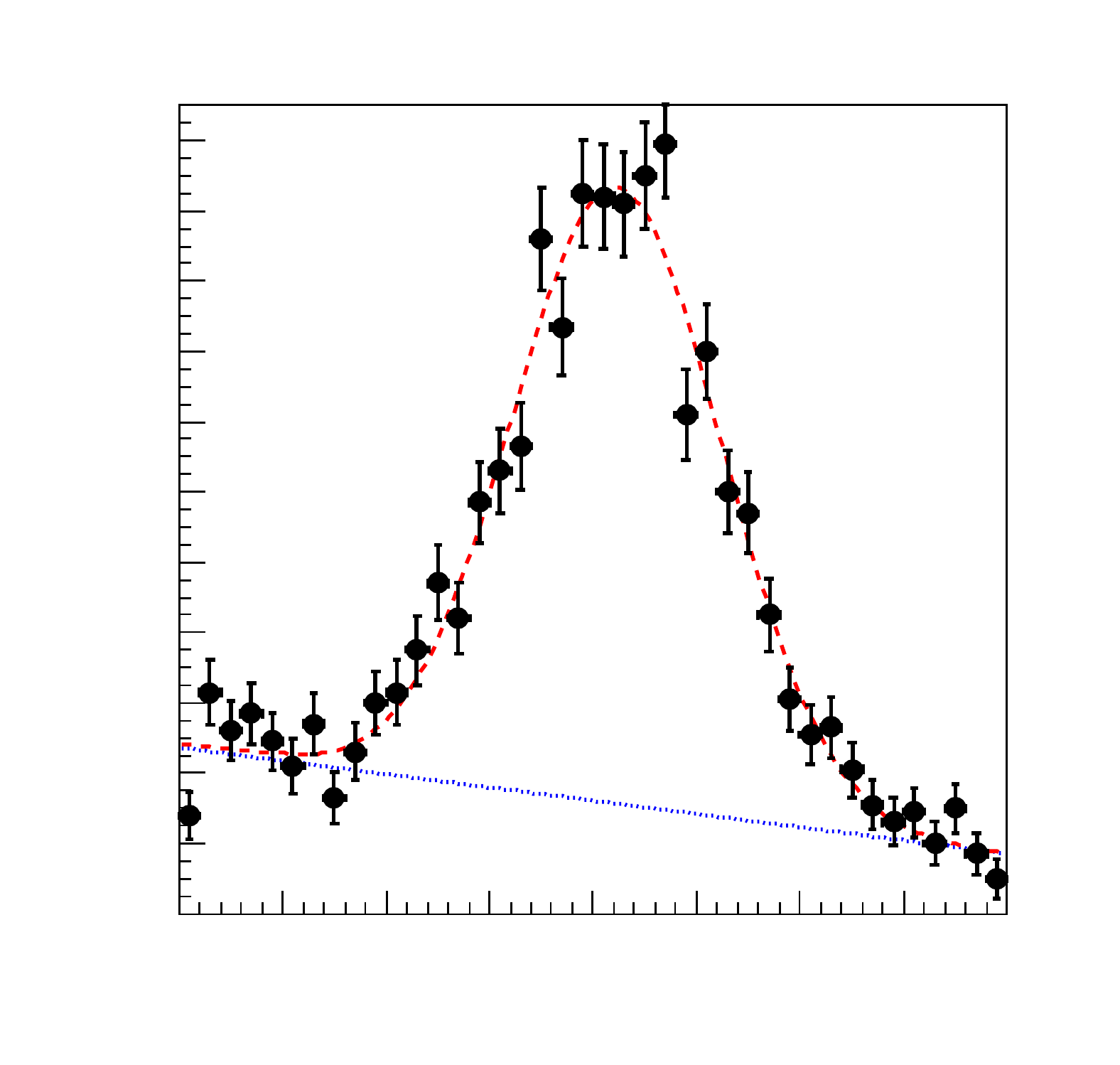} \effend
\end{subfigures}

\begin{subfigures}[p]{4}{Crystal ball fits (dashed red) with a linear
    background (dotted blue) of the ${\jpsi \to \dimu}$ data (points)
    used to determine the muon track identification efficiency, after
    requiring an identified muon. The fits are given in three bins of
    pseudo-rapidity between $2.0$ and $4.5$ and four bins of momentum
    between $0$ and $200~\gev$.\labelfig{RecEff.MuId.Pass}}
  \effbeg
  && $\quad\quad2.00 < \eta_\mu < 2.83$ 
  & $\quad\quad2.83 < \eta_\mu < 3.67$ 
  & $\quad\quad3.67 < \eta_\mu < 4.50$ \\
  \midrule
  \svg[$0 < p_\mu < 50~\gev$]{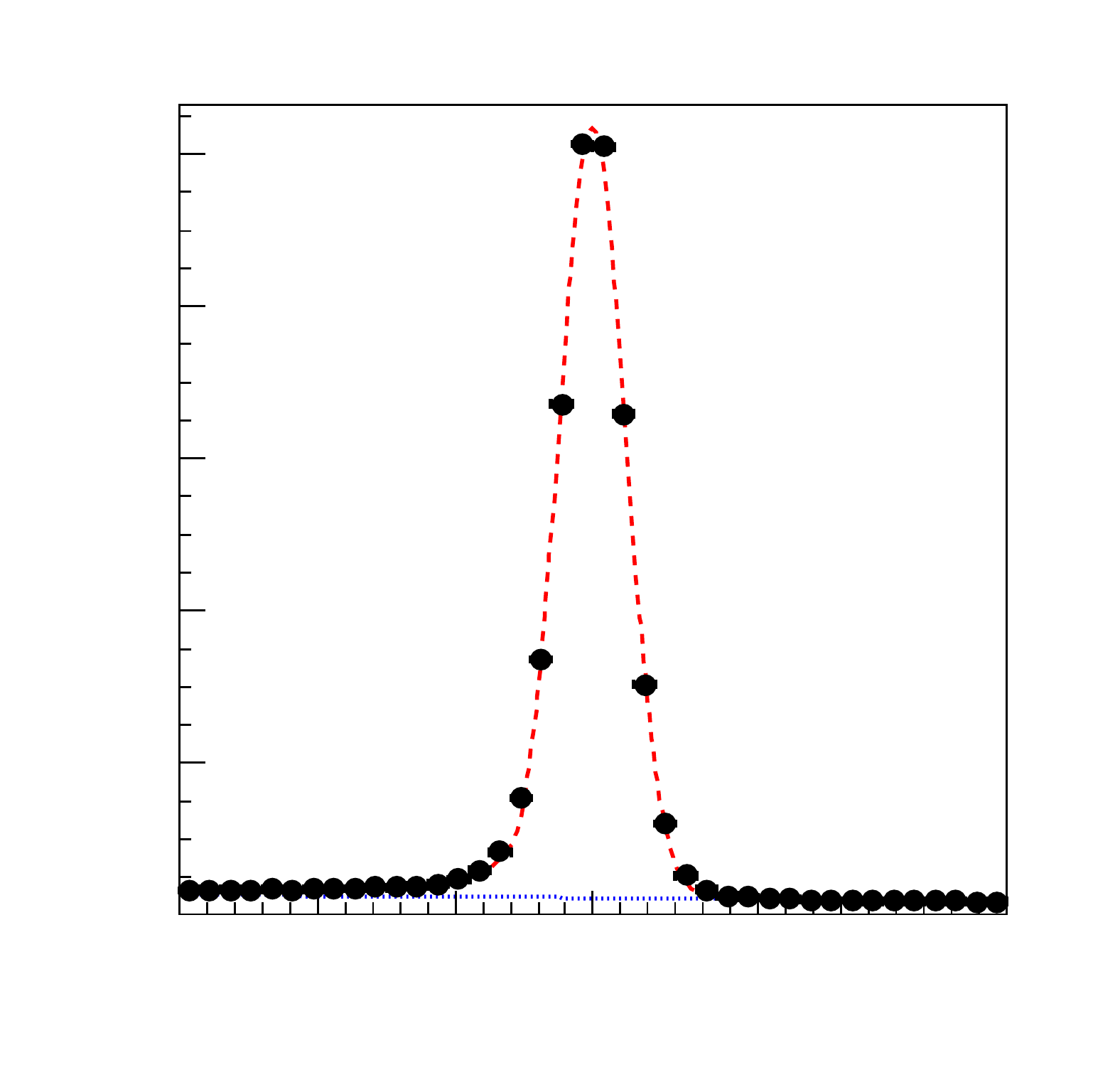}
  & \svg[2]{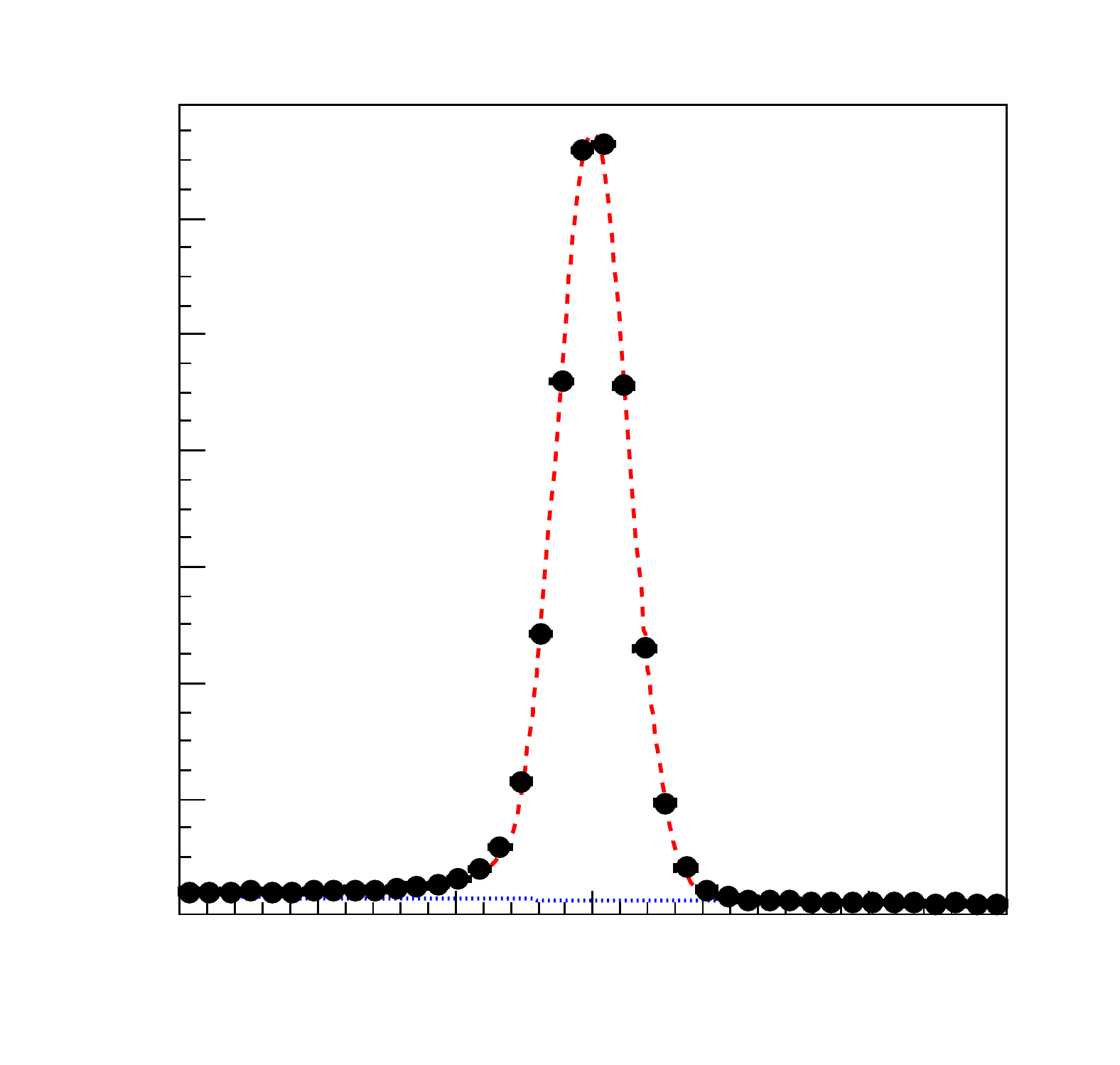}
  & \svg[2]{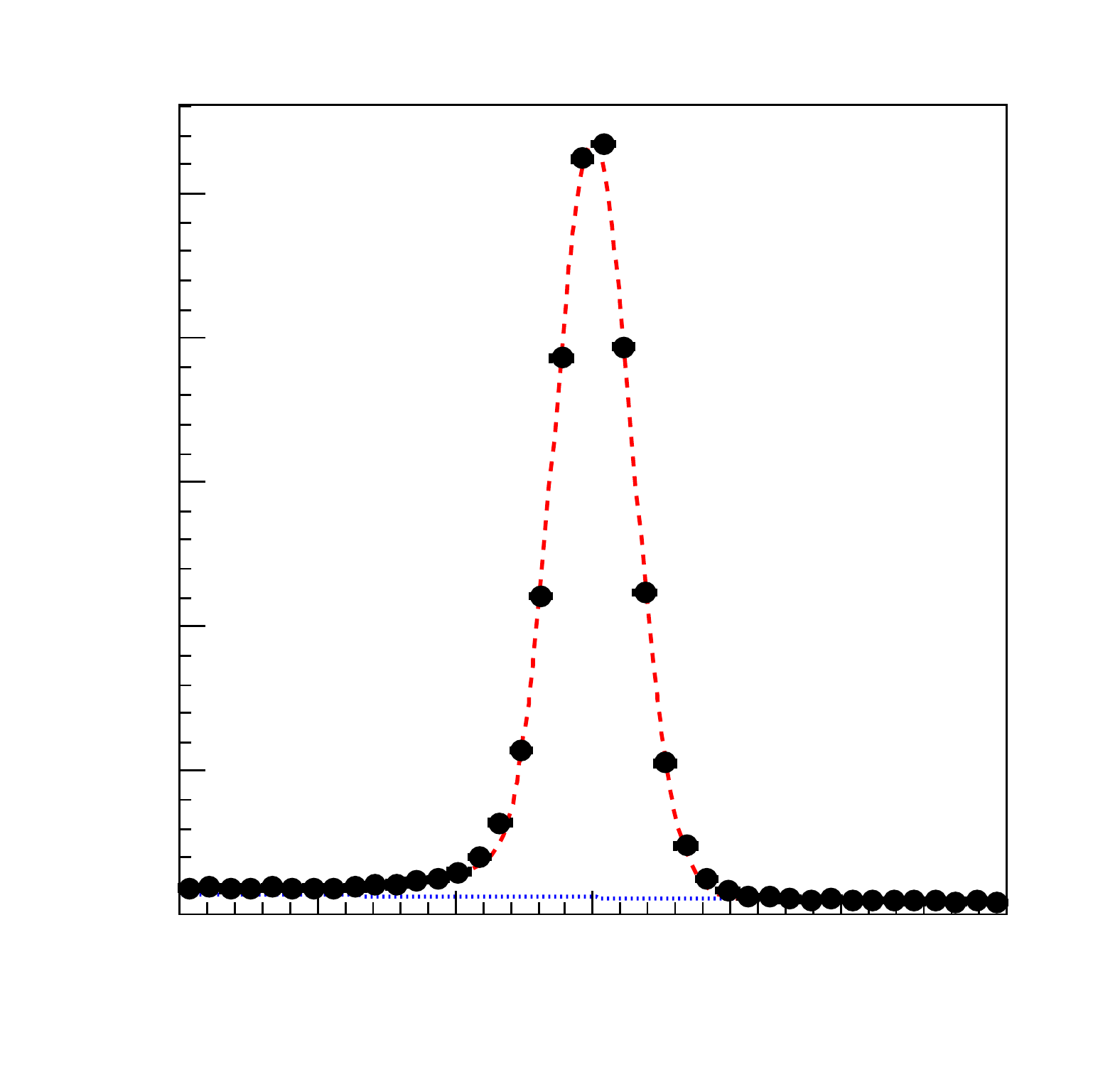} \effsep
  \svg[$50 < p_\mu < 100~\gev$]{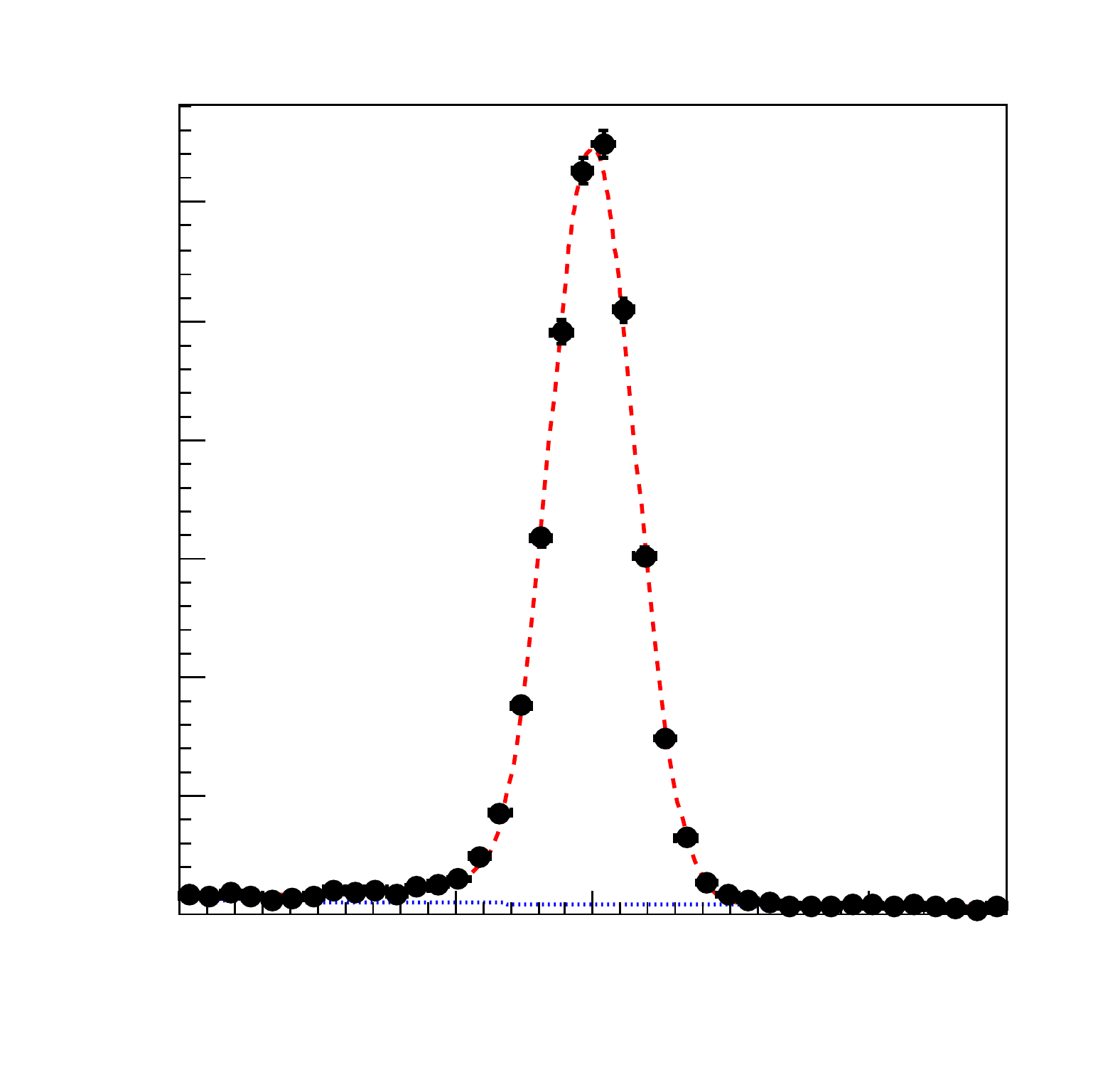}
  & \svg[2]{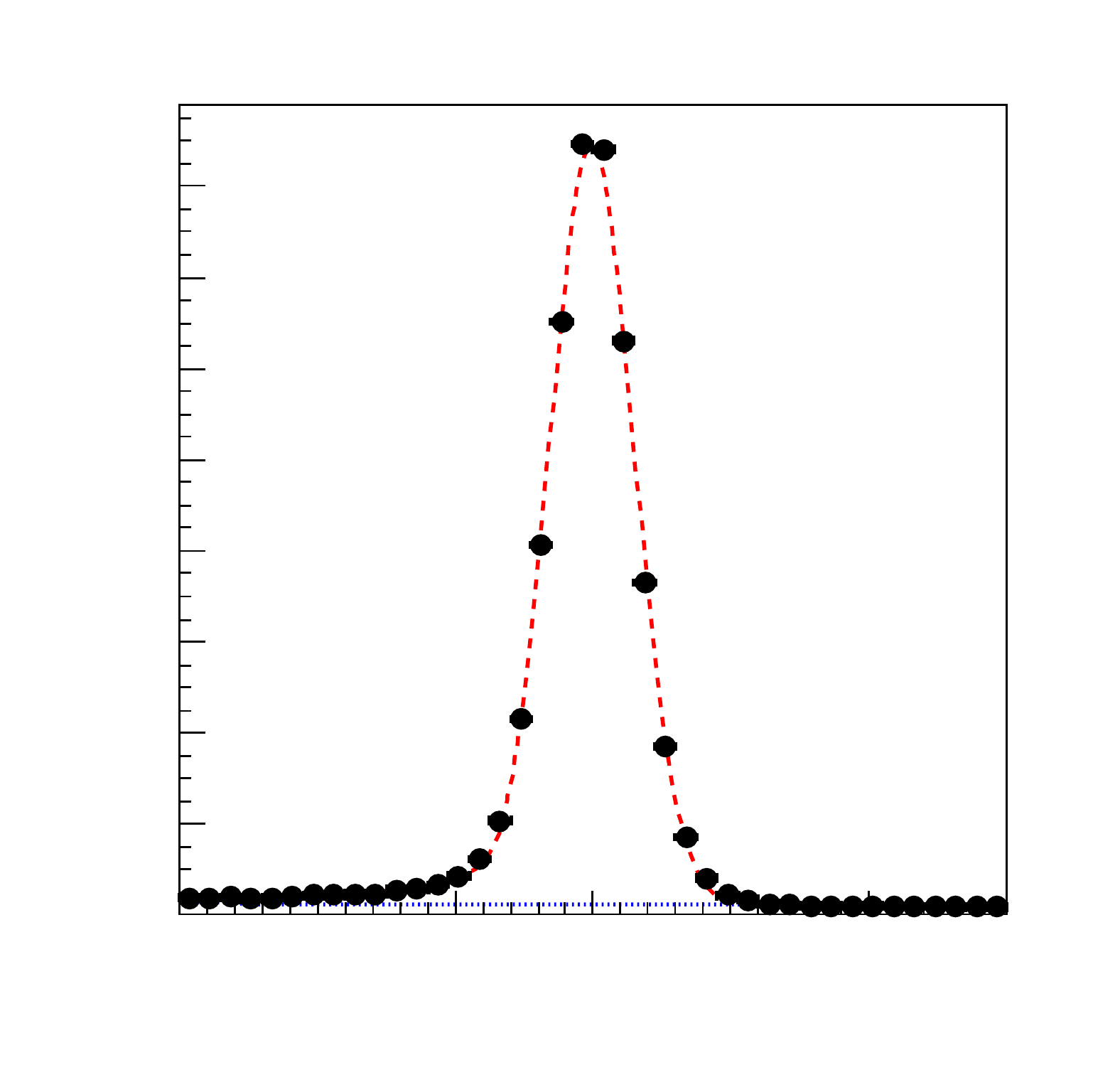}
  & \svg[2]{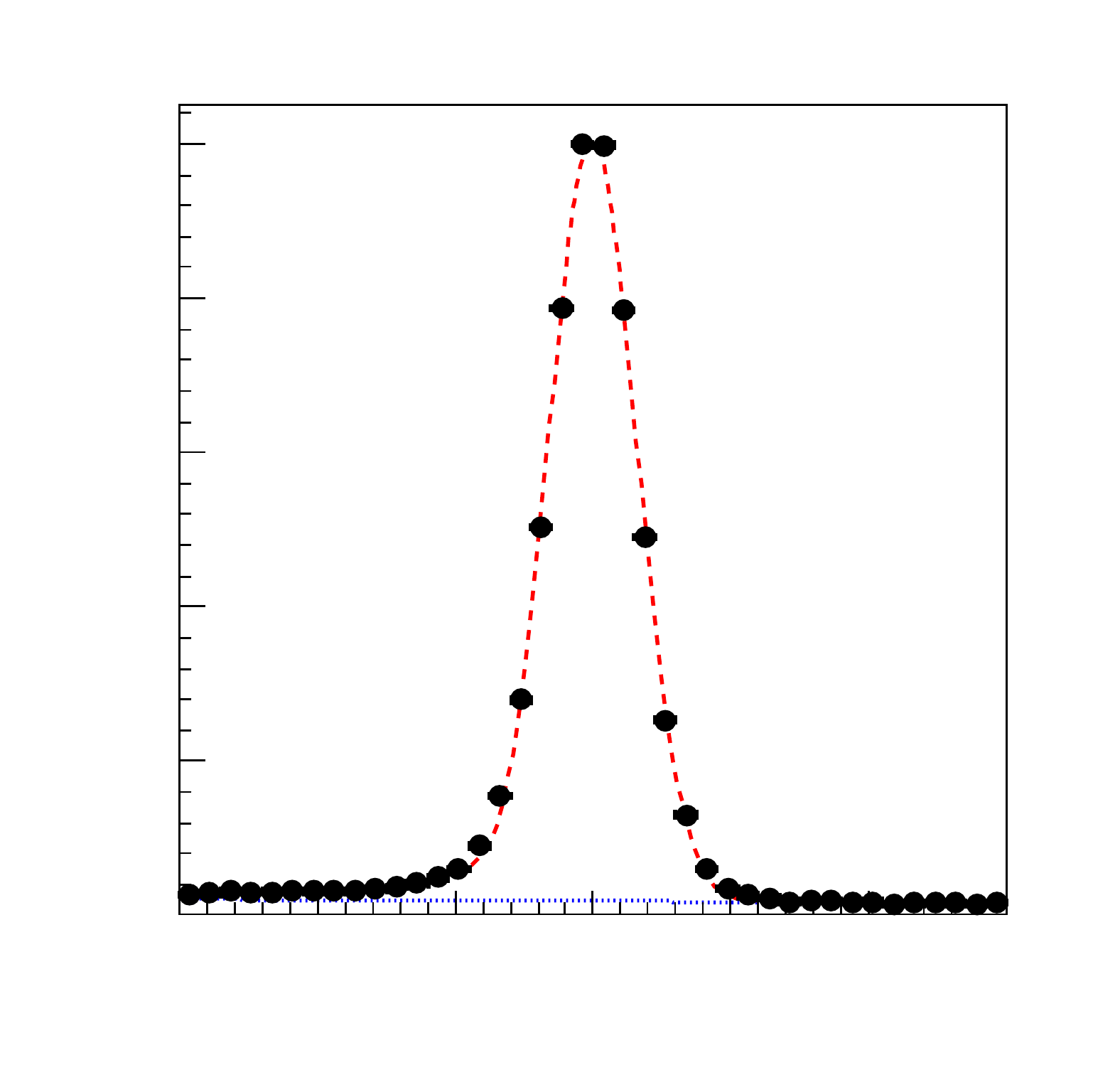} \effsep
  \svg[$100 < p_\mu < 150~\gev$]{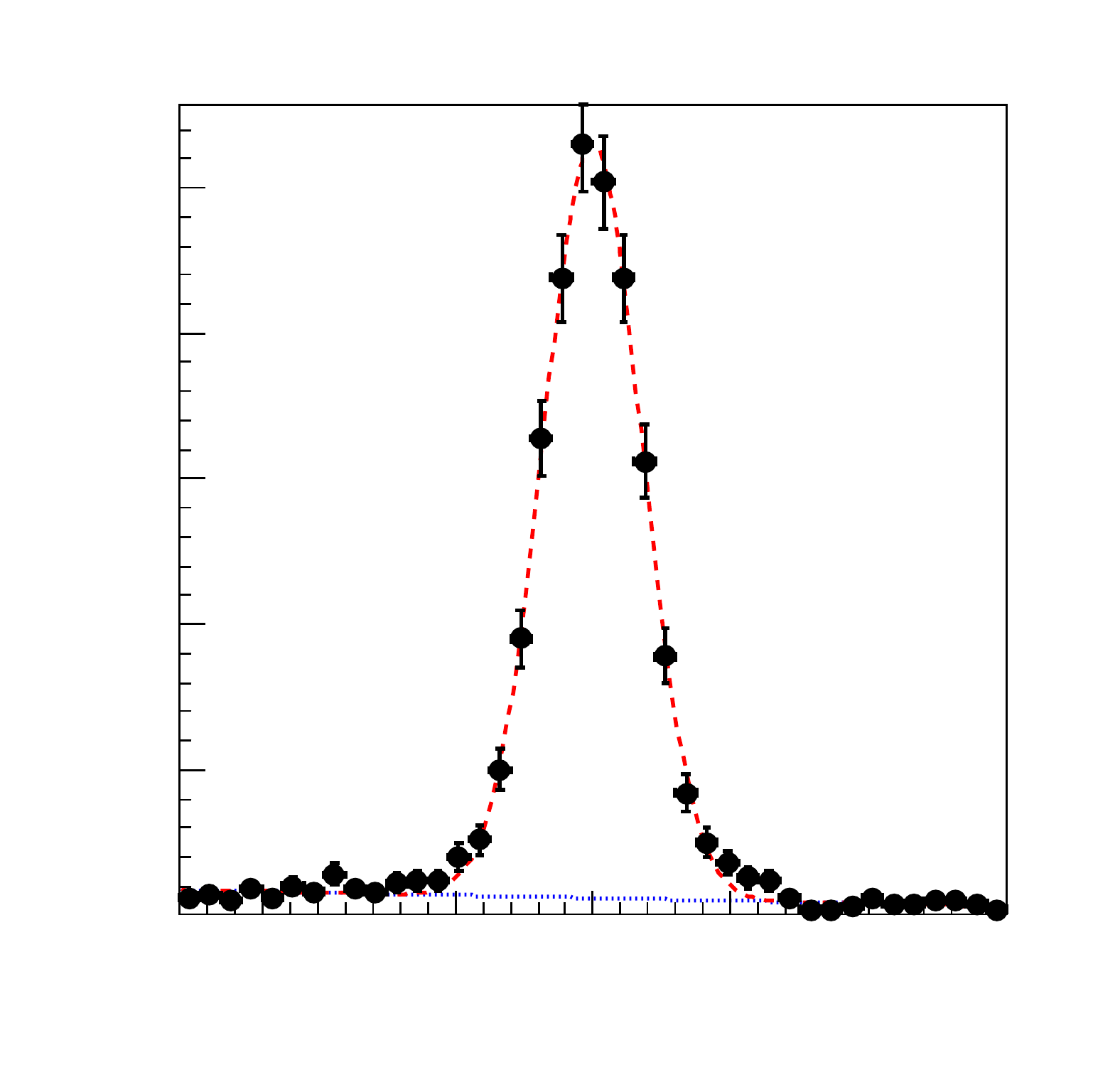}
  & \svg[2]{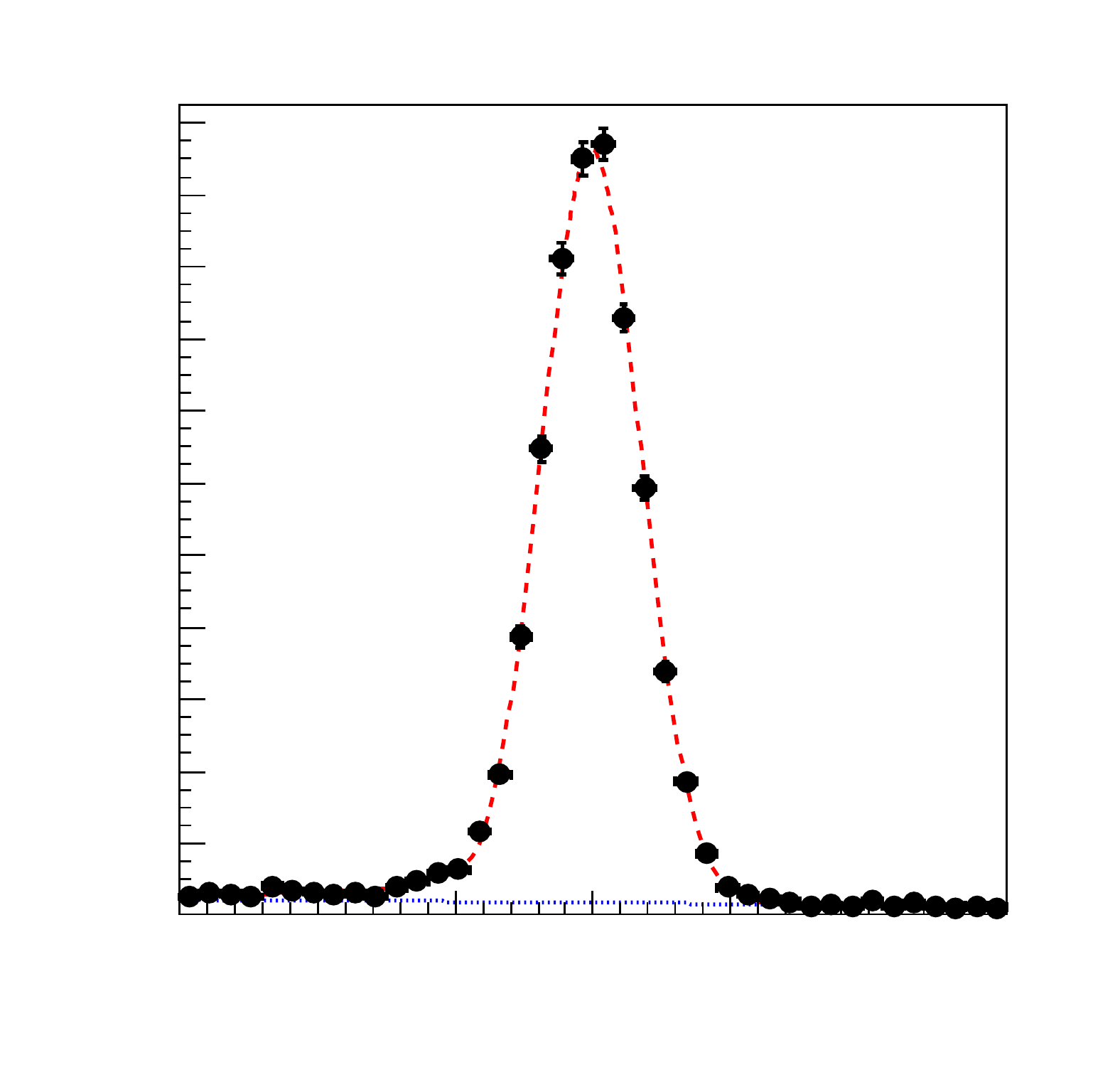}
  & \svg[2]{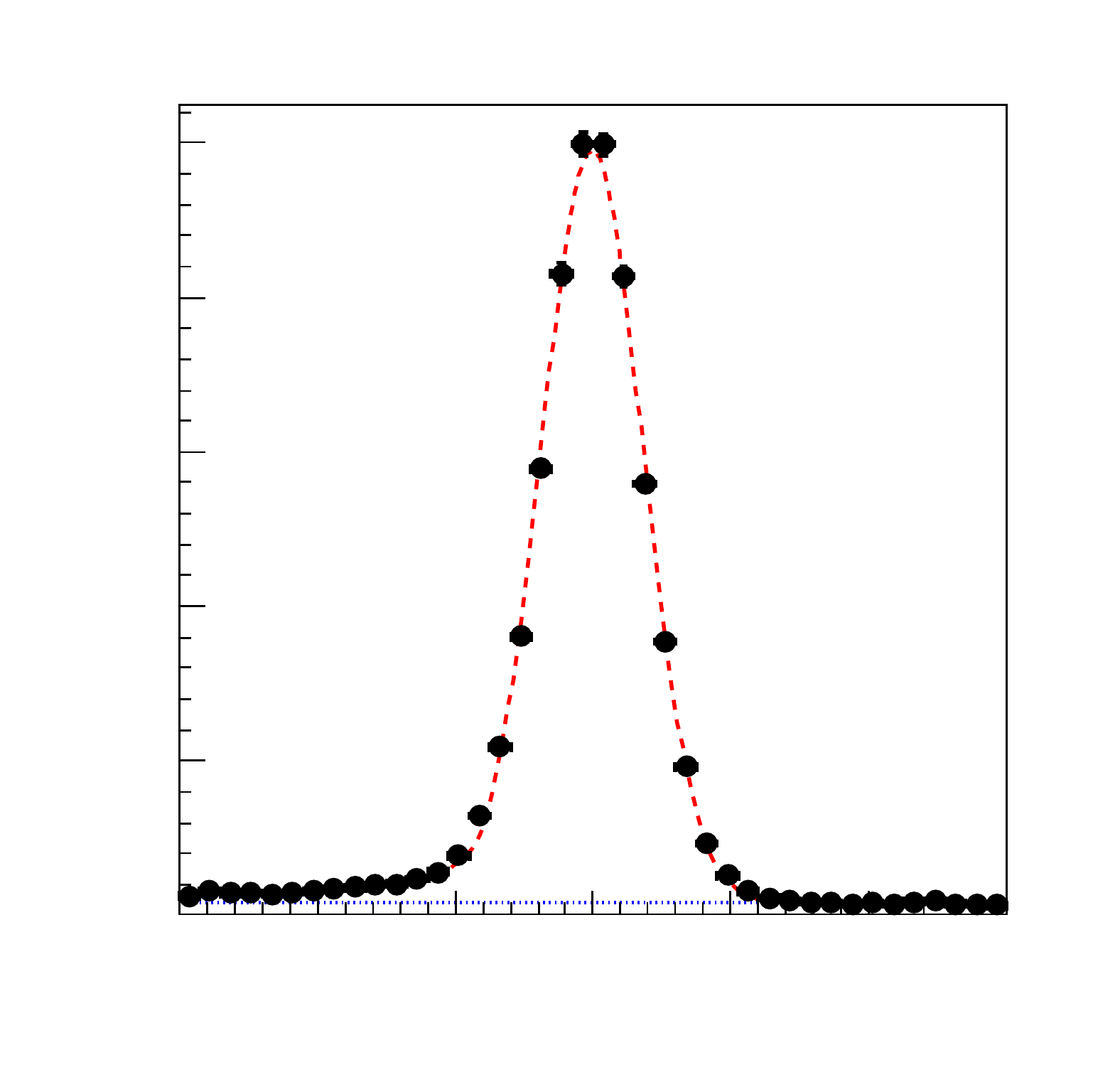} \effsep
  \svg[$150 < p_\mu < 200~\gev$]{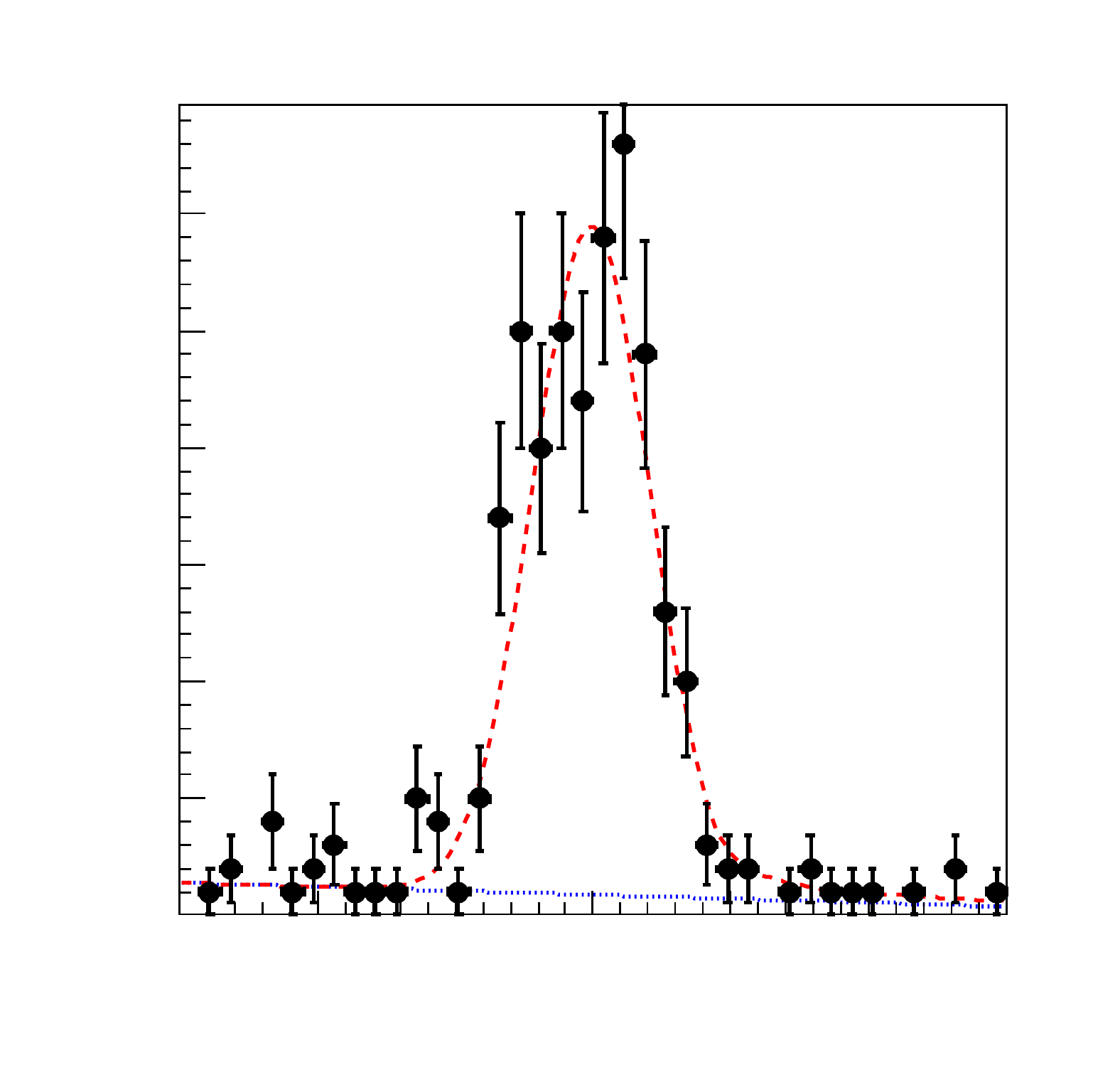}
  & \svg[2]{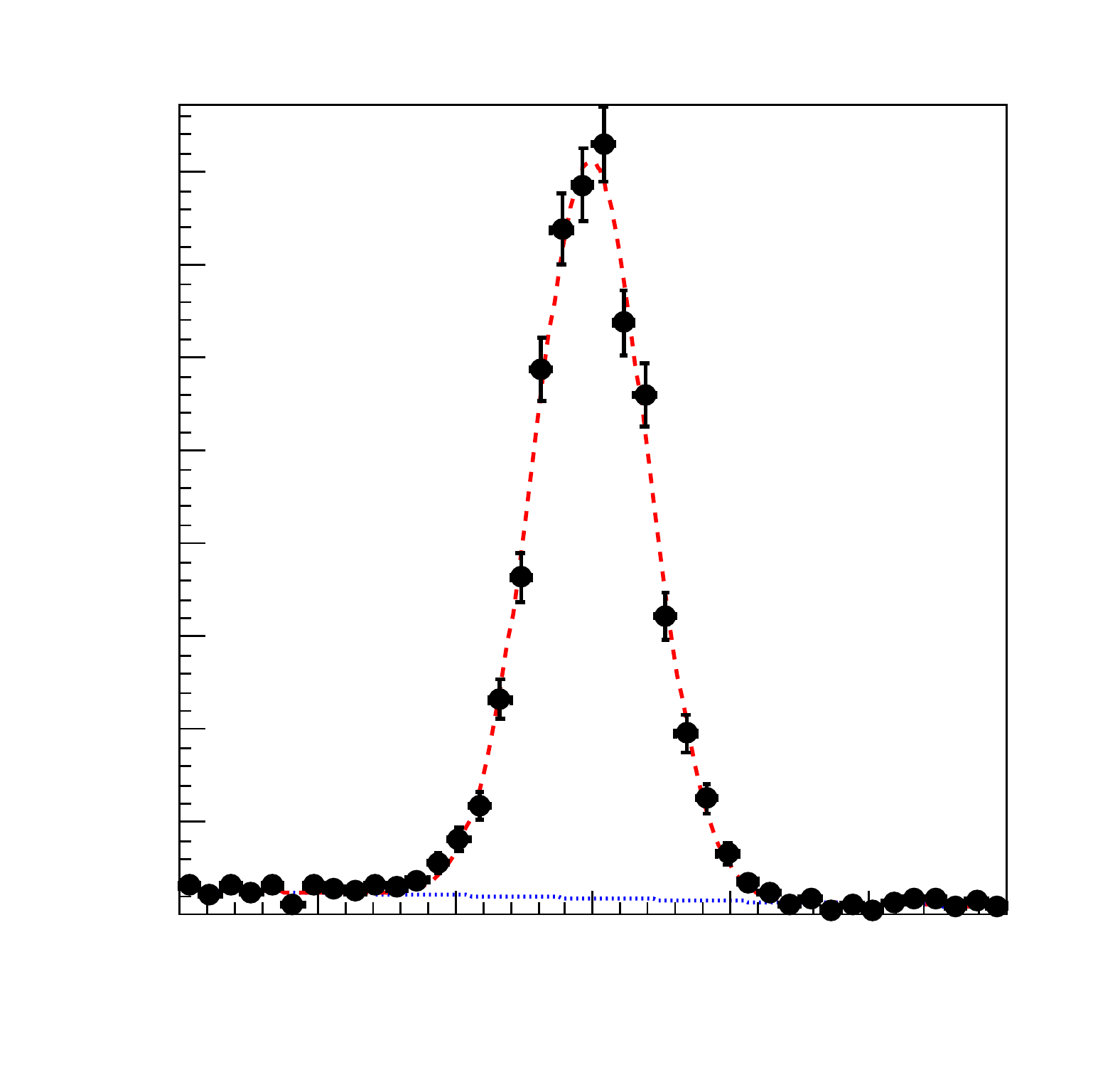}
  & \svg[2]{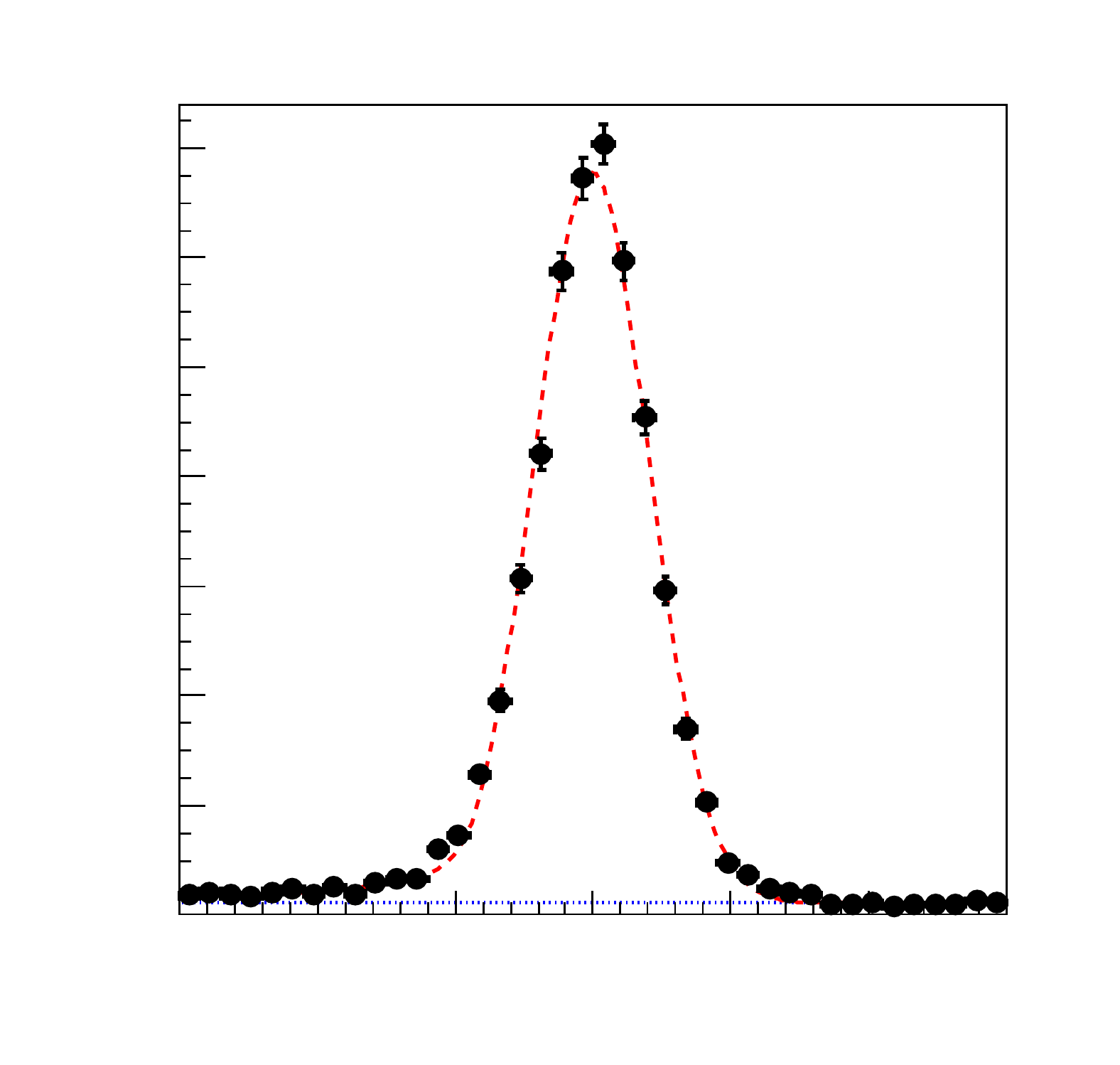} \effend
\end{subfigures}

\begin{subfigures}[p]{4}{Crystal ball fits (dashed red) with a linear
    background (dotted blue) of the ${\jpsi \to \dimu}$ data (points)
    used to determine the muon track identification efficiency, before
    requiring an identified muon. The fits are given in three bins of
    pseudo-rapidity between $2.0$ and $4.5$ and four bins of momentum
    between $0$ and $200~\gev$.\labelfig{RecEff.MuId.Total}}
  \effbeg
  && $\quad\quad2.00 < \eta_\mu < 2.83$ 
  & $\quad\quad2.83 < \eta_\mu < 3.67$ 
  & $\quad\quad3.67 < \eta_\mu < 4.50$ \\
  \midrule
  \svg[$0 < p_\mu < 50~\gev$]{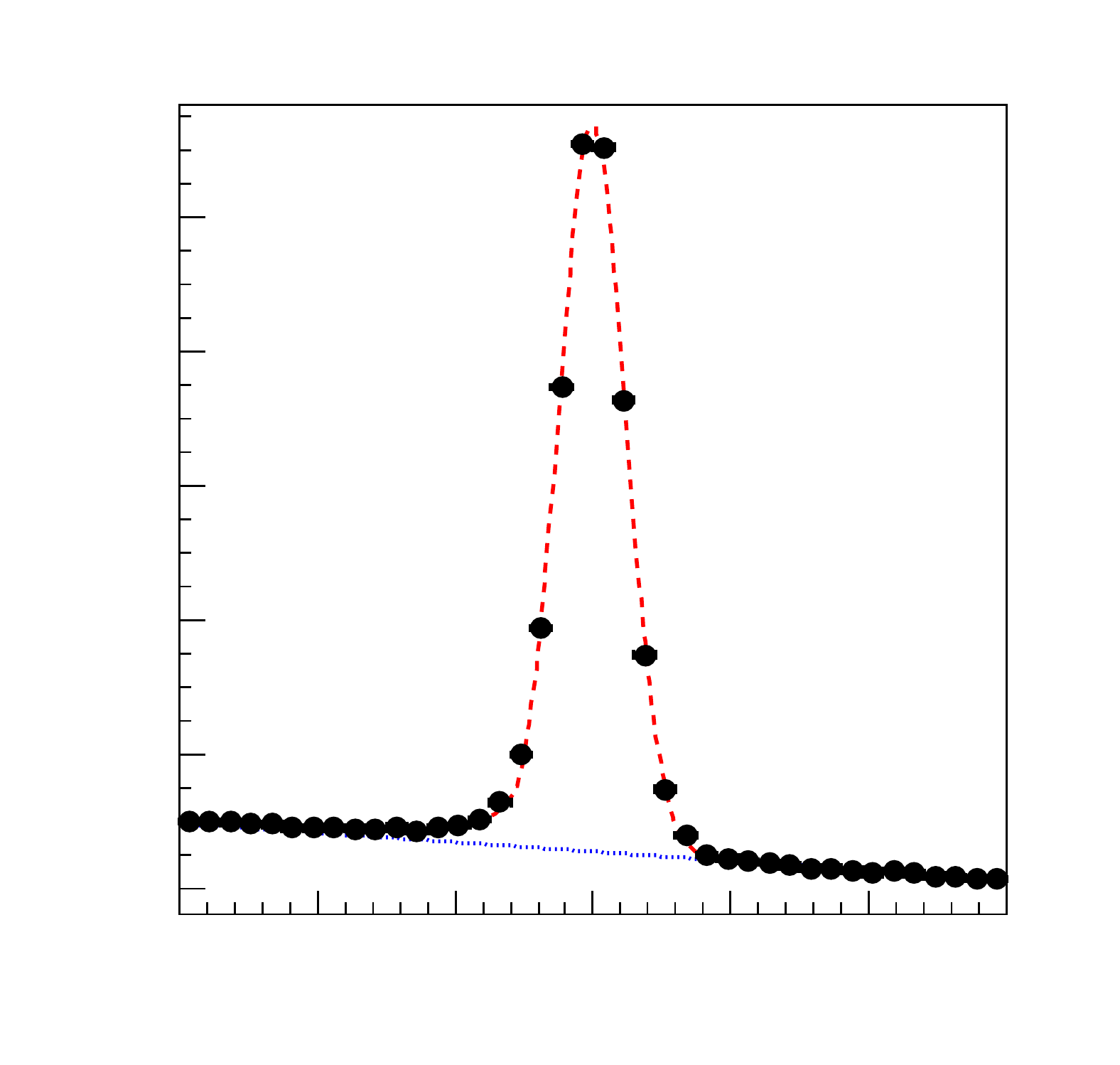}
  & \svg[2]{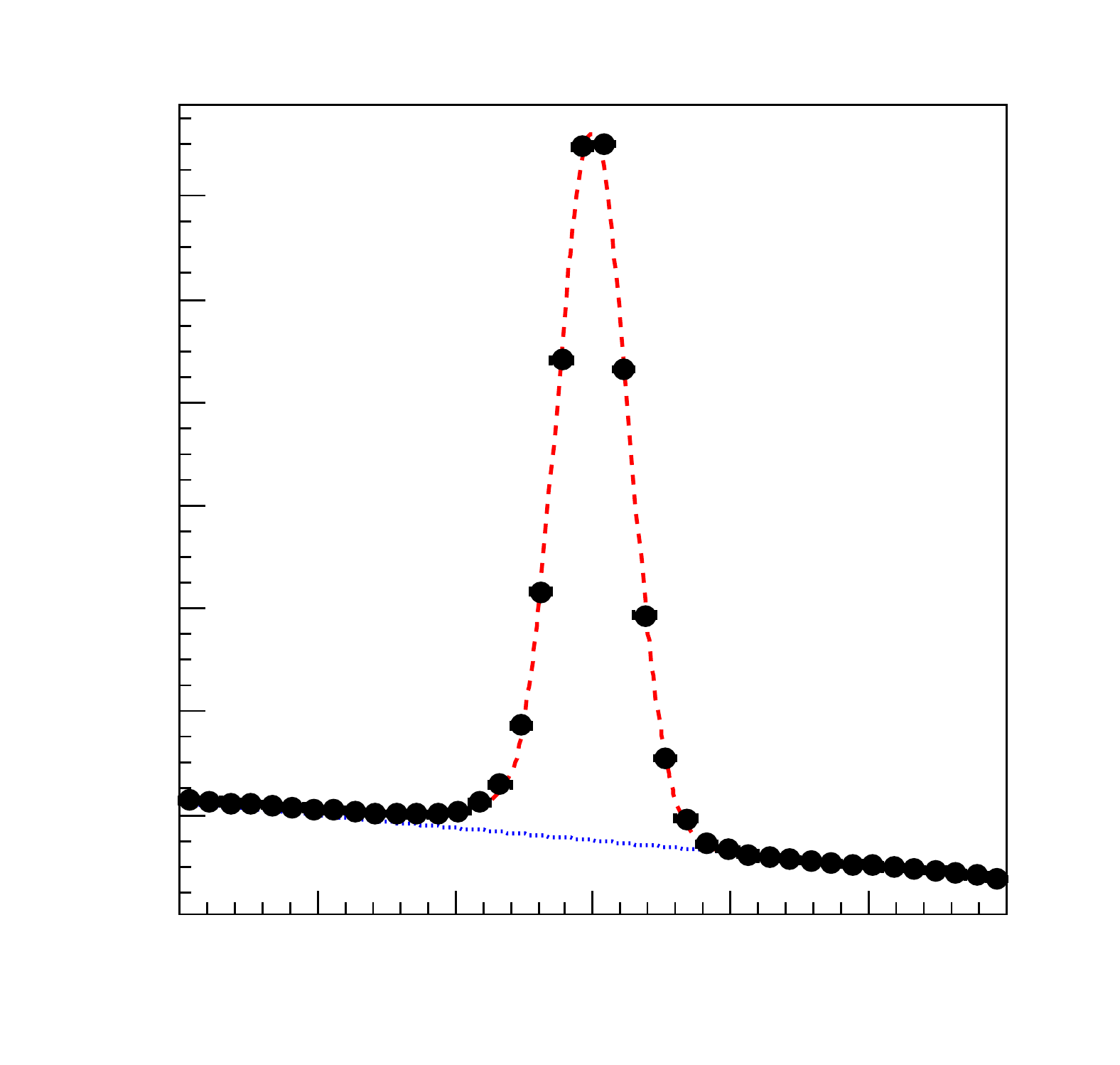}
  & \svg[2]{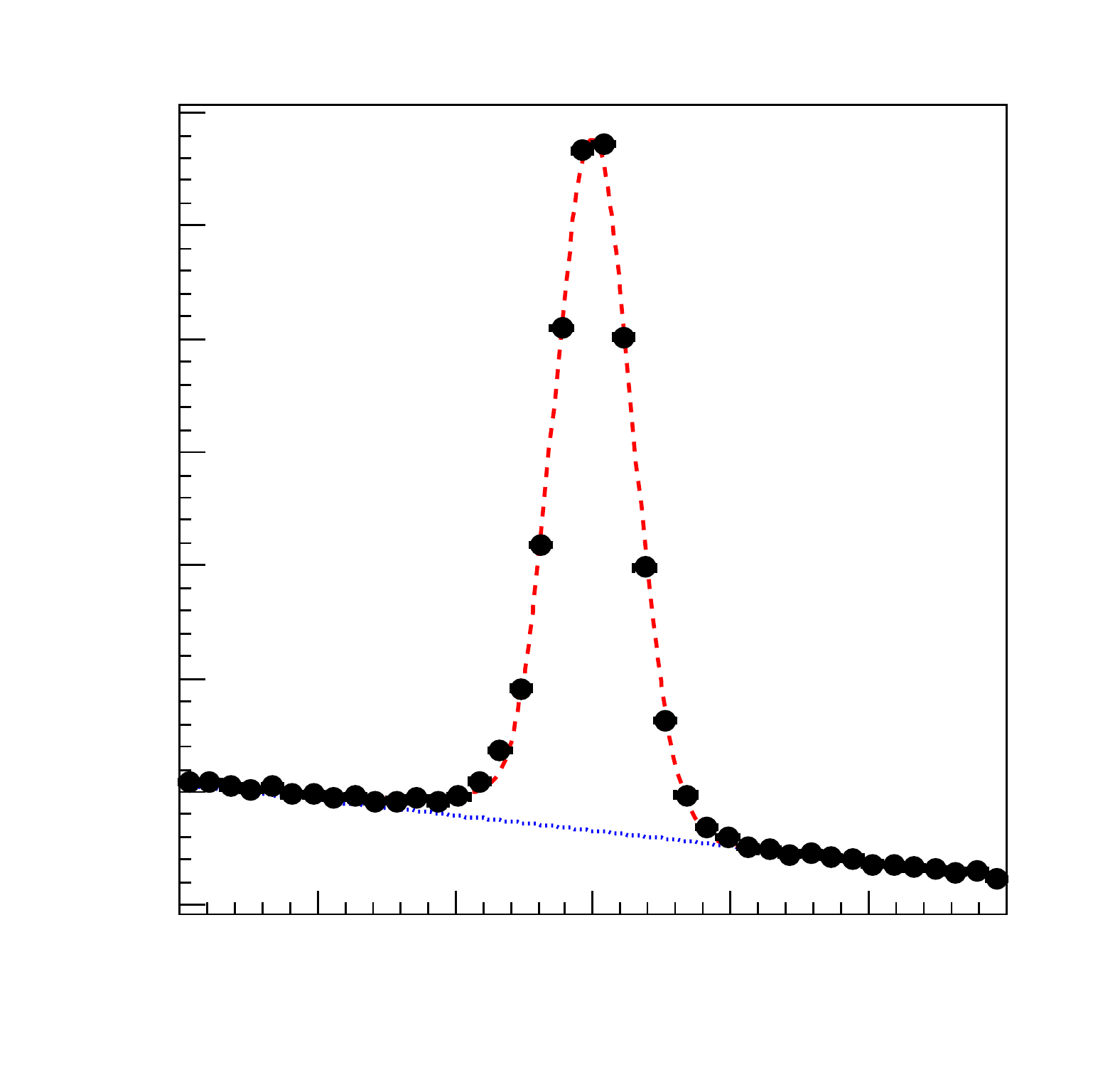} \effsep
  \svg[$50 < p_\mu < 100~\gev$]{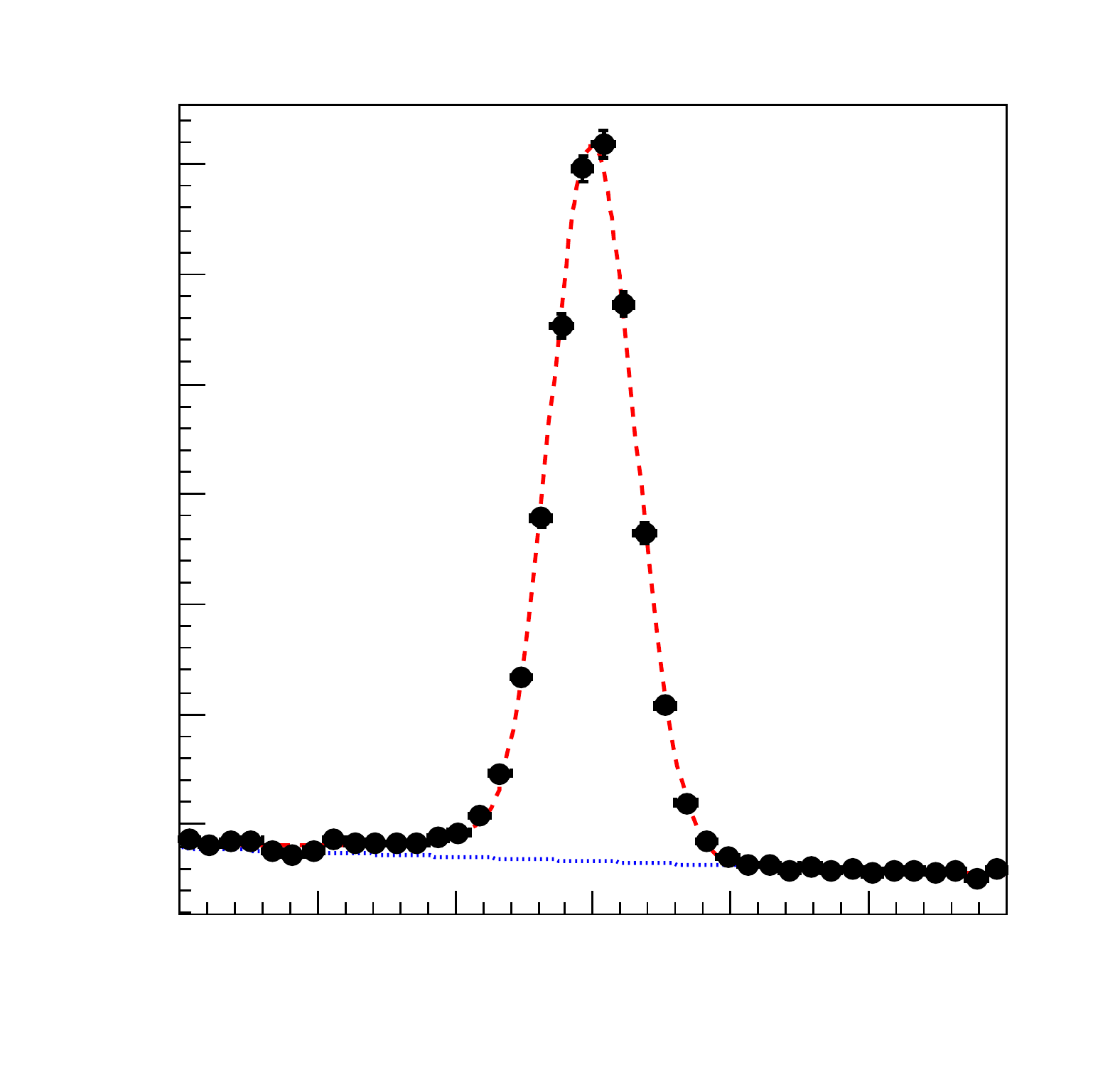}
  & \svg[2]{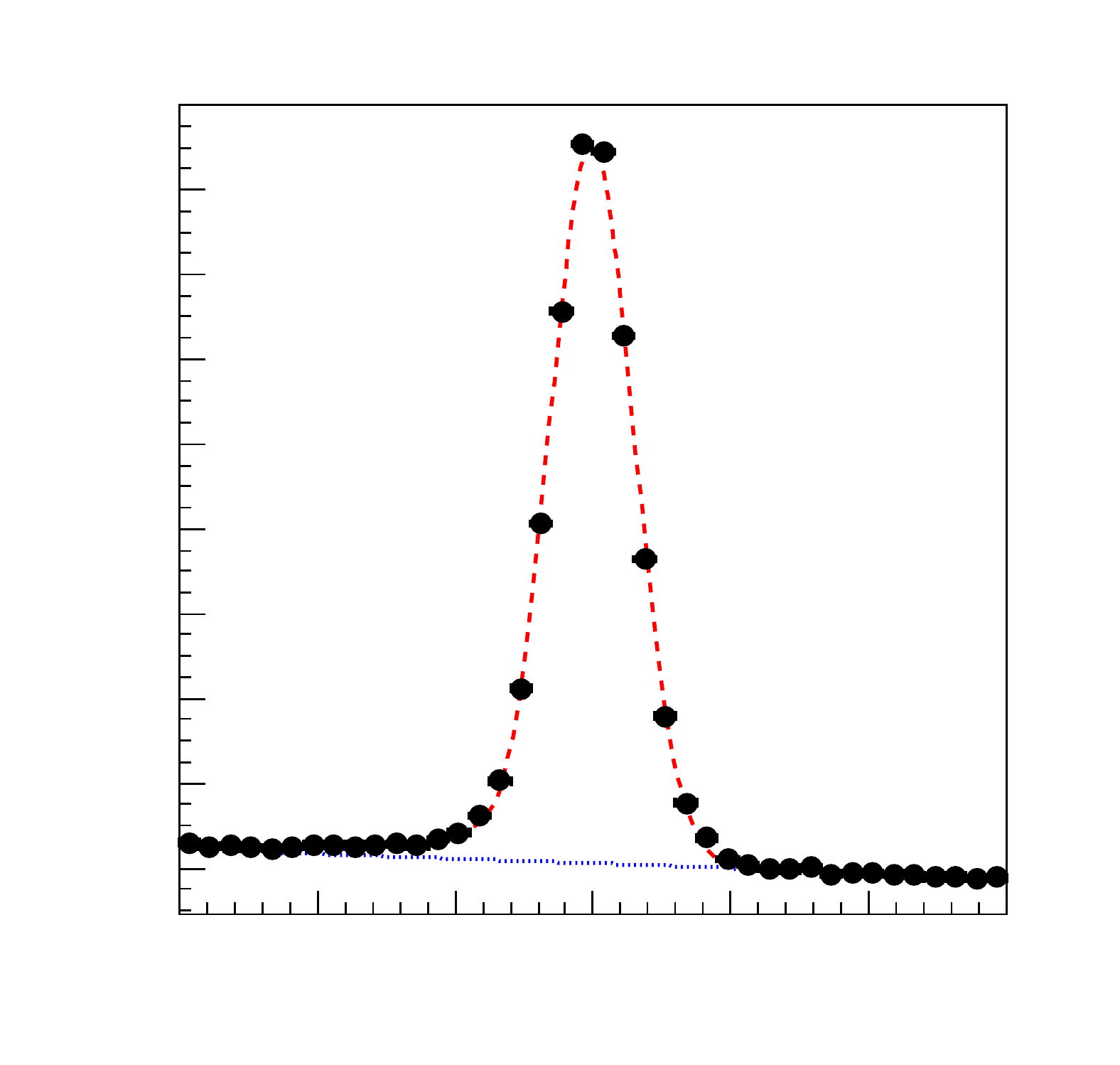}
  & \svg[2]{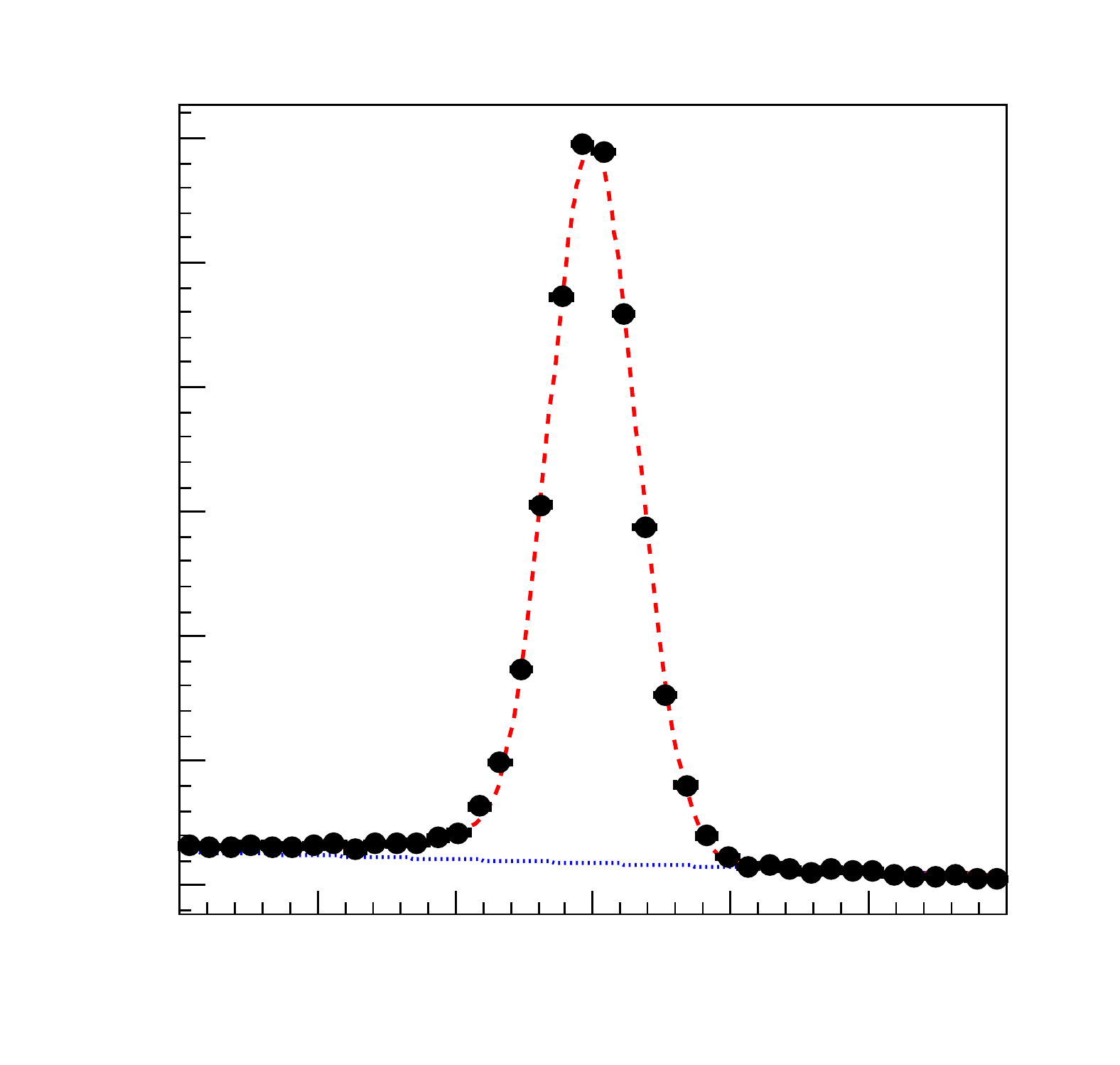} \effsep
  \svg[$100 < p_\mu < 150~\gev$]{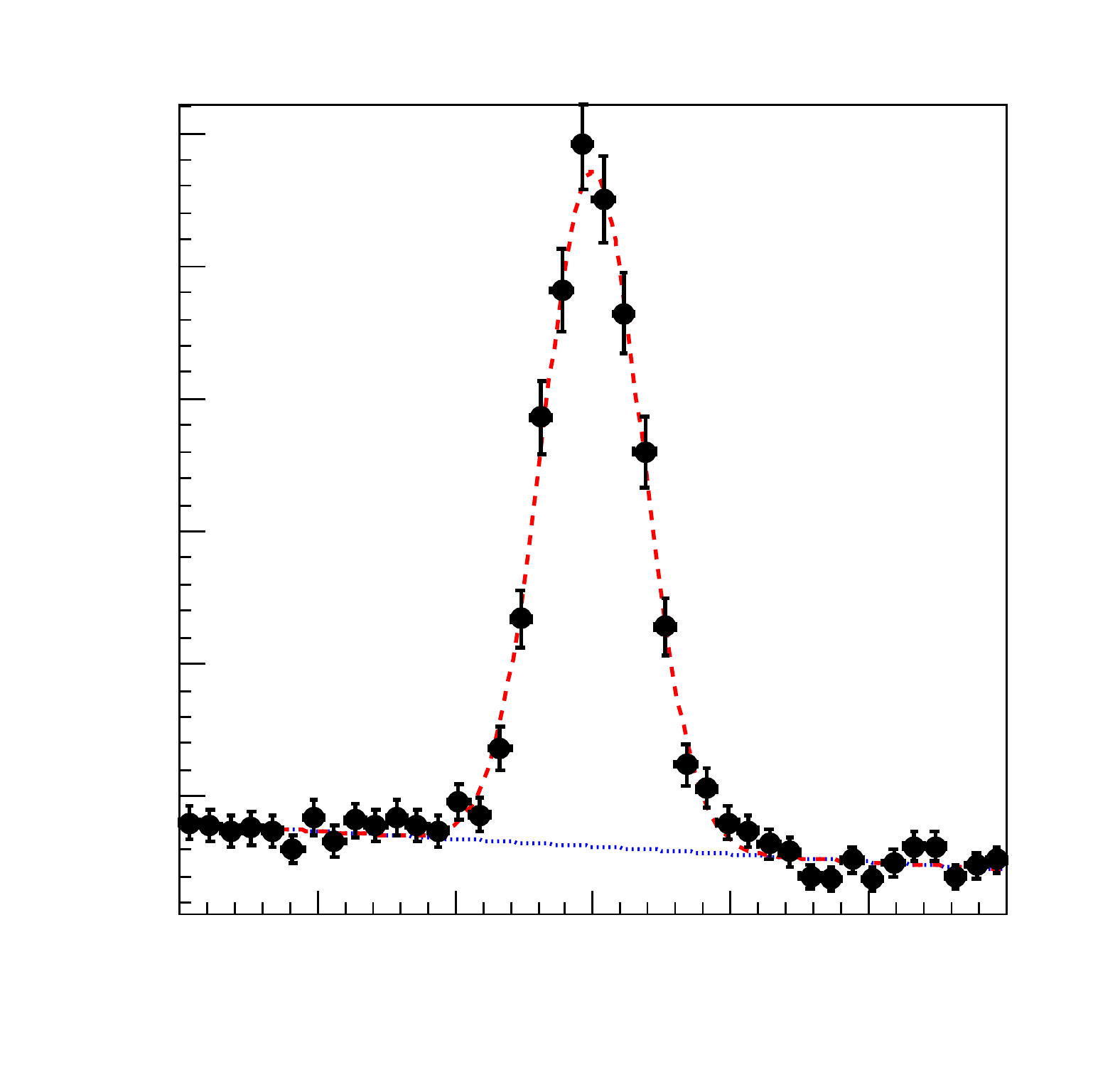}
  & \svg[2]{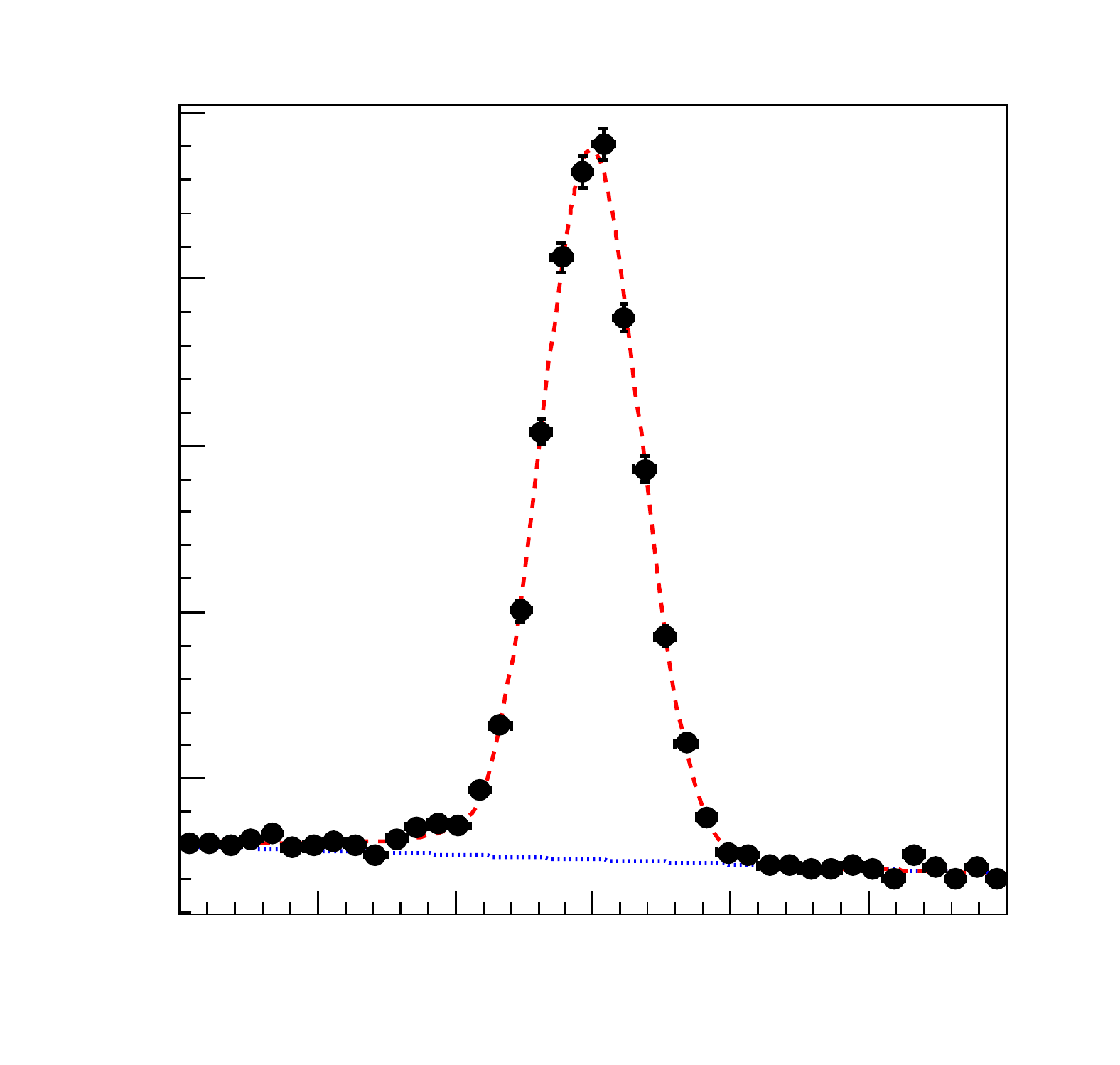}
  & \svg[2]{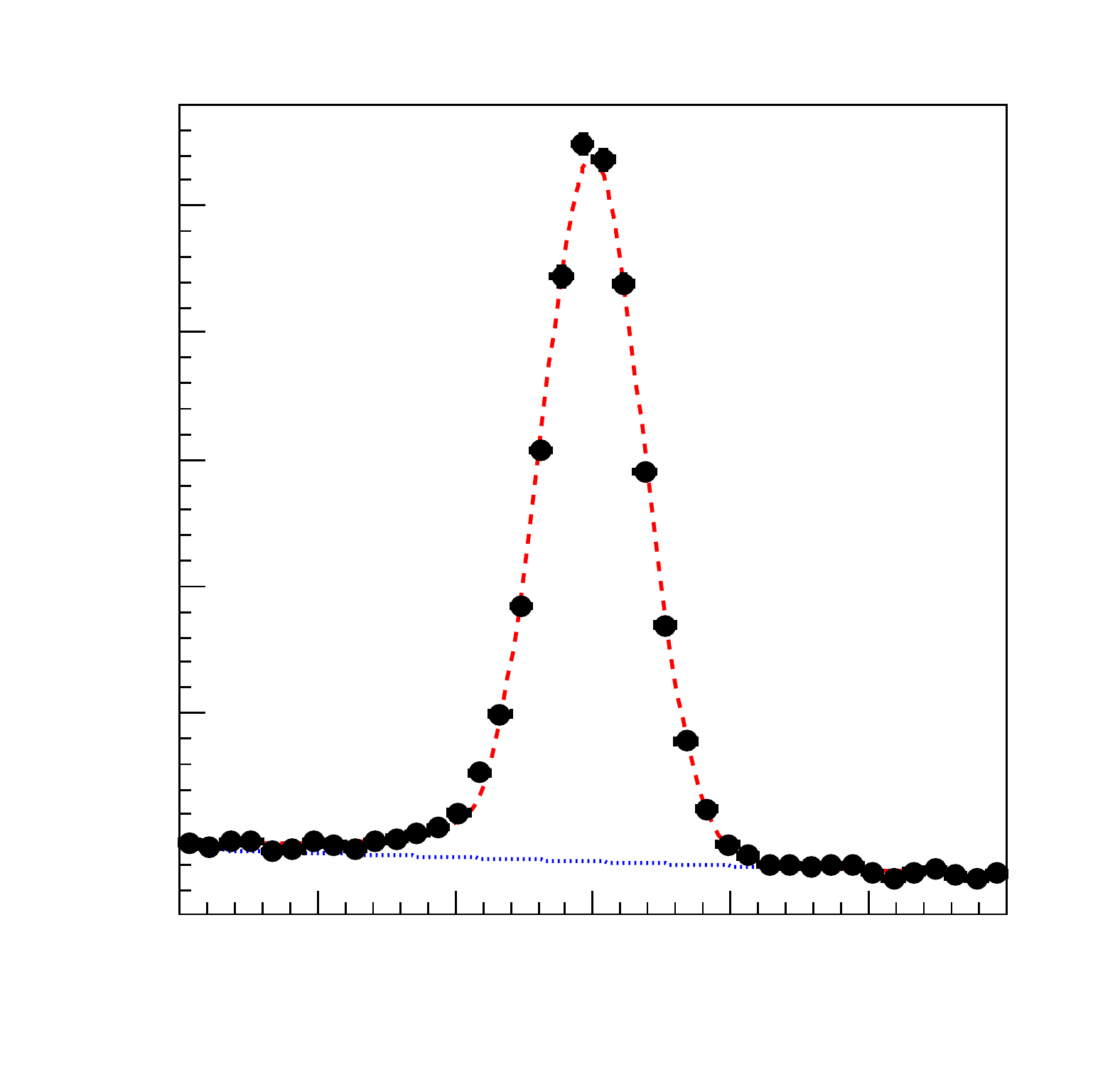} \effsep
  \svg[$150 < p_\mu < 200~\gev$]{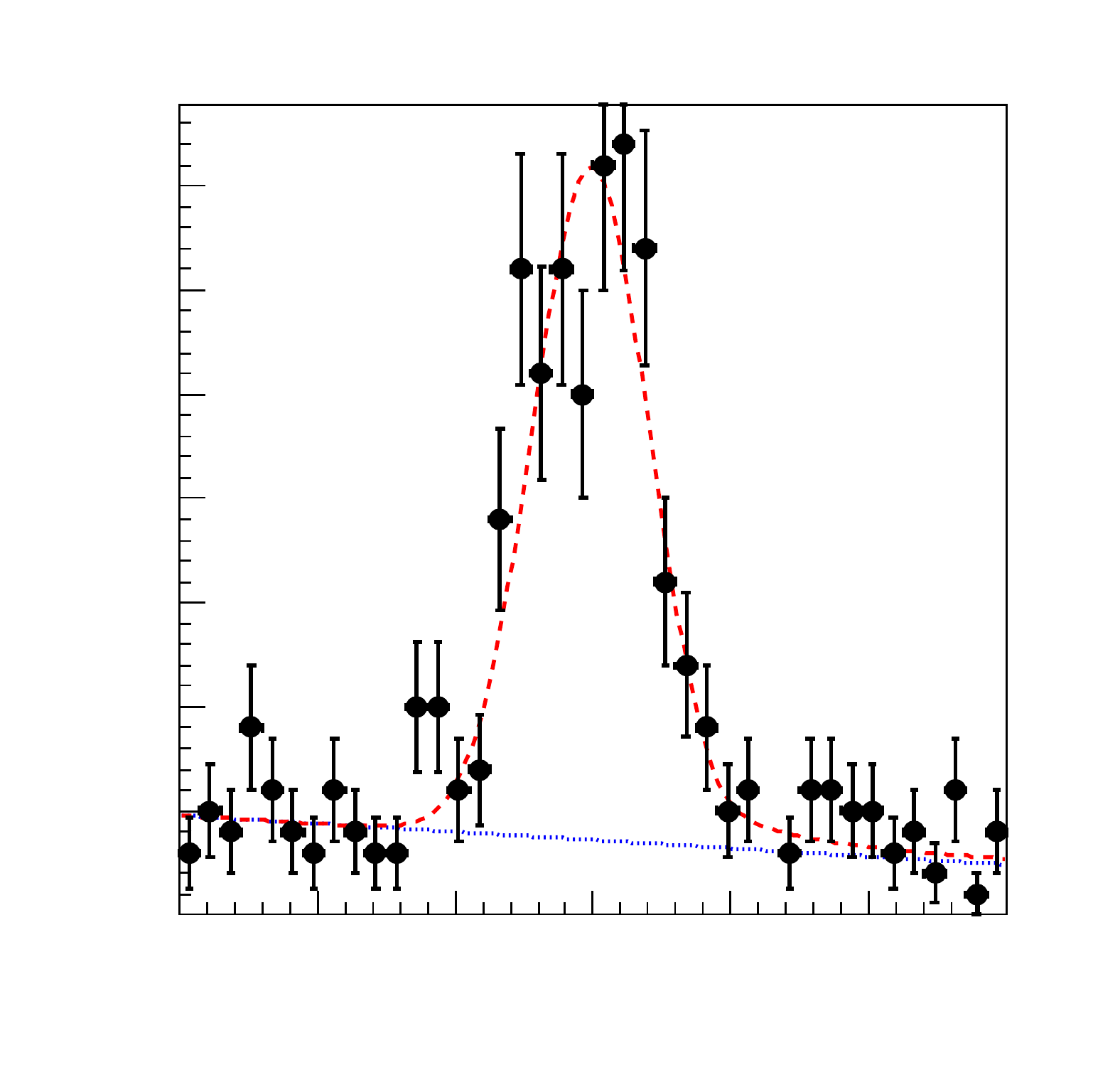}
  & \svg[2]{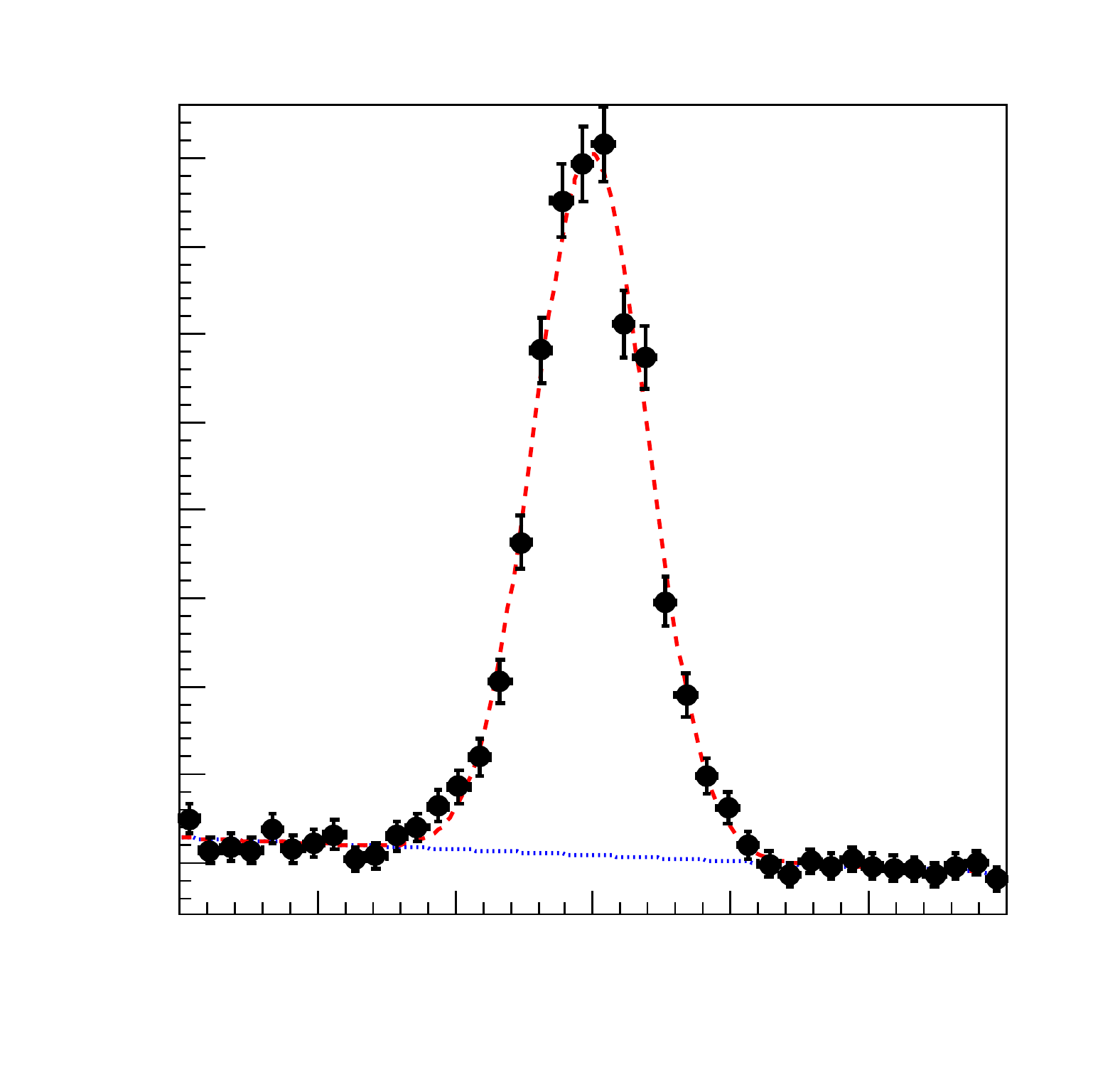}
  & \svg[2]{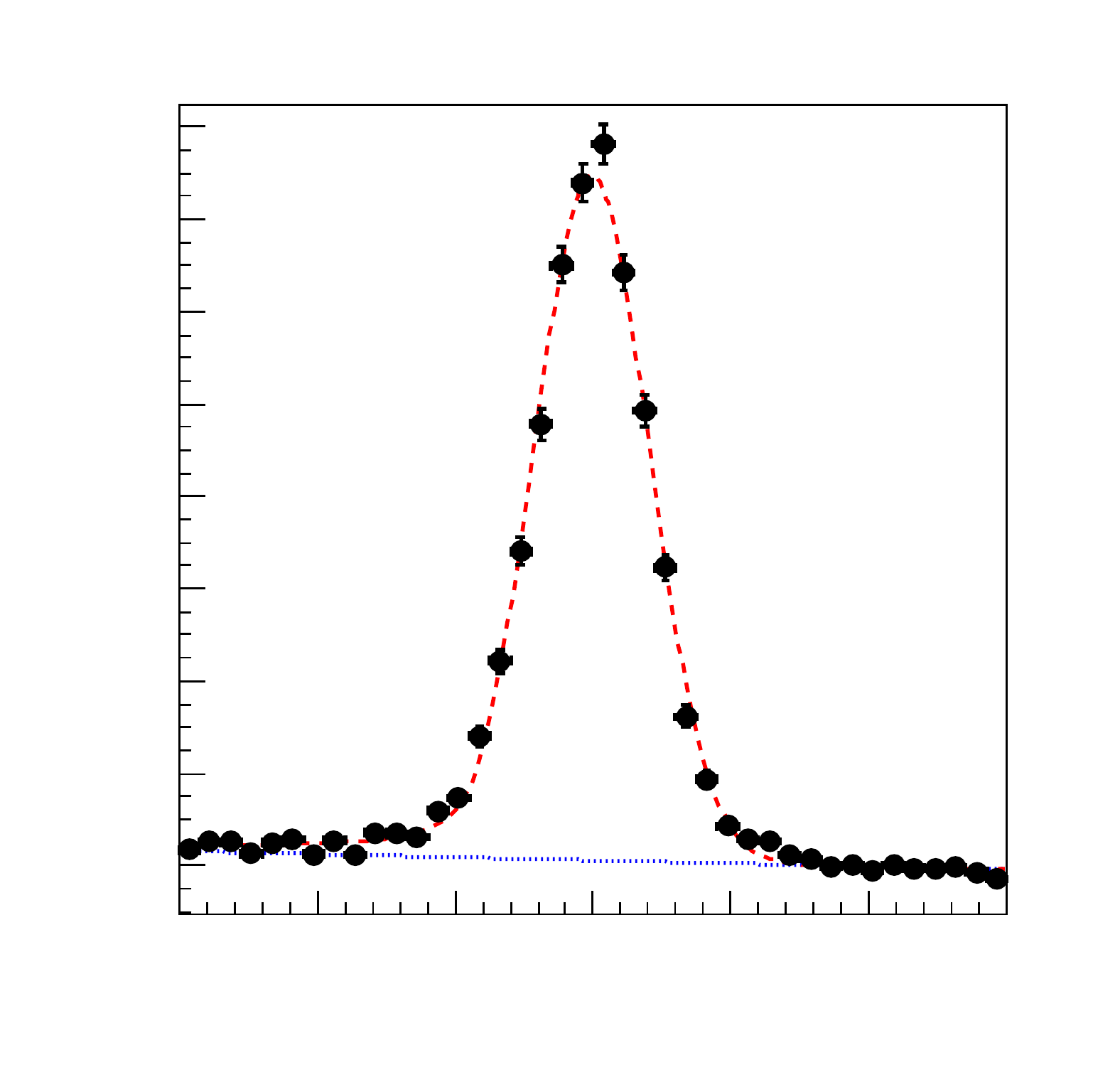} \effend
\end{subfigures}

The number of signal events in \figs{Zvr:RecEff.MuTrk.Total} through
\ref{fig:Zvr:RecEff.MuId.Pass} are extracted using two different
fitting methods. The efficiency for each bin in momentum and
pseudo-rapidity is the number of extracted signal events after
applying the selection criteria to the probe, over the number of
extracted signal events before applying the selection criteria. The
first method is performed by fitting a first degree polynomial,
\begin{equation}
  f_1(x) = P_0 + P_1 x
  \labelequ{Line}
\end{equation}
to the background on either side of the \jpsi resonance with initial
values of ${P_0 = 0}$ and ${P_1 = 0}$. For the muon track finding
efficiency, the ranges $2.70 < \m_{\dimu} < 2.95~\gev$ and $3.30 <
\m_{\dimu} < 3.70~\gev$ are fit while the ranges $2.95 < \m_{\dimu} <
3.03~\gev$ and $3.20 < \m_{\dimu} < 3.25~\gev$ are fit for the muon
identification efficiency. The number of extracted signal events is
the number of events within the distribution, less the integral of the
polynomial divided by the bin width.

The second fitting method uses four fits to converge to a final fit of
the distributions. The method begins by fitting the distributions with
\equ{Zvr:Line} and using the initial parameters ${P_0 = 0}$ and ${P_1
  = 0}$. Next, the distributions are fit with a first degree
polynomial plus a Gaussian distribution,
\begin{equation}
  f_2(x) = f_1(x) + P_2 e^{-\frac{(x - P_3)^2}{2P_4^2}}
  \labelequ{Gaussian}
\end{equation}
where the initial values for $P_0$ and $P_1$ are taken from the first
fit, ${P_2 = 0}$, ${P_3 = 3.1}$, and ${P_4 = 0.05}$. Following this,
the distributions are fit with a crystal ball
function~\cite{skwarnicki.86.1},
\begin{equation}
  f_3(x) =  f_1(x) + P_2
  \begin{cases}
    \left(\frac{P_6}{|P_5|}\right)^{P_6}
    e^{\frac{-P_5^2}{2}} \left(\frac{P_6}{|P_5|} - |P_5| - \frac{x -
        P_3}{P_4} \right)^{-P_6} & \mbox{if } \frac{x - P_3}{4} \leq -P_5 \\
    e^{-\frac{(x - P_3)^2}{2P_4^2}} & \mbox{if } \frac{x - P_3}{4} > -P_5 \\
  \end{cases}
  \labelequ{CrystalBall}
\end{equation}
where the initial values for $P_0$ through $P_4$ are taken from the
fit with \equ{Zvr:Gaussian}, ${P_5 = 1}$, and the value for $P_6$ is
fixed at $1$.

The distributions from before applying the selection criteria and
after are then fit simultaneously for each bin in momentum and
pseudo-rapidity using \equ{Zvr:CrystalBall}, where the background and
normalisation parameters $P_0$ through $P_2$ are independent between
the two distributions, but the shape parameters $P_3$ through $P_6$
are shared, with $P_6$ being allowed to vary. The number of signal
events is estimated as the integral of the final simultaneous fits
when setting the linear background to zero, ${P_1 = P_1 = 0}$, and
dividing by the bin width of the distribution.

\newsection{Combined Fit}{Fit}

The combined fit of the five cross-sections in \sec{Zed:Res} is
performed using the method of the best linear unbiased estimator from
\rfr{lyons.88.1}. The best linear unbiased estimator $\hat{x}$ is,
\begin{equation}
  \hat{x} = \vec{w} \vec{x}^\dagger
\end{equation}
where $\vec{x}$ is a row vector of the $n$ independent measurements
being combined, and $\vec{w}$ is a row vector of weights, required to
sum to unity, with an element for each measurement. The uncertainty
for $\hat{x}$ is,
\begin{equation}
  \delta_{\hat{x}}^2 = \vec{w} \mathcal{V} \vec{w}^\dagger
\end{equation}
where $\mathcal{V}$ is the $n$-by-$n$ symmetric covariance matrix for
$\vec{x}$. The weights $\vec{w}$ are found by minimising
$\delta_{\hat{x}}$ which is given by,
\begin{equation}
  \vec{w} = \left(\frac{\mathcal{V}^{-1} \vec{u}^\dagger}{\vec{u}
      \mathcal{V}^{-1} \vec{u}^\dagger} \right)^\dagger
\end{equation}
using the method of Lagrange multipliers~\cite{press.07.1} where
$\vec{u}$ is a row matrix of dimension $n$ with all entries unity and
$\mathcal{V}^{-1}$ is the inverse of the covariance matrix. The
$\chi^2$ for the estimator is given by,
\begin{equation}
  \chi^2 = \left(\hat{x}\vec{u} - \vec{x}\right) \mathcal{V}^{-1}
  \left(\hat{x}\vec{u} - \vec{x}\right)^\dagger
\end{equation}
where the number of degrees of freedom is ${n - 1}$.

For the combined measurement of \equ{Zed:Combined} the $\hat{x}$ and
 $\vec{x}$ are given by,
\begin{equation}
  \hat{x} = \sigma_{\dip \to \z \to \ditau} , ~
  \vec{x} = \sigma_{\dip \to \z \to \ditau} \begin{pmatrix}  (\mumu), 
    &  (\mue),
    &  (\emu),
    &  (\muh),
    &  (\eh)
  \end{pmatrix}
\end{equation}
where the values for the five cross-sections, $\sigma_{\dip \to \z \to
  \ditau} ~ (\tau_1\tau_2)$, are given by \equ{Zed:Categories}. The
components of the covariance matrix $\mathcal{V}$ are given by,
\begin{equation}
  \mathcal{V} = \sum_i \vec{\delta}_{\sys_i} \mathcal{C}_{\sys_i}
   \vec{\delta}_{\sys_i}^\dagger +
   \vec{\delta}_\lum \mathcal{U} \vec{\delta}_\lum^\dagger +  
   \vec{\delta}_N \mathcal{I} \vec{\delta}_N^\dagger
   \labelequ{Covariance}
\end{equation}
where numerical values for the first term are tabulated in
\tab{Zvr:Covariance.Sys}, while the second term is tabulated in
\tab{Zvr:Covariance.Lum} and the third term is tabulated in
\tab{Zvr:Covariance.Sta}. The matrix $\mathcal{C}_{\sys_i}$ is the
correlation coefficient matrix for the row vector
$\vec{\delta}_{\sys_i}$ of each systematic uncertainty $i$. The
components of these matrices for the eleven reconstruction and
selection efficiency uncertainties are given in
\tab{Zvr:Correlation}. All background uncertainties are assumed to by
uncorrelated and so their correlation coefficient matrix is the
identity matrix. The row vector $\vec{\delta}_\lum$ consists of the
luminosity uncertainty for each event category, and $\vec{\delta}_N$
is a row vector of the statistical uncertainties for the five
category. The matrix $\mathcal{U}$ is a five-by-five matrix of ones
and $\mathcal{I}$ is the five-by-five identity matrix. The covariance
matrix $\mathcal{V}$ is initially estimated for the first iteration of
the fit using the relative uncertainties of \tab{Zed:Uncertainty} and
the cross-sections of \equ{Zed:Categories}. A second iteration of the
fit is then performed using a covariance matrix determined from the
relative uncertainties of \tab{Zed:Uncertainty} and the prior best fit
value. No further iterations are required as convergence within
numerical precision is reached after this second fit.

\begin{subtables}{2}{The covariance matrices used in the best linear
    unbiased estimate decomposed into \subtab{Covariance.Sys} the
    systematic uncertainty, \subtab{Covariance.Lum}~the fully
    correlated luminosity uncertainty, and \subtab{Covariance.Sta}~the
    fully uncorrelated statistical
    uncertainty.\labeltab{Covariance}}
  \setlength{\tabcolsep}{\oldtabcolsep}
  \subfloat[]{
    \begin{tabular}[b]{r|rrrrr}
      \toprule
      & \multicolumn{5}{c}{$\displaystyle \sum_i
        \vec{\delta}_{\sys_i} \mathcal{C}_{\sys_i}
        \vec{\delta}_{\sys_i}^\dagger ~
        \left[\pb^2\right]$} \\[0.1cm]
      & \multicolumn{1}{c}{\mumu} & \multicolumn{1}{c}{\mue} 
      & \multicolumn{1}{c}{\emu}  & \multicolumn{1}{c}{\muh} 
      & \multicolumn{1}{c}{\eh}  \\
      \midrule
      \multicolumn{1}{r|}{\mumu}
      & $63.3$ & $4.9$ & $5.6$ & $6.1$ & $8.0$ \\
      \cline{2-2}
      \multicolumn{2}{r|}{\mue}
      & $15.3$ & $6.9$ & $3.5$ & $5.5$ \\
      \cline{3-3}
      \multicolumn{3}{r|}{\emu}
      & $30.4$ & $4.4$ & $21.4$ \\
      \cline{4-4}
      \multicolumn{4}{r|}{\muh}
      & $6.7$ & $7.7$ \\
      \cline{5-5}
      \multicolumn{5}{r|}{\eh}
      & $32.0$ \\
      \bottomrule
    \end{tabular}
    \labeltab{Covariance.Sys}
  }
  & \sidecaption \\
  \setlength{\tabcolsep}{\oldtabcolsep}
  \subfloat[]{
    \begin{tabular}[b]{r|rrrrr}
      \toprule
      & \multicolumn{5}{c}{$\displaystyle \vec{\delta}_\lum
        \mathcal{U} \vec{\delta}_\lum^\dagger ~
        \left[\pb^2\right]$} \\[0.1cm]
      & \multicolumn{1}{c}{\mumu} & \multicolumn{1}{c}{\mue} 
      & \multicolumn{1}{c}{\emu}  & \multicolumn{1}{c}{\muh} 
      & \multicolumn{1}{c}{\eh}  \\
      \midrule
      \multicolumn{1}{r|}{\mumu}
      & $6.4$ & $6.4$ & $6.4$ & $6.4$ & $6.4$ \\
      \cline{2-2}
      \multicolumn{2}{r|}{\mue}
      & $6.4$ & $6.4$ & $6.4$ & $6.4$ \\
      \cline{3-3}
      \multicolumn{3}{r|}{\emu}
      & $6.4$ & $6.4$ & $6.4$ \\
      \cline{4-4}
      \multicolumn{4}{r|}{\muh}
      & $6.4$ & $6.4$ \\
      \cline{5-5}
      \multicolumn{5}{r|}{\eh}
      & $6.4$ \\
      \bottomrule
    \end{tabular}
    \labeltab{Covariance.Lum}
  }
  & \setlength{\tabcolsep}{\oldtabcolsep}
  \subfloat[]{
    \begin{tabular}[b]{r|rrrrr}
      \toprule
      & \multicolumn{5}{c}{$\displaystyle \vec{\delta}_N \mathcal{I}
        \vec{\delta}_N^\dagger ~
        \left[\pb^2\right]$} \\[0.1cm]
      & \multicolumn{1}{c}{\mumu} & \multicolumn{1}{c}{\mue} 
      & \multicolumn{1}{c}{\emu}  & \multicolumn{1}{c}{\muh} 
      & \multicolumn{1}{c}{\eh}  \\
      \midrule
      \multicolumn{1}{r|}{\mumu}
      & $93.8$ & $0.0$ & $0.0$ & $0.0$ & $0.0$ \\
      \cline{2-2}
      \multicolumn{2}{r|}{\mue}
      & $26.6$ & $0.0$ & $0.0$ & $0.0$ \\
      \cline{3-3}
      \multicolumn{3}{r|}{\emu}
      & $85.0$ & $0.0$ & $0.0$ \\
      \cline{4-4}
      \multicolumn{4}{r|}{\muh}
      & $54.9$ & $0.0$ \\
      \cline{5-5}
      \multicolumn{5}{r|}{\eh}
      & $127.4$ \\
      \bottomrule
    \end{tabular}
    \labeltab{Covariance.Sta}
  } \\
\end{subtables}

\begin{table}[p]\centering
  \captionabove{Table of systematic uncertainty correlations used in
    the global fit.\labeltab{Correlation}}
  \setlength{\tabcolsep}{1.5pt}
  \begin{sideways}
    \begin{tabular}{c|cccccc|cccccc|cccccc|cccccc}
      \toprule
      & & & & & & & & & & & & & & & & & & & & \\
      & \multicolumn{6}{c|}{\mue}
      & \multicolumn{6}{c|}{\emu}  & \multicolumn{6}{c|}{\muh}
      & \multicolumn{6}{c}{\eh}  \\
      & & & & & & & & & & & & & & & & & & & & \\
      \midrule
      & & & & & & & & & & & & & & & & & & & & \\
      $\cor_\rec$
      & $\cor_\gec$     & $\cor_\trg$    & ${\cor_\trk}_1$ 
      & ${\cor_\trk}_2$ & ${\cor_\id}_1$ & ${\cor_\id}_2$
      & $\cor_\gec$     & $\cor_\trg$    & ${\cor_\trk}_1$ 
      & ${\cor_\trk}_2$ & ${\cor_\id}_1$ & ${\cor_\id}_2$
      & $\cor_\gec$     & $\cor_\trg$    & ${\cor_\trk}_1$ 
      & ${\cor_\trk}_2$ & ${\cor_\id}_1$ & ${\cor_\id}_2$
      & $\cor_\gec$     & $\cor_\trg$    & ${\cor_\trk}_1$ 
      & ${\cor_\trk}_2$ & ${\cor_\id}_1$ & ${\cor_\id}_2$ \\[0.1cm]
      $\cor_\sel$ 
      &              & $\cor_\kin$  & $\cor_\iso$ 
      & $\cor_\dphi$  & $\cor_\ips$  & $\cor_\apt$
      &              & $\cor_\kin$  & $\cor_\iso$ 
      & $\cor_\dphi$  & $\cor_\ips$  & $\cor_\apt$
      &              & $\cor_\kin$  & $\cor_\iso$ 
      & $\cor_\dphi$  & $\cor_\ips$  & $\cor_\apt$
      &              & $\cor_\kin$  & $\cor_\iso$ 
      & $\cor_\dphi$  & $\cor_\ips$  & $\cor_\apt$ \\
      & & & & & & & & & & & & & & & & & & & & \\
      \midrule
      & & & & & & & & & & & & & & & & & & & & \\
      \multirow{2}{*}{\mumu}
      & 1 & 1 & 1 & 0 & 1 & 0   & 0 & 0 & 0 & 1 & 0 & 1   
      & 1 & 1 & 1 & 1 & 1 & 0   & 0 & 0 & 0 & 1 & 0 & 0   \\[0.1cm]
      &   & 0 & 1 & 1 & 0 & 0   &   & 0 & 1 & 1 & 0 & 0  
      &   & 0 & 1 & 1 & 1 & 0   &   & 0 & 1 & 1 & 1 & 0   \\
      & & & & & & & & & & & & & & & & & & & & \\
      \midrule
      \multicolumn{7}{r|}{} & & & & & & & & & & & & & & \\
      \multicolumn{7}{r|}{\multirow{2}{*}{\mue}}
      & 0 & 0 & 0 & 0 & 0 & 0   & 1 & 1 & 1 & 0 & 1 & 0
      & 0 & 0 & 0 & 0 & 0 & 0   \\[0.1cm]
      \multicolumn{7}{r|}{}
      &   & 0 & 1 & 1 & 0 & 0   &   & 0 & 1 & 1 & 0 & 0 
      &   & 0 & 1 & 1 & 0 & 0   \\
      \multicolumn{7}{r|}{} & & & & & & & & & & & & & & \\
      \cmidrule(r){7-25}
      \multicolumn{13}{r|}{} & & & & & & & & \\
      \multicolumn{13}{r|}{\multirow{2}{*}{\emu}}
      & 0 & 0 & 0 & 1 & 0 & 0   & 1 & 1 & 1 & 1 & 1 & 0   \\[0.1cm]
      \multicolumn{13}{r|}{}
      &   & 0 & 1 & 1 & 0 & 0   &   & 0 & 1 & 1 & 0 & 0   \\
      \multicolumn{13}{r|}{} & & & & & & & & \\
      \cmidrule(r){13-25}
      \multicolumn{19}{r|}{} & & \\
      \multicolumn{19}{r|}{\multirow{2}{*}{\muh}}
      & 0 & 0 & 0 & 1 & 0 & 1 \\[0.1cm]
      \multicolumn{19}{r|}{}
      &   & 0 & 1 & 1 & 1 & 0 \\
      \multicolumn{19}{r|}{} & & \\
      \cmidrule[\heavyrulewidth](r){19-25}
    \end{tabular}
  \end{sideways}
\end{table}
\newappendix{Higgs Boson}{Hvr}

This appendix provides supplemental material for \chp{Hig}. Additional
information and plots for the \whb phenomenology of \sec{Hig:Phe} are
given in \sap{Hvr:Phe}. In \sap{Hvr:Acc} numerical values for the
acceptances and efficiencies of \sec{Hig:Nrm} are tabulated. Further
information on the statistical methods of \sec{Hig:Sta} is provided in
\sap{Hvr:Sta}.

\newsection{Higgs Phenomenology}{Phe}

Details on the masses of the \cp-even \mssm \whbs are given in
\sap{Hvr:Mass}. The methods used for calculating the branching
fractions of \sec{Hig:Br} are given in \sap{Hvr:BrSm} for the \sm \whb
and in \sap{Hvr:BrMssm} for the \mssm \whbs. The cross-section
calculations of \sec{Hig:Xs} are detailed in \sap{Hvr:XsSm} for the
\sm \whb and \sap{Hvr:XsMssm} for the \mssm \whbs.

\newsubsection{MSSM Higgs Boson Masses}{Mass}

The masses of the light and heavy \cp-even \whbs from
\fig{Hig:Mssm.Mass} are given in \fig{Hvr:Mssm.Mass}, but with
overlayed numerical values. These values are given in $17$ bins of
$10~\gev$ for the \cp-odd \whb mass and $20$ bins of $7$ for \tanb.

\begin{subfigures}{2}{Numerical values for the mass of the
    \subfig{Mssm.M.H1.Text}~light \cp-even and
    \subfig{Mssm.M.H2.Text}~heavy \cp-even \mssm \whbs as a function
    of the \cp-odd \whb mass and \tanb.\labelfig{Mssm.Mass}}
  \svgbeg \svg{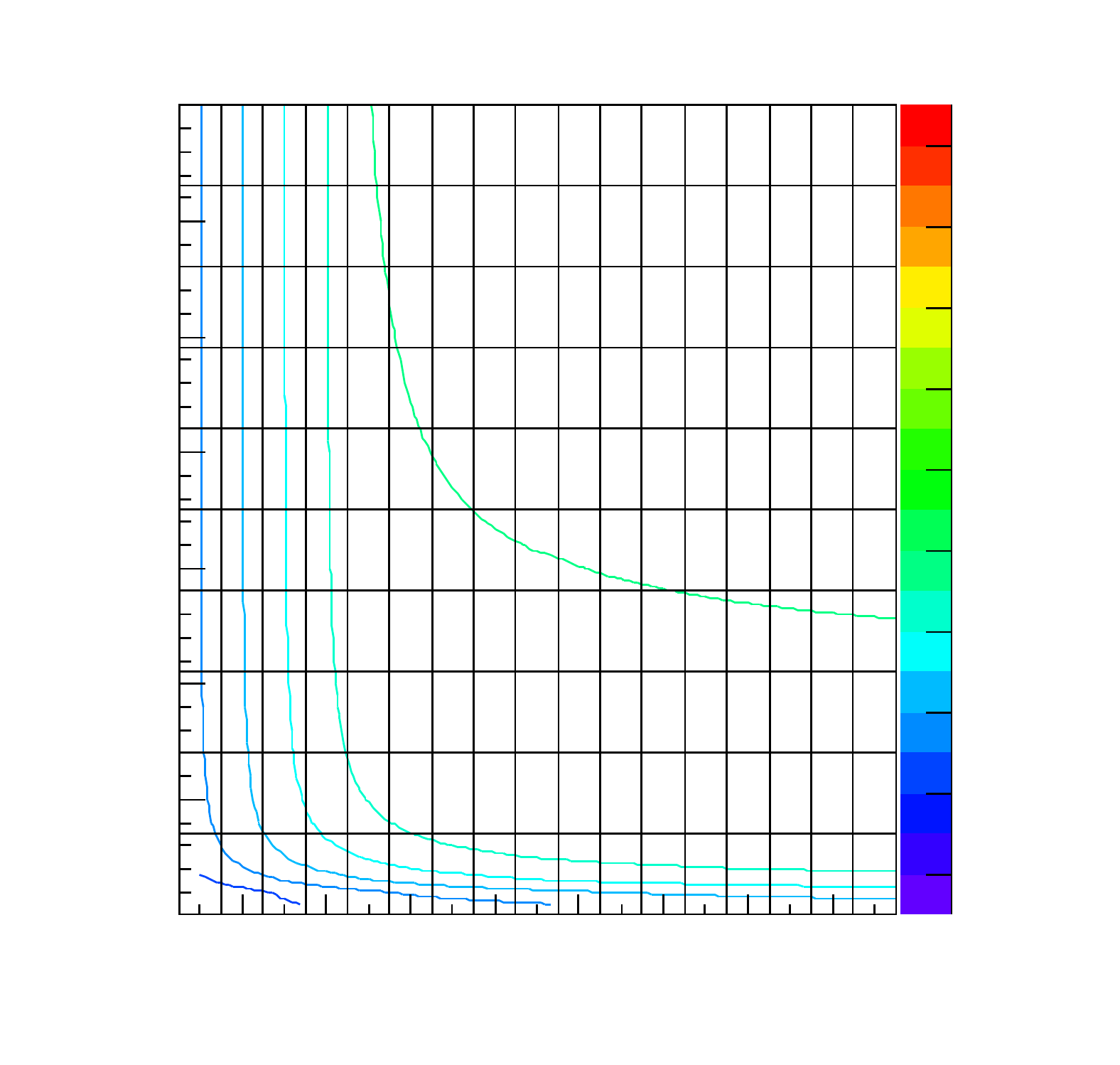} & \svg{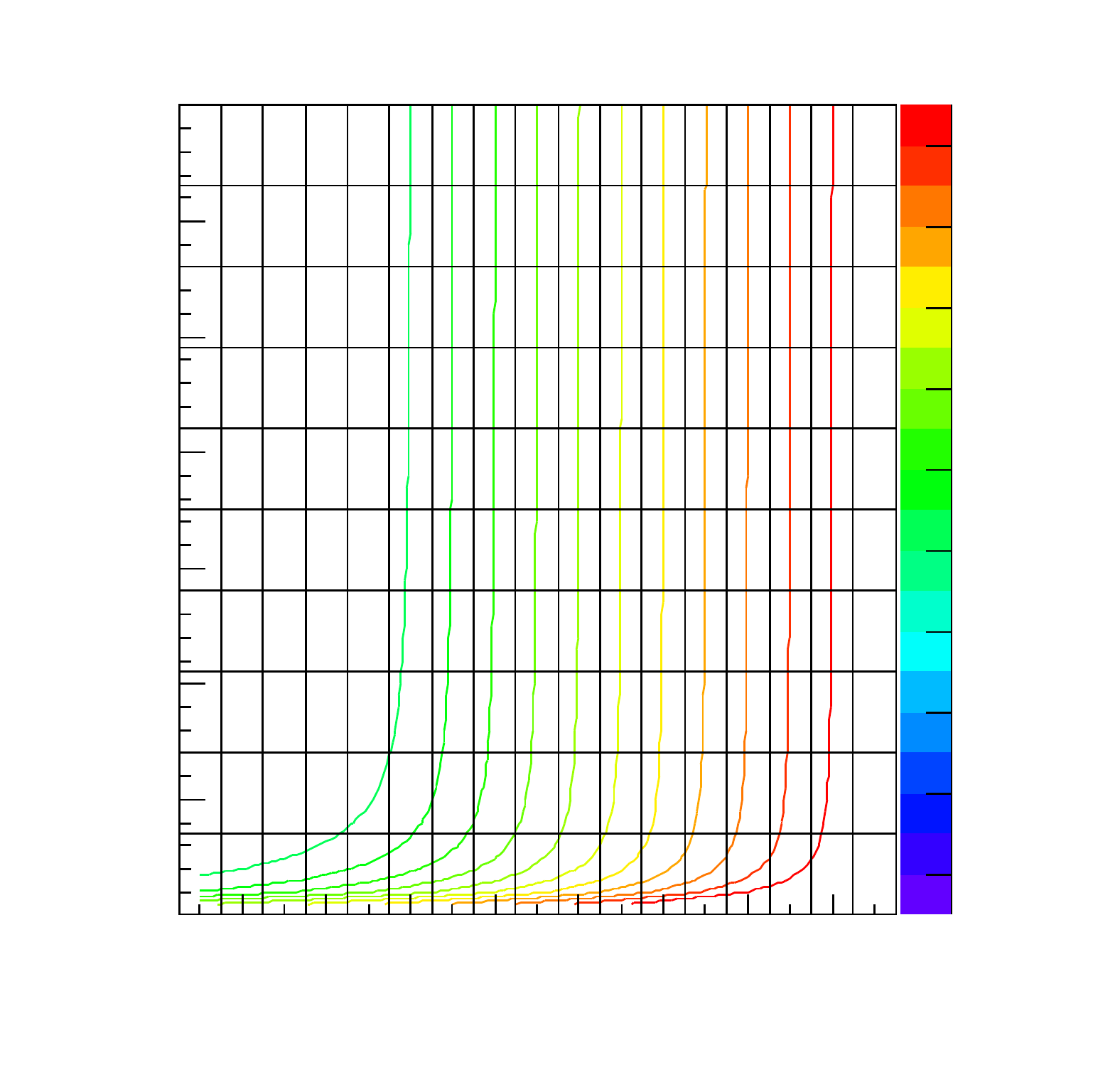} \svgend
\end{subfigures}

\newsubsection{SM Branching Fraction Calculations}{BrSm}

The \sm \whb branching fractions are calculated using the results of
\rfr{denner.11.1} and the \whb decay calculators \hdecay
\cite{\citehdecay} and \prophecy \cite{\citeprophecy}. The \whb width
is calculated as,
\begin{align*}\labelali{Sm.Width.Corrected}
  \Gamma_\hH = \Gamma_\hH^\hdecay &- \Gamma_{\hH \to \z\z}^\hdecay -
  \Gamma_{\hH \to \w\w}^\hdecay\\
  &+ \Gamma_{\hH \to \z^*\z^*}^\prophecy +
  \Gamma_{\hH \to \w^*\w^*}^\prophecy + \Gamma_{ \hH \to
    \z^*\z^*/\w^*\w^*}^\prophecy
\end{align*}
where the superscript indicates the program used to calculate the
width.

The program \hdecay calculates partial widths for all channels that
are kinematically allowed with branching fractions greater than
$10^{-4}\%$ resulting in the \whb decay width of \equ{Hig:Sm.Width.Total}
with the electron, \wuq, and \wdq pair widths excluded. For \whb
masses well above the quark pair threshold the quark pair partial
widths are calculated up to order \aS[^3], while the near threshold
widths are calculated up to order \aS[^2]. Electroweak corrections are
included at order \aE[^2] but are small in the considered \whb mass
range. For \wtq pair decays, radiative corrections at order \aS are
included and below threshold, the partial width for an on-shell and
off-shell pair is calculated as a three-body decay, $\hH \to
t\bar{t}^* \to t\bar{b}W$.

For the gauge boson combinations, the gluon pair width is calculated
up to order \aS[^2] including splittings of the gluons into heavy
quark flavours. The photon pair and \wzb and photon partial widths are
calculated including heavy fermion and \wzb loops, where no radiative
\qcd corrections are made to the quark loops as they are
small. Next-to-leading order electroweak corrections for both the
gluon and photon pair widths are applied using calculations from
\rfrs{actis.08.1} and \cite{actis.09.1}. The vector boson pair widths
are calculated in three regimes: both bosons are off-shell, one
boson is off-shell, and both bosons are on-shell. Neither interference
between the four fermion final states $\nu\nu\lep\lep$ and
$q_iq_iq_jq_j$, nor electroweak corrections are included when
calculating the vector boson pair partial widths.

Because the vector boson branching fractions are critical in the \whb
mass region considered, a more complete calculation is performed using
the decay width calculator \prophecy. Here, the vector boson partial
width is calculated explicitly for the four fermion final states by
consistently using a complex-mass scheme over the full range of masses
for the \whb at next-to-leading order. Radiative electroweak
corrections are applied on the order \aE and interference between the
four fermion states from virtual \w and \wzbs is included. In
\equ{Hvr:Sm.Width.Corrected} the full width from \hdecay is corrected
by subtracting the \hdecay vector boson partial widths and adding the
\prophecy partial width including the interference term, $\Gamma_{\hH
  \to \z^*\z^*/\w^*\w^*}^\prophecy$.

The uncertainty for the partial widths and branching fractions, given
by the coloured bands of \fig{Hig:Sm.Br} are calculated from both
parametric and theoretical uncertainty. The parametric uncertainty
incorporates the experimental uncertainty on the measurement of \aS
and the masses of the \wcq, \wbq, and \wtq masses. The theoretical
uncertainties are due to missing higher orders, and are estimated from
explicitly known electroweak and \qcd higher order calculations.

Numerical values for the \sm \whb decaying into a \wtl pair are
tabulated as a function of the \sm \whb mass in
\fig{Hvr:Sm.Br.TauTau.Text}.

\begin{subfigures}[t]{2}{Numerical values for the \sm \whb branching
    fraction into \wtl pairs as a function of the \sm \whb mass.}
  \svg[1]{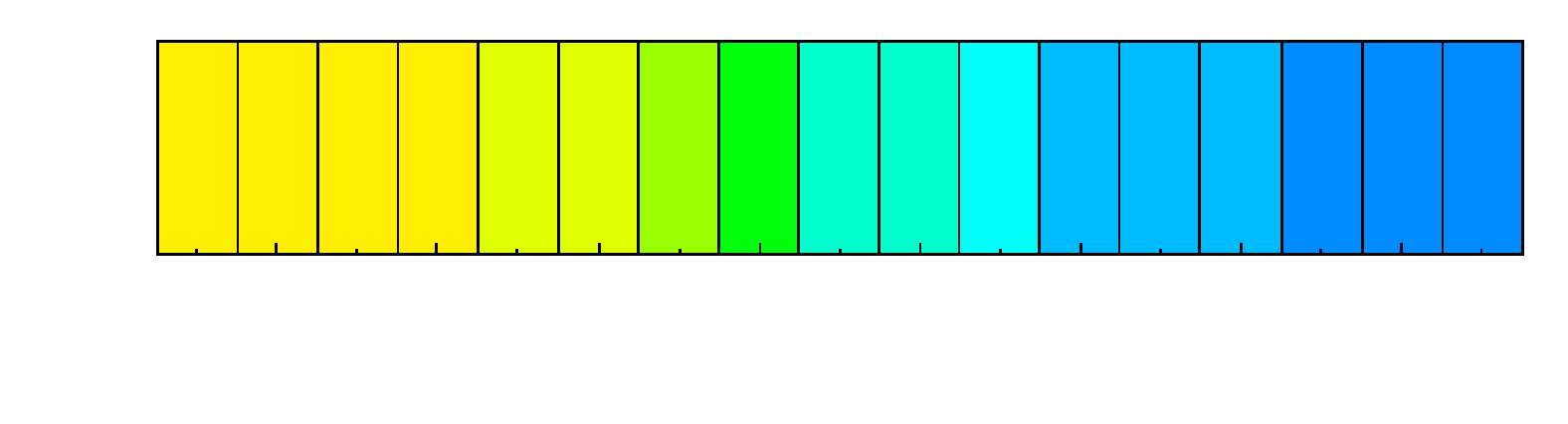} &
  \includesvg[pretex=\hspace{-0.43cm}\relsize{-3},width=0.5\columnwidth]
  {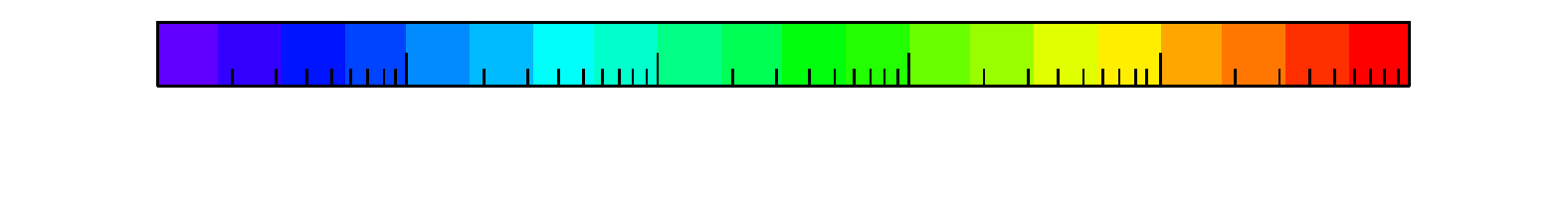} \svgend
\end{subfigures}

\newsubsection{MSSM Branching Fraction Calculations}{BrMssm}

The \wtl branching fractions of \fig{Hig:Mssm.Br} are calculated
following the recommendations of \rfr{hxswg.12.1}. The total decay
width is calculated as,
\begin{equation}
  \begin{aligned}
    \Gamma_\hz = \:&\Gamma_{\hz \to \dimu}^\feynhiggs + \Gamma_{\hz \to
      \ditau}^\feynhiggs + \Gamma_{\hz \to c\bar{c}}^\hdecay + \Gamma_{\hz \to
      b\bar{b}}^\hdecay + \Gamma_{\hz \to t\bar{t}}^\hdecay \\
    &+ \Gamma_{\hz \to gg}^\hdecay +
    \Gamma_{\hz \to \gamma\gamma}^\hdecay + \Gamma_{\hz \to
      Z\gamma}^\hdecay  + \Gamma_{\hz \to \z^*\z^*}^\prophecy +
    \Gamma_{\hz \to \w^*\w^*}^\prophecy \\
  \end{aligned}
  \labelequ{Mssm.Width.Corrected}
\end{equation}
where the superscripts indicate the decay width calculator used. The
two leptonic partial widths are calculated using \feynhiggs
\cite{\citefeynhiggs} with a full one-loop calculation. The vector
boson pair partial widths are calculated using \prophecy, just as for
the \sm, but excluding the interference terms and with the couplings
between the \whbs and the vector bosons modified using the couplings
calculated from \feynhiggs. The remaining partial widths are
calculated using \hdecay where the \whb masses and couplings are
passed to \hdecay from \feynhiggs. No uncertainties have been
calculated for the \mssm branching fractions.

Numerical values for the branching fractions of the \mssm \whbs into
\wtl pairs are tabulated as a function of the \cp-odd \whb and \tanb
in \fig{Hvr:Mssm.Br.Text}.

\begin{subfigures}{2}{Numerical values for the branching fractions, as
    a percentage, of the \subfig{Mssm.Br.H1.Text}~light \cp-even,
    \subfig{Mssm.Br.H2.Text}~heavy \cp-even, and
    \subfig{Mssm.Br.H3.Text}~\cp-odd \mssm \whbs into \wtl pairs as a
    function of the \cp-odd \whb mass and \tanb.\labelfig{Mssm.Br.Text}}
  \svgbeg
  \svg{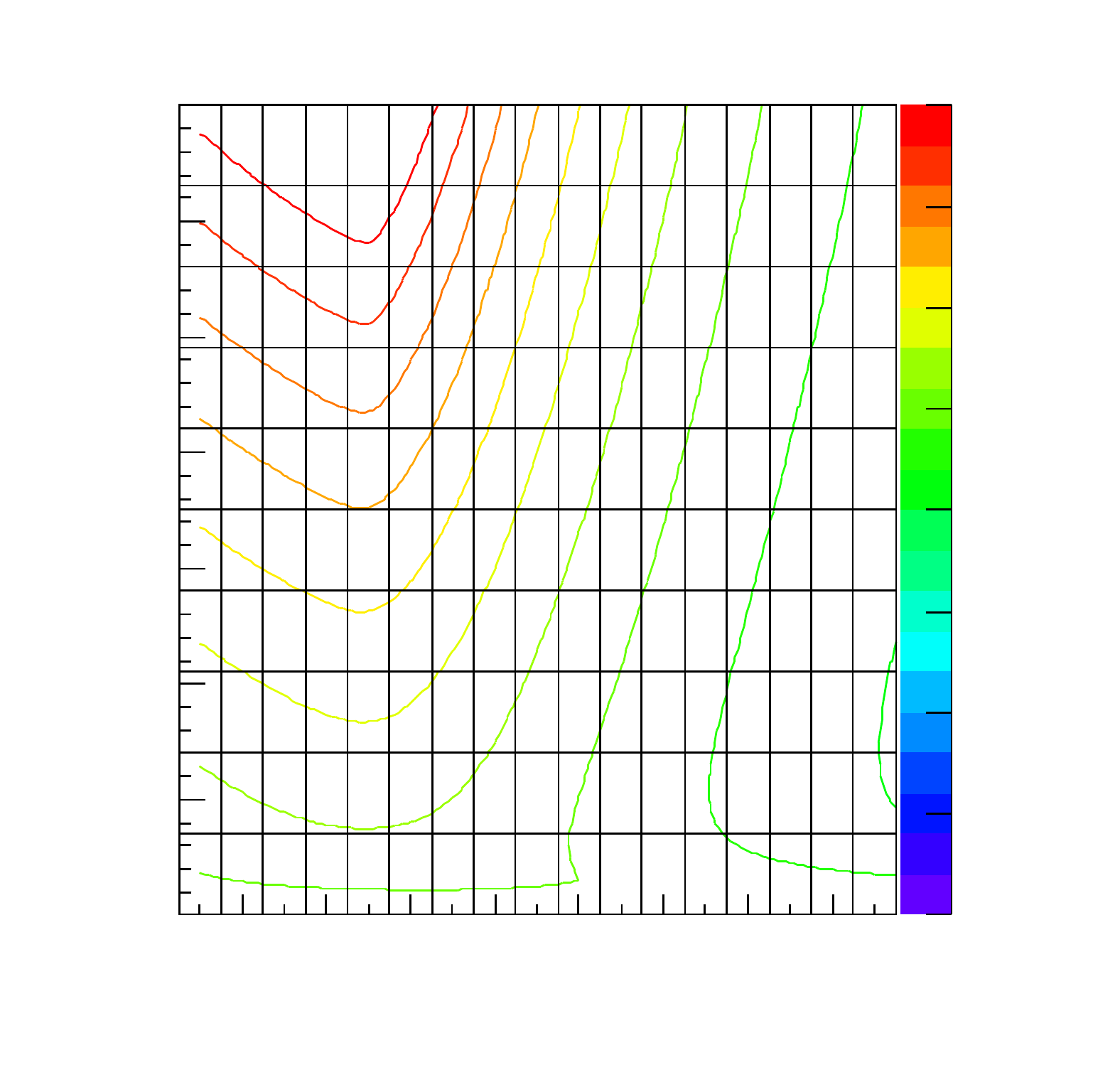} & \svg{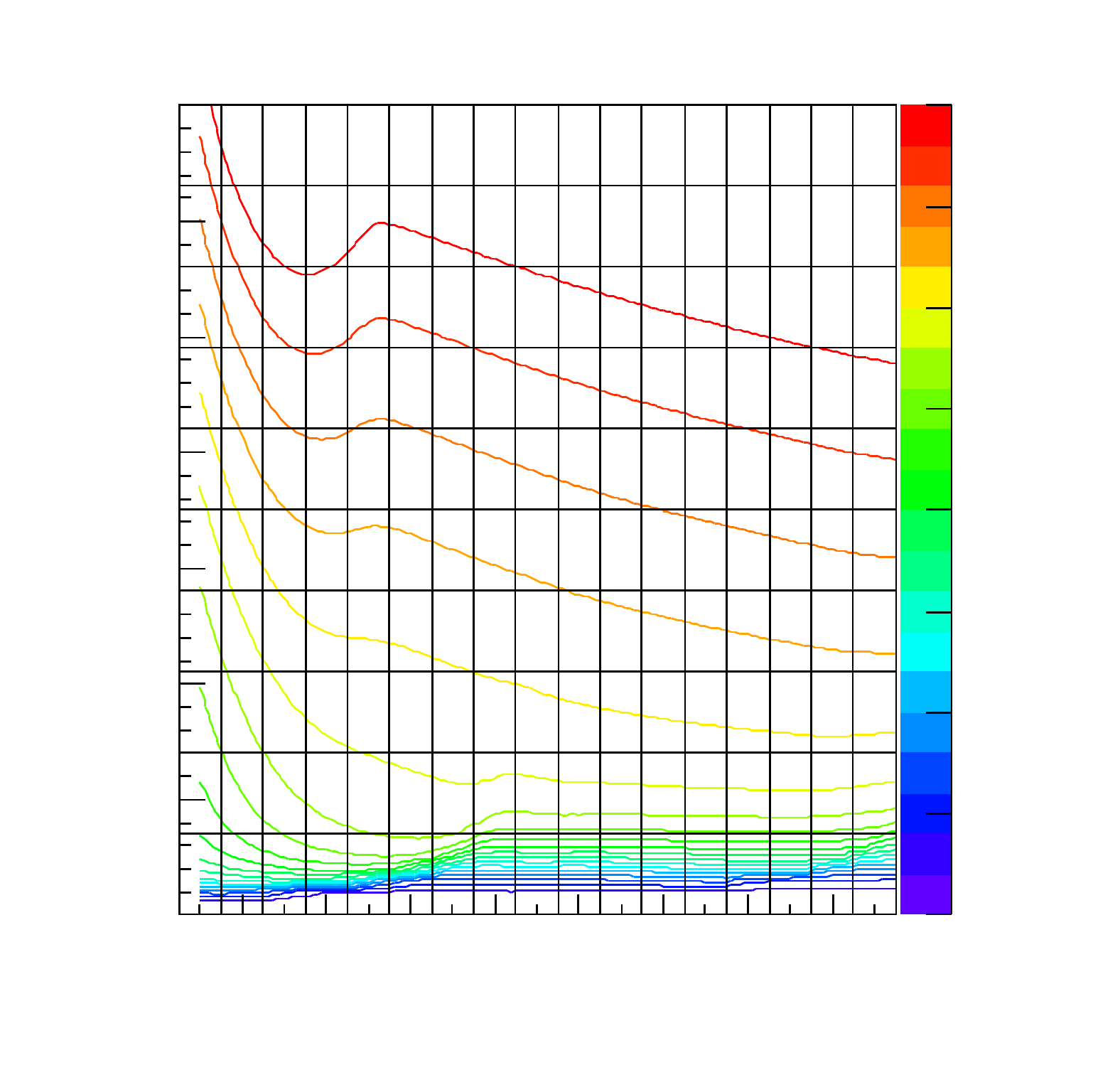} \svgend
  \sidecaption          & \svg{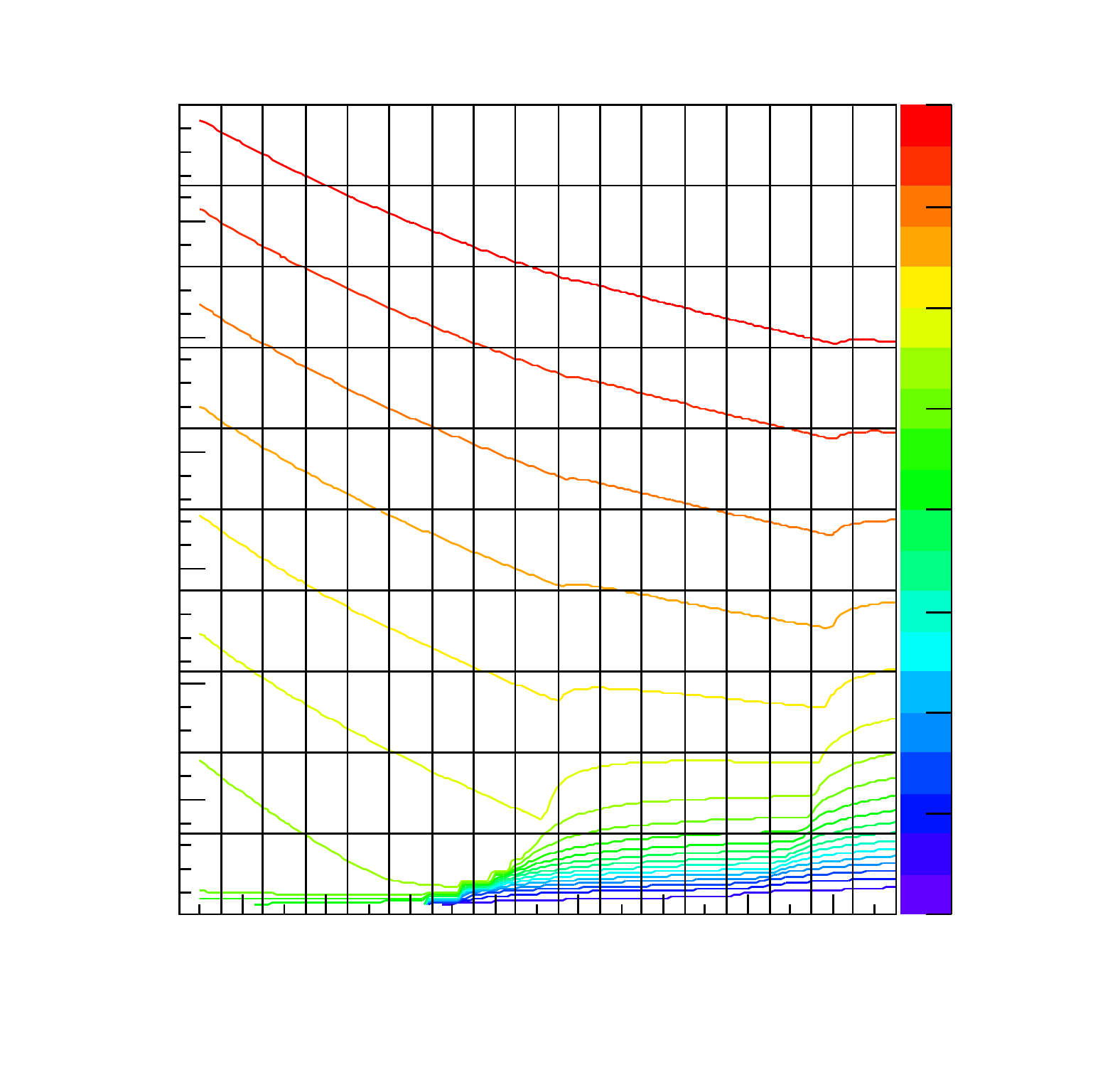} \svgend
\end{subfigures}

\newsubsection{SM Cross-Section Calculations}{XsSm}

The \ggf cross-section is determined using the
\higlu~\cite{\citehiglu} and \dfg~\cite{\citedfg} cross-section
calculators. At next-to-leading order \higlu calculates,
\begin{equation}
  \sigma_{gg \to \hH} = \sigma_{gg \to \hH}^\lo + \Delta \sigma_\mathrm{virt} +
  \Delta \sigma_{gg \to \hH g} + \Delta \sigma_{gq \to \hH q} + 
  \Delta \sigma_{qq \to \hH g}
  \labelequ{Sm.Xs.GGF}
\end{equation}
where the first term is the leading order one-loop \ggf cross-section,
the second term parametrises the infrared regularised virtual two-loop
corrections, and the three remaining terms correct for the one-leg
sub-processes, calculated with one loop. Resummation of soft gluon
contributions at next-to-leading log are then included by
\dfg. Contributions from the \wtq loop are calculated using the
large-$\m_t$ limit and added at next-to-next-to-leading log and
next-to-next-to-leading order. The result is then corrected for
electroweak contributions. The next-to-leading order corrections
increase the cross-section by approximately $80\%$ while the
next-to-next-to-leading order corrections increase the cross-section
by approximately $25\%$.

The uncertainty for the \ggf cross-section in \fig{Hig:Sm.Xs} is
indicated by the red band and is calculated from five sources. The
higher-order radiative \qcd corrections provide the largest
uncertainty and are determined by varying the factorisation and
renormalisation scales. The electroweak corrections introduce another
perturbative uncertainty, while the large-$\m_t$ approximation yields
a small uncertainty. The input masses for the \wbq and \wtq cause a
scheme dependence which is found to be small. The final source of
uncertainty is from the proton \PDF and is determined using the \mstw
\PDF sets.

The \vbf cross-section given is determined at next-to-next-to-leading
order using the \vbfh~\cite{\citevbfh} cross-section calculator. Here
the process is treated as a double deep-inelastic scattering process
where the two quark lines are assumed not to interfere and the
structure functions can be factorised. This assumption holds at not
only leading-order, but also next-to-leading order where contributions
from a single connecting colour line must conserve colour. At
next-to-next-to-leading order this assumption no longer remains valid,
but the dominant contributions can be included in the structure
functions.

At leading-order the \vbf final states can interfere with both \avp
and \aqp final states, for which the structure function factorisation
does not hold. These interference contributions, however, can be
directly calculated and are accounted for in the final cross-section
result. By kinematic arguments, this interference at next-to-leading
order and higher orders can be shown to contribute at below the
percent level, and consequently are not included. Finally, electroweak
contributions have also been included under the assumption that the
\qcd and electroweak contributions factorise.

The \avp cross-section is calculated using \vh \cite{\citevh} where
electroweak corrections are calculated at next-to-leading order and
\qcd corrections are calculated at next-to-next-to-leading
order. These two sets of corrections are assumed to factorise and were
originally calculated independently in \rfrs{ciccolini.03.1} and
\cite{brein.03.1}, respectively.

The predominant channel for \aqp is in association with a \wtq pair.
Currently, only leading order cross-section calculators are publicly
available, although full next-to-leading order cross-sections have
been calculated in \rfr{dawson.03.1}, and are found to increase the
cross-section by approximately $20\%$ at most. The results provided in
\fig{Hig:Sm.Xs} are from \rfr{hxswg.11.1} and are calculated at
next-to-leading order. Numerical values for the inclusive
cross-section of \fig{Hig:Sm.Xs} are tabulated in
\fig{Hvr:Sm.Xs.All.Text}.

\begin{subfigures}{2}{Numerical values for the inclusive \sm \whb
    cross-section as a function of the \sm \whb mass.}
  \svg[1]{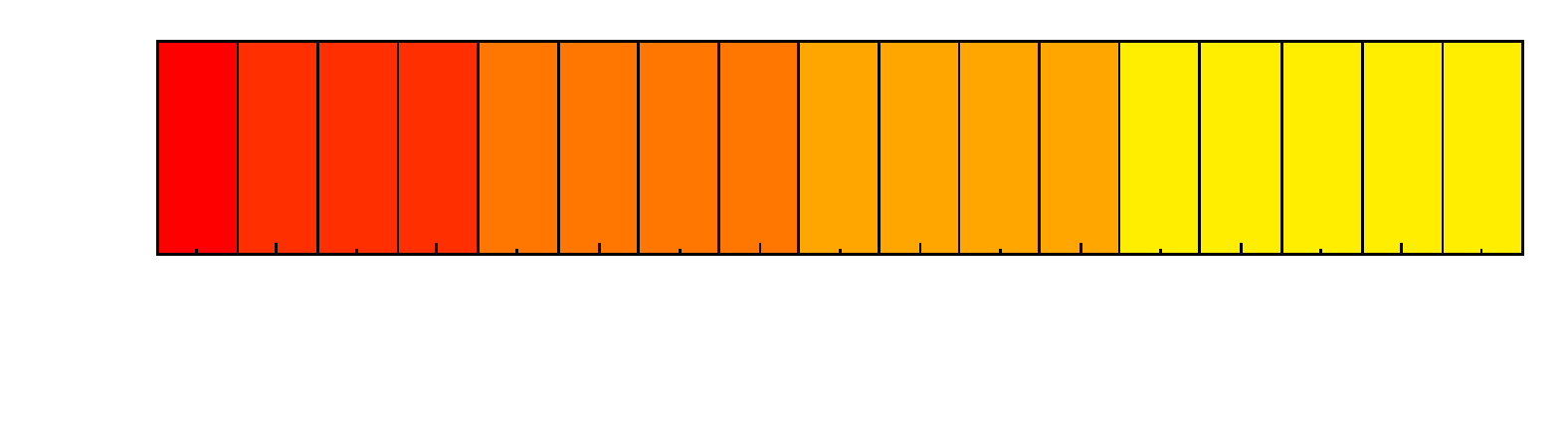} &
  \includesvg[pretex=\hspace{-0.43cm}\relsize{-3},width=0.5\columnwidth]
  {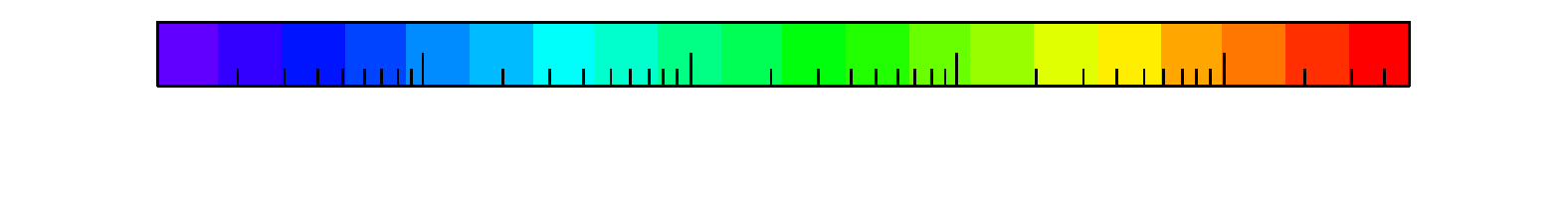} \svgend
\end{subfigures}

\newsubsection{MSSM Cross-Section Calculations}{XsMssm}

The \mssm \whb cross-sections for \ggf are determined using the
cross-section calculators \higlu and \ggh~\cite{\citeggh} with the
\mssm couplings calculated using \feynhiggs. The cross-section is
calculated as,
\begin{align*}\labelali{Mssm.Xs.GGF}
  \sigma_{gg \to \hz} =\; & \left(\frac{F_{\hz \bbbar}}{F_{\hH
        \bbbar}}\right)^2 \sigma_{gg \to \bbbar \to \hH}^\nlo +
  \left(\frac{F_{\hz \ttbar}}{F_{\hH \ttbar}}\right)^2
  \left(\sigma_{gg \to \ttbar \to \hH}^\nlo +
    \Delta\sigma_{gg \to \ttbar \to \hH}^\nnlo \right) \\
  &+ \left(\frac{F_{\hz \ttbar}F_{\hz \bbbar}}{F_{\hH \ttbar}F_{\hH
        \bbbar}}\right)\sigma_{gg \to \bbbar/\ttbar
    \to \hH}^\nlo
\end{align*}
where the pre-factors $F_{\hz q\bar{q}}$ and $F_{\hH q\bar{q}}$ are the
\mssm and \sm couplings of the \whbs with the specified quark pair,
calculated via \feynhiggs. At tree-level these couplings are given by
\figs{Thr:H12FuFu} through \ref{fig:Thr:H32FdFd} for the \mssm \whbs
and \fig{Thr:H02FF} for the \sm \whb. The next-to-leading order terms
$\sigma_{gg \to \ttbar \to \hH}^\nlo$, $\sigma_{gg \to \bbbar \to
  \hH}^\nlo$, and $\sigma_{gg \to \bbbar/\ttbar \to \hH}^\nlo$ are
calculated with \higlu for the \sm \whb using \equ{Hvr:Sm.Xs.GGF}
where the cross-section has been split into the individual \wtq, \wbq,
and interference contributions respectively.

The next-to-next-to-leading order \wtq correction term, $\Delta\sigma_{gg
  \to \ttbar \to \hH}^\nnlo$, is given by,
\begin{equation}
  \Delta\sigma_{\bbbar \to \ttbar \to \hH}^\nnlo = \left(\frac{\sigma_{gg
        \to \ttbar \to \hH}^\nnlo - \sigma_{gg \to \ttbar \to
        \hH}^\nlo}{\sigma_{gg \to \ttbar \to \hH}^\lo}\right)
  \sigma_{gg \to \ttbar \to \hH}^\lo
  \labelequ{Mssm.Xs.GGF.Delta}
\end{equation}
where all the terms are evaluated from \ggh. No electroweak
corrections are applied to the cross-sections from either
\equ{Hvr:Mssm.Xs.GGF} or \equ{Hvr:Mssm.Xs.GGF.Delta} as the \sm
couplings cannot be easily corrected to \mssm couplings. Additionally,
in \equ{Hvr:Mssm.Xs.GGF.Delta} next-to-next-to-leading log resummation
is not performed as this has not been calculated for the \cp-odd \whb
and next-to-next-to-leading log \PDF{s} are not available.

The \abp cross-section is calculated by,
\begin{equation}
  \sigma_{\bbbar \to \hz} = \left(\frac{F_{\hz \bbbar}}{F_{\hH
        \bbbar}}\right)^2 \sigma_{\bbbar \to \hH}^\nnlo
  \labelequ{Mssm.Xs.ABP}
\end{equation}
at next-to-next-to-leading order using the cross-section calculator
\bbh~\cite{\citebbh} dressed with the \mssm and \sm couplings of the
\whbs with \wbqs, $F_{\hz \bbbar}$ and $F_{\hH \bbbar}$, from
\feynhiggs. The tree-level couplings are given for the \mssm \whbs in
\figs{Thr:H12FdFd} through \ref{fig:Thr:H32FdFd} and the \sm \whb in
\fig{Thr:H02FF}. The following sub-processes for $\sigma_{\bbbar \to
  \hH}^\nnlo$ are evaluated by \bbh,
\begin{equation}
  \left. \begin{array}{lll}
      \bbbar \to \hH gg,& \bbbar \to \hH q\bar{q},& \bbbar \to \hH
      \bbbar \\
      gb \to \hH gb,& bb \to \hH bb,& bq \to \hH bq \\
      gg \to \hH \bbbar,& q\bar{q} \to \hH b\bar{b} \\
    \end{array} \right\}~\textrm{tree~level} ~
  \begin{array}{r}
    \left. \begin{array}{r}
        \bbbar \to \hH q \\ gb \to \hH b \\
      \end{array} \right\}~\textrm{one~loop} \\
    \left. \bbbar \to \hH \phantom{qq} \right\}~\textrm{two~loop} \\
  \end{array}
  \labelequ{Mssm.Xs.ABP.Breakdown}
\end{equation}
where $q$ indicates a \wuq, \wdq, \wcq, or \wsq. There is no overlap
between the one-leg sub-processes of \equ{Hvr:Sm.Xs.GGF} and
\equ{Hvr:Mssm.Xs.ABP.Breakdown} and so the \mssm \ggf cross-sections
from \equ{Hvr:Mssm.Xs.GGF} and the \abp cross-sections from
\equ{Hvr:Mssm.Xs.ABP} can be summed to determine the inclusive
cross-section without introducing double-counting.

In \fig{Hvr:Mssm.Xs.All} the inclusive cross-sections for the \mssm
\whbs, given in \fig{Hig:Mssm.Xs}, are provided with numerical values
as a function of the \cp-odd \whb mass and \tanb. The inclusive
cross-sections are separated into the \ggf component in
\fig{Hvr:Mssm.Xs.GG} and the \abp component in \fig{Hvr:Mssm.Xs.BB}.

\begin{subfigures}[p]{2}{Numerical values for the
    \subfig{Mssm.Xs.All.H1.Text}~light \cp-even,
    \subfig{Mssm.Xs.All.H2.Text}~heavy \cp-even, and
    \subfig{Mssm.Xs.All.H3.Text}~\cp-odd \mssm \whbs inclusive
    production cross-sections as a function of the \cp-odd \whb mass
    and \tanb.\labelfig{Mssm.Xs.All}}
  \svgbeg
  \svg{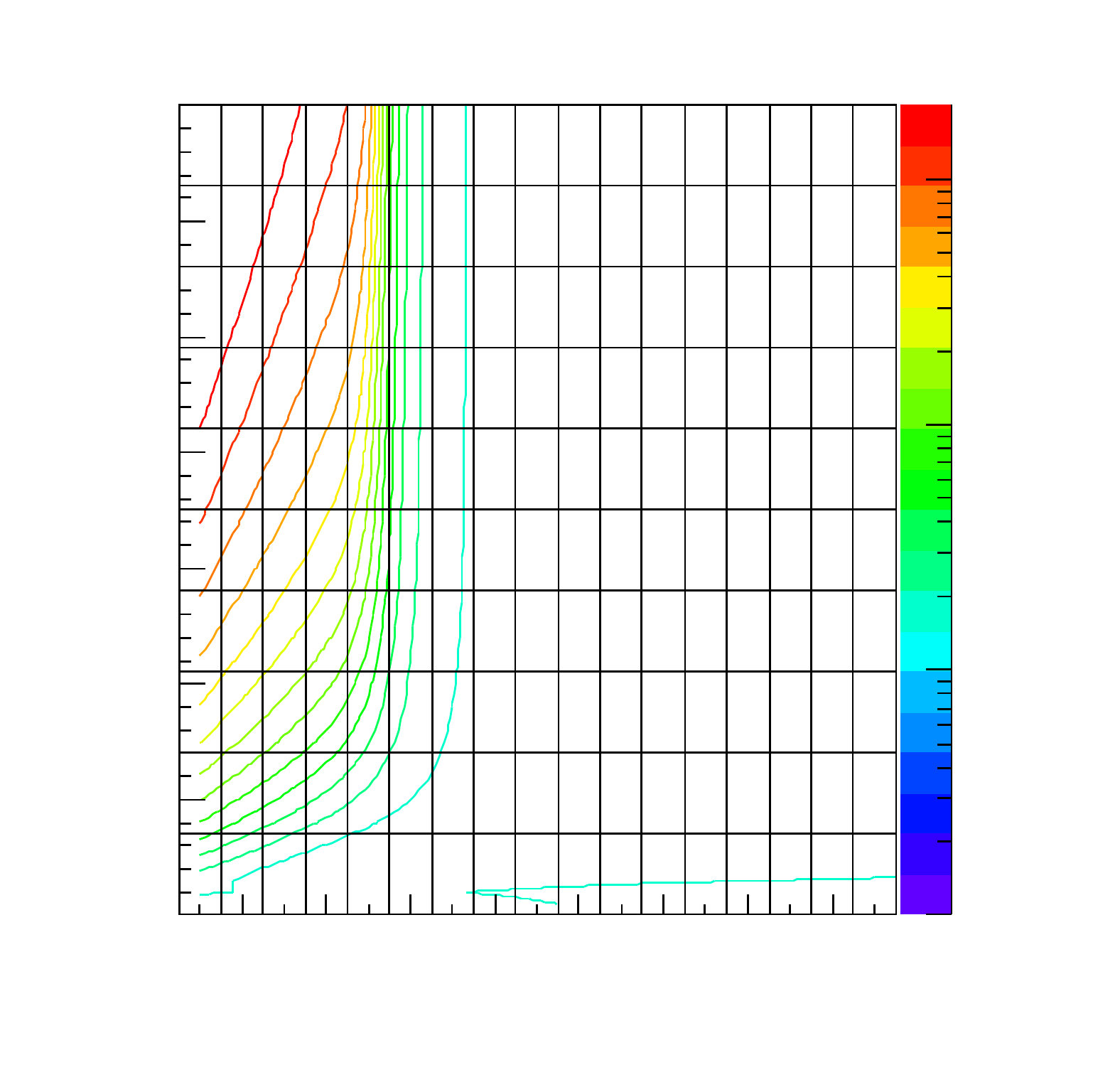} & \svg{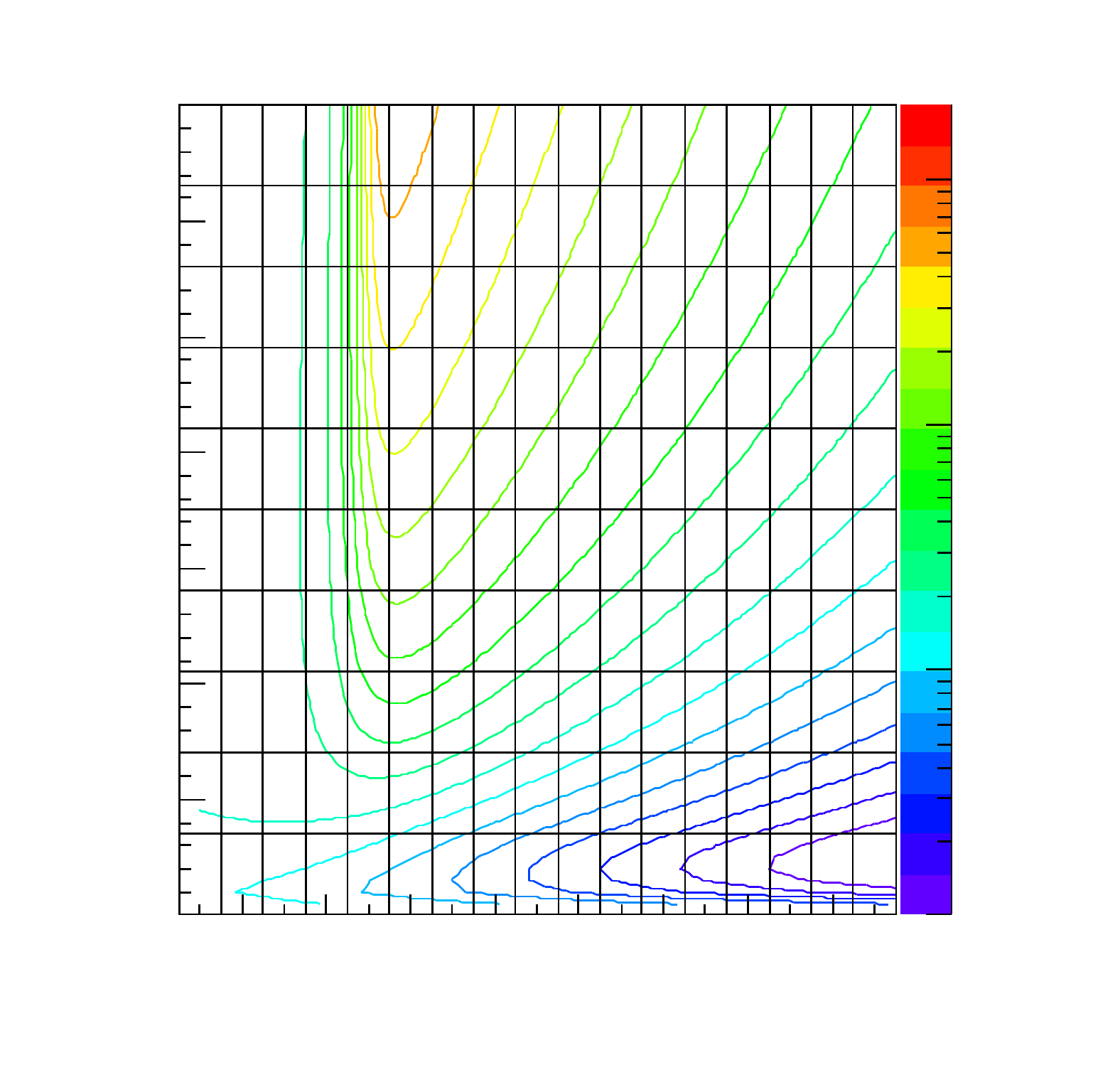} \svgend
  \sidecaption              & \svg{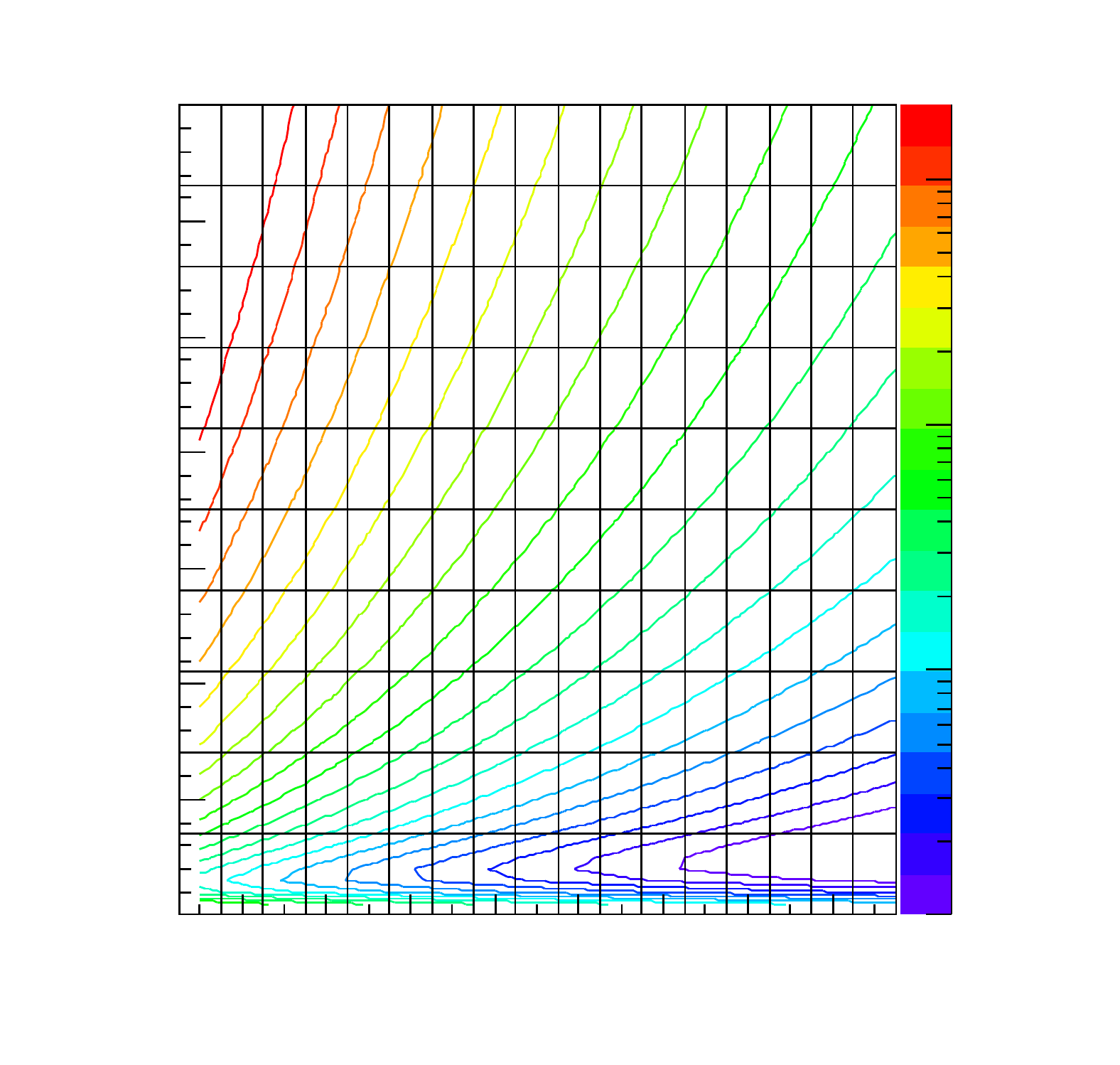} \svgend
\end{subfigures}

\begin{subfigures}[p]{2}{Coarse binnings with numerical values for the
    \subfig{Mssm.Xs.GG.H1.Text}~light \cp-even,
    \subfig{Mssm.Xs.GG.H2.Text}~heavy \cp-even, and
    \subfig{Mssm.Xs.GG.H3.Text}~\cp-odd \mssm \whbs \ggf
    cross-sections as a function of the \cp-odd \whb mass and
    \tanb.\labelfig{Mssm.Xs.GG}}
  \svgbeg
  \svg{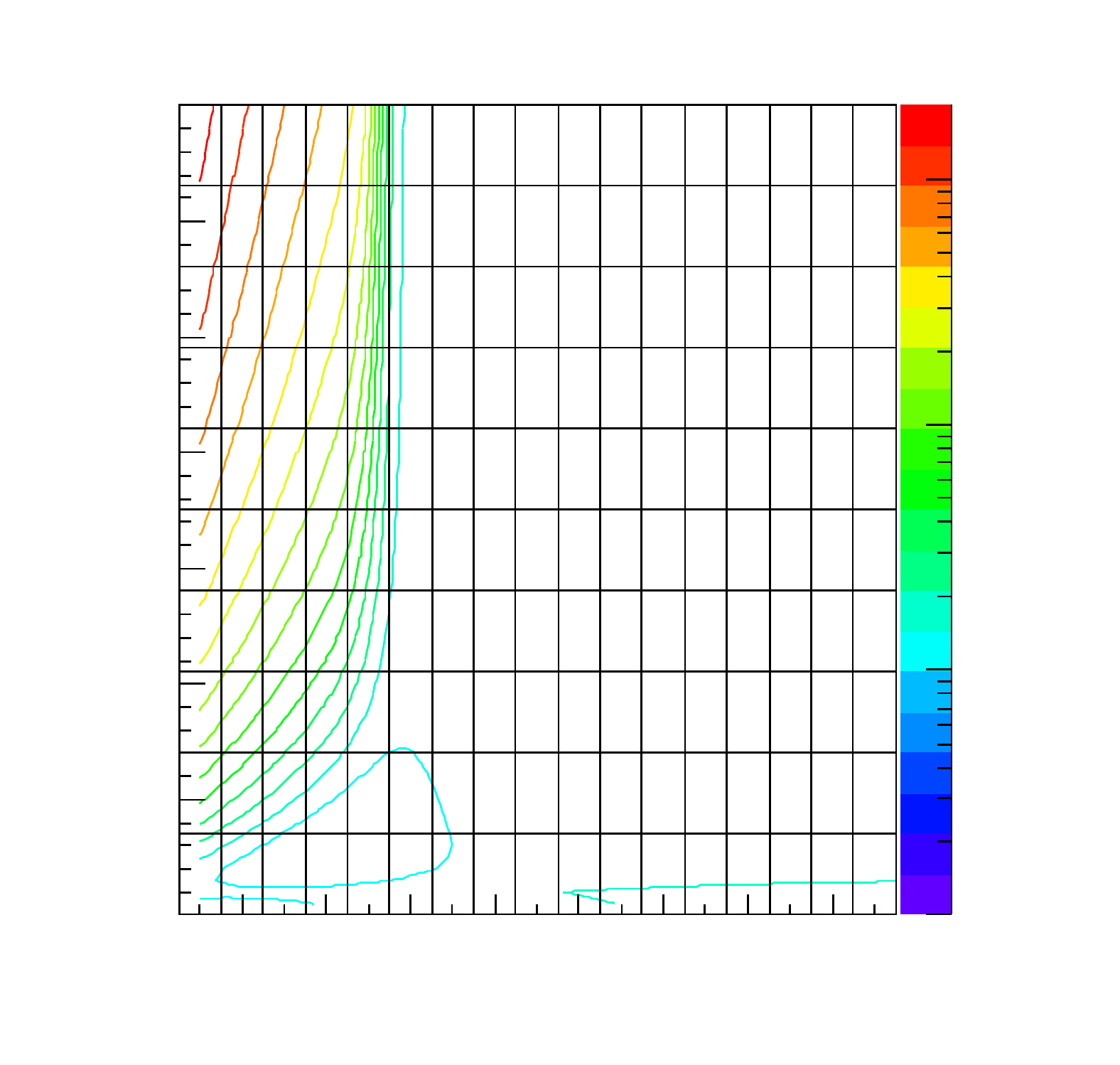} & \svg{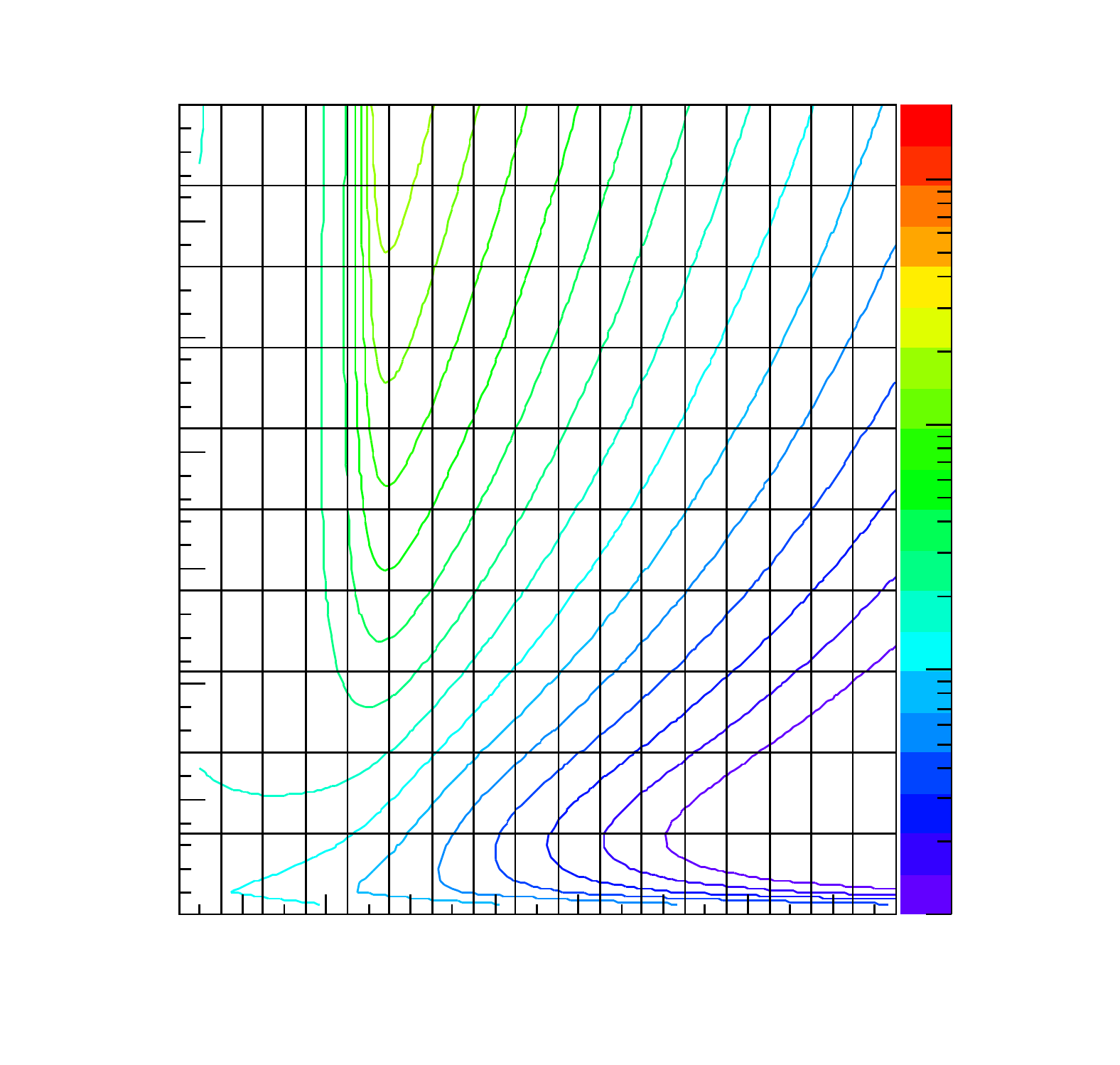} \svgend
  \sidecaption             & \svg{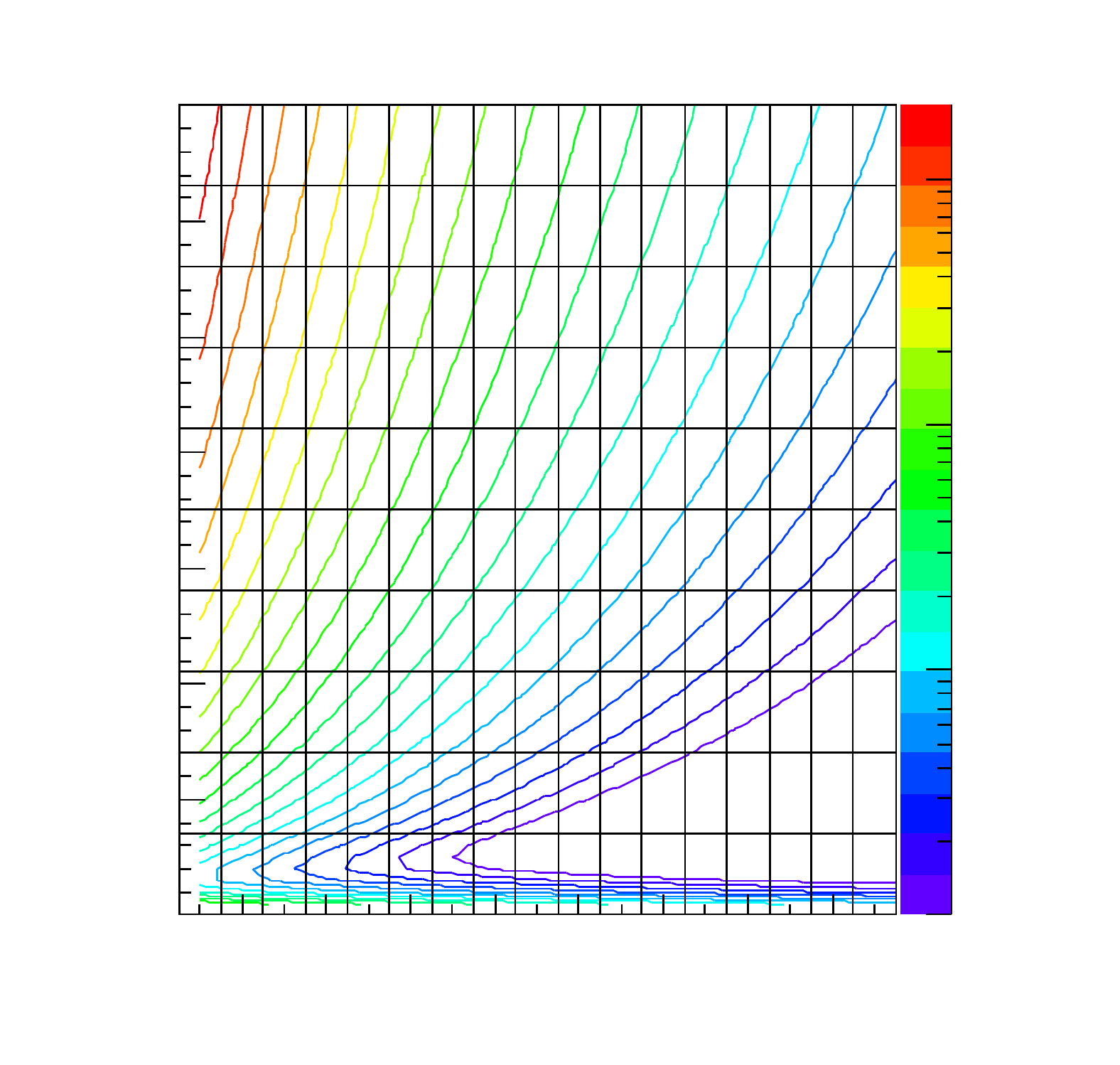} \svgend
\end{subfigures}

\begin{subfigures}[p]{2}{Coarse binnings with numerical values for the
    \subfig{Mssm.Xs.BB.H1.Text}~light \cp-even,
    \subfig{Mssm.Xs.BB.H2.Text}~heavy \cp-even, and
    \subfig{Mssm.Xs.BB.H3.Text}~\cp-odd \mssm \whbs \abp
    cross-sections as a function of the \cp-odd \whb mass and
    \tanb.\labelfig{Mssm.Xs.BB}}
  \svgbeg
  \svg{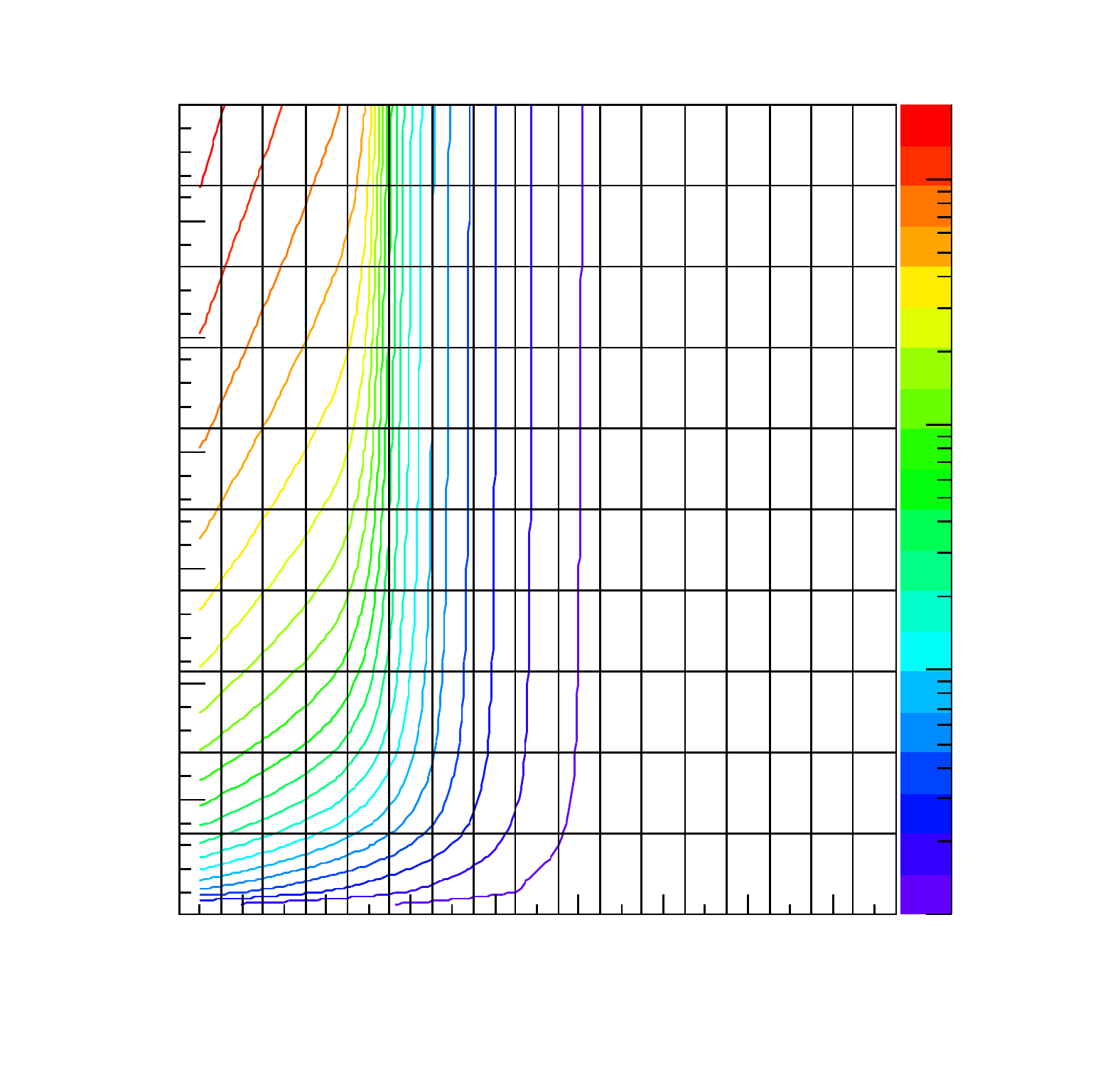} & \svg{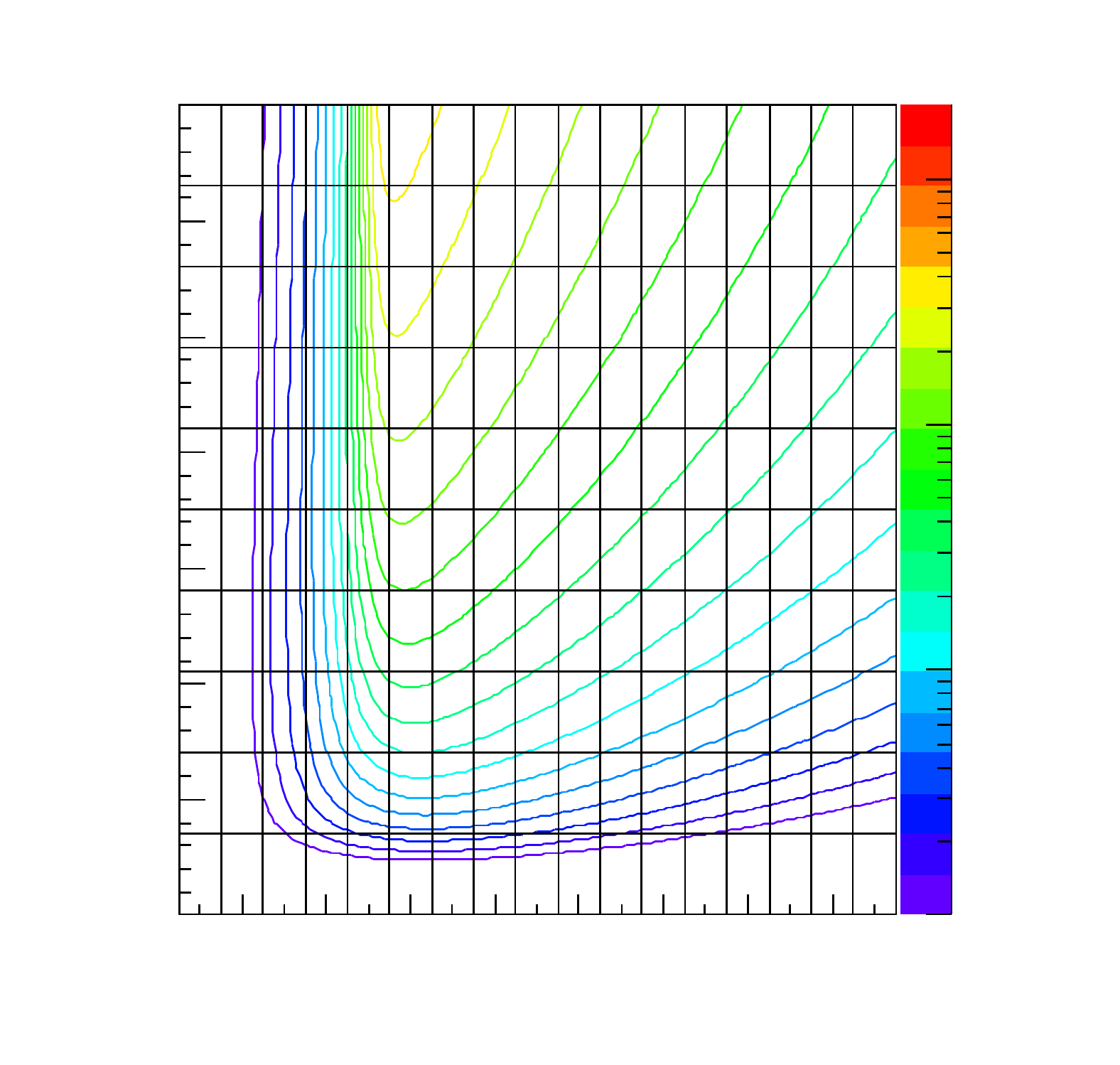} \svgend
  \sidecaption             & \svg{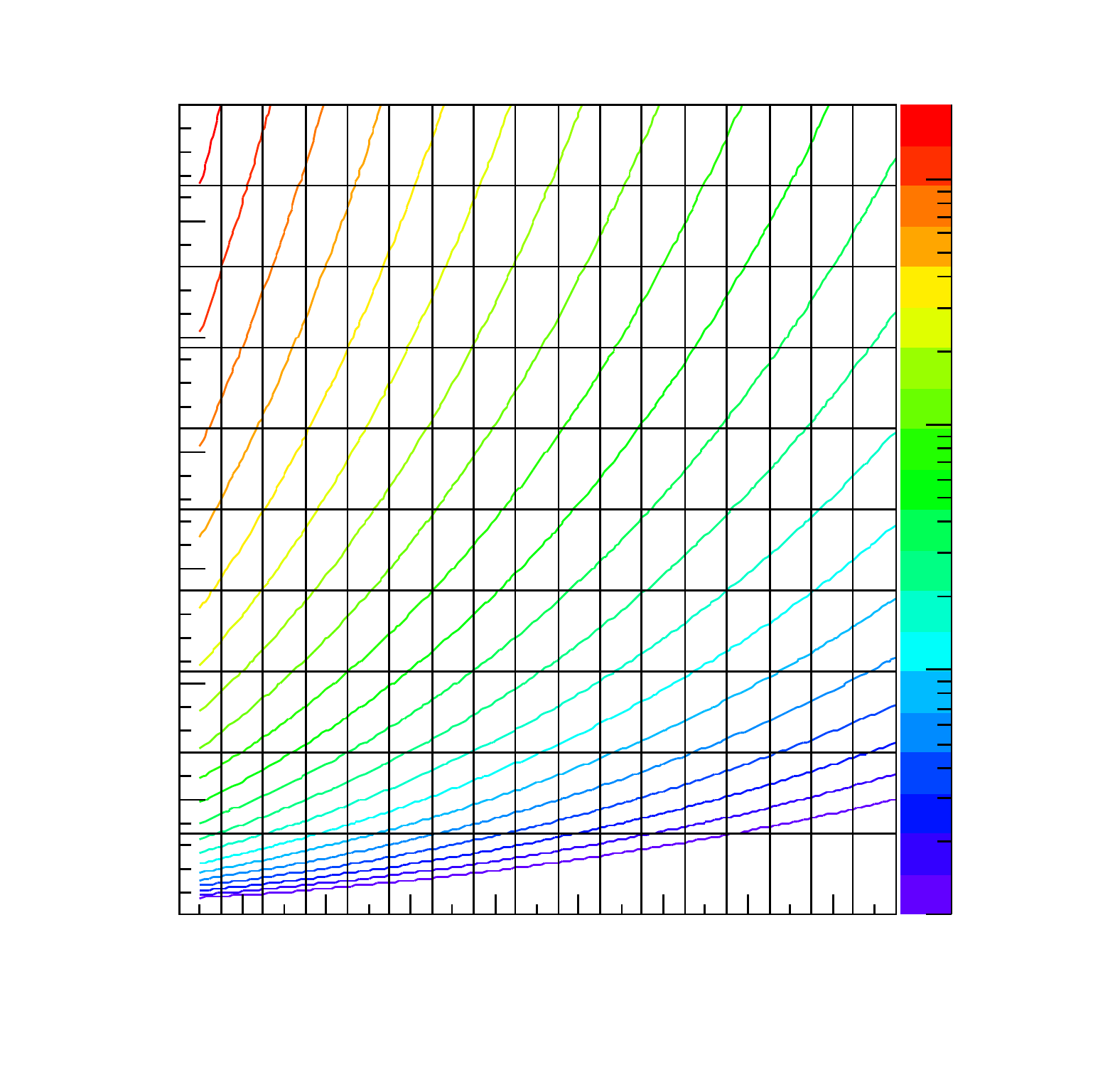} \svgend
\end{subfigures}

\newsection{Acceptances and Efficiencies}{Acc}

The numerical values for the acceptances used to calculate the number
of expected \whb events in \sec{Hig:Nrm} are given in
\fig{Hvr:Acceptance} for the five event categories as a function of
the \whb mass, $\m_\h$. In \fig{Hvr:Efficiency}, numerical values
for the efficiencies given by \equ{Hig:Efficiency} for the five event
categories are tabulated as a function of the \whb mass.

\begin{subfigures}[t]{2}{Acceptances, with numerical values, for \whbs
    decaying into the \subfig{H2TauTau2MuMu.Info.Acc.Text}~\mumu,
    \subfig{H2TauTau2MuE.Info.Acc.Text}~\mue,
    \subfig{H2TauTau2EMu.Info.Acc.Text}~\emu,
    \subfig{H2TauTau2MuPi.Info.Acc.Text}~\muh, and
    \subfig{H2TauTau2EPi.Info.Acc.Text}~\eh categories to pass the
    $\eta$ and \pt requirements of \sec{Zed:Rec} as a function of the
    \whb mass.\labelfig{Acceptance}}
  \multicolumn{2}{c}{\svg[1]{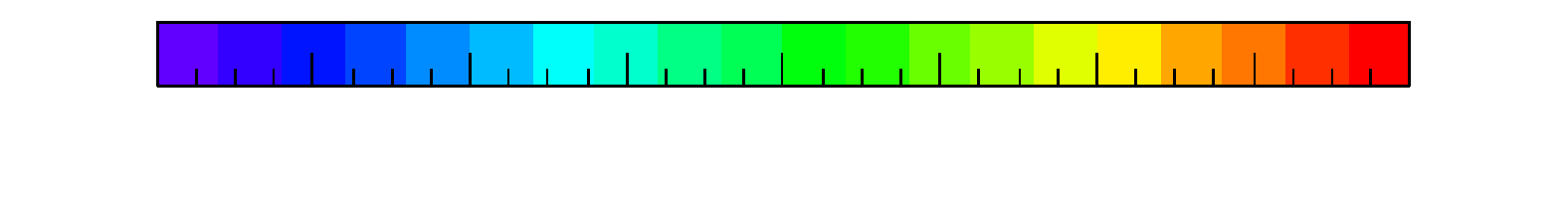}} \\
  \svg{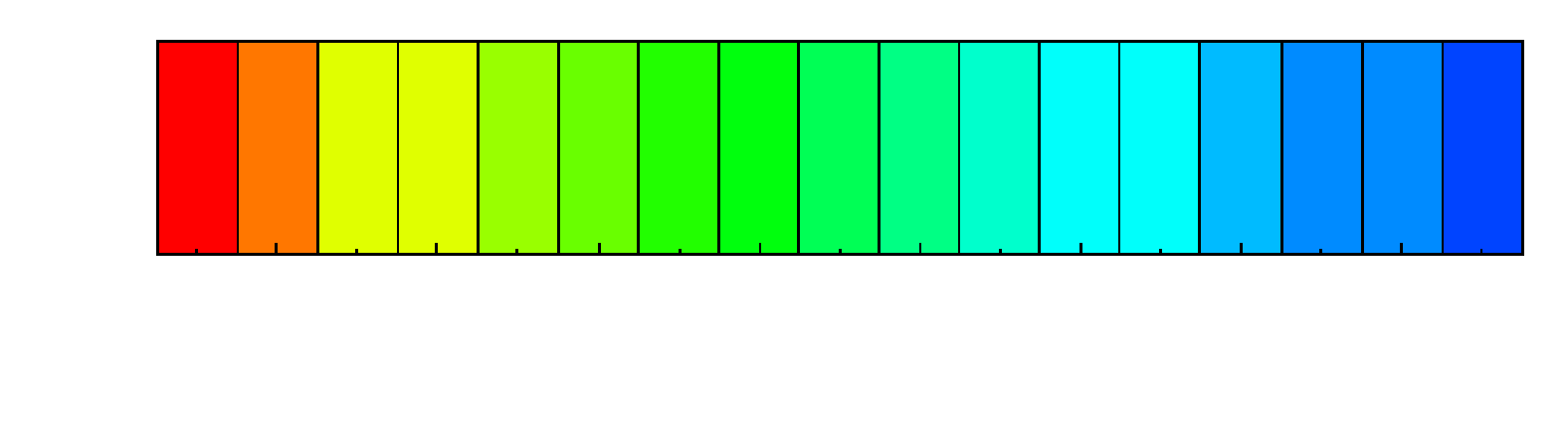} 
  & \svg{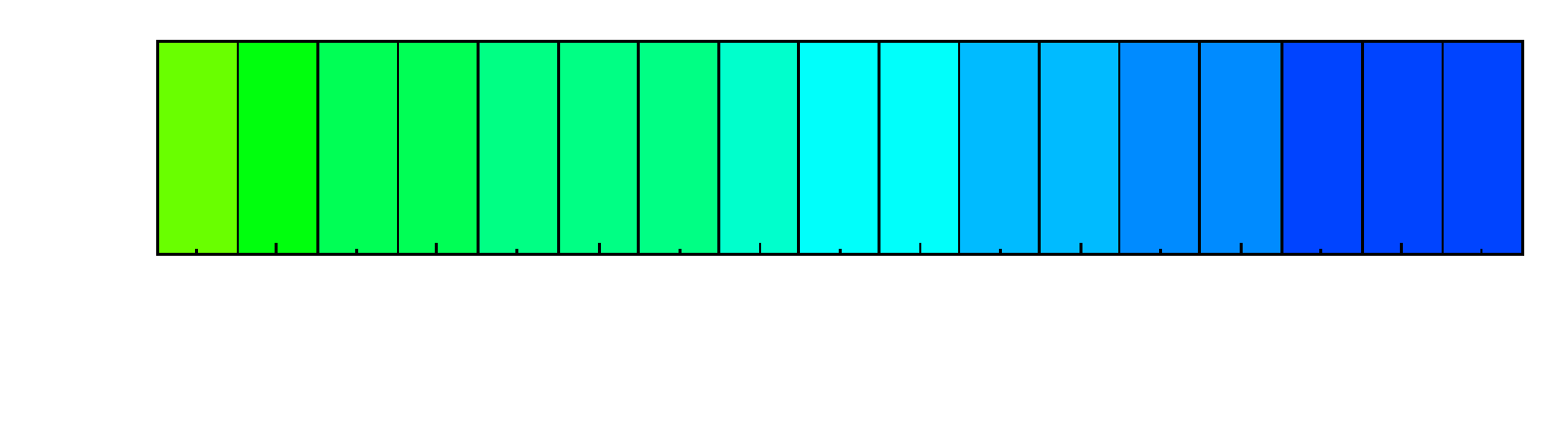} \barsep
  \svg{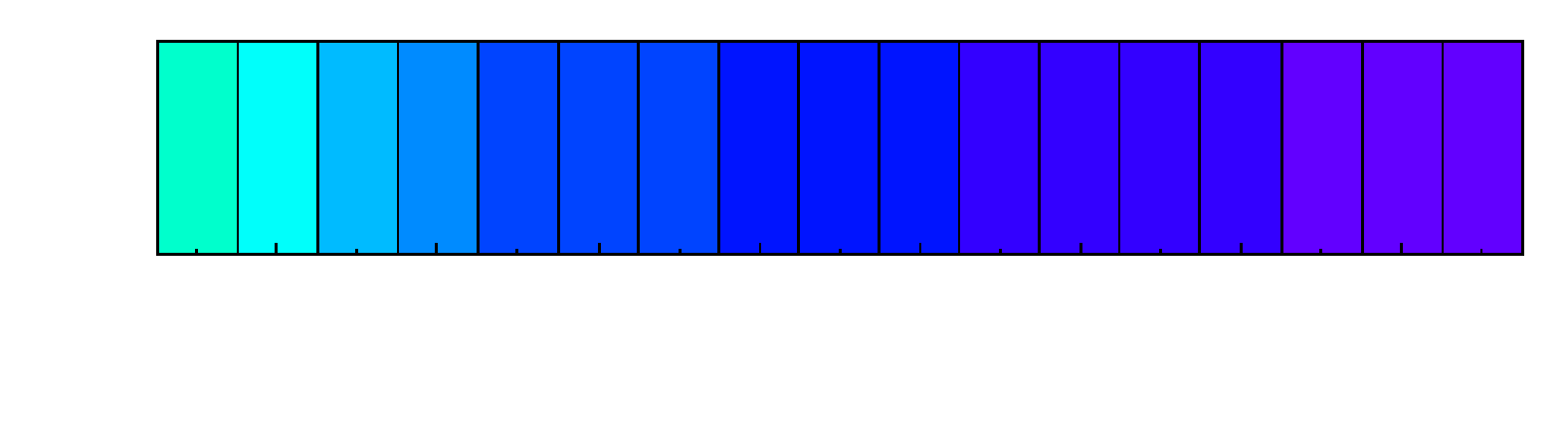} 
  & \svg{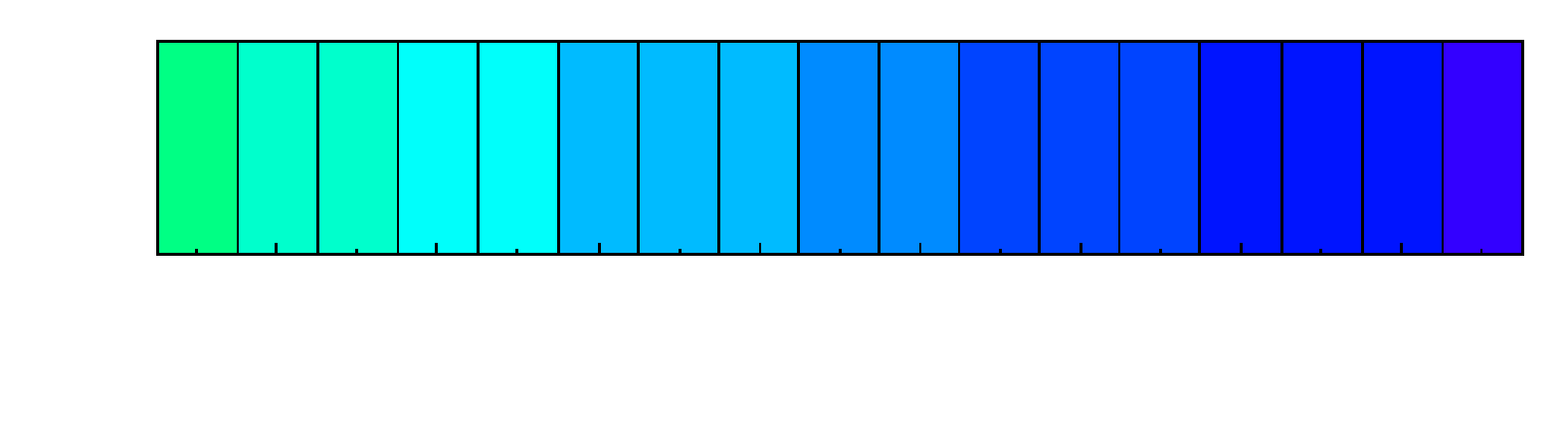} \barsep
  \svg{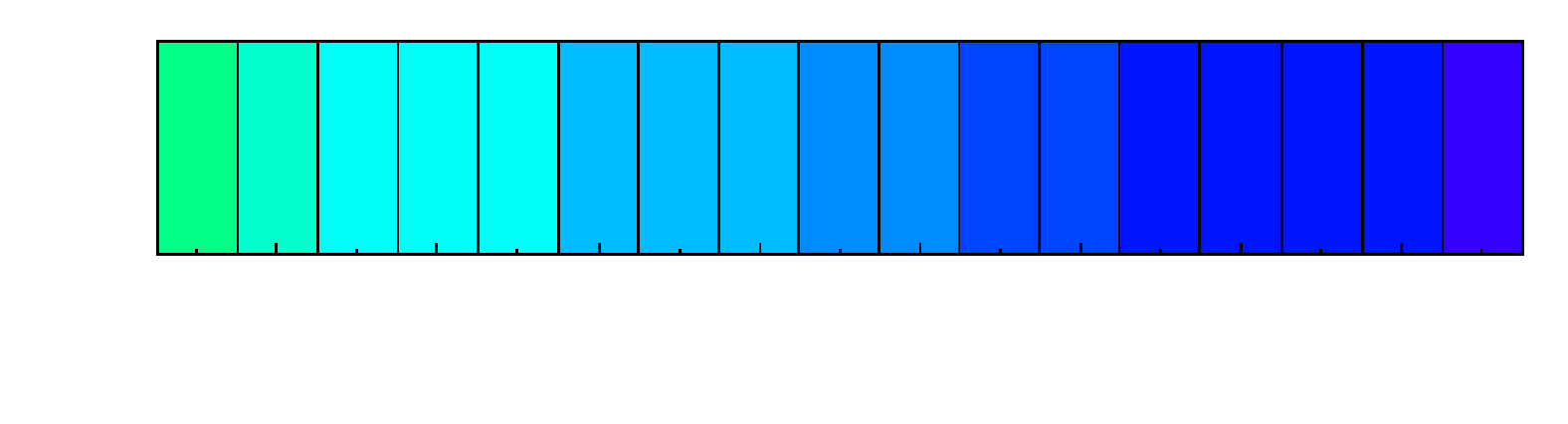} \barend
\end{subfigures}

\begin{subfigures}{2}{Efficiency corrections, with numerical values, for
    \whbs decaying into the
    \subfig{H2TauTau2MuMu.Info.Eff.Text}~\mumu,
    \subfig{H2TauTau2MuE.Info.Eff.Text}~\mue,
    \subfig{H2TauTau2EMu.Info.Eff.Text}~\emu,
    \subfig{H2TauTau2MuPi.Info.Eff.Text}~\muh, and
    \subfig{H2TauTau2EPi.Info.Eff.Text}~\eh categories as a function
    of the \whb mass. The efficiency determination is described in
    \secs{Zed:RecEff} and \ref{sec:Zed:SelEff}.\labelfig{Efficiency}}
  \multicolumn{2}{c}{\svg[1]{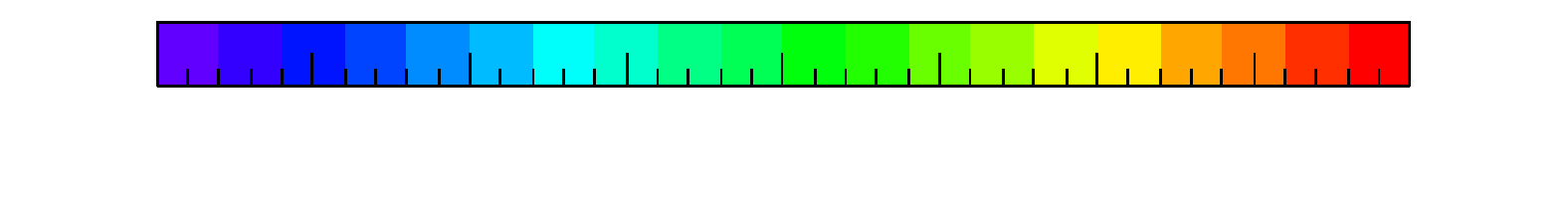}} \\
  \svg{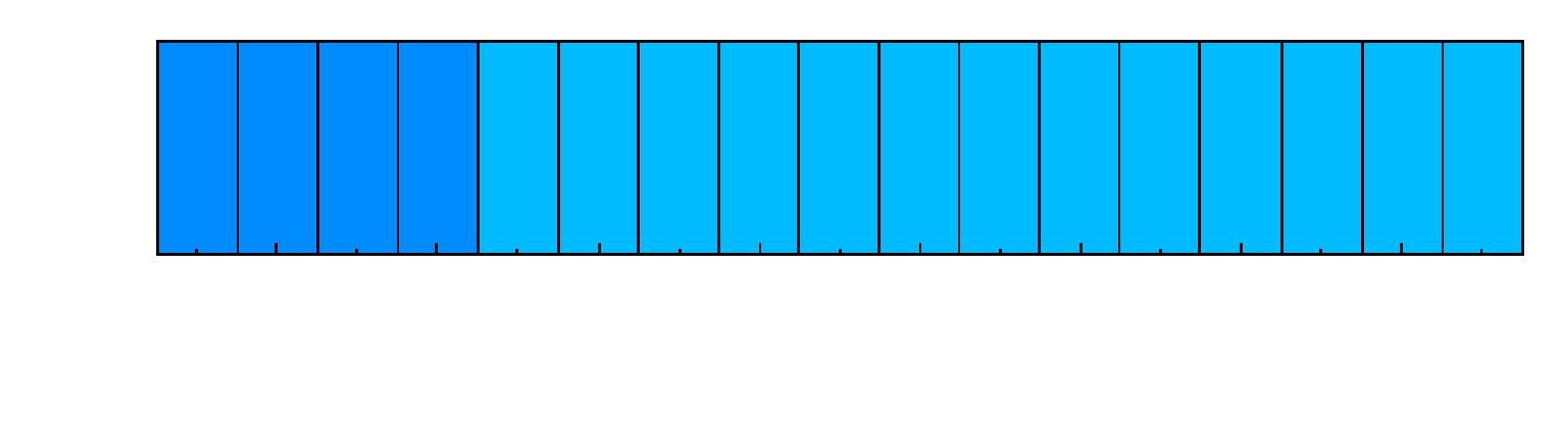} 
  & \svg{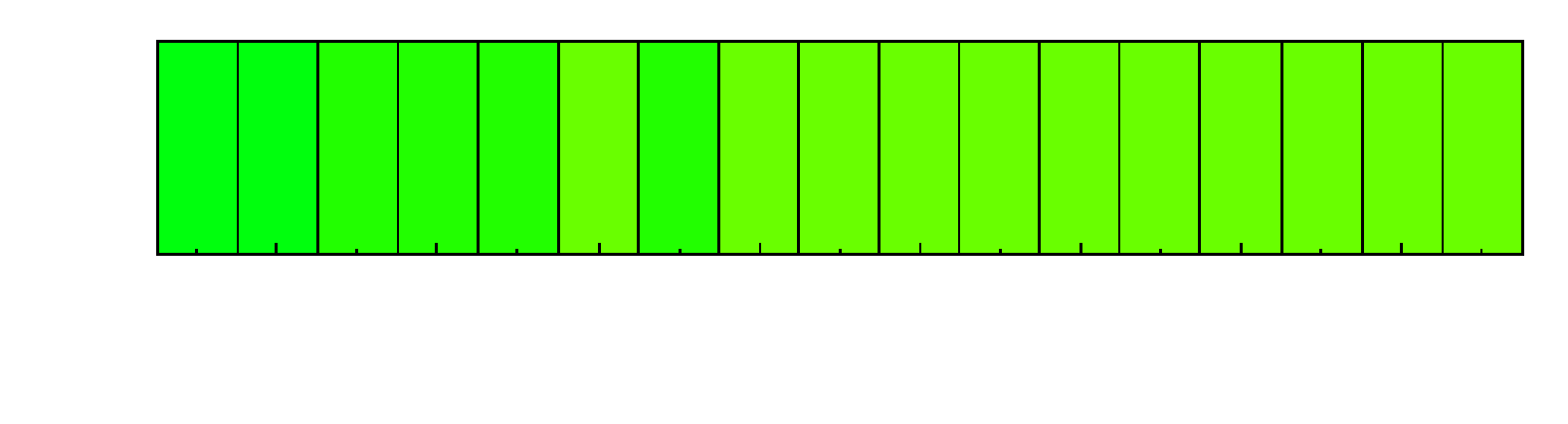} \barsep
  \svg{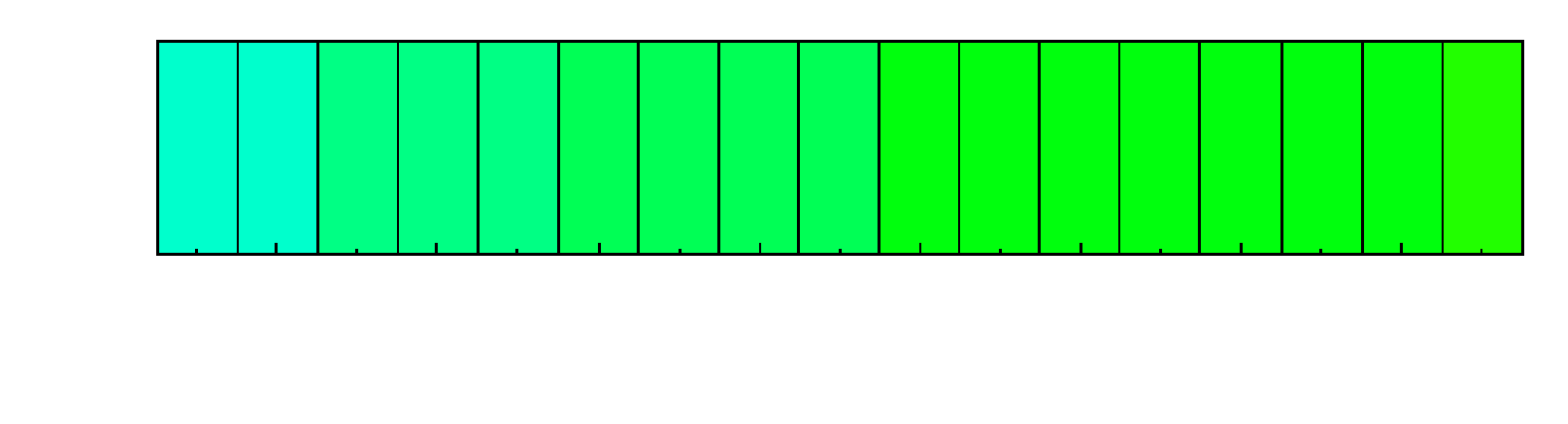} 
  & \svg{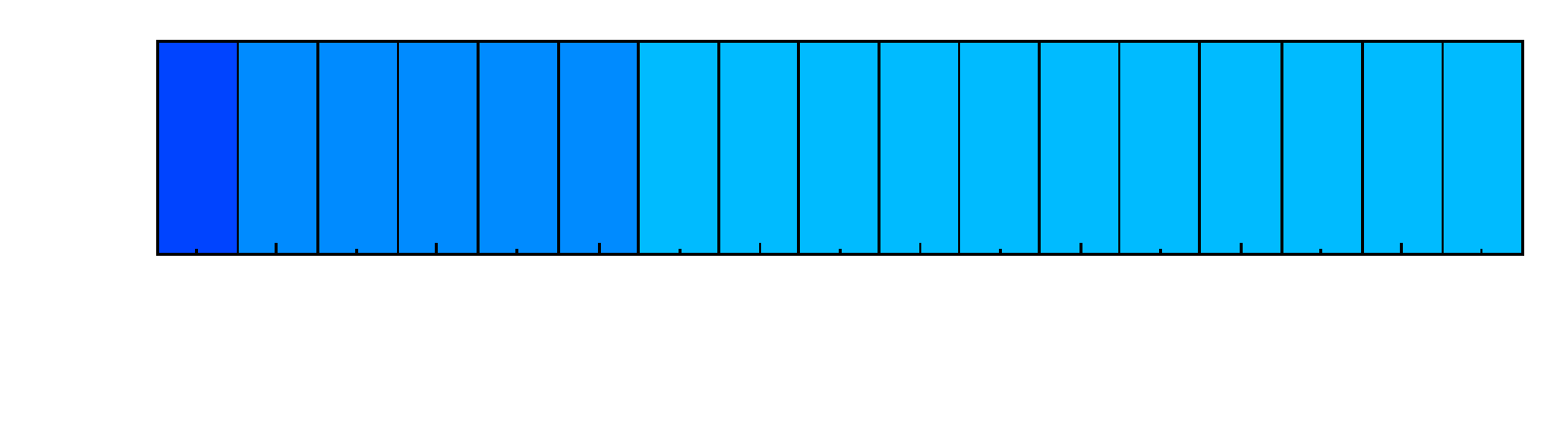} \barsep
  \svg{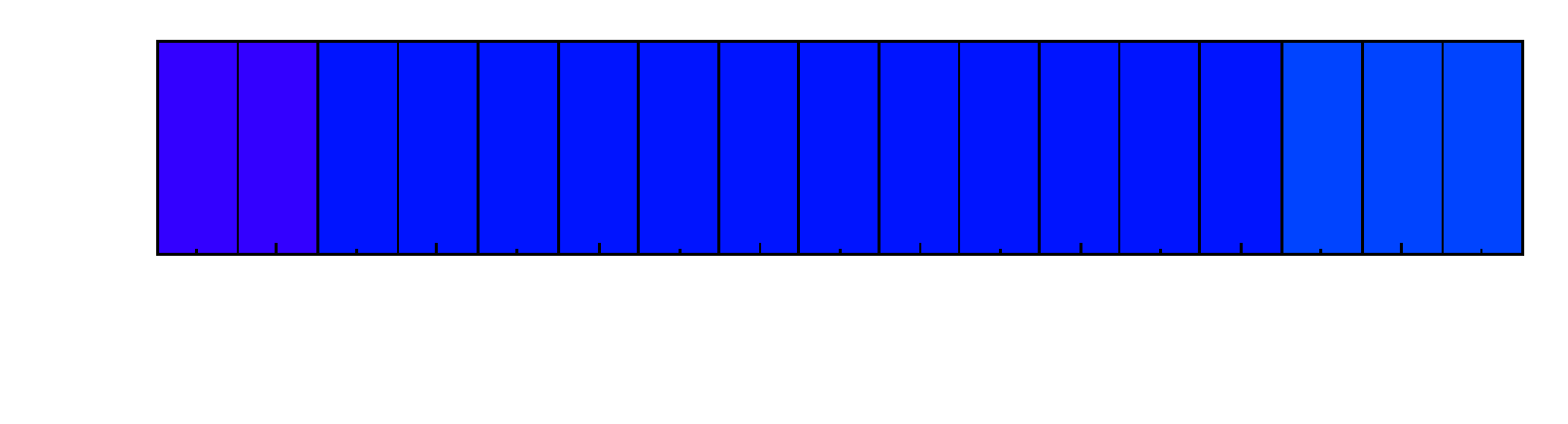} \barend
\end{subfigures}

\newsection{Statistical Methods}{Sta}

Details on the Asimov dataset of \rfr{cowan.10.1}, as well as a
derivation for \equ{Hig:Statistics.Asimov}, the extended likelihood
evaluated with an Asimov dataset, from the results \rfr{cowan.10.1} is
given in \sap{Hvr:Asi}. Further information on incorporating
systematic uncertainties into hypothesis testing using marginalisation
techniques is given in \sap{Hvr:Mar}.

\newsubsection{Asimov Dataset}{Asi}

Consider binning the $N$ events into $N_\bins$ using one or more
observables for each event, {\it e.g.} the invariant mass of the
event. The probability density function for each bin is then described
by the Poisson distribution, and so the binned likelihood function is,
\begin{align*}\labelali{Statistics.Lhs.Binned}
  \lh_b(\vars|\mu) &= \prod_i \lh_s(\var[_i]|\mu)
  = \prod_i
  \frac{({N_\bkg}_i + \mu {N_\sig}_i)^{\var[_i]} e^{-{N_\bkg}_i - \mu
      {N_\sig}_i}}{\var[_i]!} \\
  &= \prod_i
  \frac{\left(({N_\bkg} + \mu
      {N_\sig})\pdf(\vec{b}_i|\mu) \sdif{\vec{b}_i}\right)^{\var[_i]}
    e^{-({N_\bkg} + \mu
      {N_\sig})\pdf(\vec{b}_i|\mu)\sdif{\vec{b}_i}}}{\var[_i]!}
\end{align*}
where the product of the likelihoods for each bin is taken over all
$N_\bins$. The number of events in each bin is \var[_i] and the number
of expected background and signal events for each bin $i$ is given by
${N_\bkg}_i$ and ${N_\sig}_i$. In the second line, the number of
expected background and signal events for a given bin are written in
terms of the joint \pdf for the set of observables for the bin,
$\vec{b}_i$, and the width of the bin, $\dif{\vec{b}_i}$. The binned
likelihood function is dependent upon not only the number of observed
events, but also the observables of the events.

For the test statistic $q_\nu$ of \equ{Hig:Statistics.Ulr} calculated
with any likelihood function, the median is found using the definition
of the median from \equ{Hig:Statistics.Median} and the \cdf of $q_\nu$
from \equ{Hig:Statistics.Ulr.Cdf},
\begin{equation}
  \mval{q_\nu|\mu} = \left(\Phi^{-1}\left(\frac{1}{2}\right) + \frac{\nu -
      \mu}{\sigma}\right)^2 = \frac{(\nu - \mu)^2}{\sigma^2}
  \labelequ{Statistics.Median.Ulr.1}
\end{equation}
where $\sigma^2$ is the variance of the maximum likelihood estimator
$\hat{\nu}$. The binned log-likelihood function can be written as,
\begin{equation}
  \llh_b(\vars|\mu) = \ln\lh_b(\vars|\mu) = \sum_i\left(\var[_i]\ln u_i -
    u_i - x_i!\right)
\end{equation}
where the summation is over all bins, the binned likelihood function
of \equ{Hvr:Statistics.Lhs.Binned} has been used, and the change of
variable ${u_i = {N_\bkg}_i + \mu {N_\sig}_i}$ has been made. The
derivative of the binned log-likelihood function with respect to the
signal strength parameter $\mu$ is then,
\begin{equation}
  \frac{\partial \llh_b(\vec{x}|\mu)}{ \partial\mu} = \sum_i \left( \left(
  \frac{\var[_i]}{u_i} - 1\right) \frac{\partial u_i}{\partial\mu} \right)
\end{equation}
where again the summation is over all bins. If this derivative for a
given \vars is zero, then the maximum likelihood estimator $\hat{\nu}$
for \vars is $\mu$. The dataset $\asis(\mu)$ fulfils this condition
if ${\asi[_i](\mu) = u_i = {N_\bkg}_i + \mu {N_\sig}_i}$. This
representative dataset first proposed in \rfr{cowan.10.1} is named the
Asimov dataset as it was inspired by the short story of
\rfr{asimov.55.1} written by Isaac Asimov.

Calculating the test statistic $q_\nu$ using \asis[(\mu)] results in,
\begin{align*}\labelali{}
  q_\nu(\asis[(\mu)]) &= -2\left(\llh(\asis(\mu)|\nu) -
    \llh(\asis(\mu)|\mu)\right) \\
  &= \frac{(\nu - \mu)^2}{\sigma^2} \\
  &= \mval{q_\nu|\mu}
\end{align*}
for ${\mu \leq \nu}$ where \equ{Hig:Statistics.Ulr} has been used in
the first line, \equ{Hig:Statistics.Plr} in the second line, and
\equ{Hvr:Statistics.Median.Ulr.1} in the third line. Consequently, the
median test statistics parameter $q_\nu$, assuming $\mu$, can be found
by evaluating $q_\nu$ with the Asimov dataset of $\mu$. This also
provides the non-centrality parameter of $\pdf(\stat[_\nu]|\mu)$ and
$\pdf(q_\nu|\mu)$ from \equs{Hig:Statistics.Plr.Pdf} and
\ref{equ:Hig:Statistics.Ulr.Pdf}. In the example of
\fig{Hig:Statistics.Wilks} the non-centrality parameter for
$\pdf(\stat[_1]|0)$ is determined from the fit to be ${8.1 \pm 1.4}$
and found to be $7.5$ using the Asimov dataset \asis[(0)].

Using ${\asi[_i](\mu) = {N_\bkg}_i + \mu {N_\sig}_i}$ defines the
Asimov dataset for the simple and binned likelihood functions, but
must be taken to the continuous limit for the extended likelihood
function.  The extended likelihood function can be obtained by taking
the limit of the binned likelihood function as the widths of the bins
approach zero. Each bin will contain either no events or a single
event. The simple likelihood function for bin $i$ with zero events is,
\begin{equation}
  \lh_s(0|\mu) = e^{-{N_\bkg}_i - \mu{N_\sig}_i} = e^{-(N_\bkg +
    \mu N_\sig) \pdf(\vec{b}_i|\mu)\sdif{\vec{b}_i}} 
\end{equation}
where the bin centre is given by the set of observables $\vec{b}_i$ and
the width of the bin is given by $\dif{\vec{b}_i}$. Similarly,
\begin{equation}
  \begin{aligned}
    \lh_s(1|\mu) &= ({N_\bkg}_i + \mu{N_\sig}_i)e^{-{N_\bkg}_i -
      \mu{N_\sig}_i} \\
    &= \left((N_\bkg + \mu N_\sig) \pdf(\vec{b}_i|\mu)
      \sdif{\vec{b}_i}\right)
    e^{-(N_\bkg + \mu N_\sig) \pdf(\vec{b}_i|\mu)\sdif{\vec{b}_i}} \\
  \end{aligned}
\end{equation}
is the simple likelihood for a bin with one event. From these two
likelihood functions, the likelihood function for all the bins
becomes,
\begin{equation}
  \begin{aligned}
    \lh_e(\vars|\mu) &= \prod_j \left( e^{-(N_\bkg + \mu N_\sig)
        \pdf(\vec{b}_i|\mu)\sdif{\vec{b}_i}} \right) \prod_i
    \left((N_\bkg +
      \mu N_\sig) \pdf(\vec{b}_i|\mu) \sdif{\vec{b}_i} \right) \\
    &= e^{-N_\bkg - \mu N_\sig} \prod_i \big((N_\bkg +
    \mu N_\sig) \pdf(\vec{o}_i|\mu) \big) \\
  \end{aligned}
  \labelequ{Statistics.Lhs.Extended}
\end{equation}
where the first product is over all bins and the second product is
only over the $N$ bins with a single event. In the second line, the
first product has been integrated in the limit $\dif{\vec{b}_i} \to 0$
and the second product is expressed in terms of the joint \pdf for the
set of observables $\vec{o}_i$ for an event $i$. The extended
likelihood function becomes the simple likelihood function, without
the factorial denominator, when the set of observables $\vec{o}_i$ for
each event is $1$.

Evaluating the extended likelihood function for a binned
Asimov data set yields,
\begin{equation}
  \lh_e(\asis(\mu)|\nu) = e^{-N_\bkg - \nu N_\sig} \prod_i \big((N_\bkg +
  \nu N_\sig) \pdf(\vec{b}_i|\nu) \big)^{(N_\bkg + \mu
    N_\sig)\pdf(\vec{b}_i|\mu) \sdif{\vec{b}_i}}
\end{equation}
where $\vec{b}_i$ is the set of observables for each bin of $\vec{a}$
and $\dif{\vec{b}_i}$ is the width of each bin. The product is over
all bins of the Asimov dataset. However, as each bin no longer
contains a single event like the second product of
\equ{Hvr:Statistics.Lhs.Extended}, but rather ${(N_\bkg + \mu
  N_\sig)\pdf(\vec{b}_i|\mu) \sdif{\vec{b}_i}}$ events, each argument
of the product is exponentiated by this non-integer number of
events. The extended log-likelihood function can then be written as,
\begin{align*}\labelali{Statistics.Asimov}
  \llh_e(\asis(\mu)|\nu) &= \begin{aligned}[t]
    &\sum_i \ln\Big((N_\bkg + \nu
    N_\sig)\pdf(\vec{b}_i|\nu)\Big)(N_\bkg +  \mu N_\sig)
    \pdf(\vec{b}_i|\mu) \sdif{\vec{b}_i} \\
    &-N_\bkg - \nu N_\sig \\
  \end{aligned} \\
  &= \begin{aligned}[t] &\int \ln\Big((N_\bkg + \nu
    N_\sig)\pdf(\vec{o}|\nu)\Big)(N_\bkg +  \mu N_\sig)
    \pdf(\vec{o}|\mu)\sdif{\vec{o}} \\
    &-N_\bkg - \nu N_\sig \\
  \end{aligned} \\
\end{align*}
where in the first step the summation is over all bins of the Asimov
dataset and in the second step the limit has been taken as the bin
width for the Asimov dataset approaches zero. This result can then be
used to determine the Asimov test statistic for a test statistic
calculated using the extended likelihood function. When the test
statistic, Asimov or otherwise, is evaluated using a binned likelihood
function or an extended likelihood function using a \pdf from a
histogram with identical binning to the binned likelihood function,
the results from the two different likelihood function methods are
equivalent.

\newsubsection{Marginalisation}{Mar}

An alternative to the frequentist hypothesis testing of \sec{Hig:Hyp}
using \wpvs is a Bayesian approach where hypotheses are treated as
random variables. From Bayes' theorem of \rfr{bayes.64.1}, the
probability density function for the signal strength parameter $\mu$
as a function of the observed data $\vec{x}$ can be written as,
\begin{equation}
  \pdf(\mu|\vars) = \frac{\lh(\vars|\mu) \pdf(\mu)}{\int
    \lh(\vars|\mu') \pdf(\mu') \sdif{\mu'}}
  \labelequ{Statistics.Bayes}
\end{equation}
where $\pdf(\mu)$ is the prior probability density function for the
signal strength and $\lh(\vars|\mu)$ is the likelihood function for
\vars. The choice of $\pdf(\mu)$ is highly dependent upon the
analysis, although oftentimes $\pdf(\mu)$ is chosen as a divergent
uniform prior. The probability for the signal strength to be greater
than $\mu$ is then just the integral of $\pdf(\mu|\vars)$ from $\mu$
upwards, while the probability for the signal strength to be less than
$\mu$ is the integral from $\mu$ downwards.

Systematic uncertainties can be introduced into
\equ{Hvr:Statistics.Bayes} using a method similar to \rfr{helene.83.1}
by incorporating them into the \pdf{s} via nuisance parameters, \uncs,
\begin{equation}
  \pdf(\mu,\uncs|\vars) = \frac{\lh(\vars|\mu,\uncs) \pdf(\mu,\uncs)}{\int
    \lh(\vars|\mu',\uncs') \pdf(\mu') \sdif{\mu'}\sdif{\uncs'}}
  \labelequ{Statistics.Bayes.Unc}
\end{equation}
where each \unc represents a systematic uncertainty. The dependence
upon the hypothesis \pdf on the nuisance parameters can be eliminated
by,
\begin{equation}
  \pdf(\mu|\vars) = \int \pdf(\mu,\uncs|\vars) \sdif{\uncs'}
\end{equation}
where the nuisance parameters have been marginalised via
integration. If the nuisance parameters are independent of the
hypothesis, then the likelihood function can be marginalised  with
respect to \uncs,
\begin{equation}
  \lh(\vars|\mu) = \int \lh(\vars|\mu,\uncs) \pdf(\uncs) \sdif{\uncs}
\end{equation}
and used directly in \equ{Hvr:Statistics.Bayes}. The marginalised
likelihood function can also be used in the test statistics of
\equs{Hig:Statistics.Npr}, \ref{equ:Hig:Statistics.Plr}, and
\ref{equ:Hig:Statistics.Ulr}, resulting in the hybrid approach
proposed in \rfr{cousins.91.1}. The integration of the \pdf of the
observables is oftentimes not trivial and cannot be done analytically,
requiring the use of numerical methods such as Monte Carlo integration
instead.
\end{document}